\documentclass[letterpaper,12pt,titlepage,oneside,final]{book}
 
\newcommand{\href}[1]{#1} 

\usepackage[utf8]{inputenc}
\usepackage{float} 
\usepackage{scalerel}
\usepackage[normalem]{ulem}
\usepackage[english]{babel}
\usepackage{dcolumn}
\usepackage{bm}
\usepackage{blindtext}
\usepackage{verbatim}
\usepackage{relsize}
\usepackage{mathrsfs}
\usepackage{musicography}
\usepackage{amsmath}
\usepackage{blindtext}
\usepackage{cancel}
\usepackage{physics}
\usepackage{epstopdf}
\usepackage{mathtools}
\usepackage{blindtext}
\usepackage{tensor}
\usepackage{xcolor}
\usepackage[usenames,dvipsnames]{pstricks}
\usepackage{epsfig}
\usepackage{pst-grad} 
\usepackage{pst-plot} 
\usepackage{verbatim}
\usepackage{slashed}
\usepackage{dsfont}
\usepackage{upgreek}
\usepackage{faktor}
\usepackage{pagecolor}
\usepackage{caption}
\usepackage{subcaption}

\usepackage{ifthen}
\newboolean{PrintVersion}
\setboolean{PrintVersion}{false}

\usepackage{amsmath,amssymb,amstext} 
\usepackage{graphicx} 

\usepackage[pdftex,pagebackref=false]{hyperref}
\hypersetup{
    plainpages=false,       
    unicode=false,          
    pdftoolbar=true,        
    pdfmenubar=true,        
    pdffitwindow=false,     
    pdfstartview={FitH},    
    pdfnewwindow=true,      
    colorlinks=true,        
    linkcolor=blue,         
    citecolor=green,        
    filecolor=magenta,      
    urlcolor=cyan           
}
\ifthenelse{\boolean{PrintVersion}}{   
\hypersetup{	
    citecolor=black,%
    filecolor=black,%
    linkcolor=black,%
    urlcolor=black}
}{} 

\usepackage[automake,toc,abbreviations,nomain]{glossaries-extra} 
\let\origdoublepage\cleardoublepage
\newcommand{\clearemptydoublepage}{%
  \clearpage{\pagestyle{empty}\origdoublepage}}
\let\cleardoublepage\clearemptydoublepage

\newcommand{\normord}[1]{\vcentcolon\mathrel{#1}\vcentcolon}
\providecommand{\vcentcolon}{\mathrel{\mathop{:}}}

\newcommand{\openone}{\mathds{1}}
\newcommand{\M}{\mathcal{M}}
\renewcommand{\S}{\mathcal{S}}
\newcommand{\mf}{\mathsf}

\newcommand{\ii}{\mathrm{i}}

\renewcommand{\r}{\hat{\rho}}
\newcommand{\s}{\hat{\sigma}}
\newcommand{\tc}[1]{\textsc{#1}}

\DeclareMathOperator*{\sumint}{%
\mathchoice%
  {\ooalign{$\displaystyle\sum$\cr\hidewidth$\displaystyle\int$\hidewidth\cr}}
  {\ooalign{\raisebox{.14\height}{\scalebox{.7}{$\textstyle\sum$}}\cr\hidewidth$\textstyle\int$\hidewidth\cr}}
  {\ooalign{\raisebox{.2\height}{\scalebox{.6}{$\scriptstyle\sum$}}\cr$\scriptstyle\int$\cr}}
  {\ooalign{\raisebox{.2\height}{\scalebox{.6}{$\scriptstyle\sum$}}\cr$\scriptstyle\int$\cr}}
}

\definecolor{ultramarine}{RGB}{9,15,45}
\definecolor{golden}{RGB}{250,220,150}

\allowdisplaybreaks[1] 

\usepackage[automake,toc,abbreviations]{glossaries-extra} 

\newglossaryentry{computer}
{
name=computer,
description={A programmable machine that receives input data,
               stores and manipulates the data, and provides
               formatted output}
}

\newglossary*{nomenclature}{Nomenclature}
\newglossaryentry{dingledorf}
{
type=nomenclature,
name=dingledorf,
description={A person of supposed average intelligence who makes incredibly brainless misjudgments}
}

\newabbreviation{aaaaz}{AAAAZ}{American Association of Amateur Astronomers and Zoologists}

\newglossary*{symbols}{List of Symbols}
\newglossaryentry{rvec}
{
name={$\mathbf{v}$},
sort={label},
type=symbols,
description={Random vector: a location in n-dimensional Cartesian space, where each dimensional component is determined by a random process}
}
\makeglossaries

\begin{document}

\pagestyle{empty}
\pagenumbering{roman}

\begin{titlepage}
        \begin{center}
        \vspace*{1.0cm}

        \Huge
        {\bf The Information Locally Stored in Quantum Fields:\\ From Entanglement to Gravity}

        \vspace*{1.0cm}

        \normalsize
        by \\

        \vspace*{1.0cm}

        \Large
        Tales Rick Perche \\

        \vspace*{3.0cm}

        \normalsize
        A thesis\\ presented to the
        University of Waterloo \\ 
        in fulfillment of the \\
        thesis requirement for the degree of \\
        Doctor of Philosophy \\
        in \\
       Applied Mathematics (Quantum Information) \\

        \vspace*{2.0cm}

        Waterloo, Ontario, Canada, 2025 \\

        \vspace*{1.0cm}

        \copyright\ Tales Rick Perche 2025 \\
        \end{center}
\end{titlepage}

\pagestyle{plain}
\setcounter{page}{2}

\cleardoublepage 
\phantomsection

\addcontentsline{toc}{chapter}{Examining Committee}
\begin{center}\textbf{Examining Committee Membership}\end{center}
  \noindent

\vspace{-3cm}
\textcolor{white}{I wrote this PhD thesis in 6 weeks. I deeply regret starting so late, but I ended up being fairly proud of it :)}
\vspace{2cm}

The following served on the Examining Committee for this thesis. The decision of the Examining Committee is by majority vote.
  \bigskip
  
  \noindent
\begin{tabbing}
Internal-External Memberssss: \=  \kill 
External Examiner: \>  Daniel R. Terno \\ 
\> Professor, School of Mathematical and Physical Sciences,\\ \> Macquarie University \\
\end{tabbing} 
  \bigskip
  \noindent
\begin{tabbing}
Internal-External Memberssss:  \=  \kill 
Supervisors: \>Eduardo Martín-Martínez \\
\> Professor, Dept. of Applied Mathematics,\\ \>University of Waterloo \\
\\
\> David Kubiz\v nák \\
\> Professor, Dept. of Theoretical Physics,\\ \>Charles University \\
\end{tabbing}
  \bigskip
  
  \noindent
  \begin{tabbing}
Internal-External Memberssss: \=  \kill 
Internal Members: \> Achim Kempf \\
\> Professor, Dept. of Applied Mathematics,\\ \>University of Waterloo \\
\\
\> Florian Girelli \\
\> Professor, Dept. of Applied Mathematics,\\ \>University of Waterloo \\
\end{tabbing}
  \bigskip
  \bigskip
  
  \noindent
\begin{tabbing}
Internal-External Memberssss: \=  \kill 
Internal-External Member: \> Doreen Fraser \\
\> Professor, Dept. of Philosophy, \\ \>University of Waterloo \\
\end{tabbing}
  \bigskip
  

\cleardoublepage
\phantomsection    



 \addcontentsline{toc}{chapter}{Author's Declaration}
 \begin{center}\textbf{Author's Declaration}\end{center}

 \noindent
This thesis consists of material all of which I authored or co-authored: see Statement of Contributions included in the thesis. This is a true copy of the thesis, including any required final revisions, as accepted by my examiners.
 
 \noindent  
  \bigskip
  
  \noindent
I understand that my thesis may be made electronically available to the public.

\cleardoublepage
\phantomsection    

\addcontentsline{toc}{chapter}{Statement of Contributions}
\begin{center}\textbf{Statement of Contributions}\end{center}

\noindent This thesis contains results from the following works:

\begin{itemize}
    \item[(1)] \textbf{T. Rick Perche}, José Polo-Gómez, Bruno de S. L. Torres, Eduardo Martín-Martínez, ``Particle Detectors from Localized Quantum Field Theories'', \href{https://doi.org/10.1103/PhysRevD.109.045013}{Phys. Rev. D 109, 045013 (2024)}~\cite{QFTPD};
    \item[(2)] \textbf{T. Rick Perche}, ``Closed-form expressions for smeared bi-distributions of a massless scalar field: non-perturbative and asymptotic results in relativistic quantum information'', \href{https://doi.org/10.1103/PhysRevD.110.025013}{Phys. Rev. D 110, 025013 (2024)}~\cite{analytical};
    \item[(3)] \textbf{T. Rick Perche}, ``Localized non-relativistic quantum systems in curved spacetimes: a general characterization of particle detector models'', \href{https://doi.org/10.1103/PhysRevD.106.025018}{Phys. Rev. D 106, 025018 (2022)}~\cite{generalPD};
    \item[(4)] Ruhi Shah, Eduardo Martín-Martínez, \textbf{T. Rick Perche}, ``Relativistic QFT description for the interaction of a spin with a magnetic field'', \href{https://doi.org/10.1103/PhysRevD.111.044075}{Phys. Rev. D 111, 044075}~\cite{ruhi};
    \item[(5)] \textbf{T. Rick Perche}, J. P. M. Pitelli, Daniel A. T. Vanzella, ``The stress-energy tensor of an Unruh-DeWitt detector'', \href{https://doi.org/10.1103/PhysRevD.111.045004}{Phys. Rev. D 111, 045004 (2025)}~\cite{TmunuUDW};
    \item[(6)] Ivan Agullo, Béatrice Bonga, Eduardo Martín-Martínez, Sergi Nadal-Gisbert, \\\mbox{\textbf{T. Rick Perche}}, José Polo-Gómez, Patricia Ribes-Metidieri, Bruno de S. L. Torres, ``The multimode nature of spacetime entanglement in QFT'', \href{https://doi.org/10.1103/PhysRevD.111.085013}{Phys. Rev. D 111, 085013 (2025)}~\cite{patriciaAndI};
    \item[(7)] \textbf{T. Rick Perche}, José Polo-Gómez, Bruno de S. L. Torres, Eduardo Martín-Martínez, ``Fully Relativistic Entanglement Harvesting'', \href{https://doi.org/10.1103/PhysRevD.109.045018}{Phys. Rev. D 109, 045018 (2024)}~\cite{FullHarvesting};
    \item[(8)] \textbf{T. Rick Perche}, Eduardo Martín-Martínez, ``The role of quantum degrees of freedom of relativistic fields in quantum information protocols'', \href{https://doi.org/10.1103/PhysRevA.107.042612}{Phys. Rev. A 107, 042612 (2023)}~\cite{quantClass}
    \item[(9)] Eduardo Martín-Martínez, \textbf{T. Rick Perche}, ``What gravity mediated entanglement can really tell us about quantum gravity'', \href{https://doi.org/10.1103/PhysRevD.108.L101702}{Phys. Rev. D 108, L101702 (2023)}~\cite{ourBMV}
    \item[(10)] \textbf{T. Rick Perche}, Eduardo Martín-Martínez, ``The geometry of spacetime from quantum measurements'', \href{https://doi.org/10.1103/PhysRevD.105.066011}{Phys. Rev. D 105, 066011 (2022)}~\cite{geometry};
\end{itemize}
\noindent As well as results that were not yet published.

The results of manuscript (1) appear at the very end of Section~\ref{sec:LocalizedQuantumFields} and at the start of Section~\ref{sec:MoreRealisticProbes}. I was the main responsible for writing the first draft of (1) and the text was later edited by all co-authors. Parts of the text were used in the corresponding Sections.

The results of (2) appear are used in Sections~\ref{sec:UDW},~\ref{sec:GeneralEnt}, and~\ref{sec:QFTapproxQC}. I was the sole author of this manuscript, thus all contributions therein are of my authorship. Appendix~\ref{app:analytical} is also based on the text.

Manuscript (3) is presented in Section~\ref{sec:NRLQS}. As a single author, I was responsible for all the content. Appendices~\ref{app:fermi} and~\ref{app:det} are also based on content present in the manuscript.

The results of (4) appear in Section~\ref{sec:MoreRealisticProbes}. This manuscript was produced as part of Ruhi Shah's master's supervision project. I was responsible for the initial idea presented in the project, as well as coordinating it. The text was edited collaboratively between all coauthors, and Appendix~\ref{app:derivation} is my sole authorship.

The main results of (5) are the content of Section~\ref{sec:Tmunu}. I was responsible for the initial idea, the computations and for producing the initial draft. Daniel Vanzella and Jo\~ao Pitelli contributed to the text and verified the methods presented in the manuscript.

Manuscript (6) is briefly summarized in Section~\ref{sec:modeEntanglement}. This was a large collaboration resulting from a joint idea. The writing of the text was split between all coauthors, and the plot presented in Fig.~\ref{fig:LogNeg} was done by Sergi Nadal-Gisbert.

The results of manuscripts (7) appear at the start of Sections~\ref{sec:OperationallyAccessingEnt} and~\ref{sec:GeneralEnt}. I was the main responsible for the computations and writing and producing the first draft of (7) and the text was later edited by all co-authors. Parts of the text were used at the start of Section~\ref{sec:OperationallyAccessingEnt}. Appendix~\ref{app:twoQFTs} is also based on content from (7).

Results of manuscript (8) appear in Sections~\ref{sec:QCmodels} and~\ref{sec:QFTapproxQC}. I was the main responsible by the initial idea, draft, computations and plots. The text was edited collaboratively with Eduardo Mart\'in-Mart\'inez. 

A summary of the results of manuscript (9) appears in Section~\ref{sec:GME}. I was responsible for all the computations, diagrams, and for the first version of the draft. The text was edited collaboratively with Eduardo Mart\'in-Mart\'inez. Appendices~\ref{app:retGrav} and~\ref{app:FinalStates} contain computations also present in the manuscript. 

Manuscript (9) appears in Section~\ref{sec:geometryFromMeas}. I was responsible for all the computations, plots, diagrams, and for the first version of the draft. Eduardo Mart\'in-Mart\'inez proposed the original idea and helped with the editing of the manuscript.

Additional publications that resulted from my PhD, but are not included explicitly in this thesis are:
\begin{itemize}
    \item[(12)] Eduardo Martín-Martínez, \textbf{T. Rick Perche}, Bruno de S. L. Torres, ``Broken covariance of particle detector models in relativistic quantum information'', \href{https://doi.org/10.1103/PhysRevD.103.025007}{Phys. Rev. D 103, 025007 (2021)}~\cite{us2};
    \item[(13)] Bruno de S. L. Torres, \textbf{T. Rick Perche}, André G. S. Landulfo, George E. A. Matsas, ``Neutrino flavor oscillations without flavor states'', \href{https://doi.org/10.1103/PhysRevD.102.093003}{Phys. Rev. D 102, 093003 (2020)}~\cite{neutrinos};
    \item[(14)] Alvaro Ballon Bordo, David Kubiznak, \textbf{T. Rick Perche}, ``Taub-NUT solutions in conformal electrodynamics'', \href{https://doi.org/10.1016/j.physletb.2021.136312}{Phys. Lett. B 817, 136312 (2021)}~\cite{alvaro};
    \item[(15)] \textbf{T. Rick Perche}, Jonas Neuser, ``A Wavefunction Description for a Localized Quantum Particle in Curved Spacetimes'', \href{https://iopscience.iop.org/article/10.1088/1361-6382/ac103d}{Class. Quantum Grav. 38 175002 (2021)}~\cite{jonas};
    \item[(16)] I.M. Burbano, \textbf{T. Rick Perche}, Bruno de S. L. Torres, ``A path integral formulation for particle detectors: the Unruh-DeWitt model as a line defect'', \href{https://link.springer.com/article/10.1007/JHEP03(2021)076}{J. High Energy Phys. 2021, 76 (2021).}~\cite{ivan};
    \item[(17)] \textbf{T. Rick Perche}, Eduardo Martín-Martínez, ``Anti-particle detector models in QFT'', \href{https://doi.org/10.1103/PhysRevD.104.105021}{Phys. Rev. D 104, 105021 (2021)}~\cite{antiparticles};
    \item[(18)] \textbf{T. Rick Perche}, ``General features of the thermalization of particle detectors and the Unruh effect'', \href{https://doi.org/10.1103/PhysRevD.104.065001}{Phys. Rev. D 104, 065001 (2021)}~\cite{mine};
    \item[(19)] J. P. M. Pitelli, \textbf{T. Rick Perche}, ``An angular momentum based graviton detector'', \href{https://doi.org/10.1103/PhysRevD.104.065016}{Phys. Rev. D 104, 065016 (2021)}~\cite{pitelli};
    \item[(20)] \textbf{T. Rick Perche}, Caroline Lima, Eduardo Martín-Martínez, ``Harvesting entanglement from complex scalar and fermionic fields with linearly coupled particle detectors'', \href{https://doi.org/10.1103/PhysRevD.105.065016}{Phys. Rev. D 105, 065016 (2022)}~\cite{carol};
    \item[(21)] Héctor Maeso-García, \textbf{T. Rick Perche}, Eduardo Martín-Martínez, ``Entanglement harvesting: detector gap and field mass optimization'', (2022), \href{https://doi.org/10.1103/PhysRevD.106.045014}{Phys. Rev. D 106, 045014 (2022)}~\cite{hectorMass};
    \item[(22)] \textbf{T. Rick Perche}, Ahmed Shalabi, ``Spacetime curvature from ultra rapid measurements of quantum fields'', \href{https://doi.org/10.1103/PhysRevD.105.125011}{Phys. Rev. D 105, 125011 (2022)}~\cite{ahmed};
    \item[(23)] \textbf{T. Rick Perche}, Boris Ragula, Eduardo Martín-Martínez, ``Harvesting entanglement from the gravitational vacuum'', \href{https://doi.org/10.1103/PhysRevD.108.085025}{Phys. Rev. D 108, 085025 (2023)}~\cite{boris};
    \item[(24)] Finnian Gray, David Kubiznak, \textbf{T. Rick Perche}, Jaime Redondo-Yuste, ``Carrollian Motion in Magnetized Black Hole Horizons'', \href{https://doi.org/10.1103/PhysRevD.107.064009}{Phys. Rev. D 107, 064009 (2023)}~\cite{jaime};
    \item[(25)] Jiri Bicak, David Kubiznak, \textbf{T. Rick Perche}, ``Migrating Carrollian particles on magnetized black hole horizons'', \href{https://doi.org/10.1103/PhysRevD.107.104014}{Phys. Rev. D 107, 104014 (2023)}~\cite{jiri};
    \item[(26)] \textbf{T. Rick Perche}, Matheus H. Zambianco, ``Duality between amplitude and derivative coupled particle detectors in the limit of large energy gaps'', \href{https://doi.org/10.1103/PhysRevD.108.045017}{Phys. Rev. D 108, 045017 (2023)}~\cite{dualityGap};
    \item[(27)] Cameron R. D. Bunney, Leo Parry, \textbf{T. Rick Perche}, Jorma Louko, ``Ambient temperature versus ambient acceleration in the circular motion Unruh effect'', \href{https://doi.org/10.1103/PhysRevD.109.065001}{Phys. Rev. D 109, 065001 (2024)}~\cite{JormaAndI};
    \item[(28)] Philip A. LeMaitre, \textbf{T. Rick Perche}, Marius Krumm, Hans J. Briegel, ``A Universal Quantum Computer From Relativistic Motion'', \href{https://doi.org/10.48550/arXiv.2411.00105}{arXiv:2411.00105}, accepted in \textit{Phys. Rev. Lett.}~\cite{Phil};
    \item[(28)] Ahmed Shalabi, Matheus H. Zambianco, \textbf{T. Rick Perche}, ``The effect of curvature on local observables in quantum field theory'', \href{https://iopscience.iop.org/article/10.1088/1361-6382/adc534}{Class. Quantum Grav. 42 085007 (2025)}~\cite{mathahmed};
    \item[(29)] Eirini C. Telali, \textbf{T. Rick Perche}, Eduardo Martín-Martínez, ``Causality in relativistic quantum interactions without mediators'', \href{https://doi.org/10.1103/PhysRevD.111.085005}{Phys. Rev. D 111, 085005 (2025)}~\cite{eirini};
\end{itemize}

\cleardoublepage
\phantomsection    

\addcontentsline{toc}{chapter}{Abstract}
\begin{center}\textbf{Abstract}\end{center}

This thesis contains a local study of quantum field theory from fundamental, operational, and practical perspectives, with the primary goal of investigating the information that can be locally extracted from quantum fields. Central to this discussion is how the fundamental interactions of quantum fields give rise to the very objects that allow us to probe them. We approach this problem through the concept of localized quantum fields, which naturally reduce to local probes with finitely many degrees of freedom that can be accessed in realistic experiments.

Building on this detailed description of localized probes, we apply these to explore two key aspects of the information locally stored in quantum fields: entanglement and gravity. In the study of entanglement, we explore the quantification of accessible vacuum entanglement between two finite regions of spacetime. Our discussion contains both a first-principles approach based on local field degrees of freedom and an operational framework, wherein we consider the entanglement that can be harvested by coupling local probes to independent degrees of freedom of the field.

The study of entanglement in quantum field theory also leads us to classify the regimes where the quantum degrees of freedom of a field play an active role. Through the use of an effective quantum-controlled model, we show that the quantum degrees of freedom of mediating fields are only relevant in relativistic setups involving either high energies or interactions that are sufficiently localized in spacetime. In setups where these conditions are not met, a simplified effective model can accurately describe interactions while still incorporating some key relativistic elements.

Finally, we will discuss the gravitational information locally stored in quantum fields. Specifically, we will show that the correlations of quantum fields contain full information about the geometry of spacetime, and how to physically access these degrees of freedom.  While the fact that quantum fields store full gravitational information might suggest the possibility of a theory in which gravity emerges directly from quantum correlations, we speculate that gravity may instead be emergent from the entanglement in quantum field theory.

\cleardoublepage
\phantomsection    

\addcontentsline{toc}{chapter}{Acknowledgements}
\begin{center}\textbf{Acknowledgements}\end{center}


The end of my PhD came way faster than I thought it would, and it is sad that this journey has to come to an end. Nevertheless, I enjoyed my time as a PhD student very much. I was very fortunate to participate in many collaborations and visits, to organize conferences, to coordinate supervision projects, and, most importantly, to make lots of friends. I got to travel the world while doing the things that I enjoy the most and to meet others who are as excited as I am about science. None of these would be possible without the support of many people, which was arguably the most important part of this journey. 


I will start by thanking the two people who have made this possible in the first place: my supervisors \textbf{Eduardo Martín-Martínez} and \textbf{David Kubiz\v{n}ák}. I was fortunate to have two supervisors who resonated with me very well and gave me as much freedom to pursue my own research interests as they did, and I cannot think of better anyone better than these two to guide me through my PhD. Edu, I really appreciate the trust you have always put in me, and I have learned so much from you. Your care for your students is unmatched. David, your joy and passion for even the simplest things inspire everyone around you, and I was lucky to enjoy this inspiration through all of our interactions. I am also very thankful for the lovely visits to Prague and for your support in making the RQI-N conference happen. I have no words to thank you two for the support provided to me during my PhD. 

This PhD defence is also the end of a big chapter of my life, titled ``Canada''. The people that were around me during my journey are certainly what motivated me the most throughout it, and now is the time to thank them.

First, I would like to thank all current and former members of Barrio-RQI, which is certainly the best possible group that I could ever hope to be a part of. In particular, I would like to thank the following researchers, who also became very good friends during my PhD: \textbf{Kelly Wurtz} for teaching me that the best investment is usually to invest in yourself; \textbf{Boris Ragula} for being an example in owning your mistakes; \textbf{Matheus Hrabowec Zambianco} for reminding me why I do research in the first place; \textbf{José Polo-Gómez} for being an inspiring researcher and a good friend; \textbf{Erickson Tjoa} for a source of inspiration and example in academia; \textbf{Adam Teixidó-Bonfill} for showing me that listening is the most powerful skill; \textbf{Bruno de S. L. Torres} for being an example of balance between personal and academic life, as well as \textbf{Eirini Telali}, \textbf{Caroline Lima}, \textbf{Maria Papageorgiou}, \textbf{José de Ramón (Pipo)}, \textbf{Nicho Funai}, \textbf{Richard Lopp}, \textbf{Daniel Grimmer}, and \textbf{Héctor Maeso-García}.

Second, I would like to thank my current roommates, with whom I have shared unforgettable experiences in Canada: \textbf{Boris Ragula}, for bringing me into the place that I called home for so long and \textbf{Zachary Van Oosten}, for all the support, discussions and experiences we shared. I would also like to thank my former roommates \textbf{Héctor Maeso-García}, \textbf{Elisa Joly}, and \textbf{Maria Esteban} for being a part of our own sitcom, as well as \textbf{Dalila Pirvu}, and \textbf{Iván Burbano} for the support throughout the pandemic.

My stay in Canada also rendered many other friends and research colleagues that inspired me and supported me throughout this journey. I would like to thank \textbf{Ahmed Shalabi} for his unconditional support, \textbf{Evan Peters} for all the good times we shared in the Physics of Information Laboratory, \textbf{Philip LeMaitre} for introducing me to so many new experiences, as well as \textbf{Ruhi Shah}, \textbf{Everett Patterson}, \textbf{María Rosa Preciado-Rivas}, \textbf{Christopher Pollack}, \textbf{Dalila Pirvu}, \textbf{Brayden Hull}, \textbf{Cendikiawan Suryaatmadja (Diki)} for all the great time we spent together. In particular, I would like to thank Profs. \textbf{Robert Mann}, \textbf{Achim Kempf}, and \textbf{Doreen Fraser} for creating a stimulating environment for research in relativistic quantum information in Waterloo.

One of the most remarkable experiences that I had during my PhD was certainly the organization of the RQI Circuit, and it would not have been possible without the students and postdocs who collectively helped in the organization: \textbf{María Rosa Preciado-Rivas}, \textbf{José Polo-Goméz}, \textbf{Everett Patterson}, \textbf{Caroline Lima}, \textbf{Leo Parry}, \textbf{Cameron Bunney}, \textbf{Cisco Gooding}, \textbf{Anne-Catherine de la Hamette}, \textbf{Germain Tobar}, \textbf{Vasileos Fragkos}, \textbf{Jerzy Paczos}, \textbf{Dennis Rätzel}, \textbf{Ekim Hanimeli}, \textbf{Roy Barzel}, \textbf{Emanuel Schlake}, \textbf{Marian Cepok}, \textbf{Nicholas Funai}, and \textbf{Fil Simoc}. I would also like to thank \textbf{Eduardo Mart\'in-Mart\'inez} and \textbf{Jorma Louko} for the trust placed in me to organize such an ambitious event, as well as all participants and attendants of the conference.

Next, I would like to thank all members of the Gravity Laboratory group in Nottingham, who always welcomed me as if I had been a part of the group in all my visits since the summer of 2023. First, I'd like to thank Profs. \textcolor{white}{\ensuremath\heartsuit}\!\!\!\!\!\textbf{Jorma Louko} and \textbf{Silke Weinfurtner} for hosting me and cultivating such an exciting group. I am particularly thankful to Jorma, who was always very generous with his time, and who I see as a third, unofficial supervisor. Second, I would like to thank all members of the Gravity Laboratory who welcomed me and quickly became my close friends, \textcolor{white}{\ensuremath\heartsuit}\!\!\!\!\!\! \textbf{Leo Parry}, \textcolor{white}{\ensuremath\heartsuit}\!\!\!\!\!\textbf{Cameron Bunney}, \textcolor{white}{\ensuremath\heartsuit}\!\!\!\!\!\textbf{Adam Wilkinson}, \textcolor{white}{\ensuremath\heartsuit}\!\!\!\!\!\textbf{Silvia Schiattarella}, \textcolor{white}{\ensuremath\heartsuit}\!\!\!\!\!\textbf{Cisco Gooding}, \textbf{Pietro Smaniotto}, \textbf{Patrik Svan\v{c}ara}, \textbf{Sean Gregory}, \textbf{Leonardo Sollidoro}, \textbf{Vitor Barroso}, \textbf{Maciej Jarema}, \textbf{Sreelekshmi Ajithkumar (Sree)}.

I was also fortunate to spend the summer of 2024 in Vienna, and I must thank \textbf{\v{C}aslav Brukner}, \textbf{Marios Christodoulou}, \textbf{Markus Aspelmeyer} for hosting me and welcoming me into their groups. I would also like to thank \textbf{Anne-Catherine de la Hammete}, \textbf{Sara Buttler}, \textbf{Emil Broukal}, \textbf{Maria Papageorgiou}, \textbf{Daniel Vanzella}, \textbf{Robin Simmons}, \textbf{Andrea Di Biagio}, and \textbf{Ofek Bengyat} for making the stay in Vienna much more enjoyable and for the scientific discussions that we shared.

During my PhD I was also fortunate enough to visit many research groups, and I would like to thank the individual people who hosted me in each of these visits and helped me fund them: \textbf{\v{C}aslav Brukner}, who hosted me in Vienna, \textbf{Jorma Louko} and \textbf{Silke Weinfurtner}, who hosted me in Nottingham so many times, \textbf{Christopher J. Fewster}, who hosted me in York, \textbf{Sougato Bose} for hosting me in London in my three brief visits, \textbf{Charis Anastopoulos} for hosting me in Patras for a month, \textbf{Claus Laemmerzahl} for welcoming me in the Bremen Drop Tower, \textbf{Robert Jonsson} and \textbf{Guilherme Franzmann} for hosting me in Stockholm in my visits to Nordita, \textbf{Pavel Krotus} for welcoming me to Prague in all my visits, \textbf{Markus Aspelmeyer} and \textbf{Marios Christodoulou} for hosting me in Vienna in different occasions, and \textbf{Hans Briegel} for funding my very productive visits to Innsbruck.

Each article published during my PhD also has a whole story behind it. I would like to thank the people that were a part of these stories that have not yet been mentioned, many of whom became good friends through our collaborations: \textbf{Ivan Agullo}, \textbf{Alvaro Ballon-Bordo}, \textbf{Jiri Bicak},  \textbf{Béatrice Bonga}, \textbf{Hans Briegel}, \textbf{Finnian Gray}, \textbf{Marius Krumm}, \textbf{Sergi Nadal-Gisbert}, \textbf{Jonas Neuser}\textcolor{white}{\ensuremath\heartsuit}\!\!\!\!\!, \textbf{João Pitelli}, \textbf{Jaime Redondo-Yuste}, and \textbf{Patricia Ribes-Metidieri}\textcolor{white}{\ensuremath\heartsuit}\!\!\!\!\!.

 The last academics that I would like to thank are \textbf{Dra\v zen Glavan}, \textbf{Flaminia Giacomini}, \textbf{Magdalena Zych}, and \textbf{Fabio Costa}, who have helped me throughout my PhD, either directly, or indirectly, and are a source of inspiration.

I would also like to thank dear friends who have been a part of my life since before the PhD, but also throughout it, be it through stimulating discussions, hosting me for a couple of days, or sharing deep experiences: \textbf{Danilo Elias}, \textbf{Bianca Marin Moreno}, \textbf{Leonardo Bianco}, \textbf{Caio Laurenti}, \textbf{Vinicius Maciel}, \textbf{Gabriel Reis}, \textbf{Paulo Piva}, \textbf{Matheus Loures} and \textbf{Pedro Lauand}.

Last but certainly not least, I would like to thank my parents, \textbf{Enio Esteves Perche} and \textbf{Cristina Rick}, and my step-parents \textbf{Valeria Maria Rosa} and \textbf{José Luis Zavala Rubio}, who I love very dearly, and owe everything that led me here. Finally, I would like to thank my grandmother \textbf{Leani Inês Ruschel}, who always inspired me to be creative.

I thank the Vanier CGS for funding throughout my PhD as well as the David Johnston International Experience Award, the Department of Applied Mathematics and Perimeter Institute for financial aid over many of the trips during my PhD.


\cleardoublepage
\phantomsection    

\addcontentsline{toc}{chapter}{Dedication}
\begin{center}\textbf{Dedication}\end{center}

To Vó Neyta, who taught us the important saying\\

\begin{center} \textit{``Modéstia para evitar a embriaguez do triunfo''}\end{center}

\cleardoublepage
\phantomsection    

\renewcommand\contentsname{Table of Contents}
\tableofcontents
\cleardoublepage
\phantomsection    

\cleardoublepage
\phantomsection    

\addcontentsline{toc}{chapter}{List of Figures}
\listoffigures
\cleardoublepage
\phantomsection		

\addcontentsline{toc}{chapter}{List of Tables}
\listoftables
\cleardoublepage
\phantomsection		

\addcontentsline{toc}{chapter}{Preface}
{\Huge\noindent\textbf{Preface}}\\

\noindent Arguably, the most relevant aspect of a document is its target audience. This being a PhD thesis, and my primary goal being to be awarded the PhD degree, this thesis should (and mostly is) written with the committee in mind. However, I would also like for this thesis to serve as a useful reference for future researchers interested in approaching some of the tools that I have learned and developed during my PhD. To this end, this thesis contains reviews of most topics relevant to the discussion here, containing my personal view on these. In particular, while knowledge of general relativity and quantum mechanics is assumed throughout the thesis, no previous knowledge of quantum field theory is required (although it is always welcome) for reading the full content of the thesis: we will instead review all necessary content of quantum field theory in Chapter~\ref{chap:QFT}. A reader familiar with the content of Chapter~\ref{chap:QFT} is then welcome to skip it, but encouraged to check the conventions established therein.

The remaining chapters of the thesis all contain a combination of original results and reviews of relevant related research. In many cases, these are combined in the same Section for the purpose of exposition, and in some cases, the reviews themselves smoothly connect to original results, as some of the original research presented here lies exactly in the connection between known formulations of different topics. This is particularly true in Section~\ref{sec:LocalizedQuantumFields}, which connects the Fewster-Verch measurement scheme with the operational formulation of probes in quantum field theory. All parts of the thesis involving reviews of works in which I did not coauthor will be properly referenced, making clear which content is not due to my contributions.

Overall, the exposition of topics in this thesis follows a bottom-up approach, starting from more general definitions and concepts, and then particularizing to concrete applications. For instance, in Chapter~\ref{chap:meas}, we start with a fully covariant approach to measurements in quantum field theory, then reduce it to effective models, and only afterwards establish a connection with explicit physical systems. The same is true in Chapter~\ref{chap:ent}, where we first give general perspectives on the quantification of entanglement in quantum field theory, only discussing operational approaches later in the chapter.

Each Chapter in this thesis is divided into Sections containing Segments, rather than Subsections. The only exception to this rule is Section~\ref{sec:NRLQS}, which contains three Subsections, each split into their respective Segments. The Segments are identified by bold titles that summarize their main content and can be arbitrarily short. Their main goal is to clearly organize and divide the content of each section into separate subtopics. 

Finally, it is important to disclaim that any topic in theoretical physics can be described at many different levels of mathematical rigour. This is particularly relevant for the contents of this thesis, as its primary focus is deeply linked with local studies of quantum field theory, which can be formulated with arbitrary levels of mathematical rigour. For instance, in~\cite{FewsterCategory}, quantum field theory in curved spacetimes is formulated in terms of category theory. In this sense, the topics here are not presented with full mathematical rigour, as we will usually ignore topological considerations whenever they are not directly relevant to our discussions. Overall, the approach taken here uses the relevant aspects of local formulations of quantum field theory, skipping excesses in rigour that could hinder progress toward our main discussions.

\section*{Notations and Conventions}
\addcontentsline{toc}{section}{Notation and Conventions}

In this Section we will briefly summarize the main notations and conventions used throughout the thesis. Most of these are defined when first used in the manuscript, or fairly standard, but we summarize them below for convenience.

\subsubsection*{Spacetime}

Spacetime is an oriented smooth 3+1 dimensional\footnote{The only exception to this rule is in Section~\ref{sec:NRLQS}, where spacetime is taken to be $n+1$ dimensional.} Lorentzian manifold $\M$ with metric $g$, and we use the convention that timelike vector fields have negative norm (the east coast convention with signature $(-1,1,1,1)$). We usually label coordinates by $(x^0,x^1,x^2,x^3) = (t,\bm x)$, where $x^0 = t$ is a positively oriented timelike coordinate and $\bm x = (x^1,x^2,x^3) = (x^i)$ are spacelike coordinates. Greek indices run from $0$ to $3$, while Latin indices run from $1$ to $3$. Events in spacetime are denoted with serif font: $\mf x$, $\mf p$, $\mf q$. 

Given $\mathcal{O}\subset \M$, we denote its future and past domain of dependence by $D^{+}(\mathcal{O})$ and $D^{+}(\mathcal{O})$, respectively. Its domain of dependence is denoted by $D(\mathcal{O})$. We denote its causal future and causal past by $J^{+}(\mathcal{O})$ and $J^{-}(\mathcal{O})$, respectively. A Cauchy surface for a region $\mathcal{O}$ is a spacelike hypersurface such that $\mathcal{O}\subset D(\Sigma)$, and a Cauchy surface is a spatial hypersurface $\Sigma$ such that $D(\Sigma) = \M$. The causal hull of a region $\mathcal{O}$ is defined as the set $J^-(\mathcal{O})\cap J^+(\mathcal{O})$, and a set is said to be causally convex if it is equal to its causal hull. The causal complement of $\mathcal{O}$ is denoted by $\mathcal{O}' = \M\setminus (J^+(\mathcal{O})\cap J^-(\mathcal{O})).$

The Levi-Civita connection is denoted by $\nabla$, and the volume form is denoted by $\dd V$, which can be expressed as $\dd V = \sqrt{-g} \,\dd x^0\wedge... \wedge \dd x^3 \equiv \sqrt{-g} \dd^4 \mf x$, where $g$ stands for the metric determinant in a given coordinate system. The Riemann tensor is denoted by $R_{\mu\nu\alpha\beta}$ with the convention {$R_{\mu\nu\alpha\beta}v^\beta = \nabla_\mu \nabla_\nu v_\alpha - \nabla_\nu \nabla_\mu v_\alpha$}, the Ricci tensor is $R_{\mu\nu} = g^{\alpha\beta} R_{\mu\alpha\nu \beta}$, and the Ricci scalar is $R  = g^{\mu\nu}R_{\mu\nu}$. The Einstein tensor is $G_{\mu\nu} = R_{\mu\nu} - \tfrac{1}{2} R\, g_{\mu\nu}$. We use the standard notations for symmetrized and antisymmetrized indices, $A_{(\mu\nu)} = \tfrac{1}{2}(A_{\mu\nu} + A_{\nu\mu})$, $A_{[\mu\nu]} = \tfrac{1}{2}(A_{\mu\nu} - A_{\nu\mu})$.

Synge's world function is denoted $\sigma(\mf x, \mf x')$, corresponding to one-half the squared geodesic distance between events $\mf x$, $\mf x'$ that can be connected by a unique geodesic. Its derivatives with respect to its different arguments are denoted as below
\begin{equation}
    \sigma_\mu = \nabla_\mu \sigma, \quad \sigma_{\mu'} = \nabla_{\mu'}\sigma, \quad \nabla_{\mu\nu}\sigma = \sigma_{\nu\mu}, \quad \nabla_{\mu\nu'}\sigma = \sigma_{\nu'\mu}, \quad \nabla_{\mu'\nu'}\sigma = \sigma_{\nu'\mu'},
\end{equation}
as well as their natural extension for higher derivatives.

Spatial hypersurfaces are typically denoted by $\Sigma$, and their induced volume form is denoted by $\dd \Sigma$. We also denote $\dd\Sigma^\mu = n^\mu \dd \Sigma$, where $n^\mu$ is the future-pointing normalized \textit{vector} orthogonal to $\Sigma$.
Throughout the thesis, we will assume that $\M$ is globally hyperbolic so that it admits a foliation by Cauchy surfaces $\Sigma_t$, in $t\in \mathbb{R}$. In this case, the coordinate system $(t,\bm x)$ is such that $t$ parametrizes the Cauchy foliation and $\bm x$ are local coordinates in $\Sigma_t$.

All integrals are assumed to be over the entire domain where their measures is defined, e.g.
\begin{equation}
    \int\dd V f(\mf x) \equiv \int_{\M}\dd V f(\mf x),\quad\quad \int\dd \Sigma F(\bm x) \equiv \int_{\Sigma}\dd \Sigma F(\bm x).
\end{equation}
The Dirac deltas $\delta(\mf x, \mf x')$ and $\delta(\bm x, \bm x')$ are defined incorporating the measures of the spaces that the arguments belong to:
\begin{equation}
    \int \dd V \delta(\mf x, \mf x_0) f(\mf x) = f(\mf x_0), \quad\quad\int \dd \Sigma\, \delta(\bm x, \bm x_0) F(\bm x) = F(\bm x_0).
\end{equation}

When $\M$ is taken to be Minkowski spacetime, $\Sigma_t$ stands for an inertial foliation and $(t,\bm x)$ denotes an inertial coordinate system, where the metric components become $\eta_{\mu\nu} = \text{diag}(-1,1,1,1)$ and $\nabla_\mu = \partial_\mu$.

\subsubsection*{Functions, Distributions, and Bi-distributions}

The set of complex-valued smooth functions in $\M$ is denoted by $C^\infty(\M)$, and the set of smooth compactly supported functions complex functions in $\M$ is denoted $C_0^\infty(\M)$. The set of real smooth functions and real smooth compactly supported functions in $\M$ are respectively denoted by  $C^\infty(\M)_\mathbb{R}$ and $C_0^\infty(\M)_\mathbb{R}$.

A smooth function in spacetime, $f:\M\to \mathbb{C}$ (or $\mathbb{R}$) defines a unique linear functional acting in $C^\infty(\M)$:
\begin{equation}\label{eq:fofg}
    f(g) \coloneqq \int \dd V f(\mf x) g(\mf x).
\end{equation}
The symbol $f$ then stands both for the distribution $f:C^\infty(\M)\to \mathbb{C}$ and for the function $f:\M\to \mathbb{C}$: if $f$ is evaluated at an event $\mf x$, it stands for the value of the scalar function $f$, and if $f$ is evaluated a function $g$, it is understood as in Eq.~\eqref{eq:fofg}. In particular, $fg$ stands for the scalar function $f(\mf x)g(\mf x)$ and $f(g) = g(f)$ stands for its integral.

Analogously, a smooth biscalar function $A:\M\times\M\to \mathbb{C}$ defines a unique bilinear functional in $C_0^\infty(\M)$:
\begin{equation}\label{eq:Afg}
    A(f,g) \coloneqq \int \dd V \dd V' A(\mf x, \mf x') f(\mf x) g(\mf x').
\end{equation}
Moreover, the biscalar function $A(\mf x, \mf x')$ also defines a unique linear operator acting from $C_0^\infty(\M)$ to $C^\infty(\M)$:
\begin{equation}\label{eq:Af}
    Af(\mf x) \coloneqq \int \dd V' A(\mf x, \mf x') f(\mf x').
\end{equation}
The symbol $A$ then stands for the scalar function $A(\mf x, \mf x')$ when evaluated at two events in spacetime, the bi-distribution $A(f,g)$ when evaluated at two test functions, and the function $Af$, when applied to a single test function. We then have the identity $A(f,g) = f(Ag)$. Given a linear operator $L:C_0^\infty(\M)\to C^\infty(\M)$, or a bi-distribution $B:C_0^\infty(\M)\times C_0^\infty(\M) \to \mathbb{C}$, we can define formal integral kernels $L(\mf x, \mf x')$, $B(\mf x, \mf x')$ such that the identities $f(Lg) = L(f,g)$ and $B(f,g) = f(Bg)$ hold. When appropriate, we generalize the definitions of Eqs.~\eqref{eq:Afg} and~\eqref{eq:Af} for more general spaces of test functions and quotients of the space $C_0^\infty(\M)$. 

The ideas displayed in Eqs.~(\ref{eq:fofg}-\ref{eq:Af}) generalize to tensor fields of any rank: if $a$ stands for any collection of Lorentz indices, a tensor field $f^a(\mf x)$ defines a unique distribution acting in dual fields by
\begin{equation}
    f(g) = \int \dd V f^a(\mf x) g_a(\mf x).
\end{equation}
Analogously, we have a correspondence between bitensors, bi-distributions and linear operators. A bitensor $A_{aa'}(\mf x, \mf x')$ then defines the bi-distribution and linear operators
\begin{equation}
    A(f,g) = \int \dd V \dd V' f^a(\mf x) A_{aa'}(\mf x,\mf x') g^{a'}(\mf x'), \quad Af_a(\mf x) = \int \dd V' A_{aa'}(\mf x, \mf x')f^{a'}(\mf x'),
\end{equation}
also satisfying $A(f,g) = f(Ag)$.

\subsubsection*{Algebras, Operators and Hilbert Spaces}

A $\ast$-algebra (which we use synonymously with a unital $\ast$-algebra) stands for a complex vector space $\mathcal{A}$ with an associative product of elements denoted by juxtaposition and a conjugation operation ${(\,\cdot\,)}^\dagger:\mathcal{A}\to\mathcal{A}$ such that\\[2mm]
\noindent \textbf{1.} $\hat{A}(\alpha \hat{B} + \beta \hat{C}) = \alpha \hat{A}\hat{B} + \beta \hat{A}\hat{B}$ for all $\alpha,\beta\in \mathbb{C}$ and $\hat{A},\hat{B}\in \mathcal{A}$,\\[2mm]
\noindent \textbf{2.} $(\alpha\hat{A} + \beta \hat{B})^\dagger = \alpha^*\hat{A}^\dagger + \beta^*\hat{B}^\dagger$ for $\alpha,\beta\in \mathbb{C}$, $\hat{A},\hat{B}\in \mathcal{A}$, where $\alpha^*$ denotes the complex conjugate of $\alpha$,\\[2mm]
\noindent \textbf{3.} $(\hat{A}^\dagger)^\dagger = \hat{A}$ for all $\hat{A}\in \mathcal{A}$,\\[2mm]
\noindent \textbf{4.} There exists an identity element $\openone\in \mathcal{A}$ such that $\openone \hat{A} = \hat{A} \openone = \hat{A}$ for all $\hat{A}\in \mathcal{A}$. \\[2mm]
\noindent Elements of non-commutative $\ast$-algebras will be denoted with a hat.  We say that a set of elements $\{\hat{A}_i\}_i$ generates an algebra $\mathcal{A}$ if every element of the algebra can be written as a linear combination of products of elements in $\{\hat{A}_i\}_i$. If there is a suitable topology in $\mathcal{A}$, we extend linear combinations to series. Within the $\ast$-algebras, we define the commutator and anti-commutator, respectively,
\begin{equation}
    [\hat{A},\hat{B}] = \hat{A} \hat{B} - \hat{B}\hat{A}, \quad \quad 
    \{\hat{A},\hat{B}\} = \hat{A} \hat{B} + \hat{B}\hat{A}.
\end{equation}

A $\ast$-algebra morphism is a linear operation between $\ast$-algebras $\mathcal{A}_1$ and $\mathcal{A}_2$, $\Theta:\mathcal{A}_1\to\mathcal{A}_2$, such that $\Theta(\hat{A}\hat{B}) = \Theta(\hat{A})\Theta(\hat{B})$, $\Theta(\openone_1) = \openone_2$, $\Theta(\hat{A}^\dagger) = \Theta(\hat{A})^\dagger$. If $\mathcal{A}_1 = \mathcal{A}_2$ we say that $\Theta$ is an endomorphism. We say that $\Theta$ is a representation of $\mathcal{A}_1$ if the algebra $\mathcal{A}_2$ consists of operators in a Hilbert space.

Hilbert spaces will typically be denoted by $\mathscr{H}$ when they correspond to quantum systems. Pure states in $\mathscr{H}$ are normalized elements of $\mathscr{H}$ and will be denoted as with the standard Dirac notation $\ket{\psi}$, where $\bra{\psi}$ is the associated linear functional due to the Riesz representation theorem. A general state is represented by a density operator, usually denoted by $\hat{\rho}$: a positive semi-definite operator with $\text{tr}(\r) = 1$.

\subsubsection*{Symplectic Spaces}

A linear symplectic space is a vector space $\mathcal{V}\cong \mathbb{R}^{2N}$ with a non-degenerate anti-symmetric bilinear form $\bm \Omega$---the symplectic form. A basis $\{ e_{q_1},  e_{p_1},..., e_{q_N}, e_{p_N}\}$ of $\mathcal{V}$ is called symplectic if 
\begin{equation}
    \bm \Omega( e_{p_i}, e_{q_j}) = \delta_{ij}, \quad \bm \Omega( e_{q_i}, e_{q_j}) =\bm \Omega( e_{p_i}, e_{p_j}) = 0.
\end{equation}
A linear transformation $\bm S:\mathcal{V}\to\mathcal{V}$ is said to be symplectic if it preserves the symplectic form: $\bm \Omega(\bm Sv,\bm Su) = \bm \Omega(v,u)$ for all $v,u\in \mathcal{V}$. That is, if it maps symplectic bases into symplectic bases. In matrix representation, we can then write
\begin{equation}
    \bm \Omega = \bigoplus_{i=1}^N \begin{pmatrix} 0 & -1 \\ 1 & 0\end{pmatrix}, \quad \quad
    \bm \Omega^{-1} = \bigoplus_{i=1}^N \begin{pmatrix} 0 & 1 \\ -1 & 0\end{pmatrix},
\end{equation}
and a transformation is symplectic if $\bm S^\intercal \bm \Omega\bm S = \bm \Omega$, or, equivalently, if $\bm S \bm \Omega^{-1} \bm S^\intercal = \bm  \Omega^{-1}$.

The space $\mathcal{V}$ can also be seen as a linear manifold, in which case the symplectic form can be extended to act on tangent vectors. The basis $\{ e_{q_1},  e_{p_1},..., e_{q_N}, e_{p_N}\}$ induce coordinates $\xi^\alpha = \{q^1,p^1,...,q^N,p^N\}$ in $\mathcal{V}$, so that the symplectic form, seen as a 2-form, can be written as
\begin{equation}
    \bm \Omega = \frac{1}{2} \bm \Omega_{\alpha\beta} \dd\xi^\alpha\wedge \dd \xi^\beta = \sum_{i=1}^N \dd p^i \wedge \dd q^i.
\end{equation}
Any scalar function $f\in C^\infty(\mathcal{V})$ then defines a unique vector field $X_f$ through the equation 
\begin{equation}\label{eq:flowXf}
    \bm \Omega(X_f,Y) = - \dd f(Y),
\end{equation}
$X_f$ is called the Hamiltonian flow associated with $f$. Writing the components of the inverse symplectic form $\bm \Omega^{-1}$ as $\bm \Omega^{\alpha\beta}$, we can explicitly write $X_f = \Omega^{\alpha\beta}\partial_\alpha f\partial_\beta$. A function $f\in C^\infty(\mathcal{V})$ is called an observable, and the Poisson bracket between two observables is defined as
\begin{equation}\label{eq:poisson0}
    \{f,g\} = \dd f(X_g) = -\bm \Omega(X_f,X_g) = \Omega^{\alpha\beta}\partial_\alpha f \partial_\beta g.
\end{equation}
In particular, if $f$ and $g$ are linear observables defined by $f(v) = \bm \Omega(\phi_f,v)$ and $g(v) = \bm \Omega(\phi_g,v)$ for vector fields $\phi_f$ and $\phi_g$ in $\mathcal{V}$, note that
\begin{equation}\label{eq:pain}
        \{f,g\} = - \bm \Omega(\phi_f,\phi_g). 
\end{equation}

\cleardoublepage
\phantomsection		


\printglossary[type=symbols]
\cleardoublepage
\phantomsection		

\pagenumbering{arabic}

%

\addcontentsline{toc}{chapter}{Introduction}
\chapter*{Introduction}\label{chap:Intro}

For the past one hundred years, quantum field theory has stood as the most accurate framework for the description of matter. From the photons currently being absorbed by the cones and rods in your retinas to the atoms that constitute these very own cells, all known matter \textit{can} be fundamentally understood as excitations and interactions of quantum fields. However, not everything \textit{is} fully understood in terms of quantum field theory. Take a hydrogen atom, for instance. Although quantum field theory describes the electron, proton, and electromagnetic interaction individually, there is no current formulation that describes every constituent of the hydrogen atom within this framework. While one can compute corrections to effective descriptions using quantum field theory~\cite{gfactorBeier2000,gfactorIndelicato2004,gfactorTwoLoop2013,gfactor2020,gfactorExp2023}, no known state describes every part of the hydrogen atom. This matter becomes even more complicated when one realizes that describing the proton in terms of quark and gluon excitations is also not feasible---even computing the proton mass is a challenge by itself~\cite{QCD}.

Describing bound states like the hydrogen atom and proton is challenging in quantum field theory because they cannot be treated as perturbations of free theories, and non-perturbative methods are limited~\cite{nonpertComment2004,nonpertComment2021}. On the other hand, with a century of studies of quantum field theories, one could be led to think that at least we know everything that there is to know about free theories. This is not the case. For instance, there is still no effective method for computing the vacuum entanglement between two finite regions of space. Perhaps even more dramatic is the fact that a formal measurement framework for (even free) quantum field theories hadn't been formulated until 2018~\cite{FVOG} (published in 2020). These unsolved questions highlight the inherent richness and complexity of quantum fields.

It should then be clear that relying solely on quantum field theory limits our ability to describe most realistic physical systems. One way to proceed is to employ approximate descriptions---simplified models that, while practical, generally violate quantum field theory principles (typically through incompatibility with relativity). Although from a fundamental perspective, these effective models violate important fundamental laws of our universe, these incompatibilities may be acceptable if they fall below experimental precision. Ideally, one would be able to derive these effective models starting from quantum field theory, and precisely specify their regimes of validity, clearly defining the experimental conditions under which they apply.

At its core, the work presented here is motivated by the need to reconcile two competing demands. On one hand, we aim for a description of nature that faithfully adheres to the principles of relativity and quantum mechanics; on the other, we need models simple enough for explicit computations and direct experimental applications. Our focus will be to explore these connections when systems interact with a quantum field in localized regions of spacetime. Within this context, we focus on two aspects of the information locally stored in quantum fields: their entanglement structure, and the information about the background geometry of spacetime locally encoded in quantum field theory. Central to our discussion is classifying the limits where quantum field theory can be approximated by effective models and the regimes where the degrees of freedom of quantum fields play a fundamental role.

The key questions that motivate this work are: How much entanglement is there in a quantum field between two regions of spacetime? When are the quantum degrees of freedom of a field relevant? How much information about the background geometry of spacetime is encoded in quantum fields?

\definecolor{DarkGray}{RGB}{65,65,65}

\subsubsection*{Locally Probing Quantum Fields}

Quantum field theory initially emerged from efforts to reconcile quantum mechanics with special relativity. Early formulations by Dirac, Pauli, and Heisenberg successfully described many experiments but lacked an axiomatic basis. It was not until Wightman introduced a formulation based on n-point functions in 1956~\cite{WightmanOG} that the theory began to be viewed axiomatically. In 1964, Haag and Kastler established a fully axiomatic approach through what became known as the Haag-Kastler axioms~\cite{HaagOG}, framing quantum field theory as a theory of local algebras of operators associated with regions of spacetime. This local perspective eventually evolved into the field now known as algebraic quantum field theory.

This local formulation of quantum field theory not only provided a rigorous mathematical framework but also led to significant insights that influence our understanding of quantum fields, such as the Reeh-Schlieder theorem~\cite{ReehSchli}, the notion of thermality through modular Hamiltonians~\cite{Takesaki}, and the universal ultraviolet behaviour of field correlations~\cite{kayWald}. Within this algebraic framework, explicit constructions for free field theories can modelled by associating functions to observables, $f \mapsto \hat{\phi}(f)$, with each spacetime region's algebra constructed from operators $\hat{\phi}(f)$ whose support lies within that region. Although these local observables are the fundamental objects of the theory, quantum field theory was unable to, by itself, assign them an intrinsic physical meaning. It wasn't until the works of Fewster and Verch~\cite{FVOG} that the flow of information between fields during interactions was described, linking the functions $f$ to probe observables defined within the theory. Even then, the role played by the smearing functions cannot be directly related to physical measurements, as these necessarily involve probes that are either bound states of relativistic quantum fields with infinitely many degrees of freedom, or that are incompatible with the axioms of quantum field theory.

It should then be evident that the connection between effective models and fundamental descriptions becomes particularly relevant in the context of measurements. The fact that quantum field theory is the fundamental description of matter implies that even the very apparatuses that are used to measure quantum fields are themselves fundamentally quantum field theoretic objects. This forces us to choose between a fully quantum field theoretic framework for measurements---which is impractically complex for realistic setups---or the use of effective models that have some level of incompatibilities with fundamental laws, but that allow for explicit connections with physically accessible systems. Effective models also have the advantage of admitting a wider range of explicit applications, leading to an accelerated discovery of novel experimental protocols that leverage the fundamental features of quantum fields. On the other hand, it would be fair to question whether protocols discovered by making use of effective models are mere artifacts of their approximate descriptions, or whether they could be genuinely implemented and accurately described by models that respect the underlying principles that rule physics at the microscale.

It is clear that a satisfactory description of measurements in quantum field theory requires reconciling two demands. First, we need a description of nature that is as faithful as possible to the underlying principles of fundamental physics. Second, we require models that are simple enough to allow for explicit computations and direct connections with experimental setups. Ideally, we are seeking 1) a fully quantum field theoretic description of how information about a target field is encoded into a probe, that is also described as a quantum field, as well as 2) a systematic method for reducing the infinitely many degrees of freedom of the probe to a system with finite degrees of freedom that could be accessed in a realistic experiment. The Fewster-Verch framework can be used to understand the flow of information from the target field to a probe, so that the missing key of the puzzle is having an explicit method to reduce the quantum field theoretic description of quantum fields to finitely many degrees of freedom. This is essential to study the information locally stored in quantum fields.

A large portion of this thesis is devoted to bridging between foundational and practical concepts of probes in quantum field theory. Central to this connection is the concept of localized quantum fields---quantum field theories influenced by an external, classical potential that confines them in space. This model will allow us to consider localized probes described within quantum field theory that also admit a straightforward reduction to finitely many degrees of freedom, resulting in the physically accessible probes commonly known as particle detectors or Unruh-DeWitt detectors.

Once this connection is established, we explore two complementary aspects of local probes in quantum field theory. From a practical perspective, we embrace the effective Unruh-DeWitt detector models, showing how one can describe a reasonably general localized quantum system in a background curved spacetime starting from a given physical non-relativistic system. We then show that coupling this system with an external quantum field defines it as a local probe in the form of an Unruh-DeWitt detector. From a fundamental perspective, we study localized quantum fields in greater detail, providing a basis-independent notion for fields to be localized, and discussing properties of fields that are localized by physically realistic potentials. As an explicit example, we show how to describe a hydrogen atom by considering a quantum field theoretic model for the electron under the influence of an external Coulomb potential, and how it naturally reduces to a local probe of the magnetic field with two degrees of freedom corresponding to the electron spin.

As a final study of the description of local probes, we apply our findings to the study of the energy-momentum tensor of localized fields and the constraints imposed on these models by general covariance. Essentially, general covariance rules out non-dynamical localizing potentials and general relativity indirectly imposes that these potentials must be bounded. To demonstrate that these conditions still permit simple, physically consistent models, we present an analytically solvable example in which a quantum field is localized by a dynamical classical field and a perfect fluid. Overall, the goal of our studies of local probes in quantum field theory is to showcase exactly how to describe probes that both respect the fundamental properties of matter in spacetime and directly correspond to systems accessible in realistic experiments.

\subsubsection*{Entanglement in Quantum Field Theory}

The quantification of entanglement, even in finite dimensional quantum 
systems, is still a current topic of research. It should then be no surprise that this matter becomes even more complicated in the context of quantum field theories. The two main reasons for quantifying entanglement in quantum field theory are that 1) effectively, quantum fields have infinitely many quantum degrees of freedom associated with any subregion of spacetime, and 2) strictly speaking, quantum field theories do not admit a tensor product factorization associated to independent degrees of freedom. Since standard entanglement measures rely on the notion of non-separability between tensor factors, these techniques must be adapted in the context of quantum field theories. For instance, one can associate subsystems with commuting algebras of observables rather than factors in a Hilbert space.

Albeit challenging to quantify entanglement in general, a few results are known about entanglement in quantum field theory. For instance, there is robust evidence that the vacuum contains an infinite amount of entanglement between a subregion region of spacetime and its causal complement~\cite{universalEmbezzlers}, and that the entanglement entropy in this case diverges proportionally to the area of the subregion~\cite{areaLaw1993}. One of the reasons that we have results about vacuum entanglement between a region and its causal complement is that this problem reduces to computing bipartite entanglement in a pure state, which is the simplest setup for entanglement analysis. In contrast, much less is known about the entanglement between two finite subregions of spacetime, which would be an analog to the much more challenging study of entanglement in mixed bipartite states.


Despite its challenges, quantifying entanglement between two finite subregions is more physically realistic, as actual measurements of quantum fields take place in regions of finite size. The entanglement between two finite subregions of spacetime could then conceivably allow for practical applications in quantum computing and quantum cryptography. As such, one of our goals in this thesis will be to progress on the topic of quantifying the entanglement between two non-complementary regions of spacetime and how to explore methods for accessing this entanglement via localized probes. Specifically, we will discuss two complementary approaches to the study of entanglement between two finite subregions. 

The first method consists of an approach based on localized field modes. In a given spacetime region, the degrees of freedom of a Gaussian state can be represented through the expectation values of canonical pairs of smeared field operators. By considering a sufficiently large, but finite, number of modes, these degrees of freedom can be approximately encoded in a finite-dimensional covariance matrix. This representation then allows one to employ techniques of Gaussian quantum mechanics to quantify entanglement between two finite subregions of spacetime. We will consider examples of applications of this method and review the similar approaches applied to lattice field theories~\cite{KlcoUVIR,KlcoEntStrQFTI,KlcoEntStrQFTII,KlcoEntAllDist}, which suggested a series of results about the entanglement between two subregions.

The second method that we will consider relies on an operational approach, where two localized probes couple to independent degrees of freedom of the field in an attempt to extract entanglement from it. This protocol has received the name of entanglement harvesting after the modern approach presented in~\cite{Pozas-Kerstjens:2015}. Beyond reviewing the protocol, we provide a description of entanglement harvesting using probes modelled by localized quantum fields, yielding an entirely quantum field theoretic description of the protocol. This is relevant, as implementations using effective non-relativistic probes raised concerns that the protocol might be an artifact of the approximations rather than physically realistic~\cite{MaxHarvesting}. We also provide closed-form results for a commonly used setup in entanglement harvesting, which allows us to easily quantify the entanglement acquired by two detectors, the regimes in which this entanglement can be associated with the entanglement present in the field, and to consider asymptotic limits of entanglement extraction. These asymptotic results also reveal general features of the protocol, and facilitate comparisons with the broader conclusions drawn in~\cite{KlcoUVIR,KlcoEntStrQFTI,KlcoEntStrQFTII,KlcoEntAllDist}.

\subsubsection*{When are the Quantum Degrees of Freedom of Mediators Relevant?}

In a similar manner to how the internal degrees of freedom of bound states in interacting quantum field theories can be approximated by non-relativistic effective models, interactions mediated by quantum fields can be approximated by direct couplings between systems. Indeed, most non-relativistic interactions between quantum systems are prescribed as direct couplings---for example, spin-spin interactions are often modelled as $\hat{\bm \sigma}_1 \cdot \hat{\bm \sigma}_2$. This suggests the existence of a well-defined limit in which the quantum degrees of freedom of the mediators become negligible, allowing the interaction to be effectively described as a direct coupling. Identifying these regimes is essential both practically, for simplifying quantum field mediated interactions, and fundamentally, as they delineate when genuine quantum field theoretic effects manifest.


To explore the regimes where the quantum degrees of freedom of mediators are not directly relevant to interactions, we define and discuss a model for a direct-coupling interaction of quantum quantum systems formulated in terms of retarded propagators of the mediator: the quantum-controlled model. This prescription effectively defines fields whose degrees of freedom are entirely determined by the quantum systems that source them. The model is then able to incorporate some relativistic aspects of the interaction while neglecting the quantum degrees of freedom of the mediating fields. By comparing the evolution of states that interact through fully featured quantum fields with that prescribed by quantum-controlled models, we find the explicit regimes where mediators' quantum degrees of freedom are relevant. Overall, establishing the regimes of validity of the effective quantum-controlled models is an important step towards understanding when quantum field theory is necessary, as well as for connecting the fundamental descriptions of interactions with their simplified versions, that yield practical results in realistic physical systems.

An explicit example where understanding the regimes where the role of quantum degrees of freedom of mediators is relevant is in the recent experimental proposals to witness gravity-mediated entanglement~\cite{B,MV}. The goal in these proposals is to determine whether two systems that interact solely through the gravitational field can become entangled. While the experimental results can be used to infer properties of the gravitational interactions of quantum systems, one must carefully analyze the conditions under which the experiment would be able to determine whether the gravitational field has quantum degrees of freedom. By comparing the quantum field description of the gravity-mediated entanglement proposals with their quantum-controlled counterpart, we find that, under the proposed experimental parameters, both yield effectively indistinguishable results. This observation leads us to discuss a subtle point: while an observation of entanglement between the masses would rule out certain non-quantum models of gravity, it does not by itself ensure that the gravitational field has local quantum degrees of freedom. We argue that instead, one requires additional assumptions regarding locality that allow one to reach conclusions about non-classicality of gravitational degrees of freedom.

\subsubsection*{The Geometry of Spacetime from Quantum Field Theory}

The short‐distance behaviour of correlations in quantum field theory is deeply connected to the background geometry of spacetime~\cite{achim}. Specifically, the so-called Hadamard condition imposes that the two‐point correlation functions of a quantum field exhibit a universal singular structure at short distances~\cite{geometry}. This universal ultraviolet behaviour then fully encodes the information about the geodesic separation between neighbouring spacetime events, allowing one to recover the background geometry of spacetime entirely from the correlation functions of quantum fields. In fact, it has been proposed that gravity could itself be emergent from the correlations of quantum fields~\cite{achim2}.

Building on this insight, we describe an operational setup for probing spacetime geometry through measurements in quantum field theory. More precisely, using Unruh–DeWitt detectors as idealized probes, we show how one can relate correlations between sufficiently localized detectors to the correlation function of the quantum field that they couple to. By measuring these correlations, we reconstruct the spacetime metric in a coordinate system induced by a lattice of detectors that locally couple to the field.  We illustrate this recovery with examples involving detectors in various states of motion and different spacetimes, showing that the reconstruction becomes exact in the limit where the probes are perfectly localized and their separation is arbitrarily small. Overall, this example shows that not only do quantum fields store complete information about the geometry of spacetime, but that this information is physically accessible through local measurements.

After concretely showing how to recover the background geometry of spacetime from quantum measurements, we briefly discuss the possibility of rephrasing the geometry of spacetime entirely in terms of correlations of quantum fields, in an attempt to formulate a theory where gravity is emergent from quantum field theory. A preliminary analysis reveals that the main challenge in such a framework is incorporating dynamics—a problem we were unable to overcome. Instead, we argue that it might be possible to consider a theory where spacetime is emergent from the entanglement in quantum field theory, as has been proposed by different authors~\cite{VanRaamsdonkEmergent2010,ADSCFTemergent,Cao1,Cao2,emergent2021,Gui2023}.

\subsubsection*{Structure of the Thesis}

In the following six chapters, we will discuss the topics described above in detail, starting with a review of the tools of quantum field theory that will be used throughout the thesis in Chapter~\ref{chap:QFT}. Our goal in this chapter is to connect different formulations of quantum field theory, as well as setting the conventions that will be used throughout the remainder of the thesis. In Chapter~\ref{chap:meas}, we discuss local measurements in quantum field theory, and how each mode of a localized field can be approximated by an effective particle detector model. In this Chapter, we will study both localized fields and particle detector models in detail. Chapter~\ref{chap:ent} focuses on the quantification of entanglement in quantum field theory using techniques from Gaussian quantum mechanics and the entanglement harvesting protocol. In Chapter~\ref{chap:qc}, we introduce quantum-controlled models and delineate the regimes in which the quantum degrees of freedom of fields play a significant role. Chapter~\ref{chap:geometry} demonstrates that quantum fields encode complete information about the background geometry and details how local probes can access this information, also discussing the possibility of a theory where gravity emerges from quantum field correlations. Finally, Chapter~\ref{chap:conclusions} is devoted to a summary, conclusions, and future steps of the research carried on here.

Throughout this thesis, the recurring theme is the search for a harmonious interplay between the abstract principles of quantum field theory and the pragmatic models that are indispensable for real-world measurements. Whether it is through the careful construction of localized probes, the quantification of entanglement, or the effective formulation of quantum-controlled models, each part of this work aims to illuminate how effective theories emerge from, and remain consistent with, the fundamental tenets of quantum field theory, and how to use these tools to explore its essential aspects.

{\color{DarkGray}

}

\chapter{Quantum Field Theory}\label{chap:QFT}

Quantum field theory is at the root of our understanding of modern physics. Virtually everything in our Universe can be thought of in terms of excitations of quantum fields. The framework is also a century old, implying that there are too many aspects of it to fully discuss in any one document. The goal of this chapter is to provide a review of quantum field theory, focusing on its local formulation. Quantum fields are at the heart of this thesis and this material will be key in every subsequent section, so that it is only natural to start by fixing notations and conventions in quantum field theory.

During my graduate studies I found it particularly difficult to find references that would clearly connect different formulations of quantum field theory in a concise manner, and I am certainly not the first or last PhD student to face this struggle. As such, my hope is that the content of this Chapter can be useful as a reference to a younger generation of graduate students wishing to pursue similar research topics to the ones I have focused on during my PhD. In particular, this Chapter attempts to not overly focus on topological intricacies within the formulations, instead presenting the main concepts and the formal aspects that are specifically relevant for the remainder of the thesis. For a more formal perspective, the book~\cite{TheBook} is highly recommended and inspired parts of this chapter.

This chapter is organized as follows. In Section~\ref{sec:KG} we will review properties of the space of solutions of the \textit{classical} Klein-Gordon equation. It turns out that one of the main reasons for clear connections between different formulations of quantum field theory not to be ubiquitous is the fact that these rely on formulations of classical field theory, which is a topic that usually receives little attention. Section~\ref{sec:QFT} is devoted to the quantum field theory of a Klein-Gordon field, focusing on three different formulations of the theory in terms of covariant algebras of observables, bases of solutions to the Klein-Gordon equation, and algebras of observables associated to a Cauchy surface. Section~\ref{sec:Hadamard} is devoted to a brief summary of the Hadamard condition and why it is an essential additional ingredient in quantum field theories. Finally, in Section~\ref{sec:generalQFT}, we will briefly go over the formulations of quantum field theories of more general fields, such as complex scalar fields, spinor fields, electromagnetism and linearized quantum gravity.


\section{The Klein-Gordon Equation}\label{sec:KG}


In this section we will focus on the example of a classical real scalar field $\phi(\mf x)$ in a globally hyperbolic 3+1 dimensional spacetime $\mathcal{M}$. We will assume that the dynamics of the field is associated with the action
\begin{equation}
    S = \int \dd V \mathcal{L}, \quad\quad \mathcal{L} = - \frac{1}{2} \nabla_\mu \phi \nabla^\mu \phi - \frac{1}{2}V(\mf x) \phi^2,
\end{equation}
where $V(\mf x)$ is a smooth real function. In particular, $V(\mf x) = m^2$ yields a minimally coupled massive scalar field and $V(\mf x) = R(\mf x)/6$ gives us a conformally coupled scalar field. Extremizing $S$ with respect to variations of $\phi$ yields the equation of motion
\begin{equation}\label{eq:PKG}
    P\phi \coloneqq (\nabla_\mu \nabla^\mu - V(\mf x)) \phi = 0,
\end{equation}
which we will refer to as the Klein-Gordon equation.

Being a linear operator, the kernel of $P$ is a subspace of the space of smooth complex functions on $\mathcal{M}$. We denote the space of complex solutions of $P\phi = 0$ by $\S$, and the subspace of \textit{real} solutions by $\S_\mathbb{R}$, defined by the condition $\phi = \phi^*$. Although we will be looking at a real scalar field, it is convenient to describe $\S$ as a complex linear space, containing both complex and real solutions. 



\subsubsection*{The Klein-Gordon Inner Product}

Given two solutions of Eq.~\eqref{eq:PKG}, $\phi_1,\phi_2\in\S$, we define
\begin{equation}
    j_\mu(\phi_1,\phi_2) \coloneqq \phi_1^* \nabla_\mu \phi_2 - \phi_2\nabla_\mu \phi_1^*.
\end{equation}
Using the equation of motion we see that $\nabla_\mu j^\mu = 0$, so that the integral of $j^\mu$ along any Cauchy surface yields the same result provided that $\phi_1$ and $\phi_2$ decrease sufficiently fast at spatial infinity. We define
\begin{equation}
    (\phi_1,\phi_2) \coloneqq \ii \int \dd \Sigma^\mu (\phi_1^* \nabla_\mu \phi_2 - \phi_2\nabla_\mu \phi_1^*),
\end{equation}
where $\dd \Sigma^\mu = n^\mu \dd \Sigma$, with $\dd \Sigma$ being the induced volume for in the Cauchy surface and $n^\mu$ its normalized unit normal vector. One can show that the bilinear form above is sesquilinear, conjugate symmetric, and non-degenerate in $\mathcal{S}$\footnote{Indeed, if $(\phi_1,\phi_2) = 0$ for all $\phi_2\in\S$, then $\phi_1$ and $n^\mu \nabla_\mu \phi_1$ have to be zero along a Cauchy surface. Using this data as initial conditions for the Klein-Gordon equation yields $\phi_1 = 0$.}. For these reasons $(\cdot,\cdot)$ is commonly called the Klein-Gordon inner product. Importantly, the Klein-Gordon inner product is not positive definite; thus, it does not define the space $\S$ as a Hilbert space.

\subsubsection*{Bases of Solutions}

Our next goal is to find a basis for $\mathcal{S}$. Although the space of solutions lacks the neat convergence properties of a Hilbert space, it is possible to find subspaces of $\S$ where $(\cdot,\cdot)$ is positive definite, thus obtaining a Hilbert space. We then choose a maximal subspace of $\S$ where $(\cdot, \cdot)$ is positive-definite and denote it by $\S^+$. Importantly, there are infinitely many maximal subspaces of $\S$ with positive Klein-Gordon inner product, so that $\S^+$ is not unique in any sense. Given a choice of $\S^+$, the space of solutions then factors as $\S = \S^+\oplus \S^-$, where the sum is orthogonal with respect to the Klein-Gordon inner product, which is negative-definite in $\mathcal{S}^-$. That is, given $\phi\in \mathcal{S}$, it can be decomposed as
\begin{equation}
    \phi = \phi^+ + \phi^-,
\end{equation}
where $\phi^+\in\S^+$ and $\phi^-\in \S^-$ with $(\phi^+,\phi^-) = 0$. Also notice that, $- (\cdot,\cdot)$ defines $\mathcal{S}^-$ as a Hilbert space.

Given that $\mathcal{S}^+$ equipped with the Klein-Gordon inner product is a Hilbert space, we can find an orthonormal basis $\{u_{\bm k}\}_{\bm k}$, where the label $\bm k$ may be continuous, discrete, or a combination of both, depending on the spacetime geometry and $V(\mf x)$. Notice that the Klein-Gordon inner product satisfies
\begin{equation}
    (\phi_1,\phi_2) = - (\phi_1^*,\phi_2^*)^*.\label{eq:conjKGinn}
\end{equation}
Using this property we find that the set $\{u_{\bm k}^*\}_{\bm k}$ is an orthonormal basis for $\mathcal{S}^-$, so that any $\phi \in \mathcal{S}$ can be expanded as
\begin{equation}
    \phi = \phi^+ + \phi^- = \sumint_{\bm k} (u_{\bm k},\phi) u_{\bm k}   - (u_{\bm k}^*,\phi)u_{\bm k}^*.
\end{equation}
so that the set $\{u_{\bm k}, u_{\bm k}^*\}_{\bm k}$ is dense in $\S$ and orthonormal in the sense that
\begin{equation}
    (u_{\bm k}, u_{\bm k'}) = \delta(\bm k, \bm k'), \quad  (u_{\bm k}, u_{\bm k'}^*) = 0, \quad  (u_{\bm k}^*, u_{\bm k'}^*) = -\delta(\bm k, \bm k'),
\end{equation}
where $\delta(\bm k, \bm k')$ denotes the Dirac or Kronecker delta depending on whether $\bm k$ takes on continuous or discrete values.

Once the basis $\{u_{\bm k}, u_{\bm k}^*\}_{\bm k}$ has been chosen, a general real solution in $\S_\mathbb{R}$ is completely characterized by the complex coefficients $a_{\bm k} = (u_{\bm k}, \phi) = -(u_{\bm k}^*,\phi)$, and can be written as
\begin{equation}\label{eq:phiukclass}
    \phi = \sumint_{\bm k} a_{\bm k} u_{\bm k}  + a_{\bm k}^*u_{\bm k}^*.
\end{equation}
The functions $u_{\bm k}$ are usually referred to as positive frequency modes, while $u_{\bm k}^*$ are usually referred to as negative frequency modes, according to the sign of their norm defined by the Klein-Gordon inner product\footnote{The reason for the name actually comes from the fact that in static spacetimes with a static future oriented time coordinate $t$, the modes $u_{\bm k}$  with positive Klein-Gordon norm satisfy $\ii \partial_t u_{\bm k} = \omega_{\bm k} u_{\bm k}$, for adequate frequencies $\omega_{\bm k}$, while the negative frequency modes satisfy $\ii \partial_t u_{\bm k}^* = -\omega_{\bm k} u_{\bm k}^*$.}. In the decomposition~\eqref{eq:phiukclass}, the dynamics of $\phi$ in spacetime are entirely encoded in the independent modes $u_{\bm k}$, so that the coefficients $a_{\bm k}$ represent the amplitude of the field associated to each mode. In essence, the coefficients $a_{\bm k}$ encode the independent classical degrees of freedom of the field according to this mode decomposition.

\subsubsection*{Change of Basis}

One could instead have chosen a different maximal subspace of $\S$ where the Klein-Gordon inner product is positive-definite, say $\S^\bullet$, with orthonormal basis $\{v_{\bm k'}\}_{\bm k'}$. This would define the orthogonal decomposition $\S = \S^\bullet\oplus\S^{\circ}$, and a basis for $\S$ would be $\{v_{\bm k'},v_{\bm k'}^*\}_{\bm k'}$, so that any $\phi \in \S$ can also be expanded as
\begin{equation}\label{eq:KGother}
    \phi = \sumint_{\bm k'} b_{\bm k'} v_{\bm k'}  + b_{\bm k'}^*v_{\bm k'}^*,
\end{equation}
with $b_{\bm k'} = (v_{\bm k'},\phi)$. We can relate this expansion with the one of Eq.~\eqref{eq:phiukclass} by noticing that the functions $v_{\bm k'}$ are also in $\S$, so that they admit an expansion for the form
\begin{equation}\label{eq:bogoliubov}
    v_{\bm k'} =  \sumint_{\bm k} \alpha_{\bm k' \bm k}u_{\bm k} + \beta_{\bm k' \bm k}u_{\bm k}^*, \quad \alpha_{\bm k'\bm k} = (u_{\bm k},v_{\bm k'}), \quad \beta_{\bm k'\bm k} = - (u_{\bm k}^*, v_{\bm k'}),
\end{equation}
where we notice that $\alpha_{\bm k}$ and $\beta_{\bm k}$ might be complex, given that the functions $v_{\bm k'}$ themselves might (and in general will) be complex valued. Using $a_{\bm k} = (u_{\bm k},\phi)$ and Eq.~\eqref{eq:KGother}, we find that the coefficients $a_{\bm k}$ and $b_{\bm k'}$ are related by
\begin{equation}
    a_{\bm k} = \sumint_{\bm k'} \alpha_{\bm k'\bm k} b_{\bm k'} + \beta_{\bm k'\bm k}^* b_{\bm k'}^*.
\end{equation}
Notice that if $\beta_{\bm k'\bm k}$ are non-zero, the positive modes $v_{\bm k}$ will mix the positive and negative frequency modes $u_{\bm k}$ and $u_{\bm k}^*$, indicating that the subspaces $\S^+$ and $S^\bullet$ are distinct. Conversely, if $\beta_{\bm k'\bm k} = 0$ for all $\bm k$ and $\bm k'$, the basis $\{v_{\bm k'}\}_{\bm k'}$ and $\{u_{\bm k}\}_{\bm k}$ would be related by a unitary operation in $\S^+ = \S^\bullet$.

\subsubsection*{Initial Value Problem}

Finding solutions to a typical partial differential equation, such as the one defined by the Klein-Gordon equation, is usually formulated as an initial value problem. Specifically, given a Cauchy surface $\Sigma$, one considers initial conditions $\Phi,\Pi\in C^{\infty}(\Sigma)$, with the goal of solving the problem
\begin{equation}\label{eq:IVP}
    \begin{cases}
        P[\phi] = 0,\\
        \phi|_{\Sigma} = \Phi,\\
        n^\mu \nabla_\mu \phi |_\Sigma =\Pi,
    \end{cases}
\end{equation}
where $n^\mu$ denotes the future oriented normal vector to $\Sigma$. This problem can be directly solved if one has access to a basis $\{u_{\bm k}, u_{\bm k}^*\}_{\bm k}$ of the space of solutions. Indeed, expanding the solution $\phi$ as in Eq.~\eqref{eq:phiukclass}, one can obtain the coefficients $a_{\bm k}$, $a_{\bm k}^*$ by computing the Klein-Gordon inner product at the surface $\Sigma$:
\begin{equation}
    a_{\bm k} = \int \dd \Sigma^\mu (u_{\bm k}^* \partial_\mu \phi -  \phi\partial_\mu u_{\bm k}^*) =  \int \dd \Sigma (u_{\bm k}^* \Pi -  \Phi \, n^\mu \partial_\mu u_{\bm k}^*),
\end{equation}
which is entirely written in terms of the initial conditions and the basis functions.

It is convenient to define the momentum of the field $\pi(\mf x)$, such that one can write $\pi|_{\Sigma} = \Pi$ in~\eqref{eq:IVP}. To this end, consider a foliation of our globally hyperbolic spacetime by Cauchy surfaces $\Sigma_t$, where $t$ is a global future-directed timelike coordinate and define the conjugate momentum associated with this foliation as\footnote{We choose this definition for the conjugate momentum involving the induced metric in the surfaces so that it is defined as a scalar rather than as a density.}
\begin{equation}\label{eq:conjpi}
    \pi(\mf x)\sqrt{h} \coloneqq \frac{\delta S}{\delta(\partial_t\phi(\mf x))},
\end{equation}
where $h$ is the determinant of the metric induced in the surface $\Sigma_t$ for each $t$.

We will now relate the conjugate momentum with the initial conditions, as described in the initial value formulation~\eqref{eq:IVP}. We can use the foliation $\Sigma_t$ to induce a coordinate system in $\M$. We pick coordinates $(t,\bm x)$ so that for each $t$, $\bm x$ are coordinates in the surface $\Sigma_t$. Denoting the normalized field normal to $\Sigma_t$ by $n^\mu$, we can then write
\begin{equation}
    \partial_t = N n+N^i \partial_i,
\end{equation}
where $N^i = g(\partial_t,\partial_j)\delta^{ij}$. In the coordinates $(t,\bm x)$, the metric can then be written as
\begin{equation}
    g = (-N^2+N_iN^i) \dd t^2 + 2 N_i \dd x^i \dd t + h_{ij}\dd x^i \dd x^j,
\end{equation}
where, for each $t$, $h_{ij}$ is the induced metric in the surfaces $\Sigma_t$ and $N_iN^i = h_{ij}N^iN^j$. The metric determinant is then $\sqrt{-g} = N\sqrt{h}$, where $h$ is the determinant of the induced metric in the surfaces $\Sigma$, with $\dd \Sigma = \sqrt{h}\,\dd^3\bm x$. 

The metric then decomposes as the orthogonal sum 
\begin{equation}
    g_{\mu\nu} = -n_{\mu}n_\nu + h_{\mu\nu},
\end{equation}
with $h_{\mu\nu} = \delta_\mu^i\delta_\nu^j h_{ij}$. The Lagrangian can then be written as
\begin{align}
    \mathcal{L} &= \frac{1}{2} n^\mu \nabla_\mu \phi n^\nu \nabla_\nu \phi - \frac{1}{2}h^{ij}\nabla_{i}\phi\nabla_j\phi - \frac{1}{2}V(\mf x) \phi^2 \\
    &=   \frac{1}{2N^2}((\partial_t\phi)^2 - 2 \partial_t\phi N^i \partial_i\phi + N^i N^j \partial_i \phi \partial_j \phi) - \frac{1}{2}h^{ij}\nabla_i\phi\nabla_j \phi- \frac{1}{2}V(\mf x) \phi^2,
\end{align}
and the conjugate momentum is found from Eq.~\eqref{eq:conjpi}:
\begin{equation}
    \pi(\mf x) = \frac{1}{\sqrt{h}}\frac{1}{N^2}(\partial_t\phi - N^i \partial_i\phi)\sqrt{-g} = n^\mu \nabla_\mu \phi.
\end{equation}
We then see that the conjugate momentum $\pi(\mf x)$ defined in~\eqref{eq:conjpi} is precisely the normal derivative with respect to the surfaces, $n^\mu \nabla_\mu\phi$, so that $\pi|_\Sigma = \Pi$. That is, one can think of the pair $(\Phi,\Pi)\in C^{\infty}(\Sigma)$ in the initial value problem~\eqref{eq:IVP} as defining the initial value of the field and its conjugate momentum. The space of solutions $\mathcal{S}$ can then be entirely parametrized by pairs of functions $(\Phi,\Pi)\in C^{\infty}(\Sigma)$.


\subsubsection*{Phase Space}

One can also study real solutions of the Klein-Gordon equation from a phase space perspective. The first ingredient for a phase space description is then a symplectic form $\Omega$. The Klein-Gordon equation allows for the symplectic form
\begin{equation}
    \Omega(\phi_1,\phi_2) = \int \dd \Sigma^\mu(\phi_2\partial_\mu \phi_1 - \phi_1 \partial_\mu \phi_2),
\end{equation}
which is independent of the Cauchy surface due to the fact that $\nabla_\mu j^\mu(\phi_2^*,\phi_1) = 0$. The symplectic form $\Omega$ is related to the Klein-Gordon inner product through the expression
\begin{equation}
    (\phi_1,\phi_2) = \ii \Omega(\phi_2,\phi_1^*),
\end{equation}
which implies non-degeneracy and linearity, with antisymmetry being a consequence of Eq.~\eqref{eq:conjKGinn}. The symplectic form $\Omega$ then defines the space $\Gamma(\Sigma) = C_0^\infty(\Sigma)\oplus C_0^\infty(\Sigma)$\footnote{The choice of spaces $C_0^\infty(\Sigma)$ can also be replaced by more general spaces of test functions.} as an infinite dimensional symplectic manifold with symplectic form
\begin{equation}\label{eq:symplecticKG}
    \Omega(\Phi,\Pi;\Phi',\Pi') = \int \dd \Sigma (\Pi(\bm x)\Phi'(\bm x) - \Phi(\bm x) \Pi'(\bm x)).
\end{equation}

We extend the symplectic form to act in the tangent space of the symplectic manifold, where the corresponding symplectic form can be written as
\begin{equation}
    \Omega = \delta\Pi\wedge \delta \Phi.
\end{equation}
Any smooth function $O:\Gamma(\Sigma)\longrightarrow \mathbb{R}$, is then an observable of the theory and gives rise to the Hamiltonian flow
\begin{equation}
    X_O = \int \dd \Sigma \left(\frac{\delta O}{\delta \Pi} \frac{\delta}{\delta \Phi} - \frac{\delta O}{\delta \Phi}\frac{\delta}{\delta \Pi}\right).
\end{equation}
The Poisson bracket between any two observables $O(\Phi,\Pi)$ and $Q(\Phi,\Pi)$ can then be computed by
\begin{equation}
    \{O,Q\} = X_{Q}(O) = \int \dd \Sigma \left(\frac{\delta O}{\delta \Phi}\frac{\delta Q}{\delta \Pi} - \frac{\delta O}{\delta \Pi} \frac{\delta Q}{\delta \Phi}\right).
\end{equation}
Two observables $O(\Phi,\Pi)$ and $Q(\Phi,\Pi)$ will be a canonical pair if
\begin{equation}
    \{O,Q\} = 1.
\end{equation}
In this case, we say that $O$ and $Q$ define a mode of the field. This mode is entirely described within the two-dimensional symplectic submanifold parametrized by the canonical coordinates $O(\Phi,\Pi)$ and $Q(\Phi,\Pi)$. 

More generally, if $\{O_1,...,O_n\}$ and $\{Q_1,...,Q_n\}$ are observables such that
\begin{equation}
    \{O_i,O_j\} = 0, \quad \{O_i,Q_j\} = \delta_{ij}, \quad \{Q_i,Q_j\} = 0.
\end{equation}
$\{O_1,...,O_n\}$ and $\{Q_1,...,Q_n\}$ define canonical coordinates for a $2n$-dimensional symplectic submanifold, where the pairs $(O_i,Q_)$ are independent canonical pairs, representing independent degrees of freedom of solutions of the Klein-Gordon equation.

Of particular interest are linear observables of the form
\begin{equation}\label{eq:PhiPiclass}
    F(\Phi) = \int \dd \Sigma F(\bm x) \Phi(\bm x), \quad G(\Pi) = \int \dd \Sigma G(\bm x) \Pi(\bm x),
\end{equation}
whose Poison bracket becomes simply
\begin{equation}
    \{F(\Phi),G(\Pi)\} = \int \dd \Sigma F(\bm x) G(\bm x)
\end{equation}
In particular, if $\int \dd \Sigma F(\bm x) G(\bm x) = 1$, $F(\Phi)$ and $G(\Pi)$ define a canonical pair. These linear observables in the field and momentum are more often written as $\Phi(F) = F(\Phi)$ and $\Pi(G) = G(\Pi)$, corresponding to smeared field and momentum operators, where we understand $\Phi$ and $\Pi$ as distributions acting on functions defined in $C_0^\infty(\Sigma)$. These definitions allows one to derive what are commonly called the equal time canonical commutation relations for the field and conjugate momentum:
\begin{equation}\label{eq:poissonPhiPi}
    \{\Phi(F),\Pi(G)\} = \int \dd \Sigma \dd \Sigma' F(\bm x) G(\bm x') \{\Phi(\bm x), \Pi(\bm x')\} = \int \dd \Sigma F(\bm x) G(\bm x),
\end{equation}
which implies, at the distributional level, that
\begin{equation}
    \{\Phi(\bm x), \Pi(\bm x')\} = \delta(\bm x, \bm x').
\end{equation}

In this description, a general field state is a function that maps observables $O\in\Gamma$ to their expected values. That is, a state is a linear functional $\rho: C^\infty(\Gamma(\Sigma))\longrightarrow \mathbb{R}$ such that $\rho(1) = 1$ and $\rho(O)\geq 0$ whenever $O(\Phi,\Pi)\geq 0$. Any functional of this type can be written as a formal functional integral of the form
\begin{equation}
    \rho(O) = \int \text{D}\Phi \text{D}\Pi \rho(\Phi,\Pi)O(\Phi,\Pi) = \langle O \rangle_{\rho},
\end{equation}
where $\rho(\Phi,\Pi)\text{D}\Phi \text{D}\Pi$ is a measure in the phase space $\Gamma(\Sigma)$ that is normalized to $1$. This definition of a classical state is a generalization of the notion of a state being a solution to the Klein-Gordon equation that allows us to consider statistical mixtures of solutions, encoded in the measure $\rho(\Phi,\Pi)$. In particular, one can describe a single solution $(\Phi_0,\Pi_0)$ by considering a Dirac measure of the form $\rho_0(\Phi,\Pi) = \delta(\Phi-\Phi_0)\delta(\Pi-\Pi_0)$, in which case the expected value of an observable $O(\Phi,\Pi)$ becomes simply 
\begin{equation}
    \langle O\rangle_{\rho_0} = O(\Phi_0,\Pi_0).
\end{equation}

\subsubsection*{A Covariant Parametrization}

We have studied the space $\mathcal{S}$ using a basis of solutions approach and a phase space approach. The basis of solutions approach required a choice of basis $\{u_{\bm k}\}_{\bm k}$, while the phase space approach required a choice of Cauchy hypersurface $\Sigma$. We will now study a way of describing solutions to the Klein-Gordon equation in a covariant and choice independent manner using the propagators associated to the operator $P$.

The operator $P$ certainly does not admit a unique inverse, as its kernel is the space of solutions, and thus non-trivial. However, it admits two inverses $G_R$ and $G_A$ that satisfy
\begin{equation}\label{eq:GRGAcov}
    P G_R f = f, \quad \quad P G_A f = f,
\end{equation}
and are uniquely defined by the conditions $\text{supp}(G_Rf) \subset J^+(\text{supp}(f))$ and $\text{supp}(G_Af) \subset J^-(\text{supp}(f))$.

The operators $G_R$ and $G_A$ are the retarded and advanced Green's functions of $P$. Their corresponding integral kernels, $G_R(\mf x, \mf x')$ and $G_A(\mf x, \mf x')$, satisfy
\begin{align}
    (\nabla_\mu \nabla^\mu - V(\mf x))G_R(\mf x, \mf x') &= \delta(\mf x, \mf x'),\\ (\nabla_\mu \nabla^\mu - V(\mf x))G_A(\mf x, \mf x') &= \delta(\mf x, \mf x').
\end{align}
The propagator $G_R(\mf x, \mf x')$ is only non-trivial when $\mf x$ is in the causal future of $\mf x'$ ($G_R(\mf x, \mf x')$ propagates from $\mf x'$ to $\mf x$), while $G_A(\mf x, \mf x')$ is only non-trivial when $\mf x$ is in the causal past of $\mf x'$ ($G_A(\mf x, \mf x')$ propagates from $\mf x$ to $\mf x'$). As such, these Green's functions are related by $G_R(\mf x, \mf x') = G_A(\mf x', \mf x)$.

The retarded and advanced Green's functions can then be used to generate solutions to the non-homogeneous equation of motion:
\begin{equation}\label{eq:GRGAdelta}
    \phi = G_Rf\quad \text{or}\quad \phi = G_Af \quad \Rightarrow \quad  P\phi = f.
\end{equation}
Essentially, $G_Rf$ is the solution to $P\phi = f$ corresponding to the field that is created by a source $f$ and $G_Af$ is a solution that starts in the asymptotic past and is fully absorbed when it reaches the source $f$. Additionally, any function of the form $\phi = \alpha G_Rf + \beta G_Af$ with $\alpha + \beta = 1$ is a solution to $P\phi = f$, corresponding to a combination of fields that are absorbed and emitted by the source $f$.

Alternatively, a function of the form $\phi = \alpha G_Rf + \beta G_Af$ is a solution to the homogeneous equation of motion whenever $\alpha + \beta = 0$. We then define the causal propagator
\begin{equation}
    E = G_R - G_A, \quad E(f,g) = \int \dd V \dd V' f(\mf x) E(\mf x, \mf x')g(\mf x')
\end{equation}
which can be used to generate solutions of the homogeneous Klein-Gordon equation from compactly supported functions $g\in C_0^\infty(\M)$:
\begin{equation} 
    \phi = Eg \Rightarrow P\phi = 0.
\end{equation}
This is a covariant parametrization of solutions of the Klein-Gordon equation in terms of functions defined in spacetime. The map $g\mapsto Eg$ is indeed surjective in the space of solutions with compactly supported initial conditions, meaning that all solutions of the Klein-Gordon equation can be written as $Eg$ for some $g$. Moreover, the causal propagator is deeply linked to the symplectic form $\Omega$ of Eq.~\eqref{eq:symplecticKG}, as it satisfies\footnote{This equation can de derived by noticing that $\nabla^\mu(Eg \nabla_\mu(G_Rf)-G_Rf\nabla_\mu (Eg)) = f Eg$ (using Eq.~\eqref{eq:GRGAcov}) and integrating this result in the spacetime region $J^-(\Sigma)$ for a Cauchy surface placed in the causal future of the support of $f$.}
\begin{equation}\label{eq:EOmega}
    E(f,g) = - \Omega(Ef,Eg).
\end{equation}
The equation above also makes it evident that $E$ is antisymmetric, in the sense that $E(f,g) = - E(g,f)$. A particular case of this equation is when $\phi = Eg$, which gives
\begin{equation}\label{eq:phisymplecticEclass}
    \phi(f)\coloneqq \int \dd V \phi(\mf x) f(\mf x) = f(\phi) =  \Omega(\phi,Ef).
\end{equation}

The causal propagator can also be expressed in terms of a basis of solutions, $\{u_{\bm k}, u_{\bm k}^*\}$. Let $\phi = Eg$, then we can expand
\begin{align}
    \phi = Eg &= \sumint_{\bm k} (u_{\bm k},Eg)u_{\bm k} - (u_{\bm k}^*,Eg) u_{\bm k}^*\\
    &= \ii \sumint_{\bm k} \Omega(Eg,u_{\bm k}^*)u_{\bm k} - \Omega(Eg,u_{\bm k}) u_{\bm k}^*\\
    &= -\ii \sumint_{\bm k} u_{\bm k}^*(g)u_{\bm k} - u_{\bm k}(g)u_{\bm k}^*,
\end{align}
where we used Eq.~\eqref{eq:phisymplecticEclass} in the form $u_{\bm k}(g) = -\Omega(Eg,u_{\bm k})$ in the last equality. Writing the spacetime integrals of $u_{\bm k}(g)$ and $u_{\bm k}^*(g)$ explicitly, we then find
\begin{equation}
    Eg(\mf x) = \int \dd V' \frac{1}{\ii}\sumint_{\bm k} (u_{\bm k}(\mf x)u_{\bm k}^*(\mf x') - u_{\bm k}^*(\mf x)u_{\bm k}(\mf x')) g(\mf x'),    
\end{equation}
which implies that the integral kernel $E(\mf x, \mf x')$ can be written as
\begin{equation}\label{eq:Euki}
    E(\mf x, \mf x') = \frac{1}{\ii}\sumint_{\bm k} (u_{\bm k}(\mf x)u_{\bm k}^*(\mf x') - u_{\bm k}^*(\mf x)u_{\bm k}(\mf x')).
\end{equation}

Notice that the map $g\mapsto Eg$ is not injective. Indeed,  same as we have $PEh = 0$ for all $h\in C_0^\infty(\M)$, we also have $EPh = 0$, so that any two functions $f$ and $g$ such that $Eg = Ef$ differ by $Ph$ for some $h\in C_0^\infty(\M)$. Thus, the space $\S$ is instead isomorphic to the quotient space $C_\mathbb{C} \coloneqq C_0^\infty(\M)/PC_0^\infty(\M)$, with the equivalence relation $g\sim g+Ph$. The space of solutions is then entirely encoded in the space $C_\mathbb{C}$, and any solution $\phi\in \S$ is uniquely associated to the element $g\in C_\mathbb{C}$ such that $\phi = Eg$. The analogous statements hold for real solutions in $\S_\mathbb{R}$ defining $C_\mathbb{R} \coloneqq C_0^\infty(\M)_\mathbb{R}/PC_0^\infty(\M)_\mathbb{R}$. 



The space $C_\mathbb{C}$ turns out to be quite useful for defining linear observables in the symplectic space $\S$, with the symplectic form $\Omega$. Indeed, if $f:\mathcal{S}\to \mathbb{C}$ is a linear observable, it is defined by a function $f\in C_0^\infty(\M)$, with action
\begin{equation}
    f(\phi) = \int \dd V f(\mf x) \phi(\mf x).
\end{equation}
However, any observable of the form $Pf$ is trivial:
\begin{equation}
    Pf(\phi) = \int \dd V\, Pf(\mf x) \phi(\mf x) = \int \dd V f(\mf x) P\phi(\mf x) = 0,
\end{equation}
where we integrated by parts in the last equality. That is, the set of linear observables is actually identified with the set of functions in $C_\mathbb{C}$, which neglects elements of the form $Pf$. 


The symplectic form $\Omega$ then defines a Poisson bracket between any two observables of the theory. For instance, by noticing that linear observables can be written as $f(\phi) = -\Omega(Ef,\phi)$ and $g(\phi) = -\Omega(Eg,\phi)$, we quickly find the Poisson bracket between linear observables (see~\eqref{eq:pain}):
\begin{equation}
    \{f(\phi), g(\phi)\} = - \Omega(Ef,Eg) = E(f,g)
\end{equation}
Same as in our Cauchy surface description, it is common to denote linear functionals by $\phi(f) \coloneqq f(\phi)$, understood as (covariantly) smeared field observables, and giving rise to the covariant commutation relations
\begin{equation}
    \{\phi(f),\phi(g)\} = E(f,g).
\end{equation}

Indeed, any classical solution $\phi\in\mathcal{S}$ can also be seen as a distribution, in the sense that every function $\phi(\mf x)$ acts on a test function $f\in C_0^\infty(\M)$ as
\begin{equation}\label{eq:phifclass}
    \phi(f) = \int \dd V \phi(\mf x) f(\mf x).
\end{equation}
The weak solutions of the equation $P\phi = 0$ are then the distributions $\phi$ such that $\phi(Pf) = 0$ for all $f\in C_0^\infty(\M)$. When $\phi(\mf x)$ is a well-defined function, the equation $P\phi = 0$ follows from integration by parts. The quantity $\phi(f) = f(\phi)$ also has a physical meaning as the value of the observable $f$ at the solution $\phi(\mf x)$. One can then understand the function $f$ as defining the effective region of spacetime where an experimentalist has access to the field---it is impossible to measure a field (even a classical field) at a point: one always obtains an average value in a region where the measurement takes place. The expression $\phi(f)$ would then correspond to the field averaged (or smeared) in a region defined by the profile of $f(\mf x)$ (if $f$ is a real function). In this sense, when viewing solutions to the Klein-Gordon equation as the smeared fields in~\eqref{eq:phifclass}, the field is a functional that maps a `measurement apparatus' (codified by the function $f(\mf x)$) to the physically meaningful value $\phi(f)$. This is in contrast to the description of an element of the space of solutions $\S$ as an idealized field at a point, represented by $\phi(\mf x)$. We will not get into detail about how a \textit{classical} measurement apparatus can effectively access $\phi(f)$, or how it is defined by a test function $f(\mf x)$. Instead we will devote Chapter~\ref{chap:meas} to the analogous discussion in the context of quantum field theory.

A state in this context is also a linear functional that maps observables to their corresponding expectation values, $\rho:f\mapsto \rho(f)$ such that $\rho(1) = 1$ and $\rho(f)\geq 0$ when $f(\phi) \geq 0$. The interpretation of states is similar to that in our Cauchy slice phase space formulation, allowing one to compute expected values through functional integrals of the form
\begin{equation}
    \langle f\rangle_\rho \coloneqq \rho(f) = \int \text{D}\phi \rho(\phi) f(\phi),
\end{equation}
where $\rho(\phi)\text{D}\phi$ is again a formal positive measure on $\mathcal{S}$ with total measure $1$. In particular, a pure state corresponding to a standard solution $\phi_0(\mf x)$ of the Klein-Gordon equation is represented by the state $\rho_{\phi_0}(\phi) = \delta(\phi - \phi_0)$, so that 
\begin{equation}
    \langle f\rangle_{\phi_0} = f(\phi_0) = \int \dd V f(\mf x) \phi_0(\mf x).
\end{equation}

\subsubsection*{A Note About Conventions}

We should make a note at this stage regarding conventions for the differential operator $P$, the Green's functions $G_{R/A}$ and for the causal propagator, as different authors use different conventions for each of these. For instance, one can define~\cite{kasiaFewsterIntro} $P = - \nabla_\mu\nabla^\mu +V(\mf x)$ instead of $P = \nabla_\mu \nabla^\mu - V(\mf x)$ (which also depends on the metric signature), $PG_{R/A}f = -f$ instead of $PG_{R/A}f = f$ (common in electromagnetism textbooks, such as~\cite{Jackson}), and $E = G_A-G_R$ instead of $E = G_R - G_A$. Beyond these conventions, different authors might also define the Klein-Gordon inner product and the symplectic form with a relative minus sign, which typically also depend on the convention of whether $n^\mu$ in $\dd\Sigma^\mu = \dd\Sigma \, n^\mu$ stands for a future-directed normalized vector or a future-directed normalized form. Each of these choices affects countless signs throughout the formulation of classical and quantum field theory, and it is important to be consistent with these choices. Overall, when comparing results with other references, all these possible different conventions must be checked.

\section{Quantum Field Theory of a Klein-Gordon Field}\label{sec:QFT}

In this section we will describe formulations of the quantum field theory of a real scalar Klein-Gordon field. We will discuss three different approaches that can roughly be seen as 1) quantizing the covariantly smeared field operators $\phi(f)$~\eqref{eq:phifclass}, 2) quantizing the expansion coefficients $a_{\bm k}$ when the field is decomposed with respect to a given basis~\eqref{eq:phiukclass}, and 3) quantizing the field and conjugate momentum operators $\Phi(F)$, $\Pi(G)$ at a given Cauchy surface~\eqref{eq:PhiPiclass}. Specifically, we will start with the general algebraic formulation of quantum field theory in terms of $\ast$-algebras, and then move on to the effective quantization of $\phi(f)$, which will naturally lead us to the other approaches. 

\subsubsection*{Algebraic Quantum Field Theory}

In the algebraic sense, a quantum field theory can be defined as an association of open sets of spacetime to $\ast$-algebras\footnote{In some cases, it is useful to have a well-defined norm in these algebras, which requires that the local algebras are instead $\tc{C}^*$-algebras. This is mostly relevant for formal proofs in quantum field theory and we will not pursue this approach here.}, $\mathcal{O}\longmapsto \mathcal{A}(\mathcal{O})$. Specifically, the $\ast$-algebras are imposed to satisfy the following axioms:

\noindent\textbf{A1} There exists a unital $\ast$-algebra $\mathcal{A}(\M)$ with identity $\openone$ and, for each open causally convex set $\mathcal{O}\subset \M$, there exist algebras $\mathcal{A}(\mathcal{O})$ containing the unit $\openone$ that collectively generate $\mathcal{A}(\M)$.

\noindent\textbf{A2} If $\mathcal{O}_1\subset \mathcal{O}_2$, then $\mathcal{A}(\mathcal{O}_1)\subset \mathcal{A}(\mathcal{O}_2)$.

\noindent\textbf{A3} If $\mathcal{O}_1$ is spacelike separated from $\mathcal{O}_2$, then $[\mathcal{A}(\mathcal{O}_1),\mathcal{A}(\mathcal{O}_2)] = 0$.

\noindent\textbf{A4} If $\mathcal{O}_1\subset \mathcal{O}_2$ and $\mathcal{O}_1$ contains a Cauchy surface of $\mathcal{O}_2$, then $\mathcal{A}(\mathcal{O}_1) = \mathcal{A}(\mathcal{O}_2)$.

\noindent At this point, it should be mentioned that different authors often add different conditions for the association of algebras. For instance, for quantum field theories in Minkowski spacetime, it is common to include an axiom that defines Poincar\'e symmetries in the algebras. The axioms presented here are mostly inspired by~\cite{kasiaFewsterIntro} and~\cite{TheBookGoodChap}.

The elements of the $\ast$-algebras $\mathcal{A}(\mathcal{O})$ are usually referred to as the observables of the theory (even if the operators are not self-adjoint). Each of the axioms above condenses a fundamental aspect of the algebraic formulation of AQFT. \textbf{A1} represents the key aspect of this formulation, associating a local algebra to each causally convex open region of spacetime $\mathcal{O}$, so that the full algebra of the theory is nothing but the algebra that contains all the local observables. \textbf{A2} is an intuitive property that essentially states that all local operators in a subregion $\mathcal{O}_1$ are also localized in any region that contains it. \textbf{A3} is often referred to as the microcausality condition, and imposes that causally disconnected observables are independent. Finally, \textbf{A4} is related to the dynamics of the theory, stating that two regions that share the same Cauchy surface can describe the exact same observables. 

Overall, conditions \textbf{A1-A4} are usually seen as the minimal axioms satisfied by any quantum field theory\footnote{Except for minor modifications for explicit constructions, such as anti-commutation relations in \textbf{A3} for fields of half-integer spin, as we will mention in Section~\ref{sec:generalQFT}.}. In other words, it is a common conjecture that any quantum field theory admits a formulation that fulfills the four axioms above. 

While the algebras in quantum field theory can be local, the states are defined globally. A state is a linear functional $\omega: \mathcal{A}(\M)\longrightarrow \mathbb{C}$ such that $\omega(\openone) = 1$ and $\omega(\hat{A}^\dagger \hat{A}) \geq 0$. In the same spirit of our discussion in classical field theory, a state is a functional that maps observables to their expected value, generalized here for non-self adjoint operators. This can also be seen as a generalization of the notion of states as unit trace positive density operators, which allows for states to be defined even when a trace operation cannot be performed. Alternatively, if $\hat{\rho}$ is a density operator, it uniquely defines the state $\omega_{\hat{\rho}}$ by the functional relation $\omega_{\hat{\rho}}(\hat{A}) = \text{tr}(\hat{\rho}\hat{A})$. 

With this functional definition of states it is also possible to distinguish between pure and mixed states. A state is called mixed if it can be written as a convex combination of any two distinct states, and pure states are those that cannot be expressed as a convex combination of other states. 

The useful notion of operations acting on a state can be phrased in terms of operations in the algebra $\mathcal{A}(\M)$. Let $\Theta:\mathcal{A}(\M)\rightarrow \mathcal{A}(\M)$ be a $\ast$-algebra endomorphism, so that $\Theta$ is linear, $\Theta(\hat{A}\hat{B}^\dagger) = \Theta(\hat{A})\Theta(\hat{B})^\dagger$, and $\Theta(\openone) = \openone$. In this case $\Theta$ defines an operation in the algebra of observables $\mathcal{A}(\M)$ and it induces an operation in states $\omega\mapsto \tilde{\omega}$, with $\tilde{\omega}(\hat{A}) \coloneqq \omega(\Theta(\hat{A}))$. The endomorphism $\Theta$ is essentially a generalization of the typical operation $\hat{A}\mapsto \sum_i \hat{V}_i^\dagger\hat{A}\hat{V}_i$, with $\sum_i \hat{V}_i^\dagger \hat{V}_i = \openone$, which also induces the transformation $\hat{\rho} \mapsto \sum_i\hat{V}_i\hat{\rho} \hat{V}_i^\dagger$ to density operators.

One can also recover the standard description of states as vectors in a Hilbert space through the so-called GNS construction~\cite{gelfand1943imbedding,Segal1947IrreducibleRO}. In essence, given a unital $\ast$-algebra $\mathcal{A}$ and a state $\omega$, then there exists a complex Hilbert space $\mathcal{F}(\mathscr{H})$, a state $\ket{\Omega}\in\mathcal{F}(\mathscr{H})$ and a representation $\uppi_\omega$ of $\mathcal{A}$ as operators in $\mathcal{F}(\mathscr{H})$ such that
\begin{itemize}
    \item $\uppi_\omega(\mathcal{A})\ket{\Omega}$ is dense in $\mathcal{F}(\mathscr{H})$.
    \item $\omega(\hat{A}) = \langle{\Omega}|{\uppi_\omega(\hat{A})\Omega}\rangle$ for all $\hat{A}\in\mathcal{A}$.
\end{itemize}
The GNS construction allows one to represent any state as a vector in a Hilbert space which is fully generated by applications of algebra elements to the vector $\ket{\Omega}$. The fact that every state can be represented as a vector in a Hilbert space might come as a surprising feature, given that the construction works even when the state $\omega$ is mixed. However, the degrees of freedom of $\ket{\Omega}\in \mathcal{F}(\mathscr{H})$ are only in one-to-one correspondence with the degrees of freedom of $\omega$ if the representation is irreducible\footnote{Given a Hilbert space $\mathscr{H}$ and a mixed state $\hat{\rho} = \lambda\ket{\varphi}\!\!\bra{\varphi}+(1-\lambda)\ket{\psi}\!\!\bra{\psi}$ with orthogonal states $\ket{\varphi}$, $\ket{\psi}$ and $0<\lambda<1$, one can construct the~\cite{kasiaFewsterIntro} \textit{reducible} representation of the algebra $\mathcal{B}(\mathscr{H})$ in the space $\mathscr{H}\oplus\mathscr{H}$ by the association $\uppi_{\hat{\rho}}( \hat{A})= \hat{A}\oplus\hat{A}$. The vector $\ket{\varrho} = \sqrt{\lambda}\ket{\varphi}\oplus\sqrt{1-\lambda}\ket{\psi}$ then satisfies $\bra{\varrho}\uppi_{\hat{\rho}}(\mathcal{A})\ket{\varrho} = \mathcal{B}(\mathscr{H}\oplus\mathscr{H})$ and $\langle{\varrho}|{\uppi_{\hat{\rho}}(\hat{A})\varrho}\rangle = \text{tr}(\hat{\rho}\hat{A})$ for all $\hat{A}\in \mathcal{B}(\mathscr{H})$, compatible with the GNS construction. Although mathematically sound, this representation is certainly not the standard treatment employed in quantum mechanics. This is due to the fact that the representation $\uppi_{\hat{\rho}}$ is not irreducible. Alternatively, one could consider a GNS representation in a purification of the state $\hat{\rho}$, in which case the Hilbert space would contain more than only the degrees of freedom of the original state.}. Indeed, it can be shown that a GNS representation $\uppi_\omega$ of a $\ast$-algebra is irreducible\footnote{Technically, the representation is weakly irreducible if and only if the state is pure~\cite{TheBookHadamard}. If the $\ast$-algebra happens to be a $\tc{C}^*$-algebra, weak irreducibility can be replaced by irreducibility~\cite{TheBookHadamard}.} if and only if the state $\omega$ is pure~\cite{TheBookHadamard}, recovering the standard notions of quantum mechanics where states are vectors in a Hilbert space, and observables are associated with linear operators.

Importantly, the GNS representation is not guaranteed to be faithful. For a general state $\omega$, it may be the case that a GNS representation has $\uppi_\omega(\hat{A}) = 0$ for some elements $\hat{A}\neq 0$ in $\mathcal{A}(\M)$. In other words, it may be the case that some operators cannot be fully represented in $\mathcal{F}(\mathscr{H})$. On the other hand, if a GNS representation is both faithful and irreducible ($\omega$ pure), the algebra $\mathcal{A}(\M)$ can be understood as the algebra of operators acting in $\mathcal{F}(\mathscr{H})$ without any loss of information. However, even in this case, it will generally be the case that not all states acting in $\mathcal{A}(\M)$ can be represented as density operators in $\mathcal{F}(\mathscr{H})$, as we will see explicitly later on.

\subsubsection*{An Explicit Construction for a Real Scalar Field}

We have seen a general formulation of an algebraic quantum field theory. Now we can focus on the main example of this chapter: an explicit construction of the local algebras of observables for the case of a real scalar field. As we saw in Section~\ref{sec:KG}, the space of linear observables in the Klein-Gordon theory can be fully parametrized by functions in $C^\infty_0(\M)$. We will then assign each function $f\in C_0^\infty(\M)$ to a symbol $\hat{\phi}(f)$. The $\ast$-algebra $\mathcal{A}(\M)$ is generated by an identity operator $\openone$ and the symbols $\hat{\phi}(f)$, with the following identifications

\noindent \textbf{Linearity:} $\hat{\phi}(\alpha f + \beta g) \sim\alpha\hat{\phi}(f) + \beta \hat{\phi}(g)$.

\noindent \textbf{Hermiticity:} $\hat{\phi}(f)^\dagger \sim \hat{\phi}(f^*)$.

\noindent \textbf{Equations of Motion:} $\hat{\phi}(Pf) \sim 0$.

\noindent \textbf{Commutation Relations:} $[\hat{\phi}(f),\hat{\phi}(g)] \sim \ii E(f,g)$.\footnote{Notice that the commutation relations are compatible with the equations of motion condition since $E$ is a bi-solution of the homogeneous equation, that is $[\hat{\phi}(f+Ph),\hat{\phi}(g+Ph')] = E(f+Ph,g+Ph') = E(f,g)$.}

\noindent The identifications above then become equalities at the level of the $\ast$-algebra\footnote{One can also construct a representation in a $\tc{C}^*$-algebra by instead considering $f\mapsto e^{\ii \hat{\phi}(f)}$, giving rise to the so-called Weyl algebra~\cite{kasiaFewsterIntro}.}. 

The four conditions above ensure that axioms \textbf{A1-A4} are satisfied when one considers the local algebras $\mathcal{A}(\mathcal{O})$ as the algebras generated by $\hat{\phi}(f)$ with $f\in C_0^\infty(\mathcal{O})$. Indeed, properties \textbf{A1} and \textbf{A2} follow from the fact that $\mathcal{O}_1\subset\mathcal{O}_2$ implies $C_0^\infty(\mathcal{O}_1)\subset C_0^\infty(\mathcal{O}_2)$. Property \textbf{A3} follows from the fact that whenever two functions $f$ and $g$ are supported in spacelike separated regions, $E(f,g) = 0$. Finally, property \textbf{A4} is a consequence of the fact that $\hat{\phi}(Pg) = 0$, as this condition implies that the elements $\hat{\phi}(f)$ are uniquely parametrized by functions $f$ that generate different solutions to the Klein-Gordon equation. Thus, $\mathcal{A}(\mathcal{O}) = \mathcal{A}(D(\mathcal{O}))$, as initial conditions in either $\mathcal{O}$ or in its domain of dependence $D(\mathcal{O})$ can generate the same set of solutions. The condition $\hat{\phi}(Pf) = 0$ effectively makes it so that the elements of $\mathcal{A}(\mathcal{O})$ are generated by elements of $C_\mathbb{C}(\mathcal{O}) = C_0^\infty(\mathcal{O})/PC_0^\infty(\mathcal{O})$.

The field operators $\hat{\phi}(f)$ can then be understood as operator-valued distributions. Alternatively, these can be thought of as smeared field operators, a meaning better conveyed by the formal expression
\begin{equation}\label{eq:phif}
    \hat{\phi}(f) = \int \dd V \hat{\phi}(\mf x) f(\mf x).
\end{equation}
It is important to notice that $\hat{\phi}(\mf x)$ in the expression above is not a well-defined operator, it is rather used as a symbol that, in general, only makes sense when integrated against a test function, giving rise to a smeared field operator. Regardless, $\hat{\phi}(\mf x)$ is what is often called ``the quantum field'' by many authors, and it turns out to be a useful concept. For instance, it allows for the formulation of what is usually referred to as the \textit{microcausality condition}, a point-based version of axiom \textbf{A3}, stating that observables $\hat{O}_1(\mf x)$ and $\hat{O}_2(\mf x')$ must satisfy
\begin{equation}\label{eq:microcausality}
    [\hat{O}_1(\mf x),\hat{O}_2(\mf x')] = 0
\end{equation}
whenever the events $\mf x$ and $\mf x'$ are causally disconnected.

In many ways, the operator $\hat{\phi}(f)$ in~\eqref{eq:phif} is the quantum analogue of the classical smeared field $\phi(f)$ defined in Eq.~\eqref{eq:phifclass} in the previous section. Indeed, within this formulation of quantum field theory, one can still think of the function $f$ as defining the region where an experimentalist has access to a quantum field, accessing the corresponding localized observable $\hat{\phi}(f)$\footnote{This will be made explicit in Chapter~\ref{chap:meas}.}. At this stage, one can think that an experimental setup is defined by a set of test functions $\{f_1,...,f_n\}$, and that the localized observables that can be accessed in this setup are $\{\hat{\phi}(f_1),...,\hat{\phi}(f_n)\}$. This is certainly not the whole picture of measuring quantum fields, but is a valuable intuition to have at this stage.

The construction of the algebras of observables from test functions also allows one to define algebra operations from isometries. Indeed, if $\varphi:\M\mapsto\M$ is an isometry, then we can define its action on the algebra by $\varphi_*\hat{\phi}(f) = \hat{\phi}(\varphi^*f)$, inspired by the fact that
\begin{equation}
    \int \dd V \phi(\mf x) \varphi^*f(\mf x) = \int \dd V \varphi_*\phi(\mf x) f(\mf x),
\end{equation}
where $\varphi^*f(\mf x) = f(\varphi(\mf x))$, $\varphi_*\phi(\mf x) = (\varphi^{-1})^*\phi(\mf x)$. The operation $\varphi_*$ can then be extended to act in the entire algebra $\mathcal{A}(\M)$ by its action on the generators $\hat{\phi}(f)$\footnote{Notice that, in principle, $\varphi$ must be an isometry so that the operation $\hat{\phi}(\varphi^*f)$ is well defined, as we must have $\hat{\phi}(P\varphi^*f) = 0$. This happens for diffeomorphisms because $\varphi^*Pf = P\varphi^*f$, ensuring that $\varphi_*\hat{\phi}(Pf) = 0$.}. This reasoning can also be applied for one-parameter families of diffeomorphisms, inducing a one-parameter family of endomorphisms that is particularly relevant for discussions of thermality in quantum field theory\footnote{Indeed, if $\xi$ is a timelike Killing vector field and $\varphi_t$ is its flow, a thermal state (or KMS state) with inverse temperature beta with respect to $\xi$ is a state that satisfies the imaginary anti-periodicity condition $\omega(\hat{\phi}(f)\varphi_{(t+\ii \beta)*}\hat{\phi}(g)) = \omega(\varphi_{t*}\hat{\phi}(g)\hat{\phi}(f))$ as well as analiticity properties~\cite{Kubo,MartinSchwinger,KMSreview}. Although there is a rich theory of thermality in quantum field theory, we will not explicitly go into it in the thesis.}.

More general field observables, such as derivatives of the field operator, can also be defined distributionally through integration by parts in Eq.~\eqref{eq:phif}. For instance, the smeared covariant derivative of the field $\nabla_\mu\hat{\phi}$ can be defined as a distribution acting on compactly supported vector fields $j^\mu$ by the action
\begin{equation}
    \nabla\hat{\phi}(j) \coloneqq \hat{\phi}(- \nabla_\mu j^\mu) = -\int \dd V \hat{\phi}(\mf x)\nabla_\mu j^\mu(\mf x) = \int \dd V \nabla_\mu \hat{\phi}(\mf x) j^\mu(\mf x).    
\end{equation}
More general linear field operators can be defined analogously. In particular, the smeared momentum associated with a foliation with normal vector field $n$ can be written as
\begin{equation}
    \hat{\pi}(f) = \hat{\phi}(-\nabla_\mu(f\!\:n^\mu)).
\end{equation}

Non-linear field observables can be constructed using the multiplication operation in the algebra. For instance, the smeared operator $\hat{\phi}(f)^2$ is simply defined by $\hat{\phi}(f)\hat{\phi}(f)$, and can be formally written in integral form as
\begin{equation}\label{eq:phif2}
    \hat{\phi}(f)^2 = \int \dd V \dd V' f(\mf x) f(\mf x') \hat{\phi}(\mf x) \hat{\phi}(\mf x')    
\end{equation}
Although non-linear operators such as $\hat{\phi}(f)^2$ are naturally incorporated in the algebra of observables, some non-linear functions of the field are not part of the algebra at this stage. For instance, the algebra $\mathcal{A}(\M)$ does not contain an operator $\hat{\phi}^2(f)$ (not to be confused with $\hat{\phi}(f)^2$) that corresponds to
\begin{equation}\label{eq:phi2f}
    \hat{\phi}^2(f) = \int \dd V f(\mf x)\hat{\phi}(\mf x)^2.
\end{equation}
Indeed, operators that involve the square of the field $\hat{\phi}(f)$ can be understood as the irregular limit of $\hat{\phi}(f)\hat{\phi}(g)$ where $g(\mf x')\to\delta(\mf x, \mf x')$, which escape the definition of $\mathcal{A}(\M)$. Limits of this form usually yield irregular operators with divergent expected values that have to be regularized. It is common to extend the algebra to allow this type of operators, as these have an important physical significance. We won't get into details about this algebra extension~\cite{TheBookHadamard}, but we will briefly discuss how to compute and regularize expected values of operators of the form of Eq.~\eqref{eq:phi2f} in Section~\ref{sec:Hadamard}.

\subsubsection*{States and Representations}

As previously discussed, a state is a functional $\omega:\mathcal{A}(\M)\longrightarrow \mathbb{C}$ such that $\omega(\openone) = 1$ and $\omega(\hat{A}^\dagger\hat{A})\geq 0$ that maps field observables to their corresponding expected values. In the smeared operator construction we presented, the expected value of a field operator of the form $\hat{\phi}(f)$ can be thought of as an association $\omega:f\mapsto \mathbb{C}$, defining the distribution 
\begin{equation}
    \omega(\hat{\phi}(f)) = \int f(\mf x) \omega(\hat{\phi}(\mf x)),
\end{equation}
with kernel $\omega(\hat{\phi}(\mf x))$. Indeed, a general algebra element consists of sums (or series) of operators of the form $\hat{\phi}(f_1)...\hat{\phi}(f_n)$, so that the expected value of an operator in a state $\omega$ is entirely determined by the $n$-distributions
\begin{equation}
    \omega(\hat{\phi}(f_1)...\hat{\phi}(f_n)) = \int \dd V_1...\dd V_n f(\mf x_1)...f(\mf x_n) \omega(\hat{\phi}(\mf x_1)...\hat{\phi}(\mf x_n)).
\end{equation}
These $n$-distributions (or sometimes their kernels) are often referred to as the $n$-point functions of the state $\omega$, and fully determine it, by construction. It is also common to refer to the two-point function as the Wightman function, defined as the bi-distribution
\begin{equation}
    W(f,g) = \omega(\hat{\phi}(f)\hat{\phi}(g)) = \int \dd V \dd V' f(\mf x) g(\mf x')\omega(\hat{\phi}(\mf x)\hat{\phi}(\mf x')).
\end{equation}
The integral kernel of the Wightman function is often denoted by $W(\mf x, \mf x') = \omega(\hat{\phi}(\mf x)\hat{\phi}(\mf x'))$. Notice that due to the equations of motion $\hat{\phi}(f+Ph) = \hat{\phi}(f)$, we have that $W(f+Ph,g+Ph') = W(f,g)$ for all $h,h'\in C_0^\infty(\M)$. In other words, $W$ is a weak bi-solution of the Klein-Gordon equation of motion. Due to the positivity of the state, $\omega(\hat{\phi}(f^*)\hat{\phi}(f))\geq 0$, which implies that the Wightman function is a positive bi-distribution, in the sense that $W(f^*,f)\geq 0$. 

The commutation relations also imply the following relation between the Wightman function and the causal propagator:
\begin{equation}
    W(f,g) - W(g,f) = \ii E(f,g).
\end{equation}
In essence, the equation above implies that the antisymmetric part of the Wightman function of any state is entirely determined by the commutation relations of the theory so that $W(f,g)-W(g,f)$ is state independent. In other words, the Wightman function of a state is fully determined by its symmetric part. This is also made evident by considering the decomposition
\begin{equation}
    \hat{\phi}(f)\hat{\phi}(g) = \frac{1}{2}\{\hat{\phi}(f),\hat{\phi}(g)\} + \frac{\ii}{2}[\hat{\phi}(f),\hat{\phi}(g)],
\end{equation}
which implies
\begin{equation}\label{eq:WHE}
    W(f,g) = \frac{1}{2}H(f,g) + \frac{\ii}{2}E(f,g),
\end{equation}
where $H$ denotes the Hadamard distribution, explicitly defined by
\begin{equation}
    H(f,g) \coloneqq \omega(\{\hat{\phi}(f),\hat{\phi}(g)\}).
\end{equation}
The Wightman function of a state is then fully defined by the Hadamard distribution. Moreover, any symmetric weak bi-solution of the equations of motion $H$ satisfying
\begin{equation}
    |E(f,g)|^2\leq H(f,f)H(g,g)
\end{equation}
for all real compactly supported functions $f$ and $g$ defines a valid Wightman function corresponding to a state. The condition above is required from positivity of the Wightman function.

At this stage, it is also convenient to define a bi-distribution that plays a central role in interactions in quantum field theory: the Feynman propagator. The Feynman propagator evaluated at a state $\omega$ is the bi-distribution
\begin{equation}
    G_F(f,g) = \int \dd V f(\mf x) G_F(\mf x, \mf x') g(\mf x'),
\end{equation}
where for any positively time oriented time coordinate $t$, we define the kernel
\begin{equation}\label{eq:GFWxxp}
    G_F(\mf x, \mf x') = W(\mf x, \mf x') \theta(t-t') + W(\mf x', \mf x) \theta(t'-t).
\end{equation}
with the formal identification $W(\mf x, \mf x') = \omega(\hat{\phi}(\mf x)\hat{\phi}(\mf x'))$ we then see that the kernel of the Feynman propagator is a time-ordered two-point function, ordering the product of two field operators so that the operator evaluated with the larger value of temporal component always comes first. Also notice that if $\mf x$ and $\mf x'$ are spacelike separated, the operators $\hat{\phi}(\mf x)$ and $\hat{\phi}(\mf x')$ commute. Using the decomposition~\eqref{eq:WHE} we can also decompose the kernel $G_F(\mf x,\mf x')$ as
\begin{equation}\label{eq:GF}
    G_F(\mf x, \mf x') = \frac{1}{2}H(\mf x,\mf x') + \frac{\ii}{2} (G_R(\mf x, \mf x') + G_A(\mf x, \mf x')),
\end{equation}
where we used $E(\mf x, \mf x')\theta(t-t') = G_R(\mf x, \mf x')$ and $E(\mf x', \mf x)\theta(t'-t) = G_R(\mf x', \mf x) = G_A(\mf x, \mf x')$. From Eq.~\eqref{eq:GF}, we define the symmetric propagator as the bi-distribution $\Delta(f,g)$, with kernel\footnote{Notice that due to~\eqref{eq:GRGAdelta}, the symmetric propagator can be used to generate solutions to the non-homogeneous equation, as we have $P\Delta f = f$.}
\begin{equation}
    \Delta(\mf x, \mf x') = G_R(\mf x, \mf x') + G_A(\mf x, \mf x').
\end{equation}
The Feynman propagator can then be written in terms of the Hadamard distribution and the symmetric propagator as
\begin{equation}\label{eq:GFHDelta}
    G_F(f,g) = \frac{1}{2}H(f,g) + \frac{\ii}{2}\Delta(f,g).
\end{equation}
Same as the Wightman function, all the state dependence of the Feynman propagator is encoded in the Hadamard term, as $\Delta(f,g)$ is defined directly through the Green's functions associated with the equations of motion. This fact leads to the definition of the time ordered product of two linear field operators $\hat{\phi}(f)$ and $\hat{\phi}(g)$:
\begin{equation}
    T(\hat{\phi}(f)\hat{\phi}(g)) = \frac{1}{2}\{\hat{\phi}(f),\hat{\phi}(g)\} + \frac{\ii}{2}\Delta(f,g)\openone.
\end{equation}
The time ordered product can be generalized to more general field operators, but we will refrain from mentioning such generalizations, as they will not be explicitly used in the thesis.

{\color{red}
\begin{table}[h!]

\footnotesize

\centering
\color{black}

\begin{tabular}{ |c|c|c|c| } 

\hline

Distribution name & Symbol &  Alternative Form & Description  \\
\hline
\hline
{\normalsize {\color{white}O}}Retarded Green's function & $G_R$ & $\tfrac{1}{2}\Delta+\tfrac{1}{2}E$ & Retarded propagation from $\mf x'$ to $\mf x$  \\
\hline
Advanced Green's function & $G_A$ & $\tfrac{1}{2}\Delta-\tfrac{1}{2}E$ & Retarded propagation from $\mf x$ to $\mf x'$  \\
\hline
Causal propagator & $E$ &  $G_R - G_A$ & Projector into space of solutions  \\ 
\hline
Symmetric propagator & $\Delta$ &  $G_R + G_A$ & Symmetric exchange between $\mf x$ and $\mf x'$  \\ 

\hline
Hadamard function & $H$ &   & State dependent correlations\\

\hline
Wightman function & $W$ &  $\tfrac{1}{2}H + \tfrac{\ii}{2}E$ & Field's correlation function  \\
\hline
Feynman propagator & $G_F$  & $\tfrac{1}{2}H + \tfrac{\ii}{2}\Delta$ & Time-ordered correlation function \\ 
\hline
 
\end{tabular}

\caption{Distributions in quantum field theory used throughout the thesis and their descriptions.\label{tab}}
\end{table}

}

Of particular interest are states often referred to as quasifree states, in which the odd-point functions vanish, and the expected values of products of even numbers of field operators are determined by the Wightman function through Wick's theorem. A quasifree state turns out to be nothing but a zero-mean Gaussian state, as we will explicitly see when we discuss a representation of localized field modes. Thus, a quasifree state is entirely determined by it's Wightman function $W(f,g)$ (or, equivalently, by its Hadamard distribution $H(f,g)$). Also notice that each symmetric weak bi-solution of the Klein-Gordon equation uniquely defines a Hadamard distribution, and thus, a quasifree state.

One way of defining a pure quasifree state, as well as its associated GNS construction, is by decomposing the state of solutions as the orthogonal direct sum $\S = \S^+\otimes \S^-$ mentioned in the previous section. As we will see, this decomposition naturally defines a state $\omega$. Indeed, given a basis of solutions $\{u_{\bm k}, u_{\bm k}^*\}$ for $\S$ such that $\{u_{\bm k}\}_{\bm k}$ is a basis for $\S^+$, the bi-distribution defined by the kernel
\begin{equation}\label{eq:Wukuks}
    W(\mf x, \mf x') = \sumint_{\bm k}u_{\bm k}(\mf x)u_{\bm k}^*(\mf x')
\end{equation}
is positive and satisfies
\begin{align}
    W(f,g) - W(g,f) &= \int \dd V \dd V' f(\mf x) \bigg(\sumint_{\bm k}u_{\bm k}(\mf x) u_{\bm k}^*(\mf x') - u_{\bm k}^*(\mf x)u_{\bm k}(\mf x')\bigg)g(\mf x') \\
    &= \ii E(f,g),
\end{align}
so that it indeed defines a Wightman function. Therefore $W(\mf x,\mf x')$ in Eq.~\eqref{eq:Wukuks} defines a quasifree state $\omega$ through $\omega(\hat{\phi}(f)\hat{\phi}(g))$. 

We can explicitly build a GNS representation for $\omega$ by noticing that $\S^+$ is a Hilbert space with respect to the Klein-Gordon inner product. Indeed, let $K:\S\longrightarrow \S^+$ be the projection, with action
\begin{equation}
    K\phi = \sumint_{\bm k} (u_{\bm k},\phi) u_{\bm k}.
\end{equation}
Moreover, we can parametrize the set $\S^+$ by functions $f\in C_\mathbb{C}$, through $\phi = K E f$. Then the inner product between two solutions in $\S^+$ with respect to the inner product induced by the Klein-Gordon inner product is
\begin{align}
    ( K E f,K E g) &= \sumint_{\bm k} (u_{\bm k},Ef)^*(u_{\bm k},Eg) = \sumint_{\bm k} u_{\bm k}(f^*)u_{\bm k}^*(g)\\
    &=\int \dd V \dd V' f^*(\mf x) \bigg(\sumint_{\bm k} u_{\bm k}(\mf x) u_{\bm k}^*(\mf x')\bigg)g(\mf x') = W(f^*,g).
\end{align}
For convenience, from now on we denote $\mathscr{H} = \S^+$, and its states $KEF$ by $\ket{f}$, with the inner product $\braket{f}{g} = W(f^*,g)$.

Although $\mathscr{H}$ is still not the Hilbert space of the GNS construction associated to the state $\omega$, it is the first step to define it. We define the Fock space $\mathcal{F}(\mathscr{H})$ by
\begin{equation}\label{eq:Fock}
    \mathcal{F}(\mathscr{H}) = \bigoplus_{n=0}^\infty \mathscr{H}^{\odot n},
\end{equation}
where $\mathscr{H}^{\odot n} = \mathscr{H}\odot ...\odot \mathscr{H}$ is the $n$ symmetric\footnote{The symmetrization over the Fock space reflects the fact that the scalar field is bosonic so that its states are symmetric under exchanges.} tensor product of $\mathscr{H}$ with itself. The space $\mathscr{H}^{\odot 0}$ is then a one-dimensional vector space, which is spanned by a vector $\ket{\Omega}$. This turns out to be the state in the GNS construction that satisfies $\uppi_\omega(\hat{A}) = \langle \Omega | \uppi_\omega(\hat{A})\Omega\rangle$. To define the representation of $\mathcal{A}(\M)$ it is enough to define the representation of smeared field operators of the form $\hat{\phi}(f)$ which generate $\mathcal{A}(\M)$. These are defined as
\begin{equation}
    \uppi_\omega(\hat{\phi}(f)) = \hat{a}(f) + \hat{a}^\dagger(f),
\end{equation}
where $\hat{a}^\dagger$ and $\hat{a}$ are smeared creation and annihilation operators satisfying
\begin{equation}
    [\hat{a}(f),\hat{a}^\dagger(g)] = W(f,g) \openone, \quad\quad [\hat{a}(f),\hat{a}(g)] = [\hat{a}^\dagger(f),\hat{a}^\dagger(g)] = 0,
\end{equation}
with $\hat{a}(f)^\dagger = \hat{a}^\dagger(f^*)$. Their actions in vectors of $\mathcal{F}(\mathscr{H})$ is defined by 
\begin{align}
    \hat{a}^\dagger(f) \ket{\Psi} &\propto \ket{f}\odot \ket{\Psi},
\end{align}
where $\odot$ denotes the symmetrized tensor product. In particular, the conditions above imply that the state $\ket{\Omega}$ is the unique state satisfying $\hat{a}(f)\ket{\Omega} = 0$ for all $f\in C_{\mathbb{C}}$. It is also important to stress that usually $\hat{a}(f)$ and $\hat{a}^\dagger(f)$ are not elements of $\uppi_\omega(\mathcal{A}(\mathcal{O}))$ for any bounded region $\mathcal{O}$, instead being global operators in $\uppi_\omega(\mathcal{A}(\M))$\footnote{This is a consequence of the Reeh-Schlieder property, which we discuss in Chapter~\ref{chap:ent}.}.

In this GNS representation, it is possible to express the kernel $\hat{\phi}(\mf x)$ that defines the operator-valued distribution $\hat{\phi}$ as
\begin{equation}\label{eq:piomegaphi}
    \uppi_\omega(\hat{\phi}(\mf x)) = \sumint_{\bm k} \hat{a}_{\bm k} u_{\bm k}(\mf x) + \hat{a}_{\bm k}^\dagger u_{\bm k}^*(\mf x),
\end{equation}
where the creation and annihilation distributions can be written as
\begin{align}
    \hat{a}(f) = \sumint_{\bm k} u_{\bm k}(f)\hat{a}_{\bm k},\quad \quad
    \hat{a}^\dagger(f) = \sumint_{\bm k} u_{\bm k}^*(f)\hat{a}_{\bm k}^\dagger.
\end{align}
The operators $\hat{a}_{\bm k}$ and $\hat{a}_{\bm k}^\dagger$ are what many commonly refer to as the annihilation and creation operators, and they satisfy the canonical commutation relations
\begin{equation}\label{eq:CCRakakd}
    [\hat{a}_{\bm k},\hat{a}^\dagger_{\bm k'}] = \delta(\bm k, \bm k')\openone, \quad 
    [\hat{a}_{\bm k},\hat{a}_{\bm k'}] = 
    [\hat{a}^\dagger_{\bm k},\hat{a}^\dagger_{\bm k'}] = 0.
\end{equation}
It is also common to drop the symbol $\uppi_\omega$ in Eq.~\eqref{eq:piomegaphi}, and to refer to $\hat{\phi}(\mf x)$ as ``the quantum field'', keeping in mind that it only yields a well-defined operator when integrated against a test function $f$. From these expressions, it can be quickly verified that
\begin{equation}
    \bra{\Omega}\uppi_\omega(\hat{\phi}(f))\uppi_\omega(\hat{\phi}(g))\ket{\Omega} = W(f,g),
\end{equation}
where $W$ is the bisdistribution with kernel given by Eq.~\eqref{eq:Wukuks}. That is, this is the GNS representation of the state $\omega$ defined by the two-point function
\begin{equation}
    \omega(\hat{\phi}(f)\hat{\phi}(g)) = W(f,g).
\end{equation}

The representation $\uppi_\omega$ makes it so that the association $f\mapsto \uppi_\omega(\hat{\phi}(f))$ fulfills the same conditions as the association of functions in $C_\mathbb{C}$ to algebra elements $f\mapsto \hat{\phi}(f)$. In particular, the canonical commutation relations give
\begin{equation}
    [\uppi_\omega(\hat{\phi}(f)),\uppi_\omega(\hat{\phi}(g))] = \ii E(f,g).
\end{equation}
This representation is also (weakly) irreducible, implying that the state $\omega$ is pure. Moreover, GNS representations built from decompositions of the space of complex solutions, $\S = \S^+\oplus\S^-$ are typically faithful, meaning that all operators from the algebra can indeed be represented as linear operators in $\mathcal{F}(\mathscr{H})$. However, not all states in the algebra can be represented as vectors (or even density operators) in this GNS construction. Indeed, we will soon see that different choices of orthogonal decompositions of $\S$ lead to non-unitarily equivalent representations, which are able to describe different states. 

The direct sum decomposition of the Fock space~\eqref{eq:Fock} also induces a partition of the Hilbert space into eigenvectors of the so-called number operator. The number operator is defined by
\begin{equation}
    \hat{N} = \sumint_{\bm k} \hat{a}_{\bm k}^\dagger \hat{a}_{\bm k},
\end{equation}
and it is a well-defined operator in this GNS representation, with eigenvalues corresponding to each natural number, including $0$. Specifically, its eigenvectors are states $\ket{\psi_n}$ that only have components in the subspace $\mathscr{H}^{\odot n}$ in the direct sum decomposition~\eqref{eq:Fock}:
\begin{equation}
    \hat{N} \ket{\psi_n} = n \ket{\psi_n}.
\end{equation}
Equivalently, the number operator essentially counts how many operators $\hat{a}^\dagger(f)$ have to be applied to the vacuum state to produce $\ket{\psi_n}$, or, in other words, it counts the number of collective excitations that a given state has in all of its modes. For this reason, we refer to the collection of eigenvectors of $\hat{N}$ associated to the eigenvalue $n$ as $n$-excitation\footnote{A more usual name is $n$-particle states, and we will embrace it in Minkowski spacetime, when there is a clear notion of particles and a preferred vacuum.} states. Notice that every normalized state in $\mathcal{F}(\mathscr{H})$ must have a finite value for the expected number operator, by construction of the Fock space~\eqref{eq:Fock}.

In particular, a 1-excitation state is defined by a function $f\in C_\mathbb{C}$, and can be written as $\ket{f} = \hat{a}^\dagger(f) \ket{0}$. Indeed, as we saw in the construction of $\mathscr{H}$, the one-excitation space is exactly the space of positive frequency Klein-Gordon equations $\S^+$. For future reference, we note that if $\omega_f$ is the algebraic form of $\ket{f}$, then its Wightman function reads
\begin{equation}\label{eq:Wfg1part}
    \omega_f(\hat{\phi}(g)\hat{\phi}(h)) = \bra{f}\!\hat{\phi}(g)\hat{\phi}(h)\!\ket{f} = W(g,h) + W(g,f)W(f^*,h) + W(f^*,g)W(h,f).
\end{equation}

It is important to highlight that the construction discussed in this Segment is explicitly dependent on the choice of decomposition of the space of solutions into positive and negative frequencies. For instance, if one chooses a different basis for the space of solutions, say $\{v_{\bm k'}, v_{\bm k'}^*\}_{\bm k'}$ that decomposes $\S = \mathcal{S}^\bullet\oplus \S^\circ$, this would give rise to a GNS representation with respect to the quasifree state $\tilde{\omega}$, defined by the Wightman function
\begin{equation}
    \tilde{W}(f,g) = \sumint_{\bm k'} v_{\bm k'}(f)v_{\bm k'}^*(g).
\end{equation}
The associated smeared creation and annihilation operators would then be $\hat{b}(f)$ and $\hat{b}^\dagger(f)$:
\begin{equation}
    \hat{b}(f) = \sumint_{\bm k'} v_{\bm k'}(f)\hat{b}_{\bm k'},\quad 
    \hat{b}^\dagger(f) = \sumint_{\bm k'} v_{\bm k'}^*(f)\hat{b}^\dagger_{\bm k'},
\end{equation}
so that
\begin{equation}
    \uppi_{\tilde{\omega}}(\hat{\phi}(\mf x)) = \sumint_{\bm k'}v_{\bm k'}(\mf x)\hat{b}_{\bm k'} + v_{\bm k'}^*(\mf x)\hat{b}^\dagger_{\bm k'}.
\end{equation}
The state $|{\tilde{\Omega}}\rangle$ representing $\tilde{\omega}$ would then be defined by the condition $\hat{b}(f)|\tilde{\Omega}\rangle = 0$. 

If both representations are faithful and irreducible, the smeared creation and annihilation operators of each representation can be lifted to the algebra. Then, using that the modes $\{v_{\bm k'}, v_{\bm k'}^*\}_{\bm k'}$ and $\{u_{\bm k}, u_{\bm k}^*\}_{\bm k}$ are related by Eq.~\eqref{eq:bogoliubov}, one can also formally relate the creation and annihilation operators:
\begin{equation}
    \hat{a}_{\bm k} = \sumint_{\bm k'} \alpha_{\bm k' \bm k} \hat{b}_{\bm k'}+\beta^*_{\bm k' \bm k} \hat{b}^\dagger_{\bm k'}.
\end{equation}
In this case, we will have
\begin{equation}
    \tilde{\omega}(\hat{N}) = \sumint_{\bm k,\bm k'}|\beta_{\bm k'\bm k}|^2.
\end{equation}
We thus have that $\tilde{\omega}$ can be represented in as a state in the GNS representation of $\ket{\Omega}$ if and only if the integral above converges. Also notice that if any of the coefficients $\beta_{\bm k'\bm k}\neq 0$, the positive frequency spaces $\S^\bullet$ and $\S^+$ are indeed distinct. This showcases that different GNS representations are generally not unitarily equivalent, in the sense that there might be no isometry that would be able to map $\mathcal{F}(\mathscr{H})$ to $\mathcal{F}(\tilde{\mathscr{H}})$ mapping operators in one Fock space to the other and $\ket{\Omega}$ to a valid state. This example also showcases that the number operator $\hat{N}$ does not correspond to a physical observable with intrinsic meaning, instead being a relative operator that explicitly depends on the state $\omega$.

\subsubsection*{The Vacuum State and Particle States}

It is common to use the notion of a `vacuum state'' in quantum field theory. However, different authors disagree on the definitions that must be fulfilled for a state to be considered a vacuum. Arguably, the few universally agreed properties that a state must satisfy to be a vacuum state are to be a quasifree and a pure state. Unfortunately, these properties are not enough to characterize a unique state, and are fulfilled by infinitely many states, most of which do not deserve to be called a vacuum. Originally, the notion of vacuum was used for free fields in Minkowski spacetime, where there is a single state which is invariant under spacetime translations\footnote{If the Klein-Gordon differential operator $P$ is also translation invariant.}. The Minkowski vacuum also happens to be the state of minimum energy, as seen by inertial observers. Before discussing vacua in more general spacetimes and settings, let us briefly discuss the quantum field theory for a massless Klein-Gordon field in Minkowski spacetime, the Minkowski vacuum and its GNS representation.

Assume that $\M$ is Minkowski spacetime and consider a Klein-Gordon field with equation of motion $\nabla_\mu\nabla^\mu\phi = 0$, defining the theory of a free massless Klein-Gordon field. The Green's functions of the operator $P = \nabla_\mu \nabla^\mu$ can be explicitly computed in this case, and in an inertial coordinate system $(t,\bm x)$, their kernel can be written as
\begin{align}
    G_R(\mf x, \mf x') = - \frac{1}{2\pi}\delta(- (t-t')^2 + (\bm x - \bm x')^2)\theta(t-t') = -\frac{1}{4\pi|\bm x - \bm x'|}\delta(t'-t+|\bm x - \bm x'|),\nonumber\\
    G_A(\mf x, \mf x') =  - \frac{1}{2\pi}\delta(- (t-t')^2 + (\bm x - \bm x')^2)\theta(t'-t) = -\frac{1}{4\pi|\bm x - \bm x'|}\delta(t'-t-|\bm x - \bm x'|).\label{eq:GRGA}
\end{align}
From the expressions above, we see that $G_R(\mf x, \mf x') = G_A(\mf x',\mf x)$, as well as the fact that $G_R(\mf x, \mf x') = 0$ whenever $\mf x$ is in the causal past of $\mf x'$. The causal propagator can be obtained through $E(\mf x, \mf x') = G_R(\mf x, \mf x') - G_A(\mf x, \mf x')$.

The Minkowski vacuum can be defined by a specific decomposition of the space of solutions of the Klein-Gordon equation. The typical way to define this decomposition is by choosing the orthonormal basis $\{u_{\bm k},u_{\bm k}^*\}_{\bm k}$ of $\S$ defined by
\begin{equation}\label{eq:ukplanewave}
    u_{\bm k}(\mf x) = \frac{1}{(2\pi)^{3/2}}\frac{e^{\ii \mf k \cdot \mf x}}{\sqrt{2 \omega_{\bm k}}}, \quad u_{\bm k}^*(\mf x) = \frac{1}{(2\pi)^{3/2}}\frac{e^{-\ii \mf k \cdot \mf x}}{\sqrt{2 \omega_{\bm k}}},
\end{equation}
where $\bm k$ is a label in $\mathbb{R}^3$, $\mf k = (\omega_{\bm k},\bm k)$, $\omega_{\bm k} = |\bm k|$ and $\mf k \cdot \mf x = \eta_{\mu\nu}k^\mu x^\nu$. The basis $\{u_{\bm k},u_{\bm k}^*\}_{\bm k}$ of Eq.~\eqref{eq:ukplanewave} is typically referred to as the plane wave basis of solutions. One can quickly verify that this basis is orthonormal with respect to the Klein-Gordon inner product, in the sense that 
\begin{equation}
    (u_{\bm k}, u_{\bm k'}) = \delta^{(3)}(\bm k - \bm k'), \quad (u_{\bm k},u_{\bm k'}^*) = 0, \quad (u_{\bm k}^*,u_{\bm k'}^*) = - \delta^{(3)}(\bm k - \bm k').
\end{equation}
This mode decomposition then defines a unique pure quasifree state $\omega_0$ such that its Wightman function is defined by
\begin{equation}\label{eq:W0}
    W_0(\mf x, \mf x') = \int\dd^3\bm k \,u_{\bm k}(\mf x) u_{\bm k}^*(\mf x') = \lim_{\epsilon\to0^+}\frac{1}{4\pi^2}\frac{1}{-(t-t' - \ii \epsilon)^2 + (\bm x - \bm x')^2},
\end{equation}
where the regulator $\epsilon$ is necessary to ensure that the bi-distribution is well defined when evaluated at functions with overlapping support. In particular, we can decompose
\begin{equation}
    W_0(\mf x, \mf x') = \text{PV} \left(\frac{1}{4\pi^2}\frac{1}{(-(t-t')^2 + (\bm x - \bm x')^2)}\right) - \frac{\ii }{4\pi} \text{sign}(t-t') \delta(-(t-t')^2 + (\bm x - \bm x')^2),
\end{equation}
where PV denotes the principal value and $\text{sign}(u)$ is the sign function\footnote{This expression is obtained from $\lim_{\epsilon\to 0^+} \frac{1}{x\pm \ii \epsilon} = \mp \ii \pi \delta(x) + \text{PV}\frac{1}{x}$.}. The state $\omega_0$ is then invariant under Poincar\'e transformations, as the Wightman function depends only on the invariant spacetime separation between events\footnote{There are different ways of verifying this independence by considering a representation of the Poincar\'e group in the algebra, but it ends up being ultimately equivalent to the invariance of the kernel of the Wightman function.}. Also notice that using the decomposition of the Wightman in terms of the plane wave basis, we can write
\begin{equation}\label{eq:W0Fourier}
    W_0(f,g) = \int \dd^3 \bm k \,u_{\bm k}(f)u_{\bm k}^*(g) = \frac{1}{(2\pi)^3}\int \frac{\dd^3 \bm k}{{2|\bm k|}} \tilde{f}(|\bm k|, \bm k)\tilde{g}(-|\bm k|,-\bm k),
\end{equation}
where $\tilde{f}$ defines the Fourier transform of the spacetime function $f$:
\begin{equation}\label{eq:4Fourier}
    \tilde{f}(\mf k) = \int \dd^4\mf x f(\mf x) e^{\ii \mf k \cdot \mf x}
\end{equation}
with $\mf k = (\omega,\bm k)$ and $\mf k \cdot \mf x = \eta_{\mu\nu}k^\mu x^\nu$. 

The GNS representation $\uppi_0$ of the Minkowski vacuum $\omega_0$ follows the construction previously outlined, defining the space $\S^+$ as the Hilbert space $\mathscr{H}_0$ with inner product
\begin{equation}
    \braket{f}{g} = W(f^*,g),
\end{equation}
giving rise to a Fock space $\mathcal{F}(\mathscr{H}_0)$, and defining smeared creation and annihilation operators $\hat{a}$ and $\hat{a}^\dagger$. This gives rise to the familiar representation of the field operator $\hat{\phi}(\mf x)$,
\begin{equation}
    \uppi_0(\hat{\phi}(\mf x)) = \frac{1}{(2\pi)^{3/2}}\int \frac{\dd^3 \bm k}{\sqrt{2 \omega_{\bm k}}} \left(e^{\ii \mf k \cdot \mf x}\hat{a}_{\bm k} + e^{-\ii \mf k \cdot \mf x}\hat{a}_{\bm k}^\dagger\right),
\end{equation}
and the condition $\hat{a}_{\bm k}\ket{0} = 0$ defines the vector $\ket{0}$ that realizes the GNS representation of the Minkowski vacuum. This GNS construction is not only irreducible, it is also faithful, so that it can represent all operators in the algebra $\mathcal{A}(\M)$ as operators in $\mathcal{F}(\mathscr{H}_0)$.

The number operator in $\mathcal{F}(\mathscr{H}_0)$ is then given by
\begin{equation}
    \hat{N} = \int \dd^3 \bm k\, \hat{a}^\dagger_{\bm k} \hat{a}_{\bm k},
\end{equation}
As previously mentioned, its eigenvalues are natural numbers, including $0$, and any state in $\mathscr{H}_0^{\odot n}$ in the decomposition~\eqref{eq:Fock} is an eigenvector of $\hat{N}$ with eigenvalue $n$. In particular, we have $\hat{N}\ket{0} = 0$. In this GNS representation, it is common to refer to an eigenstate of the number operator $\hat{N}$ with eigenvalue $n$ as an $n$-particle state. For instance, 1 particle states are elements of $\mathscr{H}_0$, fully parametrized by functions in $C_{\mathbb{C}}$ (or, equivalently, by elements of $\S^+$), and can generally be written as
\begin{equation}
    \ket{f} = \hat{a}^\dagger(f)\ket{0} = \int \dd^3\bm k \,u_{\bm k}^*(f)\hat{a}_{\bm k}^\dagger \ket{0}.
\end{equation}
Normalization of $\ket{f}$ implies that
\begin{equation}
    \braket{f}{f} = \bra{0}\hat{a}(f^*)\hat{a}^\dagger(f)\ket{0} = W_0(f^*,f) = 1.
\end{equation}
Alternatively, writing $\varphi(\bm k) = u_{\bm k}^*(f)$, we find
\begin{equation}\label{eq:norm1part}
    \ket{f} = \int \dd^3\bm k \varphi(\bm k) \hat{a}_{\bm k}^\dagger \ket{0}, \quad \quad \braket{f}{f} = 1 \Longleftrightarrow \int \dd^3\bm k |\varphi(\bm k)|^2 = 1.
\end{equation}
Higher particle states with higher particle number can be obtained by applying multiple operators $\hat{a}^\dagger$ smeared by different functions to $\ket{0}$. 

The definition of the Minkowski vacuum describing the absence of particles, as well as the definition of particles themselves, both proved very useful for numerous different reasons. For instance, the particle number operator is deeply related to the Hamiltonian (to be discussed soon), being directly linked with resonances in inertial apparatuses that couple to the field for sufficiently long times. In these regimes one can also describe interactions in terms of particle emission and absorption, relating these with energy quanta. Similarly, an inertial system in its ground state {does not become excited} when adiabatically coupled to the field in the Minkowski vacuum, confirming the intuition that the vacuum represents the ``absence of particles''.

Although these concepts are typically useful, the examples where they apply require regimes of inertial observers coupled to the field for sufficiently long times, which is usually the regime of particle physics. This is not the regime where we want to discuss quantum field theory. Instead, we want to consider local couplings that can happen in arbitrary finite regions of spacetime. In particular, these couplings do not have to be associated with inertial observers. In fact, as we will discuss in Chapter 3, even the Minkowski vacuum produces excitations in inertial probes that couple to the field for a finite time. 

The definition of a vacuum and particles becomes even more ambiguous in more general spacetimes, where there is usually no reason to privilege a subspace of $\S$ rather than another. Given any mode decomposition $\{u_{\bm k},u_{\bm k}^*\}$ one obtains a unique state $\omega_u$ and its associated GNS construction and Fock space. Still, there is no reason to choose a basis $\{u_{\bm k},u_{\bm k}^*\}$ over any other basis $\{v_{\bm k},v_{\bm k}^*\}$, which would instead define a state $\omega_v$. And even worse, we saw that GNS representations associated with different decompositions of $\S$ are usually not unitarily equivalent, implying that it is typically not possible to represent the state $\omega_u$ in the GNS construction defined from the decomposition $\{v_{\bm k},v_{\bm k}^*\}$. Moreover, the creation and annihilation operators effectively create mode excitations when acting in the vacuum, and the modes are intrinsically global---each mode is defined by its initial conditions in a whole Cauchy surface. In summary,

\begin{center}
    \textit{``Particles are a global and relative concept, \\ill defined in a local approach to quantum field theory.''}  
\end{center}

Where does this leave us? If the notions of particles does not apply to local quantum field theory and the notion of vacuum is ambiguous, how should we approach quantum field theory? The answer to this question is actually rather simple: our goal is not to define particles or vacuua: it is to predict what happens in a physical setup. A choice-independent approach that can be taken is as follows. First identify the initial state $\omega$ of the field before an experiment is performed. This could be done by measuring different field observables which restrict the possible initial states. For instance, if one has good reason to believe that the state is quasifree, one should attempt to recover expected values such as $\omega(\{\hat{\phi}(f),\hat{\phi}(g)\})$, which entirely determine the state. Once the state is determined, one typically applies an operation to the field during an experiment. For instance, one could conceive an experiment that applies the operation $\hat{U} = e^{\ii \hat{\phi}(f)}$ to the field (for a real $f$), thus affecting field observables through $\hat{\phi}(f)\mapsto \hat{U}^\dagger\hat{\phi}(f)\hat{U}$, or, equivalently, changing the state by $\omega(\,\cdot\,)\mapsto \tilde{\omega}(\,\cdot\,) = \omega(\hat{U}^\dagger\,\cdot\,\hat{U})$. After the experiment is performed, an experimentalist would then have access to the expected value of a collection of field observables $\hat{A}_1,...,\hat{A}_n$ in the state $\tilde{\omega}$. The expected values $\omega(\hat{A}_i)$  would then be the outcome of the experiment. Overall, the notions of particles and vacua, although useful, are not necessary. All we need is to be able to describe the initial state of an experiment, as well the operations performed in it and the observables that we have access to.

\subsubsection*{Localized States and Field Modes}

When we first defined states in the context of quantum field theory, we mentioned that, unlike observables, states cannot be locally defined. This is a consequence of the fact that given a causally convex region $\mathcal{O}$, the local algebra $\mathcal{A}(\mathcal{O})$ is a type III von Neumann algebra, implying that there exist no finite rank projectors and a trace operation cannot be defined. Consequently, one cannot define a reasonable notion of state associated with local algebras $\mathcal{A}(\mathcal{O})$.

Although local states in quantum field theory are not well defined, one can still talk about local quantum degrees of freedom of a given state. This can be done by considering local canonical canonical pairs associated with the field theory. For instance, consider compactly supported functions $f_i$, $g_i$ such that $\text{supp}(f_i),\text{supp}(g_i)\subset\mathcal{O}$ for $i=1,...,N$ that satisfy
\begin{equation}
    [\hat{\phi}(f_i), \hat{\phi}(g_j)] = \ii \delta_{ij}, \quad\quad [\hat{\phi}(f_i),\hat{\phi}(f_j)] = [\hat{\phi}(g_i),\hat{\phi}(g_j)] = 0.
\end{equation}
In other words, we must have $E(f_i,g_j) = \delta_{ij}$ and $E(f_i,f_j) = E(g_i,g_j) = 0$, which define $(\hat{\phi}(f_i),\hat{\phi}(g_i))$ as independent canonical pairs. One can then create a natural mapping between the canonically conjugate observables 
\begin{equation}
    \bm{\hat\Xi} = (\hat{\phi}(f_1),\hat{\phi}(g_1),\hdots,\hat{\phi}(f_N),\hat{\phi}(g_N))^\intercal
\end{equation}
and a real phase space $\mathbb{R}^{2N}$ with symplectic form $\bm\Omega$. Picking canonical coordinates $\xi^\alpha = (q^1,p^1,...,q^N,p^N)$, we can then write
\begin{equation}\label{eq:OmegaSymplecticMatrix}
\Omega_{\alpha\beta} = \bigoplus_{j=1}^N \begin{pmatrix}
0 & -1 \\
1 & 0
\end{pmatrix}.
\end{equation}
The explicit relation between the classical phase space and the collection of canonical pairs $\hat{\bm \Xi}$ is given by the map
\begin{equation}
\hat\Xi(\bm \xi) = \Omega_{\alpha\beta}\xi^\beta\hat\Xi^\alpha,
\end{equation}
which translates the canonical commutation relations in phase space to those in the sub-algebra of operators:
\begin{equation}
    [\hat{\Xi}^\alpha,\hat{\Xi}^\beta] = \ii \Omega^{\alpha\beta} \openone,
\end{equation}
where $\Omega^{\alpha\beta}$ denotes the components of the inverse symplectic matrix. 

In the phase space formalism, the expected values of operators that are exclusively a function of $\hat{\phi}(f_i)$, $\hat{\phi}(g_i)$ can be determined by the Wigner function associated with the state. In other words, given a state $\omega$, the degrees of freedom associated to the modes $\hat{\phi}(f_i)$, $\hat{\phi}(g_i)$ can be fully described in terms of the Wigner function:
\begin{equation}
W_{\omega}(\bm\xi) = \frac{1}{(2\pi)^{2N}}\int \dd^{2N}\bm\xi'\,e^{\ii\Omega(\bm\xi,\bm\xi')} \big\langle e^{-\ii\hat\Xi(\bm\xi')} \big\rangle_\omega.
\end{equation}
The Wigner function defines a quasi-probability that can be used to compute expected values of operators that are a function of the smeared fields $\bm \Xi$. Indeed, if $\hat{A}\in\mathcal{A}(\mathcal{O})$ is any Weyl-symmetric polynomial in $\hat{\phi}(f_i)$ and $\hat{\phi}(g_i)$, $\hat{A} = A(\hat{\bm \Xi})$, the expected value of $\hat{A}$ in $\omega$ can be computed through the integral
\begin{equation}
    \omega(\hat{A}) = \int \dd^{2N} \bm \xi\, W_\omega(\bm \xi) A(\bm \xi).   
\end{equation}

This formulation becomes increasingly simplified in the case where $\omega$ is a quasifree state\footnote{This formulation can also be applied to any Gaussian state even if it has non-zero one-point function with minor changes.}, so that its Wigner function becomes a Gaussian of the form
\begin{equation}
    W(\bm \xi) = \frac{1}{\pi^N\sqrt{\det\bm \sigma}} e^{- \xi^\alpha (\bm \sigma^{-1})_{\alpha \beta} \xi_\beta},
\end{equation}
where $\bm \sigma^{-1}$ denotes the inverse of the matrix $\bm \sigma$, with components
\begin{equation}
    \sigma^{\alpha\beta} = \omega(\{\hat{\Xi}^\alpha,\hat{\Xi}^\beta\}).
\end{equation}
The covariance matrix $\bm \sigma$ then satisfies $\bm \sigma \geq \ii \bm \Omega^{-1}$ and contains all information about the quasifree state $\omega$. This formulation then reduces the problem of handling the degrees of freedom associated to the modes $\hat{\phi}(f_i)$, $\hat{\phi}(g_i)$ to a system described by Gaussian quantum mechanics.

However, it can be challenging to find functions $f_i, g_i$ that satisfy the conditions $E(f_i,g_j) = \delta_{ij}$ and $E(f_i,f_j) = E(g_i,g_j) = 0$. This task becomes simpler when one considers an equivalent representation of the quantum field theory in terms of algebras of smeared field and momentum operators along a Cauchy surface. Indeed, an alternative formulation of the quantum field theory of a scalar field can be given in terms of the association of functions $F,G\in C_0^\infty(\Sigma)$ to operators
$\hat{\Phi}(F)$ and $\hat{\Pi}(G)$
where
\begin{equation}
    [\hat{\Phi}(F),\hat{\Pi}(G)] = \ii \int \dd \Sigma F(\bm x)G(\bm x), \quad\quad [\hat{\Phi}(F),\hat{\Phi}(G)] = [\hat{\Pi}(F),\hat{\Pi}(G)] = 0,
\end{equation}
in analogy with the Poisson bracket in~\eqref{eq:poissonPhiPi}. This association, together with the commutation relations above, give rise to the $\ast$-algebra $\mathcal{A}_\Sigma(\Sigma)$, generated by complex linear combinations and products of the operators $\hat{\Phi}(F)$ and $\hat{\Pi}(G)$ and the identity $\openone$. The operators $\hat{\Phi}(F)$ and $\hat{\Pi}(G)$ can then be interpreted as smeared field and momentum operators
\begin{equation}
    \hat{\Phi}(F) = \int \dd \Sigma \hat{\Phi}(\bm x) F(\bm x), \quad \quad \hat{\Pi}(G) = \int \dd \Sigma \hat{\Pi}(\bm x) G(\bm x).
\end{equation}
The algebra $\mathcal{A}_\Sigma(\Sigma)$ turns out to be analogous to the algebra $\mathcal{A}(\M)$, being able to represent any operator in the quantum field theory. We can also define local algebras associated to subsets of the Cauchy surface $\Sigma$: if $\Sigma_\tc{a}\subset \Sigma$, we define the local algebra $\mathcal{A}_\Sigma(\Sigma_\tc{a})$ as the algebra generated by the elements $\openone$, $\hat{\Phi}(F)$, $\hat{\Pi}(G)$ with $F,G\in C_0^\infty(\Sigma_\tc{a})$

The equivalence between this Cauchy slice formulation and the algebras of operators constructed from functions $f\in C_0^\infty(\M)$ is established by Eq.~\eqref{eq:phisymplecticEclass}, extended to operator-valued distributions:
\begin{equation}\label{eq:phiPhiPi}
    \hat{\phi}(f) = \Omega(\hat{\phi},Ef) = \int \dd \Sigma (Ef\, n^\mu \nabla_\mu \hat{\phi} - \hat{\phi}\, n^\mu \nabla_\mu Ef) = - \hat{\Phi}(G) + \hat{\Pi}(F),
\end{equation}
where $F = Ef|_\Sigma, G = n^\mu \nabla_\mu Ef|_\Sigma$. Notice that the canonical commutation relations are then consistent, in the sense that any functions $f,g\in C_0^\infty(\M)$ define $F = Ef|_\Sigma, G = n^\mu \nabla_\mu Ef|_\Sigma$ and $F' = Eg|_\Sigma, G' = n^\mu \nabla_\mu Eg|_\Sigma$, so that
\begin{align}
    [\hat{\phi}(f),\hat{\phi}(g)] &= [-\hat{\Phi}(G) + \hat{\Pi}(F), - \hat{\Phi}(G') + \hat{\Pi}(F')]= -[\hat{\Pi}(F),\hat{\Phi}(G')] - [\hat{\Phi}(G),\hat{\Pi}(F')]\nonumber\\
    &= \ii \int \dd \Sigma (F(\bm x) G'(\bm x) - F'(\bm x)G(\bm x)) = \ii \Omega(Ef,Eg) = \ii E(f,g).
\end{align}
Thus, the local algebra $\mathcal{A}_\Sigma(\Sigma_\tc{a})$ is equivalent to the algebra $\mathcal{A}(D(\Sigma_\tc{a}))$, associated to the domain of dependence of $\Sigma_\tc{a}$.

Using the Cauchy surface formulation, finding canonical pairs $\hat{\Phi}(F)$ and $\hat{\Phi}(G)$ reduces to the task of finding functions $F_i$, $G_i$ in $C_0^\infty(\M)$ such that
\begin{equation}
    \int \dd \Sigma F_i(\bm x) G_j(\bm x) = \delta_{ij},
\end{equation}
or simply finding functions $F_i$ such that
\begin{equation}\label{eq:normFiFj}
    \int \dd \Sigma F_i(\bm x) F_j(\bm x) = \delta_{ij},
\end{equation}
in which case the set $\hat{\Phi}(F_i)$, $\hat{\Pi}(F_i)$ will be canonical pairs. Each of these pairs can then be mapped into covariantly smeared field observables $\hat{\phi}(f_i)$ and $\hat{\phi}(g_i)$ with $f_i$ and $g_i$ satisfying $Ef_i|_\Sigma = 0$, $n^\mu \nabla_\mu Ef_i|_\Sigma = -F_i$ and $Eg_i|_\Sigma = F_i$, $n^\mu \nabla_\mu Eg_i|_\Sigma = 0$.


\section{The Hadamard Condition}\label{sec:Hadamard}

Although we discussed that, in general, there is no unique notion of vacuum in quantum field theory, there is an argument for why one should consider a privileged state. The argument is related to expected values of operators such as $\hat{\phi}^2(f)$ in Eq.~\eqref{eq:phi2f}. As an example, let us explicitly attempt to compute the expected value of $\hat{\phi}^2(f)$ in the Minkowski vacuum. Using~\eqref{eq:phi2f} we can could write
\begin{equation}
    \omega_0(\hat{\phi}^2(f)) = \int \dd V \dd V' W_0(\mf x, \mf x')f(\mf x)\delta(\mf x,\mf x').
\end{equation}
However, we can see from Eq.~\eqref{eq:W0} that $W_0(\mf x, \mf x')$ is divergent as $\mf x'\to\mf x$, yielding a divergent result for this expected value. This is very unfortunate, given that operators of this form have important physical significance in quantum field theory. For instance, the classical Hamiltonian density for a Klein-Gordon field in inertial coordinates $(t,\bm x)$ is
\begin{equation}
    \mathcal{H}(\mf x) = \frac{1}{2}(\partial_t \phi(\mf x))^2 + \frac{1}{2}(\nabla\phi(\mf x))^2,
\end{equation}
where $\nabla$ denotes the spatial gradient along the surfaces of $t = \text{const}$. One can then write the Hamiltonian density operator in the context of quantum field theory as the operator-valued distribution with kernel
\begin{equation}
    \hat{\mathcal{H}}(\mf x) = \frac{1}{2}(\partial_t \hat{\phi}(\mf x))^2 + \frac{1}{2}(\nabla\hat{\phi}(\mf x))^2.
\end{equation}
Expected values of the form $\omega_0(\hat{\mathcal{H}}(f))$ then involve the coincidence limits of derivatives of the Wightman function~\eqref{eq:W0}, which are also divergent.

One way of handling expected values of singular operators such as $\hat{\phi}^2(\mf x)$ is by choosing a reference state, say $\omega_0$, and by instead defining the formal operator
\begin{equation}\label{eq:normordphi2f}
    \normord{\hat{\phi}^2(f)} \equiv \hat{\phi^2}(f)-\omega_0(\hat{\phi}^2(f))\openone \quad\Rightarrow\quad \omega(\normord{\hat{\phi}^2(f)}) = \omega(\hat{\phi^2}(f))-\omega_0(\hat{\phi}^2(f)).
\end{equation}
The expressions above are divergent and require explanation. If the state $\omega$ has Wightman function $W(\mf x, \mf x')$, it might be the case that $w(\mf x, \mf x') \coloneqq W(\mf x,\mf x') - W_0(\mf x, \mf x')$ is regular in the limit $\mf x'\to \mf x$. If this is the case, Eq.~\eqref{eq:normordphi2f} should be understood as
\begin{equation}
    \omega(\normord{\hat{\phi}^2(f)}) = \int \dd V w(\mf x, \mf x)f(\mf x).
\end{equation}
The operator $\normord{\hat{\phi}^2(f)}$ is called the normal ordered squared field operator. However, at this stage it is not entirely clear when (or if) the difference of Wightman functions, $w(\mf x, \mf x')$ above would be finite. However, if $\omega$ can be represented in the GNS representation of the Minkowski vacuum $\omega_0$, one can indeed show that the divergent part of $W(\mf x, \mf x')$ and $W_0(\mf x, \mf x')$ cancel in the coincidence limit of $\mf x' \to \mf x$\footnote{This is a consequence of the fact that states in $\mathcal{F}(\mathscr{H}_0)$ can be written as finite applications of the smeared creation and annihilation operators $\hat{a}^\dagger(f)$, $\hat{a}(f)$ for $f\in C_0^\infty(\M)$.}. However, as we currently stand, Eq.~\eqref{eq:normordphi2f} does not provide a well-defined value for the expectation value of $\hat{\phi}^2(f)$ in a general state. 

The solution to this issue is to impose an additional condition that must be satisfied by any state in the theory, the so-called Hadamard condition, which we state below.

\noindent\textbf{Hadamard Condition~\cite{kayWald}:} \textit{Let $t$ be a positively oriented timelike coordinate. For any $\mf x, \mf x'$ contained in a convex normal neighbourhood, the Wightman function for a Hadamard state can be written as}
\begin{equation}\label{eq:Hadamard}
    W(\mf x, \mf x') = \lim_{\epsilon \to 0^+}\frac{D^{1/2}(\mf x, \mf x')}{8 \pi^2 \sigma_{\epsilon}(\mf x, \mf x')} + v(\mf x, \mf x') \log(\sigma_\epsilon(\mf x, \mf x')/\ell^2) + w(\mf x, \mf x'),
\end{equation}
\textit{where $\sigma_\epsilon=  \sigma(\mf x, \mf x') +2 \ii \epsilon(t(\mf x) - t(\mf x')) + \epsilon^2$ is the regularized Synge's world function, $D(\mf x, \mf x') = \det(- \sigma_{\mu\nu})$ is the van Vleck-Morette determinant, $\ell$ is a parameter with units of length, and we use the convention that the branch cut for the logarithm lies along the negative real axis. Additionally, the functions $v(\mf x, \mf x')$ and $w(\mf x, \mf x')$ admit an expansion of the form}
\begin{equation}\label{eq:regvw}
    v(\mf x, \mf x') = \sum_{n=0}^\infty v_n(\mf x, \mf x') \sigma(\mf x, \mf x')^n, \quad\quad w(\mf x, \mf x') = \sum_{n=0}^\infty w_n(\mf x, \mf x') \sigma(\mf x, \mf x')^n,
\end{equation}
\textit{with all coefficients being regular in the limit $\mf x' \to \mf x$.}

\noindent The Hadamard condition essentially imposes a specific type of singularity structure for the Wightman function. Imposing that all states in the theory satisfy the Hadamard condition automatically implies that the difference between two Wightman functions is well defined in the limit $\mf x' \to \mf x$. As a matter of fact, it can be shown that the difference $W(\mf x, \mf x') - \tilde{W}(\mf x,\mf x')$ is a smooth function. It is then clear that definition~\eqref{eq:normordphi2f} can be extended to more general operators if we restrict ourselves to Hadamard states.

The Hadamard condition for the Wightman function is essentially inspired by the behaviour of the Minkowski vacuum Wightman function. Indeed, for a massive scalar field of mass $m$, one can write its Wightman function $W_m(\mf x, \mf x')$ as
\begin{equation}\label{eq:Wm}
    W_m(\mf x, \mf x') = \lim_{\epsilon\to 0^+} \frac{1}{8 \pi^2 \sigma_\epsilon(\mf x, \mf x')} + \frac{m^2}{8 \pi^2} \frac{I_1(m \sqrt{2 \sigma(\mf x, \mf x')})}{m \sqrt{2\sigma(\mf x, \mf x')}}\log(2m^2 \sigma_\epsilon(\mf x, \mf x')),
\end{equation}
where $I_1(z)$ denotes the modified Bessel function of the first kind for $\Re(z)>0$, and we consider the analytical extension of $I_1(\sqrt{z})/\sqrt{z}$. In Minkowski spacetime we can also write $\sigma_\epsilon(\mf x, \mf x') = -(t-t'-\ii \epsilon)^2 + (\bm x - \bm x')^2$ in inertial coordinates. Notice that the leading order divergence is the same as that of a massless field~\eqref{eq:W0}. The logarithm part gives both imaginary and real contributions, with the imaginary term corresponding to the modifications to the Green's propagators, making them non-zero for timelike separated events, and its real part adding the corresponding term to the Hadamard distribution. In curved spacetimes, the effect of the logarithm term in Eq.~\eqref{eq:Hadamard} is similar, resulting in violations of the Hyugen's principle (timelike propagation of massless fields\footnote{A scalar field is said to satisfy the strong Hyugen's principle if its retarded and advanced Green's functions $G_R(\mf x, \mf x')$ and $G_A(\mf x, \mf x')$ are only non-zero when $\mf x$ and $\mf x'$ are connected by null paths~\cite{RayHyugens,Hyugens1,Huygens2}.}) in more general spacetimes.

One could be tempted to interpret the term $w(\mf x, \mf x')$ in~\eqref{eq:Hadamard} as \textit{the} state dependent term. However, the parameter $\ell^2$ in the Hadamard condition is arbitrary, so upon a change $\ell \mapsto \tilde{\ell}$, we can rewrite the Hadamard condition as
\begin{equation}
    W(\mf x, \mf x') = \lim_{\epsilon \to 0^+}\frac{D^{1/2}(\mf x, \mf x')}{8 \pi^2 \sigma_{\epsilon}(\mf x, \mf x')} + v(\mf x, \mf x') \log(\sigma_\epsilon(\mf x, \mf x')/\tilde{\ell}^2) + \tilde{w}(\mf x, \mf x'),
\end{equation}
where
\begin{equation}
    \tilde{w}(\mf x, \mf x') = v(\mf x, \mf x') \text{log}(\tilde{\ell}^2/\ell^2) + w(\mf x, \mf x'),
\end{equation}
which also satisfies the condition~\eqref{eq:regvw}. Thus, the $w(\mf x, \mf x')$ term cannot be directly interpreted as the state dependent term: unless one fixes a specific parameter $\ell$ in the expression~\eqref{eq:Hadamard}, the term $w(\mf x, \mf x')$ is not uniquely defined.

The Hadamard condition is essentially the statement that regardless of the background spacetime, a quantum field has a universal UV behaviour, matching that of quantum fields in Minkowski spacetime. Alternatively, one could say that the Hadamard condition states that locally all quantum fields behave like the Minkowski vacuum. In this sense, the Hadamard condition can be seen as the quantum field theoretical version of the equivalence principle, and holds a deep connection between quantum field theory and gravity, as we will discuss further in Chapter~\ref{chap:geometry}. Moreover, it has been argued that the Hadamard condition is necessary for quantum field theory~\cite{necessityHadamard}.

From this point on, we will assume that every state in a quantum field theory satisfies the Hadamard condition. With this convention, given a reference state $\omega_0$, we define the normal order of a product of two field operators as the formal expression
\begin{equation}\label{eq:normordphi2fgen}
    \normord{\hat{\phi}^2(f)} \equiv \hat{\phi}^2(f) - \omega_0(\hat{\phi}^2(f))\openone,
\end{equation}
with the understanding that its expectation value evaluated at a given state $\omega$ is defined by
\begin{equation}
    \omega(\normord{\hat{\phi}^2(f)}) = \int \dd V \dd V'\big(W(\mf x, \mf x') - W_0(\mf x, \mf x')\big)f(\mf x)\delta(\mf x, \mf x'),
\end{equation}
where $W(\mf x, \mf x')$ is the Wightman function of $\omega$ and $W_0(\mf x, \mf x')$ is the Wightman function of the reference state $\omega_0$. The expression above is well defined due to the singularities of the Wightman functions coinciding, and the difference $W(\mf x, \mf x') - W_0(\mf x, \mf x')$ being a smooth function. One can also generalize the normal ordering for more general products of field operators, but we will not require these throughout the thesis, so we refer the interested reader to~\cite{TheBookHadamard}. 

\subsubsection*{Normal Ordering in GNS Representations}

The normal ordering simplifies significantly when we consider a GNS representation associated to a positive mode decomposition $\{u_{\bm k}\}_{\bm k}$ and use the associated state $\ket{\Omega}$ as the reference state. In this case, every operator in the GNS representation can be written in terms of creation and annihilation operators~\eqref{eq:CCRakakd} $\hat{a}_{\bm k}$ and $\hat{a}^\dagger_{\bm k}$. The normal ordering of a product of $n$ creation operators, $\hat{a}^\dagger_{\bm k_1}$,..., $\hat{a}^\dagger_{\bm k_n}$ and $m$ smeared annihilation operators $\hat{a}_{\bm k_1'}$,...,$\hat{a}_{\bm k_m'}$ (in any particular order) then gives $\hat{a}^\dagger_{\bm k_1}\cdots\hat{a}^\dagger_{\bm k_n}\hat{a}_{\bm k_1'}\cdots\hat{a}_{\bm k_m'}$. Extending the action of the normal ordering by linearity we then obtain the general expression for its action in this representation. For instance,
\begin{align}
    \normord{\hat{\phi}(\mf x)^2} \,\,&= \sumint_{\bm k, \bm k'} \!\normord{\left(u_{\bm k}(\mf x)u_{\bm k'}(\mf x)\hat{a}_{\bm k} \hat{a}_{\bm k'} + u_{\bm k}^*(\mf x)u_{\bm k'}(\mf x)\hat{a}^\dagger_{\bm k} \hat{a}_{\bm k'}+ u_{\bm k}(\mf x)u_{\bm k'}^*(\mf x)\hat{a}_{\bm k} \hat{a}^\dagger_{\bm k'}+u_{\bm k}^*(\mf x)u_{\bm k'}^*(\mf x)\hat{a}^\dagger_{\bm k} \hat{a}^\dagger_{\bm k'}\right)}\nonumber\\
    &= \sumint_{\bm k, \bm k'} \left(u_{\bm k}(\mf x)u_{\bm k'}(\mf x)\hat{a}_{\bm k} \hat{a}_{\bm k'} + u_{\bm k}^*(\mf x)u_{\bm k'}(\mf x)\hat{a}^\dagger_{\bm k} \hat{a}_{\bm k'}+ u_{\bm k}(\mf x)u_{\bm k'}^*(\mf x)\hat{a}^\dagger_{\bm k'}\hat{a}_{\bm k} +u_{\bm k}^*(\mf x)u_{\bm k'}^*(\mf x)\hat{a}^\dagger_{\bm k} \hat{a}^\dagger_{\bm k'}\right)\nonumber\\
    &= \hat{\phi}^2(\mf x) - \sumint_{\bm k} u_{\bm k}(\mf x)u_{\bm k}^*(\mf x),
\end{align}
where in the first equality, only the order of the third term was changed, and we used the commutation relations in the last equality to recover the definition~\eqref{eq:normordphi2fgen} by noticing the formal equality $\bra{0}\hat{\phi}(\mf x)^2\ket{0} = W(\mf x, \mf x)$ and~\eqref{eq:Wukuks}. This form of the normal ordering is indeed the standard definition of the normal ordering operation presented in most introductory quantum field theory textbooks (e.g.~\cite{Peskin}).

For instance, in the GNS representation of the Minkowski vacuum, we can compute the normal ordered Hamiltonian by integrating the normal ordered Hamiltonian density along a surface $t = \text{const}.$ For a free massive field of mass $m$, this gives
\begin{align}
    \normord{\hat{H}(t)} &= \frac{1}{2}\int \dd^3\bm x \normord{\left((\partial_t\hat{\phi}(\mf x))^2 + (\nabla\hat{\phi}(\mf x))^2 + m^2 \hat{\phi}(\mf x)^2\right)}\\
    &= \frac{1}{2}\int\dd^3 \bm k\,\omega_{\bm k} \normord{\left(\hat{a}_{\bm k} \hat{a}_{\bm k}^\dagger + \hat{a}_{\bm k}^\dagger\hat{a}_{\bm k}\right)} \\
    &= \int\dd^3 \bm k\,\omega_{\bm k}\,\hat{a}_{\bm k}^\dagger\hat{a}_{\bm k},\label{eq:normordH}
\end{align}
where $\omega_{\bm k} = \sqrt{\bm k^2 + m^2}$ and we used the integral representation of the Dirac delta, as well as the canonical commutation relations~\eqref{eq:CCRakakd}. Also notice that the Hamiltonian, and thus the specific energies associated to each mode, are then dependent on the reference state chosen for the normal ordering. This is perfectly fine for the Hamiltonian, as one does not typically measure the energy content of a system, but rather the difference in energy compared to other states in that system. Thus, using~\eqref{eq:normordH} to compute expected values of the Hamiltonian creates no physical issues. This is the case for many situations where the normal ordering is required. An important exception is when one tries to consider the effect of a quantum system in the gravitational field, where the total energy and momentum are responsible for sourcing gravity. 


\subsubsection*{The Hadamard Condition and General Relativity}

With the Hadamard condition, we can not only define the expected values of the operator-valued distributions $\normord{\hat{\phi}^2(f)}$ but also the expected values of unsmeared operators such as $\normord{\hat{\phi}^2(\mf x)}$. Indeed, given that $W(\mf x,\mf x') - W_0(\mf x, \mf x')$ is a smooth function in spacetime, we have that
\begin{equation}
    \omega(\normord{\hat{\phi}^2(\mf x)}) \coloneqq \lim_{\mf x'  \to \mf x} \left(W(\mf x,\mf x') - W_0(\mf x, \mf x')\right)
\end{equation}
is well defined. The same holds for any derivatives of the field:
\begin{equation}
    \omega\big(\!\!\normord{\nabla_\mu\hat{\phi}(\mf x)\nabla_\nu\hat{\phi}(\mf x)}\!\!\big) \coloneqq \lim_{\mf x'  \to \mf x} \nabla_\mu \nabla_{\nu'}\big( W(\mf x,\mf x') -  W_0(\mf x, \mf x')\big).
\end{equation}

The fact that we can define the expected values of operators defined pointwise has a significant consequence: one can write the expected value of the stress-energy tensor of the field pointwise. Indeed, the classical stress-energy tensor of a scalar field with equation of motion defined by $P = \nabla^\mu \nabla_\mu - V(\mf x)$ can be written as
\begin{equation}
    T_{\mu\nu}(\mf x) = \partial_\mu \phi(\mf x) \partial_\nu \phi(\mf x) - \tfrac{1}{2}\eta_{\mu\nu}\left(\partial_\alpha\phi(\mf x) \partial^\alpha\phi(\mf x) + V(\mf x) \phi(\mf x)\right),    
\end{equation}
so that defining $w(\mf x, \mf x') = W(\mf x, \mf x') - W_0(\mf x, \mf x')$, we can write the expected value of the normal ordered stress-energy tensor of a state $\omega$ as the smooth function
\begin{equation}
    \omega\big(\!\!\normord{\hat{T}_{\mu\nu}(\mf x)}\!\!\big) = \lim_{\mf x' \to \mf x} \left(\big(\partial_\mu \partial_{\nu'}  - \tfrac{1}{2}\eta_{\mu\nu}(\partial_\alpha\partial^{\alpha'} + V(\mf x))\big)w(\mf x, \mf x')\right).
\end{equation}
We can then write the expected value of the source term of the semiclassical Einstein's equations:
\begin{equation}\label{eq:EEsemi}
    G_{\mu\nu} = 8 \pi \ell_p^2 \,\omega\big(\!\!\normord{\hat{T}_{\mu\nu}(\mf x)}\!\!\big),
\end{equation}
where $\ell_p^2 = G$ is the Planck length. In this sense, the Hadamard condition is the key to approaching the problem of how quantum fields source gravity.

It is important to notice that the semiclassical Einstein's equations~\eqref{eq:EEsemi} do depend on a reference state (or, more generally, on a regularization scheme). Essentially, the reference state $\omega_0$ used to compute the normal ordering is postulated as a state that does not gravitate, in the sense that $\omega_0\big(\!\!\normord{\hat{T}_{\mu\nu}(\mf x)}\!\!\big)=0$. One might think that, in some cases, there might be good reasons for postulating that a particular state does not gravitate. For instance, it may seem natural to assume that the Minkowski vacuum does not generate any gravitational field. However, there is no experimental evidence that the vacuum of Minkowski spacetime does not gravitate. As a matter of fact, the large-scale structure of the Universe indicates that whichever quantum state dominates the stress-energy tensor of the Universe in the absence of matter yields an energy tensor with negative pressure. Overall, we have no unique way of prescribing the source of Einstein's equations from quantum field theory, and no clear indication of the regime of validity of semiclassical gravity\footnote{Some authors have argued that a reasonable condition for semiclassical gravity to be approximately valid would be for states such that the fluctuations in $\normord{\hat{T}_{\mu\nu}(\mf x)}$ are significantly smaller than its expected value. However, this condition fails for the reference state itself.}. Arguably, due to the challenges and ambiguities in prescribing the stress-energy tensor within quantum field theory, one could state that

\begin{center}
\textit{``quantum field theory in curved spacetimes does not uniquely prescribe how a quantum field affects gravity; it only prescribes how gravity affects quantum fields.''}
\end{center}


\section{More General Fields}\label{sec:generalQFT}

We will now briefly go over the generalizations of the discussions of this chapter for fields of higher spin and gauge fields. We will keep four main examples in mind, as these will be the most relevant ones for this thesis. The examples will be of a complex scalar field, spin 1/2 fermionic fields, electromagnetism, and linearized quantum gravity. For simplicity, we will only discuss these examples in a Minkowski background.

\subsubsection*{The Quantum Field Theory of a Complex Scalar Field}

We will now briefly mention the algebra construction associated with a complex scalar field. The theory for a complex scalar quantum field can be defined similarly to that of a real scalar field. Indeed, a classical complex field is simply a complex solution $\psi$ to the equation $P\psi = 0$, which can be derived from the extremizing the action associated with the Lagrangian
\begin{equation}
    \mathcal{L} = \nabla_\mu \psi^* \nabla^\mu \psi - V(\mf x) |\psi|^2.
\end{equation}
One can build the local algebras of observables associated to a complex field by assigning each complex function in $C_0^\infty(\M)$ to symbols $\hat{\psi}(f)$ and $\hat{\psi}^\dagger(f)$. The $\ast$-algebra $\mathcal{A}(\M)$ is generated by an identity operator $\openone$ as well as the symbols $\psi(f)$ and $\psi^\dagger(f) \coloneqq \hat{\psi}(f^*)^\dagger$, with the identifications

\noindent \textbf{Linearity:} $\hat{\psi}(\alpha f + \beta g) \sim\alpha\hat{\psi}(f) + \beta \hat{\psi}(g)$.

\noindent \textbf{Equations of Motion:} $\hat{\psi}(Pf) \sim 0$.

\noindent \textbf{Commutation Relations:} $[\hat{\psi}^\dagger(f),\hat{\psi}(g)] \sim \ii E(f,g)$,

\:\:\:\:\:\:\:\:\:\:\:\:\:\:\:\:\:\:\:\:\:\:\:\:\:\:\:\:\:\:\:\:\:\:\:\:\:\:\:\:\:\:\:\:\:\:\:\:\:\:\:\noindent $[\psi^\dagger(f),\psi^\dagger(g)] \sim 0,$

\:\:\:\:\:\:\:\:\:\:\:\:\:\:\:\:\:\:\:\:\:\:\:\:\:\:\:\:\:\:\:\:\:\:\:\:\:\:\:\:\:\:\:\:\:\:\:\:\:\:\:\noindent $[\psi(f),\psi(g)] \sim 0.$

\noindent Notice that when compared to the algebra of a real scalar field, the Hermiticity condition has been removed, and the commutation relations change. The map $f\mapsto \psi^\dagger(f)$ as defined above is also linear and represents the smeared conjugate field.

A basis of solutions of the Klein-Gordon equation $\{u_{\bm k},u_{\bm k}^*\}$ also defines unique pure quasifree state $\omega$ for a complex field theory, as well as its associated GNS representation. The state $\omega$ is defined by the Wightman function
\begin{equation}
    \omega(\hat{\psi}(f)^\dagger\hat{\psi}(g)) = \sumint_{\bm k} u_{\bm k}(f^*)u_{\bm k}(g).
\end{equation}
The analogous to the expression~\eqref{eq:piomegaphi} for the field $\hat{\psi}(\mf x)$ in this GNS representation is
\begin{equation}
    \hat{\psi}(\mf x) = \sumint_{\bm k} u_{\bm k}(\mf x) \hat{a}_{\bm k} + u_{\bm k}^*(\mf x) \hat{b}_{\bm k}^\dagger,
\end{equation}
where we now have two sets of creation and annihilation operators, satisfying the commutation relations
\begin{align}
    [\hat{a}_{\bm k},\hat{a}^\dagger_{\bm k'}] &= \delta(\bm k, \bm k')\openone, \quad 
    [\hat{a}_{\bm k},\hat{a}_{\bm k'}] = 
    [\hat{a}^\dagger_{\bm k},\hat{a}^\dagger_{\bm k'}] = 0,\\
    [\hat{b}_{\bm k},\hat{b}^\dagger_{\bm k'}] &= \delta(\bm k, \bm k')\openone, \quad \,
    [\hat{b}_{\bm k},\hat{b}_{\bm k'}] = 
    [\hat{b}^\dagger_{\bm k},\hat{b}^\dagger_{\bm k'}] = 0.
\end{align}
The state $\ket{\Omega}$ in the GNS representation of $\omega$ is then the unique state such that $\hat{a}_{\bm k}\ket{\Omega} = 0$ and $\hat{b}_{\bm k} \ket{\Omega} = 0$ for all $\bm k$.

In the context of a complex field, the operators $\hat{a}_{\bm k}$ and $\hat{a}_{\bm k}^\dagger$ are associated with creation and annihilation of particles, while the operators $\hat{b}_{\bm k}$ and $\hat{b}_{\bm k}^\dagger$ are associated with creation and annihilation of antiparticles.

\subsubsection*{The Quantum Field Theory of a Dirac Spinor}

In 3+1 Minkowski spacetime, a Dirac spinor $\psi(\mathsf{x})$ can be represented as a four-component complex field $\psi^a$, with $a \in \{1,\,2,\,3,\,4\}$, with a representation of $\text{SL}_2(\mathbb{C})$ (the universal cover of the Lorentz group SO(1,3)). A Lorentz transformation $\Lambda$ acts on spinors according to
\begin{equation}
    \psi'{}^a(\mf x) = S[\Lambda]^a_{\,\,b}\psi^b(\Lambda^{-1}\mf x) \,,
\end{equation}
where $S[\Lambda]$ is given by $S[\Lambda] = \exp(\frac{1}{2}\omega_{\mu\nu}S^{\mu\nu})$ and $S^{\mu\nu}$ are the generators of the $\text{SL}_2(\mathbb{C})$ action. Notice that for any values of $\mu,\nu\in\{0,1,2,3\}$, each of these generators is an operator in spinor space. They can be conveniently expressed in terms of the so-called gamma matrices, defined by the Clifford algebra relation
\begin{equation}
    \gamma^\mu\gamma^\nu + \gamma^\nu\gamma^\mu = -2\eta^{\mu\nu} \, ,
\end{equation}
where $\eta_{\mu\nu} = \text{diag}(-1,1,1,1)$ is the Minkowski metric in diagonal form. The generators $S^{\mu\nu}$ can then be written as \mbox{$S^{\mu\nu} = \frac{1}{4}[\gamma^\mu, \gamma^\nu]$}. In the Dirac representation the $\gamma$-matrices can be written as,
\begin{equation}\label{gammamatrices} 
\gamma^0=\begin{pmatrix}
\mathds{1}_2 & \\
& -\mathds{1}_2
\end{pmatrix}, \quad 
\gamma^i=\begin{pmatrix}
  & \sigma^i \\
-\sigma^i & 
\end{pmatrix} \, ,
\end{equation}
where $\bm \sigma = (\sigma^1,\sigma^2,\sigma^3)$ denote the Pauli matrices and $\openone_2$ is the $2\times2$ identity matrix. 

The operator $\gamma^0$ also defines a natural conjugation operation that faithfully maps spinors into their dual:
\begin{equation}
    \psi(\mf x) \mapsto \bar{\psi}(\mf x) \coloneqq \psi^\dagger(\mf x) \gamma^0.
\end{equation}
As such, every element of the spinor cotangent bundle can be uniquely written as $\bar{\psi}(\mf x)$ for its corresponding spinor $\psi(\mf x)$. The action of a co-spinor $\bar{\psi}(\mf x)$ in a spinor $\phi(\mf x)$ is denoted simply as $\bar{\psi}(\mf x)\phi(\mf x) = \psi^\dagger(\mf x)\gamma^0\phi(\mf x)$. Alternatively, we can interpret a co-spinor field as a distribution, acting on compactly supported spinor fields according to
\begin{equation}
    \bar{\psi}(\phi) \coloneqq \int \dd V \bar{\psi}(\mf x) \phi(\mf x).
\end{equation}

The dynamics of a free spinor field  $\psi(\mf x)$ of mass $m_e$ is obtained by extremizing the action associated with the \textit{real}\footnote{It is essential to have a real Lagrangian density so that operators derived from it, such as the stress-energy tensor and Hamiltonian are also real, and give rise to Hermitian operators upon quantization.} Lagrangian density
\begin{equation}
    \mathscr{L} = \frac{1}{2}\bar{\psi}(\ii\partial_\mu - m_e)\psi - \frac{1}{2} (\ii\partial_\mu\bar{\psi}\gamma^\mu + \bar{\psi}m_e)\psi,
\end{equation}
written in inertial coordinates $(t,\bm x)$. The equations of motion resulting from variation of the Lagrangian with respect to $\bar{\psi}$ and $\psi$ are, respectively,
\begin{align}
    P\psi \coloneqq (\ii\gamma^\mu\partial_\mu - m_e)\psi = 0,\\
    \bar{P}\bar{\psi} \coloneqq -(\ii\partial_\mu\bar{\psi}\gamma^\mu + \bar{\psi}m_e) = 0.
\end{align}
Denote the kernel of $P$ and $\bar{P}$ by $\mathcal{S}$ and $\bar{\mathcal{S}}$, respectively, corresponding to the space of spinor and co-spinor solutions, respectively. The conjugation operation also establishes a connection between $P$ and $\bar{P}$: $P \psi = 0 \Leftrightarrow \bar{P}\bar{\psi} = 0$, mapping $\mathcal{S}$ to $\bar{\mathcal{S}}$.

The operators $P$ and $\bar{P}$ also define unique retarded and advanced propagators, $G_R$, $G_A$, $\bar{G}_R$ and $\bar{G}_A$, satisfying
\begin{equation}
    PG_R f = P G_A f = f, \quad \quad \bar{P} \bar{G}_R \bar{f} = \bar{P} \bar{G}_A \bar{f} = \bar{f},
\end{equation}
so that ${G}_Rf$ is supported in the causal future of $\text{supp}(f)$ and ${G}_Af$ is supported in its causal past. From the retarded and advanced Green's functions, we can define the causal propagators
\begin{equation}
    E = G_R - G_A, \quad \quad \bar{E} = \bar{G}_R - \bar{G}_A,
\end{equation}
related by $\overline{E\psi} = \bar{E} \bar{\psi}$. Same as in the scalar case, the causal propagator is antisymmetric, in the sense that
\begin{equation}
    \bar{\psi}(E\phi) = - \bar{\phi}(E\psi).
\end{equation}

Given any compactly supported spinor $f$, $Ef$ is a solution to the homogeneous equation of motion. Moreover, all solutions with compactly supported initial conditions can be written as $Ef$ for some compactly supported spinor $f$. The analogous statements hold for co-spinors. At the same time, if $S_0^\infty$ denotes the set of compactly supported smooth spinors (and $\bar{S}_0^\infty$ the co-spinor analogue), the space of solutions to the Dirac equation is in one to one correspondence with the space $S_0^\infty/PS_0^\infty$, and $\bar{S}_0^\infty/\bar{P}\bar{S}_0^\infty$, while parametrizes all solutions of $\bar{P}\bar{\psi} = 0$.

Within the space of solutions of the Dirac equation, we can define a positive-definite inner product. Given a Cauchy surface $\Sigma$, we define
\begin{equation}\label{eq:spinorinnprod}
    \langle \phi,\psi\rangle \coloneqq \int \dd \Sigma_\mu \bar{\phi}\gamma^\mu \psi,
\end{equation}
and the fact that $\partial_\mu(\bar{\phi}\gamma^\mu \psi) = 0$ if $\phi$ and $\psi$ are solutions of the Dirac equation implies that the inner product~\eqref{eq:spinorinnprod} is independent of the choice of Cauchy surface. Importantly, unlike we had with the Klein-Gordon equation, the inner product defined in spinor space is \textit{positive}. The Spinor inner product satisfies the analogue of Eq.~\eqref{eq:EOmega} for spinors: 
\begin{equation}
    \bar{\phi}(E \psi) = \ii \langle E \phi,E\psi\rangle = - \bar{\psi}(E\phi).
\end{equation}

A general basis of solutions to the Dirac equation in Minkowski spacetime consists of sets of orthonormal spinors $u_{\bm p,s}(\mf x)$ and $v_{\bm p, s}(\mf x)$ labelled by $\bm p \in \mathbb{R}^3$ and $s = 1,2$ satisfying $\ii \partial_t u_{\bm p,s}(\mf x) = \omega_{\bm p}u_{\bm p,s}(\mf x)$ and $\ii \partial_t v_{\bm p,s}(\mf x) = -\omega_{\bm p}v_{\bm p,s}(\mf x)$, with $\omega_{\bm p} = \sqrt{\bm p^2 + m^2}$. Explicitly, we have
\begin{equation}
    \langle u_{\bm p,s}, u_{\bm p',s'}\rangle = \delta_{ss'} \delta^{(3)}(\bm p - \bm p'), \quad \langle v_{\bm p,s}, v_{\bm p',s'}\rangle = \delta_{ss'} \delta^{(3)}(\bm p - \bm p'), \quad \langle u_{\bm p,s}, v_{\bm p',s'}\rangle = 0.
\end{equation}
A general solution in $\mathcal{S}$ can then be written as
\begin{equation}
    \psi(\mf x) = \sum_{s=1}^2 \int \dd^3 \bm p \left(b_{\bm p,s} u_{\bm p,s}(\mf x) + c_{\bm p,s}^* v_{\bm p,s}(\mf x)\right),    
\end{equation}
where $b_{\bm p,s} = \langle u_{\bm p,s},\psi\rangle$ and $c_{\bm p,s}^* = \langle v_{\bm p,s},\psi\rangle$ are complex coefficients defined by the initial conditions in a Cauchy surface. The explicit expressions for $u_{\bm p,s}(\mf x)$ and $v_{\bm p, s}(\mf x)$, as well as multiple useful properties can be found in standard texts~\cite{Peskin}.

Similarly to the explicit construction for a real scalar field that we saw in Section~\ref{sec:QFT}, one can construct a quantum field theory for a Dirac spinor in terms of associations of compactly supported test functions to elements in an algebra. However, in this case we consider two sorts of associations: $\hat{\psi}$ and $\hat{\bar{\psi}}$, where $\hat{\psi}$ acts on co-spinors $\bar{g}\mapsto \hat{\psi}(\bar{g})$ and $\hat{\bar{\psi}}$ acts on spinors, $f\mapsto \hat{\bar{\psi}}(f)$, satisfying $\hat{\bar{\psi}}(f)^\dagger = \hat{\psi}(\bar{f})$. The algebra of operators is then defined by formal products and sums of operators $\hat{\bar{\psi}}(f)$, $\hat{\psi}(\bar{g})$ and an identity element $\openone$, satisfying the following conditions:

\noindent \textbf{Linearity:} $\hat{\psi}(\alpha \bar{f} + \beta \bar{g}) \sim\alpha\hat{\psi}(\bar{f}) + \beta \hat{\psi}(\bar{g})$,

\:\:\:\:\:\:\:\:\:\:\:\:\:\:\:\:\:\:\noindent $\hat{\bar{\psi}}(\alpha f + \beta g) \sim\alpha\hat{\bar{\psi}}(f) + \beta \hat{\bar{\psi}}(g)$.

\noindent \textbf{Equations of Motion:} $\hat{\bar{\psi}}(Pf) \sim 0$.

\:\:\:\:\:\:\:\:\:\:\:\:\:\:\:\:\:\:\:\:\:\:\:\:\:\:\:\:\:\:\:\:\:\:\:\:\:\:\:\:\:\:\:\noindent  $\hat{\psi}(\bar{P}\bar{f}) \sim 0$.

\noindent \textbf{Anticommutation Relations:} $\{\hat{\psi}(\bar{f}),\hat{\bar{\psi}}(g)\} \sim \ii \bar{f}(E g)\openone$,

\:\:\:\:\:\:\:\:\:\:\:\:\:\:\:\:\:\:\:\:\:\:\:\:\:\:\:\:\:\:\:\:\:\:\:\:\:\:\:\:\:\:\:\:\:\:\:\:\:\:\:\:\:\:\:\:\:\:\:\noindent $\{\hat{\psi}(\bar{f}),\hat{\psi}(\bar{g})\} \sim 0$,

\:\:\:\:\:\:\:\:\:\:\:\:\:\:\:\:\:\:\:\:\:\:\:\:\:\:\:\:\:\:\:\:\:\:\:\:\:\:\:\:\:\:\:\:\:\:\:\:\:\:\:\:\:\:\:\:\:\:\:\noindent $\{\hat{\bar{\psi}}(f),\hat{\bar{\psi}}(g)\} \sim 0$.

\noindent The anticommutation relations are the most distinguishing aspect of the algebra, which gives rise to the Pauli exclusion principle and ensures that self-adjoint operators of the form $\hat{\bar{\psi}}(f)\hat{\psi}(\bar{f})$ are commuting when spacelike separated. In this case one can also think of $\hat{\bar{\psi}}(f)$ and $\hat{\psi}(f)$ as smeared spinor field operators through the formal expressions
\begin{equation}
    \hat{\psi}(\bar{f}) = \int \dd V \bar{f}(\mf x)\hat{\psi}(\mf x), \quad 
    \hat{\psi}(\bar{f}) = \int \dd V \hat{\bar{\psi}}(\mf x) f(\mf x). 
\end{equation}

Finally, a basis $u_{\bm p,s}(\mf x)$ and $v_{\bm p, s}(\mf x)$ induces a GNS representation of a state $\ket{0}$ so that the field can be represented as
\begin{equation}
    \hat{\psi}(\mf x) = \sum_{s=1}^2 \int \dd^3 \bm p \left(\hat{b}_{\bm p,s} u_{\bm p,s}(\mf x) + \hat{c}_{\bm p,s} v_{\bm p,s}(\mf x)\right),    
\end{equation}
where the creation and annihilation operators $\hat{b}_{\bm p,s}, \hat{b}^\dagger_{\bm p,s}, \hat{c}_{\bm p,s}, \hat{c}^\dagger_{\bm p,s}$ satisfy the anticommutation relations
\begin{align}
    \{\hat{b}_{\bm p,s},\hat{b}^\dagger_{\bm p,s'}\} &= \delta_{ss'}\delta^{(3)}(\bm p - \bm p'), \quad \{\hat{c}_{\bm p,s},\hat{c}^\dagger_{\bm p,s'}\} = \delta_{ss'}\delta^{(3)}(\bm p - \bm p'),\\
    \{\hat{b}_{\bm p,s},\hat{b}_{\bm p,s'}\} & =\{\hat{b}_{\bm p,s},\hat{c}_{\bm p,s'}\} =\{\hat{b}_{\bm p,s},\hat{c}^\dagger_{\bm p,s'}\} =\{\hat{b}_{\bm p,s}^\dagger,\hat{c}_{\bm p,s'}\} = 0.
\end{align}
The vacuum associated to this decomposition is then defined by the condition $\hat{b}_{\bm p, s} \ket{0} = \hat{c}_{\bm p,s} \ket{0} = 0$ for all $\bm p$ and $s$. Same as in the scalar case, the operators associated to the positive frequency solutions, $\hat{b}_{\bm p,s}$ and $\hat{b}^\dagger_{\bm p,s}$ create and annihilate particles, while the operators $\hat{c}_{\bm p,s}$ and $\hat{c}^\dagger_{\bm p,s}$ create and annihilate antiparticles.

There is much more than could be said about spinor fields, but we refer the reader to the standard references~\cite{Peskin} for practical information and to~\cite{TheBookGoodChap} for a formal definition of quantum field theories of spinor fields in globally hyperbolic spacetimes. The remaining information regarding spin 1/2 fermions that will be required in this thesis will be presented as necessary.

\subsubsection*{Gauge Quantum Fields}

Gauge field theories are significantly more complex than the theories that we have explored so far. There are also many different approaches to gauge fields, and this review chapter is already too long as is. Our goal here is merely to give an overview of the quantum field theories that we will use later on, and to provide an overview of how to construct gauge-invariant algebras of observables from an algebraic perspective. For simplicity we will stick to Minkowski spacetime with inertial coordinates for this discussion.

We will focus on two theories here: electromagnetism and linearized quantum gravity. Classical electromagnetism is a gauge theory associated to the Lie group $U(1)$ for the four-potential $A_\mu(\mf x)$. The four-potential can be seen as the connection of a principal\footnote{For more about I refer the reader to Frederic Schuller's \href{https://www.youtube.com/playlist?list=PLPH7f_7ZlzxTi6kS4vCmv4ZKm9u8g5yic}{``Lectures on the Geometric Anatomy of Theoretical Physics''}.} $U(1)$ bundle associated to the $U(1)$ group action on complex fields $\psi(\mf x) \mapsto e^{-\ii Q \alpha(\mf x)} \psi(\mf x)$, where $Q$ is its associated charge. The covariant derivative associated with this gauge transformation is $D_\mu \psi = \partial_\mu \psi + \ii Q A_\mu \psi$. 

The dynamics of the electromagnetic potential are determined by the Lagrangian
\begin{equation}\label{eq:LagEM}
    \mathcal{L} = -\frac{1}{4}F^{\mu\nu}F_{\mu\nu}, \quad\quad F_{\mu\nu} = \partial_\mu A_\nu - \partial_\nu A_\mu,
\end{equation}
where $F_{\mu\nu}$ is the electromagnetic tensor, or, equivalently, the curvature in the $U(1)$ bundle. It can also be written in a basis-independent form as $\mf F = \dd \mf A$, which also makes it evident that the Lagrangian is unchanged by $\mf A \mapsto \mf A + \dd \chi$ for any smooth function $\chi(\mf x)$ (usually assumed compactly supported).

The associated equations of motion are
\begin{equation}
    PA^\nu \coloneqq \partial_\mu \partial^\mu A^\nu - \partial^\nu \partial^\mu A_\mu=\partial_\mu F^{\mu\nu} = 0,
\end{equation}
which are also invariant under the gauge transformation $A_\mu \mapsto A_\mu + \partial_\mu \chi$. This unconstrained degree of freedom in the classical equations of motion creates a series of complications when describing the degrees of freedom of the theory, resulting, for instance, in non-unique Green's functions, as well as other ambiguities. One way of approaching this issue is by identifying the gauge-invariant observables in the theory. These can be motivated by looking at the coupling of electromagnetism with a four-current $j^\mu$. 

When a charged four-current $j^\mu$ is present, its coupling to the electromagnetic field can be described at the level of the action by the additional term $-j^\mu A_\mu$ in the Lagrangian~\eqref{eq:LagEM}, resulting in the equation of motion
\begin{equation}
    \partial_\mu F^{\mu\nu} = j^\nu,
\end{equation}
which automatically implies that the four-current must satisfy the conservation equation $\nabla_\mu j^\mu = 0$. This condition then implies gauge independence of contractions of the form
\begin{equation}\label{eq:Amudist}
    A(j) = \int \dd V A_\mu(\mf x) j^\mu(\mf x).
\end{equation}
Indeed, under a gauge transformation $A_\mu \mapsto A_\mu + \partial_\mu \chi$, 
\begin{equation}
    A(j) \mapsto A(j) + \int \dd V \partial_\mu \chi(\mf x) j^\mu(\mf x) = A(j) - \int \dd V \chi(\mf x) \nabla_\mu j^\mu(\mf x) = A(j).
\end{equation}
We then define the set of conserved four-currents $\mathcal{J} = \{j^\mu\in C_0^\infty(\M): \nabla_\mu j^\mu = 0\}$, where we allow $j^\mu$ to be complex for generality. In this context, gauge invariant solutions of the equations of motion can be thought of as elements on the dual $\mathcal{J}^*$, so that a solution $A_\mu(\mf x)$ defines a unique distribution of the form of Eq.~\eqref{eq:Amudist}. In this context we can define unique retarded and advanced Green's functions satisfying $PG_Rj = j$ and $PG_A j = j$, in the sense that $G_Rj$ and $G_Aj$ are the unique elements of $\mathcal{J}^*$ satisfying the non-homogeneous equation of motion supported in the causal past and future of $\text{supp}(j)$, respectively.

From this point on, we can build a quantum theory in the usual manner by assigning symbols $\hat{A}$ that act on conserved currents in $\mathcal{J}$ by $j \mapsto \hat{A}(j)$ and, together with the identity, generate an algebra of observables satisfying the conditions

\noindent \textbf{Linearity:} $\hat{A}(\alpha j + \beta j') \sim \alpha \hat{A}(j) + \beta \hat{A}(j')$,

\noindent \textbf{Hermiticity:} $\hat{A}(j)^\dagger \sim \hat{A}(j^*)$.

\noindent \textbf{Equations of Motion:} $\hat{A}(Pf) \sim 0$.

\noindent \textbf{Commutation Relations:} $[\hat{A}(j),\hat{A}(j')] \sim \ii E(j,j')$,

\noindent where $E = G_R - G_A$ is the retarded propagator, and we allow $f$ to be any compactly support vector field in the condition $\hat{A}(Pf) = 0$. Here we again can think of the quantum field $\hat{A}_\mu(\mf x)$ as an operator-valued distribution, acting on compactly supported conserved currents according to
\begin{equation}
    \hat{A}(j) = \int \dd V \hat{A}_\mu(\mf x) j^\mu(\mf x).
\end{equation}
This construction avoids most complications regarding gauge, allowing states to be defined exactly as in Section~\ref{sec:QFT}, making the gauge invariance of the theory manifest from the construction of the algebra of observables itself.

We can use this same concept to quickly formulate the quantum theory for linearized quantum gravity. One can describe small metric fluctuations in Minkowski background by considering
\begin{equation}\label{eq:gmunuetah}
    g_{\mu\nu} = \eta_{\mu\nu} + \sqrt{8\pi}\ell_p h_{\mu\nu},
\end{equation}
where $\ell_p = \sqrt{G}$ is Planck's constant and $h_{\mu\nu}$ is a symmetric tensor. We choose these conventions for the metric perturbation $h_{\mu\nu}$ so that it has units of energy, the same as scalar and vector fields so that similar expressions to the retarded and advanced Green's functions can be used for its associated equation of motion. The Planck length then also provides a suitable expansion parameter for the metric perturbations.

The action that dictates the dynamics of the metric is the Einstein-Hilbert action, with Lagrangian $\mathcal{L} = \frac{R}{16\pi G}$, where $R$ is the Ricci scalar. Using $g_{\mu\nu}$ as in~\eqref{eq:gmunuetah}, the associated equation of motion (to leading order in $\ell_p$) becomes
\begin{equation}\label{eq:Gmunulin}
    Ph_{\mu\nu} = -\left(\partial_\alpha \partial_{(\mu} \delta^\alpha_{\nu)} - \tfrac{1}{2} \eta^{\alpha\beta}\partial_\mu \partial_\nu - \tfrac{1}{2}\delta^\alpha_\mu \delta^\beta_\nu  \partial_\sigma \partial^\sigma  - \tfrac{1}{2}\eta_{\mu\nu}(\partial^\alpha \partial^\beta - \eta^{\alpha\beta} \partial_\sigma \partial^\sigma) \right) h_{\alpha\beta} = 0.
\end{equation}
Notice that according to the definition above, $Ph_{\mu\nu} = -G_{\mu\nu}$. We choose this convention so that the Green's function for the linearized metric perturbation does not pick up a negative sign with respect to our conventions for a scalar field. As expected, the theory for the linearized metric tensor $h_{\mu\nu}$ is invariant under infinitesimal diffeomorphisms. Concretely, this implies that for any vector field $\xi^\mu$, the equations of motion are invariant under $h_{\mu\nu} \mapsto h_{\mu\nu} + \partial_\mu \xi_\nu + \partial_\nu \xi_\mu$, where $\xi_\mu$ is the vector field that generates the local diffeomorphism. 

Same as in electromagnetism, this gauge invariance imposes a conservation equation for any current that couples to $h_{\mu\nu}$. The coupling of matter with linearized metric perturbations is incorporated at the Lagrangian level by the addition of the term $- \frac{1}{2} \sqrt{8\pi} \ell_p h_{\mu\nu} T^{\mu\nu}$, derived from varying the Einstein-Hilbert action with respect to the metric. The non-homogeneous equations of motion then become
\begin{equation}\label{eq:hmunueom}
    Ph_{\mu\nu} = -\sqrt{8 \pi} \ell_p T_{\mu\nu}.
\end{equation}
The divergenceless condition of the Einstein tensor, $\nabla^\mu G_{\mu\nu} = 0$ then implies that the stress-energy tensor must also satisfy $\nabla^\mu T^{\mu\nu} = 0$. Inspired by the electromagnetic case, we define the space of divergenceless symmetric tensors of rank $(0,2)$, $\mathcal{T} = \{T_{\mu\nu}:\nabla_\mu T^{\mu\nu} = 0\}$, so that if $h_{\mu\nu}$ is a solution of $Ph_{\mu\nu} = 0$ and $T^{\mu\nu}\in \mathcal{T}$, then
\begin{equation}
    h(T) \coloneqq \int \dd V h_{\mu\nu}(\mf x) T^{\mu\nu}(\mf x)
\end{equation}
is gauge-invariant. 

Identifying solutions to the equation of motion with distributions on $\mathcal{T}$, we find retarded and advanced Green's functions $G_R$ and $G_A$, and define the causal propagator
\begin{equation}
    E_{\mu\nu\alpha'\beta'}(\mf x,\mf x') = (G_R)_{\mu\nu\alpha'\beta'}(\mf x, \mf x') - (G_A)_{\mu\nu\alpha'\beta'}(\mf x, \mf x')
\end{equation}
as usual. The explicit expressions for the retarded Green's function can be found in Appendix~\ref{app:retGrav}. We then create the association $T\in\mathcal{T}\mapsto \hat{h}(T)$. The elements $\hat{h}(T)$ and $\openone$ are then used as the generators for an algebra with the following conditions:

\noindent \textbf{Linearity:} $\hat{h}(\alpha T + \beta T') \sim \alpha \hat{h}(T) + \beta \hat{h}(T')$,

\noindent \textbf{Hermiticity:} $\hat{h}(T)^\dagger \sim \hat{h}(T^*)$.

\noindent \textbf{Equations of Motion:} $\hat{h}(Pf) \sim 0$.

\noindent \textbf{Commutation Relations:} $[\hat{h}(T),\hat{h}(T')] \sim \ii E(h,h')$,

\noindent where, same as in the electromagnetic case, we allow $f$ to be any symmetric rank 2 tensor in the condition $h(Pf) = 0$. The operators in the algebra can then be seen as the gauge invariant operators
\begin{equation}
    \hat{h}(T) = \int \dd V \hat{h}_{\mu\nu}(\mf x) T^{\mu\nu}(\mf x).
\end{equation}

%

\chapter{Locally Probing Quantum Fields}\label{chap:meas}

\vspace{-10mm}
{\tiny\textcolor{white}{If you think that this introduction for the main chapter is too short, just read the introduction, there is way more motivation there.}}
\vspace{3mm}

In the previous section, we discussed quantum field theory and defined states, observables, and expected values. However, quoting~\cite{us},
\begin{center}
    \textit{``From the retinas of our eyes to solid state sensors at the LHC, \\we never measure a quantum field other than by coupling something to it.''}   
\end{center}
\noindent The goal of this chapter is to describe how one can access expected values in quantum field theory through local probes, giving special focus to the connection between quantum field theoretic and effective descriptions of the probes.

We start with a brief review of the issues with measurements in quantum field theory in Section~\ref{sec:meas}, and then focus on local probes, starting with the description of a localized quantum field probing a free field in Section~\ref{sec:LocalizedQuantumFields}. We then simplify this description to reach the widely used Unruh-DeWitt detector model and explore its properties in Section~\ref{sec:UDW}. Section~\ref{sec:NRLQS} is devoted to connecting descriptions of non-relativistic quantum systems in curved spacetimes to particle detector models. In Section~\ref{sec:MoreRealisticProbes}, we consider more realistic probes localized by physical external potentials and study a quantum field theoretic description of a Hydrogen atom as a localized probe. In Section~\ref{sec:Tmunu}, we will discuss the stress-energy tensor of a localized field and how general covariance requires one to dynamically describe the mechanisms responsible for the localization of the probe. 

\section{Measurements in Quantum Field Theory}\label{sec:meas}

In this Section, we will discuss measurements and operations in quantum field theory, and how these can lead to an incompatibility with relativistic causality. We will also discuss the Fewster-Verch framework, a fully field theoretic description for the process of obtaining information about a quantum field by coupling it to a probe that is also described within quantum field theory.

\subsubsection*{State Updates and Causal Operations in Quantum Field Theory}

An ideal measurement in quantum theory is typically formulated in terms of a set of projectors $\hat{P}_j$ such that $\sum_j \hat{P}_j = \openone$. Each label $j$ then defines a possible outcome of the measurement, and given a pure state $\ket{\psi}$, the probability of outcome $j$ can be computed using the Born rule
\begin{equation}
    p_j = \bra{\psi}\hat{P}_j\ket{\psi}.
\end{equation}
If the outcome $j$ is obtained, one updates the state accordingly by considering $\ket{\psi}\mapsto \alpha \hat{P}_j\ket{\psi}$, where $\alpha^2 = p_j^{-1}$ is a normalization constant. Interestingly, the state update is a non-linear state operation, so that it cannot be derived from the linear dynamics of quantum theory.

This non-linearity persists when more general types of measurements are considered, such as measurements defined by operators $\hat{M}_j$ such that $\sum_j \hat{M}_j^\dagger \hat{M}_j = \openone$. Describing the state of the target system by a density operator $\hat{\rho}$, the probability of each outcome is given by
\begin{equation}
    p_j = \tr(\hat{\rho}\hat{M}_j^\dagger\hat{M}_j),
\end{equation}
and after obtaining outcome $j$ the state is updated to $\hat{\rho}_j = \alpha^2\hat{M}_j\hat{\rho}\hat{M}_j^\dagger$, again with $\alpha^2 = p_j^{-1}$. If the outcome of the experiment was not recorded, the state is updated non-selectively to $\hat{\tilde{\rho}} = \sum_j \hat{M}_j \hat{\rho}\hat{M}_j^\dagger$. Fundamentally, the lack of linearity of the state update associated with a selective measurement is a consequence of the non-deterministic aspect of quantum theory, implying that one is unable to tell which of the possible outcomes will take place until the measurement is performed. It is only after the result of the experiment is learned that one can apply the state update rule, which is not prescribed by the deterministic state evolution in quantum theory. Indeed, the so-called measurement problem in quantum mechanics~\cite{MeasBallentine,MeasZurek,MeasSchlosshauer,MeasGao} is associated with the fact that there are no clear mechanisms within the theory that determine whether a measurement takes place.

In the context of quantum field theory, state updates can be implemented in terms of algebra endomorphisms $\Theta:\mathcal{A}(\M)\rightarrow \mathcal{A}(\M)$, updating the state according to $\omega\mapsto \varpi$ such that $\varpi(\,\cdot\,) = \omega(\Theta(\,\cdot\,))$. The operation $\Theta$ would then correspond to a physical process realized in a region of spacetime. In particular, given operators $\hat{M}_j\in \mathcal{A}(\M)$ such that $\sum_j \hat{M}^\dagger_j \hat{M}_j = \openone$, the operation $\Theta(\hat{A}) = \sum_j\hat{M}_j^\dagger\hat{A}\hat{M}_j$  defines an algebra endomorphism analogous to the state update corresponding to a non-selective update. One could then import the definitions from measurements using density operators, defining a measurement by a set of operators satisfying $\sum_j \hat{M}^\dagger_j \hat{M}_j = \openone$, assigning probability $p_j = \omega(\hat{M}_j^\dagger \hat{M}_j)$ to each outcome, so that the state update after obtaining outcome $j$ would be given by $\omega_j(\,\cdot \,) = p_j^{-1}\,\omega(\hat{M}_j^\dagger\,\cdot\,\hat{M}_j)$. One issue with this reasoning is that states in quantum field theory are global in nature, so that a state update could effectively affect the expected values of observables that are arbitrarily far away from the measurement region.


\begin{figure}[h!]
    \centering
    \includegraphics[width=12cm]{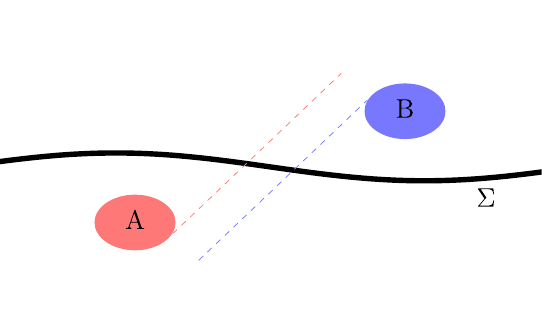}
    \caption{Schematic representation of a spacetime diagram of the setup considered by Sorkin in~\cite{Sorkin}.}
    \label{fig:sorkin}
\end{figure}

The first to notice that state updates in quantum field theory can become a real issue was Rafael Sorkin in~\cite{Sorkin}, when he showed that a projective measurement performed at a Cauchy surface could lead to faster-than-light signalling between two parties. Specifically, Sorkin considered a setup where two spacelike separated localized observers A and B interact with a field and a selective measurement is performed at a Cauchy surface $\Sigma$ in the causal future of A and causal past of B, as shown in Fig.~\ref{fig:sorkin}. He then showed that a selective measurement at $\Sigma$ could make the expected values of observables localized in B depend on the operations that were performed in A. This would imply that it is possible for A to send a signal to B, allowing for spacelike separated observers to communicate. As pointed out in~\cite{Sorkin}, any ``measurement'' that allows for such causality violations to take place must not correspond to a physical process, giving these the label of ``impossible measurements''. The fact that general state updates in quantum field theory may lead to lead to causality violations has become known as the Sorkin problem.

\begin{figure}[h!]
    \centering
    \includegraphics[width=12cm]{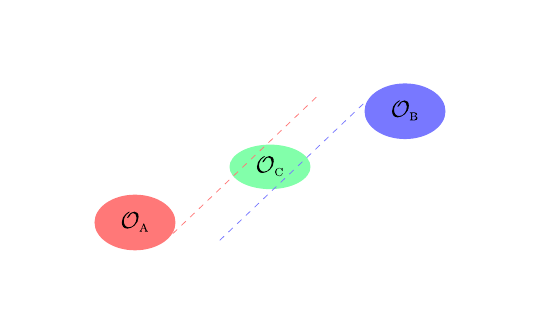}
    \caption{Schematic representation of a spacetime diagram of the setup considered in~\cite{impossible}.}
    \label{fig:impossible}
\end{figure}

It turns out that Sorkin-type problems are not exclusive to projective measurements performed in Cauchy slices. As pointed out in~\cite{impossible}, the situation initially proposed by Sorkin can be generalized to the case where one considers operations in two spacelike separated regions, $\mathcal{O}_\tc{a}$ and $\mathcal{O}_\tc{b}$, and in an additional finite region $\mathcal{O}_\tc{c}$ that is causally connected to both $\mathcal{O}_\tc{a}$ and $\mathcal{O}_\tc{b}$, as illustrated in Fig.~\ref{fig:impossible}. Essentially, some operations performed in $\mathcal{O}_\tc{c}$ might lead to expected values of observables in $\mathcal{O}_\tc{b}$ to be dependent on operations performed in $\mathcal{O}_\tc{a}$. 

For concreteness, we will consider an explicit example of operations in regions $\mathcal{O}_\tc{a}$, $\mathcal{O}_\tc{b}$ and $\mathcal{O}_\tc{c}$ that lead to Sorkin-type problems in the case of a real scalar quantum field theory. Given a self-adjoint operator $\hat{A}\in\mathcal{A}(\M)$, define the endomorphism
\begin{equation}
    \mathcal{U}_{\hat{A}}(\hat{B}) \coloneqq e^{\ii \hat{A}} \hat{B} e^{- \ii \hat{A}}.
\end{equation}
Endomorphisms of this type have been called unitary kicks associated with the observable $\hat{A}$ in~\cite{IanJubb}. One can compute the action of $\mathcal{U}_{\hat{A}}$ on smeared field operators $\hat{\phi}(g)$ through 
\begin{equation}
    \mathcal{U}_{\hat{A}}(\hat{\phi}(g)) = e^{\ii \hat{A}} \hat{\phi}(g) e^{- \ii \hat{A}} = - \ii \partial_s\left.\left(e^{\ii \hat{A}} e^{\ii s \hat{\phi}(g)} e^{- \ii \hat{A}}\right)\right|_{s=0},
\end{equation}
and from the Baker-Hausdorff-Campbell formula, we find that
\begin{equation}\label{eq:BCH}
    e^{\ii \hat{A}} e^{\ii s\hat{B}} e^{-\ii \hat{A}} = e^{\ii s \hat{B}} e^{- s [\hat{A},\hat{B}] - \tfrac{\ii s}{2}[\hat{A},[\hat{A},\hat{B}]] + ...}, 
\end{equation}
where the remaining terms in the exponent of leading order in $s$ involve more commutators of the form $[\hat{A},...,[\hat{A},\hat{B}]...]$. From this expression, we then find the action of operations of the form $\mathcal{U}_{\hat{\phi}(f)^n}$ on field operators by noticing that $[\hat{\phi}(f)^n,\hat{\phi}(g)] = n \ii E(f,g)\hat{\phi}(f)^{n-1}$, so that $[\hat{\phi}(f)^n,[\hat{\phi}(f)^n,\hat{\phi}(g)]] = 0$, and all commutators in~\eqref{eq:BCH} vanish, except for the first. This yields 
\begin{equation}
    \mathcal{U}_{\hat{\phi}(f)^n}(\hat{\phi}(g)) = \hat{\phi}(g) - n E(f,g)\hat{\phi}(f)^{n-1}.
\end{equation}

In particular, for $n=1$ we find that the unitary kick $\mathcal{U}_{\hat{\phi}(f)}$ shifts the operator $\hat{\phi}(g)$ by the classical solution $\phi_f = Ef$:
\begin{equation}
    \mathcal{U}_{\hat{\phi}(f)}(\hat{\phi}(g)) = \hat{\phi}(g) - E(f,g)\openone = \hat{\phi}(g) + \phi_f(g)
    \openone,
\end{equation}
where we obtain the last equality from $-E(f,g) = E(g,f) = g(Ef) = g(\phi_f) = \phi_f(g)$. It turns out that the operators $\mathcal{U}_{\hat{\phi}(f)}$ are causal, as shown in~\cite{IanJubb}. On the other hand,
\begin{equation}
     \mathcal{U}_{\hat{\phi}(f)^2}(\hat{\phi}(g)) = \hat{\phi}(g) - 2E(f,g)\hat{\phi}(f) = \hat{\phi}\big(g - 2 E(f,g) f\big)
\end{equation}
can lead to Sorkin-type problems if applied in the region $\mathcal{O}_\tc{c}$ in Fig.~\ref{fig:impossible}. Indeed, assume that the operation $\Theta_\tc{a} = \mathcal{U}_{\hat{\phi}(f_\tc{a})}$ is performed in region $\mathcal{O}_\tc{a}$ (with $f_\tc{a}\in C_0^\infty(\mathcal{O}_\tc{a})$), resulting in the state $\omega_\tc{a}(\,\cdot\,) = \omega(\Theta_\tc{a}(\,\cdot\,))$ and that the operation $\Theta_\tc{c} = \mathcal{U}_{\hat{\phi}(f_\tc{c})^2}$ is applied to the field at region $\mathcal{O}_\tc{c}$, with $f_\tc{c}\in C_0^\infty(\mathcal{O}_\tc{c})$. The updated state after both interactions, $\varpi$, will be given by $\varpi(\,\cdot\,) = \omega_\tc{a}(\Theta_\tc{c}(\,\cdot\,)) = \omega(\Theta_\tc{a}(\Theta_\tc{c}(\,\cdot\,)))$. The corresponding operation $\Theta_\tc{a}\circ\Theta_\tc{c}$ will then affect smeared field operators according to
\begin{align}
    \Theta_\tc{a}\circ\Theta_\tc{c}\big(\hat{\phi}(g)\big) &= \mathcal{U}_{\hat{\phi}(f_\tc{a})}\left(\mathcal{U}_{\hat{\phi}(f_\tc{c})^2}\big(\hat{\phi}(g)\big)\right) = \mathcal{U}_{\hat{\phi}(f_\tc{a})}\left(\hat{\phi}\big(g - 2 E(f_\tc{c},g) f_\tc{c}\big)\right)\\
    &= \hat{\phi}(g) - 2 E(f_\tc{c},g)\hat{\phi}(f_\tc{c}) - E(f_\tc{a},g)\openone + 2 E(f_\tc{a},f_\tc{c})E(f_\tc{c},g)\openone.
\end{align}

We can now show explicitly that the operation $\Theta_\tc{a}$, local to the region $\mathcal{O}_\tc{a}$ can affect expected values of operators localized in the region $\mathcal{O}_\tc{b}$ due to the operation $\Theta_\tc{c}$, even though $\mathcal{O}_\tc{a}$ and $\mathcal{O}_\tc{b}$ are spacetime separated. Consider $g = f_\tc{b}\in C_0^\infty(\mathcal{O}_\tc{b})$, so that $\hat{\phi}(g) = \hat{\phi}(f_\tc{b})$ is an observable in $\mathcal{O}_\tc{b}$. We then have $E(f_\tc{a},g) = E(f_\tc{a},f_\tc{b}) = 0$ due to $\mathcal{O}_\tc{a}$ and $\mathcal{O}_\tc{b}$ being causally disconnected. However, we still have
\begin{equation}
    \Theta_\tc{a}\circ\Theta_\tc{c}\big(\hat{\phi}(f_\tc{b})\big) = \hat{\phi}(f_\tc{b}) - 2 E(f_\tc{c},f_\tc{b})\hat{\phi}(f_\tc{c}) + 2 E(f_\tc{a},f_\tc{c})E(f_\tc{c},f_\tc{b})\openone.
\end{equation}
For instance, if the initial state $\omega$ is quasifree, we will then have 
\begin{equation}
    \varpi(\hat{\phi}(f_\tc{b})) = 2 E(f_\tc{a},f_\tc{c})E(f_\tc{c},f_\tc{b}),
\end{equation}
which explicitly depends on $f_\tc{a}$. We then see that if there is causal contact between $\mathcal{O}_\tc{a}$ and $\mathcal{O}_\tc{c}$ and between $\mathcal{O}_\tc{b}$ and $\mathcal{O}_\tc{c}$, the expected values of operators localized in B might depend on operations performed in A, even if the regions $\mathcal{O}_\tc{a}$ and $\mathcal{O}_\tc{b}$ are causally disconnected.

This shows that not all operations in the local algebra of observables respect causality, and some operations might lead to Sorkin-type problems. Relevant progress was made in~\cite{IanJubb} regarding the classification of causal operations in quantum field theories for free Klein-Gordon fields. However, classifying the operations that do not lead to causality violations in general quantum field theories is still an ongoing research topic. Overall, one needs to be careful when considering operations and measurements in quantum field theory, given that general local operations can lead to causality violations, implying that these cannot be realized by any physical process.

\subsubsection*{The Fewster-Verch Measurement Framework}

A possible solution to Sorkin-type problem was introduced by Fewster and Verch in a series of manuscripts describing measurements in quantum field theory~\cite{FVOG,FVGenerallyCovariant2020,FVall2023,FV2024}. Essentially, their measurement scheme considers a measurement that is induced by the interaction with another quantum field. The scheme is entirely covariant and fully prevents any Sorkin-type issues. We will now briefly describe the Fewster-Verch framework.

Consider two quantum field theories, for a field $\hat{\phi}$ that one intends to measure, and for a detector field $\hat{\phi}_\tc{d}$ that will be used to perform an effective measurement in the field $\hat{\phi}$. In principle, the field theories associated with $\hat{\phi}$ and $\hat{\phi}_\tc{d}$ are entirely decoupled, and their local algebras of observables factor as a tensor product $\mathcal{A}(\M)\otimes\mathcal{A}_\tc{d}(\M)$. However, the only way for measurements in the field $\hat{\phi}_\tc{d}$ to be able to acquire information about $\hat{\phi}$ is by considering an interaction between the two theories. We then define an, in principle completely different, quantum field theory in which the fields $\hat{\phi}$ and $\hat{\phi}_\tc{d}$ are coupled within a region contained in a compact causal diamond $\mathcal{K} = D(\mathcal{K})$. This theory is associated to an algebra of observables $\mathcal{C}(\M)$, corresponding to the coupled field theory. 

It is important to stress that, in principle, the coupled and uncoupled theories are associated to entirely different algebras. However, given the assumption of a compactly supported interaction, the algebras $\mathcal{C}(\M)$ and $\mathcal{A}(\M)\otimes \mathcal{A}_\tc{d}(\M)$ are isomorphic ``before'' and ``after'' the interaction. More precisely, defining the regions $\M^\pm = \M\backslash J^{\mp}(\mathcal{K})$, it is possible to define $\ast$-algebra isomorphisms $\gamma^\pm:\mathcal{A}(\M^\pm)\otimes \mathcal{A}_\tc{d}(\M^\pm)\rightarrow \mathcal{C}(\M^\pm)$. Indeed, in the regions $\M^\pm$ the algebra associated to the coupled theory is essentially the algebra of an uncoupled theory between $\hat{\phi}$ and $\hat{\phi}_\tc{d}$, with $\mathcal{C}(\M^\pm)$ corresponding to the algebra of observables ``after'' and ``before'' the interaction takes place\footnote{Notice that ``before'' and ``after'' do not correspond to the causal future and causal past of the interaction region, rather ``after'' stands for the complement of the causal past of the interaction region and ``before'' to the complement of its causal future.}. Moreover, due to property \textbf{A4}, we have that $\mathcal{A}(\M^\pm)\otimes \mathcal{A}_\tc{d}(\M^\pm) = \mathcal{A}(\M)\otimes \mathcal{A}_\tc{d}(\M)$ and $\mathcal{C}(\M^\pm) = \mathcal{C}(\M)$, so that the maps $\gamma^\pm$ extend to the entire coupled and uncoupled algebras. 

The isomorphisms $\gamma^\pm$ effectively allow one to work with the uncoupled algebras outside of $\mathcal{K}$. For instance, before the interaction a product state $\omega\otimes\omega_\tc{d}$ of the uncoupled algebra corresponds to a unique state $\varpi$ in $\mathcal{C}(\M)$ satisfying $(\gamma^-)^*\varpi = \omega\otimes \omega_\tc{d}$. Explicitly:
\begin{equation}
    \varpi(\gamma^-(\hat{A}\otimes\hat{B})) = \omega\otimes\omega_\tc{d}(\hat{A}\otimes\hat{B}) = \omega(\hat{A}) \omega_\tc{d}(\hat{B}) \,\,\forall\,\hat{A}\in \mathcal{A}(\M)\text{ and } \hat{B}\in\mathcal{A}_\tc{d},
\end{equation}
or equivalently, $\varpi = ((\gamma^-)^{-1})^*\omega\otimes \omega_\tc{d}$. The state $\varpi$ can then be used to describe the system before and during the interaction. Moreover, $\varpi$ is also a valid state after the interaction takes place, although it cannot be related to $\omega\otimes\omega_\tc{d}$ by the pullback through $(\gamma^-)^{-1}$. Instead, one can use $\varpi$ to compute expected values of any observable in the uncoupled theory after the interaction using $\gamma^+$, which is well defined in $\mathcal{M}^+$. For instance, one can compute the expected value of an observable $\hat{B}\in\mathcal{A}_\tc{d}(\M)$ in the uncoupled theory after the interaction by importing it to the coupled theory using the map $\gamma^+$. Specifically, the operator $\openone\otimes \hat{B}$ in $\mathcal{A}(\M)\otimes \mathcal{A}_\tc{d}(\M)$ is mapped to the operator $\hat{B}_\mathcal{C} = \gamma^+(\openone\otimes \hat{B})\in\mathcal{C}(\M)$. The expected value of $\hat{B}_\mathcal{C}$ in the state that was initially $\omega\otimes\omega_\tc{d}$ and went through the interaction is then given by
\begin{equation}
    \varpi(\hat{B}_\mathcal{C}) = \omega\otimes\omega_\tc{d}((\gamma^-)^{-1}\gamma^+(\openone \otimes \hat{B})).
\end{equation}
This naturally defines the scattering endomorphism
\begin{equation}
    \Theta \coloneqq (\gamma^-)^{-1}\gamma^+:\mathcal{A}(\M)\otimes\mathcal{A}_\tc{d}(\M)\rightarrow\mathcal{A}(\M)\otimes\mathcal{A}_\tc{d}(\M).
\end{equation}
Importantly, $\Theta$ acts trivially in local algebras located in regions that are causally disconnected from $\mathcal{K}$, which implies that this framework only prescribes non-trivial changes in the free theories in the causal future of $\mathcal{K}$. Additionally, any local interaction between quantum fields in line with this description cannot give rise to Sorkin-type problems. For a related discussion about this topic, see~\cite{impossibleImpossible}. The scattering map here is playing the role of the unitary time evolution operator associated with the interaction, described entirely in the uncoupled theory: $\Theta(\hat{A}\otimes\hat{B}) \sim \hat{U}_I^\dagger (\hat{A}\otimes\hat{B})\hat{U}_I$.

One can then define the effective final state of the detector field $\hat{\phi}_\tc{d}$ after the interaction with the field $\hat{\phi}$ (prepared in the initial state $\omega$) as the functional $\varpi_\tc{d}:\mathcal{A}_\tc{d}(\M)\rightarrow \mathbb{C}$ with action defined by
\begin{equation}\label{eq:FVuppid}
    \varpi_\tc{d}(\hat{B}) = \omega\otimes\omega_\tc{d}(\Theta(\openone \otimes \hat{B})) = \varpi(\hat{B}_\mathcal{C}).
\end{equation}
Notice that the final state of the detector explicitly depends on the initial state of the target field, $\omega$. Consequently, the expected value of $\hat{B}$ in the final state of the detector will contain information about observables and expected values of operators of the field $\hat{\phi}$. In fact, one can find the precise observable in $\mathcal{A}(\M)$ whose expectation value matches $\varpi_\tc{d}(\hat{B})$. 

We can extend the action of states in $\mathcal{A}_\tc{d}(\M)$ and in $\mathcal{A}(\M)$ to the full uncoupled algebra $\mathcal{A}(\M)\otimes\mathcal{A}_\tc{d}(\M)$ by defining their action in operators of the form $\hat{A}\otimes \hat{B}$:
\begin{align}
    \omega_{\tc{d}}(\hat{A}\otimes\hat{B}) &\coloneqq \omega_\tc{d}(\hat{B})\hat{A},\label{eq:omegadAB}\\
    \omega(\hat{A}\otimes\hat{B}) &\coloneqq \omega(\hat{A})\otimes \hat{B}\label{eq:omegaAB}
\end{align}
and extend the action to the full uncoupled algebra by linearity. The actions defined in Eqs.~\eqref{eq:omegadAB} and~\eqref{eq:omegaAB} essentially represent a partial expectation value. Then, the operator $\hat{B}_\mathcal{A} = \omega_\tc{d}((\gamma^-)^{-1}(\hat{B}_\mathcal{C})) = \omega_\tc{d}(\Theta(\openone\otimes\hat{B}))$ is such that
\begin{equation}
    \omega(\hat{B}_\mathcal{A}) = \omega(\omega_\tc{d}(\Theta(\openone\otimes\hat{B}))) = \omega\otimes\omega_\tc{d}(\Theta(\openone\otimes\hat{B})) = \varpi_{\tc{d}}(\hat{B}).
\end{equation}
In other words, the observable in $\mathcal{A}(\M)$ whose expected value in the initial state $\omega$ coincides with the expected value of $\hat{B}$ in $\varpi_\tc{d}$ is $\hat{B}_\mathcal{A} = \omega_\tc{d}(\hat{B}_\mathcal{C})$. Equivalently, a measurement of the probe observable $\hat{B}$ induces a measurement of the observable $\hat{B}_\mathcal{A} = \omega_\tc{d}(\Theta(\openone\otimes\hat{B}))$ in the field $\hat{\phi}$.

The Fewster-Verch framework can also be used to prescribe a state update for the field $\hat{\phi}$. However, as we discussed previously, the notion of state updates creates an ambiguity regarding the spacetime region where the state update should be performed. Instead, it is simpler to think of conditional probabilities for consecutive measurements, which removes this ambiguity without loss of predictive power.

Arguably, the Fewster-Verch framework does not address the question of measurements in quantum field theory---it merely prescribes an interaction-picture type of evolution for states of two quantum field theories, finding observables in each theory that are correlated. In other words, although the Fewster-Verch framework is able to correlate the expected values of a probe field to those of the target field, it does not address how one would measure the probe, which is itself also modelled by quantum field theory. In some sense, one could say that the Fewster-Verch framework reduces the measurement problem in quantum field theory to the standard measurement problem in quantum theory. Overall, the Fewster-Verch framework still relies on the fact that

\begin{center}
    \textit{``Someone, somewhere, can measure something.''}
\end{center}
With the ``something'' being an observable $\hat{B}$ in the detector field algebra after the interaction with the target field.

It is important to stress that the Fewster-Verch framework relies on the interaction between target and probe taking place in a compact region of spacetime. Indeed, the maps $\gamma^\pm$ between the uncoupled and coupled algebras and the scattering endomorphism $\Theta$ can only be defined if there are Cauchy surfaces where the uncoupled and coupled theories coincide (see \textbf{A4}). Strictly speaking, this assumption does not apply to most physical theories in our universe (e.g. the electromagnetic interaction cannot be switched off). Nevertheless, one could effectively use the Fewster-Verch framework assuming an adiabatic switching of the interaction, which one could expect to be able to describe interactions that last for sufficiently long times. 

The requirement that the interaction happens over a compactly supported region of spacetime is typically implemented by an external function that defines the interaction region. Indeed, the prototypical toy model for the application of the Fewster-Verch measurement scheme is a theory for real scalar fields $\phi$ and $\phi_\tc{d}$ defined by the Lagrangian density
\begin{equation}\label{eq:LagphiphidFV}
    \mathcal{L} = - \frac{1}{2}\partial_\mu \phi \partial^\mu \phi - \frac{1}{2}\partial_\mu \phi_\tc{d} \partial^\mu \phi_\tc{d} - \frac{m_\tc{d}^2}{2}\hat{\phi}_\tc{d}^2 - \lambda \zeta(\mf x)\phi_\tc{d}\phi.
\end{equation}
where $\lambda$ is a coupling constant with units of energy squared and $\zeta(\mf x)$ is a non-dynamical compactly supported test function\footnote{The function $\zeta(\mf x)$ could be replaced by two dynamical fields $\zeta(\mf x) = \psi_1(\mf x)\psi_2(\mf x)$ which are non-zero only in a finite region around different light-light curves that intercept at a point, so that the product $\psi_1(\mf x)\psi_2(\mf x)$ would only be non-zero in a compact causally convex region. However, if the fields $\psi_1$ and $\psi_2$ are described as quantum fields, it is not clear whether it is possible to have an uncorrelated state between the fields $\hat{\phi}$, $\hat{\phi}_\tc{d}$, $\hat{\psi}_1$ and $\hat{\psi}_2$ or to construct scattering maps that act only in the algebras of $\hat{\phi}$ and $\hat{\phi}_\tc{d}$. Alternatively, one could describe $\psi_1$ and $\psi_2$ as classical fields, in which case a semi-classical approximation would be required to determine their dynamics during the interaction.}, whose support defines the interaction region $\mathcal{K}$. In this example, it is also possible to compute the action of the scattering endomorphism in smeared field operators of the form $\hat{\phi}(f)$ and $\hat{\phi}_\tc{d}(f)$ explicitly. 

Overall, the Fewster-Verch framework prescribes the interaction between two quantum fields, a target field and a probe field, under the assumption that the interaction takes place in a compact spacetime region. The formalism uses a scattering map $\Theta$ that corresponds to the unitary evolution undergone by both fields, which provides a final state for the probe and field after the interaction takes place. It is a measurement scheme in the sense that observables in the probe field measured in its final state correspond to the expected value of an induced field operator in the target field. Given that it is entirely prescribed in terms of local interactions of relativistic quantum fields, the formalism is local and does not suffer from Sorkin-type issues. Importantly, the formalism is fairly general and can be applied to any quantum field theories defined according to \textbf{A1-A4}.

\section{Localized Quantum Fields as Probes}\label{sec:LocalizedQuantumFields}

In the Fewster-Verch formalism, the detector is prescribed as a dynamical quantum field $\hat{\phi}_\tc{d}$. However, solutions of the Klein-Gordon equation typically propagate at the speed of light, implying that the states that define the detector would also spread in space. For instance, solutions of $\nabla_\mu\nabla^\mu\phi = 0$ in Minkowski spacetime propagate only along light-like paths. A one-particle wavepacket $\ket{f} = \hat{a}^\dagger(f)\ket{0}$ is in one-to-one correspondence with the classical solution of the complex Klein-Gordon equation projected into its positive frequency part:
\begin{equation}
    Kf(\mf x) = \int \dd^3\bm k \,u^*_{\bm k}(f) u_{\bm k}(\mf x),
\end{equation}
with $u_{\bm k}(\mf x)$ being the plane waves in~\eqref{eq:ukplanewave}, which propagate along the null direction $(\omega_{\bm k}, \bm k)$. These states are not localized in a finite region of space, being more akin to a (polarization of a) wave of light, and certainly not defining a ``localized detector''.

One could instead consider localizing the state $\ket{f}$ by considering a massive field that satisfies an equation of the form $(\nabla_\mu \nabla^\mu - m^2)\phi = 0$. In fact, this was explored in~\cite{achimFlaminia}. However, even the modes $u_{\bm k}(\mf x)$ of a massive field spread throughout all the future lightcone of their support so that their support increases with time.

One way of obtaining a truly localized quantum field is by considering a field confined by an external potential. Ideally, one would even consider a field confined to a finite spatial region so that its solutions are compactly supported. An idea of this type was first considered by Unruh in his seminal paper~\cite{Unruh1976}, and we will briefly describe it here following a formalism similar to the one considered in~\cite{QFTPD}. 

\subsubsection*{Localized Quantum Fields}

Consider that the globally hyperbolic spacetime $\M$ possesses a time-like Killing vector field $\xi$ such that $\Sigma_t$ is a foliation by Cauchy surfaces which is orthogonal to $\xi$. One can then pick coordinates $(t,\bm x)$ such that $t$ is the flow of $\xi$ and $\bm x$ are coordinates in $\Sigma_t$ for each $t$. The metric can then be put in the form
\begin{equation}\label{eq:staticmetric}
    g = - \beta(\bm x)^2 \dd t^2 + h_{ij}(\bm x) \dd x^i \dd x^j,
\end{equation} 
where $h_{ij}(\bm x)$ is the induced metric in each $\Sigma_t$, being constant in time due to the fact that $t$ is a Killing flow. We then define the field $\phi_\tc{d}$ through the Lagrangian density
\begin{equation}\label{eq:Lagphid}
    \mathcal{L}_\tc{d} = - \frac{1}{2}\partial_\mu \phi_\tc{d}\partial^\mu \phi_\tc{d} - \frac{m_\tc{d}^2}{2}\phi_\tc{d}^2 - \frac{1}{2}V(\bm x)\phi_\tc{d}^2,
\end{equation}
where $V(\bm x)$ is an external potential formally defined by
\begin{equation}\label{eq:Vcompact}
    V(\bm x) = \begin{cases}
    0, \quad &\bm x \in U\\
    \infty , \quad &\bm x \notin U,
    \end{cases}
\end{equation}
with $U\subset \mathbb{R}^3$ being an open convex set. The region $U$ then defines the worldtube $T = \{\mf x = (t,\bm x): \bm x \in U\}$ and the potential $V(\bm x)$ effectively imposes Dirichlet boundary conditions for the field $\phi_\tc{d}$ at the boundary $\partial T$, with $\phi_\tc{d} = 0$ in $\M\backslash T$.

The equation of motion for the field $\phi_\tc{d}$ arising from~\eqref{eq:Lagphid} then becomes
\begin{equation}\label{eq:Pd}
    P_\tc{d}\phi_\tc{d} = (\nabla_\mu \nabla^\mu - m_\tc{d}^2 - V(\bm x))\phi_\tc{d} = 0.
\end{equation}
The differential operator $P_\tc{d}$ then defines retarded and advanced Green's functions, $E_\tc{d}^+$ and $E_\tc{d}^-$ satisfying $P_\tc{d}E_\tc{d}^\pm f = f$, with $E_\tc{d}^\pm f$ supported in $J^\pm(\text{supp}(f))$. The causal propagator is then $E_\tc{d} = E_\tc{d}^+ - E_\tc{d}^-$.

Due to the timelike symmetry that the spacetime is assumed to satisfy, the space of solutions also admits a convenient basis decomposition. In the coordinates $(t,\bm x)$ the equation of motion can be recast as
\begin{equation}
    - \partial_t^2\psi + \frac{\beta}{\sqrt{h}}\partial_i\left(\beta \sqrt{h}\, h^{ij} \partial_j \phi\right) - \beta^2(m_{\tc{d}}^2 + V(\bm x))\phi_{\tc{d}} = 0,
\end{equation}
and we can find a basis of solutions using separation of variables $u_{\bm n}(\mf x) = v_{\bm n}(t) \Phi_{\bm n}(\bm x)$, where the equation of motion can be rewritten as
\begin{equation}\label{eq:sepvar1}
    \frac{\partial_t^2v_{\bm n}(t)}{v_{\bm n}(t)} = - \omega_{\bm n}^2 = -\frac{\beta}{\sqrt{h}}\partial_i\left(\beta \sqrt{h}\, h^{ij} \partial_j \Phi_{\bm n}(\bm x)\right) + \beta^2(m_\tc{d}^2 + V(\bm x))\Phi_{\bm n}(\bm x),
\end{equation}
where $\omega_{\bm n}$ is a constant. We can define the linear differential operator 
\begin{equation}
    L \Phi = -\frac{\beta}{\sqrt{h}}\partial_i\left(\beta \sqrt{h}\, h^{ij} \partial_j \Phi\right) + \beta^2 m_\tc{d}^2 \Phi
\end{equation}
acting in smooth functions in $L^2(\Sigma_t)$ that are $0$ outside of $T\cap \Sigma_t$. The operator $L$ is self-adjoint in a suitable domain of $L^2(\Sigma_t)$ and has a discrete positive spectrum $\lambda_{\bm n}^2$ with eigenfunctions $\Phi_{\bm n}(\bm x)$. Eq.~\eqref{eq:sepvar1} then gives
\begin{equation}
    \omega_{\bm n} = |\lambda_{\bm n}| \quad \text{and} \quad v_{\bm n}(t) = e^{\pm \ii \omega_{\bm n}t},
\end{equation}
so that the basis of positive frequency solutions can be written as
\begin{equation}\label{eq:un}
    u_{\bm n}(\mf x) = e^{- \ii \omega_{\bm n} t} \Phi_{\bm n}(\bm x).
\end{equation}
Normalization according to the Klein-Gordon inner product then imposes
\begin{equation}\label{eq:normCondPhin}
    \int \dd\Sigma\, \beta(\bm x)^{-1}|\Phi_{\bm n}(\bm x)|^2 = \frac{1}{2\omega_{\bm n}}.
\end{equation}

The quantum field theory for $\hat{\phi}_\tc{d}$ can be built as in the explicit algebraic construction based on smeared field operators discussed in Section~\ref{sec:QFT}, with small modifications due to the fact that the classical solutions are supported in $T$. For instance, in the operators $\hat{\phi}_\tc{d}(f)$ will satisfy $\hat{\phi}_\tc{d}(f) = 0$ for all $f$ supported outside of $T$. Indeed, we will have $\mathcal{A}(\mathcal{O}) = \{\openone\}$ for any $\mathcal{O}$ non-overlapping with $T$, so that $\mathcal{A}(\mathcal{O}\cap T) = \mathcal{A}(\M)$ for any region $\mathcal{O}$ that contains a Cauchy surface\footnote{This association is technically not completely well defined, as we have the issue of defining $\hat{\phi}_\tc{d}(f)$ when $f$ has a support that overlaps $T$ and its complement. The natural choice $f\mapsto f|_T$ might lead to divergences due to discontinuity. In any case, provided that we remain restricted to functions with support in $T$, this construction can be used.}. Specifically, the basis $\{u_{\bm n},u_{\bm n}^*\}$ of the space of solutions gives rise to a state $\ket{0_\tc{d}}$, as well as its GNS representation in the Fock space $\mathcal{F}(\mathscr{H}_\tc{d})$, where the field operator can be written as
\begin{equation}\label{eq:GNSphid}
    \hat{\phi}_\tc{d}(\mf x) = \sum_{\bm n} \hat{a}_{\bm n} u_{\bm n}(\mf x) + \hat{a}_{\bm n}^\dagger u_{\bm n}^*(\mf x).
\end{equation}
There are relevant consequences to the fact that the modes $u_{\bm n}(\mf x)$ are discrete. First, we have that the operators $\hat{a}_{\bm n}$ and $\hat{a}_{\bm n}^\dagger$ are well defined in the algebra $\mathcal{A}_\tc{d}(\M)$ and do not necessarily need to be smeared against a test function. Indeed, denoting the causal propagator associated with the differential operator $P_\tc{d}$ in~\eqref{eq:Pd} by $E_\tc{d}$ and writing $u_{\bm n} = E_\tc{d}g_{\bm n}$ for compactly supported functions $g_{\bm n}$, we can write (using~\eqref{eq:phisymplecticEclass})
\begin{align}
    \hat{a}_{\bm n} &= (u_{\bm n},\hat{\phi}) = \ii \Omega(\hat{\phi}_\tc{d},u_{\bm n}^*) = \ii \hat{\phi}_\tc{d}(g_{\bm n}^*),\label{eq:an}\\
    \hat{a}_{\bm n}^\dagger &= - (u_{\bm n}^*,\hat{\phi}_\tc{d}) = - \ii \Omega(\hat{\phi},u_{\bm n}) = - \ii \hat{\phi}_\tc{d}(g_{\bm n}).\label{eq:and}
\end{align}
Given that the functions $g_{\bm n}$ are compactly supported in this case, the operators $\hat{a}_{\bm n} = \hat{\phi}_\tc{d}(\ii g_{\bm n}^*)$ and $\hat{a}_{\bm n}^\dagger = \hat{\phi}_\tc{d}(-\ii g_{\bm n})$ are well defined in algebra $\mathcal{A}_\tc{d}(\M)$, unlike the creation and annihilation operators associated to continuous mode solutions. The creation and annihilation operators of this confined theory also satisfy the discrete canonical commutation relations
\begin{equation}\label{eq:CCRphid}
    [\hat{a}_{\bm n},\hat{a}_{\bm n'}^\dagger] = \delta_{\bm n, \bm n'}, \quad [\hat{a}_{\bm n},\hat{a}_{\bm n'}] = [\hat{a}_{\bm n}^\dagger,\hat{a}_{\bm n'}^\dagger] = 0.
\end{equation}
In particular, the states
\begin{equation}
    \ket{\bm n} \coloneqq \hat{a}_{\bm n}^\dagger \ket{0_\tc{d}}
\end{equation}
are normalized and constitute an orthonormal basis of the one-particle Hilbert space $\mathscr{H}_\tc{d}$.

The Fock space $\mathcal{F}(\mathscr{H}_\tc{d})$ also factors in a special way when one has discrete modes. Indeed, the state $\ket{0_\tc{d}}$ factors as a tensor product of the form
\begin{equation}\label{eq:vacTensor}
    \ket{0_\tc{d}} = \bigotimes_{\bm n} \ket{0_{\bm n}},
\end{equation}
where each $\ket{0_{\bm n}}$ is defined by $\hat{a}_{\bm n}\ket{0_{\bm n}} = 0$, and represent no occupation in the mode $\bm n$. The operators $\hat{a}_{\bm n}$ and $\hat{a}_{\bm n}^\dagger$ then each act on their respective ``vacua'', $\ket{0_{\bm n}}$, with $\hat{a}_{\bm n}^\dagger$ creating excitations in the mode labelled by $\bm n$ This induces a decomposition of the Fock space $\mathcal{F}(\mathscr{H}_\tc{d})$ as a tensor product of the form
\begin{equation}\label{eq:Fockphid}
    \mathcal{F}(\mathscr{H}_\tc{d}) ={\bigotimes_{\bm n}\,} \mathscr{H}_{\bm n},
\end{equation}
where $\mathscr{H}_{\bm n}$ is simply the Hilbert space associated with each mode, spanned by states of the form $(\hat{a}_{\bm n}^\dagger)^m\ket{0_{\bm n}}$, with $m \in \mathbb{N}$, representing $m$ excitations in the mode labelled by $\bm n$. In summary, a basis for $\mathcal{F}(\mathscr{H}_\tc{d})$ is constructed with each basis element indicating how many excitations it contains in each mode. In this case, the normal ordered Hamiltonian associated to the foliation can be written as
\begin{equation}
    :\!\hat{H}\!: = \int \dd^3 \bm x \sqrt{-g}\,\normord{\hat{T}_{\mu\nu}}  n^\mu n^\nu  = \sum_{\bm n} \omega_{\bm n} \hat{a}_{\bm n}^\dagger \hat{a}_{\bm n},
\end{equation}
assigning energy $\omega_{\bm n}$ to each mode.





A physical realization of this model would correspond to a scalar field in a perfectly reflecting cavity whose motion in spacetime defines the worldtube $T$. The state $\ket{0_\tc{d}}$ represents the case where none of the modes $u_{\bm n}$ is excited, and one could argue that this is a vacuum state of the field, as seen by an observable comoving with the cavity, in the sense that it is the eigenvector of the Hamiltonian with minimum energy. Each state $\ket{\bm n}$ could be interpreted as a one-particle state within the cavity.

Notice that the facts that 1) the creation and annihilation operators $\hat{a}_{\bm n}$ and $\hat{a}_{\bm n}^\dagger$ are well defined algebra operators, 2)$\ket{\bm n} = \hat{a}_{\bm n}^\dagger \ket{0_\tc{d}}$ are normalized states, and 3) the decomposition of the Fock space as a tensor product of the Hilbert space associated to each mode, are all direct consequences of the fact that there exists a discrete orthonormal basis of solutions of the Klein-Gordon equation. As such, these results are valid whenever one considers the GNS representation of the field associated to a discrete orthonormal basis of solutions to the equations of motion. In particular, these results can hold even if the spacetime is not static.

As an explicit example in Minkowski spacetime, we consider a perfectly reflecting cubic cavity of side $d$. Considering inertial coordinates $(t,\bm x)$, define $U_d = [0,d]^3$ and the potential
\begin{equation}\label{eq:potentialCubic}
    V(\bm x) = \begin{cases}
    0, \quad & \bm x \in U_d,\\
    \infty , \quad & \bm x \notin U_d.
    \end{cases}
\end{equation}
A basis of spatial solutions of the wave equation with this potential can then be written as
\begin{equation}
    \Phi_{\bm n}(\bm x) = \frac{1}{\sqrt{2\omega_{\bm n}} }f_{n_x}(x)f_{n_y}(y)f_{n_z}(z),
\end{equation}
where the functions $f_n(u)$ are given by
\begin{equation}
    f_n(u) = \sqrt{\frac{2}{d}}\sin(\frac{\pi n u}{d}),
\end{equation}
and the corresponding eigenfrequencies are
\begin{equation}\label{eq:gapBox}
    \omega_{\bm n} = \sqrt{m^2 + \frac{\pi^2}{d^2}(n_x^2 + n_y^2 + n_z^2)},
\end{equation}
with $\bm n = (n_x,n_y,n_z)$, $n_i \in \mathbb{N}^*$.

\subsubsection*{Localized Quantum Probes}

We can now consider the case in which we use a localized field, as described above, to probe a target quantum field. For concreteness, we will stick with the example of a minimally coupled massive Klein-Gordon field under the influence of the potential~\eqref{eq:Vcompact}, giving rise to a field confined in a finite region of space\footnote{By finite region of space we mean that the restriction of $\phi(\mf x)$ to a (and thus any) Cauchy surface has compact support.}. We can now consider the quantum field theory for a target field $\hat{\phi}$ that we intend to probe through the field $\hat{\phi}_\tc{d}$. Specifically, we consider the theory defined by the Lagrangian density
\begin{equation}
    \mathcal{L} = - \frac{1}{2}\partial_\mu \phi \partial^\mu \phi - \frac{1}{2}\partial_\mu \phi_\tc{d} \partial^\mu \phi_\tc{d} - \frac{m_\tc{d}^2}{2}\phi_\tc{d}^2 - \frac{V(\bm x)}{2}\phi_\tc{d}^2 - \lambda \zeta(\mf x) \phi_\tc{d}\phi,
\end{equation}
where $\lambda$ is a real coupling constant with dimensions of energy squared and $\zeta(\mf x)$ is an adimensional\footnote{We choose $\zeta(\mf x)$ to be dimensionless so that one can keep the intuition that $\zeta(\mf x) = 1$ corresponds to the case where the fields are always coupled.} compactly supported function, defining the region of spacetime where the interaction between the fields $\phi_\tc{d}$ and $\phi$ takes place as the causal hull of $\text{supp}(\zeta)$, $\mathcal{K}$. This then configures a particular case of the Fewster-Verch measurement scheme, where we consider the detector field $\hat{\phi}_\tc{d}$ probing the free field $\hat{\phi}$. Importantly, in this case, the detector field is localized in the worldtube $T$, corresponding to a compactly supported detector.

Our next goal will be to find the final state of the detector field, $\varpi_\tc{d}$ after the interaction with the field $\hat{\phi}$. We can do this perturbatively, using the results of~\cite{FVOG}, which show that the scattering endomorphism acts in smeared field operators of the uncoupled algebra $\mathcal{A}(\M)\otimes \mathcal{A}_\tc{d}(\M)$ according to
\begin{equation}
    \Theta(\hat{\phi}(f)\otimes \openone + \openone \otimes \hat{\phi}_\tc{d}(g)) = \hat{\phi}(\tilde{f})\otimes \openone + \openone \otimes \hat{\phi}_\tc{d}(\tilde{g}),
\end{equation}
where the functions $\tilde{f}$ and $\tilde{g}$ can be computed from $f$ and $g$ using the causal propagators $E$ and $E_\tc{d}$ associated with the equations of motion of the respective fields $\hat{\phi}$ and $\hat{\phi}_\tc{d}$, as well as the total propagator associated with the interacting theory. For convenience, we will omit the tensor product with the identity in $\mathcal{A}(\M)\otimes \mathcal{A}_\tc{d}(\M)$ so that
\begin{equation}
    \Theta(\hat{\phi}(f) +  \hat{\phi}_\tc{d}(g)) = \hat{\phi}(\tilde{f}) +  \hat{\phi}_\tc{d}(\tilde{g}).
\end{equation}

Whenever $f$ and $g$ are supported in $\M^+ = \M\setminus J^-(\mathcal{K})$, the dependence of the functions $\tilde{f}$ and $\tilde{g}$ in $f$ and $g$ can also be computed perturbatively in $\lambda$. Indeed, in~\cite{FVOG}, it was found that whenever $f,g$ are supported in $\M^+$, to second order in the coupling constant, they can be written as
\begin{align}
    \tilde{f} &= f + \lambda \zeta E^-_{\tc{d}} g + \lambda^2 \zeta E_\tc{d}^-(\zeta G_Af) + \mathcal{O}(\lambda^3),\\
    \tilde{g} &= g + \lambda \zeta G_Af + \lambda^2 \zeta G_A(\zeta E_\tc{d}^- g) + \mathcal{O}(\lambda^3),
\end{align}
where $G_A$ is the advanced propagator for the field $\phi$ and $E_\tc{d}^-$ the advanced propagator for the field $\phi_\tc{d}$. In the equation above, multiplication by the function $\zeta(\mf x)$ is denoted by juxtaposition and the application of $G_A$, $E_\tc{d}^-$ to arguments involving products of functions has been denoted with parenthesis, so that $\zeta G_Af = \zeta(\mf x) G_Af(\mf x)$  and $\zeta E_\tc{d}^-g = \zeta(\mf x) E_\tc{d}^-g(\mf x)$.

We can now proceed with the Fewster-Verch prescription and compute the final state of the detector field $\varpi_\tc{d}$ perturbatively in $\lambda$. We assume that the initial state of the two fields is uncorrelated, of the form $\omega\otimes \omega_\tc{d}$. From~\eqref{eq:FVuppid}, the action of the detector field state $\varpi_\tc{d}$ in a detector observable $\hat{\phi}_\tc{d}(g)$ after the interaction can be written as
\begin{equation}
    \varpi_\tc{d}(\hat{\phi}_\tc{d}(g)) = \omega\otimes\omega_\tc{d}(\Theta(\openone\otimes\hat{\phi}(g))),
\end{equation}
where
\begin{equation}
    \Theta(\openone\otimes\hat{\phi}_\tc{d}(g)) = \hat{\phi}(\tilde{f}) +  \hat{\phi}_\tc{d}(\tilde{g})
\end{equation}
and, to second order in lambda,
\begin{align}
    \tilde{f} &= \lambda \zeta E_\tc{d}^- g + \mathcal{O}(\lambda^3)\\
    \tilde{g} &= g + \lambda^2 \zeta G_A(\zeta E_\tc{d}^- g) + \mathcal{O}(\lambda^4).
\end{align}
We then find
\begin{equation}\label{eq:uppidphig}
    \varpi_\tc{d}(\hat{\phi}_\tc{d}(g)) = \omega_\tc{d}(\hat{\phi}_\tc{d}(g)) + \lambda \omega(\hat{\phi}(\zeta E_{\tc{d}}^- g)) + \lambda^2 \omega_\tc{d}(\hat{\phi}_\tc{d}(\zeta G_A(\zeta E_\tc{d}^-g))) + \mathcal{O}(\lambda^3).
\end{equation}
The result above shows explicitly how expected values in the final state of the detector field contain information about expected values of observables of the field $\hat{\phi}$ in the state $\omega$. For instance, if one knows the value of $\omega_\tc{d}(\hat{\phi}_\tc{d}({g}))$ before the measurement takes place, the leading order in $\lambda$ corrections to this value after the interaction directly yields $\omega(\hat{\phi}(\zeta E_\tc{d}^- g))$. 

Although Eq.~\eqref{eq:uppidphig} illustrates the idea behind measurements using the Fewster-Verch framework, it is still abstract and somewhat particular---Eq.~\eqref{eq:uppidphig} gives $0$ if $\omega$ and $\omega_\tc{d}$ are quasifree. Using the compatibility of the scattering endomorphism with the algebraic structure of $\mathcal{A}(\M)\otimes \mathcal{A}_\tc{d}(\M)$, we can also compute expected values of more general operators in the detector field. In particular, let us consider the expected value of a product of operators of the form $\hat{\phi}_\tc{d}(f)\hat{\phi}_\tc{d}(g)$. When restricted to operators in $\mathcal{A}_\tc{d}(\M)$, the scattering endomorphism essentially maps
\begin{equation}\label{eq:FVphiafter}
    \hat{\phi}_\tc{d}(g) \mapsto \hat{\phi}_\tc{d}(g) + \lambda \hat{\phi}(\zeta E_{\tc{d}}^- g) + \lambda^2 \hat{\phi}_\tc{d}(\zeta G_A(\zeta E_\tc{d}^-g)) + \mathcal{O}(\lambda^3).
\end{equation}
Assuming $\omega$ and $\omega_\tc{d}$ to be quasifree states we can then apply~\eqref{eq:FVphiafter} to obtain
\begin{align}\label{eq:FVWfg}
    \varpi_\tc{d}(\hat{\phi}_\tc{d}(f)\hat{\phi}_\tc{d}(g)) = &\,\, \omega_\tc{d}(\hat{\phi}_\tc{d}(f)\hat{\phi}_\tc{d}(g)) + \lambda^2 \omega(\hat{\phi}(\zeta E_\tc{d}^- f)\hat{\phi}(\zeta E_\tc{d}^- g)) \\&+ \lambda^2 \left(\omega_\tc{d}(\hat{\phi}_\tc{d}(f)\hat{\phi}_\tc{d}(\zeta G_A(\zeta E_\tc{d}^-g))) + \omega_\tc{d}(\hat{\phi}_\tc{d}(\zeta G_A(\zeta E_\tc{d}^- f))\hat{\phi}_\tc{d}(g))\right) + \mathcal{O}(\lambda^4)\nonumber.
\end{align}


Let us now consider functions $f = -\ii g_{\bm n}$ and $g = \ii g_{\bm n}^*$, supported in the causal future of the interaction region $\mathcal{K} \supset \text{supp}(\zeta)$, such that $g_{\bm n}$ give rise to the localized modes~\eqref{eq:GNSphid} through $u_{\bm n} = E_\tc{d} g_{\bm n}$. We then have that
\begin{equation}
    \hat{\phi}_\tc{d}(f)\hat{\phi}_\tc{d}(g) = \hat{\phi}_\tc{d}(g_{\bm n})\hat{\phi}_\tc{d}(g_{\bm n}^*) = \hat{a}_{\bm n}^\dagger \hat{a}_{\bm n}
\end{equation}
is the occupation number operator associated to mode $\bm n$. Specifically, having access to the expected value of $\hat{a}_{\bm n}^\dagger \hat{a}_{\bm n}$ corresponds to having access to ``how many particles'' the field $\hat{\phi}_\tc{d}$ contains. We can now compute the expected value of the number operator after the interaction. Due to $g_{\bm n}$ being in the causal future of $\text{supp}(\zeta)$, we have that within the support of $\zeta$, $u_{\bm n} =  E_\tc{d}g_{\bm n} =- E_\tc{d}^-g_{\bm n}$, so that 
\begin{equation}\label{eq:Jesus}
    \zeta E_\tc{d}^-g_{\bm n} = -\zeta E_\tc{d}g_{\bm n} = -\zeta u_{\bm n}.
\end{equation}
Using the result above, Eq.~\eqref{eq:FVWfg} becomes
\begin{align}
    \varpi_{\tc{d}}(\hat{a}_{\bm n}^\dagger \hat{a}_{\bm n}) = &\,\,\omega_\tc{d}(\hat{a}_{\bm n}^\dagger \hat{a}_{\bm n}) + \lambda^2 \omega(\hat{\phi}(\zeta u_{\bm n})\hat{\phi}(\zeta u_{\bm n}^*)) \\
    &-\ii \lambda^2\omega_\tc{d}(\hat{a}_{\bm n}^\dagger\hat{\phi}_\tc{d}(\zeta G_A(\zeta u_{\bm n}^*))) + \ii \lambda^2\omega_\tc{d}(\hat{\phi}_\tc{d}(\zeta G_A(\zeta u_{\bm n}))\hat{a}_{\bm n}) + \mathcal{O}(\lambda^4)\\[2mm]
    =&\,\, \omega_\tc{d}(\hat{a}_{\bm n}^\dagger \hat{a}_{\bm n}) + \lambda^2 \omega(\hat{\phi}(\zeta u_{\bm n})\hat{\phi}(\zeta u_{\bm n}^*))\\[1mm]
    &- \ii  \sum_{\bm m} G_A(\zeta u_{\bm m},\zeta u_{\bm n}^*)\omega_\tc{d}(\hat{a}_{\bm n}^\dagger\hat{a}_{\bm m}) + G_A(\zeta u_{\bm m}^*,\zeta u_{\bm n}^*)\omega_\tc{d}(\hat{a}_{\bm n}^\dagger\hat{a}_{\bm m}^\dagger) 
    \\
    &+\ii \sum_{\bm m} G_A(\zeta u_{\bm m}^*,\zeta u_{\bm n})\omega_\tc{d}(\hat{a}_{\bm m}^\dagger\hat{a}_{\bm n}) + G_A(\zeta u_{\bm m},\zeta u_{\bm n})\omega_\tc{d}(\hat{a}_{\bm m}\hat{a}_{\bm n})
    + \mathcal{O}(\lambda^4),
\end{align}
where we used the expansion~\eqref{eq:GNSphid} in the last equality.

This expression considerably simplifies if the initial state $\omega_\tc{d}$ is the vacuum state of the detector, $\ket{0_\tc{d}}$, in which case $\omega_\tc{d}(\hat{a}_{\bm n} \hat{a}_{\bm m}) = \omega_\tc{d}(\hat{a}_{\bm n}^\dagger \hat{a}_{\bm m}) = \omega_\tc{d}(\hat{a}_{\bm n}^\dagger \hat{a}_{\bm m}^\dagger) = 0$, and $\varpi_\tc{d}(\hat{a}_{\bm n}^\dagger \hat{a}_{\bm n})$ gives the probability of creating a particle at the mode $u_{\bm n}$. In this case Eq.~\eqref{eq:FVWfg} yields simply
\begin{equation}\label{eq:omegadanand}
    \varpi_{\tc{d}}(\hat{a}_{\bm n}^\dagger \hat{a}_{\bm n}) = \lambda^2 \omega(\hat{\phi}(\zeta u_{\bm n})\hat{\phi}(\zeta u_{\bm n}^*)) + \mathcal{O}(\lambda^4).
\end{equation}
That is, when the field $\hat{\phi}_\tc{d}$ starts in its vacuum state, each mode $\bm n$ will have leading order probability $\lambda^2 \omega(\hat{\phi}(\zeta u_{\bm n})\hat{\phi}(\zeta u_{\bm n}^*))$ of becoming occupied after the interaction with the field $\hat{\phi}$. We can also write the mode excitation probability explicitly as a spacetime integral:
\begin{equation}\label{eq:exProbV}
    \varpi_{\tc{d}}(\hat{a}_{\bm n}^\dagger \hat{a}_{\bm n}) =  \lambda^2\int \dd V \dd V' \zeta(\mf x) \Phi_{\bm n}(\bm x)\zeta(\mf x') \Phi_{\bm n}(\bm x') e^{- \ii \omega_{\bm n}(t-t')}W(\mf x, \mf x') + \mathcal{O}(\lambda^4),
\end{equation}
where $W$ denotes the Wightman function of the quasifree state $\omega$ in the theory for the target field $\hat{\phi}$. The integral above can be seen as a time Fourier transform of the smeared two-point function. Defining $\Lambda_{\bm n}(\mf x) = \zeta(\mf x) \Phi_{\bm n}(\bm x)$ and $\Lambda_{\bm n}^\pm(\mf x) = e^{\pm \ii \omega_{\bm n}t}\Lambda_{\bm n}(\mf x)$ we can write the excitation probability of the mode $\bm n$ as
\begin{equation}
    \varpi_{\tc{d}}(\hat{a}_{\bm n}^\dagger \hat{a}_{\bm n}) = \lambda^2 W(\Lambda_{\bm n}^-,\Lambda_{\bm n}^+) + \mathcal{O}(\lambda^4).
\end{equation}

Notice that the operators that we assumed to have access to in the detector theory were defined by the compactly generating functions $g_{\bm n}$, which do not seem to have an intrinsic physical meaning on their own, but give rise to the operators $\hat{a}_{\bm n} = \hat{\phi}(\ii g_{\bm n}^*)$ and $\hat{a}_{\bm n}^\dagger = \hat{\phi}(-\ii g_{\bm n})$. On the other hand, the field operators that we ended up having access to were $\hat{\phi}(\zeta u_{\bm n})$ and $\hat{\phi}(\zeta u_{\bm n}^*)$, which are smeared by both the interaction profile $\zeta(\mf x)$ and the shape of the modes of the detector field $u_{\bm n}(\mf x)$. When we first presented the quantization of a Klein-Gordon field in Section~\ref{sec:QFT} we mentioned that we should think of smearing function $f$ in field operators of the form $\hat{\phi}(f)$ as defining the region where an experimentalist has access to the field. Here we are seeing that indeed, the \textit{shape} of a detector (and the region where it couples to the field) give rise to these smearing functions.

\subsubsection*{A More Familiar Model for the Dynamics}

The method for computing the updated state outlined above, based on the Fewster-Verch framework, allowed us to find the leading order action of the updated detector field state in any observable of its algebra. However, it would be useful to find a representation of the final detector state in terms of the GNS construction based on the state $\ket{0_\tc{d}}$. Indeed, there is a more familiar method of computing the action of the scattering endomorphism $\Theta$ on states and operators in terms of an interaction unitary time evolution operator. Unfortunately, this method is not fully well defined in terms of the algebra $\mathcal{A}(\M)\otimes\mathcal{A}_\tc{d}(\M)$, and a formal connection between it and the map $\Theta$ has not yet been fully established.

In a conventional treatment to the interaction between the fields $\hat{\phi}$ and $\hat{\phi}_\tc{d}$, one can assign the interaction Hamiltonian density
\begin{equation}
    \hat{\mathcal{H}}_I(\mf x) = \lambda \zeta(\mf x) \hat{\phi}_{\tc{d}}(\mf x)\otimes \hat{\phi}(\mf x)
\end{equation}
to the time evolution generated by the interaction. Importantly, the Hamiltonian density $\hat{\mathcal{H}}_I(\mf x)$ does not correspond to a well-defined smeared field operator in the algebra $\mathcal{A}(\M)\otimes\mathcal{A}_\tc{d}(\M)$, as it involves a direct product of fields. As a consequence, the computations that will follow are technically not fully well defined in the algebras of observables of the uncoupled theories. However, due to the fact that the modes of $\hat{\phi}_\tc{d}$ are discrete, using the GNS representation of $\hat{\phi}_\tc{d}$ associated to $\ket{0_\tc{d}}$, we can use the decomposition~\eqref{eq:GNSphid} and write
\begin{align}
    \hat{\mathcal{H}}_I(\mf x) &= \sum_{\bm n}\lambda \zeta(\mf x)(u_{\bm n}(\mf x)\hat{a}_{\bm n} + u_{\bm n}^*(\mf x) \hat{a}_{\bm n}^\dagger)\otimes \hat{\phi}(\mf x)\\
    &= \lambda\zeta(\mf x)\sum_{\bm n} \hat{a}_{\bm n} u_{\bm n}(\mf x)\hat{\phi}(\mf x) +  \hat{a}_{\bm n}^\dagger  u_{\bm n}^*(\mf x)\hat{\phi}(\mf x).
\end{align}
That is, the spacetime integral of the Hamiltonian $\hat{\mathcal{H}}_I(\mf x)$ can be written as the formal series
\begin{equation}
    \int \dd V \hat{\mathcal{H}}_I(\mf x) = \lambda\sum_{\bm n} \hat{a}_{\bm n} \hat{\phi}(\zeta u_{\bm n})+\hat{a}_{\bm n}^\dagger \hat{\phi}(\zeta u_{\bm n}^*),
\end{equation}
whose expected value in states $\omega\otimes\omega_\tc{d}$ is finite, provided that $\omega_\tc{d}$ can be represented in the GNS construction associated with $\ket{0_\tc{d}}$.

Following the prescription of time evolution in quantum mechanics, the interaction Hamiltonian density gives rise to the unitary time evolution operator
\begin{align}\label{eq:Texp}
    \hat{U}_I = \mathcal{T}\exp\left(- \ii \int \dd V \hat{\mathcal{H}}_I(\mf x)\right)\coloneqq \sum_{n=0}^\infty \hat{U}_I^{(n)},
\end{align}
where
\begin{align}
    \hat{U}_I^{(0)} &= \openone,\quad\quad
    \hat{U}_I^{(1)} = - \ii \int \dd V \hat{\mathcal{H}}_I(\mf x),\label{eq:UI1FV}\\
    \hat{U}_I^{(n)} &= (-\ii)^n \int \dd V_1 ... \dd V_n \hat{\mathcal{H}}(\mf x_1)...\hat{\mathcal{H}}(\mf x_n) \theta(t_n - t_{n-1})...\theta(t_2 - t_{1}), \text{ for } n\geq 2,
\end{align}
and $t$ is any future-directed time coordinate. Notice that because $[\hat{\mathcal{H}}_I(\mf x), \hat{\mathcal{H}}_I(\mf x')] = 0$ for causally disconnected $\mf x$ and $\mf x'$, the integral above is independent\footnote{Indeed, for $\hat{U}_I^{(2)}$, if $t$ and $s$ are two future-directed timelike coordinates, one can split integral over $\M\times \M$ into integrals over the regions $C\subset\M\times \M$ where events are causally connected and $S\subset\M\times\M$ where the events are causally separated. For $\mf x_1,\mf x_2\in C$ we have $\theta(t_2-t_1) = \theta(s_2-s_1)$, and for $\mf x_1, \mf x_2\in S$ we have $\hat{\mathcal{H}}_I(\mf x_1)\hat{\mathcal{H}}_I(\mf x_2) = \hat{\mathcal{H}}_I(\mf x_2)\hat{\mathcal{H}}_I(\mf x_1)$, so that the time ordering operation is irrelevant under the integral over the symmetric set $S$.} of the specific choice of future-directed timelike coordinate $t$. The operators of the theory then evolve according to $\hat{A} \mapsto \hat{U}_I^\dagger \hat{A} \hat{U}_I$, which induces the time evolution in the states $\omega\otimes\omega_\tc{d}(\,\cdot\,) \mapsto \omega\otimes\omega_\tc{d}(\hat{U}_I^\dagger \, \cdot \, \hat{U}_I)$. One can then identify that the relationship between the scattering morphism and the time evolution operators is given formally by $\Theta(\hat{A}) = \hat{U}_I^\dagger \hat{A} \hat{U}_I$.

The perturbative treatment for $\hat{U}_I$ indeed corresponds to the one found using the maps $\Theta$. For instance, to leading order, we have
\begin{equation}
    \hat{U}_I^\dagger \hat{\phi}_\tc{d}(g)\hat{U}_I = \hat{\phi}_\tc{d}(g) + \hat{U}_I^{(1)\dagger} \hat{\phi}_\tc{d}(g) + \hat{\phi}_\tc{d}(g) \hat{U}_I^{(1)} + \mathcal{O}(\lambda^2).
\end{equation}
We can compute the leading order effect in the detector observable $\hat{\phi}(g)$ by using~\eqref{eq:GNSphid}:
\begin{align}
    \hat{U}_I^{(1)\dagger} \hat{\phi}_\tc{d}(g) + \hat{\phi}_\tc{d}(g) \hat{U}_I^{(1)} =&\,\,\ii \lambda \sum_{\bm n \bm n'} (\hat{a}_{\bm n} \hat{\phi}(\zeta u_{\bm n})+\hat{a}_{\bm n}^\dagger \hat{\phi}(\zeta u_{\bm n}^*))(u_{\bm n'}(g) \hat{a}_{\bm n'} + u_{\bm n'}^*(g) \hat{a}_{\bm n'}^\dagger) \\&- \ii \lambda \sum_{\bm n \bm n'} (u_{\bm n'}(g) \hat{a}_{\bm n'} + u_{\bm n'}^*(g) \hat{a}_{\bm n'}^\dagger)(\hat{a}_{\bm n} \hat{\phi}(\zeta u_{\bm n})+\hat{a}_{\bm n}^\dagger \hat{\phi}(\zeta u_{\bm n}^*))\nonumber\\ = &\,\, \ii \lambda \sum_{\bm n \bm n'} (\hat{a}_{\bm n} \hat{a}_{\bm n'}^\dagger - \hat{a}_{\bm n'} \hat{a}_{\bm n})\hat{\phi}(\zeta u_{\bm n}) u_{\bm n'}^*(g) + (\hat{a}_{\bm n}^\dagger \hat{a}_{\bm n'} - \hat{a}_{\bm n'} \hat{a}_{\bm n}) \hat{\phi}(\zeta u_{\bm n}^*)u_{\bm n}(g)\nonumber\\
    =& \,\,\ii \lambda \sum_{\bm n \bm n'} \hat{\phi}(\zeta u_{\bm n}) u_{\bm n}^*(g) - \hat{\phi}(\zeta u_{\bm n}^*) u_{\bm n}(g)\nonumber\\
    =&\,\, - \lambda \hat{\phi}\left(\zeta\,\, \frac{1}{\ii}\sum_{\bm n \bm n'} u_{\bm n}^*(g) u_{\bm n} - u_{\bm n}(g) u_{\bm n}^*\right) = -\lambda \hat{\phi}(\zeta E_\tc{d}g) = \lambda \hat{\phi}(\zeta E_\tc{d}^-g),\nonumber
\end{align}
where we used the commutation relations~\eqref{eq:CCRphid} and the last equalities follow from expressing $E_\tc{d}$ in terms of the basis $u_{\bm n}$ as in~\eqref{eq:Euki} and using $\zeta E_\tc{d} g = -\zeta E_\tc{d}^- g$ for $g$ supported in $J^+(\mathcal{K})$. We then find that
\begin{equation}
    \hat{U}_I^\dagger \hat{\phi}_\tc{d}(g)\hat{U}_I = \hat{\phi}_{\tc{d}}(g) + \lambda \hat{\phi}(\zeta E_\tc{d}^-g) + \mathcal{O}(\lambda^2),
\end{equation}
matching the leading order results of~\eqref{eq:FVphiafter} for $\Theta(\hat{\phi}_\tc{d}(g))$. One can verify that this result also holds to higher orders, but we will refrain from performing these computations here for conciseness.

As a second verification, let us show that using $\hat{U}_I$ gives the result of Eq.~\eqref{eq:omegadanand} under the assumption that the initial state is of the form $\omega\otimes\omega_\tc{d}$ with $\omega$ quasifree and $\omega_\tc{d}$ corresponding to $\ket{0_\tc{d}}$. To second order in $\lambda$ we have
\begin{align}
    \omega\otimes\omega_\tc{d}(\hat{U}_\tc{I}^\dagger\hat{a}_{\bm n}^\dagger \hat{a}_{\bm n} \hat{U}_I) =&\,\, \omega_\tc{d}(\hat{a}_{\bm n}^\dagger \hat{a}_{\bm n}) + \omega_\tc{d}(\omega(\hat{U}_I^{(1)\dagger}\hat{a}_{\bm n}^\dagger \hat{a}_{\bm n} + \hat{a}_{\bm n}^\dagger \hat{a}_{\bm n}\hat{U}_I^{(1)})) \\[0.5mm]
    &\:\:\:\:\:\:\:+ \omega_\tc{d}(\omega(\hat{U}_I^{(2)\dagger}\hat{a}_{\bm n}^\dagger \hat{a}_{\bm n} + \hat{a}_{\bm n}^\dagger \hat{a}_{\bm n}\hat{U}_I^{(2)}+\hat{U}_I^{(1)\dagger}\hat{a}_{\bm n}^\dagger \hat{a}_{\bm n}\hat{U}_I^{(1)})) + \mathcal{O}(\lambda^3).\nonumber\\[2mm]
    =& \,\, \omega_\tc{d}(\omega(\hat{U}_I^{(1)\dagger}\hat{a}_{\bm n}^\dagger \hat{a}_{\bm n}\hat{U}_I^{(1)})) + \mathcal{O}(\lambda^3)\nonumber,
\end{align}
where in the last equality we used that $\omega_\tc{d}(\hat{a}_{\bm n}^\dagger\hat{A}) = \omega_\tc{d}(\hat{A}\hat{a}_{\bm n}) = 0$ for all operators $\hat{A}$ when $\omega_\tc{d}$ is the vacuum of the localized field. Expressing $\hat{U}_I^{(1)}$ as in Eq.~\eqref{eq:UI1FV} we obtain $\omega_\tc{d}(\omega(\hat{U}_I^{(1)\dagger}\hat{a}_{\bm n}^\dagger \hat{a}_{\bm n}\hat{U}_I^{(1)})) = \lambda^2\omega(\hat{\phi}(\zeta u_{\bm n})\hat{\phi}(\zeta u_{\bm n}^*))$, matching Eq.~\eqref{eq:omegadanand}.

    Overall, we can use the unitary time evolution operator to compute the final state of the detector field $\hat{\phi}_\tc{d}$. We can simplify matters even further by considering faithful irreducible GNS representations for both theories $\hat{\phi}_\tc{d}$ and $\hat{\phi}$, where the initial states of each field can be represented as density operators $\hat{\rho}_{\tc{d},0} = \ket{0_\tc{d}}\!\!\bra{0_\tc{d}}$ and $\hat{\rho}_\phi$, representing the algebraic states equivalent to $\omega_\tc{d}$ and $\omega$. In this case, one can write the detector's final state as
    \begin{equation}\label{eq:phidtrphi}
        \hat{\rho}_{\tc{d}} = \tr_{\phi}(\hat{U}_I\r_0\hat{U}_I^\dagger), \quad \quad \r_0 = \hat{\rho}_{\tc{d,0}}\otimes\hat{\rho}_\phi,
    \end{equation}
    where $\tr_\phi$ represents the partial trace with respect to the degrees of freedom of the field $\hat{\phi}$. Eq.~\eqref{eq:phidtrphi} also matches the results obtained using the scattering endomorphism $\Theta$ acting directly in the algebras of observables.
     
    We can compute the final state of the fields using the power series of the time evolution operator, giving the final state in the form:
    \begin{equation}\label{eq:rhoDyson}
        \r_{\tc{d},\phi} = \r_0 + \r^{(1)} + \r^{(2)} + \mathcal{O}(\lambda^3),
    \end{equation}
    with
    \begin{align}
        \r^{(1)} &= \hat{U}^{(1)} \r_0 + \r_0 \hat{U}^{(1)\dagger},\\
        \r^{(2)} &= \hat{U}^{(2)} \r_0 + \hat{U}^{(1)} \r_0 \hat{U}^{(1)\dagger} + \r_0 \hat{U}^{(2)\dagger}.
    \end{align}
    Notice that because the interaction Hamiltonian is linear on $\hat{\phi}(\mf x)$ and the field starts in a zero-mean Gaussian state, we have $\text{tr}(\hat{\phi}(\mf x) \hat{\rho}_\phi) = \omega(\hat{\phi}(\mf x)) = 0$, so that the term $\hat{\rho}^{(1)}$ does not contribute after the partial trace over the degrees of freedom of $\hat{\phi}(\mf x)$. The term $\r^{(2)}$ can be written as
    \begin{align}
        \r^{(2)} = \lambda^2 \int \dd V \dd V'\zeta(\mf x) \zeta(\mf x') \Big(&\hat{\phi}_\tc{d}(\mf x) \hat{\phi}(\mf x)\hat{\rho}_0 \hat{\phi}_\tc{d}(\mf x') \hat{\phi}(\mf x')\\
        &-\hat{\phi}_\tc{d}(\mf x)\hat{\phi}_\tc{d}(\mf x')  \hat{\phi}(\mf x)\hat{\phi}(\mf x')\hat{\rho}_0 \theta(t-t') \nonumber\\
        &-\hat{\rho}_0 \hat{\phi}_\tc{d}(\mf x)\hat{\phi}_\tc{d}(\mf x')  \hat{\phi}(\mf x)\hat{\phi}(\mf x')\theta(t'-t) \Big)\nonumber.
    \end{align}
    Partial tracing over the free field $\hat{\phi}(\mf x)$, and using \mbox{$\hat{\rho}_0 = \ket{0_\tc{d}}\!\!\bra{0_\tc{d}}\!\otimes\hat{\rho}_\phi$}, we then obtain
    \begin{align}
        \tr_\phi(\r^{(2)})\! =\!\lambda^2\!\! \int\! \dd V \dd V'\zeta(\mf x) \zeta(\mf x')W(\mf x, \mf x')
        \Big(&\hat{\phi}_\tc{d}(\mf x')\ket{0_\tc{d}}\!\!\bra{0_\tc{d}}\hat{\phi}_\tc{d}(\mf x) \label{eq:midComputation}\\
        &-\hat{\phi}_\tc{d}(\mf x)\hat{\phi}_\tc{d}(\mf x')  \ket{0_\tc{d}}\!\!\bra{0_\tc{d}}\theta(t-t') \nonumber\\
        &-\ket{0_\tc{d}}\!\!\bra{0_\tc{d}}\hat{\phi}_\tc{d}(\mf x)\hat{\phi}_\tc{d}(\mf x')\theta(t'-t) \Big),\nonumber
    \end{align}
    where $W(\mf x, \mf x') = \omega(\hat{\phi}(\mf x) \hat{\phi}(\mf x'))$ is the Wightman function of the field $\hat{\phi}(\mf x)$ in the state $\omega$. Equation~\eqref{eq:midComputation} then yields the leading order corrections to the state of the fields after their interaction. However, this equation is not particularly illuminating.  We can instead look at the final state of the probe field at a given mode $\bm n$. 
    

    Let us then assume that we only have access to one mode of the localized field $\hat{\phi}_\tc{d}(\mf x)$ described in the subspace $\mathscr{H}_{\bm n}$ for a given $\bm n$, which labels an eigenfrequency $\omega_{\bm n}$ and its corresponding eigenmode $u_{\bm n}(\mf x)$. Denote by $\mathscr{H}_{\bm n}^c$ the complement of this Hilbert space in the tensor product decomposition of Eq.~\eqref{eq:Fockphid}, so that the detector field's Fock space factors as $\mathcal{F}(\mathscr{H}_\tc{d}) = \mathscr{H}_{\bm n}\otimes \mathscr{H}_{\bm n}^c$.

    The  density matrix $\hat{\rho}_{\tc{d},0} = \ket{0_\tc{d}}\!\!\bra{0_\tc{d}}$ then admits the decomposition
    \begin{equation}
        \ket{0_\tc{d}}\!\!\bra{0_\tc{d}} = \bigotimes_{\bm m, \bm m'}  \ket{0_{\bm m}}\!\!\bra{0_{\bm m'}} = \hat{\rho}_{\bm n,0}\!\!\!\!\!\!\!\!\!\!\!\bigotimes_{\:\:\:\:\:\:\:(\bm m,\bm m')\neq (\bm n, \bm n)} \!\!\!\!\!\!\!\! \!\!\!\ket{0_{\bm m}^\tc{d}}\!\!\bra{0_{\bm m'}^\tc{d}},
    \end{equation}
    where $\r_{\bm n,0} = \ket{0_{\bm n}}\!\!\bra{0_{\bm n}}$.
    
    The final state that we have access to will then be given by the partial trace over of the final state of the detectors system over both the target field $\phi$ and the Hilbert space $\mathscr{H}_{\bm n}^c$:
    \begin{equation}
        \hat{\rho}_{\bm n} = \tr_{\phi,\mathscr{H}_{\bm n}^c}\left(\hat{U} \hat{\rho}_0\hat{U}^\dagger\right).
    \end{equation}
    Physically, the restriction of the field to the space $\mathscr{H}_{\bm n}$ can be realized experimentally if one only has access to excitations of the localized field with energy $\omega_{\bm n}$. For instance, consider an electromagnetic cavity which contains photodetectors that can only measure `photons' that have energy $\omega_{\bm n}$. Effectively, an experimentalist would only have access to the space $\mathscr{H}_{\bm n}$, providing physical meaning to the partial trace operation above.

    The next step is to trace the result of Eq. \eqref{eq:midComputation} over the space $\mathscr{H}_{\bm n}^c$, which we assumed to be inaccessible. To perform this computation, we write
    \begin{align}
        \hat{\phi}_\tc{d}(\mf x) = \sum_{\bm m} \hat{\phi}_{\bm m}(\mf x), \quad\quad \hat{\phi}_{\bm m}(\mf x) = \hat{a}_{\bm m} u_{\bm m}(\mf x)+\hat{a}_{\bm m}^\dagger u_{\bm m}^*(\mf x),
    \end{align}
    so that
    \begin{equation}
        \tr_{\mathscr{H}_{\bm n}^c}\left(\hat{\phi}_\tc{d}(\mf x) \hat{\phi}_\tc{d}(\mf x')\right) = \sum_{\bm m \bm m'} \tr_{\mathscr{H}_{{\bm n}}^c}\left(\hat{\phi}_{\bm m}(\mf x) \hat{\phi}_{\bm m'}(\mf x')\right).
    \end{equation}
    From Eq.~\eqref{eq:vacTensor}, the vacuum $\ket{0_\tc{d}}$ can be decomposed in terms of the ground state of each mode.
     Noticing that each $\hat{\phi}_{\bm m}(\mf x)$ term only contains one creation and one annihilation operator, we then find that for $(\bm m, \bm m') \neq (\bm n, \bm n)$,
    \begin{align}\label{eq:diagonal0D}
        \tr_{\mathscr{H}_{{\bm n}}^c}\Big(\hat{\phi}_{\bm m}(\mf x)\hat{\phi}_{\bm m'}(\mf x')\ket{0_\tc{d}}\!\!\bra{0_\tc{d}}\Big) = \delta_{\bm m ,\bm m'} u_{\bm m}(\mf x)  u_{\bm m'}^*(\mf x')\r_{{\bm n},0},
    \end{align}
    which is simply a multiple of the initial state of the mode $\bm n$. For $\bm m = \bm m' = \bm n$, the trace over $\mathscr{H}_{\bm n}^c$ simply gets rid of the components of the state in $\mathscr{H}_{\bm n}^c$, without affecting the components in $\mathscr{H}_{\bm n}$. 
    
    Using these results it is possible to trace Eq. \eqref{eq:midComputation} over the space $\mathscr{H}_{\bm n}^c$, which yields
    \begin{align}
        \!\!\!\r_{\bm n}\!&=\!\hat{\rho}_{\bm n,0} + \lambda^2 \!\int\! \dd V \dd V'\zeta(\mf x) \zeta(\mf x')W(\mf x, \mf x')
        \Big(\hat{\phi}_{\bm n}(\mf x')\r_{{\bm n},0}\hat{\phi}_{\bm n}(\mf x) \label{eq:finalRhoQFTPD}-\hat{\phi}_{\bm n}(\mf x)\hat{\phi}_{\bm n}(\mf x')  \r_{\bm n,0}\theta(t-t')\\
        &\:\:\:\:\:\:\:\:\:\:\:\:\:\:\:\:\:\:\:\:\:\:\:\:\:\:\:\:\:\:\:\:\:\:\:\:\:\:\:\:\:\:\:\:\:\:\:\:\:\:\:\:\:\:\:\:\:\:\:\:\:\:\:\:\:\:\:\:\:\:\:\:\:\:\:\:\:\:\:\:\:\:\:\:\:\:\:\:\:\:\:\:\:\:\:\:\:\:\:\:\:\:\:\:-\r_{{\bm n},0} \hat{\phi}_{\bm n}(\mf x)\hat{\phi}_{\bm n}(\mf x')\theta(t'-t) \Big)\nonumber\\
        &+\lambda^2 \r_{{\bm n},0}\sum_{\bm m\neq \bm n}\int \dd V \dd V' \zeta(\mf x) \zeta(\mf x')W(\mf x, \mf x')  u_{\bm m} (\mf x)u^*_{\bm m}(\mf x') \big( 1- \theta(t-t') - \theta(t'-t)\big) + \mathcal{O}(\lambda^4).\nonumber
    \end{align}
    Notice that the last term cancels, given that \mbox{$\theta(t-t') + \theta(t'-t) = 1$}. Defining the normalized states
    \begin{align}
        \ket{1_{\bm n}} = \hat{a}_{\bm n}^\dagger \ket{0_{\bm n}}, \quad\quad \ket{2_{\bm n}} = \frac{1}{\sqrt{2}}\hat{a}_{\bm n}^\dagger\hat{a}_{\bm n}^\dagger \ket{0_{\bm n}},
    \end{align}
    we can further write the final state $\hat{\rho}_{\bm n}$ as 
    \begin{align}
        \r_{\bm n} = \big(1 & - \lambda^2 W(\zeta u_{\bm n},\zeta u_{\bm n}^*)\big)\ket{0_{\bm n}}\!\!\bra{0_{\bm n}} + \lambda^2 W(\zeta u_{\bm n},\zeta u_{\bm n}^*)\ket{1_{\bm n}}\!\!\bra{1_{\bm n}} \\[1mm]
        &- \tfrac{\lambda^2}{\sqrt{2}} G_F(\zeta u_{\bm n}^*,\zeta u_{\bm n}^*)\ket{2_{\bm n}}\!\!\bra{0_{\bm n}} - \tfrac{\lambda^2}{\sqrt{2}} G_F(\zeta u_{\bm n},\zeta u_{\bm n})\ket{0_{\bm n}}\!\!\bra{2_{\bm n}} + \mathcal{O}\nonumber(\lambda^4),
    \end{align}
    where we used the explicit expansion~\eqref{eq:GNSphid}, as well as
    \begin{equation}
         \int \dd V \dd V' \zeta(\mf x) \zeta(\mf x') u_{\bm n}^*(\mf x)u_{\bm n}^*(\mf x') W(\mf x, \mf x') \theta(t-t') = \tfrac{1}{2}G_F(\zeta u_{\bm n}^*,\zeta u_{\bm n}^*).
    \end{equation}
    
    The Feynman propagator terms also arise in the Fewster-Verch formulation. This can be seen by noticing that
    \begin{equation}
        -\lambda^2 G_F(\zeta u_{\bm n}^*, \zeta u_{\bm n}^*) = \langle\hat{a}_{\bm n} \hat{a}_{\bm n}\rangle_{\hat{\rho}_{\bm n}}.
    \end{equation}
    Indeed, evaluating Eq.~\eqref{eq:FVWfg} at $f = g = \ii g_{\bm n}^*$ and using the expansion for the field $\hat{\phi}_\tc{d}$ in~\eqref{eq:GNSphid}, we find
    \begin{align}
        \varpi_\tc{d}(\hat{a}_{\bm n} \hat{a}_{\bm n}) &= - \lambda^2 \omega(\hat{\phi}(\zeta u_{\bm n}^*)\hat{\phi}(\zeta u_{\bm n}^*)) - \ii \lambda^2 G_A(\zeta u_{\bm n}^*,\zeta u_{\bm n}^*)\\
        &= - \lambda^2 W(\zeta u_{\bm n}^*,\zeta u_{\bm n}^*) - \ii \lambda^2 G_A(\zeta u_{\bm n}^*,\zeta u_{\bm n}^*) = - \lambda^2 G_F(\zeta u_{\bm n}^*,\zeta u_{\bm n}^*),
    \end{align}
    matching the result obtained using the time evolution operator.
    
    At this stage, it is possible to reinterpret the final result by considering the following effective scalar Hamiltonian density
    \begin{equation}\label{eq:HeffFV}
        \hat{\mathcal{H}}_{\text{eff}}(\mf x) =
        \lambda \zeta(\mf x)\hat{\phi}_{\bm n}(\mf x) \hat{\phi}(\mf x) 
        = \lambda \!\left(\Lambda_{\bm n}^-(\mf x)\hat{a}_{\bm n}  +\Lambda_{\bm n}^+(\mf x) \hat{a}_{\bm n}^\dagger \right)\hat{\phi}(\mf x),
    \end{equation}
    where we defined 
    \begin{equation}
        \Lambda_{\bm n}^-(\mf x) = \zeta(\mf x) u_{{\bm n}}(\mf x), \quad \text{and} \quad \Lambda_{\bm n}^+(\mf x) = (\Lambda_{\bm n}^-(\mf x))^*.
    \end{equation}
    The operator $\hat{\mathcal{H}}_\text{eff}(\mf x)$ only acts on the Hilbert space of the field $\hat{\phi}(\mf x)$ and on the Hilbert space $\mathscr{H}_{{\bm n}}$, which is effectively a harmonic oscillator with frequency $\omega_{\bm n}$. Defining the unitary time evolution operator
    \begin{equation}
        \hat{U}_{\text{eff}} = \mathcal{T}\exp\left(- \ii \int \dd V \hat{\mathcal{H}}_{\text{eff}}(\mf x)\right),
    \end{equation}
    it is possible to show that the leading order result for $\r_{{\bm n}}$ in Eq. \eqref{eq:finalRhoQFTPD} can be rewritten as
    \begin{equation}
        \r_{{\bm n}} = \tr_\phi\left(\hat{U}_\text{eff}(\r_{{\bm n},0}\otimes \r_\phi )\hat{U}_{\text{eff}}^\dagger\right) + \mathcal{O}(\lambda^4).
    \end{equation}
    That is, to second order in the coupling strength, it is possible to reproduce the final state of $\hat{\phi}_\tc{d}(\mf x)$ in each individual subspace $\mathscr{H}_{{\bm n}}$ by considering a linear interaction between a harmonic oscillator with the quantum field $\hat{\phi}(\mf x)$, considerably simplifying the interaction.




\section{Unruh-DeWitt Detector Models}\label{sec:UDW}

The discussion of Section~\ref{sec:meas} suggests that one can define a simplified ``detector'' model for a quantum field by restricting their attention to a specific mode of the detector field $\hat{\phi}_\tc{d}$, accessing the expected values of operators of the form $\hat{\phi}(\rho u_{\bm n})$ and $\hat{\phi}(\rho u_{\bm n}^*)$. This is the idea behind what have become known as Unruh-DeWitt detectors, after Bill Unruh~\cite{Unruh1976} and Bryce DeWitt~\cite{DeWitt}. The term Unruh-DeWitt detector has been used to describe many different models, and different authors may use it with a slightly different meaning, sometimes referring to the whole class of models or to its simplest version. Our goal in this section is to describe a general formulation of Unruh-DeWitt detectors, and to explore specific examples that will help build intuition about these models.

\subsubsection*{General Particle Detector Models}

We will now define particle detector models. Broadly speaking, a particle detector model consists of a quantum system (modelling the detector) that interacts with a quantum field in a localized spacetime region. In the description of particle detectors, one also has to provide a local timelike coordinate $\tau$ that determines the evolution of the system. 

We start with the definition of a particle detector that couples to a Lorentz scalar self-adjoint operator in a globally hyperbolic 3+1 dimensional spacetime $\M$. A particle detector of this type is defined by
\begin{enumerate}
    \item[\textbf{1.}] A positive-oriented local time coordinate $\tau$;
    \item[\textbf{2.}] A quantum system associated with a Hilbert space $\mathscr{H}_\tc{d}$ with Hamiltonian $\hat{H}_\tc{d}(\tau)$;
    \item[\textbf{3.}] A self-adjoint operator-valued compactly supported\footnote{This requirement is generally relaxed, allowing $\hat{J}(\mf x)$ to be a sufficiently localized spacetime function, rather than compactly supported.} spacetime function $\hat{J}(\mf x)$ acting on $\mathscr{H}_\tc{d}$---the detector's smeared monopole moment---supported in the region where the local coordinate $\tau$ is defined; 
    \item[\textbf{4.}] The kernel $\hat{O}(\mf x)$ of a self-adjoint Lorentz scalar operator-valued distribution \mbox{$f\mapsto \hat{O}(f)$} in the $\ast$-algebra associated with a quantum field theory;
    \item[\textbf{5.}] A coupling constant $\lambda$;
\end{enumerate}
With the structures above, we define the time-evolved monopole $\hat{\jmath}(\mf x)$ by introducing the dynamics due to the Hamiltonian $\hat{H}_\tc{d}$ to $\hat{J}(\mf x)$ through the Heisenberg equation
\begin{equation}
    \ii \pdv{\hat{\jmath}}{\tau} = [\hat{\jmath},\hat{H}_\tc{d}] + \ii \pdv{\hat{J}}{\tau},
\end{equation}
with initial condition $\hat{\jmath}|_{\tau = 0} = \hat{J}|_{\tau = 0}$. That is, we incorporate the time evolution generated by the Hamiltonian $\hat{H}_\tc{d}$ to the operator $\hat{J}(\mf x)$. The particle detector model is then defined by the interaction Hamiltonian density
\begin{equation}
    \hat{\mathcal{H}}_I(\mf x) = \lambda \hat{\jmath}(\mf x) \hat{O}(\mf x),
\end{equation}
which generates time evolution with respect to $\tau$. This simple model turns out to capture the essential features of interactions of systems with quantum fields.

Looking at conditions \textbf{1.}-\textbf{5.} one could define the interacting fields $\hat{\phi}$ and $\hat{\phi}_\tc{d}$ defined by the Lagrangian~\eqref{eq:LagphiphidFV} as a particle detector model. Indeed, one can consider \textbf{1.} $\tau = t$, where $(t,\bm x)$ are coordinates such that $t$ is the flux of a timelike Killing vector field and $\bm x$ are coordinates in the associated foliation, \textbf{2.} the Hilbert space $\mathcal{F}(\mathscr{H}_\tc{d})$ defined in~\eqref{eq:Fockphid} and the corresponding field Hamiltonian
\begin{equation}\label{eq:HIgeneralPD}
    \hat{H}_\tc{d} = \sum_{\bm n} \omega_{\bm n} \hat{a}_{\bm n}^\dagger \hat{a}_{\bm n},
\end{equation}
\textbf{3.} $\hat{J}(\mf x) = \rho(\mf x) \hat{\phi}_{\tc{d}}(0,\bm x)$, \textbf{4.} $\hat{O}(\mf x) = \hat{\phi}(\mf x)$, and \textbf{5.} the coupling constant $\lambda$. This would formally reproduce the detection model discussed in the previous section, with the caveat that the monopole operator $\hat{J}(\mf x) = \rho(\mf x) \hat{\phi}_{\tc{d}}(0,\bm x)$ is not a regular operator-valued spacetime function.

Rather than attempting to encompass the field detection models we have been discussing so far, the goal of a particle detector model, as defined here, is to allow one to consider simpler probes for quantum fields, which can be described using more standard quantum mechanics techniques. This has the advantage of simplifying the descriptions of measurements in quantum field theory, as well as allowing one to apply standard quantum information techniques to the probes. Particle detectors also facilitate the descriptions of quantum information protocols that use quantum fields as intermediates and allow one to quantify important aspects of the probes after the interaction, such as their mixedness and entanglement between different probes. As we will see, the price we will have to pay for the considerable simplification provided by particle detectors is that the operations implemented in a quantum field theory by detectors will not usually be safe from Sorkin-type problems.

Along the lines of simplifying the interaction discussed in the previous section, instead of reproducing the full interaction of the detector field $\hat{\phi}_\tc{d}$ after interacting with the field $\hat{\phi}$, one can define a particle detector model that reproduces the leading order results obtained for a given mode $\bm n$ of the detector field. Consider \textbf{1.} $\tau = t$ as in~\eqref{eq:staticmetric}, \textbf{2.} $\mathscr{H}_\tc{d} = \mathscr{H}_{\bm n}$ (defined in Eq.~\eqref{eq:Fockphid}) together with the Hamiltonian $\hat{H}_{\tc{d}} = \omega_{\bm n} \hat{a}_{\bm n}^\dagger \hat{a}_{\bm n}$, \textbf{3.} the operator
\begin{equation}\label{eq:HeisenbergPD}
    \hat{J}(\mf x) = \zeta(\mf x) (\Phi_{\bm n}(\bm x)\hat{a}_{\bm n} + \Phi_{\bm n}^*(\bm x)\hat{a}_{\bm n}^\dagger),
\end{equation}
\textbf{4.} the field operator $\hat{\phi}(\mf x)$ and $\textbf{5.}$ the coupling constant $\lambda$. With these choices, we find the interaction Hamiltonian density
\begin{equation}
    \hat{\mathcal{H}}_I(\mf x) = \lambda \zeta(\mf x) (\Phi_{\bm n}(\bm x)\hat{a}_{\bm n} e^{- \ii \omega_{\bm n} t} + \Phi_{\bm n}^*(\bm x)\hat{a}_{\bm n}^\dagger e^{\ii \omega_{\bm n} t})\hat{\phi}(\mf x),
\end{equation}
which exactly matches the effective Hamiltonian found in~\eqref{eq:HeffFV}. If the modes $\Phi_{\bm n}(\bm x)$ are real, it is typical to define the spacetime smearing function $\Lambda_{\bm n}(\mf x) = \zeta(\mf x)\Phi_{\bm n}(\bm x)$ so that the interaction Hamiltonian density becomes
\begin{equation}\label{eq:HOUDW}
    \hat{\mathcal{H}}_I(\mf x) = \lambda \Lambda_{\bm n}(\mf x)(\hat{a}_{\bm n} e^{- \ii \omega_{\bm n} t} + \hat{a}_{\bm n}^\dagger e^{\ii \omega_{\bm n} t})\hat{\phi}(\mf x).
\end{equation}
Particle detector models that can be put in the form of Eq.~\eqref{eq:HOUDW}, with $\hat{a}_{\bm n}$ and $\hat{a}_{\bm n}^\dagger$ satisfying $[\hat{a}_{\bm n},\hat{a}_{\bm n}^\dagger] = \openone$ are usually called harmonic-oscillator Unruh-DeWitt detectors\footnote{It is also usual to assume that $\lambda$ is dimensionless and $\Lambda(\mf x)$ is a spatial density in these models.}~\cite{BeiLokHOUDW1994,CharisHO2020,HODan2021,EricksonZero,MariaContinuous}.

What is commonly accepted as the simplest model of an Unruh-DeWitt detector is the two-level Unruh-DeWitt model. It can be obtained by considering any positively oriented timelike coordinate $\tau$\footnote{It is common to pick this timelike coordinate as the Fermi normal coordinate time associated to a given worldline $\mf z(\tau)$. We will discuss more about this in Section~\ref{sec:NRLQS}.}, ${\mathscr{H}}_\tc{d} = \mathbb{C}^2$, $\hat{H}_\tc{d} = \Omega\hat{\sigma}^+\hat{\sigma}^-$, and the operator $\hat{J}(\mf x) = \Lambda(\mf x) \hat{\mu}$, where $\hat{\mu} = \hat{\sigma}^+ + \hat{\sigma}^-$ is the detector's monopole moment. Here $\hat{\sigma}^\pm$ are $\mathfrak{su}(2)$ raising and lowering operators, satisfying $\hat{\sigma}^+\hat{\sigma}^- + \hat{\sigma}^-\hat{\sigma}^+ = \openone$ and $(\hat{\sigma}^\pm)^2 = 0$. The function $\Lambda(\mf x)$ is real and is usually called the spacetime smearing function. It is usually defined as having dimensions of a spatial density, making the coupling constant dimensionless. Its profile defines the region where the detector interacts with the field. The interaction Hamiltonian density then becomes
\begin{equation}\label{eq:HIUDW}
    \hat{\mathcal{H}}_I(\mf x) = \lambda \Lambda(\mf x) (e^{\ii \Omega \tau} \hat{\sigma}^+ + e^{-\ii \Omega \tau} \hat{\sigma}^-)\hat{\phi}(\mf x).
\end{equation}
The Hamiltonian $\hat{H}_\tc{d}$ defines its eigenvectors $\ket{g}$ and $\ket{e}$ as the ground and excited states, respectively, and the energy gap of the system is $\Omega$.

The two-level Unruh-DeWitt detector has become the basic tool for implementing quantum information protocols in quantum field theory. It can be thought of as the qubit of relativistic quantum information, analogous to the fundamental role played by qubits in quantum computing. As such, the two-level Unruh-DeWitt detector will be the main model that we will utilize throughout this thesis to probe quantum fields. We will discuss it in more detail briefly.

One can also generalize the notion of particle detectors to the case where a detector couples to any operator in a quantum field theory. For this definition we require not only a timelike coordinate $\tau$, but also a locally defined orthonormal frame. A fairly general particle detector can be defined by
\begin{enumerate}
    \item[\textbf{1.}] A positive-oriented local time coordinate $\tau$ and an orthonormal frame $\mf e_{\alpha}$ defined in the region where $\tau$ is defined;
    \item[\textbf{2.}] A quantum system associated with a Hilbert space $\mathscr{H}_\tc{d}(\tau)$ with Hamiltonian $\hat{H}_\tc{d}$;
    \item[\textbf{3.}] An operator-valued spacetime tensor field 
    \begin{equation}
        \hat{J}(\mf x) = \hat{J}^{\alpha_1...\alpha_n}{}_{\beta_1...\beta_m}(\mf x)\mf e_{\alpha_1}\otimes...\otimes \mf e_{\alpha_n} \otimes \mf e^{\beta_1}\otimes ... \otimes \mf e^{\beta_m},
    \end{equation} 
    where for each $\alpha_1,...\alpha_n,\beta_1,...,\beta_m$ and $\mf x$, $\hat{J}^{\alpha_1...\alpha_n}{}_{\beta_1...\beta_m}(\mf x)$ is an operator acting on $\mathscr{H}_\tc{d}$, supported in the region where the local coordinate $\tau$ is defined;
    \item[\textbf{4.}] The kernel $\hat{O}_{\alpha_1...\alpha_n}{}^{\beta_1...\beta_m}(\mf x)$ of an operator-valued distribution $f^{\alpha_1...\alpha_n}{}_{\beta_1...\beta_m}\mapsto \hat{O}(f)$ in the $\ast$-algebra associated with a quantum field theory;
    \item[\textbf{5.}] A coupling constant $\lambda$;
\end{enumerate}
We define the operator-valued tensor $\hat{\jmath}(\mf x)$, of the same rank as $\hat{J}(\mf x)$ by imposing the Heisenberg equation~\eqref{eq:HeisenbergPD} component by component in the $\mf e_\alpha$ basis. The interaction Hamiltonian density generating time evolution with respect to $\tau$ is then defined as
\begin{equation}\label{eq:PDgeneralTensor}
    \hat{\mathcal{H}}_I(\mf x) = \lambda\, \hat{\jmath}(\mf x) \cdot \hat{O}(\mf x) + \text{H.c.},
\end{equation}
where $\,\cdot\,$ denotes Lorentz contraction, and we add the Hermitian conjugate, given that neither $\hat{O}(\mf x)$ or $\hat{\jmath}(\mf x)$ need to be self-adjoint. We fix the frame $\mf e_\alpha$ here so that the quantum theory for the detector does not need to be defined relativistically. Of course, if $\hat{\jmath}(\mf x)$ can be associated with the kernel of an operator-valued distribution in a relativistic quantum field theory, the choice of frame is irrelevant.

One example of particle detector that fulfills the definition above is that of a linear complex scalar field detector, introduced in~\cite{antiparticles}. Given that it is a scalar field, one need not pick an orthonormal frame, and we can pick \textbf{1.} any positively time oriented coordinate $\tau$, \textbf{2.} $\mathscr{H}_\tc{d} = \mathbb{C}^2$ and $\hat{H}_{\tc{d}} = \Omega\hat{\sigma}^+\hat{\sigma}^-$, \textbf{3.} $\hat{J}(\mf x) = \Lambda(\mf x) \hat{\sigma}^+$, where $\Lambda(\mf x)$ is a complex spacetime smearing function, \textbf{4.} (the conjugate of) a complex scalar field $\hat{\psi}^\dagger(\mf x)$, and \textbf{5.} a coupling constant $\lambda$. The resulting interaction Hamiltonian density is
\begin{equation}
    \hat{\mathcal{H}}_I(\mf x) = \lambda (\Lambda(\mf x) e^{\ii\Omega \tau}\hat{\sigma}^+ \hat{\psi}^\dagger(\mf x) + \Lambda^*(\mf x) e^{-\ii\Omega \tau}\hat{\sigma}^- \hat{\psi}(\mf x)). 
\end{equation}
In~\cite{antiparticles} it was shown that this model can reproduce some features of the interactions of nucleons with neutrinos~\cite{neutrinos}. 

A last example that we will mention at this point is that of a two-level Unruh-DeWitt detector linearly coupled to the momentum of a scalar field. This model can be defined by picking \textbf{1.} a positively time oriented coordinate $\tau$, and a frame $\mf e_\mu$ such that $\mf e_0 = n$ is normal to the surfaces of constant $\tau$, \textbf{2.} $\mathscr{H}_\tc{d} = \mathbb{C}^2$ and $\hat{H}_{\tc{d}} = \Omega\hat{\sigma}^+\hat{\sigma}^-$, \textbf{3.} the tensor valued field operator $\hat{J}(\mf x) = \Lambda(\mf x) (\hat{\sigma}^+ + \hat{\sigma}^-)n$, \textbf{4.} the derivative of a real scalar field, $\nabla_{\mu}\hat{\phi}(\mf x)$, and \textbf{5.} a coupling constant $\lambda$. The resulting interaction Hamiltonian is then
\begin{equation}
    \hat{\mathcal{H}}_I(\mf x) =\lambda \Lambda(\mf x) (e^{\ii \Omega \tau} \hat{\sigma}^+ + e^{-\ii \Omega \tau} \hat{\sigma}^-)n^\mu \nabla_\mu\hat{\phi}(\mf x),
\end{equation}
where we can identify the momentum operator $\hat{\pi}(\mf x) = n^\mu\nabla_\mu\hat{\phi}(\mf x)$. Notice that this model could also be defined as a scalar model coupled to the scalar operator $\hat{\pi}(\mf x)$, but we mentioned it in this context for the purpose of illustrating the role of the frame $\mf e_\mu$ with a specific example.

The definition presented above is sufficiently general to describe nearly any physical interaction with a quantum field, such as the interactions of atoms with the electromagnetic field~\cite{Pozas2016,Nicho1,richard}, the interactions of nucleons with neutrinos~\cite{neutrinos,antiparticles,carol}, and the interaction of quantum systems with linearized quantum gravity~\cite{remi,pitelli,boris}.

\subsubsection*{The Dynamics of Particle Detectors}

We can now study general features of the interaction of a detector with a quantum field observable. For simplicity, we will keep our discussion to the case of detectors coupled to a scalar self-adjoint field operator $\hat{O}(\mf x)$. Having the interaction Hamiltonian density $\hat{\mathcal{H}}_I(\mf x)$ from Eq.~\eqref{eq:HIgeneralPD}, it would be natural to expect that dynamics will be given by a time evolution operator $\hat{U}_I$ of the form
\begin{equation}
    \hat{U}_I = \mathcal{T}\exp(-\ii \int \dd V \hat{\mathcal{H}}_I(\mf x)),
\end{equation}
where $\mathcal{T}\exp$ is the time ordered exponential, defined in~\eqref{eq:Texp}. However, the time ordered exponential is only independent on the prescription of time parameter if $\hat{\mathcal{H}}_I(\mf x)$ satisfies the microcausality condition $[\hat{\mathcal{H}}_I(\mf x),\hat{\mathcal{H}}_I(\mf x)] = 0$ for causally disconnected $\mf x$ and $\mf x'$. Indeed, we can write
\begin{equation}\label{eq:auxTimeOrd}
    \mathcal{T}\hat{\mathcal{H}}_I(\mf x)\hat{\mathcal{H}}_I(\mf x') = \tfrac{1}{2}\{\hat{\mathcal{H}}_I(\mf x),\hat{\mathcal{H}}_I(\mf x')\} + \tfrac{1}{2}[\hat{\mathcal{H}}_I(\mf x),\hat{\mathcal{H}}_I(\mf x')]\text{sign}(t-t').
\end{equation}
If $s$ is another positively time oriented coordinate, we have that $\text{sign}(t-t') = \text{sign}(s-s')$ whenever $\mf x$ and $\mf x'$ are timelike separated. However, for spacelike separated events, the time coordinates might not have the same order. Thus, unless $[\hat{\mathcal{H}}_I(\mf x),\hat{\mathcal{H}}_I(\mf x')] = 0$ when $\mf x$ and $\mf x'$ are spacelike separated, replacing $t-t'$ by $s-s'$ in~\eqref{eq:auxTimeOrd} will generally yield different operators. That is, the time ordering operation is only uniquely defined if $\hat{\mathcal{H}}_I(\mf x)$ satisfies the microcausality condition. 

For a particle detector with Hamiltonian density given by $\hat{\mathcal{H}}_I(\mf x)$, if $\mf x$ and $\mf x'$ are spacelike separated, we have
\begin{equation}\label{eq:HIcomm}
    [\hat{\mathcal{H}}_I(\mf x),\hat{\mathcal{H}}_I(\mf x')] = \lambda^2 [\hat{\jmath}(\mf x),\hat{\jmath}(\mf x)]\hat{O}(\mf x)\hat{O}(\mf x'),
\end{equation}
where we used that $\hat{O}(\mf x)$ satisfies the microcausality condition, as a consequence of axiom \textbf{A3}. Equation~\eqref{eq:HIcomm} shows that unless the $\hat{j}(\mf x)$ satisfies the microcausality condition, the time evolution operator cannot be defined by Eq.~\eqref{eq:Texp}.

The fact that the detector's monopole moment typically does not commute with itself at spacelike separated points is an important feature of the definitions of particle detectors given here. Indeed, if $\hat{\jmath}(\mf x)$ satisfies the microcausality condition, the association of compactly supported functions $f\mapsto\hat{\jmath}(f)$ through spacetime integration defines local algebras of observables that fulfill conditions \textbf{A1-A3} of Section~\ref{sec:QFT}\footnote{\textbf{A4} is not necessarily fulfilled, as it depends on the specific dynamics defined by $\hat{H}_\tc{d}$.}, essentially defining it as a quantum field theory. In other words, the only types of particle detectors that fulfill the microcausality condition are those defined by quantum field theories, effectively containing infinitely many degrees of freedom. Instead, a usual particle detector is a simpler system, containing at most a finite number of quantum degrees of freedom. For this reason, it is a common saying that ``particle detector models are non-relativistic''. 

Indeed, even the harmonic oscillator model, which reproduces the microcausal interaction of quantum fields to leading order, violates the microcausality condition. This is due to the fact that a mode of a localized quantum field corresponds to a quantum degree of freedom that is extended in space. Thus, the model~\eqref{eq:HIgeneralPD} couples one degree of freedom to all spacelike separated points within the support of the corresponding mode, introducing an effective interaction that couples to the field at multiple spacelike separated points simultaneously. This is also not in contradiction with the microcausality axiom \textbf{A3}, as the operators $\hat{a}_{\bm n} = \hat{\phi}(\ii g_{\bm n}^*)$ and their conjugate are localized in the whole worldtube and are not associated with an operator density. Overall, no individual mode of the field satisfies the microcausality condition, only their collective. 

We then need a specific way to describe the dynamics of a particle detector. We define the time evolution of the system with respect to the time parameter $\tau$ as
\begin{equation}
    \hat{U}_I = \mathcal{T}_\tau\exp(-\ii \int \dd V \hat{\mathcal{H}}_I(\mf x))\coloneqq \sum_{n=0}^\infty \hat{U}_I^{(n)},
\end{equation}
where $\mathcal{T}_\tau\exp$ is the time ordered exponential with respect to the time parameter $\tau$:
\begin{align}
    \hat{U}_I^{(0)} &= \openone,\quad\quad
    \hat{U}_I^{(1)} = - \ii \int \dd V \hat{\mathcal{H}}_I(\mf x),\\
    \hat{U}_I^{(n)} &= (-\ii)^n \int \dd V_1 ... \dd V_n \hat{\mathcal{H}}(\mf x_1)...\hat{\mathcal{H}}(\mf x_n) \theta(\tau_n - \tau_{n-1})...\theta(\tau_2 - \tau_{1}), \text{ for } n\geq 2.
\end{align}
It is important to stress that this is a \textit{prescription}. The fact that the time evolution operator is only defined in a given frame is a consequence of the fact that the definition of detector models allows the detectors to be non-relativistic systems that have only a single quantum degree of freedom associated with a whole spatial slice. This is in contrast to quantum fields, which effectively have degrees of freedom associated to every point of space. The time parameter $\tau$ typically defines the rest space of the system, where a non-relativistic approximation can be used to describe it. In other words, one should not expect a particle detector model to yield accurate predictions if one chooses a time parameter $\tau$ that cannot be approximated as the proper time of the detector. We will discuss this in detail in Section~\ref{sec:NRLQS}, where we will see how to describe non-relativistic quantum systems in curved spacetimes and how to describe a particle detector starting from a physical system undergoing a predetermined trajectory in spacetime.


Having a clear definition of the time evolution operator in the context of particle detectors, we can now discuss the dynamics implemented by the interaction with the field observable $\hat{O}(\mf x)$. To that end, we will assume that the initial field state $\omega$ can be represented in a given GNS representation of the corresponding field. This allows us to write the field state previous to the interaction as a density operator $\hat{\rho}_\phi$ in the Fock space of $\hat{\phi}$. We consider the detector to be in a state $\hat{\rho}_{\tc{d},0}$, such that the initial field-detector state before the interaction is given by $\r_0 = \r_{\tc{d},0}\otimes\r_{\phi}$. The final state of the detector-field system can then be written as a series expansion in the coupling constant $\lambda$:
\begin{equation}
    \r_{\tc{d},\phi} = \hat{U}_I \r_{0}\hat{U}_I^\dagger = \r_0 + \hat{U}^{(1)} \r_0 + \r_0 \hat{U}^{(1)\dagger} + \hat{U}^{(2)} \r_0 + \hat{U}^{(1)} \r_0 \hat{U}^{(1)\dagger} + \r_0 \hat{U}^{(2)\dagger} + \mathcal{O}(\lambda^3).
\end{equation}
Explicitly, we have
\begin{align}
    \hat{U}^{(1)} \r_0 &=  -\ii \lambda \int \dd V \hat{\jmath}(\mf x) \r_{\tc{d},0}\otimes\hat{O}(\mf x)\r_{\omega} =  (\r_0\hat{U}^{(1)\dagger})^\dagger,\\
    \hat{U}^{(2)} \r_0 &=  -\lambda^2 \int \dd V \dd V' \theta(\tau-\tau')\hat{\jmath}(\mf x)\hat{\jmath}(\mf x') \r_{\tc{d},0}\otimes\hat{O}(\mf x)\hat{O}(\mf x')\r_{\omega} = (\r_0\hat{U}^{(2)\dagger})^\dagger,\nonumber\\
    \hat{U}^{(1)} \r_0\hat{U}^{(1)\dagger} & =  \lambda^2 \int \dd V \dd V'\hat{\jmath}(\mf x) \r_{\tc{d},0}\hat{\jmath}(\mf x')\otimes\hat{O}(\mf x)\r_{\omega}\hat{O}(\mf x').\nonumber
\end{align}
The final state of the detector is obtained by taking the partial trace of each term in the expression above with respect to the field: $\hat{\rho}_{\tc{d}} = \tr_\phi(\r_{\tc{d},\phi})$:
\begin{align}
    \tr_\phi(\hat{U}^{(1)} \r_0) &=  -\ii \lambda \int \dd V \omega(\hat{O}(\mf x)) \hat{\jmath}(\mf x) \r_{\tc{d},0} =  (\tr_\phi(\r_0\hat{U}^{(1)\dagger}))^\dagger,\label{eq:dynUDW}\\
    \tr_\phi(\hat{U}^{(2)} \r_0) &=  -\lambda^2 \int \dd V \dd V'\omega(\hat{O}(\mf x)\hat{O}( \mf x'))\theta(\tau-\tau')\hat{\jmath}(\mf x)\hat{\jmath}(\mf x') \r_{\tc{d},0} = (\tr_\phi(\r_0\hat{U}^{(2)\dagger}))^\dagger,\nonumber\\
    \tr_\phi(\hat{U}^{(1)} \r_0\hat{U}^{(1)\dagger}) & =  \lambda^2 \int \dd V \dd V' \omega(\hat{O}(\mf x)\hat{O}( \mf x'))\hat{\jmath}(\mf x') \r_{\tc{d},0}\hat{\jmath}(\mf x),\nonumber
\end{align}
where we denoted the kernel of the distribution $\omega(\hat{O}(f))$ by $\omega(\hat{O}(\mf x))$, the kernel of the bi-distribution $\omega(\hat{O}(f)\hat{O}(g))$ by $\omega(\hat{O}(\mf x)\hat{O}( \mf x'))$ and performed a change of variables $\mf x \leftrightarrow \mf x'$ in the last equation.

The results in~\eqref{eq:dynUDW} are as far as we can go in computing the final state of the detector without further assumptions regarding the specific detector model considered. At this stage we can interpret the interaction with the quantum field as a quantum channel that maps $\hat{\rho}_{\tc{d},0}\mapsto \hat{\rho}_\tc{d}$, applying the gate $\hat{\jmath}(\mf x') \r_{\tc{d},0}\hat{\jmath}(\mf x)$ weighted by the two point function $\omega(\hat{O}(\mf x)\hat{O}( \mf x'))$, the gates $\hat{\jmath}(\mf x)\hat{\jmath}(\mf x')\r_{\tc{d},0}$ and $\r_{\tc{d},0}\hat{\jmath}(\mf x')\hat{\jmath}(\mf x)$ weighted by $\omega(\hat{O}(\mf x)\hat{O}( \mf x'))\theta(\tau-\tau')$ and the gates $\hat{\jmath}(\mf x) \r_{\tc{d},0}$ and $\r_{\tc{d},0}\hat{\jmath}(\mf x)$ weighted by the one-point function $\omega(\hat{O}(\mf x))$. Importantly, the localization of the operators $\hat{\jmath}(\mf x)$ defines the region where the one and two point functions of the observable $\hat{O}(\mf x)$ are being probed. That is, the final state of the detector will contain information regarding the expected value of products of $\hat{O}(\mf x)$ smeared against the functions that define the spacetime profile of $\hat{\jmath}(\mf x)$.



\subsubsection*{The Two-Level Unruh-DeWitt Detector}

To see a more explicit example of a particle detector, let us study the final state of a two-level Unruh-DeWitt detector after the interaction with the field. We consider the model defined in Eq.~\eqref{eq:HIUDW} of a two-level particle detector, with time-evolved monopole $\hat{\mu}(\tau) = (e^{- \ii \Omega \tau}\s^- + e^{\ii \Omega \tau}\s^+)$ coupled to the amplitude of a scalar field $\hat{O}(\mf x) = \hat{\phi}(\mf x)$. In this case, we can solve more explicitly for the dynamics of the detector, using $\hat{\jmath}(\mf x) \hat{\jmath}(\mf x') = \Lambda(\mf x)\Lambda(\mf x)e^{\ii \Omega(\tau - \tau')}\s^+\s^- + e^{-\ii \Omega (\tau - \tau')}\s^-\s^+$ in Eq.~\eqref{eq:dynUDW}:
\begin{align}\label{eq:UDWintermediate}
    &\tr_\phi(\hat{U}^{(1)} \r_0) =  \!-\ii \lambda\! \int \dd V \langle\hat{\phi}(\mf x)\rangle_\omega \Lambda(\mf x)(e^{\ii \Omega \tau} \s^+ + e^{- \ii \Omega \tau}\s^- )\r_{\tc{d},0},\\
    &\tr_\phi(\hat{U}^{(2)} \r_0) =  \!-\lambda^2 \!\!\int \dd V \dd V'W(\mf x, \mf x')\theta(\tau-\tau')\Lambda(\mf x) \Lambda(\mf x')(e^{\ii \Omega(\tau - \tau')}\s^+\s^- + e^{-\ii \Omega (\tau - \tau')}\s^-\s^+) \r_{\tc{d},0},\nonumber\\
    &\tr_\phi(\hat{U}^{(1)} \r_0\hat{U}^{(1)\dagger})  =  \!\lambda^2 \!\!\int \dd V \dd V' W(\mf x, \mf x') \Lambda(\mf x) \Lambda(\mf x') (e^{\ii \Omega \tau'} \s^+ + e^{- \ii \Omega \tau'}\s^- ) \r_{\tc{d},0}(e^{\ii \Omega \tau} \s^+ + e^{- \ii \Omega \tau}\s^- ).\nonumber
\end{align}

When recombining these terms and writing the final state for $\r_\tc{d}$, it is convenient to represent $\r_{\tc{d},0}$ and $\r_\tc{d}$ in terms of Bloch vectors, writing
\begin{align}\label{eq:BlochVector}
    \r_{\tc{d},0} = \frac{1}{2}\left(\openone + \bm a \cdot \bm{\s}\right), \quad\quad \r_{\tc{d}} = \frac{1}{2}\left(\openone + \tilde{\bm a}\cdot \bm{\s}\right), 
\end{align}
where $\bm a = (a^x,a^y,a^z)$, $\tilde{\bm a} = (\tilde{a}^x,\tilde{a}^y,\tilde{a}^z)$ with $||\bm a|| \leq 1$, $|| \tilde{\bm a}|| \leq 1$ and $\bm \s = (\s_x,\s_y,\s_z)$ is the vector of Pauli matrices. Plugging Eq~\eqref{eq:BlochVector} in~\eqref{eq:UDWintermediate} and using standard commutation relations between the sigma matrices, we find that the relationship between $\tilde{\bm a}$ and $\bm a$ is given by
\begin{align}
    \tilde{a}^x &= a^x - 2 a^z \Im(\mathcal{X}) - \big(a^x \Re(\mathcal{N} - \mathcal{K}) + a^y \Im(\mathcal{N}+\mathcal{K})\big) + \mathcal{O}(\lambda^3),\label{eq:ax}\\
    \tilde{a}^y &= a^y - 2  a^z \Re(\mathcal{X}) - \big(a^y \Re(\mathcal{N}+\mathcal{K}) - a^x \Im(\mathcal{N}-\mathcal{K})\big) + \mathcal{O}(\lambda^3),\label{eq:ay}\\
    \tilde{a}^z &= a^z + 2 (a^x\Im(\mathcal{X}) + a^y \Re(\mathcal{X})) - \big(a^z (\mathcal{L}^- + \mathcal{L}^+) - \mathcal{L}^- + \mathcal{L}^+\big) + \mathcal{O}(\lambda^3),\label{eq:az}
\end{align}
where, denoting $\Lambda^\pm(\mf x) = \Lambda(\mf x) e^{\pm \ii \Omega\tau}$, we can write
\begin{align}
    \mathcal{X} &= \lambda \omega(\hat{\phi}(\Lambda^+)) = \lambda\int \dd V \langle\hat{\phi}(\mf x)\rangle_\omega \Lambda(\mf x) e^{\ii \Omega \tau},\\
    \mathcal{L}^\pm &= \lambda^2\omega(\hat{\phi}(\Lambda^\pm)\hat{\phi}(\Lambda^\mp)) = \lambda^2\int \dd V \dd V' W(\mf x,\mf x') \Lambda(\mf x)\Lambda(\mf x')e^{\pm \ii \Omega(\tau - \tau')},\\
    \mathcal{K} &= \lambda^2\omega(\hat{\phi}(\Lambda^+)\hat{\phi}(\Lambda^+)) = \lambda^2\int \dd V \dd V' W(\mf x,\mf x') \Lambda(\mf x)\Lambda(\mf x')e^{\ii \Omega(\tau + \tau')},\\
    \mathcal{N} &= \lambda^2\int \dd V \dd V' W(\mf x,\mf x') \Lambda(\mf x)\Lambda(\mf x')e^{\ii \Omega|\tau - \tau'|},
\end{align}
and we used $e^{\ii \Omega(\tau - \tau')}\theta(\tau - \tau') +
e^{-\ii \Omega(\tau - \tau')}\theta(\tau' - \tau) = e^{\ii \Omega|\tau-\tau'|}$ to write the expression for $\mathcal{N}$. See~\cite{ruhi} for a similar derivation.

We can now analyze the detector's final state in detail. Equations~\eqref{eq:ax} and~\eqref{eq:az} for the components of the Bloch vector of the final density state $\hat{\rho}_\tc{d}$ show a clear asymmetry between the components aligned with the $z$-axis and the components orthogonal to $z$. This asymmetry is due to the fact that the qubit's internal dynamics are associated with the Hamiltonian $\hat{H}_\tc{d} = \frac{\Omega}{2}(\openone + \s_z)$. Intuitively, one can think of the qubit detector as a spin in the presence of a (quantum) magnetic field aligned with the $z$-axis. The coupling with the field $\hat{\phi}(\mf x)$ then takes place in the plane orthogonal to $z$, which can be seen by rewriting the interaction Hamiltonian~\eqref{eq:HIUDW} as
\begin{equation}
    \hat{\mathcal{H}}_I(\mf x) = \lambda \Lambda(\mf x)(\cos(\Omega \tau) \s_x - \sin(\Omega \tau) \s_y)\hat{\phi}(\mf x).
\end{equation}
One can then think of the amplitude of the scalar field (represented by $\hat{\phi}(\mf x)$) as implementing an effective magnetic field in the rotating direction $\bm n(\tau) = (\cos(\Omega \tau) , -\sin(\Omega \tau),0)$. Indeed, if one replaces $\hat{\phi}(\mf x)$ in~\eqref{eq:HIUDW} with a classical field $\phi_0(\mf x)$, the interaction becomes the same as that of a spin with a classical magnetic field of strength $\phi_0(\mf x)$ aligned with the rotating axis $\bm n(\tau)$.

The field's two-point function is also sampled within the support of $\Lambda(\mf x)$, encoded in the terms $\mathcal{L}^\pm$, $\mathcal{K}$ and $\mathcal{N}$. In particular, when the field state $\omega$ is quasifree (so that $\mathcal{X} = 0$), its excitation probability matches Eq.~\eqref{eq:exProbV}, obtained when we considered a localized field interacting with $\hat{\phi}$. Indeed, assuming that the detector starts in its ground state $\r_{\tc{d},0} = \ket{g}\!\!\bra{g}$ (described by the Bloch vector $\bm a = (0,0,-1)$), we find that $\tilde{a}^x = \tilde{a}^y = 0$ and
\begin{equation}
    \tilde{a}^z = -1 + 2\mathcal{L}^- + \mathcal{O}(\lambda^4) \quad \Rightarrow \quad  \hat{\rho}_{\tc{d}} = (1 - \mathcal{L}^-)\ket{g}\!\!\bra{g} +  \mathcal{L}^-\ket{e}\!\!\bra{e} + \mathcal{O}(\lambda^4),
\end{equation}
so that the detector's leading order excitation probability is given by
\begin{equation}
    \langle \s^+\s^-\rangle_{\hat{\rho}_\tc{d}} = \mathcal{L}^-  + \mathcal{O}(\lambda^4)= \lambda^2 W(\Lambda^-,\Lambda^+) + \mathcal{O}(\lambda^4).
\end{equation}
Noticing that the Wightman function is non-zero in $C_\mathbb{C}$, we then find that the excitation probability is always non-zero for compactly supported $\Lambda(\mf x)$ such that $\Lambda^\pm\neq Pf^\pm$. We then find that the two-level Unruh-DeWitt detector generally has a non-zero probability of becoming excited after coupling to a quantum field. Alternatively, if the detector starts in its excited state,  $\rho_{\tc{d},0} = \ket{e}\!\!\bra{e}$ (described by the Bloch vector $\bm a = (0,0,1)$), we have $\tilde{a}^x = \tilde{a}^y = 0$ and
\begin{equation}
    \tilde{a}^z = 1 - 2  \mathcal{L}^+ + \mathcal{O}(\lambda^4) \quad \Rightarrow \quad  \hat{\rho}_{\tc{d}} =(1 - \mathcal{L}^+)\ket{e}\!\!\bra{e} +  \mathcal{L}^+\ket{g}\!\!\bra{g} + \mathcal{O}(\lambda^4).
\end{equation}
Its leading order deexcitation probability is then
\begin{equation}
    \langle \s^-\s^+\rangle_{\hat{\rho}_\tc{d}}  = \mathcal{L}^+ + \mathcal{O}(\lambda^4) = \lambda^2 W(\Lambda^+,\Lambda^-) + \mathcal{O}(\lambda^4).
\end{equation}
The excitation and deexcitation probabilities are related by $\Omega \mapsto -\Omega$, as this changes the roles of the ground and excited states.

Finally, notice that we can see the explicit dependence on the choice of the time parameter $\tau$ in the time ordering operation in the observable $\mathcal{N}$, which implicitly depends on $\theta(\tau - \tau')$. Indeed, choosing another time parameter $s$ for the time ordering operation would amount to the change
\begin{equation}
    e^{\ii \Omega|\tau-\tau'|} \longmapsto e^{\ii \Omega(\tau - \tau')}\theta(s-s') + e^{-\ii \Omega(\tau - \tau')}\theta(s'-s)
\end{equation}
in the integrand of $\mathcal{N}$. From Eqs.~\eqref{eq:ax} and~\eqref{eq:ay}, we notice that the specific choice of time ordering operation is relevant to the final state of the detector (to second order in $\lambda$) whenever $a^x\neq 0$ or $a^y\neq 0$, or equivalently, whenever $\hat{\rho}_{\tc{d},0}$ does not commute with the free Hamiltonian $\hat{H}_\tc{d}$. The fact that the choice of time parameter for the prescription of the time evolution operator can be neglected when $[\hat{\rho}_{\tc{d},0},\hat{H}_\tc{d}] = 0$ has been first noted in~\cite{us2}, where this difference has also been quantified. 



\subsubsection*{Explicit Examples in Minkowski Spacetime}

We can interpret the final state of a two-level Unruh-DeWitt detector more explicitly when we consider this setup in Minkowski spacetime. Specifically, we consider inertial coordinates $(t,\bm x)$ and the real massless scalar field with equation of motion $\nabla_\mu \nabla^\mu\phi = 0$. We consider the time parameter $\tau = t$ and assume that the initial field state $\omega$ can be represented in the GNS construction associated with the Minkowski vacuum. We also assume that the spacetime smearing function $\Lambda(\mf x)$ factors as
\begin{equation}
    \Lambda(\mf x) = \chi(t) F(\bm x),
\end{equation}
so that the switching function $\chi(t)$ defines the time profile of the interaction, and the smearing function $F(\bm x)$ defines its spatial profile. Both of these are assumed to be real. In the case where the function $F(\bm x)$ is mostly supported in a finite region of space, we can think of $F(\bm x)$ as defining the shape of the detector, which is constant in the inertial frame $(t,\bm x)$, so that this setup then corresponds to an inertial two-level Unruh-DeWitt detector in Minkowski spacetime. We will discuss how to assign a state of motion and shape for detectors from physical systems in more detail in the next section when we consider non-relativistic systems in spacetime and how they give rise to particle detector models.

Given that the state $\omega$ is assumed to be in the GNS representation of the Minkowski vacuum, its Wightman function can be written as
\begin{equation}
    W(\mf x, \mf x') = W_0(\mf x, \mf x') + w(\mf x, \mf x'),
\end{equation}
where $W_0(\mf x, \mf x')$ is given by Eq.~\eqref{eq:W0} and $w(\mf x, \mf x')$ is a symmetric regular two-point function that solves the Klein-Gordon equation in both arguments. Importantly, noticing that the final state of the detector depends directly on the Wightman function $W(\mf x, \mf x')$, we see that the vacuum effects (encoded in $W_0(\mf x, \mf x')$) will always be present in the final state of the detector, regardless of the field state. In other words, the effect of the state $\omega$ on the detector always adds to the vacuum effects---the vacuum is always present.

Using the decomposition of the vacuum Wightman in terms of the plane wave basis of solutions, we can then write~\eqref{eq:W0Fourier}
\begin{align}
    W_0(\Lambda^-,\Lambda^+) &= \frac{1}{(2\pi)^3}\int \frac{\dd^3 \bm k}{2|\bm k|}\, |\tilde{\chi}(|\bm k|\,+ \,\Omega)|^2|\tilde{F}(\bm k)|^2,\label{eq:Wmp}\\
    W_0(\Lambda^+,\Lambda^+) &= \frac{1}{(2\pi)^3}\int \frac{\dd^3 \bm k}{2|\bm k|}\, \tilde{\chi}(|\bm k|\,- \,\Omega)\tilde{\chi}^*(|\bm k|\,+ \,\Omega)|\tilde{F}(\bm k)|^2,\\
    W_0(\Lambda^+,\Lambda^-) &= \frac{1}{(2\pi)^3}\int \frac{\dd^3 \bm k}{2|\bm k|}\, |\tilde{\chi}(|\bm k|\,- \,\Omega)|^2|\tilde{F}(\bm k)|^2,\label{eq:Wpm}
\end{align}
where we denote by tilde the space and time Fourier transforms compatible with the spacetime Fourier transform~\eqref{eq:4Fourier}:
\begin{align}
    \tilde{\chi}(\omega) &= \int \dd t \chi(t)e^{-\ii \omega t},\\
    \tilde{F}(\bm k) &= \int \dd^3 \bm x F(\bm x) e^{\ii \bm k \cdot \bm x},
\end{align}
so that $\tilde{\Lambda}(\omega,\bm k) = \tilde{\chi}(\omega)\tilde{F}(\bm k)$ and we used that $\chi(t)$ and $F(\bm x)$ are real, so that $\tilde{\chi}(-\omega) = \tilde{\chi}^*(\omega)$ and $\tilde{F}(-\bm k) = \tilde{F}^*(\bm k)$.

From Eqs.~\eqref{eq:Wmp} and~\eqref{eq:Wpm}, we can see that under the assumption that neither $\chi(t)$ or $F(\bm x)$ are oscillatory, the vacuum deexcitation probability is always larger than the vacuum excitation probability. Indeed, for real non-oscillatory $\chi(t)$ and $F(\bm x)$, we have that $|\tilde{\chi}(|\bm k|)|^2$ and $|\tilde{F}(\bm k)|^2$ peak at $\bm k = 0$. In turn, this implies that the integrand in~\eqref{eq:Wpm} exhibits a resonance when $|\bm k| = \Omega$, maximizing the value of $\tilde{\chi}(|\bm k| - \Omega)$. This resonance indicates that a particle detector with energy gap $\Omega$ emits wavepackets with momenta centred around $|\bm k| = \Omega$. On the other hand, the integrand in Eq.~\eqref{eq:Wmp} does not present any resonances for $\Omega>0$, showcasing the intuitive fact that the probability of exciting an inertial detector after an interaction with the vacuum is much less than the probability of deexciting it.

We can also analyze the interaction of a particle detector in its ground state with a one-particle wavepacket $\ket{f}$, defined by a function $f\in C_\mathbb{C}$,
\begin{equation}
    \ket{f} = \hat{a}^\dagger(f)\ket{0} = \int \dd^3\bm k \,u_{\bm k}^*(f) \hat{a}_{\bm k}^\dagger \ket{0}.
\end{equation}
From Eq.~\eqref{eq:Wfg1part}, we can then write the term that defines the excitation probability as  
\begin{equation}
    W_f(\Lambda^-,\Lambda^+) = W_0(\Lambda^-,\Lambda^+) + W_0(\Lambda^-,f)W_0(f^*,\Lambda^+) + W_0(f^*,\Lambda^-)W_0(\Lambda^+,f),
\end{equation}
and we can compute $W_0(\Lambda^-,f), W_0(f^*,\Lambda^+), W_0(f^*,\Lambda^-), W_0(\Lambda^+,f)$ from Eq.~\eqref{eq:W0Fourier}:
\begin{align}
    W_0(\Lambda^-,f) &= \frac{1}{(2\pi)^3}\int \frac{\dd^3\bm k}{2|\bm k|} \tilde{\chi}(|\bm k|+\Omega)\Tilde{F}(\bm k) \tilde{f}(-|\bm k|,-\bm k),\\
    W_0(f^*,\Lambda^+) &= \frac{1}{(2\pi)^3}\int \frac{\dd^3\bm k}{2|\bm k|} \tilde{f}(-|\bm k|,-\bm k) \tilde{\chi}^*(|\bm k|+\Omega)\Tilde{F}^*(\bm k),\\
    W_0(f^*,\Lambda^-) &= \frac{1}{(2\pi)^3}\int \frac{\dd^3\bm k}{2|\bm k|} \tilde{f}(-|\bm k|,-\bm k)\tilde{\chi}^*(|\bm k|-\Omega)\Tilde{F}^*(\bm k),\\
    W_0(\Lambda^+,f) &= \frac{1}{(2\pi)^3}\int \frac{\dd^3\bm k}{2|\bm k|} \tilde{\chi}(|\bm k|-\Omega)\Tilde{F}(\bm k) \tilde{f}(-|\bm k|,-\bm k),
\end{align}
where $\tilde{f}(\mf k)$ denotes the four dimensional Fourier transform defined in~\eqref{eq:4Fourier}. One can analyze the integrals above in terms of resonances between the Fourier transforms of $\tilde{f}(-|\bm k|,-\bm k)$ and $\tilde{\chi}(|\bm k|\pm\Omega)\Tilde{F}(\bm k)$, which also cause resonances in the detector's excitation probability. We will study these resonances with an explicit example later in this section.

We can make our analysis even more explicit by considering specific shapes for the interaction region defined by the functions $\chi(t)$ and $F(\bm x)$. Let
\begin{equation}\label{eq:prototype}
    \chi(t) = e^{- \frac{t^2}{2T^2}},\quad \quad
    F(\bm x) = \frac{e^{- \frac{|\bm x|^2}{2\sigma^2}}}{(2\pi\sigma^2)^{3/2}},
\end{equation}
such that $\sigma$ is a parameter with units of length that controls the spatial localization of the detector and $T$ is a parameter with units of time that controls its time duration. The normalizations of $\chi(t)$ and $F(\bm x)$ are chosen as above so that
\begin{equation}
    \lim_{T\to \infty} \chi(t) = 1, \quad \quad \lim_{\sigma \to 0} F(\bm x) = \delta^{(3)}(\bm x),
\end{equation}
so that the limit of constant interaction in time and the pointlike limit can be easily considered later on. 

Notice that the choices of Eq.~\eqref{eq:prototype} make $\Lambda(\mf x)$ not compactly supported. This does not create any divergences in our results, as $\Lambda(\mf x)$ is a sufficiently fast decaying function, and as such, the evaluation of $W$ in $\Lambda$ is well defined. However, this prevents one from considering that the detector interacts with the field in a given region of spacetime if one thinks of a ``region'' as a subset of $\M$. Nevertheless, we can still consider that the detector is localized and that its localization is defined by the profile of $\Lambda(\mf x)$, so that the detector mostly probes the field where $\Lambda(\mf x)$ has a non-negligible value. We will discuss more about the interpretations and consequences of non-compact supported spacetime smearing functions throughout the thesis, but for now, we use the choices in Eqs.~\eqref{eq:prototype} as our prototypical example of particle detectors. 

\begin{figure}[h!]
    \centering
    \includegraphics[width=11.5cm]{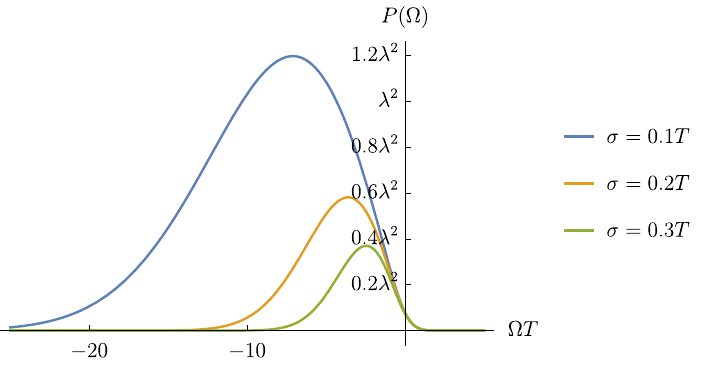}
    \caption{The two-level Unruh-DeWitt detector's leading order excitation probability as a function of $\Omega T$ for different values of $\sigma$.}
    \label{fig:pOmega}
\end{figure}

With the choices of Eq.~\eqref{eq:prototype} we can compute $W_0(\Lambda^\pm,\Lambda^\mp)$ explicitly. For convenience define
\begin{equation}
    P(\Omega) = \lambda^2 W_0(\Lambda^-,\Lambda^+) \quad\Rightarrow \quad P(-\Omega) = \lambda^2 W_0(\Lambda^+,\Lambda^-),
\end{equation}
so that $P(\Omega)$ corresponds to the detector's vacuum excitation probability and $P(-\Omega)$ to the vacuum deexcitation probability. The Fourier transforms of $F(\bm x)$ and $\chi(t)$ are then
\begin{align}
    \tilde{\chi}(\omega) = \sqrt{2\pi} T e^{- \frac{1}{2}T^2 \omega^2},\\
    \tilde{F}(\bm k) = e^{- \frac{1}{2} \sigma^2 |\bm k|^2},
\end{align}
and we can compute $P(\Omega)$ explicitly:
\begin{equation}\label{eq:POmegaproto}
    P(\Omega) = \frac{\lambda^2}{4\pi} \frac{T^2 e^{- \Omega^2 T^2}}{(T^2 + \sigma^2)}\left(1 - \frac{\sqrt{\pi}\Omega T^2}{\sqrt{T^2 + \sigma^2}}e^{\frac{\Omega^2T^4}{T^2 + \sigma^2}}\left(1-\text{erf}\left(\frac{\Omega T^2}{\sqrt{T^2 + \sigma^2}}\right)\right)\right).
\end{equation}
For concreteness, we plot $P(\Omega)$ for different values of $\sigma$ in Fig.~\ref{fig:pOmega}, where positive values of $\Omega$ correspond to the vacuum excitation probability and negative values of $\Omega$ to the deexcitation probability. We see that the excitation probability decays to $0$ as $\Omega T$ increases and that the deexcitation probability $P(-\Omega)$ has a resonance peak when $\Omega \sim 1/\sigma$. That is, when the energy gap of the detector matches the inverse size of the detector---which corresponds to the characteristic frequency of the field that resonates with the detector.

Equation~\eqref{eq:POmegaproto} also allows us to compute relevant asymptotic limits. First, one can consider the limit $\delta \to 0$, in which case the smearing function $F(\bm x)$ yields $F(\bm x) \to \delta^{(3)}(\bm x)$, defining what is commonly called a pointlike detector. Although the functions $\Lambda^\pm(\mf x)$ become singular in this limit, it is still possible to evaluate the excitation and deexcitation probabilities. Indeed, whenever $\chi(t)$ is a continuous function, the application of the Wightman function to $\Lambda^\pm(\mf x)$ is well defined, even in the pointlike limit, and we find
\begin{equation}\label{eq:poinlikeExct}
    \lim_{\sigma\to 0} P(\Omega) = \frac{\lambda^2}{4\pi} e^{- \Omega^2 T^2}\left(1 -\sqrt{\pi}\Omega T e^{\Omega^2T^2}\left(1-\text{erf}\left(\Omega T\right)\right)\right).
\end{equation}
Although the pointlike limit might yield physical results in some regimes, it is always safer to work with a regular finite-sized particle detector and take the pointlike limit at the end if convergent. For instance, Eq.~\eqref{eq:poinlikeExct} suggests that the excitation probability is unbounded as a function of $\Omega T$, as the peak of the deexcitation probability would happen at $\Omega \sim 1/\sigma$, which in the pointlike limit corresponds to $\Omega \to \infty$. 

Another relevant limit is that of long times, $T\to \infty$. This process corresponds to a usual particle detection setting, where the detector is switched on for arbitrarily long times. Even though this limit can be associated with physical processes, the limit of $T\to\infty$ of~\eqref{eq:POmegaproto} is singular. Indeed, we have
\begin{equation}
    \lim_{T\to \infty} e^{- \Omega^2 T^2}e^{\frac{\Omega^2T^4}{T^2 + \sigma^2}}\left(1-\text{erf}\left(\frac{\Omega T^2}{\sqrt{T^2 + \sigma^2}}\right)\right) = 2 e^{- \Omega^2 \sigma^2}\theta(-\Omega),
\end{equation}
so that the asymptotic limit of Eq.~\eqref{eq:POmegaproto} yields
\begin{equation}\label{eq:POmegaTinf}
    \lim_{T\to \infty} P(\Omega) \sim - \frac{\lambda^2 \Omega T e^{- \Omega^2 \sigma^2}}{2 \sqrt{\pi}}\theta(-\Omega).
\end{equation}
The expression above yields $0$ for $\Omega>0$, giving the expected result that an inertial particle detector that starts the interaction in the ground state and interacts with the vacuum for long times does not become excited. However, Eq.~\eqref{eq:POmegaTinf} yields a divergent result for the deexcitation probability. This simply tells us that this regime falls outside of the perturbative treatment we employed, as the deexcitation probability grows proportionally to $T$. Instead, it is common to compute the deexcitation rate of a detector $\mathcal{F}(-\Omega)$, defined as
\begin{equation}\label{eq:FOmega}
    \mathcal{F}(-\Omega) \coloneqq \lim_{T\to \infty} \frac{P(- \Omega)}{T} =  - \frac{\lambda^2 \Omega e^{- \Omega^2 \sigma^2}}{2 \sqrt{\pi}}\theta(-\Omega).
\end{equation}
The transition rate can be thought of as the rate that an excited detector emits particles at, and yields a finite result. Eq.~\eqref{eq:FOmega} tells us that maximal particle emission is obtained when the energy gap of the detector is of the same order as its inverse size, peaking at $\Omega = 1/2\sigma$. 


We can now study an explicit example of particle detection, where the detector interacts with a one-particle wavepacket. With the choices of~\eqref{eq:prototype} we have $W_0(\Lambda^-,f) = W_0(f^*,\Lambda^+)$ and $W_0(f^*,\Lambda^-) = W_0(\Lambda^+,f)$, so that we can define
\begin{equation}
    q(\Omega) \coloneqq  W_0(\Lambda^-,f) = W_0(f^*,\Lambda^+) \Rightarrow q(-\Omega) = W_0(f^*,\Lambda^-) = W_0(\Lambda^+,f),
\end{equation}
and defining $\mf{f}(\bm k) = \tilde{f}(-|\bm k|,-\bm k)$, we can write
\begin{equation}
    q(\Omega) = \frac{T}{(2\pi)^{\frac{5}{2}}} \int \frac{\dd^3 \bm k}{2|\bm k|} \,\mf{f}(\bm k)e^{- \frac{1}{2}\sigma^2 |\bm k|^2}e^{- \frac{1}{2}T^2 (|\bm k|+\Omega)^2} = \frac{T e^{- \frac{\Omega^2 \sigma^2}{2\alpha^2}}}{(2\pi)^{\frac{5}{2}}} \int \frac{\dd^3 \bm k}{2|\bm k|} \,\mf{f}(\bm k)e^{- \frac{1}{2}\alpha^2T^2 \left(|\bm k|+\frac{\Omega}{\alpha^2}\right)^2},
\end{equation}
with $\alpha = \sqrt{1+\sigma^2/T^2}$. We can then see that for $\Omega>0$, the terms $q(-\Omega)$ will contribute the most when $|\bm k| \sim \Omega/\alpha^2$, and the values of $\Omega$ that yields the maximum value for the integral are those that match the peaks of $\mf f(\bm k)$, also corresponding to the energy of the positive frequency classical solution corresponding to $\ket{f}$.

We can see this explicitly by choosing
\begin{equation}
    \mf f(\bm k) = N_0 e^{- \frac{1}{2}\delta^2|\bm k - \bm k_0|^2}, \quad N_0 = \frac{4 |\bm k_0|^\frac{1}{2}\delta^\frac{3}{2}\pi^\frac{3}{4}}{\sqrt{\text{erf}(|\bm k_0|\delta)}},
\end{equation}
where $N_0$ ensures $\braket{f}{f} = 1$ (see Eq.~\eqref{eq:norm1part}). Intuitively, $\ket{f}$ corresponds to a wavepacket centred at momentum $\bm k_0$, with width $1/\delta$ in momentum space. In essence, $\ket{f}$ is a localized wavepacket that moves along the direction determined by $\bm k_0$ and reaches the origin at $t=0$. The energy of $\ket{f}$ is then determined by $|\bm k_0|$, as the expected value of the normal ordered Hamiltonian~\eqref{eq:normordH} evaluated along the surface $t=0$ yields
\begin{equation}
    \bra{f}\!\normord{\hat{H}}\!\ket{f} = \frac{|\bm k_0|}{\text{erf}(|\bm k_0|\delta)}.
\end{equation}
In the case where $\delta |\bm k_0|\gg 1$ (corresponding to sharp momentum localization at $\bm k_0$), we have that its energy is simply $|\bm k_0|$. We can compute $q(\Omega)$ in closed-form:
\begin{align}
    q(\Omega) = \frac{2\pi^2N_0 e^{-\frac{\Omega^2\sigma^2}{\alpha^2}}}{\beta|\bm k_0|\delta^2} \Bigg(&e^{- \frac{\alpha^2\delta^2}{2 \beta^2}\left(|\bm k_0| + \frac{\Omega}{\alpha^2}\right)^2}\left(1 + \text{erf}\left(\frac{\delta^2 |\bm k_0| -  \Omega T^2}{\sqrt{2}\beta T}\right)\right)\label{eq:qOmega}\\
    &-e^{- \frac{\alpha^2\delta^2}{2 \beta^2}\left(|\bm k_0| - \frac{\Omega}{\alpha^2}\right)^2}\left(1 - \text{erf}\left(\frac{\delta^2 |\bm k_0| +  \Omega T^2}{\sqrt{2}\beta T}\right)\right)\Bigg)\nonumber,
\end{align}
where $\beta = \sqrt{1 + (\delta^2 + \sigma^2)/T^2}$. Although at first unhinged, we can interpret the function $q(\Omega)$ by analyzing the relationship between the variables $\Omega$, $|\bm k_0|$, $\delta$, $\sigma$ and $T$. For instance, under the assumption that $\delta |\bm k_0|\gg 1$ , we can approximate the error functions in~\eqref{eq:qOmega} by their asymptotic value at $\infty$, $\lim_{u\to\infty}\text{erf}(u) = 1$, yielding the approximation
\begin{equation}\label{eq:qAsympt}
     q(\Omega) \approx \frac{16\pi^2\pi^\frac{3}{4} e^{-\frac{\Omega^2\sigma^2}{\alpha^2}}}{\beta\sqrt{|\bm k_0|\delta}} e^{- \frac{\alpha^2\delta^2}{2 \beta^2}\left(|\bm k_0| + \frac{\Omega}{\alpha^2}\right)^2},
\end{equation}
where we also approximated $N_0 \approx 4 |\bm k_0|^\frac{1}{2}\delta^\frac{3}{2}\pi^\frac{3}{4}$. For $\Omega>0$, $q(\Omega)$ is monotonically decreasing with $|\bm k_0|$ and $\Omega$, so that $q(\Omega)$ does not present any resonances. On the other hand, we can now clearly see the peak of $q(-\Omega)$ at $|\bm k_0| = \Omega/\alpha^2$. In particular, in the case where the detector is pointlike ($\sigma \to 0$), we find $\alpha = 1$, so that resonance is achieved exactly when the energy of the wavepacket $\ket{f}$ matches the energy gap of the detector, $|\bm k_0| = \Omega$. Also notice that Eq.~\eqref{eq:qAsympt} yields finite results both in the pointlike limit ($\sigma\to 0$) and in the limit of long interaction times ($T\to \infty$). This resonance effect associated with detecting a particle of similar energy to the detector's energy gap is what is usually called a particle absorption process, justifying the name ``particle detectors''.

\subsubsection{The Gapless and Delta-Coupled Unruh-DeWitt Detectors}

Every example of local probes that we have discussed so far has been treated perturbatively. Indeed, most models cannot be solved non-perturbatively. However, there is an exception to this rule when the internal dynamics of the detector are trivial, corresponding to the so-called gapless detector models (with $\Omega = 0$). Although this case has been studied in different setups~\cite{Landulfo,nogo}, the lack of internal dynamics makes these models trivial for some relevant applications of particle detectors, as we will see in Chapter~\ref{chap:ent}. Nevertheless, they are a useful tool for extrapolating perturbative results and for studying degenerate subspaces of localized probes.

Before going through the computations with gapless detectors, let us discuss the physical interpretation of this limit of the Unruh-DeWitt model. One of the applications of particle detectors is to provide a definition of the concept of particles measured by an observer. This is done by considering a detector in a given state of motion and associating its excitations with detection of field quanta. However, it is necessary to have an energy gap to claim that energy from the field was absorbed and, thus, to make statements about ``detected particles''. Gapless detectors do not share this property and, therefore, cannot be used to discuss particle absorption/emission. Nevertheless, gapless detectors can be used to extract field correlations more precisely than gapped detectors can.

Intuitively, one can think of an Unruh-DeWitt detector as a spin-1/2 system with an energy gap $\Omega$, which is put to interact with a scalar field. The energy gap $\Omega$ in the case of a spin system can then be thought of as the result of applying an external (classical) magnetic field to the spin. In terms of the Bloch sphere, the effect of the energy gap is to add a constant rotation around the axes of the magnetic field. While this constant rotation is in place, the qubit is then put to interact with a quantum field, and the field fluctuations effectively generate another axis of rotation, also exchanging quantum information with the detector. With a gapless detector, the only ``rotation'' that takes place is due to the interaction with the quantum field. This is why, intuitively, gapless detectors can be better at extracting field correlations, as the only effect that they are sensitive to is the quantum field itself. This intuition is also aligned with the continuous variables studies of~\cite{MariaContinuous}.

Finally, notice that the lack of an energy gap does not prevent one from defining excited and ground states, as one can always consider that a magnetic field is applied before and after the interaction, but not while it is taking place. For this reason, we can maintain the notation $\{\ket{g},\ket{e}\}$ for the basis of the qubit system, and we will keep the nomenclature of ``ground'' and ``excited'' states for the eigenvectors of $ \hat{\sigma}^+\hat{\sigma}^-$, even when $\Omega = 0$.

We now consider the case where a single two-level Unruh-DeWitt detector interacts with a scalar field, with the assumption that $\Omega = 0$. In this case, the interaction Hamiltonian reduces to
\begin{equation}
    \hat{\mathcal{H}}_I(\mf x) = \lambda \Lambda(\mf x) \hat{\mu} \,\hat{\phi}(\mf x),
\end{equation}
and here we will consider that $\hat{\mu}$ is any constant operator in the qubit's Hilbert space. The fact that $\hat{\mu}$ is time-independent implies an important technical fact: we will now have the microcausality condition fulfilled by the interaction Hamiltonian $\hat{\mathcal{H}}_I(\mf x)$. Indeed,
\begin{equation}
    [\hat{\mathcal{H}}_I(\mf x), \hat{\mathcal{H}}_I(\mf x')] = \lambda^2 \Lambda(\mf x)\Lambda(\mf x') \hat{\mu}^2[\hat{\phi}(\mf x), \hat{\phi}(\mf x')],
\end{equation}
which will always commute whenever the points $\mf x$ and $\mf x'$ are spacelike separated due to the field satisfying the microcausality condition~\eqref{eq:microcausality}. This fact prevents the previously discussed incompatibilities with relativity from taking place. 

A consequence of the simple expression for the commutator $[\hat{\mathcal{H}}_I(\mf x), \hat{\mathcal{H}}_I(\mf x')]$ is that the time evolution operator for the detector and field can be solved non-perturbatively using the Magnus expansion~\cite{magnus}. In essence, due to the fact that $[[\hat{\mathcal{H}}_I(\mf x), \hat{\mathcal{H}}_I(\mf x')],\hat{\mathcal{H}}_I(\mf x'')] = 0$, we have that~\cite{magnus}
\begin{equation}
    \hat{U}_I = \mathcal{T}\exp\left(- \ii \int \dd V \hat{\mathcal{H}}_I(\mf x)\right) = e^{\hat{\Theta}_1+\hat{\Theta}_2},
\end{equation}
where
\begin{align}
    \hat{\Theta}_1 &= - \ii \int \dd V \hat{\mathcal{H}}_I(\mf x) = - \ii \lambda \hat{\mu} \hat{\phi}(\Lambda),\\
    \hat{\Theta}_2 &= - \frac{1}{2} \int \dd V \dd V' \theta(t-t')[\hat{\mathcal{H}}_I(\mf x), \hat{\mathcal{H}}_I(\mf x')] \\&= -  \ii \frac{\lambda^2}{2} \hat{\mu}^2G_R(\Lambda,\Lambda),
\end{align}
where we used that $[\hat{\phi}(\mf x), \hat{\phi}(\mf x')] = \ii E(\mf x,\mf x')$ and $\theta(t-t')E(\mf x, \mf x') = G_R(\mf x,\mf x')$. We can then write the time evolution operator for the detector-field system as
\begin{equation}
    \hat{U}_I = e^{- \ii \lambda \hat{\mu} \hat{\phi}(\Lambda)}e^{- \ii \hat{\mu}^2 \mathcal{G}},
\end{equation}
where we used that $[\hat{\mu},\hat{\mu}^2] = 0$ to separate the exponentials and we denoted $\mathcal{G} = \frac{\lambda^2}{2} G_R(\Lambda,\Lambda)$ 

We will again assume that the detector and field start in an uncorrelated state $\hat{\rho}_0 = \hat{\rho}_{\tc{d},0}\otimes \hat{\rho}_\phi$, where $\hat{\rho}_\phi$ is a representation of a quasifree state $\omega$ for the quantum field. In this case, one can compute the final state of the detector by tracing over the field's state,
\begin{align}
    \hat{\rho}_\tc{d} &= \tr_\phi\left(\hat{U}_I\hat{\rho}_0\hat{U}_I^\dagger\right) \\&= e^{- \ii \hat{\mu}^2 \mathcal{G}}\tr_\phi\left( e^{- \ii \lambda \hat{\mu} \hat{\phi}(\Lambda)}(\hat{\rho}_{\tc{d},0}\otimes \hat{\rho}_\phi )e^{\ii \lambda \hat{\mu} \hat{\phi}(\Lambda)}\right)e^{\ii \hat{\mu}^2 \mathcal{G}}\nonumber,
\end{align}
with $\mathcal{G} = \frac{\lambda^2}{2} G_R(\Lambda,\Lambda)$. We proceed with the computation assuming $\hat{\mu}^2 = \openone$ and using the identities 
\begin{align}
    e^{-\ii \lambda \hat{\mu} \hat{\phi}(\Lambda)} &= \text{cos} 
 (\lambda\hat{\phi}(\Lambda)) - \ii \hat{\mu} \,\text{sin}(\lambda \hat{\phi}(\Lambda)),\nonumber\\
    \omega(e^{\ii \hat{\phi}(f)}) &= \omega(\text{cos}(\hat{\phi}(f)) = e^{-\frac{1}{2}W(f,f)},\nonumber\\
    e^{\ii \hat{\phi}(f)}e^{\ii \hat{\phi}(g)} &= e^{\ii \hat{\phi}(f+g)}e^{\frac{\ii}{2}E(f,g)},\nonumber\\
    \omega(\text{cos}^2(\hat{\phi}(f)) &= e^{-W(f,f)}\cosh(W(f,f)),\label{eq:identities}
\end{align}
so that we find
\begin{align}
    \hat{\rho}_\tc{d} =& \omega(\text{cos}^2(\lambda \hat{\phi}(\Lambda)) e^{- \ii \hat{\mu}^2 \mathcal{G}} \hat{\rho}_{\tc{d},0} e^{\ii\hat{\mu}^2 \mathcal{G}}
    + \omega(\text{sin}^2(\lambda \hat{\phi}(\Lambda)) e^{- \ii \hat{\mu}^2 \mathcal{G}} \hat{\mu}\,\hat{\rho}_{\tc{d},0}\,\hat{\mu}e^{\ii \hat{\mu}^2 \mathcal{G}}\label{eq:finalstategapless}\\
    =& e^{- \ii \hat{\mu}^2 \mathcal{G}} \left(e^{-\xi}\cosh(\xi)\hat{\rho}_{\tc{d},0} + e^{-\xi}\sinh(\xi)\hat{\mu}\,\hat{\rho}_{\tc{d},0}\,\hat{\mu}\right)e^{\ii \hat{\mu}^2 \mathcal{G}},\nonumber
\end{align}
which establishes a quantum channel acting in the qubit with $\xi = \lambda^2 W(\Lambda,\Lambda)$. In the case of $\hat{\mu}= \hat{\sigma}^++\hat{\sigma}^-$, this quantum channel is a bit-flip channel with parameter \mbox{$p = e^{-\xi}\sinh(\xi)$}.

One can also compute the parameters $\xi$ and $\mathcal{G}$ in the case where the detector is interacting with the Minkowski vacuum of a massless scalar field, with the spacetime smearing function~\eqref{eq:prototype}
\begin{equation}
    \Lambda(\mf x) = \chi(t) F(\bm x) = e^{- \frac{t^2}{2T^2}}\frac{e^{- \frac{|\bm x|^2}{2\sigma^2}}}{(2\pi \sigma^2)^\frac{3}{2}}.
\end{equation}
We obtain:
\begin{align}
    G_R(\Lambda,\Lambda) = \frac{T/\sigma}{4\pi \alpha^2},\quad\quad
    W(\Lambda,\Lambda) = \frac{1}{4\pi \alpha^2},
\end{align}
where again $\alpha = \sqrt{1+\sigma^2/T^2}$. Notice, in particular, that the purity of the state $\hat{\rho}_\tc{d}$ is entirely determined by the parameter $\xi$. Indeed, if  $\hat{\rho}_{\tc{d},0}$ starts in a pure state, we find
\begin{equation}
    \tr(\hat{\rho}_\tc{d}^2) = e^{-2\xi}(\cosh(2\xi) + M^2\sinh(2 \xi)),
\end{equation}
 with $M^2 = \tr((\hat{\mu}\,\hat{\rho}_{\tc{d,0}})^2)$. The qubit's purity is then a decreasing function of $\xi$. Using that $\xi$ increases with $T$, we can see that the purity of the state decreases with $T$ and asymptotically reaches its minimum value when $\xi \to \lambda^2/4\pi$, unless \mbox{$[\hat{\rho}_{\tc{d},0},\hat{\mu}] = 0$}, in which case the evolution is always unitary for the detector, as it starts in an eigenstate of the Hamiltonian. We also see that the stronger the coupling, the more mixed the detector state is, as $\xi \propto \lambda^2$. The fact that the detector, in general, ends in a mixed state shows that it generally becomes entangled with the field. 

A particular case of a gapless detector is that of a delta-coupled detector, where $\Lambda(\mf x)$ is an irregular function supported along a Cauchy slice. Given a Cauchy foliation $\Sigma_\tau$, a delta-coupled detector is defined by a spacetime smearing function of the form
\begin{equation}\label{eq:pointlike}
    \Lambda(\mf x) = \eta \delta(\tau- \tau_0) F(\bm x),
\end{equation}
where $\bm x$ are coordinates in $\Sigma_{\tau_0}$ and $\eta$ is a parameter with units of time. Due to the distributional properties of the Wightman function and the retarded Green's function, even with a singular $\Lambda(\mf x)$ such as in Eq.~\eqref{eq:pointlike}, the integrals of the form $W(\Lambda,\Lambda)$ and $G_R(\Lambda,\Lambda)$ are convergent if $F(\bm x)$ is sufficiently regular. Thus, the final state of a delta-coupled detector is also given by~\eqref{eq:finalstategapless}. Also notice that if one considers internal dynamics for a delta-coupled detector, the effect of these dynamics amounts to a shift of the form $\hat{\mu} \mapsto \hat{\mu}' = \hat{U}\hat{\mu}\hat{U}^\dagger$ with unitary $\hat{U}$, so the energy gap does not amount to any significant change to the dynamics. In this sense, all delta-coupled detectors behave as if they were gapless, regardless of their internal dynamics. 

\section{Non-relativistic Quantum Systems as Particle Detectors}\label{sec:NRLQS}

Although, at this stage, we have a rather general definition of particle detectors, we have not yet explicitly discussed how one can interpret the detectors as physically realistic systems. In this Section, we will present the results of~\cite{generalPD}, which 
provide a consistent way of describing a localized non-relativistic quantum system undergoing a timelike trajectory in a background curved spacetime. Namely, using Fermi normal coordinates, it is possible to identify an inner product and canonically conjugate position and momentum operators defined in the rest space of the trajectory for each value of its proper time. This framework then naturally provides a recipe for mapping a quantum theory defined in a non-relativistic background to a theory around a timelike trajectory in curved spacetimes by reinterpreting the position and momentum operators and by introducing a local redshift factor to the Hamiltonian, which gives rise to new dynamics due to the curvature of spacetime and the acceleration of the trajectory. We then apply our formalism to particle detector models, that is, to the case where the non-relativistic quantum system is coupled to a quantum field in a curved background. This allows one to write a general definition for particle detector models, which connects the abstract definitions discussed in the previous section to physically realizable systems.

Exclusively in this Section, we will consider the more general case where spacetime is an $n+1$ dimensional manifold.

\subsection{Local Rest Frames in Curved Spacetimes}\label{sub:FNC}

The first step in describing a non-relativistic quantum system in curved spacetimes is to fix a frame such that the dynamics of the system can be approximately described using non-relativistic physics. This frame is the Fermi normal coordinate system, which we review in this Section.

\subsubsection*{Fermi Normal Coordinates}

Let $\mathcal{M}$ denote an $n+1$ dimensional spacetime with Lorentzian metric $g$, and consider a timelike trajectory $\mf z(\tau)$ in $\mathcal{M}$, parametrized by its proper time \mbox{$\tau\in(\tau_{\text{min}},\tau_{\text{max}})$}\footnote{The formalism also allows for the case where $\tau_{\text{min}} = -\infty$ and $\tau_{\text{max}} = \infty$.}. The Fermi normal coordinates around the trajectory $\mf z(\tau)$ are coordinates which are able to describe physically relevant quantities associated with an observer undergoing the trajectory. The Fermi normal coordinates are also useful because one can expand the metric components in a neighbourhood of the curve in terms of the curvature of spacetime and the trajectory's proper acceleration.

The time component of the Fermi normal coordinates is defined as the proper time of the curve, $\tau$. In order to define the spacelike coordinates, $\bm x$, we first pick an orthonormal frame $\mf{e}_\mu(\tau_0)$ in the tangent space to a given point of the curve, $T_{\mf z(\tau_0)}\mathcal{M}$ such that $e_0^\mu(\tau_0) = u^\mu(\tau_0)$ is the four-velocity of the curve. Then, we have
\begin{equation}
    g(\mf e_\mu,\mf e_\nu) = \eta_{\mu\nu},
\end{equation}
where $\eta_{\mu\nu} = \text{diag}(-1,1,1,1)$. The next step is to extend this frame along the curve $\mf z(\tau)$. To do this, we transport the vectors $\mf e_\mu(\tau_0)$ via the Fermi transport:
\begin{equation}
    \frac{\text{D}(e_\mu)^\alpha}{\dd \tau} + 2a^{[\alpha}u^{\beta]}(e_\mu)_\beta = 0, 
\end{equation}
where $\frac{\text{D}}{\dd \tau}$ denotes the covariant derivative along $\mf z(\tau)$ and $a^\mu = \frac{\text{D}u^\mu}{\dd \tau}$ is the proper acceleration of the trajectory. The Fermi transport takes into account the natural motion of the curve in order to transport vectors between different tangent spaces. Notice that because $u_\mu a^\mu = 0$, the four-velocity is always Fermi transported along the curve. Thus, Fermi transporting the frame $\mf e_\mu(\tau_0)$ along $\mf z(\tau)$ gives a frame $\mf e_\mu(\tau)$, such that $\mf e^\mu_0 = u^\mu$ for all $\tau$. This frame will be referred to as the Fermi frame.

We define the spacelike Fermi normal coordinates \mbox{$\bm x=(x^1,...,x^n)$} as follows. Let $\mathcal{N}_\mf{p}$ denote the normal neighbourhood of $\mf p$: the set of all points which can be connected to $\mf p$ by a unique geodesic. For a given $\tau$, we define the rest surface $\Sigma_\tau\subset\mathcal{N}_{\mf z(\tau)}$ as the set reached by all geodesics starting at $\mf z(\tau)$ with tangent vector orthogonal to $u^\mu$. The surfaces $\Sigma_\tau$ correspond to the local rest spaces around $\mf z (\tau)$ and define a local foliation of spacetime around the curve. Let $\mf p\in\Sigma_\tau$ for some $\tau$, then we assign coordinates $(\tau,x^1,...,x^n)$ to $\mf p$ if $\mf p = \text{exp}_{\mf z(\tau)}(x^i\mf e_i(\tau))$, where $\exp_{\mf z(\tau)}$ denotes the exponential map at the point $\mf z(\tau)$. The Fermi normal coordinates are well defined in the world tube $\mathcal{T} = \bigcup_\tau \Sigma_\tau$ around the trajectory so that any point $\mf x\in \mathcal{T}$ can be identified as $\mf x = (\tau,\bm x)$. A consequence of the definition is that the proper distance of a point $\mf x$ to the curve $\mf z(\tau)$ is given by $r = \sqrt{\delta_{ij}x^ix^j}$ so that proper distances from $\mf z(\tau)$ can be computed using the Euclidean norm of the spacelike Fermi normal coordinates. 

It is important to mention that although the time parameter of the Fermi normal coordinates is the proper time of the trajectory $\mf z (\tau)$, in general, it does not correspond to the proper time of the other trajectories defined by $\bm x = \text{const.}$ In fact, in a general curved spacetime the vector $\partial_\tau$ is not normal to the surfaces $\Sigma_\tau$, and not normalized if $\bm x \neq 0$.

It is also useful to define a local orthonormal frame associated with the Fermi normal coordinates by extending the Fermi frame to the tube $\mathcal{T}$. For a given event \mbox{$\mf x \in {\mathcal{T}}$}, we define the \emph{extended Fermi frame} $\mf e_\mu(\mf x)$ by parallel transporting the vectors $\mf e_\mu(\tau)$ along the geodesic contained in $\Sigma_\tau$ that connects $\mf z(\tau)$ to $\mf x$. This process then defines an orthonormal frame at every point within the region $\mathcal{T}$.

It is also possible to find an expression for the metric components in Fermi normal coordinates in terms of an expansion on the physical distance of a point to the curve, $r=\sqrt{\delta_{ij}x^ix^j}$. The expansion reads~\cite{poisson}
\begin{align}\label{eq:expansionFNC}
        &g_{\tau \tau}=-\left(1+a_i(\tau) x^i\right)^2-R_{0 {{i}} 0 {{j}}}(\tau)  x^{i}  x^{j} + \mathcal{O}(r^3),\nonumber\\
        &g_{\tau i}=-\frac{2}{3} R_{0 {{j}}{{i}}{{k}}}(\tau)  x^{j}  x^{k}+ \mathcal{O}(r^3),\nonumber\\
        &g_{ij}=\delta_{{{i}}{{j}}}-\frac{1}{3} R_{{{i}}{{k}}{{j}}{{l}}}(\tau)  x^{k}  x^{l}+ \mathcal{O}(r^3),
\end{align}
where $a_\mu(\tau)$ and $R_{\mu\nu\alpha\beta}(\tau)$ denote the components of acceleration and curvature in Fermi coordinates at $\mf z(\tau)$. This expansion is valid if $|\bm x|$ is sufficiently smaller than both the curvature radius of spacetime and $1/a$, where $a = \sqrt{a^\mu a_\mu}$ is the magnitude of the proper acceleration of the curve. The expansion of Eq. \eqref{eq:expansionFNC} has found many uses in the literature, such as providing a treatment for extended bodies in general relativity~\cite{DixonI,DixonII,DixonIII}, finding the energy level shift on a hydrogen atom due to curvature~\cite{parker,ParkerRevLett,ParkerHydrogen}, describing the motion of point charges in curved spacetimes~\cite{poisson}, and, more recently, describing localized non-relativistic systems in curved spacetimes~\cite{us,jonas,mine,theguy,ahmed,Nico}.

\subsubsection*{The Fermi bound}\label{sub:fermibound}

In this Segment, we define, estimate, and discuss a quantity with units of length, which we name the \emph{Fermi bound}. The Fermi bound is essentially the maximum radius that a system centred at the curve $\mf z(\tau)$ can have in order to be completely described in terms of Fermi normal coordinates.

\begin{figure}[h!]
    \centering
    \includegraphics[width=8cm]{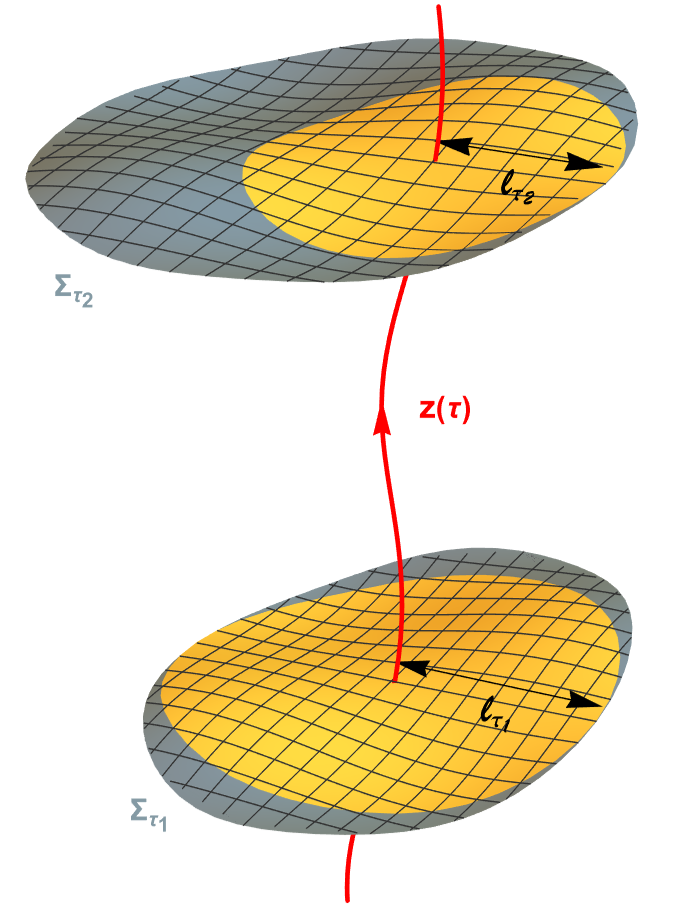}
    \caption{Schematic representation of the region delimited by the $\tau$-Fermi bound (in yellow) within each constant $\tau$ surface $\Sigma_\tau$ (in gray).}
    \label{fig:FermiScheme}
\end{figure}

We first define the $\tau$-Fermi bound. Consider the set of spacelike geodesics which connect $\mf z(\tau)$ to the boundary of $\Sigma_\tau$. The $\tau$-Fermi bound $\ell_\tau$ is defined as the minimum proper length of maximally extended geodesics in this set. In essence, it is the largest radius that a spacelike ball $B\subset T_{\mf z(\tau)}\mathcal{M}$ orthogonal to $u^\mu(\tau)$ can have so that $\exp_{\mf z(\tau)}(B)\subset\Sigma_\tau$. Thus, any system defined in $\Sigma_\tau$ which is centred at $\mf z(\tau)$ and contained in a ball with a proper radius smaller than $\ell_\tau$ can be entirely described using Fermi normal coordinates. There are two parameters that control the size of $\ell_\tau$. The curve's acceleration effectively bends the surfaces $\Sigma_\tau$ so that some geodesics overlap after a length of $1/a$, even in flat spacetimes. Meanwhile, spacetime may be positively curved, which makes nearby geodesics converge so that they overlap after a certain distance. Overall, $\ell_\tau$ is controlled by the curve's acceleration and the curvature of spacetime. A schematic representation of the region delimited by the $\tau$-Fermi bound within each rest space can be found in Fig. \ref{fig:FermiScheme}.

The \emph{Fermi bound} $\ell$ is defined as the infimum of the \mbox{$\tau$-Fermi} bounds. That is,
\begin{equation}
    \ell = \inf_\tau \,\,\ell_\tau.
\end{equation} 
Each of the $\tau$-Fermi bounds defines a bound for the size of a system in $\Sigma_\tau$, which can be described in terms of spacelike Fermi normal coordinates. Thus, the Fermi-bound is a bound for the size of a system centred at the curve $\mf z(\tau)$, which can be entirely described by Fermi normal coordinates at all times. The Fermi bound also defines a world tube around the trajectory, where systems that can be entirely described in Fermi normal coordinates may have support. This tube is defined as the region spanned by all geodesics contained in $\Sigma_\tau$ which have proper length smaller than the Fermi bound $\ell$ for each $\tau$.

\begin{figure}[h]
    \centering
    \includegraphics[width=10cm]{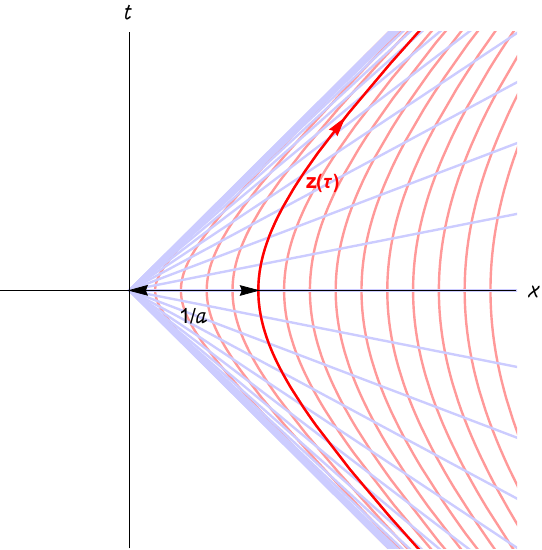}
    \caption{Fermi normal coordinates for a uniformly accelerated trajectory in Minkowski spacetime.}
    \label{fig:acceleration}
\end{figure}

An illustrative example of Fermi normal coordinates and the Fermi bound can be obtained for a uniformly accelerated trajectory in Minkowski spacetime. Consider inertial coordinates $(t,x,y,z)$ and a uniformly accelerated observer undergoing a trajectory $\mf z(\tau) = (\frac{1}{a}\cosh(a\tau),\frac{1}{a}\sinh(a\tau),0,0)$. Then, the Fermi normal coordinates around $\mf z(\tau)$ are the Rindler coordinates $(\tau,\bm x)$, with $\bm x = (X,y,z)$. The metric in these coordinates reads
\begin{equation}
    g = -(1+aX)^2 \dd\tau^2 + dX^2 + dy^2 + dz^2.
\end{equation}
From this expression, one can see that the metric becomes degenerate at $X = -1/a$, which corresponds to the events of the form $(0,0,y,z)$ in inertial coordinates. These are also the events where the Fermi normal coordinates break down. Given that the proper distance is given by the Euclidean distance in the spacelike Fermi normal coordinates, we see that the Fermi bound for a uniformly accelerated trajectory is $\ell = 1/a$. The Fermi normal coordinates and the Fermi bound around a uniformly accelerated trajectory in Minkowski spacetime are depicted in Fig. \ref{fig:acceleration}. 

More generally, in Appendix \ref{app:fermi} we show that under the conditions where the expansion of Eq. \eqref{eq:expansionFNC} is valid, it is possible to estimate the Fermi bound from below by
\begin{equation}\label{eq:approximationFermi}
    \ell\gtrsim  \inf_\tau \left(\frac{1}{a+\sqrt{\lambda_R}}\right),
\end{equation}
where $a = \sqrt{a_\mu a^\mu}$ is the norm of the four acceleration of $\mf z(\tau)$ and $\lambda_R$ is the largest positive eigenvalue of the operator $-R_{0i0j}$, if there are any. This estimate can be useful for providing bounds for the regime of validity of frameworks which use Fermi normal coordinates. Notice that for the case of uniformly accelerated trajectories in Rindler space, this estimate is exact.

\subsection{Non-relativistic quantum systems in curved spacetimes}\label{sec:NRQS}

We now have the tools that will allow us to write a framework that allows one to describe a localized non-relativistic quantum system in curved spacetimes. We will work with systems that can be described by a wavefunction in an $n$ dimensional space and by internal degrees of freedom defined in a finite dimensional space.

\subsubsection*{A single particle in non-relativistic quantum mechanics}\label{sub:nonrel}

To later obtain a generalization to curved spacetimes, let us start by considering a system that can be described in terms of a wavefunction in a non-relativistic setup. We will assume that the system can be described in a Hilbert space {$\mathscr{H} = \mathscr{H}_\textsc{x}\otimes\mathscr{H}_\textsc{s}$}, where $\mathscr{H}_\textsc{x}$ is associated with the position degrees of freedom of the particle (its wavefunction), and $\mathscr{H}_\textsc{s}$ is a finite dimensional Hilbert space associated with its additional internal degrees of freedom (for instance its spin). Then, the canonical variables associated to the position degrees of freedom in $\mathscr{H}_\textsc{x}$ are the position and momentum operators, $\hat{x}^i$ and $\hat{p}_j$. These satisfy the commutation relations
\begin{equation}\label{eq:xpcomm}
    \comm{\hat{x}^i}{\hat{p}_j} = \ii \delta^i_j\openone.
\end{equation}
When translating this description to curved spacetimes, it will be useful to work in the position representation of such system. Let $\ket{\bm x}$ denote the (non-normalizable) eigenstates of $\hat{\bm x} = (\hat{x}^1,...,\hat{x}^n)$, and let $\ket{s}$ be any basis for $\mathscr{H}_\textsc{s}$. Then any state $\ket{\psi}$ can be written as
\begin{equation}
    \ket{\psi} = \sum_s \int \dd^{n} \bm x\, \braket{\bm x,s}{\psi}\,\ket{\bm x,s}.
\end{equation}
We define $\psi^s(\bm x) = \braket{\bm x,s}{\psi}$ as the wavefunction representation of $\ket{\psi}$ in the basis $\ket{s}$. Normalization of the state $\ket{\psi}$ then implies
\begin{align}
    \braket{\psi}{\psi} &= \sum_{s,s'} \int \dd^n \bm x\,\dd^n \bm x' \,(\psi^{s'}(\bm x'))^*\psi^s(\bm x) \braket{\bm x'}{\bm x}\braket{s'}{s}\nonumber\\
    &= \sum_{s,s'} \int \dd^n \bm x(\psi^{s'}(\bm x))^*\psi^s(\bm x) \delta_{s's}\nonumber\\
    &= \int \dd^n \bm x \,\psi_s^*(\bm x) \psi^s(\bm x)=1,
\end{align}
where we denote $\psi_s(\bm x)=\delta_{s's}\psi^{s'}(\bm x)$, and we used Einstein's summation convention from the second to third lines. That is, the components $\psi^s(\bm x)$ can be seen as elements of $L^2(\mathbb{R}^n)$. In fact, we have $\mathscr{H}_\textsc{x}\cong L^2(\mathbb{R}^n)$, where the isomorphism is $\ket{\psi}\longmapsto \psi(\bm x) = \braket{\bm x}{\psi}$. In the space $L^2(\mathbb{R}^n)$, the position operator acts in the wavefunctions as multiplication,
\begin{equation}\label{eq:x}
    \bra{\bm x}\hat{x}^i \ket{\psi} = x^i \psi(\bm x),
\end{equation}
and the momentum operator acts according to
\begin{equation}\label{eq:p}
    \bra{\bm x} \hat{p}_j \ket{\psi} = - \ii \partial_j \psi(\bm x).
\end{equation}
From Eqs. \eqref{eq:x} and \eqref{eq:p}, it is clear that the commutation relations of Eq. \eqref{eq:xpcomm} are satisfied. 

We assume that the dynamics of the system are prescribed by a self-adjoint Hamiltonian $\hat{H}(\hat{\bm x},\hat{\bm p}, \{\hat{s}_i\},t)$, where  $\{\hat{s}_i\}$ denotes any collection of operators acting in $\mathscr{H}_\textsc{s}$ and $t$ denotes a possible external time dependence on the Hamiltonian. This Hamiltonian then generates unitary time evolution according to Schr\"odinger's equation,
\begin{equation}
    \ii\dv{}{t}\ket{\psi(t)} = \hat{H}(t)\ket{\psi(t)},
\end{equation}
where we have omitted the dependence of $\hat{H}$ in $\hat{\bm x}$, $\hat{\bm p}$ and $\{\hat{s}_i\}$ to lighten the notation. Equivalently, one can write the time evolved state in terms of the time evolution operator $\hat{U}(t,t_0)$, defined by $\ket{\psi(t)} = \hat{U}(t,t_0)\ket{\psi(t_0)}$ so that Schr\"odinger's equation gives
\begin{equation}\label{eq:SchU}
    \ii\dv{}{t}\hat{U}(t,t_0) = \hat{H}(t)\hat{U}(t,t_0),
\end{equation}
which can be shown to define a unitary operator $\hat{U}(t,t_0)$. In fact, the solution to Eq. \eqref{eq:SchU} reads
\begin{equation}
    \hat{U}(t,t_0) = \mathcal{T}\text{exp}\left(-\ii \int \dd t \hat{H}(t)\right),
\end{equation}
where $\mathcal{T}\exp$ denotes the time ordered exponential. 

Finally, an important remark needs to be made regarding the system's Hamiltonian. Given that we will later provide a framework that allows one to approximately describe the non-relativistic system in a general relativistic setup, it will be important to consider the \emph{total} energy of the system, which takes into consideration its rest mass. This essentially amounts to adding a term $m c^2 \,\openone$ to the non-relativistic Hamiltonian\footnote{we reintroduced the factor of $c$ here for clarity.}, where $m$ denotes the rest mass of the system. For instance, a particle of mass $m$ under the influence of a potential $V(\bm x)$ should be associated to the Hamiltonian
\begin{equation}\label{eq:HO}
    \hat{H} = m + \frac{\hat{\bm p}^2}{2m} + V(\hat{\bm x}).
\end{equation}
Notice that the introduction of the rest mass does not influence the dynamics of the system, given that it amounts to an overall shift in the energy levels.


\subsubsection*{A localized non-relativistic quantum system in curved spacetimes}\label{sub:main}

We now provide a framework for describing the localized quantum system from the previous Segment undergoing a timelike trajectory $\mf z(\tau)$ in a given $(n+1)$ dimensional background spacetime $\mathcal{M}$. We assume that the quantum system is localized in space at each instant of time. This corresponds to the assumption that there exists a timelike curve $\mf z(\tau)$, parametrizing the events around which the quantum system is localized. For convenience, we will refer to $\mf z(\tau)$ as the trajectory of the system. We then assign Fermi normal coordinates $(\tau,\bm x)$ around the curve $\mf z(\tau)$, so that the local rest spaces of the system are the surfaces $\Sigma_\tau$, defined by constant values of the $\tau$ coordinate. Our goal here is to define the Hilbert space for the wavefunctions at each value of the time coordinate $\tau$ as $L^2(\Sigma_\tau)$ with a suitable integration measure. However, there are some important remarks that have to be considered. 

First, we notice that $\Sigma_\tau$ is only locally defined and does not extend past the normal neighbourhood of $\mf z(\tau)$. This means that if one wishes to consider wavefunctions completely defined in $L^2(\Sigma_\tau)$, then one must consider functions which are defined in a finite-sized box or, equivalently, that the potential which traps the system is infinite outside of a region centred at $\bm x = 0$ with radius smaller than the Fermi bound $\ell$. This condition can be relaxed if the potential is strong enough so that the effective localization of the system is mostly within a radius $\ell$ of the trajectory. Within this relaxed assumption, one loses information about the ``tails of the wavefunction''. Nevertheless, if these tails can be assumed negligible compared to the values of the wavefunction in the region where the surfaces $\Sigma_\tau$ are well defined, the description is approximately valid. Overall, we will call the assumption that the wavefunction is completely localized in each of the $\Sigma_\tau$ surfaces the assumption of \emph{Fermi localization}, and the assumption that the wavefunctions are approximately localized within the $\Sigma_\tau$ surfaces will be called \emph{approximate Fermi localization}. It is important to note that a physical (finite) trapping potential cannot produce wavefunctions that are Fermi localized, only approximately Fermi localized.

Second, it is important to mention that the formalism developed here is not a fundamental description and cannot be valid for a system with arbitrarily high energies. In fact, an important assumption for our model is that the non-relativistic energy of the system is sufficiently small compared to its rest energy. This can also be formulated as the assumption that $\sqrt{\langle \hat{\bm p}^2\rangle}$ is small compared to the system's rest mass $m$, or, in other words, that the system's average velocity $\sqrt{\langle\hat{\bm v}^2\rangle}$ is small compared to the speed of light. As we will discuss, this assumption will ensure that the dynamics introduced by the motion of the system and the spacetime curvature reduce to corrections previously found in the literature in similar setups.

The first step to formulating our description for a non-relativistic system in curved spacetimes is to appropriately determine the inner product in $L^2(\Sigma_\tau)$. The natural choice is to define the inner product as the integral with respect to the measure of the surfaces $\Sigma_\tau$. That is, for $\psi(\bm x)$ and $\phi(\bm x)$ defined in $\Sigma_\tau$,
\begin{equation}\label{eq:innprod}
    (\psi,\phi)_\tau \equiv \int_{\Sigma_\tau} \dd \Sigma \, \psi^*(\bm x) \phi(\bm x),
\end{equation}
where $\dd \Sigma = \sqrt{g_\Sigma(\tau,\bm x)}\,\dd^n \bm x$ is the invariant volume measure in $\Sigma_\tau$, with $g_\Sigma$ being the determinant of the induced metric in the rest surfaces $\Sigma_\tau$. Here we see that the assumption of Fermi localization ensures that the wavefunctions above are well defined within the surface and can be integrated in Eq. \eqref{eq:innprod}. Approximate Fermi localization then ensures that only the tails of the wavefunctions are neglected in Eq. \eqref{eq:innprod}. More than a geometrical and natural inner product, $(\psi,\phi)_\tau$ defined above is obtained when one considers the reduction of Dirac spinors to wavefunctions defined in local rest spaces, as was done in \cite{jonas}. We will also see that the inner product of Eq. \eqref{eq:innprod} allows one to define a consistent quantum theory, with self-adjoint canonically conjugate position and momentum operators for the system.

Under the assumption of Fermi localization, one can then find position and momentum operators defined in terms of their actions in wavefunctions $\psi(\bm x)\in L^2(\Sigma_\tau)$. We define the position operator $\hat{\bm x} = \hat{x}^i \mathsf{e}_i$, where $\mf{e}_i$ denotes the extended Fermi frame and the $\hat{x}^i$ are defined through their action on wavefunctions as
\begin{equation}\label{eq:posOp}
    \hat{x}^i:\psi(\bm x)\longmapsto x^i \psi(\bm x).
\end{equation}
The equation above is simply the generalization of the idea that each component of the position operator multiplies the wavefunction by the components of $\bm x$, here in Fermi normal coordinates. Physically, the definition of Eq. \eqref{eq:posOp} is justified by the fact that in Fermi normal coordinates, $|\bm x|=\sqrt{\delta_{ij}x^ix^j}$ corresponds to the proper distance between a point and the center of the curve. It is also important to mention that the components of the position operator defined above are also self-adjoint with respect to the inner product defined in Eq. \eqref{eq:innprod}.

The momentum operator in the position representation can also be defined by its action on wavefunctions \mbox{$\psi(\bm x)\in\Sigma_\tau$}. We define its components by
\begin{equation}\label{eq:pOp}
    \hat{p}_j:\psi(\bm x) \longmapsto  \frac{- \ii}{(g_\Sigma)^\frac{1}{4}}\pdv{}{x^j} \left((g_\Sigma)^\frac{1}{4}\psi(\bm x)\right).
\end{equation}
Although the factors of $1/4$ may seem out of place at first glance, they are necessary for the $\hat{p}_j$ operators to be self-adjoint with respect to the inner product of Eq. \eqref{eq:innprod}. In fact,
\begin{align}
    (\psi,\hat{p}_j\phi)_\tau &=  \int\dd^n\bm x \sqrt{g_{\Sigma}}\,\psi^*(\bm x) \frac{-\ii}{(g_\Sigma)^\frac{1}{4}}\partial_j \left((g_\Sigma)^\frac{1}{4}\phi(\bm x)\right)\nonumber\\
    &= -\ii \int \dd^n\bm x(g_\Sigma)^\frac{1}{4} \psi^*(\bm x) \partial_j \left((g_\Sigma)^\frac{1}{4}\phi(\bm x)\right)\nonumber\\
    &= \ii \int \dd^n\bm x\partial_j\left((g_\Sigma)^\frac{1}{4} \psi^*(\bm x)\right) (g_\Sigma)^\frac{1}{4}\phi(\bm x)\nonumber\\
    &= \int \dd^n\bm x\left(-\ii\partial_j\left((g_\Sigma)^\frac{1}{4} \psi\right)\right)^* (g_\Sigma)^\frac{1}{4}\phi(\bm x)\nonumber\\
    &= \int \dd^n\bm x\sqrt{g_{\Sigma}}\left(\frac{-\ii}{(g_\Sigma)^\frac{1}{4}}\partial_j\left((g_\Sigma)^\frac{1}{4} \psi(\bm x)\right)\right)^* \!\!\phi(\bm x) \nonumber\\
    &= (\hat{p}_j \psi,\phi)_\tau,
\end{align}
where we have integrated by parts in the third equality and the boundary terms vanish under the assumption of Fermi localization\footnote{Under the assumption of approximate Fermi localization, one would then obtain boundary terms of the same order as the terms lost in the inner product, yielding approximately self-adjoint $\hat{p}_j$ operators. In order to obtain a fully consistent framework, one can then effectively truncate approximately Fermi localized wavefunctions so that their dynamics can be approximated by that of compactly supported wavefunctions.}. Not only are the momentum operator components defined in Eq. \eqref{eq:pOp} self-adjoint, but they also satisfy the canonical commutation relations with the position operator of Eq. \eqref{eq:posOp}:
\begin{equation}
    \hat{x}^i \hat{p}_j - \hat{p}_j \hat{x}^i : \psi(\bm x) \longmapsto \ii \delta^i_j\psi(\bm x).
\end{equation}
Thus, the $\hat{x}^i$ and $\hat{p}_j$ operators defined here are valid generalizations of the position and momentum operators of quantum mechanics. Indeed, given any non-relativistic quantum system described in terms of its position and momentum operators $\hat{\bm x}$ and $\hat{\bm p}$, one can describe it around a trajectory $\mf z(\tau)$ for a fixed $\tau$ by considering its wavefunction to be in $\Sigma_\tau$, and interpreting the corresponding position and momentum operators as Eqs. \eqref{eq:posOp} and \eqref{eq:pOp}. We remark that canonically conjugated position and momentum operators fully define a quantum theory for a wavefunction at a given time, so that the description provided here is indeed enough for describing the position degrees of freedom of the quantum state around the trajectory $\mf z(\tau)$.

To fully describe the system in curved spacetimes, one also requires to describe the additional internal degrees of freedom of the system contained in the collection of operators $\{\hat{s}_i\}$ in this more general setup. Given that these degrees of freedom are internal to the particle, we describe them in the same Hilbert space $\mathscr{H}_\textsc{s}$, with no modifications to the inner product when one goes to curved spacetimes. Notice, however, that although their description will not change, it might be necessary to introduce different dynamics for the $\{\hat{s}_i\}$ operators in order to describe their evolution in curved spacetimes\footnote{For instance, the dynamics of the spin of particles can be affected by spacetime curvature in a non-trivial way, as discussed in~\cite{parker,jonas,theguy,Nico}.}. This will be further discussed when we consider time evolution in this framework. 

The treatment given so far allows one to describe a non-relativistic quantum system in curved spacetimes at a given surface $\Sigma_\tau$. However, we have not yet mentioned how time evolution can be implemented in this description. In other words, we have yet to describe a Hamiltonian formulation in this setup. First, notice that at each value of the time parameter $\tau$, the wavefunctions are defined in a \emph{different} Hilbert space $L^2(\Sigma_\tau)$. This adds extra complications when writing Schr\"odinger's equation, as one cannot differentiate states with respect to time via a limit of infinitesimal differences, since $\psi(\tau_0+\delta\tau,\bm x)$ and $\psi(\tau_0,\bm x)$ are defined in different Hilbert spaces. To make sense of Schr\"odinger's equation in this setup, we must locally extend the wavefunctions defined in $\Sigma_{\tau_0}$, so that we obtain a function $\psi(\tau,\bm x)$ for $\tau\in[\tau_0,\tau_0 + \varepsilon)$ for a small $\varepsilon>0$. It is then possible to compare its values at different $\tau$'s so that differentiation can be performed. This essentially amounts to differentiation of a scalar function locally defined in spacetime with respect to the time parameter $\tau$.

At this stage, one could naively think that given a Hamiltonian $\hat{H}(\hat{\bm x},\hat{\bm p},\{\hat{s}_i\},t)$ for a quantum particle in a non-relativistic setup, it is enough to replace its dependence on $\hat{\bm x}$, $\hat{\bm p}$ and $\{\hat{s}_i\}$ as described previously, together with the replacement $t\longmapsto\tau$ in order to write Schr\"odinger's equation. However, an important missing ingredient also has to be considered: redshift. As mentioned in Subsection \ref{sub:FNC}, the time parameter $\tau$ only corresponds to the proper time of an observer along the curve $\mf z(\tau)$, but not locally around it. This implies that the time evolution at each point of space should contain a redshift factor associated with the time dilation of the foliation defined by the $\Sigma_\tau$ surfaces. In Appendix \ref{app:det}, we compute the corresponding redshift factor. It is given by
\begin{equation}\label{eq:redshift}
    \gamma(\tau,\bm x) = \abs{g_{\tau\tau} - g_{\tau i} g_{\tau j} \bar{g}^{ij}}^{\frac{1}{2}},
\end{equation}
where $\bar{g}^{ij}$ denotes the inverse of the induced metric in the $\Sigma_\tau$ surfaces. 

In a classical system, one would take this redshift factor into account by multiplying the local Hamiltonian of the system by $\gamma(\tau,{\bm x})$, giving rise to the effective Hamiltonian $\gamma(\tau,{\bm x})\hat{H}({\bm x},{\bm p},\{{s}_i\},\tau)$. In a quantum setup, one would then be tempted to describe the Hamiltonian as $\gamma(\tau,\hat{\bm x})\hat{H}(\hat{\bm x},\hat{\bm p},\{\hat{s}_i\},\tau)$, promoting the space dependence in $\gamma(\tau,\bm x)$ to the position operator $\hat{\bm x}$. However, this product will not necessarily be self-adjoint due to the dependence of $\hat{H}$ on $\hat{\bm p}$. To obtain a self-adjoint Hamiltonian, one could then use the Weyl quantization prescription~\cite{Weyl1927} for the Hamiltonian $\gamma(\tau,\hat{\bm x})\hat{H}(\hat{\bm x},\hat{\bm p},\{\hat{s}_i\},\tau)$, or use the Moyal product~\cite{moyal}. A simpler way to handle the self-adjointness problem is to define the Hamiltonian via a symmetrization as
\begin{equation}\label{eq:Hcurved}
    \hat{\mathsf{H}}(\hat{\bm x},\hat{\bm p},\{\hat{s}_i\},\tau) = \frac{1}{2}\left(\gamma(\tau,\hat{\bm x})\hat{H}(\hat{\bm x},\hat{\bm p},\{\hat{s}_i\},\tau) + \text{H.c.}\right).
\end{equation}
Although different quantization methods might, in principle, give different Hamiltonians, in Subsection \ref{sub:coincidence} we will argue that the Hamiltonian can be well approximated by \mbox{$\hat{H}(\hat{\bm x},\hat{\bm p},\{\hat{s}_i\},\tau)+ma_i(\tau)\hat{x}^i + \frac{m}{2}R_{0i0j}(\mf z(\tau))\hat{x}^i\hat{x}^j$}, where $m$ is the rest mass of the system. Moreover, this approximate correction is also independent of ordering ambiguities and yields the same result for any quantization prescription chosen. 

Having the system's Hamiltonian properly prescribed, we are at the stage where we can write Schr\"odinger's equation. We first write Schr\"odinger's equation for a system with no extra spin degrees of freedom. That is, in the case where the Hilbert space $\mathscr{H}_\textsc{s}$ is trivial, and the system can be entirely described by its wavefunction. In this case, Schr\"odinger's equation can be written simply as
\begin{equation}
    \ii\pdv{}{\tau}\psi(\tau,\bm x) =\hat{\mathsf{H}}(\hat{\bm x},\hat{\bm p},\tau)\psi(\tau,\bm x),
\end{equation}
where the $\hat{\bm x}$ and $\hat{\bm p}$ operators act in $\psi(\tau,\bm x)$ according to Eqs. \eqref{eq:posOp} and \eqref{eq:pOp}. We remark that although it might look like the newly introduced dynamics and the extra factors in the differential operator $\hat{\bm p}$ give rise to a much more complicated differential equation, one can instead use the commutation relations between $\hat{\bm x}$ and $\hat{\bm p}$ in order to find the solutions to Schr\"odinger's equation (as is usually done with the quantum harmonic oscillator, for instance).

To write Schr\"odinger's equation when the system also has internal degrees of freedom in $\mathscr{H}_\textsc{s}$, we write states in the Dirac notation, with \mbox{$\ket{\psi}\in \mathscr{H}^{(\tau)}_{\textsc{x}}\otimes \mathscr{H}_\textsc{s}$}, where $\mathscr{H}^{(\tau)}_{\textsc{x}}\cong L^2(\Sigma_\tau)$ for each $\tau$. In this context, the position eigenvectors are $\ket{\bm x}$ such that $\psi^s(\tau,\bm x) = \braket{\bm x,s}{\psi(\tau)}$ and a decomposition of the identity in the position basis can be written as
\begin{equation}\label{eq:idDecomp}
    \openone = \sum_s \int \dd \Sigma \ket{\bm x,s}\!\!\bra{\bm x,s}.
\end{equation}
In terms of Dirac's notation, we can then write Schr\"odinger's equation as
\begin{equation}
    \ii \dv{}{\tau}\ket{\psi} = \hat{\mathscr{H}}(\tau) \ket{\psi},
\end{equation}
where the $\tau$ differentiation in the position spaces is understood via the local extension of the wavefunctions, and we omitted the dependence of $\hat{\mathscr{H}}$ in $\hat{\bm x}$, $\hat{\bm p}$ and $\{\hat{s}_i\}$ for simplicity. This equation also defines unitary operators $\hat{U}(\tau,\tau_0)$ by $\hat{U}(\tau,\tau_0)\ket{\psi(\tau_0)} = \ket{\psi(\tau)}$\footnote{It is important to keep in mind that this family of unitary operators acts in different Hilbert spaces. That is,
\begin{align}
    \hat{U}(\tau,\tau_0): \mathscr{H}^{(\tau_0)}_{\textsc{x}}\otimes \mathscr{H}_\textsc{s}&\longrightarrow  \mathscr{H}^{(\tau)}_{\textsc{x}}\otimes \mathscr{H}_\textsc{s}\\
    \:\:\:\:\ket{\psi(\tau_0)}&\longmapsto \ket{\psi(\tau)},\nonumber
\end{align}
where $\mathscr{H}^{(\tau_0)}_{\textsc{x}}$ and $\mathscr{H}^{(\tau)}_{\textsc{x}}$ denote the $L^2$ spaces at $\Sigma_{\tau_0}$ and $\Sigma_\tau$, respectively.}.

Finally, we comment on the possible need to perform additional changes to the Hamiltonian $\hat{H}$, apart from the redshift factor and the replacement of the position and momentum operators for their definitions in Eqs. \eqref{eq:posOp} and \eqref{eq:pOp}. These additional changes could come from interactions of the other internal degrees of freedom of the system (encoded in the $\{\hat{s}_i\}$ operators) with curvature and acceleration. Although it is not possible to give a general recipe for adapting general operators to curved spacetimes, the framework provided here can accommodate these changes in each case with minor modifications. For instance, {in~\cite{jonas,theguy,Nico}}, a fermionic particle in curved spacetimes is described, and the coupling of its spin with curvature is obtained. This could be implemented here by adding terms to the Hamiltonian of Eq. \eqref{eq:Hcurved} corresponding to this interaction.

\begin{figure}[h!]
    \centering
    \includegraphics[width=8cm]{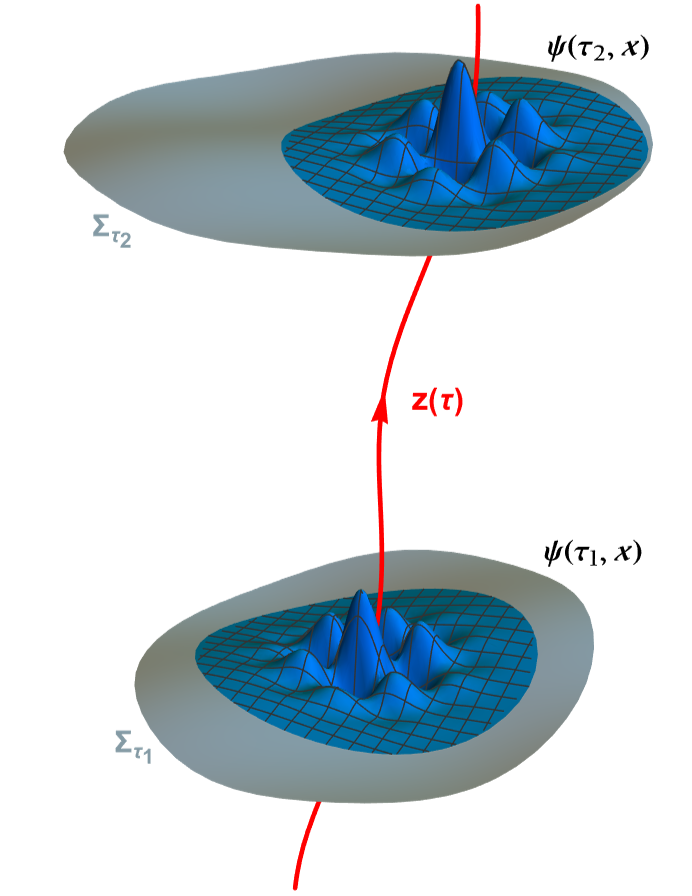}
    \caption{Schematic representation of the model for localized non-relativistic systems in curved spacetimes, with wavefunctions defined in the local rests paces.}
    \label{fig:WavefunctionScheme}
\end{figure}

Overall, in this Segment, we completed one of the main goals of this Section: we provided a consistent description for a localized non-relativistic quantum system in curved spacetimes. A schematic representation of the model obtained is displayed in Fig. \ref{fig:WavefunctionScheme}. Throughout the remainder of the Section, we will discuss consequences and applications of this formalism.

\subsubsection*{Discussion of the regime of validity of the model}\label{sub:regime}

In this Section we discuss the regime of validity of the framework presented in Subsection \ref{sub:main}, and discuss the compatibility of the model with the framework of general relativity.

First, we remark once again that for the formalism presented in the Segment above to be applied, the non-relativistic quantum system must be sufficiently localized within the Fermi bound of the trajectory, $\ell$. For instance, the formulation works for systems which are compactly supported with support contained in a sphere of radius smaller than $\ell$. However, in order to perfectly trap a quantum particle, one requires an infinite potential, which is an unphysical assumption. If the potential is finite but strong enough to approximately Fermi localize the system, the formalism can still be applied by neglecting the tails of the wavefunction outside of a sphere of radius $\ell$. Although one expects to lose some information about the system, the loss due to this approximation can be controlled by analyzing the tails of the wavefunctions associated with the relevant states of the system. 

It is possible to quantify the regime of validity of the theory using the estimate for the Fermi bound of Eq. \eqref{eq:approximationFermi}. Consider a non-relativistic quantum theory, which we wish to describe around a trajectory $\mf z(\tau)$ in curved spacetimes. We will work under the assumption that the system is \emph{strongly supported} within a region of radius $R(\varepsilon)$. The notion of strong support has been used by Eduardo Mart\'in-Mart\'inez to characterize functions that decay quickly with some characteristic scale since 2015~\cite{Pozas-Kerstjens:2015}. In the context of our framework, a state of a quantum field will be said to be strongly supported within a region of radius $R(\varepsilon)$ if the expected value of a set of observables of interest can be computed to precision $\varepsilon$ by performing spatial integrals of the wavefunctions in a spatial ball of radius $R(\varepsilon)$ centred at $\mf z(\tau)$ for each $\tau$. The collection of observables of interest will explicitly depend on the system under consideration and on the observables that are relevant to the predictions one wishes to compute in each setup.

Within the assumption that the relevant states of the system have strong support within a region of radius $R(\varepsilon)$, the condition for our framework to be applicable is $R(\varepsilon)<\ell$, and the errors in the description are controlled by the parameter $\varepsilon$. Assuming the system to be sufficiently localized with respect to the curvature of spacetime and the trajectory's acceleration, one can then use the estimate of Eq. \eqref{eq:approximationFermi}, which gives the approximate condition
\begin{equation}
    R(\varepsilon)<\frac{1}{a + \sqrt{\lambda_R}},
\end{equation}
where $\lambda_R$ is the largest positive eigenvalue of $-R_{0i0j}$ and $a$ denotes the maximum acceleration of the system along its motion. If the trapping potential which localizes the system is sufficiently strong, a good estimate for the localization of the system is $\sqrt{\langle\hat{\bm x}^2\rangle}$ so that the framework can be applied when $\sqrt{\langle\hat{\bm x}^2\rangle}<1/(a+\sqrt{\lambda_R})$. However, different systems might require different methods of estimating their localization depending on the specific shape of the trapping potential.

Another important assumption for our setup is that the energy of the system is non-relativistic (that is, sufficiently smaller than the system's rest energy). Although this assumption is not explicitly required in order to construct the formalism, it is important so that the prescribed Hamiltonian of Eq. \eqref{eq:Hcurved} is able to accurately predict the relativistic corrections to the internal dynamics of the system. In fact, we will later show that, under this assumption, the framework yields the most relevant relativistic corrections for the description of wavefunctions in curved spacetimes found in~\cite{jonas}. Moreover, we remark again that other relativistic modifications to the prescription of Eq. \eqref{eq:Hcurved} might be required, which take into account the relationship of the internal degrees of freedom of the system with the geometry of spacetime and the trajectory's motion.

We now comment on the relationship between the non-relativistic quantum theory and the framework of general relativity. In our model, we assumed that spacetime is not affected by the quantum system. This assumption is reasonable, provided that the stresses and energy of the quantum system are small enough, in which case their effect on the spacetime metric can be neglected. This a usual assumption even when considering relativistic quantum theories in curved spacetimes, so it is natural to expect that this assumption also has to be made in our treatment. Moreover, considering the effect of superpositions of quantum systems in a background spacetime has been argued to lead to superpositions of spacetimes~\cite{flaminiaSuperpositionSpacetimes2022}, but these fall beyond the scope of our discussions.

Finally, we comment on the relationship of our model with causality and what is missed in this treatment. The formalism developed here allows one to consider the effect of the local geometry of a curved spacetime in the dynamics of a quantum system locally via the dependence on the position operator $\hat{\bm x}$. Although this interaction affects the system's position degree of freedom locally, in a non-relativistic setup, this effect changes a single degree of freedom (the system's wavefunction) at each value of $\tau$. In essence, this means that the local effect of curvature in each portion of the system affects it instantaneously in its own frame, implying that information within this quantum system propagates acausally. This is expected from any non-relativistic quantum theory and is also the case for our framework. The consequences of this causality violation have been carefully studied in the literature in similar setups (see e.g.~\cite{us2,PipoFTL,BrunoDan}). An expected consequence of this violation is the ability of non-relativistic systems to signal between spacelike separated points, violating causality at the order of their effective size. For instance, when coupling these systems to a quantum field, one would find that the operations performed on the field would, in general, lead to the Sorkin-type problems discussed in Section~\ref{sec:meas}. We remark that no physical system should be able to signal faster than light, and this can be seen as another bound for the regime of validity of our framework, similar to what happens with specific non-relativistic quantum systems in curved spacetimes ~\cite{eduardo2015,us2,PipoFTL,mariaPipoNew}.

Overall, we conclude that the framework presented in this Section can be employed to accurately describe non-relativistic quantum systems provided that 1) the system is sufficiently localized with respect to the Fermi bound, 2) its non-relativistic energy is sufficiently smaller than its rest energy, and 3) one is not interested in its uses in communication protocols between regions which are spacelike separated by proper distances which are of the order of the size of the system. 





\subsubsection*{The coincidence limit and first order corrections}\label{sub:coincidence}

We will now describe the first order corrections emerging due to the motion of the system and spacetime curvature due to the addition of the redshift factor in Eq. \eqref{eq:redshift}.

In~\cite{jonas}, a non-relativistic quantum formalism has been presented for the approximate description of a localized fermionic particle in curved spacetimes. The authors then traced over the spin degrees of freedom of the particle, obtaining the free Hamiltonian for a localized non-relativistic wavefunction added to corrections in terms of curvature and acceleration. For systems such that their non-relativistic energy is much smaller than their rest mass, the most relevant correction terms found were $ma_i \hat{x}^i$ and $\frac{m}{2}R_{0i0j}\hat{x}^i\hat{x}^j$. {These corrections were also found in~\cite{gravDetec,roura,theguy,Nico}, among others.} We will now show that to first order in curvature and acceleration, these corrections are also the ones provided by the formalism presented in Subsection \ref{sub:main}. 

The new dynamics introduced by the formalism we presented are due to the introduction of the redshift factor in the original Hamiltonian, which localizes the quantum system. Using the expansions of Eq. \eqref{eq:expansionFNC}, we can write the redshift factor as
\begin{equation}
    \gamma(\tau,\bm x) = 1 + a_i x^i + \frac{1}{2} R_{0i0j}x^ix^j + \mathcal{O}(r^3).
\end{equation}
Considering only the first order corrections in Eq. \eqref{eq:Hcurved} due to spacetime curvature and acceleration (which are also the corrections to second order in the system's localization), one obtains the approximate Hamiltonian
\begin{equation}\label{eq:Kwurtz}
    \hat{\mathsf{H}}(\tau) \approx \hat{H}(\tau) + \frac{1}{2}\left(\left(a_i \hat{x}^i + \tfrac{1}{2} R_{0i0j}\hat{x}^i\hat{x}^j\right)\hat{H}(\tau) + \text{H.c.}\right).
\end{equation}
Notice that the Hamiltonian $\hat{H}(\tau)$ can be split into two contributions: the rest mass of the system and its non-relativistic energy, associated with an operator $\hat{H}_{\textsc{nr}}(\tau)$ which contains its kinetic and potential energies,
\begin{equation}
    \hat{H}(\tau)  = m +\hat{H}_{\textsc{nr}}(\tau).
\end{equation}
The assumption that the system's degrees of freedom are non-relativistic then implies that its non-relativistic energy is much smaller than its rest mass, $\langle\hat{H}_{\textsc{nr}}(\tau)\rangle\ll m$. In particular, this implies that the terms of the form $\left(a_i \hat{x}^i + \tfrac{1}{2} R_{0i0j}\hat{x}^i\hat{x}^j\right)\hat{H}_{\textsc{nr}}(\tau)$ in Eq. \eqref{eq:Kwurtz} contribute much less than the terms which involve the rest mass of the system. In fact, reintroducing units of $c$, the term $a_i x^i$ picks up a factor of $1/c^2$, while the factor of $R_{0i0j}\hat{x}^i\hat{x}^j$ is of the order of the ratio of the size of the system by the curvature radius of spacetime. It is then safe to neglect these terms in most non-relativistic setups. 

Under these approximations, we can write
\begin{equation}
    \left(a_i \hat{x}^i + \tfrac{1}{2} R_{0i0j}\hat{x}^i\hat{x}^j\right)\!\hat{H}(\tau)\approx m\!\left(a_i \hat{x}^i + \tfrac{1}{2} R_{0i0j}\hat{x}^i\hat{x}^j\right),
\end{equation}
so that the Hamiltonian that promotes time evolution with respect to the time parameter $\tau$ can be written as
\begin{equation}\label{eq:approxH}
    \hat{\mathsf{H}}(\tau)\approx \hat{H}(\tau)+m  a_i(\tau) \hat{x}^i + \frac{m}{2}R_{0i0j}(\tau)\hat{x}^i \hat{x}^j.
\end{equation}
Notice that the correction terms arise from the product of the rest mass term $m\openone$ with the redshift factor. Due to the fact that the rest mass term in the Hamiltonian is proportional to the identity and that $\gamma(\tau,\hat{\bm x})$ only depends on $\bm x$, the leading order relativistic corrections to the Hamiltonian are independent of the ordering ambiguities that might show up (See Eq. \eqref{eq:Hcurved} and related discussions). The Hamiltonian of Eq. \eqref{eq:approxH} also precisely matches the coupling of a localized system with curvature used in~\cite{gravDetec,pitelli} and the most relevant corrections found in~\cite{jonas}. However, in~\cite{jonas}, other corrections of the order of acceleration and curvature times the system's non-relativistic energy are also found. These extra corrections are not naturally taken into account in our formalism, and can be neglected if the non-relativistic rest energy of the system is sufficiently smaller than its rest mass. If, on the other hand, one wishes to consider these corrections, it is enough to add them to the prescription of the Hamiltonian in Eq. \eqref{eq:Hcurved}.

Eq. \eqref{eq:approxH} gives a Hamiltonian that is naturally self-adjoint under this non-relativistic approximation. Eq. \eqref{eq:approxH} also gives a simple expression for $\hat{\mathsf{H}}(\tau)$, with a quadratic correction to the Hamiltonian $\hat{H}(\tau)$. This approximation is valid provided that the energy of the system is much smaller than $mc^2$ and that the system is more localized than both the curvature radius of spacetime and than $c^2/a$, where we reintroduced the factors of $c$ in order to make the limits of validity more explicit.

\subsubsection*{A uniformly accelerated quantum harmonic oscillator in constant curvature spacetimes}\label{sub:example}

In this Segment we exemplify the formalism developed in this section to a harmonic oscillator with uniform proper acceleration in a constant curvature spacetime. For convenience, we will assume {the spatial dimension to be} $n=3$ in this example.

The Riemann curvature tensor in a constant curvature spacetime takes the shape
\begin{equation}
    R_{\mu\nu\alpha\beta} = \alpha (g_{\mu\alpha}g_{\nu \beta} - g_{\mu\beta}g_{\nu\alpha}),
\end{equation}
where $\alpha$ is a constant, related to the Ricci curvature by $\alpha = R/12$. The sign of $R$ determines whether spacetime is positively ($R>0$) or negatively ($R<0$) curved, corresponding to deSitter and anti-deSitter spacetimes, respectively. In particular, in Fermi normal coordinates associated with a trajectory $\mf{z}(\tau)$, we obtain
\begin{equation}\label{eq:Rconst}
    R_{0i0j}(\mf z(\tau)) = -\alpha\delta_{ij},
\end{equation}
so that $R_{0i0j}\hat{x}^i \hat{x}^j = -\alpha \hat{\bm x}^2$, which is the relevant quantity for the redshift factor $\gamma(\tau,\hat{\bm x})$ that influences the Hamiltonian of the system.

The Hamiltonian for a quantum harmonic oscillator with frequency $\omega$ and mass $m$ can then be written as Eq. \eqref{eq:HO} by taking $V(\bm x) = m \omega^2 \bm x^2/2$. That is,
\begin{equation}
    \hat{H} = m + \frac{\hat{\bm p}^2}{2m} + \frac{m \omega^2}{2}\hat{\bm x}^2.
\end{equation}
To describe the system in a background curved spacetime, we must then have that the particle is localized in a region sufficiently smaller than the Fermi bound. A quantum harmonic oscillator is localized in a region of the order of $\sqrt{\langle\hat{\bm x}^2\rangle}=\kappa/\sqrt{m\omega}$, where $\kappa=\sqrt{2\langle\hat{n}\rangle+1}$. We then use the approximation derived in Appendix \ref{app:fermi} for the Fermi bound, $\ell \gtrsim 1/(a+\sqrt{\lambda_R})$, where $\lambda_R$ is the largest negative eigenvalue of $R_{0i0j}$. Using Eq. \eqref{eq:Rconst}, we find that $\lambda_R = \text{max}(0,-\alpha)$. Thus, the formalism developed in this section can be applied if the relevant states for the setup satisfy
\begin{equation}
    \frac{\kappa}{\sqrt{m \omega}}<\frac{1}{a + \sqrt{|\alpha|}}.
\end{equation}
If the spacetime is positively curved, we can then describe the system if the frequency of the harmonic oscillator is larger than $a^2\kappa^2/m$. If spacetime is negatively curved, curvature reduces the Fermi bound so that one must have a frequency larger than $(a+ \sqrt{R/12})^2\kappa^2/m$. In either case, we will assume that the frequency of our harmonic oscillator is large enough so that we can employ our formalism.

Employing the approximation of Eq. \eqref{eq:approxH}, we then obtain the Hamiltonian of the system
\begin{equation}
    \hat{\mathsf{H}} = m + m\, \bm a \cdot \hat{\bm x} -\frac{m \alpha}{2} \hat{\bm x}^2 + \frac{\hat{\bm p}^2}{2m} + \frac{m\omega^2}{2}\hat{\bm x}^2.
\end{equation}
Given that all terms in the Hamiltonian above are quadratic, it still defines a quantum harmonic oscillator. In fact, $\hat{\mathsf{H}}$ can be rewritten as
\begin{equation}
    \hat{\mathsf{H}} = m -\frac{m \bm a^2}{2(\omega^2 - \alpha)}+ \frac{\hat{\bm p}^2}{2m} + \frac{m(\omega^2-\alpha)}{2}\left(\hat{\bm x}+\frac{\bm a}{\omega^2 - \alpha}\right)^2.
\end{equation}
From the equation above, we can see that the resulting theory is a harmonic oscillator with frequency $\omega'$ given by $\omega'^2 = \omega^2 -\alpha$, with wavefunctions shifted by the vector $- \bm a /\omega'^2$. The acceleration of the system and curvature of spacetime also shift the energy of the ground state by $-m\bm a^2/2\omega'^2$.

Notice that if the curvature of spacetime is too large (with $R>0$), then it is possible that the effective potential generated by deSitter curvature is larger than the trapping potential of the oscillator, which would result in the particle accelerating away from the center of the coordinate system. In this regime, the wavefunction of the system cannot be considered to be localized, and the condition of Fermi localization breaks down, in which case this formalism would not be suitable for its description. 

On the other hand, anti-deSitter spacetime creates an effective trapping potential for the particle and increases the frequency of the oscillator. This phenomenon has also been seen for a localized fermionic system in~\cite{jonas}, where it was also found that the curvature of anti-deSitter spacetime can be responsible for trapping the particle even when $\omega = 0$.

\subsection{Application to Particle Detector Models}\label{sec:detectors}

We finally have all the tools to relate the framework presented in Section \ref{sec:NRQS} to the formalism of particle detector models. In this Section we formulate a general notion of a non-relativistic particle detector based on the coupling of a given non-relativistic quantum system with a quantum field.

\subsubsection*{General non-relativistic particle detector models}\label{sub:generalPDs}


To describe a particle detector model from the framework presented in Section \ref{sec:NRQS}, we require two extra ingredients apart from the non-relativistic quantum system: a quantum field theory and an interaction between the quantum field and the localized system. In this Section, we will refer to the non-relativistic quantum system as ``the detector''.


The interaction of the detector with a field is prescribed in terms of an interaction Hamiltonian which couples to an operator-valued distribution of a quantum field theory, say $\hat{O}^b(\mf x)$, where $b$ stands for any collection of Lorentz indices. To produce a scalar interaction Hamiltonian, one must have an operator in the detector's Hilbert space, which is a tensor of the same rank as that of $\hat{O}^b(\mf x)$. We define the tensor operator as $\hat{\mu}(\tau) = \hat{\mu}^b(\tau)\mf{E}_b$, where $\hat{\mu}^b(\tau)$ is an operator in $\mathscr{H}_\textsc{x}^{(\tau)}\otimes\mathscr{H}_\textsc{s}$ and $\mf E_b$ denotes the orthonormal frame for tensors of the same rank as $\hat{O}^b(\mf x)$ built from the extended Fermi frame $\mf e_\mu$. We further assume $\hat{\mu}^b(\tau)$ to be only a function of the operators $\hat{\bm x}$, $\hat{\bm p}$ and $\{\hat{s}_i\}$ and of the time parameter $\tau$. For convenience, we will work in the interaction picture from now on, so that $\hat{\mu}^b(\tau)$ includes the free time evolution associated with the detector's free Hamiltonian $\hat{\mathsf{H}}(\tau)$. Then, the interaction Hamiltonian is prescribed in the interaction picture as
\begin{equation}\label{eq:generalPD}
    \hat{H}_I(\tau) = \lambda \gamma(\tau,\hat{\bm x})\hat{\mu}^\dagger_b(\tau)\hat{O}^b(\tau,\hat{\bm x}) + \text{H.c.},
\end{equation}
where $\hat{O}^b(\tau,{\bm x})$ denote{s} the components of the operator $\hat{O}^b(\mf x)$ in the frame $\mf{E}_b$\footnote{We consider $\hat{O}^b(\mf x)$ and $\hat{\mu}^b(\tau)$ to be written in the orthonormal frame $\mf E_b$ in order to avoid unnecessary metric prescriptions in the contraction.} evaluated in Fermi normal coordinates around the curve $\mf z(\tau)$, { and the replacement of the dependence in the (classical) coordinates $\bm x$ by the quantum position operator $\hat{\bm x}$ formally means} 
\begin{equation}\label{eq:defOb}
    {\hat{O}^b(\tau,\hat{\bm x}) \coloneqq  \int
\dd\Sigma \,\hat{O}^b (\tau , \bm{x}) \ket{\bm{x}}\!\!\bra{\bm{x}}_{\tau},}
\end{equation}
{where the subscript $\tau$ in $\ket{\bm x}\!\!\bra{\bm x}_\tau$ denotes time evolution with respect to the detector's free Hamiltonian: $\ket{\bm x}\!\!\bra{\bm x}_\tau = \hat{U}^\dagger(\tau)\ket{\bm x}\!\!\bra{\bm x}\hat{U}(\tau)$ with $\hat{U} = \mathcal{T}\exp(-\ii \int^\tau \hat{\mathsf{H}}(\tau')\dd \tau')$.} In Eq. \eqref{eq:generalPD}, $\gamma(\tau,\hat{\bm x})$ denotes the redshift factor of Eq. \eqref{eq:redshift} and $\hat{\mu}_b^\dagger(\tau)$ denotes the dual field to $\hat{\mu}^b(\tau)$. 

We can show that this system satisfies the general definition presented in Section~\ref{sec:UDW} whenever the interaction Hamiltonian is diagonal in position basis. That is, whenever $\bra{\bm x'}\hat{\mu}_b^\dagger(\tau)\ket{\bm x} = \hat{\mu}_b^\dagger(\tau,\bm x)\delta(\bm x, \bm x')$, where $\hat{\mu}_b^\dagger(\tau,\bm x)$ acts only on $\mathscr{H}_\tc{s}$. This happens whenever $\hat{\mu}_b^\dagger(\tau)$ is only a function of the position operator $\hat{\bm x}$. Indeed, in this case we can insert position identities~\eqref{eq:defOb} in the interaction Hamiltonian~\eqref{eq:generalPD} to recast it as
\begin{equation}
    \hat{\mathsf{H}}_I(\tau) = \lambda \int \dd \Sigma \gamma(\tau,\bm x)\mu^\dagger_b(\tau,\bm x)\hat{O}^b(\tau,\hat{\bm x})\ket{\bm x}\!\!\bra{\bm x}_\tau + \text{H.c.},
\end{equation}
where we notice that $\sqrt{-g} \dd^n \bm x = \gamma(\tau,\bm x) \dd \Sigma$ in Fermi normal coordinates. The interaction Hamiltonian density is then 
\begin{equation}
    \hat{\mathsf{H}}_I(\mf x) = \lambda \mu^\dagger_b(\tau,\bm x)\hat{O}^b(\mf x) + \text{H.c.},
\end{equation}
identifying $\hat{\mu}_b^\dagger(\tau,\bm x)$ as the operator valued tensor $\hat{j}(\mf x)$ in Eq.~\eqref{eq:PDgeneralTensor}.

We can further expand the Hamiltonian by assuming that there are no internal degrees of freedom associated with $\mathscr{H}_\textsc{s}$ and that the detector's free Hamiltonian $\hat{\mathsf{H}}$ is independent of $\tau$ and has discrete energy eigenvectors, $\ket{\psi_n}$ with energy eigenvalues $E_n$ with $\hat{\mathsf{H}}\ket{\psi_n} = E_n \ket{\psi_n}$. The eigenfunctions are defined as $\psi_n(\tau,\bm x) = \braket{\bm x}{\psi_n(\tau)} = e^{-\ii E_n \tau}\psi_n(\bm x)$, where we write the wavefunction at $\tau = 0$ as $\psi_n(\bm x)$. In the eigenbasis of the free Hamiltonian, the interaction Hamiltonian reads
\begin{align}
    \hat{H}_I(\tau) &= \lambda \sum_{nm}\int \dd^n \bm x \sqrt{-g}\, \psi_n^*(\bm x)\psi_m(\bm x)e^{i\Omega_{nm}\tau}\!\mu_b^*(\mf x) \hat{O}^b(\mf x)\ket{\psi_n}\!\!\bra{\psi_m}+\text{H.c.},
\end{align}
where $\Omega_{nm} = E_n - E_m$ is the energy gap between the states labelled by $n$ and $m$. Then, in order to draw a better comparison with the previous models in the literature (See e.g. \cite{us,us2,mine}), we define the spacetime smearing tensors $(\Lambda_{nm})^b(\mf x) = \psi_n(\bm x)\psi_m^*(\bm x)\mu^b(\mf x)$. We can then write the Interaction Hamiltonian as
\begin{align}\label{eq:Henergybasis}
    \hat{H}_I(\tau)\! &= \!\lambda\! \sum_{nm}\!\!\int \!\!\dd^n \bm x \sqrt{-g}\,(\Lambda_{nm})_b^*(\mf x) \hat{O}^b(\mf x)e^{\ii\Omega_{nm}\tau}\!\ket{\psi_n}\!\!\bra{\psi_m}+\text{H.c.},
\end{align}
which is a generalization of the original result connecting wavefunctions and particle detectors in~\cite{Unruh-Wald}. The integrand in the expression above can be identified as the Hamiltonian density. In fact, most recent studies that consider finite-sized particle detectors in curved spacetimes (for instance~\cite{us,us2,PipoFTL,antiparticles,mine,ericksonBH,carol,geometry,ahmed,ericksonKen}) prescribe the interaction of the detector with the field in terms of the Hamiltonian density in order to highlight the locality of the theory. In the approach presented here, locality is implemented in terms of the dependence on the position operator of the non-relativistic quantum system.

Overall, the model of Eq. \eqref{eq:generalPD} for the interaction of a localized non-relativistic quantum system with a quantum field represents the most general interaction between a non-relativistic quantum system localized around a trajectory and an operator in a quantum field theory. The considerations about the description of non-relativistic quantum systems in curved spacetimes from Section \ref{sec:NRQS} (including its regimes of validity and covariance of the model) also apply to the general particle detector models presented here and naturally impose a limit for the regime of validity for these models. 

Although the model of Eq. \eqref{eq:generalPD} is very general, and as we will see, can recover many models in the literature, it is not able to implement some features of specific models in the literature, especially when it comes to delocalization of the center of mass of detectors, which was considered in~\cite{achimDelocalized,achimDelocalizedHarvesting}, for instance. This delocalization would amount to describing the curve $\mf z(\tau)$ quantum mechanically, which would require slight changes in our formalism.

As we will show in the examples below, standard particle detector models used in the literature can be recovered from the model of Eq. \eqref{eq:generalPD} by choosing an appropriate quantum system, together with the detector and field operator that mediate the interaction.

\subsubsection{The Scalar Unruh-DeWitt model} 

The simplest scalar Unruh-DeWitt model found in the literature consists of a two-level system coupled to a real scalar quantum field $\hat{\phi}(\mf x)$ according to the interaction Hamiltonian density of Eq,~\eqref{eq:HIUDW}.
The interaction of Eq. \eqref{eq:HIUDW} can be recovered from the general model of Eq. \eqref{eq:generalPD} from any non-relativistic quantum system by restricting it to two levels and neglecting the terms of the interaction which commute with the detector's free Hamiltonian. Consider a localized quantum system that is entirely described by its position degrees of freedom and has ($\tau$ independent) discrete energy levels $E_n$ with eigenstates $\ket{\psi_n}$. We prescribe the interaction with the quantum field by the Hamiltonian
\begin{equation}\label{eq:intermediaryUDW}
    \hat{H}(\tau) = \lambda\gamma(\tau,\hat{\bm x}) f(\tau,\hat{\bm x}) \hat{\phi}(\tau,\hat{\bm x}),
\end{equation}
where $f(\mf x) = f(\tau,{\bm x})$ is any real scalar function evaluated in Fermi normal coordinates. We then identify the operator $\hat{\mu}(\tau) = f(\tau,\hat{\bm x})$ and the field operator \mbox{$\hat{O}(\mf x) = \hat{\phi}(\mf x)$}, so that Eq. \eqref{eq:intermediaryUDW} reads
\begin{align}
    \hat{H}_I(\tau) &= \lambda \sum_{nm}\int \dd^n \bm x \sqrt{-g}\,\Lambda_{mn}(\mf x) \hat{\phi}(\mf x)e^{i\Omega_{nm}\tau}\ket{\psi_n}\!\!\bra{\psi_m},
\end{align}
with $\Omega_{nm} = E_n - E_m$ and $\Lambda_{nm}(\mf x) = \psi_n^*(\bm x)\psi_m(\bm x) f(\tau,\bm x)$.

Restricting this system to two energy levels with an energy gap $\Omega$, say $n = g$ and $m = e$, we obtain the Hamiltonian
\begin{align}
    \hat{H}_I(\tau) = \lambda \int  \dd^n \bm x \sqrt{-g}\,\hat{\phi}(\mf x)
    &\Big(\Lambda_{eg}(\mf x)e^{\ii\Omega\tau}\ket{\psi_e}\!\!\bra{\psi_g} +\Lambda_{ge}(\mf x)e^{-\ii\Omega\tau} \ket{\psi_g}\!\!\bra{\psi_e}\nonumber\\
    &\:\:\:\:\:\:\:\:\:\:\:+\Lambda_{gg}(\mf x)\ket{\psi_g}\!\!\bra{\psi_g}+\Lambda_{ee}(\mf x)\ket{\psi_e}\!\!\bra{\psi_e}\Big),\nonumber
\end{align}
where we split the terms that do not commute with the free Hamiltonian in the first line and the terms that do in the second line. As previously mentioned, to recover the Unruh-DeWitt model, we neglect the terms that commute with the detector's free Hamiltonian. This is a reasonable assumption if one is mostly interested in the excitation and de-excitation of the detector, as these terms do not contribute to energy level excitations to leading order in perturbation theory. In order to recover the Unruh-DeWitt model, we assume that \mbox{$\psi_g(\bm x)\psi^*_e(\bm x) = \psi^*_g(\bm x)\psi_e(\bm x)$}, so that \mbox{$\Lambda(\mf x) = \Lambda_{ge}(\mf x) = \Lambda_{eg}(\mf x)$}, and the interaction Hamiltonian reads
\begin{align}
    \hat{H}_I(\tau) \!=& \lambda\!\! \int\!\! \dd^n \bm x \sqrt{-g} \Lambda(\mf x)\!\!\left(e^{\ii\Omega\tau}\!\!\:\!\ket{\psi_e}\!\!\bra{\psi_g} \!+\!e^{\!-\ii\Omega\tau} \!\:\!\!\ket{\psi_g}\!\!\bra{\psi_e}\right)\!\hat{\phi}(\mf x).
\end{align}
Denoting $\hat{\sigma}^+ = \ket{\psi_e}\!\!\bra{\psi_g}$ and $\hat{\sigma}^- = \ket{\psi_g}\!\!\bra{\psi_e}$, one then identifies the exact same Hamiltonian density from Eq. \eqref{eq:HIUDW}.

It is worth mentioning that instead of neglecting the part of the interaction Hamiltonian which commutes with the detector's free Hamiltonian, one could alternatively model an auxiliary internal degree of freedom for the detector in $\mathscr{H}_\textsc{s} = \mathbb{C}^2$ in order to recover the Unruh-DeWitt model. Then, by adding an energy gap for states $\ket{0},\ket{1}\in\mathbb{C}^2$ one could choose the ground and excited states as $\ket{\psi_g'} = \ket{\psi_g,0}$ and $\ket{\psi_e'} = \ket{\psi_e,1}$ and \mbox{$\hat{\mu}(\tau,\hat{\bm x}) = g(\hat{\bm x})(\hat{\sigma}^+(\tau) + \hat{\sigma}^-(\tau))$}. These choices, together with the reduction to the two dimensional subspace spanned by $\ket{\psi_g'}$ and $\ket{\psi_e'}$, reduce the model of Eq. \eqref{eq:generalPD} to Eq. \eqref{eq:HIUDW} exactly. Thus, we have argued how the general particle detector model from Eq. \eqref{eq:generalPD} can be used to recover the most used particle detector model in the literature by picking a quantum system which yields the corresponding spacetime smearing function $\Lambda(\mf x)$.

Finally, we mention that one could choose any other scalar operator for $\hat{O}(\mf x)$, such as $:\!\!\hat{\phi}^2(\mf x)\!\!:\,$, which would give detector models studied in~\cite{eduAchimBosonFermion,Sachs1,sachs2018entanglement}, for instance. Moreover, the quantum field theory that this detector couples to can be more general than a scalar field theory. For instance, for a spinor field $\hat{\psi}(\mf x)$, the operator $\hat{O}(\mf x)$ can be chosen as $\hat{O}(\mf x) \!= \,:\!\!\hat{\bar{\psi}}(\mf x) \hat{\psi}(\mf x)\!\!:\,$, which would recover other models studied in~\cite{eduAchimBosonFermion,Sachs1,sachs2018entanglement}, for instance. A generalization of the reduction presented here can also be carried naturally for the case of complex scalar fields, recovering the scalar models of~\cite{neutrinos,antiparticles,carol}. Overall, the model of Eq. \eqref{eq:generalPD} can be used to recover any coupling of a non-relativistic systems with a scalar quantum field, or with a scalar operator in a more general quantum field theory.




\subsubsection{The light-matter interaction} 

To recover models of particle detectors based on the light-matter interaction, and to extend these to curved spacetimes, we consider our non-relativistic quantum system to be a hydrogen atom. That is, its free Hamiltonian is prescribed as
\begin{equation}\label{eq:atom}
    \hat{H}_\textrm{atom} = \frac{\hat{\bm p}^2}{2 m_e}-\frac{e^2}{4\pi |\hat{\bm x}|},
\end{equation}
where $m_e$ is the reduced mass of the electron, and $e$ is its fundamental charge. This Hamiltonian admits bound states $\ket{\psi_{nlm}}$ labelled by three quantum numbers, with $n\in \mathbb{N}$, $0\leq l \leq n-1$, $-l\leq m \leq l$, so that \mbox{$\hat{H}_\textrm{atom}\ket{\psi_{nlm}} = E_n \ket{\psi_{nlm}}$}, where $E_n = - \alpha^2 m_e/2n^2$ and $\alpha$ denotes the fine structure constant.

The interaction of an atom with a background quantum electromagnetic field then defines a particle detector model, where a localized non-relativistic quantum system (the atom) couples to a quantum field (the electromagnetic field). Although there are different ways of prescribing this interaction~\cite{Nicho1,richard}, here we will focus on the so-called dipole interaction, where, for inertial motion in flat spacetimes, the interaction Hamiltonian can be written as
\begin{equation}\label{eq:dipole}
    \hat{H}_I(t) = -e\,\hat{\bm x}\cdot \hat{\bm E}(t,\hat{\bm x}),
\end{equation}
where $\hat{\bm E}(t,\bm x)$ is the electric field in the frame of the atom. It is important to mention that the Hamiltonian of Eq. \eqref{eq:dipole} only accurately models the interaction of an atom with the electromagnetic field in some regimes\footnote{The interested reader can check~\cite{Nicho1,richard} for more detailed discussions.}. The Hamiltonians from Eqs. \eqref{eq:atom} and \eqref{eq:dipole}, together with a description of the electric field, then determine a particle detector model for an inertial atom in flat spacetime.

In order to consider this system undergoing a trajectory $\mf z(\tau)$ in curved spacetimes, we must first check the conditions for the framework presented in Section \ref{sec:NRQS} to be applicable. That is, we must have that the system is more localized than the Fermi bound of the curve and that its non-relativistic energy is sufficiently smaller than its rest energy. For an electron in an atom, we have $|E|\sim 13,6\text{eV}$ and $m_e \approx 0.5\text{MeV}$, so this approximation is valid. In order to address the localization of the atom, we use that an atom's extension in space can be approximately bounded by $1/|E|$, where $E$ is the average energy of the state considered. Assuming the Hydrogen atom to be in the state $\ket{\psi_{nlm}}$, and using the estimate for the Fermi bound of Eq. \eqref{eq:approximationFermi}, we find that a regime where the formalism can be used for this description is when
\begin{equation}
    \frac{1}{|E_n|} = \frac{2 n^2}{\alpha^2 m_e} < \frac{1}{a + \sqrt{\lambda_R}}.
\end{equation}
That is, an atom is well described undergoing an accelerated motion in curved spacetimes if its proper acceleration added to the square root of the largest eigenvalue of $-R_{0i0j}$ is smaller than $\alpha^2 m_e/2n^2$. For instance, for a hydrogen atom in its ground state undergoing uniformly accelerated motion in a flat spacetime, we obtain that its acceleration must be smaller than $10^{25}\text{m}/\text{s}^2$. This acceleration is large enough for probing an Unruh temperature as large as $10^6 \textrm{K}$, for instance.

Under the Fermi localization assumption, our framework then allows us to describe an atom undergoing a trajectory $\mf z(\tau)$ in a curved spacetime. We promote the $\hat{\bm x}$ and $\hat{\bm p}$ operators according to Eqs. \eqref{eq:posOp} and \eqref{eq:pOp}, and take into account the redshift factor discussed in Eq. \eqref{eq:redshift} in the Hamiltonian, which introduces new dynamics to the atom due to its acceleration and due to the curvature of spacetime. These can be approximated by Eq. \eqref{eq:approxH}.

In order to describe the coupling of the atom with the field, one must prescribe the interaction of Eq. \eqref{eq:dipole} covariantly, evaluating it in Fermi normal coordinates, and replacing $\bm x\longmapsto \hat{\bm x}$. We start by discussing the operator $\hat{\bm E}(\tau,\bm x)$. The effective electric field seen by an observer with four-velocity $v^\mu$ can be written as
\begin{equation}
    E^\mu(\mf x) = F^{\mu\nu}(\mf x)v_\nu.
\end{equation}
$E^\mu(\mf x)$ is then always a spacelike vector orthogonal to $v^\mu$, so that the electric field is defined in the rest space associated to the observer. The electric field associated to observers that move along constant $\bm x$ curves in Fermi normal coordinates can then be written as $E^\mu(\mf x) = F^{\mu\nu}(\mf x)u_\nu(\mf x)$, where $u^\mu(\mf x) = (\partial_\tau)^\mu/\sqrt{|g_{\tau\tau}|}$ is the four-velocity of observers moving along lines defined by $\bm x = \text{const}.$ The position vector $\bm x$ in the dipole interaction can then be associated with the position vector in Fermi normal coordinates, $\bm x = x^i \mf{e}_i$. Thus, the classical interaction Hamiltonian between the atom and the field can be written as
\begin{equation}
    {H}_I(\tau) = -e \gamma(\tau,{\bm x}){x}^i {E}_i(\tau,{\bm x}),
\end{equation}
where we added the classical redshift from Eq. \eqref{eq:redshift} and the contraction $x^i E_i$ naturally generalizes the dot product from Eq. \eqref{eq:dipole}. The quantum Hamiltonian then reads
\begin{equation}
    \hat{H}_I(\tau) = -e \gamma(\tau,\hat{\bm x})\hat{x}^\mu {E}_\mu(\tau,\hat{\bm x}),
\end{equation}
where $x^\mu = (0,x^i)$. Notice that this Hamiltonian is self-adjoint because it is only a function of the position operator $\hat{\bm x}$. In terms of the $\hat{\mu}$ operator defined in Eq. \eqref{eq:generalPD}, this particle detector model couples to the electromagnetic field strength, $F_{\mu\nu}(\mf x)$, so that the $\hat{\mu}^{\mu\nu}(\tau)$ operator that defines it is \mbox{$\hat{\mu}^{\mu\nu}(\tau,\hat{\bm x}) = \hat{x}^\mu u^\nu(\tau,\hat{\bm x})$} and the coupling constant is given by $\lambda = - e$. Thus, we have shown how to describe the light-matter interaction in curved spacetimes as a particular case of the particle detector model of Eq. \eqref{eq:generalPD}.

\section{More Realistic Localized Probes}\label{sec:MoreRealisticProbes}

When we introduced localized probes for quantum fields, we started our discussion by considering a real scalar quantum field that is confined to a finite region of space in a globally hyperbolic spacetime. Although this model was useful for gaining insight into how to apply the Fewster-Verch framework in an explicit example of measuring a quantum field, it is not an entirely physical model, as one cannot physically realize an infinite trapping potential. As a matter of fact, when we considered particle detectors in Section~\ref{sec:UDW}, we quickly gave up on the strict requirement of compactly supported probes for our applications. A more realistic model would consider a bounded confining potential, in which case the field modes would not be compactly supported, and there would exist an energy threshold after which the field modes become continuous. The goal of this section is to consider more realistic field probes, starting with a discussion of localized fields that are not compactly supported, then discussing the mixed spectrum of quantum fields, and finally analyzing a quantum field theoretic description of a hydrogen atom, and how to reduce this description to a probe of the magnetic field.

\subsubsection*{Non Compactly Supported Localized Fields}

Let us start by considering a generalization of the compactly supported field described in Section~\ref{sec:LocalizedQuantumFields}. We consider a static globally hyperbolic spacetime $\M$ such that the metric can be decomposed as in Eq.~\eqref{eq:staticmetric} in coordinates $(t,\bm x)$, where $t$ corresponds to the flux of a timelike Killing vector field and $\bm x$ are \textit{global} coordinates in the Cauchy surface orthogonal to the flow. We consider a real scalar quantum field $\hat{\phi}_\tc{d}$ with dynamics associated to the Lagrangian
\begin{equation}
    \mathcal{L}_\tc{d} = - \frac{1}{2}\nabla_\mu \phi_\tc{d}\nabla^\mu \phi_\tc{d} - \frac{m_\tc{d}^2}{2}\phi_\tc{d}^2 - \frac{1}{2}V(\bm x)\phi_\tc{d}^2,
\end{equation}
where the function $V(\mf x)$ is a smooth confining potential such that $m_\tc{d}^2 + V(\bm x)>0$. Being confining here means that there exists a timelike curve $\mf z(\tau)$ such that $V(\mf x) \to \infty$ as $\sigma(\mf z(\tau),\mf x)\to \infty$ for all $\tau$. That is, a confining potential is one such that $V(\mf x)\to \infty$ as $\mf x$ goes to spatial infinity. 

The equations of motion for $\phi_\tc{d}$ take the same shape as Eq.~\eqref{eq:Pd} and a basis of solutions can be found by separation of variables as in~\eqref{eq:sepvar1}. The relevant differential operator for the separation of variables in this case is
\begin{equation}
    L \Phi = -\frac{\beta}{\sqrt{h}}\partial_i\left(\beta \sqrt{h}\, h^{ij} \partial_j \Phi\right) + \beta^2 (m_\tc{d}^2 + V(\bm x))\Phi
\end{equation}
acting in a subdomain of $L^2(\Sigma_t)$ where it is self-adjoint. The fact that $V(\bm x)$ is confining then implies that the spectrum of $L$ is discrete, giving rise to positive\footnote{This fact comes from the condition that $V(\mf x)$ is positive is a consequence of the fact that $m_\tc{d}^2 + V(\bm x)>0$. If this condition was violated, it could give rise to unstable field modes with imaginary frequency.} eigenvalues $\lambda_{\bm n}^2$ and eigenfunctions $\Phi_{\bm n}(\bm x)$, so that one can write the basis of positive frequency solutions as in Eq.~\eqref{eq:un}:
\begin{equation}
    u_{\bm n}(\mf x) = e^{- \ii \omega_{\bm n}t} \Phi_{\bm n}(\bm x), 
\end{equation}
where $\Phi_{\bm n}$ satisfies the normalization condition~\eqref{eq:normCondPhin} and $\omega_{\bm n} = |\lambda_{\bm n}|$.

The confining potential $V(\mf x)$ essentially implies that the functions $u_{\bm n}(\mf x)$ are localized in space, in the sense that $u_{\bm n}|_\Sigma \in L^2(\Sigma)$ for any Cauchy surface $\Sigma$. The quantization of this theory can then be obtained by the standard procedure outlined in Section~\ref{sec:LocalizedQuantumFields}, and one can find a GNS representation associated with the state defined by the positive modes $\{u_{\bm n}\}_{\bm n}$, where the field operator can be written as
\begin{equation}\label{eq:phidexp}
    \hat{\phi}_\tc{d}(\mf x) = \sum_{\bm n} u_{\bm n}(\mf x) \hat{a}_{\bm n} + u_{\bm n}^*(\mf x) \hat{a}_{\bm n}^\dagger.
\end{equation}

The creation and annihilation operators $\hat{a}_{\bm n}$ and $\hat{a}_{\bm n}^\dagger$ then satisfy the discrete commutation relations~\eqref{eq:CCRphid}, so that the tensor product factorization of the Fock space as a tensor product also holds~\eqref{eq:Fockphid}. If $\ket{0_\tc{d}}$ is the vacuum state associated with the mode decomposition $\{u_{\bm n},u_{\bm n}^*\}_{\bm n}$, then the states obtained by repeated applications of the creation operators are also normalizable. In essence, most results valid for a compactly supported field trapped by an infinite potential are still valid in this non-compactly supported case, with the exception that here we have regular local algebras of observables that can properly map any function $f\in C_0^\infty(\M)$ to operators $\hat{\phi}(f)$. 

Even though the mode functions $u_{\bm n}$ are not compactly supported in this case, the operators $\hat{a}_{\bm n}$ and $\hat{a}_{\bm n}^\dagger$ can still be seen as regular operators in the $\ast$-algebra, similar to what we had in the compactly supported case. Indeed, as we mentioned, one can extend the algebra by considering smeared field operators acting on more general test functions of sufficient rapid decay. We can indeed find suitable functions $g_{\bm n}$ such that $u_{\bm n} = E_{\tc{d}}g_{\bm n}$, so that the relations $\hat{a}_{\bm n} = \hat{\phi}(\ii g_{\bm n}^*)$ and $\hat{a}_{\bm n}^\dagger = \hat{\phi}(- \ii g_{\bm n})$ in~\eqref{eq:an},~\eqref{eq:and} still hold, and the creation and annihilation operators associated to each mode are indeed well defined operators in the $\ast$-algebra.

The field $\hat{\phi}_\tc{d}$ is localized in the sense that each of its modes is localized around the minima of the potential $V(\mf x)$. It would be better to have a basis-independent way of checking whether a quantum field is localized. One definition for a field to be localized could be given as follows. Given an event $\mf x_0$, denote its normal neighbourhood by $N_{\mf x_0}$ and consider an orthonormal basis $e_{\alpha}$ at $T_{\mf x_0}\M$ such that $e_0 \propto \partial_t$ in the static coordinates $(t,\bm x)$. Denote the Riemann normal coordinates centered at $\mf x_0$ constructed from the basis $e_\alpha$ by $y^\alpha$, such that an event $\mf x\in N_{\mf x_0}$ has coordinates $y^\alpha$ if $\mf x = \exp_{\mf x_0}(y^\alpha e_\alpha)$. Define
\begin{equation}
    \ell_{\mf x_0} = \sup(\{r:\exp_{\mf x_0}(y^\alpha e_\alpha)\in N_{\mf x_0}\,\,\forall y^\alpha \text{ such that } |\delta_{\alpha\beta}y^\alpha y^\beta| < r\}),\quad \text{ and } \quad \ell_{\tc{r}} = \inf_{\mf x_0} \ell_{\mf x_0}.
\end{equation} 
We will assume that $\ell_\tc{r}\neq 0$. Given a compactly supported function $f(y^\alpha)$ in $\mathbb{R}^4$ with support compact contained in $\{(y^\alpha)\in \mathbb{R}^4: |\delta_{\alpha\beta} y^\alpha y^\beta | < \ell_\tc{r}\}$, define ${f}_{\mf x_0}(\mf x) = f(y^\alpha(\mf x))$, the pushforward of $f$ by the Riemann normal coordinates centered at $\mf x_0$. We can then say that a state $\omega_\tc{d}$ is localized with respect to a foliation $\Sigma_t$ if for each $\mf y \in \M$ and $n\in \mathbb{N}$, and each fixed real smooth compactly supported function $f$, 
\begin{equation}
    \lim_{\sigma(\mf y, \mf x_0)\to \infty}\omega_\tc{d}(\hat{\phi}_\tc{d}({f}_{\mf x_0})^n) = 0,\label{eq:locCond}
\end{equation}
where the limit is taken keeping $\mf x_0$ in the same surface $\Sigma_t$ as $\mf y$. If every state in a quantum field theory is localized with respect to a given foliation, then one could say that the field is localized.

The idea behind this definition lies in the fact that the smeared field operators $\hat{\phi}({f}_{\mf x_0})$ can be thought of as the field operators that an experimentalist comoving with the foliation $\Sigma_t$ would have access to if probing the field in spacetime regions of size smaller than $\ell_\tc{r}$. If, for a fixed shape $f$, all the expected values of measurements vanish when $f$ is taken to spatial infinity, one could then say that the field is localized. We show that the condition~\eqref{eq:locCond} is satisfied by Gaussian states that can be represented in the GNS representation~\eqref{eq:phidexp}\footnote{this result is valid whenever the mode functions $\Phi_{\bm n}(\bm x)$ asymptotically decay exponentially with the proper distance between events.} in Appendix~\ref{app:locCond}. Also notice that the localization condition~\eqref{eq:locCond} is immediately satisfied by compactly supported localized fields, and it is violated by any field with translation invariance, such as a free Klein-Gordon field with equation of motion $\nabla_\mu \nabla^\mu\phi - m^2\phi = 0$ in Minkowski spacetime. Importantly, notice that the condition~\eqref{eq:locCond} does not violate the Hadamard condition, as the functions $f_{\mf x_0}$ are kept fixed in the limit. In essence, the convergence of ~\eqref{eq:locCond} is not uniform.

For an explicit example of a non-compactly supported field, we can consider a field in Minkowski spacetime under the influence of a quadratic potential. Specifically, we consider inertial coordinates $(t,\bm x)$ and a the time-independent potential
\begin{equation}
    V(\bm x) = \frac{|\bm x|^2}{2\ell^4}.
\end{equation}
The parameter $\ell$ has dimensions of length and controls the strength of the potential, with smaller values of $\ell$ corresponding to stronger potentials. The equations of motion for the field then become
\begin{equation}
    \left(\partial_\mu \partial^\mu - m_\tc{d}^2 - \frac{|\bm x|^2}{\ell^4}\right) \phi(\mf x) = 0.
\end{equation}
The differential operator $L$, in this case, is
\begin{equation}\label{eq:Lgen}
    L = - \nabla^2 + m_{\tc{d}}^2 + \frac{|\bm x|^2}{\ell^4},
\end{equation}
with eigenfunctions labelled by $n_x,n_y,n_z\in \mathbb{N}$ given by
\begin{equation}
    \Phi_{\bm n}(\bm x) = \frac{1}{\sqrt{2\omega_{\bm n}}}f_{n_x}(x)f_{n_y}(y)f_{n_z}(z), \,\,\text{ where }\,\, f_n(u) = \frac{1}{\sqrt{2^n n!}}\frac{e^{-\frac{u^2}{2\ell^2}}}{\pi^\frac{1}{4} \sqrt{\ell}}H_n(u/\ell),
\end{equation}
and $H_n(u)$ denote the Hermite polynomials. The eigenfrequencies $\omega_{\bm n}$ are characterized by the eigenvalues of $L$, and are explicitly given by
\begin{equation}\label{eq:gapHO}
    \omega_{\bm n} = \sqrt{m_\tc{d}^2+ \frac{2}{\ell^2}\left(n_x + n_y + n_z +\frac{3}{2}\right)}.
\end{equation}
The quantization of the field using the basis of solutions $u_{\bm n}(\mf x)$ yields a field operator of the form~\eqref{eq:phidexp}, where the index $\bm n$ is given by $\bm n = (n_x,n_y,n_z)$ with $n_i \in \mathbb{N}$.

We can analyze the spatial localization of the modes of the quantum field in this example. We see that all modes are exponentially decaying as $e^{-|\bm x|^2/2\ell^2}$. That is, the parameter $\ell$ which is inversely related to the strength of the potential $V(\bm x)$ is also related to the spatial localization of the field. However, highly energetic modes will be less localized, as it is well known that the region where the Hermite polynomials are non-negligible grows as $\sqrt{2n + 1}$. One way of seeing that this field is localized is by noticing that each of the relevant modes for a given physical scenario is exponentially localized.

\subsubsection*{Fields with Discrete and Continuum Spectrum}

While a compactly supported field is unphysical due to its infinite potential outside of a region, a potential that goes to infinity at spatial infinity is an idealization, as all known potentials become at most constant at infinity: all laboratories have a finite size and finite energy. That is, one would expect only a discrete (or even finite) number of sufficiently low energy modes to be localized, while the remaining modes would be scattering states. 

In the context of a static spacetime and the decomposition of the equation of motion in terms of the self-adjoint operator $L$ in~\eqref{eq:Lgen}, this is to say that $L$ must have a mixture of discrete and continuous eigenvalues, with the discrete eigenvalues being below a certain threshold. Let us then assume that $L$ possesses a discrete set of eigenvalues $\lambda_{\bm n}^2$, with eigenfunctions $\Phi_{\bm n}(\bm x)$ parametrized by the discrete index $\bm n$, in addition to a continuous spectrum parametrized by $\bm k\in \mathbb{R}^3$, $\nu_{\bm k}^2$, with generalized eigenfunctions $\Upphi_{\bm k}(\bm x)$, so that $\lambda_{\bm n}^2<\nu_{\bm k}^2$ for all $\bm n$ and $\bm k$. One then obtains a basis of positive frequency solutions of the equations of motion given by $\{u_{\bm n},U_{\bm k}\}_{\bm n, \bm k}$, where
\begin{equation}
    u_{\bm n} = e^{- \ii \omega_{\bm n} t}\Phi_{\bm n}(\bm x), \quad \quad U_{\bm k}(\mf x) = e^{- \ii w_{\bm k}t}\Upphi_{\bm k}(\bm x),
\end{equation}
where $\omega_{\bm n} = |\lambda_{\bm n}|$ and $w_{\bm k} = |\nu_{\bm k}|$. Normalization with respect to the Klein-Gordon inner product then implies
\begin{equation}
    \int \dd \Sigma \,\beta^{-1}\Phi_{\bm n}^*(\bm x) \Phi_{\bm n'}(\mf x) = \frac{1}{2\omega_{\bm n}}\delta_{\bm n, \bm n'}, \quad \quad \int \dd \Sigma \,\beta^{-1} \Upphi_{\bm k}^*(\bm x) \Upphi_{\bm k'}(\mf x) = \frac{1}{2w_{\bm k}}\delta^{(\bm k)}(\bm k - \bm k').
\end{equation}

The basis of positive frequency solutions $\{u_{\bm n},U_{\bm k}\}_{\bm n, \bm k}$ defines a vacuum state $\ket{0_\tc{d}}$ as well as its GNS representation, where the field operator can be written as
\begin{equation}
\begin{aligned}
    \hat{\phi}_\tc{d}(\mf x) =& \sum_{\bm n} \left(u_{\bm n}(\mf x) \hat{a}_{\bm n} + u_{\bm n}^*(\mf x) \hat{a}^\dagger_{\bm n}\right) + \int \dd^3 \bm k \left(U_{\bm k}(\mf x) \hat{a}_{\bm k} +U_{\bm k}^*(\mf x) \hat{a}^\dagger_{\bm k} \right){,}
\end{aligned}
 \label{decomposition}
\end{equation}
where the creation and annihilation operators labelled by discrete indices $\bm n$ satisfy discrete commutation relations, and the creation and annihilation operators associated with continuous indices satisfy continuous commutation relations. Consequently, the operators $\hat{a}_{\bm n}$ and $\hat{a}_{\bm n}^\dagger$ are well defined within the algebra and give rise to normalized states $\ket{\bm n} = \hat{a}_{\bm n}^\dagger\ket{0_\tc{d}}$. This is certainly not the case for the operators $\hat{a}_{\bm k}$ and $\hat{a}_{\bm k}^\dagger$, which have similar behaviour to the creation and annihilation operators associated to the Minkowski vacuum.

The mix of continuous and discrete basis of solutions makes it so that the the Fock space $\mathcal{F}(\mathscr{H}_\tc{d})$ does not factor entirely as a tensor product of the different field modes, as we had in Eq.~\eqref{eq:Fockphid}. Instead, 
the vacuum state factors as
\begin{equation}
    \ket{0_\tc{d}} = \left(\bigotimes_{\bm n} \ket{0_{\bm n}}\right)\otimes \ket{0_{\text{cont}}},
\end{equation}
where $\ket{0_{\bm n}}$ are the zero occupation states in each mode defined by $a_{\bm n}\ket{0_{\bm n}} = 0$ and $\ket{0_\text{cont}}$ is the state defined by $\hat{a}_{\bm k}\ket{0_{\text{cont}}}$ for all continuous labels $\bm k$. The Fock space then factors as
\begin{equation}
    \mathcal{F}(\mathscr{H}_\tc{d}) = \left(\bigotimes_{{\bm n}} \mathscr{H}_{\bm n}\right)\otimes \mathcal{F}(\mathscr{H}_\text{cont}).
\end{equation}

The resulting quantum field $\hat{\phi}_\tc{d}$ is then certainly not a localized field, as it contains modes that are not integrable in $L^2(\Sigma_t)$. Indeed, due to the continuous modes not being integrable in $\Sigma_t$, the localization condition~\eqref{eq:locCond} is violated. However, this GNS representation induces a natural decomposition of the field into discrete and continuous modes:
\begin{equation}
    \hat{\phi}_\tc{d}(\mf x) = \hat{\phi}_{\text{loc}}(\mf x) + \hat{\phi}_\text{cont}(\mf x),
\end{equation}
with
\begin{equation}
    \hat{\phi}_{\text{loc}}(\mf x) = \sum_{\bm n} \left(u_{\bm n}(\mf x) \hat{a}_{\bm n} + u_{\bm n}^*(\mf x) \hat{a}^\dagger_{\bm n}\right), \quad \quad \hat{\phi}_{\text{cont}}(\mf x) \int \dd^3 \bm k \left(U_{\bm k}(\mf x) \hat{a}_{\bm k} +U_{\bm k}^*(\mf x) \hat{a}^\dagger_{\bm k} \right).
\end{equation}
While neither $\hat{\phi}_\text{loc}$ or $\hat{\phi}_\text{cont}$ are quantum fields individually, states of this quantum field theory satisfy the localization condition~\eqref{eq:locCond} with the replacement $\hat{\phi}_\tc{d}\mapsto \hat{\phi}_\text{loc}$\footnote{The proof of Appendix~\ref{app:locCond} only depends on the fact that the modes $\Phi_{\bm n}(\bm x)$ decay sufficiently fast, so that the conclusion still holds for $\hat{\phi}_\text{loc}$.}. In other words, observables associated with $\hat{\phi}_\text{loc}$ are localized in space, while operators associated with $\hat{\phi}_\text{cont}$ are not.

One can then consider the discrete part of $\hat{\phi}_\tc{d}$ to be a localized probe, while $\hat{\phi}_\text{cont}$ corresponds to propagating modes. Indeed, the modes $U_{\bm k}$ correspond to scattering states that can be produced if the field $\hat{\phi}_\tc{d}$ acquires sufficient energy. One can draw an analogy with an electron bound to an atom interacting with an external electromagnetic pulse: if the energy of the pulse is sufficiently small, the electron will remain bound to the atom but with a different energy level, described by a bound state of the form $\ket{\bm n}$. On the other hand, a highly energetic electromagnetic pulse may cause the electron to escape the atom instead, making it more accurately represented by a wavepacket supported involving a continuous sum over states of the form $\hat{a}^\dagger_{\bm k}\ket{0_\tc{d}}$.

\subsubsection*{A Realistic Example: The Hydrogen Atom as a Localized Quantum Field}

We will now describe a realistic example using the tools discussed so far: an electron bound to a hydrogen atom. One can describe a hydrogen atom as an electron field under the influence of an external Coulomb potential. Although the Coulomb potential is not smooth (or bounded from below), it admits well-defined discrete field modes with energies $\omega_{\bm n}<0$ and continuous modes for $w_{\bm k}\geq 0$, allowing one to describe it with the tools of the previous sections. In order to find the discrete modes explicitly, we will consider this example in Minkowski spacetime using inertial coordinates $(t,\bm x)$. The content of the next Segments has been discussed in~\cite{ruhi}.

One can couple a Dirac spinor to electromagnetism by considering the U(1) gauge transformation associated with the charge $- q$,
\begin{equation}
    \psi(\mf x) \mapsto e^{\ii q\alpha(\mf x)} \psi(\mf x),
\end{equation}
which is generated by the electromagnetic four-potential $A = A_\mu \text{d}x^\mu$. The corresponding Lagrangian density for a Dirac spinor $\psi$ minimally coupled to an electromagnetic four-potential $A_\mu$ is
\begin{equation}\label{QED-lagrangian-density}
    \mathscr{L} = \bar{\psi}(\ii\gamma^\mu D_\mu - m_e)\psi  - \frac{1}{4} F_{\mu\nu}F^{\mu\nu}\, .
\end{equation}
Here $\bar{\psi} \equiv\psi^\dagger \gamma^0$, $D_\mu \equiv\partial_\mu - \ii q A_\mu$ is the covariant derivative with respect to the U(1) gauge transformation,  and \mbox{$F_{\mu\nu} = \partial_\mu A_\nu - \partial_\nu A_\mu$} is the electromagnetic field strength tensor.

To describe the electron field in an atom, we model the nucleus as a non-dynamical static point charge with four-current density $j^\mu(\mf x)$. The full Lagrangian for the theory then becomes
\begin{equation}\label{QED-lagrangian-density2}
    \mathscr{L} = \bar{\psi}(\ii\gamma^\mu D_\mu - m_e)\psi  - \frac{1}{4} F_{\mu\nu}F^{\mu\nu} - j^\mu A_\mu\, .
\end{equation}
The corresponding equations of motion are
\begin{align}
    \partial^\mu F_{\mu\nu} = j_\nu -q \bar{\psi}\gamma_\nu \psi,\\
    (\ii \slashed{\partial} - m_e)\psi = -q\slashed{A} \psi.
\end{align}
The solutions for the four-potential $A_\mu$ can be written as a sum of the solutions to the free part $A_\mu^{(\text{free})}$, and the part sourced by the nucleus $A^{(\text{atom})}_\mu$,
\begin{align}\label{eq:Adecomp}
    A_\mu = A_\mu^{(\text{free})} + A^{(\text{atom})}_\mu \,, 
\end{align}
which satisfy
\begin{align}
    \partial^\mu F^{(\text{atom})}_{\mu\nu} &= j_\nu,\\
    \partial^\mu F^{(\text{free})}_{\mu\nu} &= -q \bar{\psi}\gamma_\nu \psi. 
\end{align}

We assume that the nucleus is at rest with respect to the inertial frame $(t,\bm x)$ and prescribe
\begin{equation}\label{eq:nucleusJmu}
    j^\mu(\mf x) = Q u^\mu \delta^{(3)}(\bm x ),
\end{equation}
where $u^\mu = (1,0,0,0)$ is the four-velocity of the nucleus and $Q = {qZ}$ is its charge. The pointlike charge then sources a Coulomb potential $A^{(\text{atom})}_\mu(r)$, where $r = |\bm x|$ is the radial coordinate centered at the nucleus, $r = \sqrt{x^i x_i}$. In the Coulomb gauge, the potential is given by
\begin{equation}\label{atom-potential}
    A_{\mu}^{(\text{atom})}(t, \bm{x}) = 
        -  \frac{Q}{r} u_\mu.
\end{equation}

The electron orbitals can be found by solving the equations of motion for the $\psi(\mf x)$ field only considering the Coulomb potential sourced by the nucleus: 
\begin{equation}\label{Dirac-eqn-atom}
    0=(\ii\slashed{\partial} +  q\slashed{A}^{(\text{atom})}(\mathsf{x}) - m_e)\psi(\mathsf{x}) \, .
\end{equation}
To obtain the basis of solutions for the electron, we can split the time and space components of the equation in this inertial frame, 
\begin{equation}\label{Dirac-eqn-atom-split}
    \ii\partial_0\psi(\mathsf{x}) = (-\ii\gamma^0\gamma^i\partial_i - \gamma^0m_e - qA_0(r))\psi(\mathsf{x}) \, .
\end{equation}
A general solution can be found by looking for static solutions, which are eigenfunctions of the Hamiltonian
\begin{equation}\label{H-Dirac-atom}
    \mathsf{H}_{\text{atom}} = (-\ii\gamma^0\gamma^i\partial_i + \gamma^0m_e - qA_0(r)) \, .
\end{equation}
That is, we look for solutions to the following time-independent Dirac equation,
\begin{equation}\label{TIDE-atom}
    \mathsf{H}_{\text{atom}}\psi(\bm{x}) = E\psi(\bm{x}) \, .
\end{equation}
The Hamiltonian $\mathsf{H}_\text{atom}$ is analogous to the familiar Hamiltonian of a Schrodinger hydrogen atom. In this case, the Hamiltonian acts on a spinor-valued field, and its eigenfunctions correspond to the \emph{classical} static solutions. 

The solutions to this eigenvalue problem can be classified in terms of eigenvalues of a set of operators (acting on classical spinor fields) that commute with $\mathsf{H}_\text{atom}$. The following operators are relevant to this problem: 
\begin{align}\label{Dirac-atom-operators}
    \bm{\mf{J}} &= \bm{\mf{L}} + {\bm \Upsigma}\,, \quad\quad {\bm{\mf{L}}} = - \ii \bm{r} \times \bm \nabla \, , \quad\quad {\bm \Upsigma} = \begin{pmatrix}
        \bm \sigma & 0 \\
        0 & \bm \sigma
    \end{pmatrix} \,,
\end{align}
corresponding to total angular momentum ($\bm{\mf{J}}$), orbital angular momentum ($\bm{\mf{L}}$), and spin ($\bm{\Upsigma}$). In addition, we define the parity operator,
\begin{equation}\label{Dirac-parity-operator}
    \mf{P}\psi(t, \bm{x}) = \gamma^0\psi(t, -\bm{x}) \, .
\end{equation} 
Both the parity operator and the total angular momentum operator commute with the Hamiltonian $\mf{H}_{\text{atom}}$, as well as $\mf{J}_z$,  
\begin{equation}\label{Dirac-atom-commute}
    [\bm{\mf{J}}, \mf{H}_{\text{atom}}] = [\mf{J}_z, \mf{H}_{\text{atom}}] = [\mf{P}, \mf{H}_{\text{atom}}] = 0 \,.
\end{equation}
Together with $\mf{H}_\text{atom}$, their eigenvalues are enough to label all bound stationary solutions of Dirac's hydrogen atom. Notice that
although the total angular momentum is conserved, the orbital angular momentum ($\bm{\mf{L}}$) and spin ($\bm \Upsigma$) are not. The solutions of the equation of motion can, therefore, be labelled by four quantum numbers, $n$, $j$, $m$, and $p$, defined by
\begin{align}
    \mf{J}^2\psi_{njmp} &= j(j+1)\psi_{njmp} \, , \\
    \mf{J}_z\psi_{njmp} &= m\psi_{njmp} \, ,  \\
    \quad\quad\quad\quad\quad\quad\quad\mf{P}\psi_{njmp} &= p\psi_{njmp} \, , \\ \quad\quad\quad\quad\mf{H}_{\text{atom}}\psi_{njmp} &= E_{nj}\psi_{njmp}\label{TIDE},
\end{align}
where 
\begin{equation}\label{Energies-dirac-atom}
    E_{nj} =\frac{m_e}{\sqrt{1+\frac{(Z \alpha)^2}{\left(n-j-\frac{1}{2}+\sqrt{\left(j+\frac{1}{2}\right)^2-(Z \alpha)^2}\right)^2}}}
\end{equation}
and $\alpha$ is the fine structure constant. The quantum numbers $n$, $j$, $m$, and $p$ take the discrete values
\begin{align}\label{Dirac-atom-eigenvalues}
    n &= 1,\,2\,,...\,,\\
    j &= \frac{1}{2},\,...\,,\,n-\frac{1}{2}\,, \\
    m &= -j, -(j-1), ..., j-1, j \, , \\
    p &= \begin{cases}
        +1, \text{ if } j = n - 1/2,\\
        \pm 1, \text{ if } j \neq n - 1/2        
    \end{cases} .
\end{align}
Notice that, unlike the case of Schrodinger's hydrogen atom, in Dirac's atom, the energy levels depend on the total orbital quantum number $j$.

As with any time-independent spherically symmetric external potential $A_0(r)$, the four-component Dirac eigenfunctions $\psi_{njmp}$ can be split into two two-component bispinors,
\begin{equation}\label{psi-sol-atom}
    \psi_{njmp}(\bm x) = \begin{pmatrix}
        g_{nj}(r)\Omega_{jml}(\theta,\phi)\\
        f_{nj}(r)\Omega_{jml'}(\theta,\phi)
    \end{pmatrix} \, .
\end{equation}
Here the functions $g_{nj}(r)$ and $f_{nj}(r)$ define the effective localization lengthscale of the modes and $\Omega_{jml}$ are the spinor spherical harmonics~\cite{RQMgreiner, spinorsphericalharmonics}, where $l$ labels orbital angular momentum according to $\bm{\mf{L}}^2\Omega_{jml} = l(l+1)\Omega_{jml}$. In Eq.~\eqref{psi-sol-atom} $l$ and $l'$ can take the values of $j \pm \tfrac{1}{2}$. $l$ and $l'$ are related by the parity eigenvalue $p = \pm 1$ through $l' - l =  p$.  

The operator $\mf{H}_\text{atom}$ also possesses a continuous spectrum for $E\geq m_e$, representing scattering states. These can be labelled by a continuous parameter $k\geq 0$ and the quantum numbers $j,m,p$. The corresponding eigenfunctions can be classified by the sign of the eigenvalues of the operator $\ii \partial_t$. We will notate them as $u_{kjmp}(\bm x)$ for positive frequencies (electron states) and $v_{kjmp}(\bm x)$ for negative frequencies (positron states). The explicit expressions for the scattering solutions can be found in~\cite{RQMgreiner}.

We can then write a general solution of Eq.~\eqref{Dirac-eqn-atom} as a mode expansion in terms of the solutions $\psi_{\bm N}(\bm x)$, as well as a continuous set of unbound solutions $u_{\bm k}(\bm x)$ and $v_{\bm k}(\bm x)$, where we use the multi-indices $\bm N = (n,j,m,p)$, $\bm k = (k,j,m,p)$ and $k$ is a continuous parameter. Overall, a general solution of the equation of motion can be written in terms of coefficients $b_{\bm N}$, {${b}_{\bm k}$} and {${c}_{\bm k}$} as
\begin{align}\label{Dirac-solution}
    \psi(\mf x) = \sum_{\bm N} b_{\bm N} e^{-\ii E_{\bm N}t}\psi_{\bm N}(\bm{x})
     + \sumint_{\bm k} \left( b_{\bm k} e^{- \ii E_{\bm k} t} u_{\bm k}(\bm x) + c_{\bm k}^* e^{\ii E_{\bm k} t} v_{\bm k}(\bm x)\right).
\end{align}

The classical theory for the field $\psi(\mf x)$ under the influence of a Coulomb potential can then be used to define a quantum field theory for the field $\hat{\psi}(\mf x)$, as discussed in Section~\ref{sec:generalQFT}. This quantum field theory will then have a unique quasifree state $\omega_0$ invariant under time translations with respect to $u^\mu$, defined by the mode decomposition of Eq.~\eqref{Dirac-solution}. This state will then give rise to a GNS representation, defining a Fock space for the electron states. Effectively, this process corresponds to promoting the coefficients in~\eqref{Dirac-solution} to operators, \mbox{$b_{\bm N}\rightarrow\hat{b}_{\bm N}$}, \mbox{$b_{\bm k}\rightarrow\hat{b}_{\bm k}$}, \mbox{$c_{\bm k}\rightarrow\hat{c}_{\bm k}$}. The operators $\hat{b}_{\bm N},\, \hat{b}_{\bm k},\, \hat{c}_{\bm k}$ are the annihilation operators and $\hat{b}^\dagger_{\bm N},\, \hat{b}^\dagger_{\bm k},\, \hat{c}^\dagger_{\bm k}$ are the corresponding creation operators. These operators are defined by the canonical anti-commutation relations,
\begin{align}
    \{\hat{b}_{\bm N}, \hat{b}^\dagger_{\bm N'}\} &= \delta_{\bm{NN}'},\label{eq:commNN}\\
    \{\hat{b}_{\bm k}, \hat{b}^\dagger_{\bm k'}\} &= \delta(\bm k,\bm k'),\\
    \{\hat{c}_{\bm k}, \hat{c}^\dagger_{\bm k'}\} &= \delta(\bm k,\bm k').
\end{align} 
Equation~\eqref{eq:commNN} reflects the fact that the discrete modes $\psi_{\bm N}(\bm x)$ are orthonormal according to the Dirac inner product
\begin{equation}
    (\psi_{\bm N}, \psi_{\bm N'}) = \int \dd^3 \bm x \bar{\psi}_{\bm N}(\bm x) \gamma^0 \psi_{\bm N'}(\bm x)  = \delta_{\bm N \bm N'}.
\end{equation}
The creation and annihilation operators define the vacuum $\ket{0}$ associated to $\omega_0$ by $\hat{b}_{\bm N} \ket{0} = \hat{b}_{\bm k} \ket{0} = \hat{c}_{\bm k} \ket{0} = 0$ for all $\bm N$ and $\bm k$. The electron quantum field can then be written as
\begin{align}\label{psi-general}
    \hat{\psi}(\mf x) = &\sum_{\bm N} \hat{b}_{\bm N} e^{-\ii E_{\bm N}t}\psi_{\bm N}(\bm{x})+ \sumint_{\bm k} \left( \hat{b}_{\bm k} e^{- \ii E_{\bm k} t} u_{\bm k}(\bm x) + \hat{c}^\dagger_{\bm k} e^{\ii E_{\bm k} t} v_{\bm k}(\bm x)\right).
\end{align}
In the context of this quantum field theory, the Hamiltonian density that prescribes the dynamics of the electron field under the influence of the Coulomb potential is
\begin{equation}
    \hat{\mathcal{H}}_{\text{atom}}(\mf x) = \hat{\bar{\psi}}(\mf x)(-\ii\gamma^i\partial_i + m_e + q \slashed{A}^{\text(atom)})\hat{\psi}(\mf{x}),
\end{equation}
not to be confused with the operator $\mf{H}_\text{atom}$, which is a differential operator that acts in classical solutions of the equation of motion. The corresponding Hamiltonian of the quantum theory for the fermionic field under the influence of the Coulomb potential can be found by regularizing\footnote{Same as in the scalar field case, the Hamiltonian density $\hat{\mathcal{H}}_\text{atom}(\mf x)$ is quadratic in the field operator $\hat{\psi}(\mf x)$, so that it requires to be normal ordered with respect to a given reference state. We pick the vacuum $\ket{0}$ for the normal ordering.} and integrating the Hamiltonian density along a spatial slice $t = \text{const.}$ We regularize the Hamiltonian by subtracting its expected value in the vacuum $\ket{0}$ defined by the expansion of Eq.~\eqref{psi-general}:
\begin{align}\label{H-atom-quantized}
    &\normord{\hat{H}_{\text{atom}}} = \int \dd^3 \bm x\, \Big(\hat{\mathcal{H}}_{\text{atom}}(\mf x) - \bra{0}\hat{\mathcal{H}}_{\text{atom}}(\mf x)\ket{0} \Big) \nonumber \\
    &= \sum_{\bm N} E_{\bm N}\hat{b}^\dagger_{\bm N}\hat{b}_{\bm N} + \sumint_{\bm k} E_{\bm{k}} (\hat{b}_{\bm k}^\dagger\hat{b}_{\bm k} + \hat{c}_{\bm k}^\dagger\hat{c}_{\bm k}).
\end{align}
The eigenstates of the Hamiltonian can be constructed by repeated applications of the creation operators on the vacuum state $\ket{0}$, and these eigenstates span the field's Fock space, $\mathcal{F}_\psi$. The one-particle states of the theory are
\begin{align}
    \ket{\bm N} &= \hat{b}_{\bm N}^\dagger \ket{0},\\
    \ket{\bm k,+} &= \hat{b}_{\bm k}^\dagger \ket{0},\\
    \ket{\bm k,-} &= \hat{c}_{\bm k}^\dagger \ket{0}.
\end{align}
Notice that the states $\ket{\bm k,\pm}$ are not normalizable, as they correspond to the continuous spectrum of $\hat{{H}}_\text{atom}$. Therefore, the scattering states are not physical, although they form a useful basis for the study of scattering processes. On the other hand, the bound states $\ket{\bm N}$ are localized physical states, satisfying
\begin{equation}\label{eq:wearenotparticlephysicists}
    \braket{\bm N}{\bm N'} = \delta_{\bm N\bm N'}.
\end{equation}
These states correspond to electrons bound to the hydrogen atom with quantum numbers $\bm N = (n,j,m,p)$.

While Eq.~\eqref{psi-general} allows one to describe an electron in terms of a fully featured quantum field, it does not describe aspects of the electron-proton interaction. For instance, the hyperfine splitting in the electron energy levels comes from the interaction of the spins of the proton and electron through the magnetic field, which is not present in the description above\footnote{It is possible to effectively implement the hyperfine splitting by considering an additional $\mathbb{C}^2$ quantum degree of freedom corresponding to the proton spin and introducing an appropriate dipole coupling with the field $\psi(\mf x)$~\cite{RQMgreiner}.}. 

The lack of the description for the degrees of freedom of the proton in this model also implies that the bound states in the hydrogen atom are electron states. This is unlike the Schr\"odinger atom description, where one defines both centre of mass and internal degrees of freedom for the system, mixing the electron and proton wavefunctions~\cite{richard}. That is, the bound states of a Schr\"odinger atom do not correspond to the electron degrees of freedom per s\'e, but to a combination of the electron and proton systems. In order to obtain this feature in a quantum field theoretic description, one would need to consider both the electron and proton as fermionic fields, which would require more sophisticated bi-spinor techniques~\cite{bispinorQED} and falls beyond our current goals.

{
Finally, notice that although the electron field is fully relativistic, this model clearly privileges the reference frame of the nucleus. This implies that the solutions of the electronic field are not Poincar\'e invariant, instead only being invariant under time translations in the direction of the nucleus four-velocity $u^\mu$ and rotations around the origin in the nucleus' rest space. However, the lack of Poincar\'e covariance is to be expected: no localized system can be invariant under translations or arbitrary boosts. On the other hand, symmetry under arbitrary transformations generated by the Poincar\'e group can be restored by also applying these transformations to the nucleus ($j^\mu(\mf x)$) and Coulomb field. In other words, the field representation of Eq.~\eqref{psi-general} is valid in any inertial frame where the nucleus current density takes the shape of Eq.~\eqref{eq:nucleusJmu}. 
}

Our next goal throughout this section will be to describe the coupling of the spin of an electron in a hydrogen atom to an external electromagnetic field. We will do so by first restricting the spinor field to a single spherically symmetric orbital, and then considering its coupling with electromagnetism and recasting it in a familiar manner.

\subsubsection*{Reducing a Relativistic Atom to a Spin 1/2 System }\label{sec:C2reduction}

We will now reduce the quantum field description of the Hydrogen atom presented above to an effective two-level system. To do this, we restrict the quantum numbers of the fermionic field~\eqref{Dirac-eqn-atom} to two degrees of freedom corresponding to the one-particle sector of the $s$ orbital of an atom. 

In the usual Schr\"odinger description of a Hydrogen atom, the $s$ orbitals are defined by the vanishing of the quantum number associated with orbital angular momentum ($l=0$). However, in the Dirac description, the orbital angular momentum does not define a quantum number, as it does not commute with the atom Hamiltonian. Instead, in this description, the $s$ orbitals correspond to the quantum numbers $j=  1/2$ and $p = +1$\footnote{To compare the orbitals defined by a Dirac hydrogen-like atom with the ones in the Schr\"odinger description, it is necessary to employ non-relativistic approximations. These are usually done through a Foldy-Wouthuysen~\cite{FoldyWou,RQMgreiner} transformation. In this framework, the top two components of the Dirac spinor, which have well-defined orbital angular momentum, are dominant. This implies that, in the non-relativistic approximation, the Dirac spinor describing a state with $j=\frac{1}{2}$ and $l=0$ has positive parity since the dominant top component has $l=0$ and the bottom component has $l'=1$, yielding $p = l'-l = +1$.}. The two degrees of freedom in a given $s$ orbital are encoded in the magnetic quantum number $m$, which can take the values $m = \pm\frac{1}{2}$. For instance, the $1s$ orbital corresponds to the subspace defined by the quantum numbers $n=1$ and $j=1/2$, which imply $p=+1$.

To reduce the quantum field description to a specific $s$ orbital, we fix the quantum number $n = n_0$. For convenience, we introduce the following notation for the operators and states associated with the quantum numbers $(n,j,m,p) =(n_0,\frac{1}{2},\pm\tfrac{1}{2},+1)$:
\begin{equation}
    \hat{b}_\uparrow \coloneqq \hat{b}_{n_0,\frac{1}{2},\frac{1}{2},+1},\quad\quad
    \hat{b}_\downarrow \coloneqq \hat{b}_{n_0,\frac{1}{2},-\frac{1}{2},+1}.
\end{equation}
\vspace{-10mm}

\begin{equation}
\begin{aligned}\label{uparrowdownarrowstates}
    \ket{\uparrow}\, &\coloneqq  \ket{n_0,\tfrac{1}{2},\tfrac{1}{2},+1} = \hat{b}^\dagger_{\uparrow}|0\rangle ,\\
    \ket{\downarrow}\, &\coloneqq  \ket{n_0,\tfrac{1}{2},-\tfrac{1}{2},+1} = \hat{b}^\dagger_{\downarrow}|0\rangle  .
    \end{aligned}
\end{equation}
The fact that the modes of the field $\hat{\psi}(\mf x)$ are discrete implies that the states $\ket{\uparrow}$ and $\ket{\downarrow}$ are normalized (see Eq.~\eqref{eq:wearenotparticlephysicists}). $\ket{\uparrow}$ and $\ket{\downarrow}$ form an orthonormal basis for the two dimensional subspace that they span, $\mathscr{H}_\text{s}\cong \mathbb{C}^2$. Within this subspace, we define ladder operators $\hat{\sigma}_+$ and $\hat{\sigma}_-$ as
\begin{align}
    \hat{\sigma}_+ &\coloneqq \ket{\uparrow}\!\bra{\downarrow}\,, \quad \hat{\sigma}_- \coloneqq \ket{\downarrow}\!\bra{\uparrow} \,,
\end{align}
and the operators 
\begin{align}
    \hat{\sigma}_x &\coloneqq \hat{\sigma}_+ + \hat{\sigma}_- \,,\\
    \hat{\sigma}_y &\coloneqq -\ii(\hat{\sigma}_+ - \hat{\sigma}_-), \,\\
    \hat{\sigma}_z &\coloneqq \hat{\sigma}_+ \hat{\sigma}_- - \hat{\sigma}_-\hat{\sigma}_+,
\end{align}
which satisfy the $\mathfrak{su}(2)$ algebra commutation relations
\begin{equation}
    [\hat{\sigma}_i, \hat{\sigma}_j] = 2\ii\epsilon_{ij}^{\,\,\,\,k}\,\hat{\sigma}_{k} \, ,
\end{equation}
so that they act as the Pauli operators in $\mathscr{H}_\text{s}$ and their matrix representation in the basis $\{\ket{\uparrow},\ket{\downarrow}\}$ takes the form
\begin{equation}
    \hat{\sigma}_x = \begin{pmatrix}
        0&&1\\
        1&&0
    \end{pmatrix} ,\, \hat{\sigma}_y =\begin{pmatrix}
        0&&-\ii\\
        \ii&&0
    \end{pmatrix},\,
    \hat{\sigma}_z  = \begin{pmatrix}
        1&&0\\
        0&&-1
    \end{pmatrix}\,.
\end{equation}

We define the projector into the subspace $\mathscr{H}_\tc{s}$,
\begin{equation}
    \hat{P}_{\text{s}} = \ket{\uparrow}\!\bra{\uparrow}+\ket{\downarrow}\!\bra{\downarrow}. \, 
\end{equation}
The projector $\hat{P}_\text{s}$ can be used to reduce operators acting on the Fock space $\mathcal{F}_\psi$ to $\mathscr{H}_\text{s}\cong \mathbb{C}^2$. For instance, consider an operator acting on the Fock space $\mathcal{F}_\psi$ of the form, 
\begin{equation}\label{generaloperatorM}
    \hat{M}(\mf x) = \hat{\bar{\psi}}(\mf{x})O(\mf{x})\hat{\psi}(\mf{x})\,,
\end{equation}
where $O(\mf{x})$ is an operator that acts on (the classical) spinor space. Let ${M_{\bm{N}\bm{N}'}}(\mf{x}) \coloneqq \bar{\psi}_{\bm{N}}(\mf{x}) O(\mf{x})\psi_{\bm{N}'}(\mf{x})$, where $\bm{N}, \, \bm{N}'$ are labels for the  quantum numbers $(n,\,j,\, m,\,p)$ introduced in the previous section. The operator $\hat{M}(\mf x)$ can be generally written as
\begin{align}\label{Moperator}
    \hat{M}(\mf x) &= \sum_{\bm{N},\,\bm{N}'} M_{\bm{N}\bm{N}'}(\mf{x})\hat{b}^\dagger_{\bm{N}}\hat{b}_{\bm{N}'}\,.
\end{align}
The action of the projector $\hat{P}_\text{s}$ on the operators $\hat{b}^\dagger_{\bm{N}}\hat{b}_{\bm{N}'}$ in this expansion can be written in the basis $\{\ket{\uparrow},\ket{\downarrow}\}$ as 
\begin{equation}
    \hat{P}_\text{s}\hat{b}_{\bm N}^\dagger \hat{b}_{\bm N'} \hat{P}_\text{s} = \sum_{m \in \{\uparrow, \downarrow\}} \sum_{ m' \in \{\uparrow, \downarrow\}} \delta_{m \bm{N}} \delta_{m' \bm{N}'}\ket{m}\!\bra{m'}\, ,
\end{equation}
so that applying the projector $\hat{P}_\text{s}$ on Eq.~\eqref{generaloperatorM} yields
\begin{align}
    \hat{M}_{\text{s}}(\mf x) \coloneqq \hat{P}_\text{s}\hat{M}(\mf x)\hat{P}_\text{s} &= M_{\uparrow\uparrow}(\mf{x})\ket{\uparrow}\!\bra{\uparrow} + M_{\uparrow\downarrow}(\mf{x})\ket{\uparrow}\!\bra{\downarrow} \nonumber\\*
    &+ M_{\downarrow\uparrow}(\mf{x})\ket{\downarrow}\!\bra{\uparrow} + M_{\downarrow\downarrow}(\mf{x})\ket{\downarrow}\!\bra{\downarrow}\,,
\end{align}
with matrix representation
\begin{equation}
    \hat{M}_\text{s}(\mf x) = \begin{pmatrix}
        M_{\uparrow\uparrow}(\mf{x})&&M_{\uparrow\downarrow}(\mf{x})\\
        M_{\downarrow\uparrow}(\mf{x})&&M_{\downarrow\downarrow}(\mf{x})
    \end{pmatrix}\, .
\end{equation}
From this point on, we use the subindex s to indicate projection to $\mathscr{H}_\text{s}$.

As an example, consider the second-quantized Hamiltonian (\ref{H-atom-quantized}). We can build its projection on the subspace $\mathscr{H}_\text{s}$, \mbox{$\hat{H}_{\text{atom,}\text{s}}=\hat{P}_\text{s} \hat{H}_{\text{atom}}\hat{P}_\text{s}$}, given by
\begin{equation}\label{H-atom-lzero}
    \hat{H}_{\text{atom,}\text{s}} = \sum_{m = \pm \frac{1}{2}} E_{m}\hat{b}_{m}^\dagger\hat{b}_m = E_{\uparrow}\hat{b}_{\uparrow}^\dagger\hat{b}_\uparrow + E_{\downarrow}\hat{b}_{\downarrow}^\dagger\hat{b}_\downarrow \,.
\end{equation}
Notice that since we are not considering any hyperfine interactions or any external electromagnetic fields, spins up and down have the same energy, $E_{\uparrow} = E_\downarrow$, yielding a $\hat{H}_{\text{atom,}\text{s}}$ proportional to the identity in $\mathscr{H}_\text{s}$. Defining $E_0 \coloneqq E_{\uparrow} = E_\downarrow$, a matrix representation for the atom Hamiltonian in this subspace is then 
\begin{equation}
    \hat{H}_{\text{atom,}\text{s}} = \begin{pmatrix}
        E_0&&0\\0&&E_0
    \end{pmatrix} = E_0\hat{\mathds{1}} \, .
\end{equation}

Reducing the electron field theory to the two-dimensional $\mathscr{H}_\text{s}$ subspace considerably simplifies its dynamics by disregarding particle creation effects and transitions to higher energy modes. Although the states $\ket{\uparrow}$ and $\ket{\downarrow}$ still represent mode excitations of a relativistic field, it is important to remark that the projector $\hat{P}_\text{s}$ is non-local in spacetime. This is because projectors of the form $\ket{\uparrow}\!\bra{\uparrow}$, $\ket{\downarrow}\!\bra{\downarrow}$ are intrinsically non-local, leading to covariance and causality violations which are controlled by the size of the localization of the field modes. This phenomenon has been discussed in Section~\ref{sec:LocalizedQuantumFields}, and is simply an example of the general fact that restricting a quantum field theory to modes of energies below a certain cutoff in a given frame introduces causality violations~\cite{polchinski1999effective, BurgessEffectiveFieldTheory,Bruno}. Nevertheless, the reduction of the field theory to the subspace $\mathscr{H}_\text{s}$ is justified whenever one considers processes that effectively take place at low energy and affect only the $s$ orbital, such as the interaction of a spin in an $s$ orbital with an external electromagnetic field that does not produce mode excitations. It is in this regime where one obtains the leading order relativistic corrections to the hydrogen-like atom~\cite{Dirac-electron-theory,RQMgreiner}.

\subsubsection*{An Electron Coupled to an External Electromagnetic Field}\label{sec:coupleExtMag}

As an example of the application of the reduction method introduced in the previous section, we derive the Zeeman Hamiltonian 
\begin{equation}
    \hat{H}_I = - \gamma\, \hat{\bm \sigma}\cdot \bm B,\label{eq:ZeemanBase}
\end{equation} 
for the interaction of a spin with an external electromagnetic field starting from the quantum field theoretic description of the electron. The coupling of a fermionic field with electromagnetism is encoded in the Lagrangian density of Eq.~\eqref{QED-lagrangian-density}, which contains the interaction term between $\psi(\mf x)$ and an external electromagnetic field $A^{\text{(ext)}}_\mu(\mf x)$ (see Eq.~\eqref{eq:Adecomp}). For simplicity, for the remainder of this Section, we will denote the external field simply by $A_\mu(\mf x)$. The interaction Lagrangian density can then be written as
\begin{equation}
    \mathscr{L}_I(\mf x) = q\bar{\psi}(\mf x) \slashed{A}(\mf x)\psi(\mf x).
\end{equation}
In the quantum field theory description of $\hat{\psi}(\mf x)$, we can then write the associated Hamiltonian density as
\begin{equation}\label{h-int-density}
    \hat{\mathcal{H}}_I(\mf x) = - 
 q\hat{\bar{\psi}}(\mf x) \slashed{A}(\mf x)\hat{\psi}(\mf x).
\end{equation}
The interaction Hamiltonian due to the external field $A_{\mu}(\mf x)$ is obtained by integrating Eq.~\eqref{h-int-density} over the slice $t = \text{const}.$,
\begin{equation}\label{H-interaction}
    \hat{H}_{\text{ext}}(t) = -  q\int \dd^3 \bm x \,\hat{\bar{\psi}}(\mf{x})\slashed{A}(\mf x)\hat{\psi}(\mf{x})\,,
\end{equation}
where the quantum field $\hat{\psi}(\mf x)$ is given by Eq.~\eqref{psi-general}. The interaction Hamiltonian $\hat{H}_\text{ext}$ can be projected to the $\mathscr{H}_\text{s}$ subspace using the projector $\hat{P}_\text{s}$ from the previous section. It reads
\begin{align}\label{zeeman-non-reduced}
    &\hat{H}_I(t) \coloneqq \hat{P}_\text{s}\hat{H}_{\text{ext}}(t)\hat{P}_\text{s} \\
    &= -  q\sum_{m \in \{\uparrow, \downarrow\}} \sum_{ m' \in \{\uparrow, \downarrow\}}  \int \dd^3 \bm x\, \hat{\bar{\psi}}_m(\bm x)\slashed{A}(\mf x)\hat{{\psi}}_{m'}(\bm x)\ket{m}\bra{m'}\nonumber,
\end{align}
where the time dependence of the modes $\psi_{\uparrow}$ and $\psi_{\downarrow}$ cancels, as $E_\uparrow = E_\downarrow$. 
The states $\psi_\uparrow(\bm x)$ and $\psi_\downarrow(\bm x)$ are solutions to the time-independent Diac equation, and can be explicitly written as
\begin{align}
    \psi_\uparrow(\bm x) = \ii\begin{pmatrix}
    g(r)\frac{1}{2\sqrt{\pi}}\\
    0\\
    -\ii f(r)\frac{z}{2\sqrt{\pi}r}\\
    -\ii f(r)\frac{x+\ii y}{2\sqrt{\pi}r}
\end{pmatrix} \, ,\label{Psi-lzero-up}\quad\quad
    \psi_\downarrow(\bm x) = \ii\begin{pmatrix}
    0\\
    g(r)\frac{1}{2\sqrt{\pi}}\\
    -\ii f(r)\frac{x-\ii y}{2\sqrt{\pi}r}\\
    \ii f(r)\frac{z}{2\sqrt{\pi}r}]\\
\end{pmatrix} \, ,
\end{align}
where $\bm x = (x,y,z)$ and $r = |\bm x|$. Importantly, the mode functions $f(r)$ are significantly smaller in magnitude than the mode functions $g(r)$, with $f(r)/g(r) = \mathcal{O}(\alpha)$. For instance, when $n_0 = 1$, the radial functions $g$ and $f$ are
\begin{align}
    g(r) &= k_1a_0^{-3/2}e^{-Z r/a_0}(r/a_0)^{\beta-1}\label{g} \, ,\\*
    f(r) &= k_2a_0^{-3/2}e^{-Z r/a_0}(r/a_0)^{\beta-1}\label{f}\, ,
\end{align}
with $a_0$, $\beta$ $k_1$ and $k_2$ defined as
\begin{align}
    a_0 &= \frac{1}{m_e\alpha}\, , 
  &&&  \beta &= \sqrt{1 - Z^2\alpha^2}\label{a0beta} \, ,\\
    k_1 &= 2^\beta Z^{\beta + \tfrac{1}{2}} \sqrt{\frac{1+\beta}{\Gamma(1+2\beta)}} \, ,  &&&
    k_2 &= -k_1\sqrt{\frac{1-\beta}{1+\beta}},\label{k1k2}
\end{align}
 where $a_0$ is the Bohr radius. In this example one can see that the functions $f(r)$ are significantly smaller than $g(r)$ by noticing that  $k_2 = \mathcal{O}(\alpha)$, while $k_1$ is  $\mathcal{O}(1)$.

Substituting the solutions~\eqref{Psi-lzero-up} in the interaction Hamiltonian density (\ref{zeeman-non-reduced}), we obtain
\begin{align}\label{H-interaction-simple}
    \hat{H}_I(t) &=  q \int \dd^3\bm x\, \frac{f(r)g(r)}{2\pi r}\hat{\bm{\sigma}}\cdot(\bm{x}\times \bm{A}(t, \bm{x}))\, .
\end{align}
It is possible to rewrite $\hat{H}_I(t)$ as a smeared version of the Zeeman Hamiltonian~\eqref{eq:ZeemanBase} added to a boundary term by using the following vector identities 
\begin{align}
    \bm \nabla \psi {\times} \bm{v} &= \bm \nabla \times (\psi \bm{v} ) - \psi \bm \nabla {\times} \bm{v} \label{prop2}\, ,\\
    \bm u \cdot (\bm \nabla\times \bm v) & = (\bm \nabla \times \bm u) \cdot \bm v - \bm \nabla \cdot( \bm u \times \bm v) \label{prop3}\, .
\end{align}
Let us first define the useful function $\upphi(r)$ in terms of the radial functions $f(r)$ and $g(r)$ by the conditions
\begin{equation}\label{Phi-def}
    \bm{\nabla}\upphi(r) = \frac{f(r)g(r)}{r}\bm{x}\, ,\quad \lim_{r\to\infty} \upphi(r) = 0.
\end{equation}
Using $\bm \nabla \upphi(r) = \partial_r\upphi(r)\bm{x}/r$, the unique solution to Eq.~\eqref{Phi-def} can be written as\footnote{For instance, when $n_0 = 1$ we can write $\upphi(r)$ explicitly in terms of the incomplete gamma function $\Gamma(s,t)$: 
\begin{align}\label{Phi}
    \upphi(r) = \frac{2\sqrt{1-\beta^2}}{a_0^2 \Gamma(2\beta+1)} \, \Gamma(2\beta - 1, 2r/a_0) \, .
\end{align}},
\begin{equation}\label{eq:phir}
    \upphi(r) = -\int_r^\infty \dd r\, f(r)g(r) .
\end{equation}

We can then recast the interaction Hamiltonian as
\begin{equation}
    \hat{H}_I(t) = \frac{q}{2\pi} \int \dd^3\bm x\, \hat{\bm{\sigma}}\cdot(\nabla\upphi\times \bm{A}(t, \bm{x}))\, .
\end{equation}
Using (\ref{prop2}) we obtain a dependence on $\bm B = \nabla\times \bm A$:
\begin{align}
    \hat{\bm{\sigma}}\cdot(\bm{\nabla}\upphi \times \bm{A})&=  \hat{\bm{\sigma}}\cdot(\bm{\nabla}\times(\upphi\bm{A}) - \upphi\bm{\nabla}\times \bm{A})\nonumber\\
    &= - \upphi\,\hat{\bm{\sigma}}\cdot\bm{B} +  \hat{\bm{\sigma}}\cdot(\bm{\nabla} \times (\upphi\bm{A})).\label{eq:RuhiIsAStar}
\end{align}
where the term $\hat{\bm{\sigma}}\cdot(\bm{\nabla} \times (\upphi\bm{A}))$ can be recast as a boundary term using~\eqref{prop3}:
\begin{align}
    \hat{\bm{\sigma}}\cdot(\bm{\nabla} \times (\upphi\bm{A})) &=   - \bm{\nabla}\cdot(\hat{\bm{\sigma}} \times (\upphi\bm{A})) + (\bm{\nabla}\times\hat{\bm{\sigma}})\cdot \upphi\bm{A}\nonumber\\    
    &=  - \bm{\nabla}\cdot(\hat{\bm{\sigma}} \times (\upphi\bm{A})) \, ,
\end{align}
where we used $\nabla\times\hat{\bm \sigma} = 0$, given that $\hat{\bm \sigma}$ is independent of $\bm x$. The boundary term can be safely neglected due to the fact that $\upphi(r)$ decays exponentially as $r$ increases. 

Plugging the result of Eq.~\eqref{eq:RuhiIsAStar} in the Hamiltonian~\eqref{H-interaction-simple} and neglecting the boundary term, we obtain
\begin{equation}\label{H-interaction-zeeman}
    \hat{H}_I(t) = - \frac{q}{2\pi} \int d^3\bm x\, \upphi(r)\, \hat{\bm{\sigma}}\cdot\bm{B}(t, \bm{x})\, .
\end{equation}
The interaction Hamiltonian above can be seen as a generalization of the Zeeman effect, which takes into account that the magnetic field that couples to the spin is smeared by the function $\upphi(r)$.

The Zeeman interaction in its familiar form can be obtained by assuming that the magnetic field is approximately homogeneous within the localization of the atom in the same spirit as in the dipole approximation~\cite{ScullyBook}. Indeed, if $\bm B(t,\bm x) = \bm B(t)$, the radial integral of $\upphi(r)$ factors out. We compute this integral in Appendix~\ref{app:derivation}, resulting in
\begin{equation}\label{hflatzeemanexact}
\hat{H}_I(t) =  -\frac{q}{2m_e}\hat{\bm{\sigma}}\cdot \bm{B}(t)\left(1 -\frac{4}{3}\int_0^\infty \!\dd r \, r^2 f^2(r)\right).
\end{equation}
Using the fact that $f(r) = \mathcal{O}(\alpha)$, and defining the spin operator $\hat{\bm S} = \frac{\hbar}{2} \hat{\bm \sigma}$, we find
\begin{equation}\label{eq:HintB}
    \hat{H}_I(t) = - \frac{q}{m_e} \hat{\bm{S}}\cdot \bm{B}(t) + \mathcal{O}(\alpha^2) \, .
\end{equation}
To leading order in the fine-structure constant $\alpha$, this is the Zeeman Hamiltonian for the ground state splitting of a Hydrogen atom. The higher order corrections in $\alpha$ are defined by the specific shape of the mode functions $f(r)$. Equation~\eqref{H-interaction-zeeman} effectively gives the corrections to the electron $g$-factor due to the fact that the electron is localized in an atomic orbital. This effect has been first noted by Breit in the case of hydrogen-like atoms in~\cite{Breit1928}.

Notice that the leading order corrections in $\alpha$ to the Zeeman Hamiltonian presented above are of order $\mathcal{O}(\alpha^2)$, while it is well known that corrections from QED interactions to the electron $g$-factor are of first order in $\alpha$. To introduce these QED corrections, it would be enough to incorporate the renormalized interactions obtained from higher loop QED considerations~\cite{gfactorBeier2000,gfactorIndelicato2004,gfactorTwoLoop2013,gfactor2020,gfactorExp2023}.

Finally, notice that the reduction of the QED interaction Hamiltonian to Eq.~\eqref{H-interaction-zeeman} presented above is also valid when the electron field is under the influence of any spherically symmetric electric potential $A_0(r)$. This can be seen by noticing that localized mode solutions with quantum numbers $j=1/2$ and $p=+1$ take the form of Eq.~\eqref{Psi-lzero-up} with different radial functions $f(r)$ and $g(r)$ determined by $A_0(r)$. This would allow one to consider more general models for the nucleus, such as incorporating a finite size, and Eq.~\eqref{hflatzeemanexact} would still describe the coupling of a spin with an external electromagnetic field as well as the corrections arising from the different shapes of the electron orbitals.

Effectively, when considering a quantum magnetic field $\hat{\bm B}(\mf x)$, our results have shown explicitly how to reduce a fully relativistic theory for the electron to a two-level particle detector model coupled to the magnetic field. Indeed, this model has been studied in detail in~\cite{ruhi}, but we will not focus on its specific properties here. The main point of this example is to showcase that the connection between fundamental and operational perspectives of measurements in quantum field theory is not merely theoretical, but also applies to physically realistic systems that are typically used in experimental setups.

\section{The Stress-Energy Tensor of Localized Probes}\label{sec:Tmunu}

In this Section, we will discuss a fundamental aspect of the description of localized quantum fields: their stress-energy tensor. We will discuss the necessity to dynamically describe the localizing potential in these theories and present a model for a localizing potential that interacts semiclassically with a quantum field, localizing it in space. The results of this section are based on~\cite{TmunuUDW}.

\subsubsection*{External Potentials Break General Covariance.}

The classical stress-energy tensor associated with the field $\phi_\tc{d}(\mf x)$,  whose equation of motion is defined by the Lagrangian given by Eq. \eqref{eq:Lagphid} is
\begin{equation}\begin{aligned}
    T_{\mu\nu} = \partial_\mu \phi_{\tc{d}}\partial_\nu \phi_{\tc{d}}-\frac{1}{2}g_{\mu\nu}\Big(\partial_\alpha \phi_{\tc{d}}\partial^\alpha \phi_{\tc{d}} + m_{\tc{d}}^2 \phi_\tc{d}^2 + V(\mf x) \phi_\tc{d}^2\Big).
    \label{TmunuVx}
\end{aligned}\end{equation}
By writing the expression above for the quantum field $\hat{\phi}_\tc{d}$ and taking expected values of the operator-valued stress-energy tensor, $\langle \hat{T}_{\mu\nu}\rangle$ (with the appropriate renormalization methods~\cite{birrell_davies,HadamardRenormalization2008}){,} one could claim that this is how the field $\phi_\tc{d}$ gravitates. However, there is an important matter which is not addressed in this description: what is the physical origin of the potential $V(\mf x)$? Whatever is the physical system that sources the potential $V(\mf x)$, both the source and the energy associated with $V(\mf x)$ will contribute non-negligibly to the total stress-energy tensor in the spacetime. 
Moreover, if the source of the localization potential $V({\mf x})$ is not taken into account, the field theory associated to
Eq.~(\ref{TmunuVx}) irredeemably breaks general covariance. Indeed, the stress-energy tensor of Eq.~\eqref{TmunuVx} is not covariantly conserved:
\begin{equation}
    \partial^\mu T_{\mu\nu} = \partial_\nu \phi_\tc{d}(\partial^\mu \partial_\mu - m_\tc{d}^2 - V(\mf x))\phi_\tc{d} - \frac{1}{2}\phi_\tc{d}^2\partial_\nu V(\mf x),
\end{equation}
which yields $\partial^\mu T_{\mu\nu} = - \frac{1}{2}\phi_\tc{d}^2\partial_\nu V(\mf x)$ on shell. That is, the only potentials that would produce a theory fully compatible with general covariance are constant, being unable to localize the field modes of $\phi_\tc{d}$.

One option would be to consider that $V(\mf x)$ is generated by another scalar field. In that case, $V(\mf x)$ would be replaced by $V(\phi_\tc{c}(\mf x))$ and, for any fixed solution for the field $\phi_\tc{c}$, the potential which would affect $\phi_\tc{d}$ would be a function of $\phi_\tc{c}(\mf x)$. In that case, the full Lagrangian of the theory would be 
\begin{equation}
\begin{aligned}
    \mathcal{L} =& - \frac{1}{2} \partial_\mu\phi_{\tc{d}}\partial^\mu\phi_{\tc{d}} - \frac{m_\tc{d}^2}{2} \phi_{\tc{d}}^2 -\frac{V(\phi_\tc{c})}{2}\phi^2_\tc{d}
    - \frac{1}{2} \partial_\mu\phi_\tc{c}\partial^\mu \phi_\tc{c} - \frac{m_\tc{c}^2}{2}\phi_\tc{c}^2
\end{aligned}
\end{equation}
and the associated stress-energy tensor would be
\begin{equation}
\begin{aligned}
    T_{\mu\nu} =& \partial_\mu \phi_{\tc{d}}\partial_\nu \phi_{\tc{d}}+\partial_\mu \phi_{\tc{c}}\partial_\nu \phi_{\tc{c}}-\frac{1}{2}g_{\mu\nu}\Big(\partial_\alpha \phi_{\tc{d}}\partial^\alpha \phi_{\tc{d}} + m_{\tc{d}}^2 \phi_\tc{d}^2 + V(\phi_\tc{c})\phi_\tc{d}^2+ \partial_\alpha \phi_{\tc{c}}\partial^\alpha \phi_{\tc{c}} + m_{\tc{c}}^2 \phi_\tc{c}^2\Big){,}
\end{aligned}
\end{equation}
which is covariantly conserved on shell. However, the solutions of the Klein-Gordon equation for $\phi_\tc{c}$ will generally not be confined to a finite region of space and will propagate away. One way to prevent this would be to find a fine tuned solitonic solution for the fields $\phi_\tc{d}$ and $\phi_\tc{c}$. Unfortunately, this would imply that small changes to the system (such as when $\phi_\tc{d}$ interacts with an external field) would likely break the bound system, leading to the fields $\phi_\tc{c}$ and $\phi_\tc{d}$ propagating away.

An alternative way of keeping the field $\phi_\tc{c}$ localized in space would be to consider an external potential $V_\tc{c}(\bm x)$ which localizes it. This would amount to adding a term of the form $- \frac{1}{2}V_\tc{c}(\bm x) \phi_\tc{c}^2$ to the Lagrangian, which would allow for a bound solution for $\phi_\tc{c}$. One would then naturally wonder what is the physical system that sources the potential $V_\tc{c}$ and what are its contributions to the stress-energy tensor. One could, of course, introduce yet another scalar field $\phi_\tc{b}$ and to add another interaction of the form $V(\phi_\tc{b})\phi_\tc{c}^2$, but this would quickly lead us to a rabbit hole, where one would always need to add another localized field to source the effective potential that localizes the previous one.

A possible alternative solution to this puzzle (as we will show throughout this Section) is to introduce matter with two degrees of freedom---a perfect fluid, for instance. One degree of freedom would be responsible for the localization of the field $\phi_{\tc{c}}$ regardless of the state of the field $\phi_\tc{d}$, and the other one would depend on the equations of motion and a boundary condition. As we will see, the additional matter with two degrees of freedom would allow a stable solitonic solution for the fields $\phi_\tc{c}$ and $\phi_\tc{d}$, maintaining the shape of the solution $\phi_\tc{c}$ (and thus of the localizing potential $V(\phi_\tc{c})$), regardless of the state of $\phi_\tc{d}$.

\subsubsection*{A Comment on Localization}

Before presenting an explicit example of a general covariant localized quantum field, let us discuss the implications of the fact that the external potential must be incorporated in the stress-energy tensor of the theory to compactly supported fields. Strictly speaking, one can only have modes of the field that are compactly supported in the limit where the external potential goes to infinity outside of a given region. In this case, the associated stress-energy tensor would also diverge pointwise. This implies that, at the very least, the energy density associated with this system would be infinite. Taking the Hoop conjecture~\cite{hoopConj,HoopReview} seriously, we conclude that
\begin{center}
    \textit{``A general covariant model for a perfectly localized field is a black hole.''}
\end{center}
\noindent This fact implies that all general covariant formulations of localized probes in quantum field theory must have infinite support in space. At the same time, it does not seem reasonable to assume that it is possible to perfectly control the mechanisms responsible for localizing the interactions of quantum fields. This would imply the field observables $\hat{\phi}(f)$ with compactly supported $f\in C_0^\infty(\mathcal{O})$ could technically not be probed, as the functions $f$ are defined by the spaces of the probes and their interaction regions. Instead, we would be restricted to accessing only operators $\hat{\phi}(f)$ that do not belong to any local algebra (only to the global $\mathcal{A}(\M)$), with non-compactly supported functions $f$. In this case, one must use the profile of the functions $f$ to determine effective regions where one has access to the field, determining how much access one has to each region by the value of $|f(\mf x)|$ within it. In essence, this suggests that a better notion of ``regions'' in quantum field theory could be given in terms of functions of fast decay, rather than by subsets of spacetime. This would allow one to incorporate the fact that accessing the field in a sharp spacetime region is an idealized scenario while maintaining some notion of localization. A concrete implementation of this idea, as well as its implications to fundamental concepts, such as causality, will appear in a future work.

\subsubsection*{A Consistent Semiclassical Description for a Particle Detector}
\label{sec:model}

A description of a localized quantum field which takes into account the physical system that localizes it can be formulated in terms of a Lorentz invariant action in Minkowski spacetime. The full Lagrangian depending on the scalar field $\phi_\tc{d}$,  a complex\footnote{We chose a complex-valued field so that a time-independent potential and energy-momentum distribution (which basically depends on $|\psi_\tc{c}(\mf x)|^2$)  can be obtained when $\psi_\tc{c}$ is in a stationary state.} field $\psi_\tc{c}$ that produces an effective potential responsible for the localization of $\phi_{\tc{d}}$, and on the fluid configuration  is given by
\begin{equation}
\begin{aligned}
    \mathcal{L} = &- \frac{1}{2} \partial_\mu\phi_{\tc{d}}\partial^\mu\phi_{\tc{d}} - \frac{m_\tc{d}^2}{2} \phi_{\tc{d}}^2 - \frac{\alpha}{2} |\psi_{\tc{c}}|^2\phi_{\tc{d}}^2\\
    &- \partial_\mu\psi_\tc{c}^*\partial^\mu \psi_\tc{c} - m_\tc{c}^2|\psi_\tc{c}|^2 - V_{\tc{c}}(|\psi_\tc{c}|^2)+(1- \mu  |\psi_\tc{c}|^2) \mathcal{L}^{\textrm{fluid}},
    \label{full lagrangian}
\end{aligned}
\end{equation}
where $\mu$ is a coupling constant with units of squared length, $\alpha$ is a dimensionless constant, $m_\tc{d}$ and $m_\tc{c}$ are the masses of the fields $\phi_\tc{d}$ and $\psi_\tc{c}$ and $V_\tc{c}(|\psi_\tc{c}|^2)$ is a self-interaction term for the field $\psi_\tc{c}$.  

The role of the Lagrangian $\mathcal{L}^{\textrm{fluid}}$ in this description is two-fold. It gives rise to the energy-momentum of the fluid and appears explicitly on the equations of motion for $\psi_\tc{c}$ due to the non-minimal coupling between the fluid and the field $\psi_{\tc{c}}$. In this way, the exact form of the on-shell Lagrangian $\mathcal{L}^{\textrm{fluid}}$ turns out to be essential. There are several possible on-shell real Lagrangians (all of them giving rise to the same energy-momentum tensor). The most common options are $\mathcal{L}^\textrm{fluid}=P$~\cite{Schutz1970} and $\mathcal{L}^\textrm{fluid}=-\rho$~\cite{Brown1993}, where $P$ and $\rho$ are the proper pressure and the proper energy density of the fluid. The transition between these two Lagrangians is made through the addition of a surface integral in the fluid action, i.e., the Lagrangian is modified by a total derivative term~\cite{Brown1993}. This clearly affects its on-shell value without affecting the equations of motion. Also, by considering the fluid as constituted by particles with fixed rest mass and structure (solitons), the average on-shell Lagrangian turns out to be of the form~\cite{Avelino2018} 
\begin{equation}
    \mathcal{L}^{\textrm{fluid}} = T^{\textrm{fluid}} = -\rho+3P,
\end{equation}
i.e., the trace of the stress-energy tensor 
\begin{equation}
    T^\textrm{fluid}_{\mu\nu}  =  (\rho + P)u_\mu u_\nu+ P g_{\mu\nu}.
\end{equation}
In all cases, for the fluid to be modelled by a collection of particles, the equation of state $w=p/\rho$ must satisfy $0\leq w\leq 1/3$. 

The equations of motion for the fields $\phi_\tc{d}$ and $\psi_\tc{c}$ are given  by
\begin{subequations}
      \begin{align}
        & (\Box-m_\tc{d}^2 - \alpha|\psi_\tc{c}|^2)\phi_\tc{d}  = 0,\label{eqmotion1}\\
        &\left(\Box-m_\tc{c}^2- F_\tc{c}(|\psi_\tc{c}|^2) -\mu  \mathcal{L}^{\textrm{fluid}}- \frac{\alpha}{2}\phi_\tc{d}^2\right) \psi_{\tc{c}} = 0,\label{eqmotion2}
      \end{align}
      \label{eqmotion}
    \end{subequations}
where we defined
\begin{equation}
    F_\tc{c}(|\psi_\tc{c}|^2) = \pdv{V_{\tc{c}}}{|\psi_{\tc{c}}|^2}.
\end{equation}
Notice that Eq.~(\ref{eqmotion2}) shows explicitly how the Lagrangian of the fluid $\mathcal{L}^{\textrm{fluid}}$ affects the equations of motion for $\psi_\tc{c}$.

The equations of motion for a perfect fluid minimally coupled to gravity are equivalent to $\nabla^\mu T_{\mu\nu}^{\textrm{fluid}}=0$ along with the conservation of particle number~\cite{Schutz1970}. However, in our case, the fluid is also coupled to matter fields. Hence the equations of motion for the fluid turns out to be given by the conservation of the full stress-energy tensor in the spacetime, i.e.,  $\nabla_\mu T^{\mu\nu}=0$, where 

\begin{equation}
\begin{aligned}\label{eq:fullTmunu}
    T_{\mu\nu} \equiv& - \frac{2}{\sqrt{-g}} \frac{\delta (\sqrt{-g}\mathcal{L})}{\delta g^{\mu\nu}}\\
    =& \partial_\mu \phi_{\tc{d}}\partial_\nu \phi_{\tc{d}}+2\Re(\partial_\mu \psi_{\tc{c}}^*\partial_\nu \psi_{\tc{c}}) -\frac{1}{2}g_{\mu\nu}\Big(\partial_\alpha \phi_{\tc{d}}\partial^\alpha \phi_{\tc{d}} + m_{\tc{d}}^2 \phi_\tc{d}^2 + \alpha |\psi_\tc{c}|^2 \phi_\tc{d}^2\\
    &+ 2\partial_\alpha \psi_{\tc{c}}^*\partial^\alpha \psi_{\tc{c}} + 2m_{\tc{c}}^2 |\psi_\tc{c}|^2 + 2 V_\tc{c}(|\psi_\tc{c}|^2)\Big)+(1-\mu |\psi_\tc{c}|^2)T^\textrm{fluid}_{\mu\nu}.
\end{aligned}
\end{equation}
We can then compute its divergence,

\begin{equation}
\begin{aligned}
    \partial^\mu T_{\mu\nu} =&  \left[\Big(\Box-m_\tc{d}^2 - \alpha|\psi_\tc{c}|^2 \Big)\phi_{\tc{d}}\right]\partial_\nu \phi_{\tc{d}} \\&+2\Re\left\{\left[\Big(\Box-m_\tc{c}^2  -F_\tc{c}(|\psi_\tc{c}|^2)- \mu  \mathcal{L}^\textrm{fluid}- \tfrac{\alpha}{2}\phi_\tc{d}^2\Big)\psi_{\tc{c}}\right]\partial_\nu \psi_{\tc{c}}^*\right\}
    \\&+ (1-\mu|\psi_\tc{c}|^2) \partial^\mu T^\textrm{fluid}_{\mu\nu} -\mu T^\textrm{fluid}_{\mu\nu} \partial^\mu |\psi_\tc{c}|^2 + \mu  \mathcal{L}^\textrm{fluid} \partial_\nu|\psi_\tc{c}|^2,
\end{aligned}
\label{diverce}
\end{equation}
where we added and subtracted $ \mu  \mathcal{L}^\textrm{fluid} \partial_\nu|\psi_\tc{c}|^2$ to explicitly factor the equation of motion for the field $\psi_\tc{c}$. Using the equations of motion \eqref{eqmotion}, the first and second lines in Eq.~(\ref{diverce}) vanish, and we see that the divergencelessness of $T_{\mu\nu}$ is ensured provided that the perfect fluid stress-energy tensor satisfies
\begin{equation}\label{eq:Tmunurequirement}
\begin{aligned}
    (1-\mu|\psi_\tc{c}|^2) \partial^\mu T^\textrm{fluid}_{\mu\nu} &-\mu T^\textrm{fluid}_{\mu\nu} \partial^\mu |\psi_\tc{c}|^2 + \mu  \mathcal{L}^{\textrm{fluid}} \partial_\nu|\psi_\tc{c}|^2 = 0,
\end{aligned}\end{equation}
which turns into a differential equation for $u_\mu$, $\rho$ and $P$.

We are interested in using the field $\psi_\tc{c}$ to source a time-independent potential for the field $\phi_\tc{d}$. This can be obtained if $\psi_\tc{c}$ is of the form
\begin{equation}\label{eq:ansatzpsic}
    \psi_\tc{c}(\mf x) = e^{\ii \omega_\tc{c} t}\Psi_\tc{c}(\bm x).
\end{equation}
Below we will analyze a particular case which allows this ansatz for $\psi_\tc{c}$ to be a solution to the equations of motion.

We will now analyze the case where the field $\phi_\tc{d}$ is such that $\phi_\tc{d}^2(\mf x) = g(\bm x)$ is time-independent in a given inertial frame. In the semiclassical context $\phi_\tc{d}^2(\mf x)$ would be replaced by the renormalized expected value of $\hat{\phi}_\tc{d}^2$,  $\langle \normord{\hat{\phi}_\tc{d}^2(\mf x)}\rangle$, which is time-independent whenever $\hat{\phi}_\tc{d}$ is in an eigenstate of its Hamiltonian (e.g., if $\alpha = 0$ and $\hat{\phi}_\tc{d}$ were in its vacuum state, we would have $g(\bm x) = 0$).  In this case, the equation of motion for $\psi_\tc{c}(\mf x) = e^{\ii \omega_\tc{c} t}\Psi_\tc{c}(\bm x)$ reads
\begin{equation}\begin{aligned}
    \Big(\omega_\tc{c}^2 - m_\tc{c}^2& + \nabla^2 - f(\bm x)\Big)\Psi_\tc{c}(\bm x) = 0,
    \end{aligned}
\end{equation}
where we defined
\begin{equation}\label{eq:f}
    f(\bm x) = \mu \mathcal{L}^\textrm{fluid} + \frac{\alpha}{2}g(\bm x) + F_\tc{c}(\bm x), \quad \quad F_\tc{c}(\bm x) = \pdv{V_\tc{c}}{|\Psi_\tc{c}|^2}.
\end{equation}
Notice that once $g(\bm x)$ is fixed and the potential $V_\tc{c}$ is chosen, the fluid Lagrangian completely determines $f(\bm x)$. Thus, $\Psi_\tc{c}(\bm x)$ is an eigenfunction of the operator \mbox{$-\nabla^2 + f(\bm x)$}. Recalling that we are interested in localized solutions, we should
look for eigenfunctions with negative eigenvalues:
\begin{equation}
    (-\nabla^2 + f(\bm x)) \Psi_\tc{c}(\bm x)  = -\lambda_\tc{c}^2 \Psi_\tc{c}(\bm x){.}
    \label{potential f}
\end{equation}
Hence, Eq. \eqref{eq:ansatzpsic} is a stationary localized 
solution to the equations of motion provided that
\begin{equation}
    \omega_\tc{c}^2 = m_\tc{c}^2 - \lambda_\tc{c}^2,
    \label{frequency}
\end{equation}
with $m_\tc{c}^2 > \lambda_\tc{c}^2$. 

Due to the fact that $|\psi_\tc{c}(\mf x)|^2 = |\Psi_\tc{c}(\bm x)|^2$ is time-independent in this case, it is natural to impose that the fluid described by $T^\textrm{fluid}_{\mu\nu}$ undergoes motion in the $\partial_t$ direction, and that both $\rho$ and $P$ are time-independent. We then find that the $0-$th component of Eq.~(\ref{eq:Tmunurequirement}) is trivial, while
\begin{equation}
    \partial^\mu T^\textrm{fluid}_{\mu i } = \partial_i  P
\end{equation}
and Eq. \eqref{eq:Tmunurequirement} becomes a differential equation for $P$, 
\begin{equation}
\begin{aligned}
    (1-\mu |\Psi_\tc{c}|^2)\partial_i P - \mu  \partial_i|\Psi_\tc{c}|^2 P+(f(\bm x) -\tfrac{\alpha}{2}g(\bm x) -F_\tc{c}(\bm x))\partial_i |\Psi_\tc{c}|^2 = 0
\end{aligned}
\label{eq for P}
\end{equation}
where $f$, $g$, and $|\Psi_\tc{c}|^2$ are given.

Finding a solution $P$ for the above equation also gives us the proper energy density of the fluid ($\rho$) through the equation $\mu \mathcal{L}^\textrm{fluid} + \frac{\alpha}{2}g(\bm x) +F_\tc{c}\left(|\Psi_\tc{c}|^2\right)= f(\bm x)$. In particular, the proper energy density of the fluid depends on the choice of the on-shell Lagrangian $\mathcal{L}^\textrm{fluid}$ (i.e., the choice of the non-minimal coupling between the fluid and $\psi_\tc{c}$).

We consider two on-shell Lagrangians given by \mbox{$\mathcal{L}^\textrm{fluid}=-\rho+3\eta P$} so that \mbox{$\eta=0$} for the choice \mbox{$\mathcal{L}^\textrm{fluid}=-\rho$} and $\eta=1$ for \mbox{$\mathcal{L}^\textrm{fluid}=T^{\textrm{fluid}}=-\rho+3P$}\footnote{We will not consider Lagrangians that are independent of $\rho$ (such as $\mathcal{L}^\textrm{fluid} = P$), so that the energy density can be determined from $\mathcal{L}^\textrm{fluid}$.}.  We can then find the equation of state of the fluid, relating $\rho$ and $P$, in terms of the functions $g(\bm x)$, $F_\tc{c}(\bm x)$ and $f(\bm x)$: 
\begin{equation}
    \rho = 3 \eta P + \frac{\alpha}{2 \mu}g(\bm x) + \frac{F_{\tc{c}}(\bm x)}{\mu}  - \frac{f(\bm x)}{\mu}.
    \label{eq for rho}
\end{equation}

For a given time-independent configuration of the field $\phi_\tc{d}$, this solution is stable and satisfies $\partial^\mu T_{\mu\nu} = 0$ in the whole spacetime. In particular, notice that the function $f(\bm x)$ is independent of the choice of $g(\bm x)$, so that different $g$'s yield different pressures $P$ and proper energy densities $\rho$ for the fluid. Also notice that whether $T^{\textrm{fluid}}_{\mu\nu}$ satisfies energy conditions or not will explicitly depend on the choices of $f(\bm x)$, $g(\bm x)$ and $\Psi_\tc{c}(\bm x)$. Ideally, to ensure localization of the whole system (in the sense that $T_{\mu\nu}$ goes to zero at spatial infinity), one would require that both $\rho$ and $P$ go to zero at infinity, so that the integration constant arising in Eq. \eqref{eq for P} is not arbitrary.

\subsubsection*{An Explicit Example of a Localized Quantum Field}
\label{sec:example}

In this Section, we will present a concrete example of a realization of the time-independent model constructed above. The first step is to pick the state of the classical field, which will source the external potential for the quantum field $\phi_\tc{d}(\mf x)$. The explicit example we will construct will be spherically symmetric, so we use spherical coordinates $(t,r,\theta,\phi)$ and choose the state for the field $\psi_\tc{c}(\mf x)$ to be
\begin{equation}
    \psi_\tc{c}(\mf x) = \frac{1}{\ell}e^{-\ii \omega_\tc{c} t}\sech(\frac{r}{\ell}),
    \label{psic}
\end{equation}
so that the effective potential generated by $\psi_\tc{c}(\mf x)$ on $\phi_\tc{d}$ is given by
\begin{equation}
    \alpha|\psi_\tc{c}|^2 = \frac{\alpha}{\ell^2} \sech^2\left(\frac{r}{\ell}\right).
\end{equation}
This implies that the equation of motion for the field $\phi_\tc{d}$ becomes
\begin{equation}
    \left(\Box-m_\tc{d}^2 - \frac{\alpha}{\ell^2}\sech^2\left(\tfrac{r}{\ell}\right)\right)\phi_\tc{d}  = 0.
\end{equation}
We also find that
\begin{equation}
    \nabla^2 \Psi_\tc{c}(\bm x)  = \left(\frac{1}{\ell^2} -\frac{2}{\ell^2} \sech^2\left(\tfrac{r}{\ell}\right) - \frac{2}{\ell^2} \frac{\tanh(\tfrac{r}{\ell})}{\tfrac{r}{\ell}}\right)\Psi_\tc{c}(\bm x),
    \label{eqpsic}
\end{equation}
so that it has the shape of the eigenvalue equation~\eqref{potential f} with
\begin{align}
    f(\bm x) &= f(r) = -\frac{2}{\ell^2} \sech^2(\tfrac{r}{\ell}) - \frac{2}{\ell^2} \frac{\tanh(\tfrac{r}{\ell})}{r/\ell},
\end{align}
and the corresponding eigenvalue $\lambda_{\tc{c}} = 1/\ell$, so that \mbox{$(\nabla^2 - f(\bm x)) \Psi_\tc{c}= \lambda_{\tc{c}}\Psi_\tc{c}$}. We also find the frequency $\omega_\tc{c}$ from Eq. \eqref{frequency}
\begin{equation}
    \omega_{\tc{c}} = \sqrt{m_\tc{c}^2 - \frac{1}{\ell^2}}.
\end{equation}
For convenience, we assume that $\langle:\!\hat{\phi}_\tc{d}^2(\mf x)\!:\rangle = g(\mf x) = 0$ here so that we can pick $\mathcal{L}^\textrm{fluid}$ and the potential $V_\tc{c}(|\psi_\tc{c}|^2)$ as
\begin{equation}
    V_\tc{c}(|\psi_\tc{c}|^2) = -(|\psi_\tc{c}|^2)^2, \quad\quad \mathcal{L}^{\textrm{fluid}} =-\frac{2}{\mu\ell^2} \frac{\tanh(\tfrac{r}{\ell})}{r/\ell},
\end{equation}
so that $F_\tc{c}(\bm x)$ is given by
\begin{equation}
    F_\tc{c}(\bm x) = -2|\psi_\tc{c}(\mf x)|^2 = -\frac{2}{\ell^2}\sech^2(\tfrac{r}{\ell}),
\end{equation}
and Eq.~\eqref{eq:f} is satisfied. Notice that both $\mathcal{L}^{\text{fluid}}$ and $V_\tc{c}(\bm x)$ are both smooth bounded functions in space. Equation \eqref{potential f} then holds provided that $\psi_\tc{c}(\mf x)$ is given by Eq. \eqref{psic}. 

We now solve Eq.~\eqref{eq for P} for the pressure $P$ and Eq.~\eqref{eq for rho} for the energy density $\rho$. Due to spherical symmetry and time invariance of the system, we have $P = P(r)$ and $\rho = \rho(r)$. The differential equation for $P$ becomes
\begin{equation}
    P'(r)  + \frac{2 \mu \tanh(\tfrac{r}{\ell})P(r)}{\ell(-\mu + \ell^2\cosh^2(\tfrac{r}{\ell}))}  = \frac{4 \tanh^2(\tfrac{r}{\ell})}{r\ell^2( - \mu+\ell^2 \cosh^2(\tfrac{r}{\ell}))}.
\end{equation}
The solution for $P(r)$ can be written as
\begin{align}\label{eq:P}
    P(r)  &= \frac{1}{\ell^2\left(\ell^2 -\mu \sech^2(\tfrac{r}{\ell})\right)}\int_r^\infty dr' G(r'),
\end{align}
where
\begin{align}
    G(r) \label{eq:Gofr}
    &=\frac{4 \sech^2(\tfrac{r}{\ell})\tanh^2(\tfrac{r}{\ell})}{r},
\end{align}
and we picked the integration constant such that $\lim_{r\to \infty} P(r) = 0$. Unfortunately, no known closed expression for the integral of $G(r)$ is known. The energy density can be found by using Eq.~\eqref{eq for rho}, which yields
\begin{equation}\label{eq:rho}
    \rho(r) = 3\eta  P(r) + \frac{2}{\mu\ell^2}\frac{\tanh(\tfrac{r}{\ell})}{r/\ell}. 
\end{equation}

In order to have a concrete model, we will also pick a value for the constant $\mu$. However, not all values for $\mu$ will yield a physical model. Notice that for $\mu  >\ell^2$, the solution for $P(r)$ is divergent at $r = \ell \,\text{arcsech}(\ell^2/|\mu|)$. Provided that $\mu <\ell^2$, the solution for $P(r)$ is smooth, positive and decreasing. It is also important to ensure that the energy density of the fluid is positive. First, notice that for large $r$, $\rho(r)$ behaves as $\frac{2}{\mu \ell r}$, so that we must have $\mu>0$ to ensure $\rho(r)>0$. Given that $P(r)$ is positive and smooth for $\mu<\ell^2$, $\mu \in (0,\ell^2)$ ensures that $\rho(r)>0$ for all $r$.

\begin{figure}[htb]
    \centering
    \includegraphics[width=11cm]{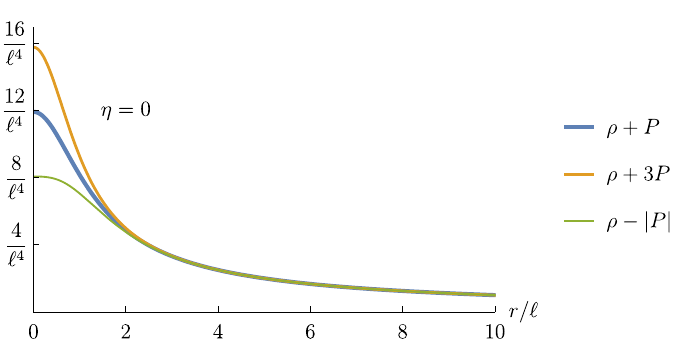}
    \caption{$\rho+P$, $\rho+3P$, $\rho-|P|$ as a function of $r/\ell$ for $\eta=0$ and $\mu=\ell^2/5$.}
    \label{fig1}
\end{figure}

\begin{figure}[htb]
    \centering
    \includegraphics[width=11cm]{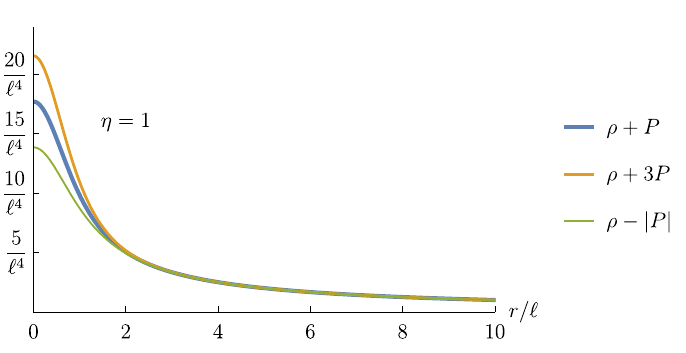}
    \caption{$\rho+P$, $\rho+3P$, $\rho-|P|$ as a function of $r/\ell$ for $\eta=1$ and $\mu=\ell^2/5$.}
    \label{fig2}
\end{figure}

One can go a step further and demand that the null, weak, strong and dominant energy conditions are satisfied by the fluid, imposing that $\rho+P>0$, $\rho+3P>0$ and $\rho-|P|>0$. Given that both $\rho$ and $P$ are positive, the only condition that adds extra constraints is $\rho-|P|>0$. We find that $\rho-|P|>0$ if $2/\mu > 3 (\eta-1) P(0)$, and $P(0)$ can be computed in closed form~\cite{mathematica}:
\begin{align}
    P(0) = \frac{4 \left(4 \log(A) - 40 \zeta'(-3) - \frac{1}{3} - \frac{4}{45}\log(2)\right)}{\ell^2(\ell^2 - \mu)}\equiv \frac{g_0}{\ell^2(\ell^2  - \mu)}
\end{align}
where $\zeta$ denotes Riemann's Zeta function and $A$ is the Glaisher constant, yielding $g_0 \approx 1.53971$. We then find that $\rho(r) - |P(r)|>0$ if either $\eta>\frac{1}{3}$, or if $\mu < \frac{\ell^2}{1+(1-3\eta) g_0/2}$. 
For instance, if $\eta=0$ we have that $\rho(r) - |P(r)|$ will only be positive if $0<\mu\lesssim 0.565017 \ell^2$. In Figs.~\ref{fig1} and~\ref{fig2} we plot $\rho+P$, $\rho+3P$ and $\rho-|P|$ for values of the constant $\mu$ that respect the energy conditions.
\begin{figure}[htb]
    \centering
    \includegraphics[width=11cm]{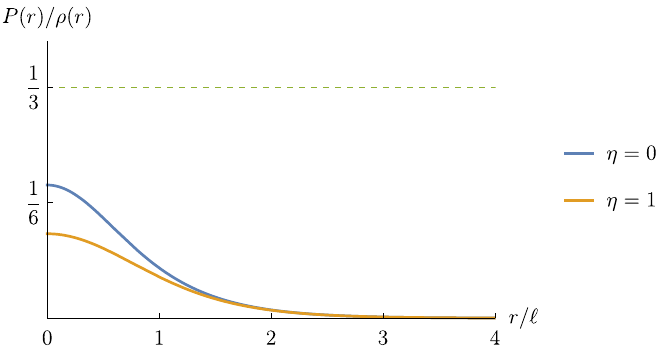}
    \caption{$w:=P/\rho$ for $\eta=0$ and $\eta=1$ with the choice $\mu = \ell^2/5$.}
    \label{figw}
\end{figure}

Fig.~\ref{figw} shows that $0<w:=P/\rho<1/3$ for both $\eta=0$ and $\eta=1$. This is especially important for the case $\eta=1$, where the fluid is constituted by particles with fixed mass and structure. In any case, the fluid is composed of non-exotic matter content. 
Finally, as expected, both quantities are localized around the origin within a lengthscale characterized by the parameter $\ell$.

With these classical solutions for the fields $\psi_\tc{c}(\mf x)$, and for the fluid parameters $\rho(r)$ and $P(r)$, the field $\phi_{\tc{d}}$ experiences the effective potential 
\begin{equation}
    V(r) = \frac{\alpha}{\ell^2} \sech^2\left(\frac{r}{\ell}\right).
\end{equation}
From here on, we pick $\alpha = -6$. The equation of motion for the field $\phi_\tc{d}$ can then be solved by separation of variables. The operator
\begin{equation}
    L = - \nabla^2 - \frac{6}{\ell^2}\sech^2(\tfrac{r}{\ell}) 
\end{equation}
possesses one eigenfunction with negative eigenvalue and a continuous spectrum in $[0,\infty)$. The Klein-Gordon normalized eigenfunction of $L$ and its respective eigenvalue are
\begin{align}
    \Phi_1(r) &= \sqrt{\frac{3}{8\pi \ell\omega_\tc{d}}}\frac{\tanh(\tfrac{r}{\ell})}{r \cosh(\tfrac{r}{\ell})}, &&& \mu_1 &= -\frac{1}{\ell^2}.\label{eq:Phi1}
\end{align}
We label the generalized eigenfunctions in the continuous spectrum of $L$ by $v_{klm}(\bm x)$, where $k^2/\ell^2$ is the corresponding eigenvalue, and $l,m$ are the usual angular momentum labels.

The corresponding quantum field $\hat{\phi}_\tc{d}$ can then be written as
\begin{align}
    \hat{\phi}_\tc{d}(\mf x) =& \left(e^{- \ii \omega_{1} t}\Phi_{1}(r) \hat{a}_{1} + e^{\ii \omega_{1} t}\Phi^*_{1}(r) \hat{a}^\dagger_{1}\right)\\
    &\!\!\!\!\!\!\!\!\!\!\!\!\!\!\!\!\!+ \sum_{l,m}\int\dd k \left(e^{- \ii w_{k} t}v_{klm}(\bm x) \hat{b}_{klm} +e^{\ii w_{k} t}v^*_{klm}(\bm x) \hat{b}^\dagger_{klm} \right),\nonumber
\end{align}
where $\omega_1 = \sqrt{m_\tc{d}^2 - \tfrac{1}{\ell^2}}$ and $w_{k} = \sqrt{m_{\tc{d}}^2 + \frac{k^2}{\ell^2}}$. Notice that the modes $v_{klm}(\bm x)$ are not localized, as they do not belong to $L^2(\mathbb{R}^3)$, as they are associated with scattering states.

\subsubsection*{The Stress-Energy Tensor of a Particle Detector}
\label{sec:tmunuexample}

We now describe the stress-energy tensor of a Unruh-DeWitt detector that matches the description of Section~\ref{sec:example}. The energy-tensor for this configuration can be written as Eq.~\eqref{eq:fullTmunu}, with the replacement $\phi_\tc{d}(\mf x)\mapsto \hat{\phi}_\tc{d}(\mf x)$, giving rise to an operator-valued energy momentum tensor $\hat{T}_{\mu\nu}$. One can then obtain a classical stress-energy tensor by taking the expected value of the normal ordered energy tensor
\begin{equation}
    \langle\normord{\hat{T}_{\mu\nu}}\rangle_{\hat{\rho}_\tc{d}} = \langle\hat{T}_{\mu\nu}\rangle_{\hat{\rho}_\tc{d}} - \bra{0_\tc{d}}\hat{T}_{\mu\nu}\ket{0_\tc{d}},
\end{equation}
where $\hat{\rho}_\tc{d}$ denotes the state of the detector field $\hat{\phi}_\tc{d}$ and we choose to use as a reference state the vacuum\footnote{In order to obtain a finite value for the stress-energy tensor, one requires to choose a renormalization scheme. We choose the reference state $\ket{0_\tc{d}}$ for convenience, but another natural choice would be to consider a subtraction using the Minkowski vacuum.} of the field $\hat{\phi}_\tc{d}(\mf x)$. For convenience, we define
\begin{align}
    \hat{T}_{\mu\nu}^{\phi_\tc{d}} &=  \partial_\mu \hat{\phi}_{\tc{d}}\partial_\nu \hat{\phi}_{\tc{d}} -\frac{1}{2}g_{\mu\nu}\!\left(\partial_\alpha \hat{\phi}_{\tc{d}}\partial^\alpha \hat{\phi}_{\tc{d}} + m_{\tc{d}}^2 \hat{\phi}_\tc{d}^2\right)\!, \\
    \hat{T}_{\mu\nu}^{\phi_\tc{d}\psi_\tc{c}} &= -\frac{1}{2}g_{\mu\nu}\alpha |\psi_\tc{c}|^2 \hat{\phi}_\tc{d}^2,\\
    T_{\mu\nu}^{\psi_\tc{c}} &= 2\Re(\partial_\mu \psi_{\tc{c}}^*\partial_\nu \psi_{\tc{c}})-g_{\mu\nu}\left(\partial_\alpha \psi_{\tc{c}}^*\partial^\alpha \psi_{\tc{c}} + m_{\tc{c}}^2 |\psi_\tc{c}|^2  +  V_\tc{c}(|\psi_\tc{c}|^2)\right),\nonumber\\
    T_{\mu\nu}^{\psi_\tc{c}\text{fluid}} &=  -\mu |\psi_\tc{c}|^2T^\textrm{fluid}_{\mu\nu},
\end{align}
so that the stress-energy tensor of the detector can be expressed as the sum
\begin{equation}
    \hat{T}_{\mu\nu} = \hat{T}_{\mu\nu}^{\phi_\tc{d}} + \hat{T}_{\mu\nu}^{\phi_\tc{d}\psi_\tc{c}} + T_{\mu\nu}^{\psi_\tc{c}} + T_{\mu\nu}^{\psi_\tc{c}\text{fluid}} + T^\textrm{fluid}_{\mu\nu}.
\end{equation}
Notice that only $\hat{T}_{\mu\nu}^{\phi_\tc{d}}$ and $\hat{T}_{\mu\nu}^{\phi_\tc{d}\psi_\tc{c}}$ are operator valued. However, all terms that contain dependence on the fluid indirectly depend on the state of the field $\hat{\phi}_\tc{d}$ through the function $g(\bm x)$.

As an explicit example, we consider a harmonic oscillator Unruh-DeWitt detector interacting with a free massless quantum field according to the interaction Hamiltonian
\begin{equation}
    \hat{\mathcal{H}}(\mf x) = \lambda \Lambda(\mf x)(e^{\ii \Omega t} \hat{a}^\dagger + e^{- \ii \Omega t}\hat{a})\hat{\phi}(\mf x).
\end{equation}
We consider the detector to have the spacetime smearing function
\begin{equation}
    \Lambda(\mf x) = e^{-\frac{t^2}{2T^2}} \sqrt{\frac{3}{8\pi \ell\omega_{\tc{d}}}}\frac{\tanh(\tfrac{r}{\ell})}{r \cosh(\tfrac{r}{\ell})}.
\end{equation}
This spacetime smearing function corresponds to a detector modelled by the bound mode $\Phi_1(r)$ in Eq.~\eqref{eq:Phi1} and a switching function $\zeta(\mf x) = e^{-\frac{t^2}{2 T^2}}$ that controls the time profile of the interaction with the parameter $T$. This detector is well modelled (to leading order) by the field $\hat{\phi}_\tc{d}$, interacting with the field $\hat{\phi}$ when $\Omega = \omega_1 = \sqrt{m_\tc{d}^2 - \tfrac{1}{\ell^2}}$, as discussed in Section~\ref{sec:LocalizedQuantumFields}.

\begin{figure}[htb]
    \centering
    \includegraphics[width=11cm]{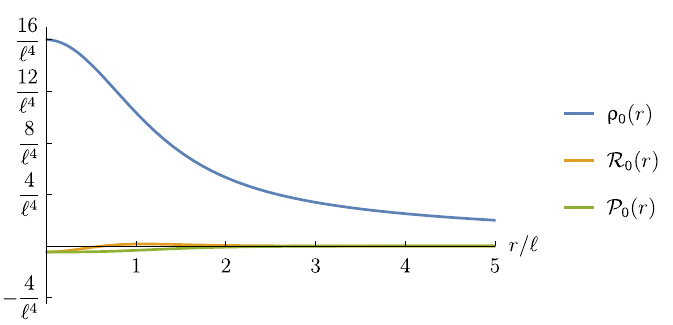}
    \caption{$\uprho_0(r), \mathcal{R}_0(r),$ and $\mathcal{P}_0(r)$ for $\eta=0$, $\mu=\ell^2/5$, $m_\tc{c} = \frac{2}{\ell}$.}
    \label{Tmunu0}
\end{figure}

The detector is modelled by the combination of the field $\psi_\tc{c}$, the fluid, and the field $\hat{\phi}_\tc{d}$, which models its internal dynamics. When the detector is in its ground state, the field $\hat{\phi}_\tc{d}$ is then in its vacuum state,  $\ket{0_\tc{d}}$. We assume that the free field $\hat{\phi}$ is in a state denoted by $\hat{\rho}_\phi$. The stress-energy tensor of the detector is then a combination of the stress-energy tensor of the quantum field $\hat{\phi}_\tc{d}$, the classical field $\psi_\tc{c}$, and the perfect fluid. In the vacuum state $\ket{0_\tc{d}}$, we have $\langle: \!\hat{T}_{\mu\nu}^{\phi_\tc{d}}\!:\rangle = \langle:\!\hat{T}_{\mu\nu}^{\phi_\tc{d}\psi_\tc{c}}\!:\rangle = 0$, so that the stress-energy tensor $T_{\mu\nu}^{\ket{0_\tc{d}}} = \bra{0_\tc{d}} \normord{\hat{T}_{\mu\nu}}\ket{0_\tc{d}}$ can be written as
\begin{equation}
    T_{\mu\nu}^{\ket{0_\tc{d}}} = T_{\mu\nu}^{\psi_\tc{c}} + T_{\mu\nu}^{\psi_\tc{c}\text{fluid}} + T^\textrm{fluid}_{\mu\nu},
\end{equation}
with $\rho$ given by Eq.~\eqref{eq:rho}, $P$ given by Eq.~\eqref{eq:P} and $\psi_\tc{c}(\mf x)$ given by Eq.~\eqref{psic}. This results in a $T_{\mu\nu}$ of the form
\begin{equation}
    T_{\mu\nu}^{\ket{0_\tc{d}}} = \uprho_0(r)u_\mu u_\nu + \mathcal{R}_0(r)r_\mu r_\nu + \mathcal{P}_0(r) \Omega_{\mu\nu},\label{eq:Tmunu0}
\end{equation}
where $\mathsf{u} = \mathsf{e}_t$, $\mathsf{r} = \mathsf{e}_r$, $\Omega = \mathsf{e}_\theta\otimes \mathsf{e}_\theta + \mathsf{e}_\phi\otimes \mathsf{e}_\phi$ in the normalized spherical frame $\mf{e}_t = \partial_t$, $\mf{e}_r = \partial_r$, $\mf{e}_\theta = \frac{1}{r}\partial_\theta$, $\mf{e}_\phi = \frac{1}{r \sin\theta}\partial_\phi$. The energy tensor is then diagonal in spherical coordinates, and $\uprho_0(r) = T^{\mu\nu}_0u_\mu u_\nu$ corresponds to the energy density in this frame, $\mathcal{R}_0(r)  = T^{\mu\nu}r_\mu r_\nu$ is the radial pressure, and $\mathcal{P}_0(r) = T^{\mu\nu} \Omega_{\mu\nu}$ is the pressure in the angular directions. Their explicit expressions are
\begin{align}
    \uprho_0(r) &= \frac{2\sech^2(\tfrac{r}{\ell})}{\ell^2}m_\tc{c}^2  + \left(1 - \frac{\mu \sech^2(\tfrac{r}{\ell})}{\ell^2}\right)\rho(r),\\
    \mathcal{R}_0(r) &= - \frac{2\sech^4(\tfrac{r}{\ell})}{\ell^4}+ \left(1 - \frac{\mu \sech^2(\tfrac{r}{\ell})}{\ell^2}\right)P(r),\nonumber\\
    \mathcal{P}_0(r) & = -\frac{2\sech^2(\tfrac{r}{\ell})}{\ell^4} + \left(1 - \frac{\mu \sech^2(\tfrac{r}{\ell})}{\ell^2}\right)P(r) \nonumber,
\end{align}
where $\rho(r)$ and $P(r)$ are the energy density and pressure of the perfect fluid. The plots of $\uprho_0(r), \mathcal{R}_0(r),$ and $\mathcal{P}_0(r)$ can be found in Fig.~\ref{Tmunu0}. We see that the radial and angular pressures assume negative values, but it is simple to check that all energy conditions are verified. Notice that these results are independent of any specific property of the field $\hat{\phi}_\tc{d}$, as they correspond only to the system responsible for generating the trapping potential (which depends on $m_\tc{c}$, $\ell$, and the parameters of the fluid).

\begin{figure}[htb]
    \centering
    \includegraphics[width=11.5cm]{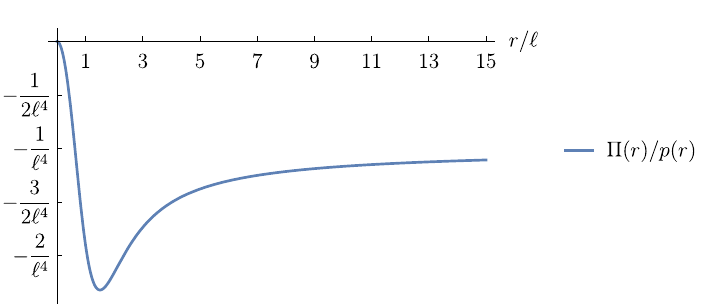}
    \caption{The pressure deviator $\Pi(r)/p(r)$ for $\eta = 0$, $\mu = \ell^2/5$, $m_\tc{c} = \frac{2}{\ell}$.}
    \label{deviator}
\end{figure}

Given the pressures $\mathcal{R}_0(r)$ and $\mathcal{P}_0(r)$ we can calculate the pressure deviator $\Pi(r)$~\cite{landau}. This quantity is defined as the traceless part of the the spatial components of the energy momentum tensor. It measures the difference from the matter content described by $\uprho_0(r)$,  $\mathcal{R}_0(r)$ and $\mathcal{P}_0(r)$ and a perfect fluid modelled by a gas of particles. It can be calculated through the Landau decomposition~\cite{landau}
\begin{equation}
    T_{\mu\nu} = \rho u_\mu u_\nu + (p(r) +\Pi(r))r_\mu r_\nu + (p(r) - \tfrac{1}{2}\Pi(r))\Omega_{\mu\nu},
\end{equation}
where
\begin{align}
    \Pi(r) &= \frac{2}{3}\left(\mathcal{R}_0(r) - \mathcal{P}_0(r)\right) = \frac{4}{3}\frac{\sech^2(\tfrac{r}{\ell})\tanh^2(\tfrac{r}{\ell})}{\ell^4},\nonumber\\
    p(r) &= \frac{\mathcal{R}_0(r) + 2 \mathcal{P}_0(r)}{3}.
\end{align} 
In Fig.~\ref{deviator}, we plot $\Pi(r)/p(r)$ as a function of $r/l$. We observe that the radial pressure approaches zero as $r \to \infty$, indicating that $T_{\mu\nu}$ does not represent a perfect fluid, even at infinity.

\begin{figure}[htb]
    \centering
    \includegraphics[width=10cm]{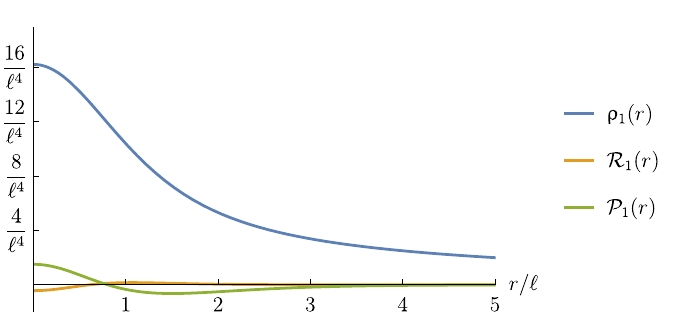}
    \caption{$\uprho_1(r), \mathcal{R}_1(r),$ and $\mathcal{P}_1(r)$ for $\eta=0$, $\mu=\ell^2/5$, $m_\tc{c} = \frac{2}{\ell}$, and $m_\tc{d} = \frac{5}{\ell}$.}
    \label{Tmunu1}
\end{figure}

The detector's first excited state is \mbox{$\ket{1_\tc{d}} = \hat{a}_1^\dagger \ket{0_\tc{d}}$}. To compute the stress-energy tensor when the detector is in this configuration, one could perform the same procedure as that of Section~\ref{sec:example}, using $g(\bm x)$ as the renormalized expected value of $\hat{\phi}_\tc{d}^2$, so that the fluid absorbs the dependence on $g(\bm x)$. We have
\begin{align}
    g(\bm x) &= \langle 1_\tc{d} | \!:\!\hat{\phi}_\tc{d}^2(\mf x)\!:\!| 1_\tc{d} \rangle =\langle 1_\tc{d} | \hat{\phi}_\tc{d}^2(\mf x)| 1_\tc{d} \rangle - \langle 0_\tc{d} | \hat{\phi}_\tc{d}^2(\mf x)| 0_\tc{d} \rangle= \frac{6\csch^4(\tfrac{2r}{\ell})\sinh^6(\tfrac{r}{l})}{\pi r^2 \omega_\tc{d} \ell}.
\end{align}
The energy density and pressure of the perfect fluid are then changed to $\rho_1(r)$ and $P_1(r)$, given explicitly by
\begin{equation}
    P_1(r) = \frac{1}{\ell^2 - \mu \sech^2(\tfrac{r}{\ell})}\int_r^\infty G_1(r),
\end{equation}
where $G_1(r) = G(r) + \Delta G(r)$ with $G(r)$ defined in Eq.~\eqref{eq:Gofr}, and 
\begin{align}
    \Delta G(r) &= -\frac{9 \tanh^3(\tfrac{r}{\ell})\sech^4(\tfrac{r}{\ell})}{4 \pi r^2 \ell^2 \omega_\tc{d}}.
\end{align}
The energy density will then be given by
\begin{equation}
    \rho_1(r) = 3 \eta P_1(r) - \mathcal{L}^{\text{fluid}},
\end{equation}
where
\begin{align}
    \mathcal{L}^\text{fluid} =-\frac{2}{\mu\ell^2} \frac{\tanh(\tfrac{r}{\ell})}{r/\ell} - \frac{\alpha}{2\mu}g(\bm x).
\end{align}
Computing the expected value of the renormalized stress-energy densities $\langle\normord{\hat{T}_{\mu\nu}^{\phi_\tc{d}}}\rangle$ and $\langle\normord{\hat{T}_{\mu\nu}^{\phi_\tc{d}\psi_\tc{c}}}\rangle$, one obtains a stress-energy tensor of the form
\begin{equation}
    T^{\ket{1_\tc{d}}}_{\mu\nu} = \uprho_1(r)u_\mu u_\nu + \mathcal{R}_1(r)r_\mu r_\nu + \mathcal{P}_1(r) \Omega_{\mu\nu},
\end{equation}
where $\uprho_1(r)$, $\mathcal{R}_1(r)$, and $\mathcal{P}_1(r)$ play the same role as $\uprho_0(r)$, $\mathcal{R}_0(r)$, and $\mathcal{P}_0(r)$ in Eq.~\eqref{Tmunu0}. However, their expressions are cumbersome and do not provide any important insight. The plots of these quantities (when the detector is in its excited state) are displayed in Fig.~\ref{Tmunu1}.

\subsubsection*{Final Remarks}
 
  We end this Section by stressing that the model for a general covariant t localized quantum field discussed above is a toy model in the sense that, at least in principle, it does not correspond to any known physical system. The model also has drawbacks, such as the fact that it does not have a total finite energy (due to the $1/r$ decay in $\rho(r)$), as well as the fact that this localized field technically does not allow one to dynamically switch the interaction with an external field. This means that this model is still not capable of fully describing the process of locally probing a quantum field $\hat{\phi}$. However, this is an example of a localized field with a well-defined stress-energy tensor and shows that it is possible to conceive a general covariant localized probe that fulfills many properties that are desired for physical systems.

%

\chapter{Entanglement in Quantum Field Theory}\label{chap:ent}

Entanglement is typically seen as a key feature of quantum theories, and its applications range from interpretations in quantum foundations to a resource in quantum computing. However, its properties are still a topic of current research, even in finite dimensional systems. Specifically, there are very few explicit results about entanglement in quantum field theory.

The goal of this chapter is to discuss entanglement in quantum field theory, focusing on how to quantify it and how to probe it. We will start with a brief review of entanglement in Section~\ref{sec:EntRev}, and we will discuss the challenges and general results related to entanglement in quantum field theory in Section~\ref{sec:entQFThard}. In Section~\ref{sec:modeEntanglement}, we will discuss methods of quantifying entanglement between two finite regions of spacetime in quantum field theory. In Section~\ref{sec:OperationallyAccessingEnt}, we will discuss how one can use the local probes discussed in Chapter~\ref{chap:meas} to attempt to extract entanglement from a quantum field, in a protocol that has become known as entanglement harvesting. In Section~\ref{sec:GeneralEnt}, we will discuss some general results of the protocol and how they relate to some results of Section~\ref{sec:modeEntanglement}.

\section{A Very Brief Review of Entanglement}\label{sec:EntRev}

Entanglement in quantum mechanics is typically introduced through the notion of separability in a tensor-product structure. For instance, consider two subsystems A and B with Hilbert spaces $\mathscr{H}_\tc{a}$ and $\mathscr{H}_\tc{b}$, so that the total Hilbert space of the composite system is $\mathscr{H} = \mathscr{H}_\tc{a} \otimes \mathscr{H}_\tc{b}$. A pure state $\ket{\psi}\in \mathscr{H}$ is then said to be separable if there exist states $\ket{\psi_\tc{a}}\in\mathscr{H}_\tc{a}$ and $\ket{\psi_\tc{b}}\in\mathscr{H}_\tc{b}$ such that
\begin{equation}
    \ket{\psi} = \ket{\psi_\tc{a}}\otimes \ket{\psi_\tc{b}}.
\end{equation}
In this case $\ket{\psi_\tc{a}}$ fully determines the outcomes of any measurement of observables in $\mathscr{H}_\tc{a}$ and $\ket{\psi_\tc{b}}$ determines the outcomes of measurements in $\mathscr{H}_\tc{b}$. Pure separable states are those that do not possess any correlations between systems A and B, in the sense that $\langle \hat{A}\otimes\hat{B}\rangle_\psi = \langle \hat{A}\rangle_{\psi_\tc{a}}\langle\hat{B}\rangle_{\psi_\tc{b}}$. A pure state is said to be entangled if it is not separable, in which case there are no local pure states in $\mathscr{H}_\tc{a}$ and $\mathscr{H}_\tc{b}$ that fully determine the outcome of local measurements in A or B. A typical example of an entangled state is when $\mathscr{H}_\tc{a} = \mathscr{H}_\tc{b} = \mathbb{C}^2$ with bases $\{\ket{0},\ket{1}\}$ and
\begin{equation}\label{eq:EPR}
    \ket{\psi} = \frac{1}{\sqrt{2}}\left(\ket{00} + \ket{11}\right).
\end{equation}

A standard way of quantifying entanglement in pure bipartite systems is by computing how much information is lost about $\ket{\psi}$ when one only considers local measurements in $\mathscr{H}_\tc{a}$ or in $\mathscr{H}_\tc{b}$. For instance, $\ket{\psi}$ in~\eqref{eq:EPR} is a maximally entangled state, as $\r_\tc{a} = \tr_\tc{a}(\ket{\psi}\!\!\bra{\psi}) = \r_\tc{b} = \tr_\tc{b}(\ket{\psi}\!\!\bra{\psi}) = \frac{1}{2}\openone$. The fact that the reduced states of $\ket{\psi}$ in $\mathscr{H}_\tc{a}$ and $\mathscr{H}_\tc{b}$ are maximally mixed indicates that the partial states contain no information about $\ket{\psi}$.

This lost information can be computed through the entanglement entropy
\begin{equation}
    S_E(\ket{\psi}\!\!\bra{\psi}) \coloneqq S(\hat{\rho}_\tc{a}) - S(\ket{\psi}\!\!\bra{\psi}) = S(\hat{\rho}_\tc{a}) = S(\hat{\rho}_\tc{b}),
\end{equation}
where $\hat{\rho}_\tc{a} =  \tr_\tc{b}(\ket{\psi}\!\!\bra{\psi})$, $\hat{\rho}_\tc{b} =  \tr_\tc{a}(\ket{\psi}\!\!\bra{\psi})$ are the partial states of $\ket{\psi}$ with respect to A and B, and $S$ denotes the von Neuman entropy
\begin{equation}
    S(\hat{\rho}) = - \tr(\hat{\rho}\log\hat{\rho}),
\end{equation}
which can be intuitively seen as a quantifier of the uncertainty associated with $\hat{\rho}$, being $0$ for pure states. The entanglement entropy is then $0$ for any separable pure state, indicating that $\ket{\psi}$ is entangled whenever $S_E(\ket{\psi}\!\!\bra{\psi})\neq 0$, and the maximum value of the entanglement entropy is $\log(n)$, where $n = \min(\text{dim}(\mathscr{H}_\tc{a}),\text{dim}(\mathscr{H}_\tc{b}))$. In particular, if $\text{dim}(\mathscr{H}_\tc{a})=\text{dim}(\mathscr{H}_\tc{b})$, a pure state that maximizes the entanglement entropy is called a maximally entangled state (e.g. the state $\ket{\psi}$ in~\eqref{eq:EPR}).

The discussion of entanglement becomes slightly more involved when one considers mixed states. This is due to the fact that mixed states might be composed of classical mixtures of unentangled states, which can contain classical correlations between the partial states in A and B. Overall, a bipartite mixed state is separable if it can be written as
\begin{equation}
    \hat{\rho} = \sum_i p_i \r_\tc{a}^i \otimes \r_\tc{b}^i,
\end{equation}
where $\r_\tc{a}^i$ and $\r_\tc{b}^i$ are states in $\mathscr{H}_\tc{a}$ and $\mathscr{H}_\tc{b}$. A non-separable state is then said to be entangled. In this more general case, $S_E(\hat{\rho})$ is not enough to quantify whether $\r$ is entangled or not, as the partial states $\r_\tc{a}$ and $\r_\tc{b}$ might not be enough to fully describe the state $\r$ even if it is separable, due to classical correlations. 

Nevertheless, one can still use some notion of the entanglement entropy to quantify the entanglement of mixed states through the entanglement of formation. The entanglement of formation is defined as
\begin{equation}\label{eq:formation}
    E_F(\hat{\rho}) = \min_{p_i,\ket{\psi_i}}\sum_i p_i S_E(\ket{\psi_i}\!\!\bra{\psi_i}),
\end{equation}
where the minimum is taken over all $p_i\geq 0$ and pure states $\ket{\psi_i}$ such that
\begin{equation}
    \hat{\rho} = \sum_i p_i \ket{\psi_i}\!\!\bra{\psi_i}.
\end{equation}
In other words, the entanglement of formation corresponds to the minimum entanglement entropy of decompositions of $\hat{\rho}$ in terms of pure states. It gives the minimum amount of entanglement that the pure states must have to compose $\hat{\rho}$. Unfortunately, the optimization of Eq.~\eqref{eq:formation} makes the entanglement of formation very challenging to compute in practice, even in low dimensional quantum systems.

Another way of quantifying entanglement in mixed states is the distillable entanglement, which codifies how many maximally entangled pairs can be extracted from $n$ copies of $\hat{\rho}$. Essentially,  one can distinguish between classical and quantum correlations by classifying correlations that can and cannot increase through local operations and classical communication (LOCC)~\cite{monotones,LOCC}. The entanglement between two systems then cannot increase under LOCC, in the sense that no sequence of local operations acting on the individual systems can increase the entanglement between them, even if these operations depend on classical parameters related to the other system. In this context, the distillable entanglement of $\hat{\rho}$ is defined as the asymptotic rate of maximally entangled states that can be extracted from $n$ copies of $\hat{\rho}$. Needless to say, the distillable entanglement is also not computationally friendly. Interestingly, one can show that the distillable entanglement is upper bounded by the entanglement of formation, implying that not all bipartite entanglement in a state can be distilled. 

One practical way of quantifying entanglement is through the negativity. Essentially, the negativity relies on the fact that the operation of transposition is positive, but not completely positive. In other words, the partial transposition operation $\r_\tc{a}\otimes \r_\tc{b}\mapsto \r_\tc{a}\otimes\r_\tc{b}^\intercal$ (and extended by linearity) can map a density operator $\r$ to an operator that is not positive. We denote the partial transpose with respect to B by $\r^{\intercal_\tc{b}}$. The fact that partial transposition always maps separable states into positive operators implies that states with negative partial transpose must be entangled. This is known as Peres' criterion, after~\cite{Peres,HorodeckiPeres}. Moreover, one can use the negative eigenvalues of the partial transpose to quantify entanglement. Denoting by $\sigma_\tc{b}^-(\r)$, the set of negative eigenvalues of $\r^{\intercal_\tc{b}}$, the negativity of a bipartite density operator $\r$ is defined as~\cite{negativityOG}
\begin{equation}\label{eq:negativity}
    \mathcal{N}(\r) = \sum_{\lambda\in \sigma_\tc{b}^-(\r)}\!\! |\lambda| \,\,= \frac{||\r^{\intercal_\tc{b}}||_{1} - 1}{2},
\end{equation}
where $|| \, \cdot \, ||_\tc{1}$ denotes the trace norm $||\hat{A}||_1 = \text{tr}\sqrt{\hat{A}^\dagger\hat{A}}$. It is also convenient to define the logarithmic negativity
\begin{equation}
    E_\mathcal{N} = \log\left(2\mathcal{N} + 1\right) = \log(||\r^\intercal_\tc{b}||_1),
\end{equation}
which also yields the maximum of $\log(n)$ for maximally entangled states when $\text{dim}(\mathscr{H}_\tc{a})=\text{dim}(\mathscr{H}_\tc{b}) = n$. It can then be shown that the negativity does not increase under LOCC~\cite{Plenio} and that the logarithmic negativity is an upper bound to the distillable entanglement. However, the negativity is not always a faithful measure of entanglement, in the sense that not all entangled states have zero negativity. Indeed, in Hilbert spaces $\mathscr{H}_\tc{a}$ and $\mathscr{H}_\tc{b}$ with dimensions greater than $2$, not all entangled states lead to non-positive partial transposes. However, in the particular case where $\mathscr{H}_\tc{b} = \mathscr{H}_\tc{a} = \mathbb{C}^2$ (or whenever $\text{dim}(\mathscr{H}_\tc{a})+ \text{dim}(\mathscr{H}_\tc{b}) \leq 5$), the negativity is a faithful entanglement measure. 

If even in bipartite systems quantifying entanglement can become tricky, it should come as no surprise that entanglement in systems with more than two parties is even more challenging. Indeed, systems with more than two parties can have multiple different kinds of entanglement between the different parties~\cite{twoQubitInequivEnt,twoQubitInequivEntCabello}. This is perhaps better exemplified with the $\ket{GHZ}$ and $\ket{W}$ states in a three-qubit system:
\begin{align}\label{eq:GHZW}
    \ket{W} = \frac{1}{\sqrt{3}}(\ket{100} + \ket{010} + \ket{001}),\quad\quad\quad
    \ket{GHZ} = \frac{1}{\sqrt{2}}(\ket{000} + \ket{111}).
\end{align}
The $\ket{W}$ state essentially corresponds to a case where each pair of qubits is equally entangled with each other. Indeed, {$\mathcal{N}(\hat{\rho}^W_\tc{ab}) = \mathcal{N}(\hat{\rho}^W_\tc{bc}) = \mathcal{N}(\hat{\rho}^W_\tc{ac}) = \frac{1}{6}(\sqrt{5}-1)$, and $\ket{W}$ is the tripartite state that maximizes the negativity between all pairs.} On the other hand, $\ket{GHZ}$ is such that neither of the reduced bipartite systems is individually entangled with each other, as $\r^{GHZ}_{\tc{ab}} = \r^{GHZ}_{\tc{bc}} = \r^{GHZ}_{\tc{ac}} = \frac{1}{2}(\ket{00}\!\!\bra{00}+\ket{11}\!\!\bra{11})$ are separable. In a way, the state $\ket{GHZ}$ contains entanglement that involves all parties simultaneously, and if any qubit is lost, all information about the state is lost. However, performing a local measurement in one of the parties associated to the eigenspaces of $\hat{\sigma}_x$ in the state $\ket{GHZ}$ yields maximally entangled states for the remaining parties---this fact will be important later on when we discuss recent progress regarding entanglement in quantum field theory. 

As the number of parties becomes larger, new types of entanglement arise, making the classification of entanglement in $n$-partite systems still an active topic of research~\cite{entQuantFiniteDim2013,EntQuantNature2016,MarceloTripartite2019}.

\section{Challenges of Quantifying Entanglement in Quantum Field Theory}\label{sec:entQFThard}

Given the challenges of quantifying entanglement even in finite dimensional Hilbert spaces with a finite number of parties, it is to be expected that the quantifying entanglement in the context of quantum field theory makes the situation even worse. A way of seeing why this is the case is by thinking of the analogy between a massive real scalar quantum field and a lattice of interacting quantum harmonic oscillators~\cite{areaLaw1993}, where one assigns one oscillator degree of freedom to each point of space---quantum field theory can be thought of as a limit of an infinite-partite system with infinite dimensional Hilbert spaces. Overall, the worst-case scenario for entanglement analysis.

In many instances, it is not suitable to think of quantum fields in terms of lattices of harmonic oscillators attached to each point of space; instead, quantum field theory is an association of algebras of observables to regions of spacetime. However, in this context, the standard tools for studying entanglement do not apply. In fact, in this case, it is not even clear what one means by quantifying entanglement in a quantum field theory, as one does not have a tensor product decomposition, or even different states associated with different parties. Although we only have one state and no tensor product decomposition, we can still find methods to effectively assign independent quantum degrees of freedom to spacelike separated regions of spacetime, and thus compute the entanglement between these regions. For the purposes of this chapter,  ``quantifying entanglement in quantum field theory'' will refer to quantifying entanglement in \textit{one} state between two regions of spacetime. We start by discussing some known results about entanglement of the Minkowski vacuum in two causally complementary regions.


\subsubsection*{Vacuum Entanglement between Complementary Regions}

Although the standard techniques of quantification of entanglement cannot be straightforwardly applied in quantum field theory, there are strong arguments that states in quantum field theory are highly entangled. In particular, the Minkowski vacuum is argued to contain an infinite amount of entanglement, as we will discuss here. 

A simple argument for why the Minkowski vacuum must contain entanglement between complementary regions of spacetime is the fact that its correlation function $\omega_0(\hat{\phi}(f)\hat{\phi}(g))$ is non-degenerate. This implies that degrees of freedom associated with any two spacetime regions are always correlated. While correlations cannot be associated with entanglement in general (they might be classical correlations), we note that the Minkowski vacuum is a pure state. This implies that any correlations between independent degrees of freedom that fully describe the state are not due to classical mixtures, and must necessarily be quantum in nature. In particular, this implies that the vacuum must contain entanglement between a region $\mathcal{O}$ and its causal complement $\mathcal{O}' = \M\setminus(J^+(\mathcal{O})\cup J^-(\mathcal{O}))$. While this informal argument can be used as evidence that the vacuum is entangled, we can refine it by analyzing the Reeh-Schlieder theorem:

\noindent \textbf{Reeh-Schlieder Theorem~\cite{ReehSchli,HaagOG}:}\textit{Given a causally convex bounded set $\mathcal{O}$, let $\uppi_0$ be the GNS representation of the Minkowski vacuum $\ket{0}$. Then $\ket{0}$ is both cyclic and separating for $\uppi_0(\mathcal{A}(\mathcal{O}))$.}

In the theorem above, we say that a vector $\ket{\Psi}$ in a Hilbert space $\mathscr{H}$ is cyclic with respect to a subalgebra $\mathcal{A}$ of linear operators on $\mathscr{H}$ if $\hat{A} \ket{\Psi} = 0 \Rightarrow \hat{A} = 0$ for all $\hat{A} \in \mathcal{A}$. We say that $\ket{\psi}$ is separating for $\mathcal{A}$ if the set $\mathcal{A}\ket{\Psi}$ is dense in $\mathscr{H}$. The Reeh-Schlieder theorem then implies that given any bounded region $\mathcal{O}$, 1) there are no \textit{local} operators in $\mathcal{A}(\mathcal{O})$ that annihilate the vacuum, and 2) that every state in $\mathcal{F}(\mathscr{H}_0)$ can effectively be produced by applying \textit{local} operators in $\mathcal{A}(\mathcal{O})$ to the vacuum.

The Reeh-Schlieder theorem then implies that although the local algebra $\mathcal{A}(\mathcal{O})$ only captures a subset of the degrees of freedom of the quantum field theory, its action on the vacuum is sufficient to effectively produce any state. Noticing that states in quantum field theories are globally defined, this implies that there are localized operators in $\mathcal{A}(\mathcal{O})$ that can produce field excitations in regions arbitrarily far away from $\mathcal{O}$ when applied to the vacuum. At first glance, this fact might look incompatible with relativistic causality, but it is important to stress that general operators in $\mathcal{A}(\mathcal{O})$ cannot simply be applied to the vacuum. Indeed, only unitary operators could potentially arise from well-defined operations in the algebra whose action in $\ket{\Omega}$ produces another pure state, and even so, not all unitary operators correspond to physical processes, as discussed in Section~\ref{sec:meas}. Instead, one should see the cyclicity of the vacuum as evidence that it contains entanglement between any two regions.

To understand why one can argue that a cyclic and separating state is entangled, it is instructive to consider what these concepts entail in the case of a finite dimensional bipartite system. Let $\mathscr{H}_1$ and $\mathscr{H}_2$ be two Hilbert spaces of dimension $n$ and $\mathscr{H} = \mathscr{H}_1 \otimes \mathscr{H}_2$. In this context, a vector $\ket{\psi}\in\mathscr{H}$ is cyclic for the algebra of linear operators of the form $\hat{A}_1\otimes \openone_2$ if and only if $\ket{\psi}$ has maximal Schmidt rank, in the sense that there exist bases $\{\ket{i_1}\}_{i=1}^n$ and $\{\ket{i_2}\}_{i=1}^n$ of $\mathscr{H}_1$ and $\mathscr{H}_2$ such that
\begin{equation}\label{eq:Schmidt}
    \ket{\psi} = \sum_{i = 1}^n \psi_i\ket{i_1i_2},
\end{equation}
with all $\psi_i\neq0$. It is then clear that the action of all linear operators acting on $\mathscr{H}_1$, $\hat{A}_1 = \sum_{jk} A_{jk} \ket{j_1}\!\!\bra{k_1}$, can generate any state of $\mathscr{H}$ by acting on $\ket{\psi}$, as
\begin{equation}
    \hat{A}_1\ket{\psi} = \sum_{ij} A_{ji}\psi_i\ket{j_1i_2}.
\end{equation}
The equation above also shows that, in this finite dimensional case, any cyclic state is also separating, as $\hat{A}\ket{\psi} = 0$ implies $A_{ji}\psi_i = 0$ for all $i,j$, leading to $\hat{A} = 0$ from the fact that $\psi_i\neq 0$. The decomposition of Eq.~\eqref{eq:Schmidt} also automatically implies that $\ket{\psi}$ has non-zero entanglement entropy. This discussion also generalizes to infinite dimensional separable Hilbert spaces (with bounded operators $\hat{A}_1$).  

In the more general context of quantum field theory, where one cannot decompose the Fock state as a tensor product corresponding to different regions of spacetime, the statement that the vacuum is cyclic and separating can then be seen as a generalization of the concept of vectors of maximal Schmidt rank, indicating that the Minkowski vacuum contains a high amount of entanglement between a bounded region and its causal complement. Indeed, there are good arguments for why the vacuum entanglement between a region and its causal complement is not only high, but infinite. For instance, in~\cite{universalEmbezzlers} it was shown that by considering arbitrary states $\ket{\varphi_\tc{a}}$ and $\ket{\varphi_\tc{b}}$ in any two Hilbert spaces $\mathscr{H}_\tc{a}$ and $\mathscr{H}_\tc{b}$, an arbitrary target state $\ket{\Psi}\in \mathscr{H}_\tc{a}\otimes\mathscr{H}_\tc{b}$, and $\epsilon>0$, it is possible to find unitaries $\hat{U}_\tc{a}\in \mathcal{A}(\mathcal{O})\otimes\mathcal{B}(\mathscr{H}_\tc{a})$ and  $\hat{U}_\tc{b}\in \mathcal{A}(\mathcal{O}')\otimes\mathcal{B}(\mathscr{H}_\tc{b})$, such that
\begin{equation}
    \left|\left|\hat{U}_\tc{a}\hat{U}_\tc{b}(\ket{0}\otimes \ket{\varphi_\tc{a}}\otimes \ket{\varphi_\tc{b}}) - \ket{0}\otimes \ket{\Psi}\right|\right| < \epsilon.
\end{equation}
In particular, this result implies that it is possible to find local unitaries that produce any entangled state in $\mathscr{H}_\tc{a}\otimes\mathscr{H}_\tc{b}$ to arbitrary precision by coupling system A to the vacuum in region $\mathcal{O}$ and system B to the vacuum in $\mathcal{O}'$. Unfortunately, the results of~\cite{universalEmbezzlers} are not constructive, so it does provide an explicit form for the unitaries $\hat{U}_\tc{a}$ and $\hat{U}_\tc{b}$.

\subsubsection*{Quantifying Entanglement Between Two {Non-Complementary} Regions}

The entanglement between a region and its complement is, in many ways, physically inaccessible: when probing any features of quantum fields, we are always restricted to local measurements at finite energies. While the entanglement between complementary regions is infinite, analyses of entanglement entropy between a region and its complement in lattice field theories show that most of the entanglement is localized at the boundary of the regions and that the entanglement entropy is given by $A/4\epsilon^2$, where $A$ is the area of contact between the regions\footnote{Specifically, $A$ is the area of the spatial region $\Sigma_{\mathcal{O}}$ (contained in a Cauchy surface) such that $D(\Sigma_{\mathcal{O}})$ is the smallest causal diamond that contains the causal hull of $\mathcal{O}$.} and $\epsilon$ is the lattice separation, corresponding to a UV cutoff~\cite{areaLaw1993,AreaLawsCirac2008,areaLawReview2010,Kelly}. This quantity diverges as $\epsilon \to 0^+$, confirming that the entanglement between a region and its causal complement is infinite. However, it also gives very little useful information about the physically accessible entanglement between two regions---one cannot realistically access the degrees of freedom of a quantum field in sharp regions, and in this example, the infinitely thin overlap between the regions is the relevant region where the entanglement is localized.

For the remainder of this Chapter, we will then focus on quantifying entanglement between two non-complementary regions, when there is a non-zero distance between them. Not only will this get rid of the ``infinite entanglement'' between the regions, but it will also yield entanglement that is, at least in principle, physically accessible. However, this introduces a series of challenges in the quantification of entanglement. Indeed, while there is plenty of evidence that the Minkowski vacuum contains entanglement between a spacetime region and its causal complement, these arguments do not immediately imply that the vacuum contains entanglement between two non-complementary regions. In this case the Reeh-Schlieder does not provide much useful information, as being cyclic and separating only evidences entanglement between bipartitions. The theorem of~\cite{universalEmbezzlers} also does not apply unless $\mathcal{O}'$ is the causal complement of $\mathcal{O}$. Finally, the simple argument that the vacuum contains correlations between any two regions also does not imply that two finite regions are entangled with each other, given that the state that represents the degrees of freedom of each region cannot be assumed to be pure\footnote{As a matter of fact, the reduced state to two regions cannot even be formally defined due to the type III nature of local algebras.}.

In summary, given two causally disconnected regions $\mathcal{O}_\tc{a}$ and $\mathcal{O}_\tc{b}$ with respective causal complements $\mathcal{O}_\tc{a}'$ and $\mathcal{O}_\tc{b}'$, the arguments that apply to complementary regions only ensure that there is entanglement between degrees of freedom in $\mathcal{O}_\tc{a}$ and $\mathcal{O}_\tc{a}'$ (resp. B). Even though $\mathcal{O}_\tc{b}\subset\mathcal{O}_\tc{a}'$, it is not a guarantee that the field degrees of freedom of $\mathcal{O}_\tc{a}$ and $\mathcal{O}_\tc{b}$ are entangled. This can be seen by noticing that the vacuum state of Minkowski spacetime is a pure state that can be fully described by its degrees of freedom in $\mathcal{O}_\tc{a}$, $\mathcal{O}_\tc{b}$, and $\mathcal{O}_\text{ext}=(\mathcal{O}_\tc{a}\cup\mathcal{O}_\tc{b})'$. This means that the problem of quantifying the entanglement between the two regions $\mathcal{O}_\tc{a}$ and $\mathcal{O}_\tc{b}$ can be intuitively thought in terms of quantifying tripartite entanglement of a pure state, or, in terms of quantifying bipartite entanglement with mixed states. Both of these cases are much more challenging than the case of pure bipartite states.

Throughout the remainder of the chapter, we will focus on two ways of quantifying entanglement between two non-complementary regions in quantum field theory. The first approach will be to identify field degrees of freedom localized in two regions and to attempt to quantify the entanglement between them. The second approach will be to use localized probes to attempt to extract entanglement from a quantum field, inferring entanglement in the field from the entanglement acquired by the probes.

\section{Field Entanglement between Localized Modes}\label{sec:modeEntanglement}

In this Section, we will discuss an approach for quantifying entanglement between two finite regions of a quantum field theory by analyzing entanglement between degrees of freedom localized in each region. Specifically, we will consider degrees of freedom associated with canonical pairs in the respective regions, allowing us to use tools of Gaussian quantum mechanics for computing the entanglement between local field modes explicitly. This section is based on the results of~\cite{patriciaAndI}, but we will also summarize the methods discussed in~\cite{KlcoUVIR,KlcoEntStrQFTI,KlcoEntStrQFTII,KlcoEntAllDist}, which have yielded important results regarding entanglement in quantum field theory.

\subsubsection*{Local Degrees of Freedom Associated to Two Regions in QFT}

One way of quantifying entanglement between two spacetime regions $\mathcal{O}_\tc{a}$ and $\mathcal{O}_\tc{b}$ is by utilizing the phase space quantum mechanics techniques described in Section~\ref{sec:QFT}. Let us consider the explicit case where the regions $\mathcal{O}_\tc{a}$ and $\mathcal{O}_\tc{b}$ are causal diamonds\footnote{In this case, If $\tilde{\mathcal{O}}_\tc{a}$ and $\tilde{\mathcal{O}}_\tc{b}$ are any sets contained in $\mathcal{O}_\tc{a}$ and  $\mathcal{O}_\tc{b}$, respectively, their associated algebras can be fully represented in $\mathcal{A}(\mathcal{O}_\tc{a})$ and $\mathcal{A}(\mathcal{O}_\tc{b})$ by \textbf{A2}.}. To ensure that the degrees of freedom in the regions $\mathcal{O}_\tc{a}$ and $\mathcal{O}_\tc{b}$ are independent, we assume that the two causal diamonds are spacelike separated. 

One can then find a Cauchy surface $\Sigma$ that overlaps $\mathcal{O}_\tc{a}$ and $\mathcal{O}_\tc{b}$, such that $\Sigma_\tc{a}\subset\Sigma$ is a Cauchy surface for $\mathcal{O}_\tc{a}$ and $\Sigma_\tc{b}\subset\Sigma$ is a Cauchy surface for $\mathcal{O}_\tc{b}$. Canonical modes in each of these regions can be defined by considering sets of functions $F_{\tc{a},i}$, $G_{\tc{a},i}$ in $C_0^\infty(\Sigma_\tc{a})$ and $F_{\tc{b},i}$, $G_{\tc{b},i}$ in $C_0^\infty(\Sigma_\tc{b})$ such that
\begin{equation}
    \int \dd\Sigma F_{\tc{a},i}(\bm x) G_{\tc{a},j}(\bm x) = \delta_{ij}, \quad 
    \int \dd\Sigma F_{\tc{b},i}(\bm x) G_{\tc{b},j}(\bm x) = \delta_{ij},
\end{equation}
giving rise to the independent canonical pairs $(\hat{\Phi}(F_{\tc{a},i}),\hat{\Pi}(G_{\tc{a},i}))$ and $(\hat{\Phi}(F_{\tc{b},i}), \hat{\Pi}(G_{\tc{b},i}))$. 

If the sets of functions are maximal linearly independent sets\footnote{By a maximal linear independent set we mean that the only $F\in C_0^\infty(\Sigma_\tc{a})$ satisfying $F_{\tc{a},i}(F) = G_{\tc{a},i}(F) = 0 \,\,\,\,\,\forall\, i$ is $F(\bm x) = 0$, and the analogous statement for B.} in $C_0^\infty(\Sigma_\tc{a})$ and $C_0^\infty(\Sigma_\tc{b})$, it is then possible to fully represent the degrees of freedom of the field in the regions $\mathcal{O}_\tc{a}$ and $\mathcal{O}_\tc{b}$ in terms of the field and momentum operators smeared against the functions $F_{\tc{a},i}$, $G_{\tc{a},i}$, $F_{\tc{b},i}$, $G_{\tc{b},i}$. In this case, the sets of modes $(\hat{\Phi}(F_{\tc{a},i})$, $\hat{\Pi}(G_{\tc{a},i}))$ fully encompass all degrees of freedom of $\mathcal{O}_\tc{a}$ (respectively, for B). One way of finding such a maximal linearly independent set would be by considering $F_{\tc{a},i} = G_{\tc{a},i}$ as an orthonormal basis of smooth compactly supported functions in the \textit{real} Hilbert space $L^2(\Sigma_\tc{a})$, defining the complete set of modes $(\hat{\Phi}(F_{\tc{a},i}), \hat{\Pi}(F_{\tc{a},i}))$. With the analogous procedure for B, we would obtain the complete set of field modes $(\hat{\Phi}(F_{\tc{b},i}), \hat{\Pi}(F_{\tc{b},i}))$, which are independent of the modes in A. Although this approach would fully represent the degrees of freedom of a quantum field, the standard techniques of Gaussian quantum mechanics would not be applicable in a straightforward manner to this infinite dimensional case.

We can instead look for a finite, but sufficiently large, number of modes $(\hat{\Phi}(F_{\tc{a},i}), \hat{\Pi}(F_{\tc{a},i}))$ and $(\hat{\Phi}(F_{\tc{b},i}), \hat{\Pi}(F_{\tc{b},i}))$, with $i=1,...N$ in each region. This would allow us to describe a quasifree state $\omega$ within the familiar domain of finite dimensional Gaussian quantum mechanics, where there are simple and effective techniques to quantify entanglement between independent degrees of freedom.

Having $2N + 2N = 4N$ degrees of freedom associated to the collection of modes in A and B, we split the classical phase space where they can be represented as a direct sum $\mathbb{R}^{4N} = \mathbb{R}^{2N}\oplus \mathbb{R}^{2N}$ with the $4N$ dimensional symplectic form $\bm \Omega$ (analogous to Eq.~\eqref{eq:OmegaSymplecticMatrix}), with the first factor associated associated to operators defined in the region $\mathcal{O}_\tc{a}$ and the second factor associated to operators in $\mathcal{O}_\tc{b}$, and canonical coordinates $\xi^\alpha = (q_\tc{a}^1,p_{\tc{a}}^1,...,q_\tc{a}^N,p_{\tc{a}}^N,q_\tc{b}^1,p_{\tc{b}}^1,...,q_\tc{b}^N,p_{\tc{b}}^N)$. The association $\hat{\Xi}(\bm \xi) = \Omega_{\alpha\beta}\xi^\beta \hat{\Xi}^\alpha$ then creates the correspondence 
\begin{align}
    q_\tc{a}^i&\mapsto \hat{\Phi}(F_{\tc{a},i}), \quad\quad q_\tc{b}^i\mapsto \hat{\Phi}(F_{\tc{b},i}),\\
    p_\tc{a}^i&\mapsto \hat{\Pi}(F_{\tc{a},i}),  \quad\quad p_\tc{b}^i\mapsto \hat{\Pi}(F_{\tc{a},i}),
\end{align}
representing the canonical modes in a phase space.

The covariance matrix $\bm \sigma$ of a quasifree state $\omega$ then factors as
\begin{equation}
    \bm \sigma  = \begin{pmatrix}\bm \sigma_\tc{a} & \bm \eta \\ \bm \eta^\intercal & \bm \sigma_\tc{b}\end{pmatrix},
\end{equation}
where $\bm \sigma_\tc{a}$ and $\bm \sigma_\tc{b}$ are the covariance matrix associated to the modes in A and B individually, and $\bm \eta$ is a matrix of correlations, defined by the blocks
\begin{equation}
    \eta^{ij} = \begin{pmatrix}\langle \{\hat{\Phi}(F_{\tc{a},i}),\hat{\Phi}(F_{\tc{b},j})\}\rangle_\omega & \langle \{\hat{\Phi}(F_{\tc{a},i}),\hat{\Pi}(F_{\tc{b},j})\}\rangle_\omega\\
    \langle \{\hat{\Pi}(F_{\tc{a},i}),\hat{\Phi}(F_{\tc{b},j})\}\rangle_\omega & \langle \{\hat{\Pi}(F_{\tc{a},i)},\hat{\Pi}(F_{\tc{b},j})\}\rangle_\omega\end{pmatrix}.
\end{equation}

Regarding the entanglement analysis, the simplification brought by the restriction to a finite number of modes gives a rather simple quantification of bipartite entanglement. It turns out that {\color{black} for Gaussian bisymmetric states\footnote{These are bipartite Gaussian states that are invariant under internal permutations of modes within either side of the partition.} 
the Peres-Horodecki separability criterion is not only a sufficient but also a necessary condition~\cite{Simon2000, Serafini2005}. This means that in these Gaussian scenarios, separable states are exactly those with a positive partial transpose (PPT), and therefore, the negativity and the logarithmic negativity are \textit{faithful} entanglement monotones. 

We can compute the negativity associated to the representation of the state $\omega$ by defining the covariance matrix of the partial transpose with respect to B, $\bm\sigma^\Gamma$, as the result of reversing the sign of the momenta associated with system B: 
\begin{equation}
\bm\sigma^\Gamma = (\openone_{\textsc a} \oplus \bm T_\textsc{b}) \bm\sigma (\openone_{\textsc a} \oplus \bm T_\textsc{b}), 
\end{equation}
where
\begin{equation}
\bm T_\textsc{b} = \bigoplus_{j=1}^{N}\begin{pmatrix}
    1 & 0 \\
    0 & -1
\end{pmatrix}.
\end{equation}}
The negativity is then entirely determined by the symplectic spectrum\footnote{Notice that the symplectic eigenvalue is composed of pairs $\{\pm \nu_j\}_j$. When we refer to the symplectic spectrum, we will mean the positive symplectic spectrum, consisting only of $\{\nu_j\}_j$.} of ${\bm\sigma}^\Gamma$, $\{{\nu}_1^\Gamma,\hdots,{\nu}_{N}^\Gamma\}$:
\begin{equation}
    E_{\mathcal{N}} = \sum_{j=1}^{N} \max\big(0,-\log_2\nu_j^\Gamma\big).\label{eq:logNeg}
\end{equation}
Moreover, the symplectic spectrum of ${\bm\sigma}^\Gamma$, is simply given by the absolute value of the eigenvalues of $\ii \bm \Omega \bm \sigma^\Gamma$, and can therefore be computed in a straightforward manner. From Eq.~\eqref{eq:logNeg} we see that the modes in A and B will be entangled if and only if $\bm{\sigma}^\Gamma$ has at least one symplectic eigenvalue strictly below 1, i.e., if and only if the condition $\bm\sigma^\Gamma \geq \ii \bm\Omega^{-1}$ is violated, in which case $\bm \sigma^\Gamma$ would not define a state.

Having the tools to quantify entanglement in quantum field theory, we can move forward with applying these techniques to specific examples. As the tools discussed here suggest, there are no general expressions for the entanglement between two localized regions in quantum field theory, and most of what is known applies only to specific examples, focusing mostly on the Minkowski vacuum of a real scalar quantum field. 


\definecolor{LightBlue}{RGB}{100,200,230}


\subsubsection*{Entanglement between two Spacetime Regions}

We will now focus on quantifying entanglement of the Minkowski vacuum of a real scalar field, $\omega_0\leftrightarrow \ket{0}$. The setups that will be discussed here were first used in this context in~\cite{ubiquitous}, where the authors studied entanglement between two modes $(\hat{\Phi}(F_\tc{a}),\hat{\Pi}(F_\tc{a}))$ and $(\hat{\Phi}(F_\tc{b}),\hat{\Pi}(F_\tc{b}))$, defined by non-overlapping spherically symmetric spatial smearing functions $F_\tc{a}$ and $F_\tc{b}$ (among other more general examples). The functions $F_\tc{a}(\bm x)$ and $F_\tc{b}(\bm x)$ were defined along a spatial slice $t = 0$ in inertial coordinates $(t,\bm x)$. The functions $F_\tc{a}$ and $F_\tc{b}$ considered were spherically symmetric\footnote{The studies in~\cite{ubiquitous} were conducted in spacetimes of different dimensions, but we will focus on the case of $3+1$ Minkowski spacetime.} with no overlapping support, prescribed as
\begin{equation}\label{eq:FAFB}
    F_\tc{a}(\bm x) = F^{(\delta)}(\bm x - \bm x_\tc{a}), \quad \quad F_\tc{b}(\bm x) = F^{(\delta)}(\bm x - \bm x_\tc{b}),
\end{equation}
with
\begin{equation}\label{Eq: smearing family}
    F^{(\delta)}(\bm x) = N_\delta \left(  1 - \tfrac{|\bm x|^2}{R^2} \right)^{\!\delta} \, \theta\left( 1- |\bm x|/R \right),
\end{equation}
where $|\bm x_\tc{a} - \bm x_{\tc{b}}|>2R$, $\delta\geq 1$ and
\begin{equation}\label{Eq: A_delta}
N_\delta = \frac{1}{\pi^{\frac{3}{4}} R^\frac{3}{2}}\sqrt{\frac{\Gamma(\tfrac{5}{2}+2\delta)}{\Gamma(1+2\delta)}},
\end{equation}
ensuring normalization according to~\eqref{eq:normFiFj}. The functions $F_\tc{a}(\bm x)$ and $F_\tc{b}(\bm x)$ are then compactly supported within the sphere of radius $R$ centered at $\bm x_\tc{a}$ and $\bm x_\tc{b}$, respectively. Although not of class $C^\infty$, these functions possess $\lfloor \delta\rfloor$ derivatives, peaking at $\bm x = \bm x_\tc{a}$ and $\bm x = \bm x_\tc{b}$ respectively, and decaying to $0$ at the boundary of the spheres where they are supported. The functions $F^{(\delta)}(\bm x)$ are particularly convenient for the Gaussian quantum mechanics approach, as the expected values of the smeared field operators can be found in closed form~\cite{ubiquitous}. 

Although there is reasonable evidence that the Minkowski vacuum is a highly entangled state, the studies of~\cite{ubiquitous} found that no such modes $(\hat{\Phi}(F_\tc{a}),\hat{\Pi}(F_\tc{a}))$, $(\hat{\Phi}(F_\tc{b}),\hat{\Pi}(F_\tc{b}))$ are ever entangled in $3+1$ dimensional Minkowski spacetime. In other words, although the vacuum is a highly entangled state, entanglement in quantum field theory is not as ubiquitous as one might have expected.

\begin{figure}[h!]
    \centering
    \includegraphics[width=14cm]{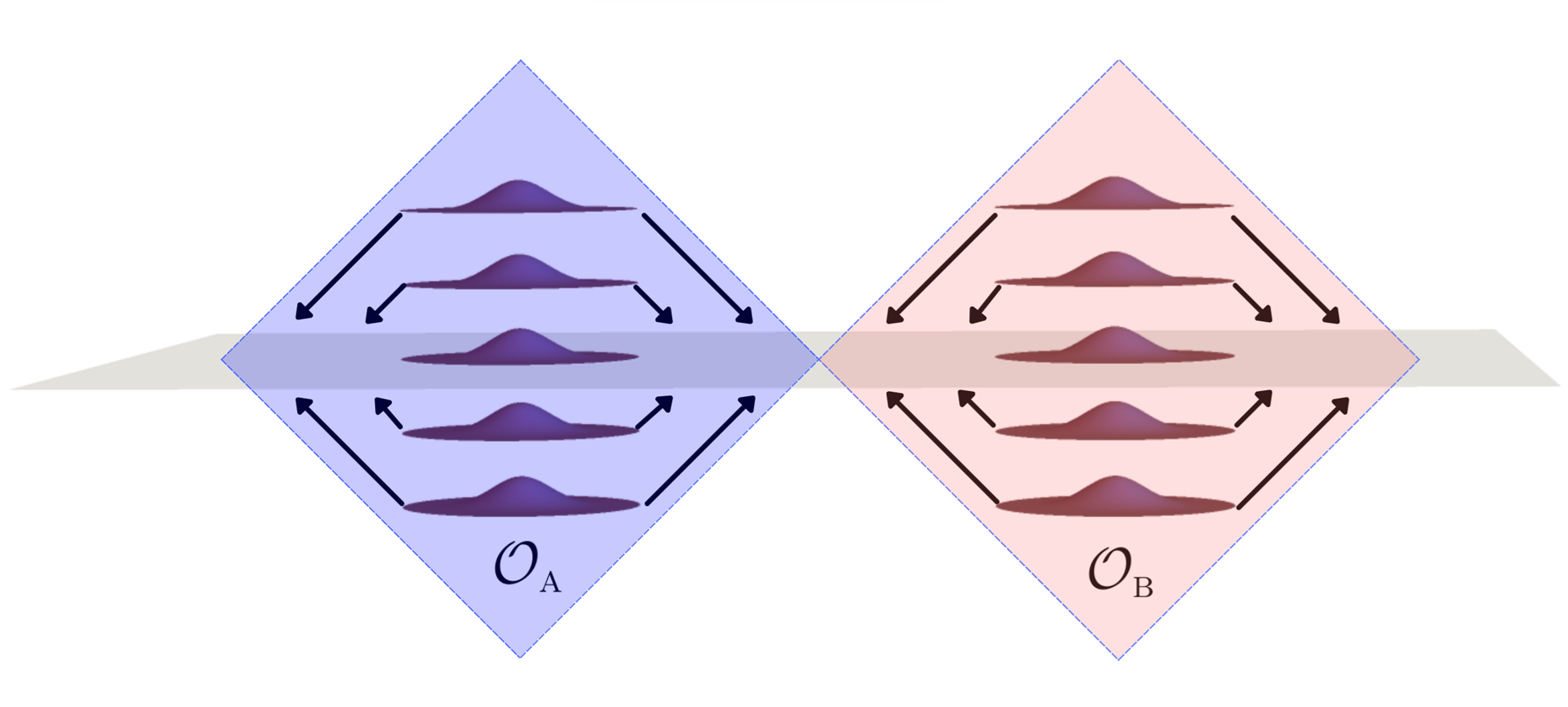}
    \caption{Schematic representation of the localization of the spatial modes in spacelike separated regions $\mathcal{O}_\tc{a}$ and $\mathcal{O}_\tc{b}$. The arrows indicate the operation of moving the modes by solving the equations of motion corresponding to each initial condition to find the corresponding position and momentum of each mode in the surface $t=0$.}
    \label{fig:setupEntGauss}
\end{figure}

A natural generalization of the example studied in~\cite{ubiquitous} would be to instead consider many different field modes localized in a given region of spacetime. An attempt in this direction was considered in the collaboration~\cite{patriciaAndI}, where we studied entanglement between multiple sets of modes in two spacelike separated regions. Specifically, we considered the modes that correspond to initial conditions $F_\tc{a}(\bm x)$ and $F_\tc{b}(\bm x)$ at different times\footnote{The motivation for this choice of modes in~\cite{patriciaAndI} stemmed from the fact that two spacelike separated particle detectors coupled to the Minkowski vacuum for finite times can become entangled. We will discuss this idea in more detail in the next Segment.}. That is, if $\Sigma_{t_0}$ denotes the Cauchy surface defined by $t = t_0$, the modes considered correspond to the smeared field operators
\begin{align}
    \hat{\Phi}_{t_i}(F_\tc{a}) = \int_{\Sigma_{t_i}} \dd \Sigma \hat{\Phi}(\bm x) F_{\tc{a}}(\bm x), \quad \quad \hat{\Pi}_{t_i}(F_\tc{a}) = \int_{\Sigma_{t_i}} \dd \Sigma \hat{\Pi}(\bm x) F_{\tc{a}}(\bm x),\\
    \hat{\Phi}_{t_i}(F_\tc{b}) = \int_{\Sigma_{t_i}} \dd \Sigma \hat{\Phi}(\bm x) F_{\tc{b}}(\bm x), \quad \quad \hat{\Pi}_{t_i}(F_\tc{b}) = \int_{\Sigma_{t_i}} \dd \Sigma \hat{\Pi}(\bm x) F_{\tc{b}}(\bm x),
\end{align}
for multiple values of $t_i$, corresponding to different Cauchy surfaces and $F_\tc{a}(\bm x)$, $F_\tc{b}(\bm x)$ given by Eq.~\eqref{eq:FAFB}. Importantly, we assume that the values of $t_i$ are picked so that there exist two spacelike separated causal diamonds $\mathcal{O}_\tc{a}$ and $\mathcal{O}_{\tc{b}}$ that contain $\text{supp}(F_\tc{a})\cap\Sigma_{t_i}$ and $\text{supp}(F_\tc{b})\cap\Sigma_{t_i}$ for all $t_i$, respectively for A and B (see Fig.~\ref{fig:setupEntGauss}).

By choosing field and momentum operators at different times, we have that for each $t_i$, the modes $(\hat{\Phi}_{t_i}(F_\tc{a}),\hat{\Pi}_{t_i}(F_\tc{a}))$, $(\hat{\Phi}_{t_i}(F_\tc{b}),\hat{\Pi}_{t_i}(F_\tc{b}))$ belong to the algebra $\mathcal{A}(\Sigma_{t_i})$. In particular, for $t_i\neq t_j$, the pairs  $(\hat{\Phi}_{t_i}(F_\tc{a}),\hat{\Pi}_{t_i}(F_\tc{a}))$ and $(\hat{\Phi}_{t_j}(F_\tc{a}),\hat{\Pi}_{t_j}(F_\tc{a}))$ belong to different algebras. We can still compare them both by noticing that each of the operators $\hat{\Phi}(F_{\tc{a}/\tc{b},i})$ and $\hat{\Pi}(F_{\tc{a}/\tc{b},i})$ correspond to covariantly smeared field operators $\hat{\phi}(f_{\tc{a}/\tc{b},i})$, $\hat{\phi}(g_{\tc{a}/\tc{b},i})$ through~\eqref{eq:phiPhiPi}, where 
\begin{align}
    Ef_{\tc{a}/\tc{b},i}|_{\Sigma_{t_i}} &= 0, &&&  Eg_{\tc{a}/\tc{b},i}|_{\Sigma_{t_i}} &= F_{\tc{a}/\tc{b}}, \\
    n^\mu \nabla_\mu Ef_{\tc{a}/\tc{b},i}|_{\Sigma_{t_i}} &= - F_{\tc{a}/\tc{b}}, &&& n^\mu \nabla_\mu Eg_{\tc{a}/\tc{b},i}|_{\Sigma_{t_i}} &= 0. 
\end{align}
Consequently, we have $\hat{\phi}(f_{\tc{a},i}),\hat{\phi}(g_{\tc{a},j})\in \mathcal{A}(\mathcal{O}_\tc{a})$ and $\hat{\phi}(f_{\tc{b},i}),\hat{\phi}(g_{\tc{b},j})\in \mathcal{A}(\mathcal{O}_\tc{b})$, so that the operators associated with the modes labelled by  A and B commute. However, this choice of modes has the unfortunate consequence that different pairs within each region, say $(\hat{\phi}(f_{\tc{a},i})$, $\hat{\phi}(g_{\tc{a},i}))$ and $(\hat{\phi}(f_{\tc{a},j})$, $\hat{\phi}(g_{\tc{a},j}))$, might not be commuting. This can be mapped to our standard Gaussian quantum mechanics formulation by performing a symplectic Gram-Schmidt procedure\footnote{The Gram-Schmidt procedure can become computationally challenging depending on the number of modes considered.}, which results in independent modes within each region. The explicit computation regarding characterization of the modes can be found in~\cite{patriciaAndI}\footnote{Notice that in~\cite{patriciaAndI} the modes were represented in a single algebra $\mathcal{A}(\Sigma)$, associated to a single Cauchy surface, rather than in the covariant algebra $\mathcal{A}(\M)$. These methods are, of course, completely equivalent.}, and the Gram-Schmidt algorithm can be found in, e.g.~\cite{KlcoEntStrQFTI}.

\begin{figure}[h!]
    \centering
    \includegraphics[width=0.7\textwidth]{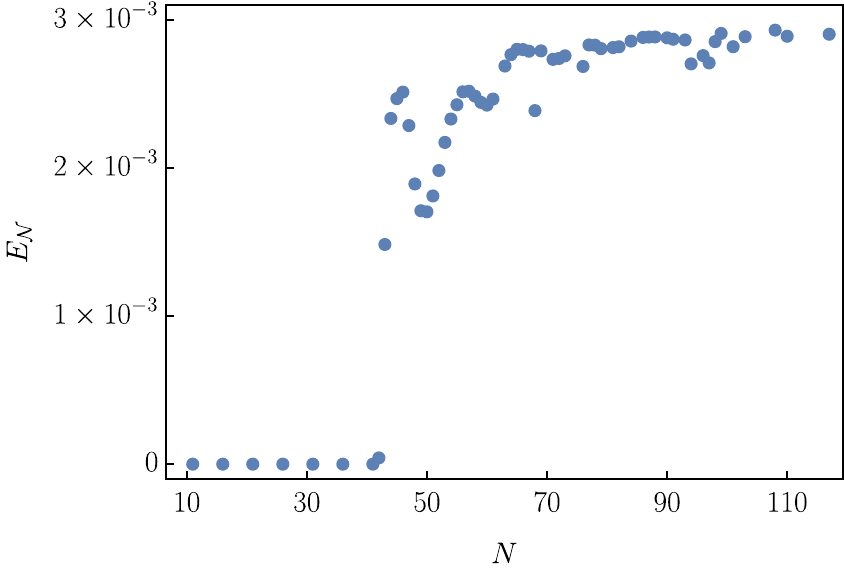}
    \caption{{Logarithmic negativity quantifying the entanglement between the two sets of $N$ modes associated with detectors A and B, as a function $N$, when the separation between the centres of the regions A and B is $|\bm x_\textsc{a}-\bm x_\textsc{b}| = T + 2R$. }}
    \label{fig:LogNeg}
\end{figure}

The specific setup used in~\cite{patriciaAndI} considered $N$ Cauchy surfaces, parametrized by
\begin{equation}
    t_i = - \frac{T}{2} + \frac{i}{N-1} T,
\end{equation}
with $i\in \{0,1,...,N-1\}$, the shape functions in~\eqref{eq:FAFB} with the parameter $\delta = 2$ when the separation between the causal diamonds $\mathcal{O}_\tc{a}$ and $\mathcal{O}_\tc{b}$ is minimal: $|\bm x_\tc{a} - \bm x_\tc{b}| = T + 2R$. The logarithmic negativity as a function of the number of modes considered is displayed in Fig.~\ref{fig:LogNeg}. The plot shows that the entanglement between the two sets of modes in each region is zero unless sufficiently many modes are considered (equivalently, unless $N$ is sufficiently large). We also see that as different modes are considered (with different choices of $t_i$), the negativity does not monotonically increase; instead, it first oscillates and then asymptotes to a constant value. This is indicative that the different choices of local modes in A and B contain different components of whichever modes are responsible for most of the entanglement between the two regions---simply increasing the number of modes while changing all of them does not guarantee that the most entangled modes between the two regions can be more faithfully represented.

The result displayed in Fig.~\ref{fig:LogNeg} demonstrates that there is indeed vacuum entanglement between two causally disjoint regions of spacetime. Also notice that due to the fact that we only considered specific types of modes, the logarithmic negativity displayed in Fig.~\ref{fig:LogNeg} is merely a lower bound for the entanglement between the two regions. 

\subsubsection*{To Which Field Modes do Probes have Access to?}

We have now explicitly seen that the vacuum of a quantum field theory possesses entanglement between two finite spacelike separated regions. However, this does not immediately imply that one can access this resource. As we discussed in Chapter~\ref{chap:meas}, one accesses a quantum field through interactions with localized probes, but it is not yet clear which specific set of modes a probe has access to.

For instance, consider an inertial two-level Unruh-DeWitt detector in Minkowski spacetime, defined by the interaction Hamiltonian density
\begin{equation}\label{eq:HIUDWchiF}
    \hat{\mathcal{H}}(\mf x) = \lambda \Lambda(\mf x) (e^{\ii \Omega t}\hat{\sigma}^+ + e^{- \ii \Omega t}\s^-)\hat{\phi}(\mf x), \quad \Lambda(\mf x) = \chi(t)F(\bm x),
\end{equation}
where $\chi(t)$ is the switching function and $F(\bm x)$ is the smearing function, written in inertial coordinates $(t,\bm x)$. For convenience, we also assume that $F$ is a real positive function, normalized in $L^2(\mathbb{R}^3)$. The interaction Hamiltonian associated to the interaction of the detector with the field can be obtained by integration in the spatial variables $\bm x$:
\begin{equation}
    \hat{H}_I(t) = \lambda \chi(t) (e^{\ii \Omega t}\hat{\sigma}^+ + e^{- \ii \Omega t}\s^-) \int \dd^3 \bm x F(\bm x) \hat{\phi}(\mf x).
\end{equation}
We can then identify the operator
\begin{equation}
    \hat{\Phi}_t(F) = \int \dd^3 \bm x F(\bm x)\hat{\phi}(t, \bm x),
\end{equation}
which naturally shows up in the interaction Hamiltonian. 

The detector then directly couples to the operators defined by $\hat{\Phi}_t(F)$ (with $t$ in the support of $\chi(t)$) in the algebra $\mathcal{A}(\mathcal{O})$, where $\mathcal{O}$ is a causal diamond that contains the support of $\Lambda(\mf x)$. The set $\hat{\Phi}_t(F)$ for $t\in \text{supp}(\chi(t))$ can then be mapped to a single Cauchy surface, say $t=0$, where each operator $\hat{\Phi}_t(F)$ in general corresponds to a mixture of smeared field and momentum operators. Alternatively, this uncountable set of field operators can be represented in the algebra $\mathcal{A}(\mathcal{O})$. Interestingly, the detector modelled by~\eqref{eq:HIUDWchiF} does not directly couple to the field's conjugate momentum $\hat{\Pi}_t(F)$ at each surface. Instead, an inertial particle detector is naturally associated with the inertial time coordinate $t$, and the operator $\hat{\Pi}_t(F)$ appears in the relationship between $\hat{H}_I(t+\delta t)$ and $\hat{H}_I(t)$:
\begin{equation}
    \frac{\hat{H}_I(t+\delta t) - \hat{H}_I(t)}{\delta t} = \lambda \partial_t(\Lambda^+(\mf x) \hat{\sigma}^+ + \Lambda^-(\mf x) \hat{\sigma}^-)\hat{\Phi}_t(F) + \lambda(\Lambda^+(\mf x) \hat{\sigma}^+ +\Lambda^-(\mf x) \hat{\sigma}^-)\hat{\Pi}_t(F) + \mathcal{O}(\delta t).
\end{equation}
One could then argue that the operator $\hat{\Pi}_t(F)$ is indirectly probed by an Unruh-DeWitt detector. Overall, to describe the dynamics of an inertial detector with the field, it would also be a natural choice to select canonical pairs of the form $(\hat{\Phi}_t(F),\hat{\Pi}_t(F))$ whenever $F$ is a real function normalized in $L^2(\mathbb{R}^3)$.


Indeed, the original motivation in~\cite{patriciaAndI} for considering modes with constant shape in different time slices was to attempt to quantify the entanglement between modes that two particle detectors with spatial shape $F(\bm x)$ couple to, and compare it with the entanglement that can be acquired by the probes. In this sense, the logarithmic negativity displayed in Fig.~\ref{fig:LogNeg} can also be seen as an upper bound to the entanglement that can be acquired by detectors that couple to the field with spacetime smearing functions $\chi(t)F_\tc{a}(\bm x)$ and $\chi(t)F_\tc{b}(\bm x)$. Indeed, in~\cite{patriciaAndI}, it was shown that the entanglement that can be acquired by detectors in this setup was orders of magnitude smaller than the field entanglement between the field modes considered in Fig.~\ref{fig:LogNeg}. We will discuss more about the entanglement that can be acquired by two detectors coupled to the field in Section~\ref{sec:OperationallyAccessingEnt}.


\subsubsection*{Recent Progress and Future Steps}

We conclude this section by mentioning recent results regarding entanglement in quantum field theory due to the contributions of Natalie Klco's group. We will briefly summarize the techniques presented in~\cite{KlcoEntStrQFTI} and the results obtained in~\cite{KlcoUVIR,KlcoEntStrQFTII,KlcoEntAllDist}, as these are particularly relevant for the next steps of research related to vacuum entanglement in quantum field theory, as well as for ideas discussed in the next chapter.

The references~\cite{KlcoUVIR,KlcoEntStrQFTI,KlcoEntStrQFTII,KlcoEntAllDist} all take a slightly different approach to entanglement in quantum field theory, by exploiting the fact that a real scalar quantum field can be approximated by a lattice of coupled harmonic oscillators. Within this lattice, the oscillators at each site determine canonical modes of the lattice field theory, which can then be represented in a finite dimensional phase space. Although the degrees of freedom represented in Klco's approach are different from the ones that we have discussed so far, the techniques presented in~\cite{KlcoEntStrQFTI} and later applied in~\cite{KlcoUVIR,KlcoEntStrQFTII,KlcoEntAllDist} can be directly translated to our setup, where the modes are defined by smeared field and momentum operators. For this reason, we will present these techniques in the context of our setup.

First we notice that in the Minkowski vacuum $\omega_0$, given a flat Cauchy surface $\Sigma$ and any functions $F,G\in C_0^\infty(\Sigma)$,
\begin{equation}
    \omega_0(\{\hat{\Phi}(F),\hat{\Pi}(G)\}) = 0.
\end{equation}
This result implies that given any set of independent modes $(\hat{\Phi}(F_{i}),\hat{\Pi}(F_{i}))$, (with $i\in\{1,...,N\}$), the covariance matrix associated with $\omega_0$ takes the form
\begin{equation}\label{eq:sigma0}
    \bm  \sigma_{0} = \begin{pmatrix}G_{11} & 0 & G_{12} & 0 & \dots & G_{1N} & 0\\
    0 & H_{11} & 0 & H_{12} & \dots & 0 & H_{1N}\\
    G_{21} & 0 & G_{22} & 0 & \dots & G_{2N} & 0\\
    0 & H_{21} & 0 & H_{22} & \dots & 0 & H_{2N}\\
    \vdots & \vdots & \vdots & \vdots &\ddots & \vdots  & \vdots\\
    G_{N1} & 0 & G_{N2} & 0 & \dots & G_{NN} & 0\\
    0 & H_{1N} & 0 & H_{2N} & \dots & 0 & H_{NN}
    \end{pmatrix},
\end{equation}
where
\begin{align}\label{eq:GijHij}
    G_{ij} = \omega_0(\{\hat{\Phi}(F_i),\hat{\Phi}(F_j)\}),\quad \quad H_{ij} = \omega_0(\{\hat{\Pi}(F_i),\hat{\Pi}(F_j)\}).
\end{align}
In other words, if $\bm P_\Phi: \mathbb{R}^{N}\to \mathbb{R}^{2N}$ and $\bm P_\Pi: \mathbb{R}^{N}\to \mathbb{R}^{2N}$ are linear operators defined by
\begin{equation}\label{eq:PPhiPPi}
    \bm P_\Phi(q^1,...,q^N) = (q^1,0,q^2,0,...,q^N,0), \quad
    \bm P_\Pi(p^1,...,p^N) = (0,p^1,0,p^2,...,0,p^N),
\end{equation}
we can write
\begin{equation}
    \bm \sigma_0 = \bm P_\Phi \bm G \bm P_\Phi^\intercal + \bm P_\Pi \bm H \bm P_\Pi^\intercal,
\end{equation}
where $(\bm G)_{ij} = G_{ij}$ and $(\bm H)_{ij}=  H_{ij}$. It turns out that many other operations in $\bm \sigma_0$ also decompose in terms of $\bm G$ and $\bm H$. For instance, the (positive) symplectic spectrum of $\bm \sigma_0$ is given by the eigenvalues of $\sqrt{\bm G \bm H}$, which also coincide with the eigenvalues of $\sqrt{\bm H\bm G}$. 

This factorization becomes particularly useful when one wishes to check entanglement between two parties. Indeed, consider commuting modes $(\hat{\Phi}(F_{\tc{a},i}),\hat{\Pi}(F_{\tc{a},i}))$ and $(\hat{\Phi}(F_{\tc{b},i}),\hat{\Pi}(F_{\tc{b},i}))$ for $i\in\{1,...,N\}$, associated to commuting algebras $\mathcal{A}(\mathcal{O}_\tc{a})$ and $\mathcal{A}(\mathcal{O}_\tc{b})$. Then the covariance matrix associated with the Minkowski vacuum also factors as~\eqref{eq:sigma0}, now a $4N\times4N$ matrix, with $\bm G$ and $\bm H$ containing the correlations of the field and momentum operators between all modes labelled A and B. The operators $\bm P_\Phi$ and $\bm P_\Pi$ naturally generalize to this higher dimensional space. In this case, the partial transpose with respect to B can be written as
\begin{equation}
    \bm \sigma_0^\Gamma = \bm P_\Phi \bm G \bm P_\Phi^\intercal + \bm P_\Pi \bm H^\Gamma \bm P_\Pi^\intercal,
\end{equation}
where $\bm H^\Gamma = \bm C_\tc{b} \bm H \bm C_\tc{b}$ and $\bm C_\tc{b}$ is given by
\begin{equation}
    \bm C_\tc{b} = \begin{pmatrix}\openone_\tc{a} & 0\\ 0 & -\openone_\tc{b}\end{pmatrix},
\end{equation}
effectively mapping $\hat{\Pi}(F_{\tc{b},i})\mapsto -\hat{\Pi}(F_{\tc{b},i})$. Thus, the symplectic eigenvalues of $\bm \sigma_0^\Gamma$ are given by the eigenvalues of $\sqrt{\bm G\bm H^\Gamma}$. Moreover, the operators $\bm G$ and $\bm H^\Gamma$ can be used to find sets of modes that fully encode the negativity between systems A and B, as follows.

Let $\{\nu_j^\Gamma\}_{j=1,...,2N}$ be the symplectic eigenvalues of $\bm \sigma_0^\Gamma$, ordered in increasing order, so that the eigenvalues of $\bm H^\Gamma \bm G$ and $\bm G \bm H^\Gamma$ are $(\nu_j^\Gamma)^2$. The eigenvectors of $\bm H^\Gamma \bm G$ and $\bm G \bm H^\Gamma$ can then be used to characterize the modes associated to the eigenvalues of $\bm \sigma_0^\Gamma$ that are smaller than one. The modes associated with eigenvalues smaller than one will then correspond to each \textit{negativity core}. Let $\mf v_{\Phi,j}$, $\mf v_{\Pi,j}\in \mathbb{R}^{2N}$ be the eigenvectors
\begin{equation}
    \bm H^\Gamma \bm G \mf v_{\Phi,j} = (\nu_j^\Gamma)^2\mf v_{\Phi,j}, \quad \bm G \bm H^\Gamma \mf v_{\Pi,j} = (\nu_j^\Gamma)^2\mf  v_{\Pi,j}.
\end{equation}
Each pair of vectors $(\bm P_\Phi\mf v_{\Phi,j},\bm P_\Pi\mf v_{\Pi,j})$ in the symplectic space $\mathbb{R}^{4N}$ is then associated to an eigenvalue $\nu_j^\Gamma$ of $\sqrt{\bm G \bm H^\Gamma}$, corresponding to possible negative symplectic eigenvalues of the partially transposed covariance matrix. We can use the vectors $\mf v_{\Phi,j}$ and $\mf v_{\Pi,j}$ to build a symplectic operation that is local in A and B and maps each pair $(\hat{\Phi}(F_{\tc{a}/\tc{b},j}),\hat{\Pi}(F_{\tc{a}/\tc{b},j}))$ to a corresponding canonical pair of operators that are associated to a negativity core.

To build the symplectic transformation, the convention used in~\cite{KlcoEntStrQFTI} normalizes the vectors $\mf v_{\Phi,j}$, $\mf v_{\Pi,j}$ by imposing the conditions
\begin{equation}
    \mf v_{\Phi,j}^\intercal\bm G \mf v_{\Phi,j} = \nu_j^\Gamma, \quad \mf v_{\Pi,j}^\intercal\bm H^\Gamma \mf v_{\Pi,j} = \nu_j^\Gamma, \quad\sum_{j=1}^{N} \mf v_{\Phi,j}\mf v_{\Pi,j}^\intercal = \openone_{N}.
\end{equation}
For convenience we define $\bm P_\tc{a}$ and $\bm P_\tc{b}$ as the projectors in A and B, so that $\openone = \bm P_\tc{a}\oplus \bm P_\tc{b}$. We then construct the basis, local in A, $\{\bm P_\tc{a} \bm P_\Phi\mf v_{\Phi,j},\bm P_\tc{a} \bm P_\Pi \mf v_{\Pi,j}\}$. The basis $\{\bm P_\tc{a} \bm P_\Phi\mf v_{\Phi,j},\bm P_\tc{a} \bm P_\Pi \mf v_{\Pi,j}\}$ is in general not symplectic. This can be fixed by applying the symplectic Gram-Schmidt procedure to it, producing a basis $\{\mf u_{\Phi,j}^\tc{a},\mf u^\tc{a}_{\Pi,j}\}$. We then define ${\bm S}_\tc{a}$:
\begin{equation}
     {{\bm S}}_\tc{a}\bm P_\tc{a} \mf e_{\Phi,j} = \bm P_{\tc{a}}\bm P_\Phi \mf v_{\Phi,j},\quad
     {\bm S}_\tc{a}\bm P_{\tc{a}}\mf e_{\Pi,j} = \bm P_{\tc{a}}\bm P_\Pi \mf v_{\Pi,j},
\end{equation}
where $\{\bm P_{\tc{a}}\mf e_{\Phi,j},\bm P_{\tc{a}}\mf e_{\Phi,j}\}$ is the basis associated with the modes $(\hat{\Phi}(F_{\tc{a},j}),\hat{\Pi}(F_{\tc{b},j}))$.  The matrix representation of $\bm S_\tc{a}$ in the basis $\{\bm P_{\tc{a}}\mf e_{\Phi,j},\bm P_{\tc{a}}\mf e_{\Phi,j}\}$ is then
\begin{equation}
    \bm S_\tc{a} = \begin{pmatrix}
        (\bm P_\Phi \mf u_{\Phi,1}^\tc{a})^\intercal\\
        (\bm P_\Pi \mf u_{\Pi,1}^\tc{a})^\intercal\\
        \vdots\\
        (\bm P_\Phi \mf u_{\Phi,N}^\tc{a})^\intercal\\
        (\bm P_\Pi \mf u_{\Pi,N}^\tc{a})^\intercal
    \end{pmatrix}.
\end{equation}
Finally, we define the transformation $\bm S_\tc{b}$ by reversing the modes in $\bm S_\tc{a}$:
\begin{equation}
    \bm S_\tc{b} = \bm R \bm S_\tc{a} \bm R, \quad R = \begin{pmatrix}
        0 & \dots & 0 & \openone_2\\[1mm]
   \dots & \ddots & \openone_2 & 0\\[1mm]
   0 & \openone_2 & \ddots & \vdots\\[1mm]
   \openone_2 & 0 &\cdots  & 0
    \end{pmatrix}.
\end{equation}
The symplectic transformation that maps the covariance matrix to the modes associated to the negativity is then
\begin{equation}
    \bm S = \bm S_\tc{a}\oplus \bm S_\tc{b},
\end{equation}
so that the covariance matrix $\bm \sigma_0' = \bm S \bm \sigma \bm S^\intercal$ represents the vacuum in this ``negativity basis''. Within $\bm \sigma_0'$, the first canonical mode in A is entangled with the last mode of B, and contribute to the total logarithmic negativity with $- \log(\nu_j^\Gamma)$. Overall, the canonical pair $j$ (with $1\leq j\leq 2N$) in A will be entangled with the canonical pair $2N+1-j$ in B, and these contribute additively to the logarithmic negativity whenever $\nu_j^\Gamma<1$. This method allows one to fully classify which modes contribute to the non-bound entanglement between the modes, as well as how much each mode contributes to the negativity. 

Applying these techniques to lattice field theories, Prof. Klco's group was able to show numerous results about entanglement in quantum field theory (extending their results to the limit of the continuum). We will summarize some of the most relevant conclusions obtained in their recent works below.

\noindent\textbf{Exponential Decay of Entanglement:} In~\cite{KlcoUVIR,KlcoEntAllDist}, it was argued that the entanglement between two spacetime regions decays exponentially with the distance between them, even when the field is massless with polynomially decaying field correlations.

\noindent\textbf{UV-IR Connection:} Also in~\cite{KlcoEntAllDist}, it was argued that vacuum entanglement between two regions exists, even when the regions are arbitrarily separated. Moreover, the further apart the regions are, the more energy the local modes that maximize the negativity must have. This fact was called the UV-IR connection~\cite{KlcoUVIR}\footnote{The name stems from the fact that when the separation between the regions is large (the field modes between them are in the IR range), the local modes that encode the entanglement between the regions must have energies in the UV range.}. 

\noindent\textbf{GHZ-Type Entanglement:} In~\cite{KlcoEntStrQFTII} it was shown that it is possible to recover a polynomial decay (for massless fields) in the entanglement between two causally disconnected regions, provided that one performs a selective measurement to the field in the region complementary to the two regions of interest. This result indicates that the vacuum entanglement behaves similarly to the GHZ state: when one considers entanglement between only two regions, the remaining degrees of freedom of the field are effectively traced out, creating mixedness in the local modes. However, when a local measurement is performed in the complementary region, one can recover entanglement between two regions that behaves like the field's correlations. This behaviour for the entanglement between two regions and their complement is similar to that observed for a GHZ state~\eqref{eq:GHZW}, indicating that vacuum entanglement is genuinely multipartite, indiscriminately entangling all neighbouring regions. 

Overall, the tools developed in~\cite{KlcoEntStrQFTI} and discussed above give a clear pathway to study entanglement in quantum field theory and suggest natural steps forward. For instance, applying Klco's techniques to the approach of localized field modes would allow one to find the specific sets of local modes $\hat{\phi}(f_{\tc{a}/\tc{b},i})$, $\hat{\phi}(g_{\tc{a}/\tc{b},i})$ that contain complete information about the negativity between two finite spacetime regions. This could also be used to indicate the specific local probes that should be utilized to extract entanglement from the vacuum. These are topics that are currently being studied and have the potential to provide insight into our understanding of vacuum entanglement and even lead to practical applications.

\section{Operationally Accessing the Entanglement in QFT}\label{sec:OperationallyAccessingEnt}

An alternative way of quantifying the entanglement in quantum field theory is by quantifying the entanglement that can be acquired by probes that couple to a field. The concept of utilizing localized probes to access entanglement in quantum field theory was first considered by Valentini in 1991~\cite{Valentini1991}, later studied by Reznik and collaborators in the 2000's~\cite{Reznik2003}, and a modern approach was introduced by Pozas-Kerstjens and Mart\'in-Mart\'inez in~\cite{Pozas-Kerstjens:2015}, when the protocol took the name of \textit{entanglement harvesting}. The typical protocol of entanglement harvesting considers two particle detectors that couple to a localized quantum field in an attempt to extract entanglement from the field. 

Throughout the last decade entanglement harvesting has been studied with two-level Unruh-DeWitt detectors in a variety of spacetimes, considering different states of motion for the detectors and different states for a real scalar quantum field~\cite{Reznik2005,Retzker2005,Reznik2007,Salton:2014jaa,Pozas-Kerstjens:2015,Ng1,Henderson2019,bandlimitedHarv2020,ampEntBH2020,HarvestingDelocalized,ericksonNew,HarvestingAccelerationRobb,twist2022,cisco2023harvesting,SchwarzchildHarvestingWellDone}. The protocol has also been studied with more general detector models that couple to different quantum fields, studying entanglement harvesting from the electromagnetic~\cite{Pozas2016}, neutrino~\cite{carol}, gravitational fields~\cite{boris}, among other generalizations~\cite{HarvestingSuperposed,threeHarvesting2022,tripartiteBHarvesting}. In this Section, we will describe the protocol of entanglement harvesting, starting with the formulation first presented in~\cite{FullHarvesting}, which uses two localized quantum fields as probes, and later presenting the simplified formulation in terms of particle detectors, which has become the standard approach to the protocol.

\subsubsection*{Two Localized Probes Coupled to a Quantum Field}

    We will now consider two localized real scalar quantum fields $\hat{\phi}_\tc{a}(\mf x)$ and $\hat{\phi}_\tc{b}(\mf x)$ in $3+1$ dimensional Minkowski spacetime, under the influence of confining potentials $V_\tc{a}(\bm x)$ and $V_\tc{b}(\bm x)$ coupled to a real massless scalar quantum field $\hat{\phi}$. The theory of the three field $\hat{\phi}_\tc{a}$, $\hat{\phi}_\tc{b}$, and $\hat{\phi}$ is described by the Lagrangian
    \begin{align}
        \mathcal{L} = -\tfrac{1}{2} \partial_\mu \phi\partial^\mu \phi -\tfrac{1}{2} \partial_\mu \phi_\tc{a}\partial^\mu \phi_\tc{a} - \tfrac{1}{2}(m_\tc{d}^2 + V_\tc{a}(\bm x))&\phi_\tc{a}^2-\tfrac{1}{2} \partial_\mu \phi_\tc{b}\partial^\mu \phi_\tc{b}- \tfrac{1}{2}(m_\tc{b}^2 + V_\tc{b}(\bm x))\phi_\tc{b}^2 \nonumber\\[2pt]
        &- \lambda \zeta_\tc{a}(\mf x) \phi_\tc{a} \phi - \lambda \zeta_\tc{b}(\mf x) \phi_\tc{b} \phi,
    \end{align}
    where $\zeta_\tc{a}(\mf x)$ and $\zeta_{\tc{b}}(\mf x)$ are spacetime smearing functions that are localized in spacetime. For convenience we assume for now that $\zeta_\tc{a}(\mf x)$ and $\zeta_{\tc{b}}(\mf x)$ are compactly supported in causal diamonds $\mathcal{O}_\tc{a}$ and $\mathcal{O}_\tc{b}$. As a consequence of the confining potentials, both fields will have discrete modes $u_{\bm n_\tc{a}}(\mf x) = e^{- \ii \omega_{\bm n_\tc{a}}t} \Phi_{\bm n_{\tc{a}}}(\bm x)$, $u_{\bm n_\tc{b}}(\mf x) = e^{- \ii \omega_{\bm n_\tc{b}}t} \Phi_{\bm n_{\tc{b}}}(\bm x)$, labelled by the discrete indices $\bm n_\tc{a}$ and $\bm n_{\tc{b}}$ and admitting expansions of the form~\eqref{eq:phidexp}. This gives rise to the vacuum states $\ket{0_\tc{a}}$ and $\ket{0_\tc{b}}$ for each field. 
    
    The fields then interact linearly with a free Klein-Gordon field $\hat{\phi}(\mf x)$, so that the interaction Hamiltonian density of the interacting theory can be written as
    \begin{equation}\label{eq:HI2fields}
        \hat{\mathcal{H}}_I(\mf x) = \underbrace{\lambda \zeta_\tc{a}(\mf x)\hat{\phi}(\mf x)\hat{\phi}_\tc{a}(\mf x)}_{\text{\normalsize  $\hat{\mathcal{H}}_{I,\tc{a}}(\mf x)$}} + \underbrace{\lambda\zeta_\tc{b}(\mf x)\hat\phi(\mf x) \hat{\phi}_\tc{b}(\mf x)}_{\text{\normalsize  $\hat{\mathcal{H}}_{I,\tc{b}}(\mf x)$}}.
    \end{equation}
    By picking initial states for the system of the three fields $\hat{\phi}_\tc{a}(\mf x)$, $\hat{\phi}_\tc{b}(\mf x)$, and $\hat{\phi}(\mf x)$, one can then compute the final state of the probe fields by applying the time evolution operator
    \begin{equation}
        \hat{U}_I = \mathcal{T}\exp\left(-\ii \int \dd V \hat{\mathcal{H}}_I(\mf x)\right) = \mathcal{T}\exp\left(- \ii \int \dd V (\hat{\mathcal{H}}_{I,\tc{a}}(\mf x) + \hat{\mathcal{H}}_{I,\tc{b}}(\mf x))\right)
    \end{equation}
    and tracing over $\hat\phi$, analogous to the calculation performed in Section~\ref{sec:LocalizedQuantumFields}. Our goal is to quantify the entanglement in the final probes state after the interaction with the field. 
    
    The computation of the final state of the probe fields $\hat{\phi}_\tc{a}$ and $\hat{\phi}_\tc{b}$ follows steps analogous to~\eqref{eq:rhoDyson}. We consider the initial state 
    \begin{equation}\label{eq:rho02fields}
        \hat{\rho}_0 = \ket{0_\tc{a}}\!\!\bra{0_\tc{a}} \otimes \ket{0_\tc{b}}\!\!\bra{0_\tc{b}} \otimes \hat{\rho}_\phi,
    \end{equation}
    where $\hat{\rho}_\phi$ is a zero mean Gaussian state for the field $\hat{\phi}(\mf x)$ in a suitable GNS representation. The calculation is straightforward, but tedious, and was first performed in this context in~\cite{FullHarvesting}. We present these explicit computations in Appendix~\ref{app:twoQFTs}. In summary, to leading order in $\lambda$, each mode $\bm n_\tc{a}$ and $\bm n_\tc{b}$ evolves independently, according to the interaction Hamiltonian densities
    \begin{equation}\label{eq:hIABQFT}
        \hat{\mathcal{H}}_{\bm n_\tc{a},\text{eff}}(\mf x) = \lambda \hat{Q}_{{}_{\bm n_\tc{a}}}^{\tc{a}}\!(\mf x) \hat{\phi}(\mf x), \quad \quad  \hat{\mathcal{H}}_{\bm n_\tc{b},\text{eff}}(\mf x) = \lambda \hat{Q}_{{}_{\bm n_\tc{b}}}^{\tc{b}}\!(\mf x) \hat{\phi}(\mf x),
    \end{equation}
    where
    \begin{align}
        \hat{Q}_{{}_{\bm n_\tc{a}}}^{\tc{a}}\!(\mf x) = \Lambda_\tc{a}(\mf x) e^{- \ii \omega_{\bm n_\tc{a}} t}\hat{a}^{\tc{a}}_{{}_{\bm n_\tc{a}}}+\Lambda^*_\tc{a}(\mf x) e^{\ii \omega_{\bm n_\tc{a}} t}\hat{a}^{\tc{a}\dagger}_{{}_{\bm n_\tc{a}}},\quad\quad \Lambda_\tc{a}(\mf x) &\coloneqq \zeta_\tc{a}(\mf x) \Phi^{\tc{a}}_{{}_{\bm n_\tc{a}}}(\mf x),\\
        \hat{Q}_{{}_{\bm n_\tc{b}}}^\tc{b}\!(\mf x) = \Lambda_\tc{b}(\mf x)e^{- \ii \omega_{\bm n_\tc{b}} t} \hat{a}^{\tc{b}}_{{}_{\bm n_\tc{b}}}+\Lambda^\ast_\tc{b}(\mf x) e^{\ii \omega_{\bm n_\tc{b}} t}\hat{a}^{\tc{b}\dagger}_{{}_{\bm n_\tc{b}}},\quad \quad
        \Lambda_\tc{b}(\mf x) &\coloneqq \zeta_\tc{b}(\mf x) \Phi^{\tc{b}}_{{}_{\bm n_\tc{b}}}(\mf x).
    \end{align}
    with $\hat{a}^{\tc{a}}_{{}_{\bm n_\tc{a}}}$, $\hat{a}^{\tc{a}\dagger}_{{}_{\bm n_\tc{a}}}$, $\hat{a}^{\tc{b}}_{{}_{\bm n_\tc{b}}}$, and $\hat{a}^{\tc{b}\dagger}_{{}_{\bm n_\tc{b}}}$ being the creation and annihilation operators associated with excitations in the modes $\bm n_\tc{a}$ and $\bm n_\tc{b}$ for the respective fields $\hat{\phi}_\tc{a}(\mf x)$ and $\hat{\phi}_\tc{b}(\mf x)$. This shows that when the localized fields start the interaction in their vacuum state, each individual pair of modes of localized quantum fields effectively behaves as particle detector models to leading order in perturbation theory. In this case, where the probe fields are scalar fields, they specifically correspond to harmonic oscillator particle detectors.

    Let us then focus on two specific modes labelled by $\bm n_\tc{a}$ and $\bm n_\tc{b}$, and define the normalized states
    \begin{align}
        \ket{1_\tc{a}} = \hat{a}_{\bm n_\tc{a}}^\dagger \ket{0_{\bm n_\tc{a}}}, \quad\quad 
        \ket{2_\tc{a}} = \frac{1}{\sqrt{2}}\hat{a}_{\bm n_\tc{a}}^\dagger\hat{a}_{\bm n_\tc{a}}^\dagger \ket{0_{\bm n_\tc{a}}},\\
        \ket{1_\tc{b}} = \hat{a}_{\bm n_\tc{b}}^\dagger \ket{0_{\bm n_\tc{b}}}, \quad\quad 
        \ket{2_\tc{b}} = \frac{1}{\sqrt{2}}\hat{a}_{\bm n_\tc{b}}^\dagger\hat{a}_{\bm n_\tc{b}}^\dagger \ket{0_{\bm n_\tc{b}}}.
    \end{align}
    Let $\hat{\rho}_\tc{ab}$ denote the final state of the modes $\bm n_\tc{a}$ and $\bm n_\tc{b}$, obtained using the initial state~\eqref{eq:rho02fields}. To leading order in perturbation theory, $\hat{\rho}_\tc{ab}$ only acquires components in the subspace spanned by $\{\ket{0_\tc{a}},\ket{1_\tc{a}},\ket{2_\tc{a}}\}\otimes\{\ket{0_\tc{b}},\ket{1_\tc{b}},\ket{2_\tc{b}}\}$. In this basis we can write the final density operator in matrix form as

\begin{align}\label{eq:rhoAB}
    \r_\tc{ab} = \begin{pmatrix} {\text{\textbf{M}}} & 0_{7\times2} \\ 0_{2\times7} & 0_{2\times2}
\end{pmatrix} +\mathcal{O}(\lambda^4),  
\end{align} 
where
\begin{align}
\text{\textbf{M}} = \begin{pmatrix}
       1- \mathcal{L}_\tc{aa}^- - \mathcal{L}_\tc{bb}^-  & 0 & \mathcal{K}_\tc{b}^* & 0 & \mathcal{M}^* & 0 & \mathcal{K}_{\tc{a}}^*\\
        0 & \mathcal{L}_{\textsc{bb}}^- & 0 & \mathcal{L}_{\textsc{ab}}^- & 0& 0& 0\\
        \mathcal{K}_\tc{b} & 0 & 0 & 0 & 0 & 0& 0\\
        0 & (\mathcal{L}_{\textsc{ab}}^-)^*  & 0 & \mathcal{L}_{\textsc{aa}}^-  & 0& 0& 0\\
        \mathcal{M} & 0 & 0 & 0 & 0& 0& 0\\
        0 & 0 & 0 & 0 & 0& 0& 0\\
        \mathcal{K}_{\tc{a}} & 0 & 0 & 0 & 0& 0 & 0
    \end{pmatrix},
\end{align}
\begin{align}
    \mathcal{L}_{\tc{ij}}^- &= \lambda^2 \int \dd V \dd V' \Lambda_\tc{i}(\mf x) \Lambda_\tc{j}(\mf x') e^{- \ii (\omega_{\bm n_\tc{i}} t-\omega_{\bm n_\tc{j}}t')} W(\mf x,\mf x') = \lambda^2 W(\Lambda_\tc{i}^-,\Lambda_\tc{j}^+),\label{eq:Lij}\\
    \mathcal{K}_{\tc{i}} &= - \tfrac{\lambda^2}{\sqrt{2}} \int \dd V \dd V' \Lambda_\tc{i}(\mf x) \Lambda_\tc{i}(\mf x') e^{\ii \omega_{\bm n_\tc{i}}(t+t')} G_F(\mf x,\mf x') = - \tfrac{\lambda^2}{\sqrt{2}} G_F(\Lambda_\tc{i}^+,\Lambda_\tc{i}^+),\label{eq:Kij}\\
    \mathcal{M} &= -\lambda^2 \int \dd V \dd V' \Lambda_\tc{a}(\mf x) \Lambda_\tc{b}(\mf x') e^{\ii (\omega_{\bm n_\tc{a}}t+\omega_{\bm n_\tc{b}}t')} G_F(\mf x,\mf x') = -\lambda^2G_F(\Lambda_\tc{a}^+,\Lambda_\tc{b}^+) ,\label{eq:M}
\end{align}
where $\Lambda_\tc{i}^\pm (\mf x) = e^{\pm \ii \omega_{\bm n_\tc{i}t}}\Lambda_\tc{i}(\mf x)$, and $W$ and $G_F$ denote the Wightman function and Feynman propagator of the field $\hat{\phi}$ in the state $\hat{\rho}_\phi$. In the expressions above, $\tc{I},\tc{J} \in \{\tc{A},\tc{B}\}$. The $\mathcal{L}_\tc{aa}^-$, $\mathcal{L}_\tc{bb}^-$, $\mathcal{K}_\tc{a}$ and $\mathcal{K}_\tc{b}$ terms are local to each probe, while the terms $\mathcal{L}_{\tc{ab}}$, and $\mathcal{M}$ correspond to the correlations acquired by the two probes. Also notice that $\mathcal{L}_{\tc{aa}}^-$ and $\mathcal{L}_{\tc{bb}}^-$ correspond to the excitation probability of the individual field modes $\bm n_\tc{a}$ and $\bm n_\tc{b}$.

We can then quantify the entanglement present in the final state of the detectors (if any). Noticing that the final state in Eq.~\eqref{eq:rhoAB} is a mixed state (as each field mode becomes entangled with the field $\hat{\phi}$), we pick the negativity~\eqref{eq:negativity} as an entanglement quantifier. For the state given in Eq.~\eqref{eq:rhoAB}, it reads
\begin{align}\label{eq:neg}
    \mathcal{N}(\r_\tc{ab}) &= \max\left(0,\sqrt{|\mathcal{M}|^2 + \left(\tfrac{\mathcal{L}^-_\tc{aa} - \mathcal{L}^-_\tc{bb}}{2}\right)^2} - \tfrac{\mathcal{L}^-_{\tc{aa}} + \mathcal{L}^-_\tc{bb}}{2}\right) + \mathcal{O}(\lambda^4).
\end{align}
Moreover, if the detectors' local terms are the same (i.e., $\mathcal{L}^-_\tc{aa} = \mathcal{L}^-_\tc{bb} = \mathcal{L}^-$), Eq. \eqref{eq:neg} simplifies to \mbox{$\mathcal{N}(\r_\tc{ab}) = \max(0,|\mathcal{M}| - \mathcal{L}^-)$}. Overall, the entanglement in the state of the detectors is a competition between the non-local $\mathcal{M}$ term and the local noise terms $\mathcal{L}^-_\tc{aa}$ and $\mathcal{L}^-_\tc{bb}$\footnote{The argument that entanglement at leading order is always a competition between local noise and the correlation term $\mathcal{M}$  can be made even if the detectors are not identical since we can bound $\mathcal{N}$ using
 \begin{align}\label{eq:negab2fields}
    \mathcal{N}(\r_\tc{ab})\leq \sqrt{|\mathcal{M}|^2+\frac{(\mathcal{L}^-_{\textsc{aa}} - \mathcal{L}^-_{\textsc{bb}})^2}{4}} - \frac{\mathcal{L}^-_{\textsc{aa}} + \mathcal{L}^-_{\textsc{bb}}}{2} \leq |\mathcal{M}|+\frac{|\mathcal{L}^-_{\textsc{aa}} - \mathcal{L}^-_{\textsc{bb}}|}{{2}} - \frac{\mathcal{L}^-_{\textsc{aa}} + \mathcal{L}^-_{\textsc{bb}}}{2} \leq |\mathcal{M}|,
\end{align}
so that we obtain $\mathcal{N}(\r_\tc{ab}) \leq |\mathcal{M}|$. This argument was taken from~\cite{hectorMass}.}.

    Interestingly, since at leading order in perturbation theory, different modes of the probe fields do not interact with each other, each pair of modes may acquire some amount of entanglement, and not only the two modes labelled by $\bm n_\tc{a}$ and $\bm n_\tc{b}$. This means that the amount of entanglement that the two localized quantum fields can acquire is lower bounded by Eq.~\eqref{eq:neg}.
    
    Notice that the computations Appendix~\ref{app:twoQFTs} also indirectly show that the analogy between particle detectors and modes of localized quantum field theories presented in Section~\ref{sec:LocalizedQuantumFields} also holds when multiple localized quantum fields are considered. In particular, this implies that the results for the final detector state and negativity immediately carry on to the case of two harmonic oscillator detectors coupled to a scalar quantum field.

    \subsubsection*{Entanglement Harvesting vs. Communication}

    It should be no surprise that two systems coupled to a quantum field can become entangled. For instance, two spins interact electromagnetically through the magnetic field sourced by each, resulting in an effective interaction Hamiltonian that directly couples their spins and is well known to generate entanglement. However, the mechanism that allows the spins to become entangled in this case is \textit{communication} through the field, which requires the systems to be causally connected. In this example with spins, both probes are coupled to the same field degrees of freedom, and this type of scenario cannot be used to infer entanglement between different field degrees of freedom.

    To extract entanglement from the field, the detectors must not be coupled to the same field degrees of freedom. In other words, the entanglement between the probes that couple to the field in regions $\mathcal{O}_\tc{a}$ and $\mathcal{O}_\tc{b}$ can only be associated with field entanglement between the respective regions if the algebras $\mathcal{A}(\mathcal{O}_\tc{a})$ and $\mathcal{A}(\mathcal{O}_\tc{b})$ commute. For instance, the algebras will commute whenever $\mathcal{O}_\tc{a}$ and $\mathcal{O}_\tc{b}$ are spacelike separated\footnote{However, in the case of a massless field in Minkowski spacetime, any two regions that are not connected by lightlike geodesics also fulfill this condition, allowing for timelike entanglement harvesting~\cite{pastfuture}.}. When $[\mathcal{A}(\mathcal{O}_\tc{a}),\mathcal{A}(\mathcal{O}_\tc{b})] = 0$, one can then fully associate any entanglement acquired by the probes to entanglement that was previously present in the field between regions $\mathcal{O}_\tc{a}$ and $\mathcal{O}_\tc{b}$, configuring the protocol that has become known as 
    \textit{entanglement harvesting}.

    At this stage notice that if the supports of $\zeta_\tc{a}(\mf x)$ and $\zeta_\tc{b}(\mf x)$ are spacelike separated, the interaction Hamiltonian densities $\hat{\mathcal{H}}_{I,\tc{a}}(\mf x) + \hat{\mathcal{H}}_{I,\tc{b}}(\mf x)$ commute. This implies that the unitary time evolution operator factors as
    \begin{equation}
        \hat{U}_I = \hat{U}_{\tc{a},\phi}\hat{U}_{\tc{b},\phi}, \quad\quad\quad \hat{U}_{\tc{a}/\tc{b},\phi} = \mathcal{T}\exp\left(-\ii \int \dd V \hat{\mathcal{H}}_{I,\tc{a}/\tc{b}}(\mf x)\right).
    \end{equation}
    Although the total unitary time evolution operator $\hat{U}_I$ factors as a product acting locally in A and B when the interaction regions are spacelike separated, this does not ensure that the final state of probes A and B will be separable once we trace over the fields degrees of freedom. Essentially, the unitaries $\hat{U}_{\tc{a},\phi}$ and $\hat{U}_{\tc{b},\phi}$ will entangle the degrees of freedom of A (resp. B) to the degrees of freedom of the field $\hat{\phi}$ in the coupling region defined by the profile of $\zeta_\tc{a}(\mf x)$ (resp. B). If these local degrees of freedom of $\hat{\phi}$ are sufficiently entangled between regions A and B, it is then possible that the final state of the fields A and B will be entangled after the interaction. In other words, entanglement harvesting can be understood as an entanglement swap operation, where the probes A and B attempt to extract entanglement between the degrees of freedom of the field $\hat{\phi}$ in the regions defined by the support of $\zeta_\tc{a}(\mf x)$ and $\zeta_{\tc{b}}(\mf x)$.

    However, as discussed in Section~\ref{sec:Tmunu}, compact support is an idealization, and any realistic description for the modes of a quantum field will necessarily extend itself through all space. Whatever mechanism is responsible for localizing the interaction between the probes and field will also not be able to fully suppress the interactions. This is to say that in the example of the previous Segment, the assumption that the interaction profiles $\zeta_\tc{a}(\mf x)$ and $\zeta_\tc{b}(\mf x)$ are compactly supported is an idealization. Instead, it is likely that in any realistic scenario, both detectors will couple to the whole algebra of observables of the target field $\mathcal{A}(\M)$, and that the spacetime smearing functions $\Lambda_\tc{i}(\mf x)$ will be non-zero throughout all spacetime. 
    
    We then need a method to classify which entanglement acquired by the probes is genuinely harvested, and which entanglement is acquired through communication. This classification was first done in~\cite{ericksonNew}, where the negativity of Eq.~\eqref{eq:negab2fields} was split into two distinct terms, corresponding to entanglement through communication and entanglement extraction from the field. We will take a slightly different (but equivalent) approach here. The classification of the different contributions to entanglement between the probes can be done in terms of the propagators involved in the entanglement between the detectors, which can be split into state dependent and state independent parts (see~\eqref{eq:WHE} and~\eqref{eq:GFHDelta}). Specifically, the $\mathcal{M}$ term in Eq.~\eqref{eq:negab2fields} can be decomposed in terms of the Hadamard distribution and the symmetric propagator:
    \begin{equation}
        \mathcal{M} =  - \lambda^2 G_F(\Lambda_\tc{a}^+,\Lambda_\tc{b}^+) = - \frac{\lambda^2}{2}H(\Lambda_\tc{a}^+,\Lambda_\tc{b}^+) - \frac{\ii \lambda^2}{2} \Delta(\Lambda_\tc{a}^+,\Lambda_\tc{b}^+). 
    \end{equation}
    The decomposition above separated the state dependent terms in the propagator $\mathcal{M}$ (encoded in $H$) and the state independent terms (encoded in $\Delta$). Only the term $H(\Lambda_\tc{a}^+,\Lambda_\tc{b}^+)$ can encode entanglement in the field, given that $\Delta(\Lambda_\tc{a}^+,\Lambda_\tc{b}^+)$ is state independent: it represents the entanglement that is acquired between the probes when they are causally connected via symmetric exchanges through the field. Also notice that whenever the supports of $\Lambda_\tc{a}(\mf x)$ and $\Lambda_\tc{b}(\mf x)$ are spacelike separated, $\Delta(\Lambda_\tc{a}^+,\Lambda_\tc{b}^+) = 0$. We can extend this concept to the case where two detectors interact with the field in non-compactly supported regions. Essentially, entanglement acquired by the probes can be mostly tied to the entanglement in the field $\hat{\phi}$ whenever $\Delta(f_\tc{a},f_\tc{b})$, $E(f_\tc{a},f_\tc{b})$ are negligible in comparison to $\mathcal{N}(\hat{\rho}_\tc{ab})$ for all functions $f_\tc{a}$ related to detector A and functions $f_\tc{b}$ related to detector B. Equivalently,  if $\tilde{\mathcal{N}}(\hat{\rho}_\tc{ab})$ denotes the negativity of final state of two probes setting $G_R(f_\tc{a},f_\tc{b})\mapsto 0$, and $G_R(f_\tc{b},f_\tc{a})\mapsto 0$, the condition so that the entanglement between the probes can be traced back to the field is that $\mathcal{N}(\r_\tc{ab}) \approx \tilde{\mathcal{N}}(\r_\tc{ab})$.

    For instance, in the setup of the previous Segment, we have
    \begin{equation}
        \tilde{\mathcal{N}}(\hat{\rho}_\tc{ab}) = \max\left(0,\sqrt{\frac{1}{4}|H(\Lambda_{\tc{a}}^+,\Lambda_{\tc{b}}^+)|^2 + \left(\tfrac{\mathcal{L}^-_\tc{aa} - \mathcal{L}^-_\tc{bb}}{2}\right)^2} - \tfrac{\mathcal{L}^-_{\tc{aa}} + \mathcal{L}^-_\tc{bb}}{2}\right) + \mathcal{O}(\lambda^4),
    \end{equation}
    and we will have $\tilde{\mathcal{N}}(\hat{\rho}_\tc{ab}) \approx {\mathcal{N}}(\hat{\rho}_\tc{ab})$ if $\tfrac{1}{2}|\Delta(\Lambda^+_\tc{a},\Lambda^+_\tc{b})| \ll \mathcal{N}(\r_\tc{ab})$. This condition can be fulfilled by considering systems that are sufficiently separated in space (or in time, for a massless field in Minkowski spacetime).

    \subsubsection*{Explicit Examples}

    For concreteness, we now present specific examples of entanglement harvesting using two localized quantum field theories. Specifically, we consider the lowest energy modes of 1) fields under the influence of quadratic potentials and 2) fields in cubic boxes with Dirichlet boundary conditions, as described in Sections~\ref{sec:LocalizedQuantumFields} and~\ref{sec:MoreRealisticProbes}. We will see that, indeed, it is possible for these localized quantum fields to extract entanglement from the vacuum of a free Klein-Gordon field in $3+1$ Minkowski spacetime.

    As our first example of fully relativistic entanglement harvesting, we consider two localized quantum fields in Minkowski spacetime under the influence of potentials $V_\tc{a}(\bm x) = |\bm x|^2/2\ell^4$ and $V_\tc{b}(\bm x) = V_\tc{a}(\bm x - \bm L)$, where $L = |\bm L|$ denotes the proper distance between the centers of the trapping potentials. In essence, the two quantum fields are identical, apart from a spatial shift in the potentials that confine them. Under these assumptions, the energy levels of each field take the form of Eq.~\eqref{eq:gapHO}, with their lowest energy levels being $\omega_{\bm 0_\tc{a}} = \omega_{\bm 0_\tc{b}} = \sqrt{m_\tc{d}^2 +   3/\ell^2}$. 
    
    Both fields will interact with a free scalar field $\hat{\phi}(\mf x)$ according to the interaction Hamiltonian of Eq. \eqref{eq:hIABQFT}, where the functions $\zeta_\tc{a}(\mf x)$ and $\zeta_{\tc{b}}(\mf x)$ will be prescribed as
    \begin{equation}\label{eq:switching}
        \zeta_\tc{a}(\mf x) = \zeta_\tc{b}(\mf x) = e^{-\frac{{\pi} t^2}{2T^2}}.
    \end{equation}
    This corresponds to interactions that are adiabatically switched on and peak at $t = 0$. The effective time of the switching is controlled by the timescale $T$. The reason that we consider $\zeta_\tc{a}(\mf x) = \zeta_\tc{b}(\mf x)$ independent of the spatial coordinates is that the effective spacetime region where the localized fields interact with $\hat{\phi}(\mf x)$ is defined by the product of $\zeta_\tc{a}(\mf x)$ with the mode localization of the fields. The spatial localization of the modes together with the time localization of $\zeta_\tc{a}(\mf x)$ then gives an overall interaction which is localized in spacetime for each mode. 

    We consider the three fields to start in their respective vacua, $\ket{0_\tc{a}}\otimes \ket{0_\tc{b}} \otimes \ket{0}$, and we assume that we only have access to the localized fields' mode excitations with the lowest energy, $\omega_{\bm 0_\tc{a}} = \omega_{\bm 0_\tc{b}} \equiv \Omega$. The negativity of the final state of the probes, $r_\tc{ab}$, takes the same form of Eq.~\eqref{eq:neg}, and in the case where the excitation probabilities are the same (as we are considering here), it becomes
    \begin{equation}
        \mathcal{N}(\r_\tc{d}) = \max(0, |\mathcal{M}| - \mathcal{L}^-) + \mathcal{O}(\lambda^4),
    \end{equation}
    where the $\mathcal{L}^-$ and $\mathcal{M}$ terms are given by
    \begin{align}
        \mathcal{L}^- &= \lambda^2 \int \dd V \dd V' \Lambda_\tc{a}(\mf x)\Lambda_\tc{a}(\mf x') e^{- \ii \Omega(t-t')} W(\mf x,\mf x') =\lambda^2 W(\Lambda_\tc{a}^-, \Lambda_\tc{a}^+),\label{eq:LQFTA}\\
        &= \lambda^2 \int \dd V \dd V' \Lambda_\tc{b}(\mf x)\Lambda_\tc{b}(\mf x') e^{- \ii\Omega (t-t')} W(\mf x,\mf x')=\lambda^2 W(\Lambda_\tc{b}^-, \Lambda_\tc{b}^+),\nonumber\\
        \mathcal{M} &= -\lambda^2 \int \dd V \dd V' \Lambda_\tc{a}(\mf x)\Lambda_\tc{b}(\mf x') e^{\ii \Omega(t+t')} G_F(\mf x,\mf x') = -\lambda^2 G_F(\Lambda_\tc{a}^+, \Lambda_\tc{b}^+),\nonumber
    \end{align}
    with $\Omega = \sqrt{m^2 + 3/\ell^2}$ and the spacetime smearing functions are given by
    \begin{align}
        \Lambda_\tc{a}(\mf x) &= \zeta_\tc{a}(\mf x) \Phi^{\tc{a}}_{\bm 0_\tc{a}}\!(\bm x) = e^{-\frac{\pi t^2}{2T^2}} \!\!\left(\frac{1}{\pi\ell^2}\right)^{\frac{3}{4}}\!\!\!\frac{e^{-\frac{|\bm x|^2}{2\ell^2}}}{\left(m^2 + \tfrac{3}{\ell^2}\right)^{1/4}},\nonumber\\
        \Lambda_\tc{b}(\mf x) &= \zeta_\tc{b}(\mf x) \Phi^{\tc{b}}_{\bm 0_\tc{b}}\!(\bm x) = \Lambda_\tc{a}(t,\bm x - \bm L).\label{eq:smearingsQFTHarvesting}
    \end{align}
    We then see that the effective size of the interaction region can be estimated by looking at the standard deviation of the space dependent Gaussian function in Eq.~\eqref{eq:smearingsQFTHarvesting}. In this case, the spatial size of the interaction region can be estimated to be $\sigma \sim \ell$, so that smaller values of the parameter $\ell$ that defines the confining potential correspond to more localized detectors.

    \begin{figure}[h!]
        \centering
        \includegraphics[width=12cm]{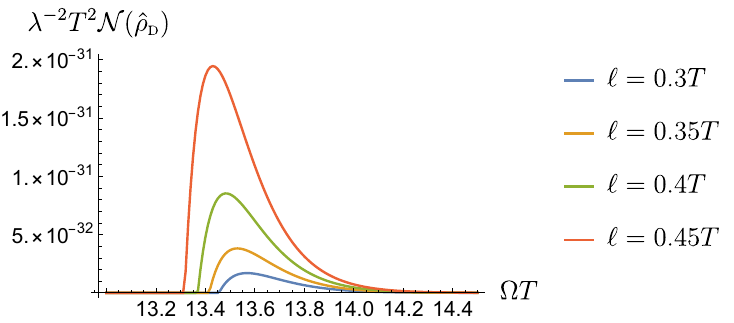}
        \caption{The negativity of the state of two localized quantum fields confined by a quadratic potential when restricted to their lowest energy after interacting with a massless scalar field. The negativity is plotted as a function of the energy of the modes $\Omega = \sqrt{m^2 + 3/\ell^2}$. The time duration of the interaction $T$ is used as a scale. The separation between the detectors' interaction regions for these plots is $L = 5 T$.}
        \label{fig:quadratic}
    \end{figure}
    
   We focus on the case where the interaction regions are approximately spacelike separated\footnote{Even though the Gaussian tails of the switching are technically infinitely long, using these switchings is effectively equivalent to considering compactly supported switchings.}, so that entanglement acquired by the localized modes via communication can be neglected. For this reason, we consider $L = 5T$ in the specific example that we explore here, where we verified that $\tfrac{1}{2}\Delta(\Lambda_\tc{a}^+,\Lambda_\tc{b}^+)$ is $6$ orders of magnitude smaller than $\mathcal{N}(\hat{\rho}_\tc{ab}$. In Fig.~\ref{fig:quadratic}, we plot the entanglement acquired by the localized modes as a function of their energy gap. The plot is what is expected for the behaviour of entanglement harvesting in the Minkowski vacuum, where there is a threshold in the energy gap below which no entanglement can be extracted. For $\Omega T$ above this threshold, the entanglement peaks and quickly decays. 

    \begin{figure}[h!]
        \centering
        \includegraphics[width=12cm]{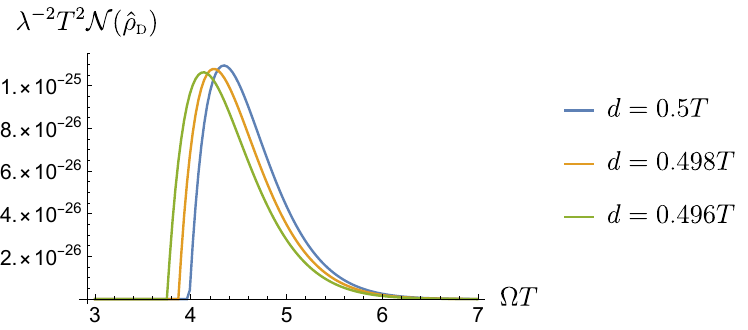}
        \caption{The negativity of the state of two localized quantum fields in boxes of sides $d$ when restricted to their lowest energy mode after spacelike interaction with a massless scalar field. The negativity is plotted as a function of the energy of the localized mode, $\Omega = \sqrt{m^2 + 3\pi/d^2}$. The time duration of the interaction $T$ is used as a scale. The separation between the detectors' interaction regions for these plots is $L = 4.5 T$.}
        \label{fig:box}
    \end{figure}

    As a second example, in Fig.~\ref{fig:box}, we consider the case where two massive fields in cubic cavities of size length $d$ with Dirichlet boundary conditions interact with a free massless scalar field. The box localization was discussed in Section~\ref{sec:LocalizedQuantumFields}. We consider the same choices of $\zeta_\tc{a}(\mf x)$ and $\zeta_\tc{b}(\mf x)$  as in Eq. \eqref{eq:switching}. We also restrict the localized fields to the lowest energy mode $\bm 1_\tc{a} = \bm 1_{\tc b} = (1,1,1)$, with energy $\omega^\tc{a}_{\bm 1_\tc{a}} = \omega^{\tc{b}}_{\bm 1_{\tc{b}}} = \sqrt{m^2 + 3\pi^2/d^2}$. To ensure that communication between the detectors is negligible, we pick $d \sim 0.5 T$ and consider the distance between the cavities to be given by $L = 5T$. The negativity in this case can be seen in Fig.~\ref{fig:box} as a function of the energy gap of the $\bm 1_{\tc{a}}$, $\bm 1_{\tc{b}}$ modes of the fields. The behaviour of the negativity is similar to most cases of entanglement harvesting in spacelike separated regions. We see more entanglement in this setup due to the smaller choice of $L$, which can be taken in this case because the communication between the detectors is naturally smaller due to the compact support of the modes in space.

\subsubsection*{Entanglement Harvesting with Unruh-DeWitt Detectors}

We will now review the protocol of entanglement harvesting in a context similar to the one discussed in~\cite{Pozas-Kerstjens:2015}, which could be considered the standard modern description of the protocol. This description uses two two-level Unruh-DeWitt detectors coupled to the amplitude of a real scalar quantum field. We will describe the protocol in a general globally hyperbolic spacetime, and later restrict it to inertial detectors in Minkowski spacetime for explicit examples. 

As discussed in Section~\ref{sec:UDW}, the first step to define an Unruh-DeWitt detector is to define a timelike coordinate $\tau$ that will be used to prescribe time evolution. With two detectors, one must pick a single timelike coordinate $\tau$ to prescribe the evolution of \text{both} detectors. This is because the unitary time evolution operator will generally involve time ordered interactions that involve products of spacetime smearing functions associated to each detector, and a single notion of time ordering must picked. For instance, one could define two two-level Unruh-DeWitt detectors by defining two trajectories $\mf z_\tc{a}(\tau_\tc{a})$ and $\mf z_\tc{b}(\tau_\tc{b})$ with respective energy gaps $\Omega_\tc{a}$ and $\Omega_\tc{b}$, such that the spacetime smearing functions $\Lambda_\tc{a}(\mf x)$ and $\Lambda_\tc{b}(\mf x)$ are supported the regions where the Fermi normal coordinates associated with each trajectory is defined. The interaction Hamiltonian density for the interaction of the two detectors with a real scalar quantum field $\hat{\phi}$ can then be written as
\begin{equation}\label{eq:HItwoUDWs}
    \hat{\mathcal{H}}_I(\mf x) = \lambda \Lambda_\tc{a}(\mf x) (e^{\ii \Omega_\tc{a} \tau_\tc{a}}\hat{\sigma}_\tc{a}^+ + e^{-\ii \Omega_\tc{a} \tau_\tc{a}}\hat{\sigma}_\tc{a}^-)\hat{\phi}(\mf x)+\lambda \Lambda_\tc{b}(\mf x) (e^{\ii \Omega_\tc{b} \tau_\tc{b}}\hat{\sigma}_\tc{b}^+ + e^{-\ii \Omega_\tc{b} \tau_\tc{b}}\hat{\sigma}_\tc{b}^-)\hat{\phi}(\mf x)
\end{equation}
acting in $\mathscr{H}_\tc{a}\otimes\mathscr{H}_\tc{b}\otimes \mathcal{F}(\mathscr{H})$ with $\mathscr{H}_\tc{a} = \mathscr{H}_\tc{b} = \mathbb{C}^2$ and $\mathcal{F}(\mathscr{H})$ being the Fock space of the field $\hat{\phi}$ in a suitable GNS representation. To fully define the model, we will also assume that it is possible to find a time coordinate $\tau$ that coincides with both $\tau_\tc{a}$ in the support of $\Lambda_\tc{a}(\mf x)$ and $\tau_\tc{b}$ in the support of $\Lambda_\tc{b}(\mf x)$. This is generally possible in globally hyperbolic spacetimes if the supports of $\Lambda_\tc{a}(\mf x)$ and $\Lambda_\tc{b}(\mf x)$ do not overlap by choosing appropriate time parametrizations such that the events $\mf z_\tc{a}(\tau_\tc{a} = 0)$ and $\mf z_\tc{b}(\tau_\tc{b} = 0)$ are spacelike separated.

We can compute the final state of the detectors after the interaction by applying the time evolution operator
\begin{equation}
    \hat{U}_I = \mathcal{T}_\tau\exp(-\ii \int \dd V \hat{\mathcal{H}}(\mf x)).
\end{equation}
For simplicity, we will consider the case where the detectors start in their respective ground states and the initial state of the field is a quasifree state $\omega$, represented as the density operator $\hat{\rho}_\phi$, so that the initial state of the complete system is $\hat{\rho}_0 = \ket{g_\tc{a}}\!\!\bra{g_\tc{a}}\otimes \ket{g_\tc{b}}\!\!\bra{g_\tc{b}}\otimes \hat{\rho}_\phi$. To leading order in $\lambda$, the final state of the two detector system can be written as
\begin{equation}\label{eq:UDWabFinal}
    \hat{\rho}_\tc{ab} = \begin{pmatrix}
        1 - \mathcal{L}_\tc{aa}^- - \mathcal{L}_\tc{bb}^- & 0 & 0 & \mathcal{M}^*\\
        0 & \mathcal{L}_\tc{bb}^- & \mathcal{L}_{\tc{ab}}^- & 0\\
        0 & (\mathcal{L}_\tc{ab}^-)^* & \mathcal{L}_\tc{aa}^- & 0\\
        \mathcal{M} & 0 & 0 & 0
    \end{pmatrix} + \mathcal{O}(\lambda^2)
\end{equation}
in the basis $\{\ket{g_\tc{a}},\ket{e_\tc{a}}\}\otimes \{\ket{g_\tc{b}},\ket{e_\tc{b}}\}$. The expressions for $\mathcal{M}$ and $\mathcal{L}_{\tc{ij}}^-$ are analogous to~\eqref{eq:Lij} and~\eqref{eq:M}, with $\mathcal{M} = - \lambda^2 G_F(\Lambda_\tc{a}^+,\Lambda_\tc{b}^+)$ and $\mathcal{L}_{\tc{ij}}^- = \lambda^2 W(\Lambda_\tc{i}^-,\Lambda_\tc{j}^+)$ and $\Lambda_\tc{i}^\pm(\mf x) = \Lambda(\mf x) e^{\pm \ii \Omega_\tc{i}\tau}$. In this case, the leading order negativity between the two states is also given by Eq.~\eqref{eq:negab2fields}, and our previous analysis of entanglement harvesting applies equally to the case of two two-level Unruh-DeWitt detectors that start their interaction in the ground state.

Notice that when the detectors start in their ground state, the final state of detectors system does not depend explicitly on the time ordering with respect to $\tau$, so that the results in this setup are covariant to leading order in $\lambda$. Indeed, in~\cite{us2}, it was shown that even when multiple detectors are considered, if their initial state commutes with their collective free Hamiltonians, their final state is independent of the choice of time parameter that prescribes the evolution. More generally, in~\cite{hectorCov}, it was shown that the negativity of the final two detectors state is independent of the time ordering operation for any choice of pure state (although the final state might not be). Specifically, it was shown that considering the initial state $\hat{\rho}_o = \ket{\psi_\tc{a}}\!\!\bra{\psi_\tc{a}}\otimes \ket{\psi_\tc{b}}\!\!\bra{\psi_\tc{b}}\otimes \hat{\rho}_\phi$ with
\begin{equation}
    \ket{\psi}_\tc{a} = \cos \alpha_\tc{a} \ket{g_\tc{a}} - e^{\ii \beta_\tc{a}}\sin\alpha_\tc{a} \ket{e_\tc{a}}, \quad \quad 
    \ket{\psi}_\tc{b} = \cos \alpha_\tc{b} \ket{g_\tc{b}} - e^{\ii \beta_\tc{b}}\sin\alpha_\tc{b} \ket{e_\tc{b}},
\end{equation}
the final state of the detectors can be written as
\begin{equation}
    \hat{\rho}_\tc{ab} =  \begin{pmatrix}
        1 - \mathcal{L}_\tc{aa}^\text{gen} - \mathcal{L}_\tc{bb}^\text{gen} & \mathcal{X}^* & \mathcal{Y}^* & (\mathcal{M}^\text{gen})^*\\
        \mathcal{X} & \mathcal{L}_\tc{bb}^\text{gen} & \mathcal{L}_{\tc{ab}}^\text{gen} & 0\\
        \mathcal{Y} & (\mathcal{L}_\tc{ab}^\text{gen})^* & \mathcal{L}_\tc{aa}^\text{gen} & 0\\
        \mathcal{M}^\text{gen} & 0 & 0 & 0
    \end{pmatrix} + \mathcal{O}(\lambda^2),
\end{equation}
where 
\begin{align}
    \mathcal{L}_{\tc{ij}}^\text{gen} &= \lambda^2 W\left(\cos^2\alpha_\tc{j}\,\Lambda_\tc{i}^- - e^{- 2 \ii \beta_\tc{j}}\sin^2\alpha_\tc{j} \,\Lambda_\tc{i}^+, \cos^2 \alpha_\tc{i} \,\Lambda_\tc{j}^+ - e^{2 \ii \beta_\tc{i}} \sin^2 \alpha_\tc{i} \,\Lambda_\tc{j}^-\right),\\
    \mathcal{M}^\text{gen} &= \lambda^2 G_F\left(\cos^2 \alpha_\tc{a} \,\Lambda_\tc{a}^+ - e^{2 \ii \beta_\tc{a}} \sin^2 \alpha_{\tc{a}} \,\Lambda_\tc{a}^-, \cos^2 \alpha_\tc{b} \,\Lambda_\tc{b}^+ - e^{2 \ii \beta_\tc{b}}\sin^2 \alpha_\tc{b}\,\Lambda_\tc{b}^-\right),
\end{align}
and the $\mathcal{X}$ and $\mathcal{Y}$ terms explicitly depend on the choice of time parameter $\tau$~\cite{hectorCov}. The leading order negativity of the state above then reads
\begin{align}\label{eq:neggen}
    \mathcal{N}(\r_\tc{ab}) &= \max\left(0,\sqrt{|\mathcal{M}^\text{gen}|^2 + \left(\tfrac{\mathcal{L}^\text{gen}_\tc{aa} - \mathcal{L}^\text{gen}_\tc{bb}}{2}\right)^2} - \tfrac{\mathcal{L}^\text{gen}_{\tc{aa}} + \mathcal{L}^\text{gen}_\tc{bb}}{2}\right) + \mathcal{O}(\lambda^4).
\end{align}
The fact that the expression above depends only on $\mathcal{M}^\text{gen}$ and $\mathcal{L}_{\tc{ij}}^\text{gen}$ then shows that the leading order entanglement between the detectors is indeed independent of the specific choice of time parameter $\tau$.

\subsubsection*{An Explicit Example with two-level Unruh-DeWitt detectors}

We now study an explicit example of entanglement harvesting with two-level Unruh-DeWitt detectors. We will consider the case where the detectors are inertial in Minkowski spacetime, defined by the prototypical spacetime smearing functions
\begin{align}\label{eq:prototypeAB}
    \Lambda_\tc{a}(\mf x) = e^{- \frac{t^2}{2 T^2}} \frac{e^{-\frac{|\bm x|^2}{2 \sigma^2}}}{(2\pi \sigma^2)^\frac{3}{2}},\quad \quad
    \Lambda_\tc{b}(\mf x) = e^{- \frac{(t-t_0)^2}{2 T^2}} \frac{e^{-\frac{|\bm x-\bm L|^2}{2 \sigma^2}}}{(2\pi \sigma^2)^\frac{3}{2}},
\end{align}
written in inertial coordinates $(t,\bm x)$. This choice defines the interaction regions of the detectors to be spacetime Gaussians of spatial width $\sigma$ and effective time duration controlled by the parameter $T$. The interaction regions are shifted in space by $\bm L$ and in time by $t_0$ with respect to the inertial frame. Due to the fact that the spacetime smearing functions of Eq.~\eqref{eq:prototypeAB} differ only by spacetime translations, we find that $\mathcal{L}_{\tc{aa}}^{-} = \mathcal{L}_{\tc{bb}}^{-} = \mathcal{L}$. This scenario has been studied multiple times in the literature (see e.g.~\cite{Pozas-Kerstjens:2015,Pozas2016,ericksonNew,hectorMass,Ng1}). However, it was only in the $\sigma \to 0$ limit that analytical results were found for the relevant smeared bi-distributions necessary to compute the negativity. While the $\mathcal{L}$ term can be evaluated analytically and is given by Eq.~\eqref{eq:Wpm}, to numerically evaluate $G(\Lambda_\tc{a}^+,\Lambda_\tc{b}^+)$ effectively, one usually writes, in momentum space~\cite{Pozas-Kerstjens:2015},
\begin{align}\label{eq:GABk}
    G(\Lambda_\tc{a}^+,\Lambda_\tc{b}^+) = \frac{\lambda^2 T^2e^{\ii \Omega t_0}}{4 \pi }&\!\int \dd |\bm k|\,|\bm k| e^{- |\bm k|^2 \sigma^2} e^{-(\Omega^2 + |\bm k|^2)T^2} \text{sinc}(|\bm k||\bm L|)\\
    &\times \Bigg(e^{-\ii |\bm k| t_0}\text{erfc}\left(\ii |\bm k|T -\frac{t_0}{2T}\right)+e^{\ii |\bm k| t_0}\text{erfc}\left(\ii |\bm k|T +\frac{t_0}{2T}\right)\Bigg),\nonumber
\end{align}
which, up to this point, could not be solved in terms of elementary functions.

On the other hand, the results of~\cite{analytical} (that we display in Appendix~\ref{app:analytical}) give the exact value of the relevant propagators in this case. We find
\begin{align}
    G(\Lambda_\tc{a}^+,\Lambda_\tc{b}^+) =
    &\frac{\lambda^2 T^2e^{-\Omega^2T^2}e^{\ii \Omega t_0}}{8 \sqrt{\pi} |\bm L| \sqrt{T^2+\sigma^2}}\Bigg(e^{- \frac{(|\bm L| + t_0)^2}{4(T^2 + \sigma^2)}}\text{erfi}\left(\frac{|\bm L| +t_0}{2\sqrt{T^2 + \sigma^2}}\right)+e^{- \frac{(|\bm L| - t_0)^2}{4(T^2 + \sigma^2)}}\text{erfi}\left(\frac{|\bm L| -t_0}{2\sqrt{T^2 + \sigma^2}}\right)\nonumber\\
    &\:\:\:\:-\ii\left(e^{-\frac{(|\bm L| + t_0)^2}{4 (T^2 + \sigma^2)}}\text{erf}\left(\frac{|\bm L|T^2 - t_0\sigma^2 }{2 T \sigma\sqrt{T^{2} + \sigma^{2}}}\right)+e^{-\frac{(|\bm L| - t_0)^2}{4 (T^2 + \sigma^2)}} \text{erf}\left(\frac{|\bm L|T^2 + t_0\sigma^2 }{2 T \sigma\sqrt{T^{2} + \sigma^{2}}}\right)\right)\Bigg).\label{eq:GAB}
\end{align}
We have also checked that the numerical integration of Eq.~\eqref{eq:GABk} matches the results of Eq.~\eqref{eq:GAB} for numerous parameters $|\bm L|$, $t_0$, $T$, $\sigma$ and $\Omega$. Combining the equation above with Eq.~\eqref{eq:Wpm}, one then finds a closed-form analytical expression for the negativity. In particular, for the case where there is no time separation between the interactions, $t_0 = 0$, we obtain, with $\alpha = \sqrt{1 + \sigma^2/T^2}$,{\footnotesize
\begin{equation}
    \mathcal{N}(\hat{\rho}_\tc{ab}) = \frac{\lambda^2 e^{-\Omega^2 T^2}}{4\pi \alpha^2 } \left(\sqrt{\pi}\alpha e^{- \frac{|\bm L|^2}{4\alpha^2T^2}}\frac{T}{|\bm L| }\sqrt{\text{erf}\left(\frac{|\bm L|}{2 \alpha \sigma}\right)^2 + \text{erfi}\left(\frac{|\bm L|}{2\alpha T}\right)^2}  + \frac{\sqrt{\pi}\Omega T}{\alpha}e^{\frac{\Omega^2T^2}{\alpha^2}}\text{erfc}\left(\frac{\Omega T}{\alpha}\right)- 1 \right),\label{eq:negAnal}
\end{equation}}
whenever the quantity above is positive. 

We can now explicitly check for which values of the parameters $L$, $T$, $\sigma$ and $\Omega$ the communication between the detectors is negligible. The imaginary part of the Feynman propagator at $t_0 = 0$ reads
\begin{equation}\label{eq:DAB}
    \tfrac{1}{2}|\Delta(\Lambda_\tc{a}^+,\Lambda_\tc{b}^+)| = \frac{e^{-\Omega^2 T^2}}{4\alpha \sqrt{\pi}} \frac{T}{|\bm L|}e^{- \frac{|\bm L|^2}{4\alpha^2T^2}}\text{erf}\left(\frac{|\bm L|}{2 \alpha \sigma}\right).
\end{equation}
We then see that the $\text{erf}$ term in Eq.~\eqref{eq:negAnal} comes exclusively from the signalling between the detectors. Eq.~\eqref{eq:DAB} can then be used to estimate the signalling between the detectors, so we are looking for situations that configure entanglement harvesting, in which $\mathcal{N}(\hat{\rho}_\tc{ab})>0$ and $\frac{1}{2}|\Delta(\Lambda_\tc{a}^+,\Lambda_\tc{b}^+)|\ll\mathcal{N}(\hat{\rho}_\tc{ab})$. From the expression above, one also confirms the exponential decay of the signalling between two Gaussian detectors, which was seen in~\cite{mariaPipoNew}.

\begin{figure}[h!]
    \centering
    \includegraphics[width=12cm]{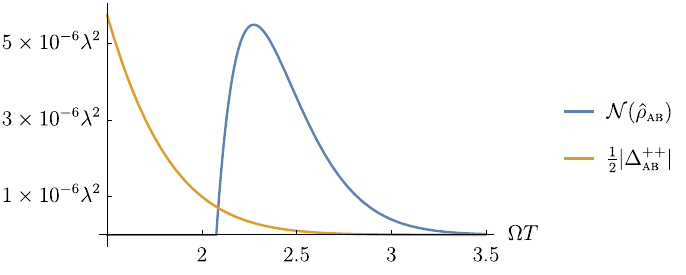}
    \caption{The negativity and signalling contribution for two detectors interacting with a massless scalar field in Gaussian spacetime regions separated by $|\bm L| = 5 T$, with detector sizes $\sigma = 0.01T$.}
    \label{fig:n}
\end{figure}

In Fig.~\ref{fig:n}, we plot the negativity and the signalling contribution as a function of the detectors' energy gap for $\sigma = 0.01 T$ when the detectors are separated by a distance of $|\bm L| = 5 T$. We find that the negativity becomes an order of magnitude larger than the signalling from the moment at which it peaks, and this ratio continues to increase as $\Omega T$ increases.In this setup, one can safely associate the entanglement between the detectors to field entanglement whenever $\Omega T > L/2T$.

\section{General Results of Entanglement Harvesting}\label{sec:GeneralEnt}

In this final Section we will review general results regarding entanglement harvesting, and discuss how the observed entanglement that can be harvested by localized probes relates to the general results about entanglement in quantum field theory discussed in Section~\ref{sec:modeEntanglement}.

\subsubsection*{No-go Theorem for Entanglement Extraction}

As we have seen, two causally disconnected detectors can end up entangled through their interaction with a field. Indeed, many different scenarios were found where spacelike separated detectors can extract entanglement from a quantum field (see e.g.~\cite{Reznik2005,Retzker2005,Reznik2007,Salton:2014jaa,Pozas-Kerstjens:2015,Ng1,Henderson2019,bandlimitedHarv2020,ampEntBH2020,HarvestingDelocalized,ericksonNew,HarvestingAccelerationRobb,twist2022,cisco2023harvesting,SchwarzchildHarvestingWellDone}). On the other hand, there are also many regimes where spacelike entanglement cannot be harvested. In particular, there are families of such scenarios where the harvesting of entanglement is forbidden that are covered by a \textit{no-go theorem}~\cite{nogo}, which we briefly summarize below.

When two detectors interact with the field in spacelike separated regions, we can factor the unitary time evolution $\hat{U}_I$ as 
\begin{equation}
    \hat{U}_I  = \hat{U}_{\tc{a}\phi}\hat{U}_{\tc{b}\phi}= \hat{U}_{\tc{b}\phi}\hat{U}_{\tc{a}\phi},
\end{equation}
where $\hat{U}_{\tc{a}\phi}$ acts only on detector A and on the field, and $\hat{U}_{\tc{b}\phi}$ only acts on B and on the field. We can factor $\hat{U}_I$ as such because field operators smeared against the interaction regions A and B commute (as the supports of $\Lambda_\tc{a}$ and $\Lambda_\tc{b}$ are spacelike separated and the field satisfies the microcausality condition), and because observables that act on detector A commute with observables of detector B. Notice that because the unitary $\hat{U}_I$ factors as the product of two unitaries that act on systems A and B separately, $\hat{U}_I$ is unable to directly couple A and B. However, as we previously mentioned, the field degrees of freedom supported in the interaction regions A and B can be entangled, which might allow entanglement to be exchanged between the field in regions A and B and the detectors.

The no-go theorem in~\cite{nogo} points out specific cases where the unitary time evolution prescribed by the interaction with the field can be written as a simple-generated unitary. That is, when $\hat{U}_\tc{a} = e^{- \ii \hat{m}_\tc{a}\otimes \hat{X}_\tc{a}}$ or $\hat{U}_\tc{b} = e^{- \ii \hat{m}_\tc{b}\otimes \hat{X}_\tc{b}}$ for operators $\hat{m}_{\tc{a}}$ and $\hat{m}_{\tc{b}}$ that act in the respective detectors Hilbert spaces and field observables $\hat{X}_\tc{a}$ and $\hat{X}_\tc{b}$ localized in each detector's coupling regions. In this case, it is possible to show that at least one of the commuting quantum channels implemented in each detector is an entanglement breaking channel, implying that the two detectors end up in a separable state after their interaction with the field. In these cases, entanglement harvesting is not possible.

The two notable cases where $\hat{U}_\tc{a}$ and $\hat{U}_\tc{b}$ are simple-generated unitaries are the case of two-level gapless detectors and the particular case of two-level delta-coupled detectors. 

In the case of gapless detectors, we have $\Omega_\tc{a} = \Omega_\tc{b} = 0$ so that the free evolution of the detectors' monopole moments is trivial: $\hat{\mu}_\tc{i}(t) = \hat{\mu}_\tc{i}$. In this case, $[[\hat{\mathcal{H}}_I(\mf x),\hat{\mathcal{H}}_I(\mf x')],\hat{\mathcal{H}}_I(\mf x'')] = 0$, so that the unitary time evolution operator for each detector can be computed using the Magnus expansion~\cite{HarvestingQueNemLouko,Landulfo}:
\begin{equation}
    \hat{U}_{\tc{a},\phi} = e^{\ii\varphi_\textsc{a}}\,e^{- \ii \lambda \hat{\mu}_\tc{a} \hat{\phi}(\Lambda_\tc{a})}, \quad\quad 
    \hat{U}_{\tc{b},\phi} = e^{\ii\varphi_\textsc{b}}\,e^{- \ii \lambda \hat{\mu}_\tc{b} \hat{\phi}(\Lambda_\tc{b}) },
\end{equation}
for real phases $\varphi_\textsc{a},\varphi_\textsc{b}$. The unitaries $\hat{U}_{\tc{a},\phi}$ and $\hat{U}_{\tc{b},\phi}$  are then simple-generated unitaries, thus implementing an entanglement breaking channel, as explicitly discussed in~\cite{nogo}.

As we discussed in Section~\ref{sec:UDW}, the delta-coupled case is a particular case of gapless detectors. As such, it is not possible to harvest entanglement with two spacelike separated delta-coupled detectors. Explicitly, in the delta coupled limit, the spacetime smearing functions can be written as $\Lambda_\tc{i}(\mf x) = \eta \delta(t-t_\tc{i}) F_\tc{i}(\bm x)$, where $t_\tc{i}$ are the times at which the (sudden) couplings happen. $\eta$ is a parameter with dimensions of time, and $F_\tc{i}(\bm x)$ is a smearing function that defines the spatial profile of the interaction regions. We then have 

\begin{equation}
    \hat{U}_{\tc{a},\phi} = e^{- \ii \lambda \eta \hat{\mu}_\tc{a}(t_\tc{a}) \hat{\Phi}_{t_\tc{a}}(F_\tc{a})}, \quad \hat{U}_{\tc{b},\phi} = e^{- \ii \lambda \eta \hat{\mu}_\tc{b}(t_\tc{b}) \hat{\Phi}_{t_\tc{b}}(F_\tc{b})},
\end{equation}
which are simple-generated. Notice, however, that it is possible to harvest entanglement if the detectors' coupling is given by \textit{multiple} sudden interactions, that is, when the coupling is described by a linear combination of a sufficiently large number of terms, each represented by a delta coupling~\cite{JoseEdu2024Nonperturbative}.

Overall, the no-go theorem implies that the detectors must have non-trivial internal dynamics to be able to extract entanglement from a quantum field, and that detectors must couple to the field for a sufficiently long time to become entangled with each other. In particular, the no-go theorems also imply that our techniques for solving the dynamics of detectors non-perturbatively cannot be applied to the entanglement harvesting protocol. This fact significantly limits our ability to study the protocol in more general settings, such as the limit of large coupling constants, which could provide significant optimizations to entanglement harvesting.

\subsubsection*{The Behaviour of Entanglement Harvesting with Distance and the UV-IR Connection}

Although obtaining general results about the entanglement harvesting protocol is challenging, explicit examples might be used to confirm the general results about the behaviour of vacuum entanglement in quantum field theory discussed in Section~\ref{sec:modeEntanglement}. In particular, we will briefly show that the example of entanglement harvesting in Minkowski spacetime~\eqref{eq:negAnal} confirms the exponential decay of entanglement with distance, as well as the fact that there is entanglement between any two regions, and that entanglement harvesting also displays a UV-IR connection. These three results involve the asymptotic limit of $|\bm L|\to \infty$ of the setup, which we will now focus on.

Having the analytical expression for the negativity allows one to estimate the behaviour of the entanglement that can be harvested by detectors that couple to a massless field in the limit of large $|\bm L|$. This is done by considering the asymptotic expansion of Eq.~\eqref{eq:GAB}. One finds
\begin{equation}
    |G(\Lambda_\tc{a}^+,\Lambda_\tc{b}^+)| = \frac{\lambda^2e^{- \Omega^2 T^2}}{2\pi} \left(\frac{T^2}{|\bm L|^2} + \frac{2T^2(T^2+\sigma^2)}{|\bm L|^4}+ \mathcal{O}\left(|\bm L|^{-6}\right)\right)\!.
\end{equation}
Notice, however, that the term above must be larger than the local vacuum excitations of the detectors to ensure that any entanglement can be harvested at all. The vacuum noise is independent of $|\bm L|$, but, as has been shown first in~\cite{hectorMass}, there is always a large enough value of the energy gap $\Omega$, which allows the Gaussian detectors to harvest entanglement. To see which value of $\Omega$ maximizes the negativity, we differentiate Eq.~\eqref{eq:neg} assuming $\sigma\ll T$, and set the result to zero, which yields the same asymptotic result for $\Omega$ that was first found in~\cite{hectorMass}: $\Omega T \sim | \bm L|/(2T)$. 

In other words, the value of $\Omega$ that maximizes the entanglement that can be extracted by the detectors is proportional to the distance between them. This can be seen as the analogue of the UV-IR correspondence for the entanglement harvesting setup, showcasing that the detectors must have arbitrarily high energies to be able to extract entanglement from regions asymptotically separated in space.

Having the optimal value of $\Omega$ allows us to evaluate the negativity at said energy gap, and to consider the asymptotic limit of the resulting expression. We find that for $\Omega T \sim | \bm L|/(2T)$, the negativity is positive, and has the asymptotic expression
\begin{equation}\label{eq:negAnalAsympt}
    \mathcal{N}(\hat{\rho}_{\tc{ab}}) = \frac{4\lambda^2 e^{-\frac{|\bm L|^2}{4 T^2}}}{\pi }\frac{T^4}{|\bm L|^4} + \mathcal{O}\left(\frac{T^6}{|\bm L|^6}e^{-\frac{|\bm L|^2}{4 T^2}}\right).
\end{equation}
The fact that the negativity is positive confirms that two detectors can be used to extract entanglement from any two arbitrarily separated regions of a quantum field. In this limit, of course, the communication between the detectors is negligible, as they are too far away to signal to each other\footnote{Indeed, the signalling term, in this case, decays as $e^{- |\bm L|^2/2T^2}$.}. This result then confirms that not only are any two independent regions of a quantum field are entangled, but that this entanglement is accessible to localized probes.

Finally, Eq.~\eqref{eq:negAnalAsympt} also gives the asymptotic behaviour of the entanglement between the two regions. The maximum entanglement that can be harvested by a pair of inertial detectors that couple to a massless field in Gaussian spacetime regions decays as $\ell^{-4}e^{-\ell^2/4}$. This decay is certainly faster than a simple exponential decay, was argued in~\cite{KlcoUVIR}, but, as we previously discussed, the entanglement that a pair of detectors can extract is upper bounded by the entanglement present in the field, so it is no surprise that the decay of the negativity in~\eqref{eq:negAnalAsympt} is faster than exponential. This faster decay rate of entanglement with the distance also showcases how unoptimized the specific setups described in Section~\ref{sec:OperationallyAccessingEnt} are, and begs the question of what the optimal setup for the protocol of entanglement harvesting would be.

\textcolor{white}{I sincerely considered adding two more sections here about entanglement harvesting from complex scalar, fermionic and gravitational fields. I guess the thesis already turned up too long.}

%

\definecolor{QC}{RGB}{150,120,100}

\chapter{When is Quantum Field Theory Necessary?}\label{chap:qc}

    Although quantum field theory is the most accurate description of matter, many physical setups do not require a full description in terms of quantum fields. An example of this fact is the effective description of external potentials as classical fields that we employed in our discussions of localized fields. In this Chapter, we will discuss an effective theory that allows one to describe two systems that interact via a quantum field that neglects the degrees of freedom of the field mediating the interaction while maintaining some relativistic aspects of the interaction. The explicit model that we will discuss here will be referred to as the quantum-controlled model. It was introduced in~\cite{quantClass} in an attempt to understand the roles that quantum degrees of freedom of a quantum field actively play in relativistic quantum information protocols, such as entanglement harvesting. We will define the model in Section~\ref{sec:QCmodels}, and study the limit in which it approximates interactions through quantum fields in Section~\ref{sec:QFTapproxQC}. As an application of the model, in Section~\ref{sec:GME} we will use the quantum-controlled model to describe the recent proposals of gravity mediated entanglement experiments introduced in~\cite{B,MV}.

\section{Quantum-Controlled Models}\label{sec:QCmodels}

    A simple but pedagogical example is the interaction of two spins (labelled by A and B), usually modelled by the J-coupling:
    \begin{equation}\label{eq:Jcoupling}
        \hat{H}_I(t) = -J \hat{\bm \sigma}_\tc{a}(t)\cdot \hat{\bm \sigma}_\tc{b}(t),
    \end{equation}
    where $\hat{\bm \sigma}_\tc{a}$ and $\hat{\bm \sigma}_\tc{b}$ are the sigma vectors associated to each system. This is a direct coupling between the spins, which certainly does not take into account relativistic aspects of the interaction. However, physical couplings compatible with relativistic principles must be local. 
    
    The interaction between spins~\eqref{eq:Jcoupling} turns out to be a consequence of the fact that each of the spins locally couple to the magnetic field. Indeed, in Section~\ref{sec:MoreRealisticProbes}, we discussed a quantum field theoretic description for an atom and how this description naturally gives rise to a coupling of effective spin degrees of freedom with an external quantum magnetic field. Effectively, a spin at $\bm x_0$ couples to the magnetic field according to the interaction
    \begin{equation}\label{eq:sigmadotBhat}
        \hat{H}_I(t) = - \gamma\, \hat{\bm \sigma}(t) \cdot \hat{\bm B}(t,\bm x_0). 
    \end{equation}
    One way of understanding the J-coupling locally (but without invoking the quantum degrees of freedom of the magnetic field) is to consider that each spin couples locally to the magnetic field at their respective locations $\bm x_\tc{a}$ and $\bm x_\tc{b}$:
    \begin{equation}\label{eq:HintAB}
        \hat{H}_{I,\tc{a}}(t) = - \gamma \hat{\bm \sigma}_\tc{a}(t)\cdot\hat{\bm B}(t,\bm x_\tc{a}),\, \quad \quad 
        \hat{H}_{I,\tc{b}}(t) = - \gamma \hat{\bm \sigma}_\tc{b}(t)\cdot\hat{\bm B}(t,\bm x_\tc{b}), 
    \end{equation}
    where $\hat{\bm B}(t,\bm x)$ denotes the magnetic field at each point of spacetime. Due to the local couplings~\eqref{eq:HintAB}, each spin sources a magnetic field associated with a magnetic dipole $-\gamma\hat{\bm \sigma}_\tc{a/b}$. Using a non-relativistic approach, we can write 
    \begin{align}
        \hat{\bm B}_\tc{a}(t,\bm x) &= \frac{\gamma}{4 \pi |\bm x - \bm x_\tc{a}|^3}\left(\hat{\bm \sigma}_\tc{a} - 3  (\hat{\bm\sigma}_\tc{a}\cdot\big(\bm x \!-\! \bm x_\tc{a})\big)\tfrac{\bm x - \bm x_\tc{a}}{|\bm x - \bm x_\tc{a}|^2}\right),\label{eq:BaBb}\\
        \hat{\bm B}_\tc{b}(t,\bm x) &= \frac{\gamma}{4 \pi |\bm x - \bm x_\tc{b}|^3}\left(\hat{\bm \sigma}_\tc{b} - 3  \big(\hat{\bm\sigma}_\tc{b}\cdot(\bm x \!-\! \bm x_\tc{b})\big)\tfrac{\bm x - \bm x_\tc{b}}{|\bm x - \bm x_\tc{b}|^2}\right).\nonumber
    \end{align}
    One can then recover a more general form of the J-coupling by replacing $\hat{\bm B}$ in Eq.~\eqref{eq:sigmadotBhat} by the magnetic field sourced by the spins, $\hat{\bm B}\mapsto \hat{\bm B}_\tc{a}+\hat{\bm B}_\tc{b}$, ignoring the self-interaction terms:
    \begin{equation}
        \hat{H}_I(t) = \tfrac{1}{2}(\hat{H}_{\tc{a},I}(t) + \hat{H}_{\tc{a},I}(t)) = - \frac{\gamma^2}{2\pi |\bm r_{\tc{ab}}|^3}\left(\hat{\bm \sigma}_\tc{a}(t)\cdot \hat{\bm \sigma}_\tc{b}(t) - \tfrac{3}{|\bm r_\tc{ab}|^2}  \big(\hat{\bm\sigma}_\tc{a}(t)\cdot\bm r_{\tc{ab}}\big)\big(\hat{\bm\sigma}_\tc{b}(t)\cdot\bm r_{\tc{ab}}\big)\right),
    \end{equation}
    where we defined $\bm r_\tc{ab} = \bm x_\tc{a} - \bm x_\tc{b}$. The added factor of $1/2$ arises due to the energy stored in the magnetic field itself. This extra factor has been discussed in more detail in the Appendix of~\cite{eirini} for the case of a scalar field and the discussion naturally generalizes to linear couplings in more general field theories, such as electromagnetism.
    
    Notice that the magnetic fields $\hat{\bm B}_\tc{a/b}$ in~\eqref{eq:BaBb} are not quantum fields in the typical sense. In fact, the ``fields'' defined in Eq.~\eqref{eq:BaBb} do not have any independent degrees of freedom: they instead propagate the degrees of freedom of A and B to all points of spacetime in a non-relativistic manner. In this sense, the degrees of freedom of the ``field'' $\hat{B}$ are entirely determined by the quantum sources A and B. 

    The model presented above for the non-relativistic interaction of spins through a magnetic field is what inspires the quantum-controlled model (qc-model), which we present below.

\subsubsection*{The Quantum Controlled Model}

Consider a globally hyperbolic spacetime $\M$ and two non-relativistic quantum systems A and B described with respect to the trajectories $\mf z_\tc{a}(\tau_\tc{a})$ and $\mf z_\tc{b}(\tau_\tc{b})$ according to the formulation presented in Section~\ref{sec:NRLQS}. Here $\tau_\tc{a}$ and $\tau_\tc{b}$ denote their respective Fermi normal coordinate times, and we assume each system to be defined in non-overlapping worldtubes around $\mf z_\tc{a}$ and $\mf z_\tc{b}$, so that we can consider a single global timelike coordinate $\tau$ such that $\tau = \tau_\tc{a}$ within the support of system A and $\tau = \tau_\tc{b}$ along the support of system B. Also consider a free \textit{classical} field theory for a tensor field $\phi$ with equation of motion $P\phi = 0$, and a (also tensor-valued) real observable $O(\mf x) = L\phi(\mf x)$, where $L$ is a linear operator. 

We assume that the classical field couples to systems A and B through couplings of the form $\lambda \hat{\jmath}_\tc{a}(\mf x) \cdot O(\mf x)$ and $\lambda \hat{\jmath}_\tc{b}(\mf x) \cdot O(\mf x)$, where $\hat{\jmath}_\tc{a}(\mf x)$ and $\hat{\jmath}_\tc{b}(\mf x)$ are (ideally compactly supported) tensor-valued self-adjoint operator currents\footnote{In the context of the description provided in Section~\ref{sec:NRLQS}, each component of the operator-valued currents in their respective extended Fermi frame would be written as
\begin{equation}
    \hat{\jmath}_\tc{a}^a(\mf x) = \int \dd\Sigma_\tc{a} f_\tc{a}^a(\mf x) \ket{\bm x_\tc{a}}\!\!\bra{\bm x_\tc{a}}, \quad \quad 
    \hat{\jmath}_\tc{b}^a(\mf x) = \int \dd\Sigma_\tc{b} f_\tc{b}^a(\mf x) \ket{\bm x_\tc{b}}\!\!\bra{\bm x_\tc{b}},
\end{equation}
where $f_\tc{a}^a(\mf x)$ and $f_\tc{b}^a(\mf x)$ are scalar functions in spacetime for each Lorentz index $a$.} that incorporate the free dynamics of systems A and B (that is, satisfying Eq.~\eqref{eq:HeisenbergPD}). The coupling of each system with the classical field is then associated with the respective Hamiltonian densities\footnote{One can generalize these interactions to the case where the operator currents are not self-adjoint and $O(\mf x)$, but we will restrict ourselves to the real self-adjoint case for simplicity.}
\begin{equation}
    \hat{\mathcal{H}}_{\tc{a},I}(\mf x) = \lambda \hat{\jmath}_\tc{a}(\mf x) \cdot O(\mf x),\quad \quad \hat{\mathcal{H}}_{\tc{b},I}(\mf x) = \lambda \hat{\jmath}_\tc{b}(\mf x) \cdot O(\mf x).
\end{equation}
Using the Hamiltonian density above, one can find the effect of the classical operator $O(\mf x)$ in systems A and B by applying the time evolution associated with each of the local Hamiltonian densities. In a sense, this is a classical version of a particle detector model, where the field $\phi(\mf x)$ is classical.

If the field were coupled to a classical current $\hat{\jmath}(\mf x)$ with a coupling of the form $j(\mf x)\cdot O(\mf x)$, one would be able to find the backreaction in the field $\phi$, by solving the equations of motion
\begin{equation}
    P\phi = L^* j,
\end{equation}
where $L^*$ is the linear operator defined by
\begin{equation}
    \int \dd V L\phi(\mf x) \cdot j(\mf x) = \int \phi(\mf x) \cdot L^*j(\mf x).
\end{equation}
To leading order in the coupling constant, the effect of the source $j(\mf x)$ on the field can be computed through the field's retarded Green's function, $G_R$ by
\begin{equation}
    \phi(\mf x) = G_R L^* j(\mf x),
\end{equation}
In particular, the leading order linear operator $O(\mf x)$ sourced by systems A and B can be written as
\begin{equation}
    O(\mf x) =  \tilde{G}_Rj(\mf x)\coloneqq L G_RL^*j(\mf x).
\end{equation}
Essentially, $j(\mf x)$ is the effective source for the operator $O(\mf x)$, and $\tilde{G}_R = LG_RL^*$ is its effective Green's function. For instance, in the case of the coupling of spins previously discussed, $\phi$ corresponds to the electromagnetic potential $A_\mu$, $j$ corresponds to a magnetic dipole, $L = \bm\nabla\times$ is the curl so that $L^*j$ yields the four-current associated with the dipole. The operator $\tilde{G}_R$ applied to $j$ then gives Amp\'ere's law.

In the context of the interactions of the quantum systems A and B with the field $\phi$, we then define the quantum-controlled observables
\begin{align}
    \hat{O}_\tc{a}(\mf x) &= \int \dd V' \tilde{G}_R(\mf x,\mf x') \cdot \hat{\jmath}_\tc{a}(\mf x'),\label{eq:OaOb}\\
    \hat{O}_\tc{b}(\mf x) &= \int \dd V' \tilde{G}_R(\mf x, \mf x') \cdot \hat{\jmath}_\tc{b}(\mf x').\nonumber
\end{align}
These are essentially a generalization of the quantum-controlled magnetic fields sourced by the spins A and B in~\eqref{eq:BaBb}, incorporating the retarded propagation of the field $\phi$. Notice that we have not considered the coupling constant $\lambda$ in the definitions for the qc-fields in~\eqref{eq:OaOb}. This will be convenient for writing the interaction Hamiltonian density in the qc-model.

The qc-model for the interaction between the spins is then defined by the interaction Hamiltonian density
\begin{align}
    \hat{\mathcal{H}}_{\text{qc}}(\mf x) &= \frac{\lambda^2}{2} \left(\hat{\jmath}_\tc{a}(\mf x)\cdot \hat{O}_\tc{b}(\mf x) + \hat{\jmath}_\tc{b}(\mf x)\cdot \hat{O}_\tc{a}(\mf x) \right)\\
    &= \frac{\lambda^2}{2} \int \dd V' \left(\hat{\jmath}_\tc{a}(\mf x)\cdot \tilde{G}_R(\mf x, \mf x')\cdot \hat{\jmath}_\tc{b}(\mf x') + \hat{\jmath}_\tc{b}(\mf x)\cdot \tilde{G}_R(\mf x, \mf x')\cdot \hat{\jmath}_\tc{a}(\mf x') \right).\label{eq:qcHIgen}
\end{align}
The quantum-controlled model then considers a direct coupling between systems A and B that ignores self-interactions and respects the causal propagation of signals imposed by the relativistic description of the background spacetime. For instance, if $\hat{\jmath}_\tc{a}(\mf x)$ and $\hat{\jmath}_\tc{b}(\mf x)$ are spacelike separated, $\hat{\mathcal{H}}_\tc{qc}(\mf x)$ identically vanishes, and no interaction between A and B takes place. However, the qc-model is not fully compatible with relativistic causality, as was discussed in detail in~\cite{eirini}. We will briefly mention more about this incompatibility later in this chapter.

Notice that the time evolution of systems A and B generated by the qc-model is \textit{unitary}, reflecting the fact that only systems A and B participate in the interaction. This is unlike the case where the interaction is mediated by a field with quantum degrees of freedom, which also becomes entangled with the sources. The qc-interaction is also second order in the coupling constant. This is due to the fact that it directly couples the currents, which are each proportional to the coupling constant. The interaction then incorporates the dynamics of the classical field $\phi$ through the retarded propagators, but it does not incorporate its degrees of freedom, instead propagating the sources themselves. Also notice that the qc-interaction Hamiltonian density~\eqref{eq:qcHIgen} satisfies
\begin{equation}
    \int \dd V \hat{\mathcal{H}}_\text{qc}(\mf x) = \frac{\lambda^2}{2}\int \dd V \dd V' \hat{\jmath}_\tc{a}(\mf x)\cdot \tilde{\Delta}(\mf x, \mf x')\cdot \hat{\jmath}_\tc{b}(\mf x'), 
\end{equation}
where $\tilde{\Delta}(\mf x, \mf x') = \tilde{G}_R(\mf x, \mf x') + \tilde{G}_A(\mf x, \mf x') = \tilde{G}_R(\mf x, \mf x') + \tilde{G}_R(\mf x', \mf x)$ is the symmetric propagator associated with the observable $O(\mf x)$. This implies that the leading order results from the qc-model depend exclusively on the symmetric propagator. 

At this stage, the qc-model is an ad-hoc description, as we have not yet shown that it can approximate interactions mediated by quantum fields. The goal of Section~\ref{sec:QFTapproxQC} will be to address this point with a concrete example where the qc-model is analogous to a two-level Unruh-DeWitt detector.

\subsubsection*{The Quantum-Classical Analogue of a Two-Level Unruh-DeWitt Detector}

We will now describe the qc-model for the interaction of two two-level systems via a scalar field. This model is in many ways analogous to the interaction of two two-level Unruh-DeWitt detectors. For simplicity, we will also restrict this example to a real massless scalar field in Minkowski spacetime and assume the qubits to undergo comoving inertial trajectories. We associate a Hilbert space $\mathcal{H}_\tc{i}\cong  \mathbb{C}^2$ to each system and consider their free dynamics to be implemented by the free Hamiltonians
\begin{equation}
    \hat{H}_\textsc{a} = \Omega \hat{\sigma}^+_\textsc{a}\hat{\sigma}^-_\textsc{a}, \quad 
    \hat{H}_\textsc{b} = \Omega \hat{\sigma}^+_\textsc{b}\hat{\sigma}^-_\textsc{b},
\end{equation}
where we are assuming for simplicity that the qubits have the same energy gap $\Omega$. 

The interactions with the field are prescribed by picking the observable $O(\mf x)$\footnote{In this case the operator $L$ is trivial and we have $\tilde{G}_R = G_R$, $\tilde{\Delta} = \Delta$.} as $\lambda\hat{\jmath}_\tc{a}(\mf x) \phi(\mf x)$ and $\lambda\hat{\jmath}_\tc{b}(\mf x)\phi(\mf x)$, where
\begin{equation}
    \hat{\jmath}_\tc{a}(\mf x) = \Lambda_\tc{a}(\mf x) \hat{\mu}_\tc{a}(t), \quad 
    \hat{\jmath}_\tc{a}(\mf x) = \Lambda_\tc{b}(\mf x) \hat{\mu}_\tc{b}(t), \quad \text{ with } \quad \hat{\mu}_\tc{i}(t) = e^{\ii \Omega t}\hat{\sigma}^+_\tc{i}+e^{-\ii \Omega t}\hat{\sigma}^-_\tc{i},
\end{equation}
where $(t,\bm x)$ are inertial coordinates, making this model analogous to an inertial two-level Unruh-DeWitt model. The qc-interaction Hamiltonian density for the systems can then be written as
\begin{align}\label{eq:classHpointlike}
    \hat{\mathcal{H}}_\text{qc}(\mf x) \!= \!\frac{\lambda^2}{2}\!\!\int &\dd V' \Big(\Lambda_\textsc{a}(\mf x)\Lambda_\textsc{b}(\mf x')\hat{\mu}_\textsc{a}(t)\hat{\mu}_\textsc{b}(t')G_R(\mf x, \mf x')+\Lambda_\textsc{b}(\mf x)\Lambda_\textsc{a}(\mf x')\hat{\mu}_\textsc{b}(t)\hat{\mu}_\textsc{a}(t')G_R(\mf x,\mf x')\Big).
\end{align}

The dynamics of the pair A, B are determined by the interaction unitary time evolution operator, prescribed with respect to the inertial time parameter $t$:
\begin{align}\label{eq:UIclass}
    \hat{U} &= \mathcal{T}_t\exp\left(-\ii \int \dd V  \hat{\mathcal{H}}_\text{qc}(\mf x)\right) = \openone - \ii \int \dd V  \hat{\mathcal{H}}_\text{qc}(\mf x) + \mathcal{O}(\lambda^4).\nonumber
\end{align}
In the basis $\{\ket{g_\textsc{a}g_\textsc{b}},\ket{g_\textsc{a}e_\textsc{b}},\ket{e_\textsc{a}g_\textsc{b}},\ket{e_\textsc{a}e_\textsc{b}}\}$, such that $\hat{\sigma}_\textsc{i}^+ \ket{g_\textsc{i}} = \ket {e_\textsc{i}}$ and $\hat{\sigma}_\textsc{i}^+ \ket{e_\textsc{i}}= 0$, this unitary can be written to leading order in the coupling constant as
\begin{equation}\label{eq:classU}
    \hat{U} = \begin{pmatrix}
        0 & 0 & 0 & -\mathcal{M}_{\textsc{c}}^*\\
        0 & 0 & -\mathcal{N}^*_{\textsc{c}} & 0 \\
        0 & \mathcal{N}_{\textsc{c}} & 0 & 0\\
        \mathcal{M}_{\textsc{c}} & 0 & 0 & 0
    \end{pmatrix} + \mathcal{O}(\lambda^4),
\end{equation}
where
\begin{align}
    \mathcal{M}_\textsc{c} = -  \frac{\ii \lambda^2}{2} \int \dd V\dd V' e^{\ii \Omega(t+t')}\Lambda_\textsc{a}(\mf x)\Lambda_\textsc{b}(\mf x')\Delta(\mf x, \mf x') = -  \frac{\ii\lambda^2}{2}\Delta(\Lambda_\tc{a}^+,\Lambda_\tc{b}^+),\nonumber \\
    \mathcal{N}_\textsc{c} = -\frac{\ii\lambda^2}{2} \int \dd V\dd V' e^{\ii \Omega(t-t')}\Lambda_\textsc{a}(\mf x)\Lambda_\textsc{b}(\mf x')\Delta(\mf x, \mf x') = - \frac{\ii\lambda^2}{2}\Delta(\Lambda_\tc{a}^+,\Lambda_\tc{b}^-),\label{eq:McNc}
\end{align}
where $\Lambda_\tc{i} =  e^{\pm\ii \Omega t}\Lambda_\tc{i}(\mf x)$, as usual.

If the two systems start in their ground state, \mbox{$\hat{\rho}_0 = \ket{g_\textsc{a}}\!\!\bra{g_\textsc{a}}\otimes\ket{g_\textsc{b}}\!\!\bra{g_\textsc{b}}$}, then, after the interaction, the state of the quantum systems is given by $\hat{\rho} = \hat{U}\hat{\rho}_0 \hat{U}^\dagger$. To fourth order in the coupling $\lambda$, we obtain
\begin{equation}\label{eq:rhoClass}
    \hat{\rho}_{\textsc{c}} = \begin{pmatrix}
        1 - |\mathcal{M}_{\textsc{c}}|^2 & 0 & 0 & \mathcal{M}_{\textsc{c}}^*\\
        0 & 0 & 0 & 0 \\
        0 & 0 & 0 & 0\\
        \mathcal{M}_{\textsc{c}} & 0 & 0 & |\mathcal{M}_{\textsc{c}}|^2
    \end{pmatrix} + \mathcal{O}(\lambda^6).
\end{equation}
We again note that $\hat{\rho}_{\textsc{c}}$ above is a pure state due to the fact that the evolution of two quantum systems interacting through a {qc-field} is unitary. Indeed, the final state of the systems can also be written as $\hat{\rho}_{\textsc{c}} = \ket{\psi}\!\!\bra{\psi}$, where the state vector $\ket{\psi}$ is given by
\begin{equation}
    \ket{\psi} = \frac{\ket{g_\tc{a}g_{\tc{b}}} + \mathcal{M}_{\textsc{c}}\ket{e_\tc{a}e_{\tc{b}}}}{\sqrt{1 + |\mathcal{M}_{\textsc{c}}|^2}} + \mathcal{O}(\lambda^6)
\end{equation}
to fourth order in $\lambda$.

\section{When can Quantum Field Theory be Approximated by a QC Model?}\label{sec:QFTapproxQC}

Having defined quantum-controlled models and studied the dynamics in a particular example, we now have the tools to compare a qc-interaction with interactions mediated by quantum fields. In this Section, we will consider explicit examples that allow us to conclude the regimes where one can approximate interactions mediated by quantum fields by the simpler unitary evolution provided by the qc-model. This comparison can be made in a straightforward fashion when one analyzes the entanglement that can be acquired in each description, which we will analyze below. For studies of different relativistic quantum information protocols, we refer the reader to~\cite{quantClass}.

\subsubsection*{Entanglement in QC-Models}

We will now consider two pointlike two-level quantum systems directly coupled according to the quantum-controlled model described in Section \ref{sec:QCmodels}. We have seen that if the qubits start in the ground state, the final state of the system is given by Eq. \eqref{eq:rhoClass}. In particular, the relevant matrix elements of the final state are proportional to the $\mathcal{M}_\tc{c}$ term, which is given by an integral of the symmetric propagator $\Delta(\mf x,\mf x')$, which automatically implies that if the interactions of qubits $\tc{A}$ and $\tc{B}$ are causally disconnected, then the qubits's state is unaffected. In particular, for a massless field in a spacetime that respects the strong Huygens's principle~\cite{Hyugens1,RayHyugens,Huygens2}, $G_R(\mf x, \mf x')$ and $G_A(\mf x,\mf x')$ are only non-zero when $\mf x$ and $\mf x'$ are lightlike separated, and there is no ``leakage'' of the propagators inside the lightcone. In this case, two qubits can only affect each other when interacting via a qc-field if their interactions are at some point lightlike separated. 

It is possible to quantify the entanglement acquired by the qubits via communication through the propagation of the {qc-field}. In order to quantify the entanglement of the final state of Eq. \eqref{eq:rhoClass}, and as usual, we choose the negativity as an entanglement quantifier. The partial transpose of the state of Eq. \eqref{eq:rhoClass} has a single negative eigenvalue, $-|\mathcal{M}_\tc{c}|$, so that its negativity reads, to leading order in $\lambda$,
\begin{equation}\label{eq:NegC}
    \mf{N}(\hat{\rho}_\tc{c}) = |\mathcal{M}_\tc{c}| = \frac{\lambda^2}{2}\Delta(\Lambda_\tc{a}^+,\Lambda_\tc{b}^+).
\end{equation}
From the expression above, we also confirm that when the qubits' region of interaction are not causally connected, they will also not be entangled (see the definition of $\mathcal{M}_\tc{c}$ in Eq. \eqref{eq:McNc}). Also notice that the negativity in this case precisely matches the quantifier of entanglement acquired via communication discussed in Section~\ref{sec:OperationallyAccessingEnt}.

\begin{figure}[h!]
    \centering
    \includegraphics[width=10cm]{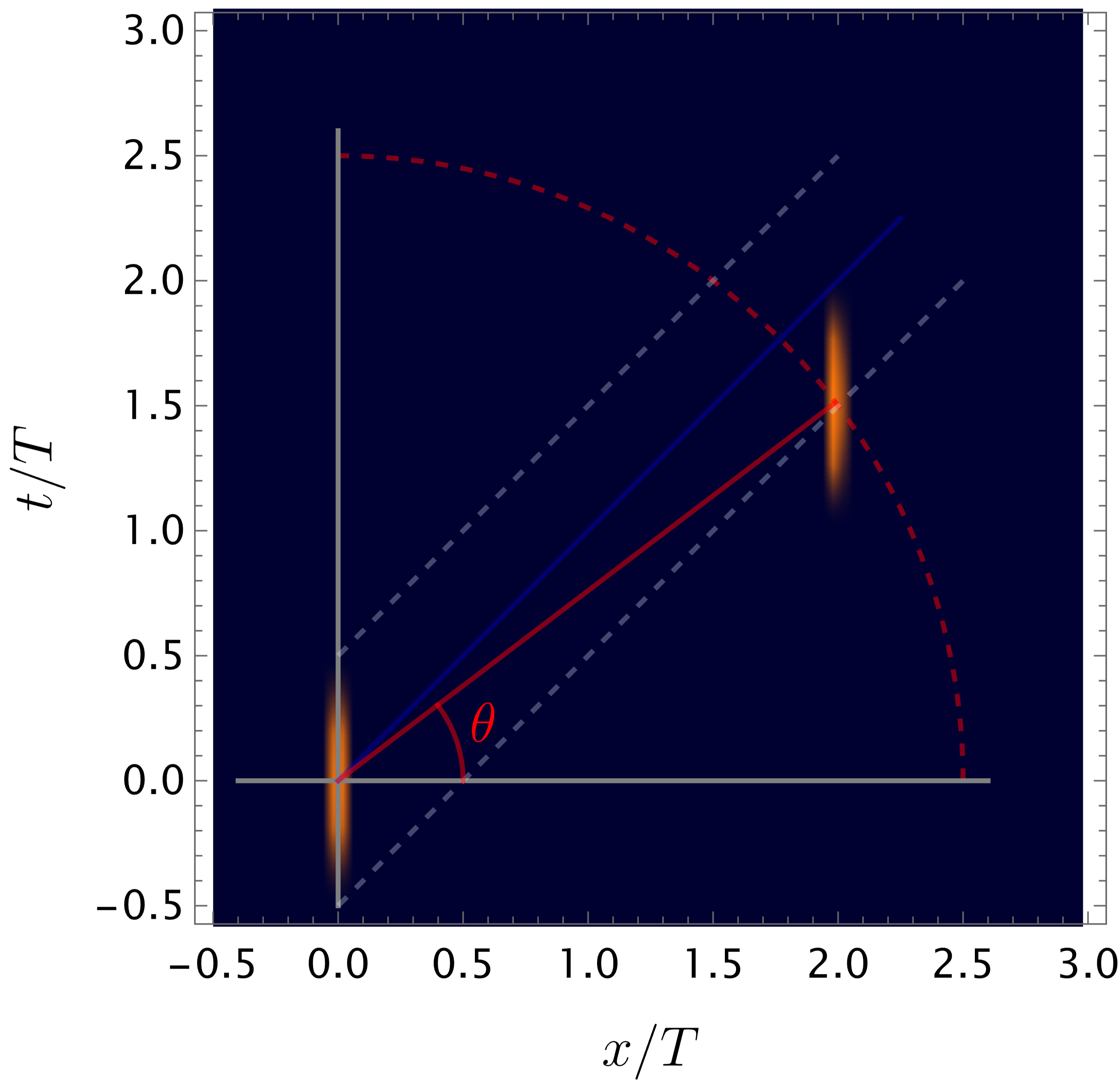}
    \caption{Setup for the configuration of the detectors as a function of the angle $\theta$.}
    \label{fig:SetupTheta}
\end{figure}

We consider a concrete example with a massless real scalar field in Minkowski spacetime and a specific spatial and temporal profile for the qubits, which undergo inertial comoving trajectories separated by a distance $L = |\bm L|$, where $\bm L$ is the separation vector between them. We will first prescribe the spacetime smearing functions as
\begin{align}
    \Lambda_\tc{a}(\mf x) &= \chi(t) \delta^{(3)}(\bm x),\label{eq:LambdaA}\\
    \Lambda_\tc{b}(\mf x) &= \chi(t-t_0) \delta^{(3)}(\bm x - \bm L),\label{eq:LambdaB}
\end{align}
where
\begin{equation}\label{eq:chiGaussian}
    \chi(t) = e^{-t^2/T^2}.
\end{equation}
With these choices, the systems are pointlike, $t_0$ is the time delay between the switchings, and $T$ controls the time duration of the interactions. The interaction of qubit $\tc{A}$ is centred at the origin of the coordinate system, and the interaction of qubit $\tc{B}$ is centred at the event $(t_0,\bm L)$. This choice makes the interaction non-compactly supported. Same as in the previous example we considered with detectors coupled to quantum fields, the main consequence of this choice is that, in principle, the qubits will always be in causal contact due to the tails of the Gaussians. However, 99.9999\% of the area of the switching function is concentrated in an interval of width $7T$ centred at the Gaussian peak. We then define the interval $[t_m -3.5T,t_m+3.5T]$ as the strong support of the Gaussian, where $t_m$ is its peak value. As we have discussed, signalling outside this region will be negligible compared to the effect of the interaction when the strong supports are lightlike separated.

\begin{figure}[h]
    \centering
    \includegraphics[width=10cm]{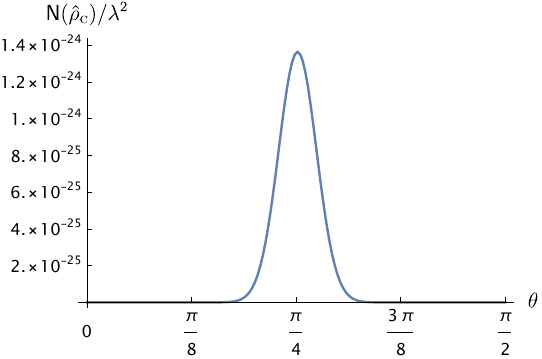}
    \caption{Plot of the negativity in the qubits state as a function of the angle $\theta$ for $\Omega T = 10$, $L = 10T$ for the Gaussian switching functions.}
    \label{fig:ClassicalNegativityGaussianTheta}
\end{figure}

We will analyze the negativity acquired by the qubits as we position system $\tc{B}$ around different events of the form $(L\sin(\theta),L\cos(\theta),0,0)$ parametrized by the parameter $\theta\in(0,\pi/2)$, as shown in Fig. \ref{fig:SetupTheta}. In Fig. \ref{fig:ClassicalNegativityGaussianTheta} we plot the negativity in the qubits state as a function of $\theta$. We consider the distance between the systems to be $L = 10T$, which ensures that the strong support of the Gaussians is spacelike separated. As we can see, there is no entanglement until a certain value of $\theta$, where the qubits stop being effectively spacelike separated. We then see a peak of the negativity when the interaction regions are lightlike separated at $\theta = \pi/4$. Notice that the plot is not completely symmetric with respect to the $\theta = \pi/4$ axis because the interaction regions are smeared in time, which slightly breaks the symmetry. 

\subsubsection*{Entanglement through quantum fields}\label{sub:entQuant}

We can now consider the entangling protocol outlined above in the case where the qubits are coupled to a real scalar \emph{quantum} field, that is, when the qubits are two-level Unruh-DeWitt detectors. This essentially defines an entanglement harvesting protocol, where we also allow the detectors to be causally connected. This is not an issue, as our goal here is not to extract entanglement from a quantum field but rather to obtain results that can be compared to the qc-interaction.

Considering two inertial Unruh-DeWitt detectors initially in their ground states coupled to the vacuum state of a real scalar quantum field, the final state of the detectors to leading order will be given by Eq.~\eqref{eq:UDWabFinal}, and the negativity of their final state will be given by Eq.~\eqref{eq:neg}. Specifically, for detectors with spacetime smearing functions related by a spacetime translation interacting with the vacuum of Minkowski spacetime, the leading order negativity of the final state $\hat{\rho}_\tc{d}$ is given by the simplified expression
\begin{equation}\label{eq:Nquant}
    \mf{N}(\hat{\rho}_\tc{d}) = \max(0,|\mathcal{M}| - \mathcal{L}^-),
\end{equation}
where $\mathcal{L}^- \equiv \mathcal{L}_{\tc{aa}}^- = \mathcal{L}_\tc{bb}^-$ as a consequence of the detectors being identical. As discussed in Section~\ref{sec:OperationallyAccessingEnt}, the negativity reflects the competition between the non-local terms arising from $\mathcal{M} = \lambda^2 G_F(\Lambda_\tc{a}^+,\Lambda_\tc{b}^+)$ and the local noise term $\mathcal{L} = \lambda^2 W(\Lambda_\tc{a}^-,\Lambda_\tc{a}^+)= \lambda^2 W(\Lambda_\tc{b}^-,\Lambda_\tc{b}^+)$. This local `vacuum' noise term is present only in the case where the field is quantum, as can be seen comparing Eqs.~\eqref{eq:NegC} and~\eqref{eq:Nquant}.

\begin{figure}[h]
    \centering
    \includegraphics[width=10cm]{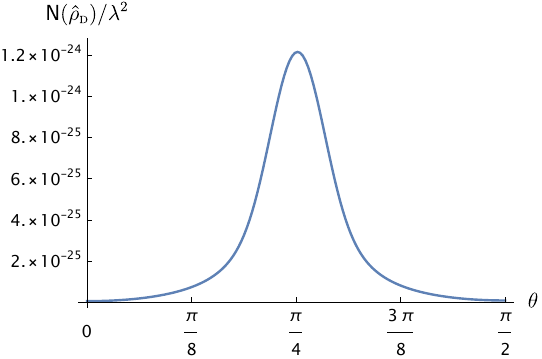}
    \caption{Plot of the negativity in the detectors state as a function of the angle $\theta$ for $\Omega T = 10$, $L = t_0 = 10T$ for Gaussian switching functions.}
    \label{fig:QuantumNegativityGaussianTheta}
\end{figure}

In order to draw a fair comparison between the {quantum-controlled and the truly quantum} models, we consider the same choice of spacetime smearing function of Eqs. \eqref{eq:LambdaA} and \eqref{eq:LambdaB}, and plot the negativity as a function of $\theta$ in Fig.~\ref{fig:QuantumNegativityGaussianTheta}. Unlike the quantum-controlled case, and aligned with the discussions of Section~\ref{sec:OperationallyAccessingEnt}, here we have that even when the {detectors' interaction regions} are fully spacelike separated ($\theta\approx 0$), it is still possible for them to become entangled. This is precisely the entanglement that is extracted from the field, and not acquired by the detectors via communication. Namely, this is a feature of the protocols of entanglement harvesting that explicitly depends on the field's quantum degrees of freedom. We also see a peak when the interaction regions are lightlike separated due to the entanglement that the detectors acquire through communication via the field.

\subsubsection*{When are the quantum degrees of freedom of the field negligible?}

As we discussed, a legitimate question that can be asked about the quantum-controlled model is whether it can reproduce the phenomenology of the fully quantum model in some regimes. If the qc-model is to hold any physical value, it should indeed be able to reproduce the same physics as the fully quantum model in the regimes where the quantum features of the field do not play any relevant role. To answer this question, we will now compare the two models, giving special attention to the regimes where the quantum field model can be well approximated as a qc-model. 

We choose to do this comparison for the the study of the entanglement acquired by two detectors when they interact with the field. The question of whether a model where the field is not fully quantum can predict that two systems that interact with the field get entangled is certainly relevant~\cite{MVPRD2020}, and this comparison showcases the differences that appear in other more general protocols when considering two quantum systems communicating through a quantum field.

The scales relevant for addressing the regimes where qc-fields can approximate quantum fields are the detectors' spatial separation $L$, their time separation, $t_0$, their energy gap $\Omega$ and the time of their switching, $T$. The relevant dimensionless parameters are then $L/T$, $t_0/T$ and $\Omega T$. We already saw that as $L/T$ increases past $t_0/T$ and as $t_0/T$ increases past $L/T$, the entanglement acquired by the detectors decreases (the further from light contact, the less entanglement between the detectors there will be in both models), so that the optimal rate $L/t_0$ is approximately $1$ making the detectors approximately lightlike separated. We also saw that the quantum field case can feature entanglement even when the {detectors' interaction regions} are spacelike separated, which is impossible in the {quantum-controlled} case. {In this sense, one of the conditions that is required for the quantum field to reduce to {the quantum-controlled model} is that the detectors have to be causally connected}. This imposes a restriction on the parameters $L/T$ and $t_0/T$. The study that remains to be conducted is what are the conditions over $\Omega T$ which allow the {fully featured} quantum field case to be well modelled by the {quantum-controlled} field scenario.

A main difference between {the cases where the field has quantum degrees of freedom or not} is the fact that {fully featured} quantum fields can produce local noise excitations in the detectors. This local noise is a consequence of the detectors becoming entangled with the field itself, which decoheres their state. The decoherence results in a decrease in the entanglement between the detectors, as they share part of the entanglement with the field. This can be seen in Eq. \eqref{eq:Nquant} for the negativity of the detectors, where we see that the vacuum noise $\mathcal{L}^-$ contributes negatively to the entanglement acquired by them. It is then clear that a condition so that the {true} quantum case can be mimicked by the {quantum-controlled case} is that the $\mathcal{L}^-$ term is much smaller than the nonlocal $\mathcal{M}$ term. This condition can be achieved if $\Omega T\gg 1$ (see Fig.~\ref{fig:pOmega}), or, in other words, in the limit where the interaction time is much larger than the characteristic time scale of the detectors.

\begin{figure}[h!]
    \centering
    \begin{subfigure}{.5\textwidth}
  \centering
  \includegraphics[width=7.5cm]{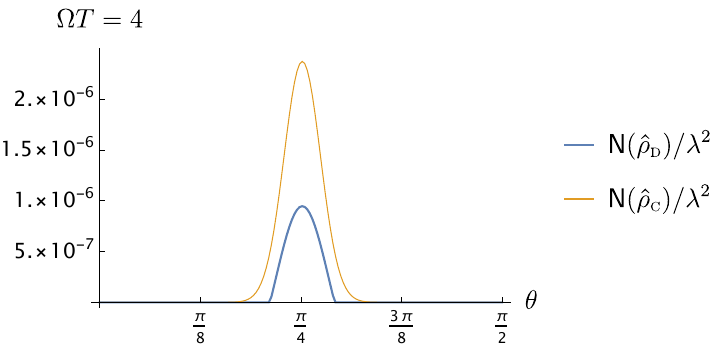}
  \includegraphics[width=7.5cm]{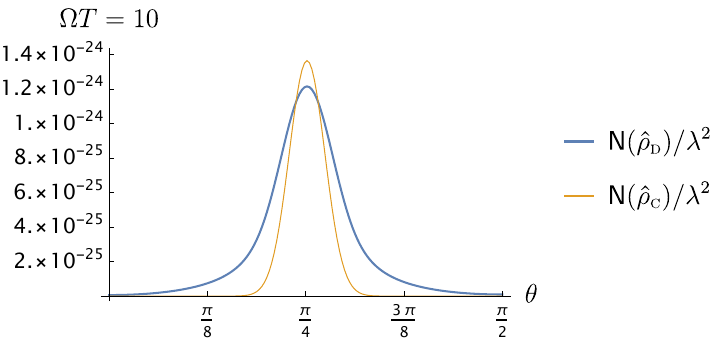}
  \caption*{}
\end{subfigure}%
\begin{subfigure}{.5\textwidth}
  \centering
  \includegraphics[width=7.5cm]{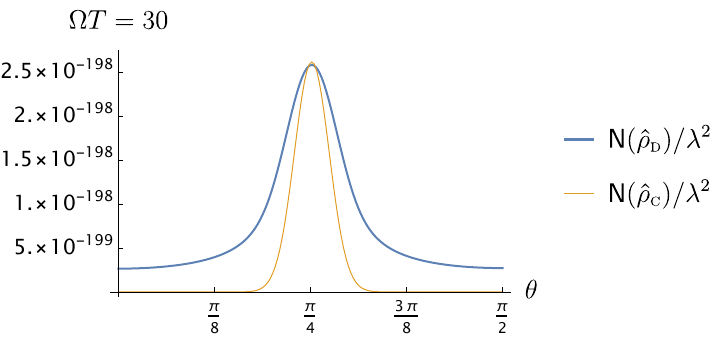}
  \includegraphics[width=7.5cm]{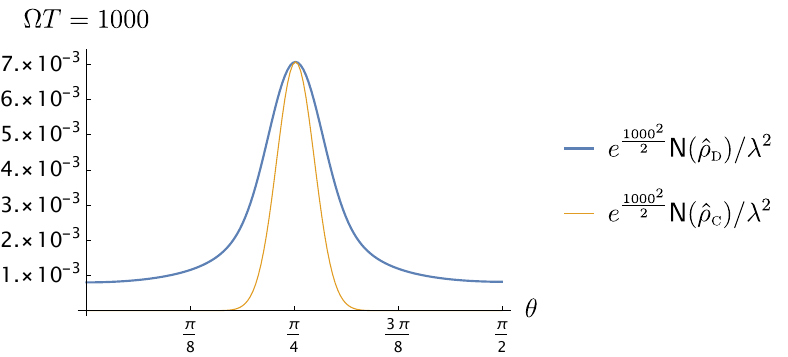}
  \caption*{}
\end{subfigure}
    \caption{Plot of the negativity in the detectors state for both the classical and quantum cases as a function of the angle $\theta$ for $L = t_0 = 10T$, and multiple values of $\Omega T$ for Gaussian supported switching functions.}
    \label{fig:VaryingOmegaClassicalQuantumGaussian}
\end{figure}

Another condition for the quantum field and {qc-field} models to behave similarly is that $\mathcal{M} \approx \mathcal{M}_\tc{c}$. Noticing that
\begin{equation}
    \mathcal{M} = \lambda^2G_F(\Lambda_\tc{a}^+,\Lambda_\tc{b}^+) = \tfrac{\lambda^2}{2}H(\Lambda_\tc{a}^+,\Lambda_\tc{b}^+)+\tfrac{\ii\lambda^2}{2}\Delta(\Lambda_\tc{a}^+,\Lambda_\tc{b}^+) = \tfrac{\lambda^2}{2}H(\Lambda_\tc{a}^+,\Lambda_\tc{b}^+) + \mathcal{M}_\tc{c}
\end{equation}
we see that the condition $\mathcal{M} \approx \mathcal{M}_\tc{c}$ is equivalent to the statement that the imaginary part of the propagator contributes significantly more to the $\mathcal{M}$ term than its real part. In Fig. \ref{fig:VaryingOmegaClassicalQuantumGaussian} we show plots for the negativity as a function of $\theta$ for the setup of Fig. \ref{fig:SetupTheta} for different values of $\Omega T$ considering both the pointlike spacetime smearing function with Gaussian switching of Eq. \eqref{eq:chiGaussian}. We see that when the {detectors' interaction regions} are lightlike connected, it is possible to get more entanglement between the detectors when their interaction is via the {qc-field} than when the detectors interact via the quantum field. As we mentioned earlier, this is due to the local noise, which decreases the entanglement acquired by the detectors when they interact with a quantum field. That is, although we always have $|\mathcal{M}| \geq |\mathcal{M}_\tc{c}|$, we also have $\mathcal{L}>0$, which allows the negativity in the {quantum-controlled} case to surpass that of the quantum case when $\mathcal{L}$ is comparable to $|\mathcal{M}|$. In Fig.~\ref{fig:VaryingOmegaClassicalQuantumGaussian}, we also see that under the assumption that $\Omega T\gg 1$, the {quantum field and qc-field} models give similar predictions when the {detectors' interaction regions} are causally connected ($\theta \approx \pi/4$). 

Overall, we can conclude that the {fully} quantum case can be well modelled by the {quantum-controlled} case only if three conditions are satisfied: 1) the systems involved in the protocol must be causally connected ($T\gg L$), 2) the interaction time with the mediating field has to be much larger than the characteristic time scale of the detectors ($T \gg 1/\Omega$), and 3) the interactions with the field have to be sufficiently weak ($\lambda\ll 1)$. {The third condition is necessary to avoid the major discrepancies between classical and quantum physics that take place for high energies, {which are not implemented simply by the retarded Green's function of the classical field theory for $\phi$}.} 

Finally, notice that if the three conditions above are satisfied, then the density operator of Eq. \eqref{eq:UDWabFinal} obtained in the fully quantum case reduces to the density operator obtained in the {qc-model} in Eq. \eqref{eq:rhoClass}. Indeed, assuming the three conditions, we have $|\mathcal{L}_{\tc{ij}}|\leq \mathcal{L}\ll |\mathcal{M}|$. Then, to leading order in $\lambda$, we have
\begin{align}
    \hat{\rho}_\tc{d} = |\mathcal{M}|&\!\begin{pmatrix}
    (1 - 2{\mathcal{L}^-})/{|\mathcal{M}|} & 0 & 0& \!\!{\mathcal{M}^*}/{|\mathcal{M}|}\\
    0 & \!\!{\mathcal{L}^-}/{|\mathcal{M}|} & {\mathcal{L}_{\tc{ab}}}/{|\mathcal{M}|} & 0 \\ 
    0 & \!\!{(\mathcal{L}_\tc{ab})^*}/{|\mathcal{M}|} & {\mathcal{L}^-}/{|\mathcal{M}|} & 0\\
    {\mathcal{M}}/{|\mathcal{M}|} & 0& 0& 0\end{pmatrix}\nonumber\approx \begin{pmatrix}
    1  & 0 & 0 & {\mathcal{M}^*}\\
    0 & 0 & 0 & 0 \\ 
    0 & 0 & 0 & 0\\
    {\mathcal{M}} & 0& 0& 0\end{pmatrix},
\end{align}
which is the leading order result from Eq. \eqref{eq:rhoClass}. That is, the three assumptions discussed above ensure that the classical model can be used to approximate the interaction with the quantum field.

\subsubsection*{Two gapless detectors interacting with a scalar quantum field}

For a second comparison between the models, we consider the cases where the qubits are gapless, $\Omega = 0$. In this case both the Unruh-DeWitt and the quantum-controlled models can be solved non-perturbatively, yielding a more clear comparison. 

We start by considering two gapless Unruh-DeWitt detectors, so the total interaction Hamiltonian density is given by
\begin{equation}
    \hat{\mathcal{H}}_I(\mf x) = \lambda \hat{\mu}_\tc{a} \Lambda_\tc{a}(\mf x) \hat{\phi}(\mf x) + \lambda \hat{\mu}_\tc{b} \Lambda_\tc{b}(\mf x) \hat{\phi}(\mf x),
\end{equation}
In this case a similar method to that used for one gapless detector can also be applied. This case has been studied in~\cite{Landulfo} under the assumption that the interaction of one of the detectors happens before the other. We will not make this assumption here and will instead obtain results for arbitrary interaction regions for the two detectors, as was done in~\cite{analytical}.

The Magnus expansion can be used to compute the time evolution operator. We find
\begin{equation}
    \hat{U}_I = e^{\hat{\Theta}_1+\hat{\Theta}_2},
\end{equation}
where
\begin{align}
    \hat{\Theta}_1 &= - \ii \int \dd V \hat{\mathcal{H}}_I(\mf x) = - \ii \lambda \hat{\mu}_\tc{a} \hat{\phi}(\Lambda_\tc{a})- \ii \lambda \hat{\mu}_\tc{b} \hat{\phi}(\Lambda_\tc{b}),\label{eq:TH2}\\
    \hat{\Theta}_2 &= - \frac{1}{2} \int \dd V \dd V' \theta(t-t')[\hat{\mathcal{H}}_I(\mf x), \hat{\mathcal{H}}_I(\mf x')] \\&= - \frac{\lambda^2}{2}\int \dd V \dd V' \theta(t-t')[\hat{\phi}(\mf x), \hat{\phi}(\mf x')]\Big(\hat{\mu}_\tc{a}^2 \Lambda_\tc{a}(\mf x)\Lambda_\tc{a}(\mf x')+\hat{\mu}_\tc{a}\hat{\mu}_\tc{b} \Lambda_\tc{a}(\mf x)\Lambda_\tc{b}(\mf x')\nonumber\\
    &\:\:\:\:\:\:\:\:\:\:\:\:\:\:\:\:\:\:\:\:\:\:\:\:\:\:\:\:\:\:\:\:\:\:\:\:\:\:\:\:\:\:\:\:\:\:\:\:\:\:\:\:\:\:\:\:\:\:\:\:\:\:\:\:\:\:\:\:\:\:\:\:+\hat{\mu}_\tc{b}\hat{\mu}_\tc{a} \Lambda_\tc{b}(\mf x)\Lambda_\tc{a}(\mf x')+\hat{\mu}_\tc{b}^2 \Lambda_\tc{b}(\mf x)\Lambda_\tc{b}(\mf x')\Big)\nonumber\\
    &= - \frac{\ii \lambda^2}{2}\Big(\hat{\mu}_\tc{a}^2 G_R(\Lambda_\tc{a},\Lambda_\tc{a})+\hat{\mu}_\tc{b}^2 G_R(\Lambda_\tc{b},\Lambda_\tc{b})+\hat{\mu}_\tc{a}\hat{\mu}_\tc{b} \left(G_R(\Lambda_\tc{a},\Lambda_\tc{b})+ G_R(\Lambda_\tc{b},\Lambda_\tc{a})\right)\Big),\nonumber
\end{align}
where we used that $\theta(t-t')[\hat{\phi}(\mf x),\hat{\phi}(\mf x')] = \ii G_R(\mf x, \mf x')$, and that the operators $\hat{\mu}_\tc{a}$ and $\hat{\mu}_\tc{b}$ commute. We then denote $\Delta_\tc{ab} = \lambda^2\Delta(\Lambda_\tc{a},\Lambda_\tc{b})$, $\mathcal{G}_\tc{a} = \frac{\lambda^2}{2}G_R(\Lambda_\tc{a},\Lambda_\tc{a})$, and $\mathcal{G}_\tc{b} = \frac{\lambda^2}{2}G_R(\Lambda_\tc{b},\Lambda_\tc{b})$, so that Eq.~\eqref{eq:TH2} allows us to write
\begin{equation}
    \hat{\Theta}_2 = - \ii \hat{\mu}_\tc{a}^2\mathcal{G}_\tc{a}- \ii\hat{\mu}_\tc{b}^2 \mathcal{G}_\tc{b} -  \tfrac{\ii}{2}\hat{\mu}_\tc{a}\hat{\mu}_\tc{b}\Delta_\tc{ab}.
\end{equation}
The fact that the commutator $[\hat{\mathcal{H}}_I(\mf x), \hat{\mathcal{H}}_I(\mf x')]$ commutes with $\mathcal{H}_I(\mf x'')$ implies that only $\hat{\Theta}_1$ and $\hat{\Theta}_2$ are non-zero in the Magnus expansion so that the unitary time evolution operator reads
\begin{equation}\label{eq:UI2g}
    \hat{U}_I = e^{- \ii \lambda\hat{\mu}_\tc{a}\hat{\phi}(\Lambda_\tc{a}) - \ii \lambda\hat{\mu}_\tc{b} \hat{\phi}(\Lambda_\tc{b})} e^{- \ii \hat{\mu}_\tc{a}^2\mathcal{G}_\tc{a}}e^{- \ii\hat{\mu}_\tc{b}^2 \mathcal{G}_\tc{b}}e^{-  \tfrac{\ii}{2}\hat{\mu}_\tc{a}\hat{\mu}_\tc{b}\Delta_\tc{ab}},
\end{equation}
where we used that $[\hat{\Theta}_1,\hat{\Theta}_2] = 0$ to separate the exponentials. One can also use the Baker-Campbell-Hausdorff formula in order to factor $\hat{U}_I$ as
\begin{align}
    \hat{U}_I &= e^{- \ii \lambda\hat{\mu}_\tc{a}\hat{\phi}(\Lambda_\tc{a})}e^{- \ii \lambda\hat{\mu}_\tc{b} \hat{\phi}(\Lambda_\tc{b})} e^{- \ii \hat{\mu}_\tc{a}^2\mathcal{G}_\tc{a}- \ii\hat{\mu}_\tc{b}^2 \mathcal{G}_\tc{b}}e^{-  \tfrac{\ii}{2}\hat{\mu}_\tc{a}\hat{\mu}_\tc{b}(\Delta_\tc{ab}-E_{\tc{ab}})}\nonumber\\
    &= e^{- \ii \lambda\hat{\mu}_\tc{b} \hat{\phi}(\Lambda_\tc{b})} e^{- \ii \lambda\hat{\mu}_\tc{a}\hat{\phi}(\Lambda_\tc{a})}e^{- \ii \hat{\mu}_\tc{a}^2\mathcal{G}_\tc{a}- \ii\hat{\mu}_\tc{b}^2 \mathcal{G}_\tc{b}}e^{-  \tfrac{\ii}{2}\hat{\mu}_\tc{a}\hat{\mu}_\tc{b}(\Delta_\tc{ab}+E_{\tc{ab}})},\label{eq:BCH2UDW}
\end{align}
where $E_{\tc{ab}} = \lambda^2 E(\Lambda_\tc{a},\Lambda_\tc{b})$, and we have
\begin{align}
    \tfrac{1}{2}(\Delta_{\tc{ab}} - E_\tc{ab}) &= \lambda^2G_A(\Lambda_\tc{a},\Lambda_\tc{b}) = \lambda^2G_R(\Lambda_\tc{b},\Lambda_\tc{a}),\\
    \tfrac{1}{2}(\Delta_{\tc{ab}} + E_\tc{ab}) &= \lambda^2G_R(\Lambda_\tc{a},\Lambda_\tc{b}) = \lambda^2G_A(\Lambda_\tc{b},\Lambda_\tc{a}).
\end{align}

To compute the final state of the two detectors after tracing the field, it is more convenient to work with the expression from Eq. \eqref{eq:UI2g}. Assume that the initial state of the detectors-field systems is $\hat{\rho}_0 = \hat{\rho}_{\tc{ab},0}\otimes \hat{\rho}_\omega$, where, as usual, $\hat{\rho}_\omega$ is the representation of a quasifree state $\omega$ in the quantum field theory. We further assume that $\hat{\mu}_\tc{a}^2 = \hat{\mu}_\tc{b}^2 = \openone$, so that the effect of the local unitaries $e^{- \ii \hat{\mu}_\tc{a}^2\mathcal{G}_\tc{a}}e^{- \ii\hat{\mu}_\tc{b}^2 \mathcal{G}_\tc{b}}$ becomes negligible. Given that the unitary $e^{-  \tfrac{\ii}{2}\hat{\mu}_\tc{a}\hat{\mu}_\tc{b}\Delta_\tc{ab}}$ commutes with the field-dependent term, we can separate the action of the unitary
\begin{equation}
    \hat{U}_\phi = e^{- \ii \lambda\hat{\mu}_\tc{a}\hat{\phi}(\Lambda_\tc{a}) - \ii \lambda\hat{\mu}_\tc{b} \hat{\phi}(\Lambda_\tc{b})}
\end{equation}
from the rest. To proceed with the computations, we denote the eigenstate of $\hat{\mu}_\tc{a}$ and $\hat{\mu}_\tc{b}$ by $\ket{\pm_\tc{a}}$ and $\ket{\pm_\tc{b}}$, so that
\begin{align}
    \hat{U}_\phi \hat{\rho}_0\hat{U}_\phi^\dagger &= e^{- \ii \lambda\hat{\mu}_\tc{a}\hat{\phi}(\Lambda_\tc{a}) - \ii \lambda\hat{\mu}_\tc{b} \hat{\phi}(\Lambda_\tc{b})}\hat{\rho}_{0}e^{\ii \lambda\hat{\mu}_\tc{a}\hat{\phi}(\Lambda_\tc{a}) + \ii \lambda\hat{\mu}_\tc{b} \hat{\phi}(\Lambda_\tc{b})}\nonumber\\
    &=\!\!\!\!\sum_{{\substack{{}_{\mu_\tc{a},\mu_\tc{b} = \pm}\\{}_{\mu_\tc{a}',\mu_\tc{b}' = \pm}}}} \!\!e^{- \ii \lambda\hat{\phi}({\mu}_\tc{a}\Lambda_\tc{a}+{\mu}_\tc{b}\Lambda_\tc{b})}\hat{\rho}_\omega e^{\ii \lambda\hat{\phi}({\mu}_\tc{a}'\Lambda_\tc{a}+{\mu}_\tc{b}'\Lambda_\tc{b})}\nonumber\,\bra{\mu_\tc{a}\mu_{\tc{b}}}\hat{\rho}_{\tc{ab},0}\ket{\mu_\tc{a}'\mu_{\tc{b}}'}  \ket{\mu_\tc{a}\mu_{\tc{b}}}\!\!\bra{\mu_\tc{a}'\mu_{\tc{b}}'}.
\end{align}
The next step is to trace over the field to obtain the state $\hat{\sigma}_\tc{ab} = \tr_\phi(\hat{U}_\phi \hat{\rho}_0\hat{U}_\phi^\dagger)$. We find
\begin{align}
    &\hat{\sigma}_\tc{ab} 
    =\!\!\!\!\sum_{{\substack{{}_{\mu_\tc{a},\mu_\tc{b} = \pm}\\{}_{\mu_\tc{a}',\mu_\tc{b}' = \pm}}}} \!\!\omega\!\left(e^{\ii \lambda\hat{\phi}({\mu}_\tc{a}'\Lambda_\tc{a}+{\mu}_\tc{b}'\Lambda_\tc{b})}e^{- \ii \lambda\hat{\phi}({\mu}_\tc{a}\Lambda_\tc{a}+{\mu}_\tc{b}\Lambda_\tc{b})}\right)\,\bra{\mu_\tc{a}\mu_{\tc{b}}}\hat{\rho}_{\tc{ab},0}\ket{\mu_\tc{a}'\mu_{\tc{b}}'}  \ket{\mu_\tc{a}\mu_{\tc{b}}}\!\!\bra{\mu_\tc{a}'\mu_{\tc{b}}'}\nonumber\\[10pt]
    &=\!\!\!\!\sum_{{\substack{{}_{\mu_\tc{a},\mu_\tc{b} = \pm}\\{}_{\mu_\tc{a}',\mu_\tc{b}' = \pm}}}} \!\!\! e^{\frac{\ii\lambda^2}{2}E(\mu_\tc{a}'\Lambda_\tc{a} + \mu_{\tc{b}}'\Lambda_\tc{b},\mu_\tc{a}\Lambda_\tc{a} + \mu_{\tc{b}}\Lambda_\tc{b})- \frac{\lambda^2}{2}||(\mu_{\tc{a}} - \mu_{\tc{a}}')\Lambda_{\tc{a}} + (\mu_{\tc{b}} - \mu_{\tc{b}}')\Lambda_{\tc{b}}||^2}\nonumber\,\bra{\mu_\tc{a}\mu_{\tc{b}}}\hat{\rho}_{\tc{ab},0}\ket{\mu_\tc{a}'\mu_{\tc{b}}'}  \ket{\mu_\tc{a}\mu_{\tc{b}}}\!\!\bra{\mu_\tc{a}'\mu_{\tc{b}}'},
\end{align}
where we used
\begin{equation}
    \omega\left(e^{\ii \lambda \hat{\phi}(f)}e^{\ii \lambda \hat{\phi}(g)}\right) = e^{-\frac{\ii \lambda^2}{2}E(f,g) - \frac{\lambda^2}{2} W(f+g,f+g)},
\end{equation}
and we denoted $||f||^2 = W(f,f)$. We can now incorporate the unitary $e^{-  \tfrac{\ii}{2}\hat{\mu}_\tc{a}\hat{\mu}_\tc{b}\Delta_\tc{ab}}$ again so that the final state of the detectors is given by
\begin{equation}
    \hat{\rho}_\tc{ab} = e^{-  \tfrac{\ii}{2}\hat{\mu}_\tc{a}\hat{\mu}_\tc{b}\Delta_\tc{ab}} \hat{\sigma}_\tc{ab}e^{\tfrac{\ii}{2}\hat{\mu}_\tc{a}\hat{\mu}_\tc{b}\Delta_\tc{ab}}.\label{eq:sab}
\end{equation} 
Once again, notice that the final state of the qubits is entirely given in terms of bi-distributions of the quantum field smeared against $\Lambda_\tc{a}(\mf x)$ and $\Lambda_{\tc b}(\mf x')$. 

Finally, we write $\rho_{ij}$, $i,j = 1,...,4$ for the components of $\hat{\rho}_{\tc{ab},0}$ in the basis $\{\ket{+_\tc{a}+_\tc{b}},\ket{+_\tc{a}-_\tc{b}}$ $,\ket{-_\tc{a}+_\tc{b}},\ket{-_\tc{a}-_\tc{b}}\}$, so that the final state of the detectors state can be written (in this same basis) as
\begin{equation}\label{eq:rhoABgapless}
{\footnotesize
    \hat{\rho}_\tc{ab} = \begin{pmatrix}
        \rho_{11} & e^{-2W_{\tc{bb}} + \ii(E_\tc{ab} - \Delta_{\tc{ab}})}\rho_{12} & e^{-2W_{\tc{aa}} - \ii(E_\tc{ab} + \Delta_{\tc{ab}})}\rho_{13} & e^{-2(W_{\tc{aa}}+W_{\tc{bb}}+H_{\tc{ab}})}\rho_{14} \\
        e^{-2W_{\tc{bb}} - \ii(E_\tc{ab} - \Delta_{\tc{ab}})}\rho_{21} & \rho_{22} & e^{-2(W_{\tc{aa}}+W_{\tc{bb}}-H_{\tc{ab}})}\rho_{23} & e^{-2W_{\tc{aa}} + \ii(E_\tc{ab} + \Delta_{\tc{ab}})}\rho_{24} \\
        e^{-2W_{\tc{aa}} + \ii(E_\tc{ab} + \Delta_{\tc{ab}})}\rho_{31} & e^{-2(W_{\tc{aa}}+W_{\tc{bb}}-H_{\tc{ab}})}\rho_{32} & \rho_{33} & e^{-2W_{\tc{bb}} - \ii(E_\tc{ab} - \Delta_{\tc{ab}})}\rho_{34} \\
        e^{-2(W_{\tc{aa}}+W_{\tc{bb}}+H_{\tc{ab}})}\rho_{41} & e^{-2W_{\tc{aa}} - \ii(E_\tc{ab} + \Delta_{\tc{ab}})}\rho_{42} & e^{-2W_{\tc{bb}} + \ii(E_\tc{ab} - \Delta_{\tc{ab}})}\rho_{43} & \rho_{44} 
    \end{pmatrix}.}
\end{equation}
For each of the bi-distributions $W, H, E$, and $\Delta$, we use the convention $A_{\tc{ab}} = \lambda^2 A(\Lambda_\tc{a},\Lambda_\tc{b})$. 

\subsubsection*{The Gapless Quantum-Controlled Model}

The quantum-controlled version of this gapless model is defined by the interaction Hamiltonian density
\begin{equation}
    \hat{\mathcal{H}}_\tc{qc}(\mf x) = \frac{\lambda^2}{2} \hat{\mu}_\tc{a} \hat{\mu}_\tc{b} \int \dd V'  \left(\Lambda_\tc{a}(\mf x)\Lambda_\tc{b}(\mf x') G_R(\mf x, \mf x') + \Lambda_\tc{b}(\mf x)\Lambda_\tc{a}(\mf x') G_R(\mf x, \mf x')\right).
\end{equation}
We then have that, in the gapless case, the Hamiltonian density commutes with itself at different times, so the unitary time evolution operator can be computed as the exponential
\begin{equation}
    \hat{U}_\tc{c} = \exp(-\ii \int \dd V\hat{\mathcal{H}}_\tc{qc}(\mf x) ) =  e^{- \frac{\ii}{2} \hat{\mu}_\tc{a}\hat{\mu}_\tc{b}\Delta_{\tc{ab}}}.
\end{equation}
Writing again $\rho_{ij}$, $i,j = 1,...,4$ for the components of $\hat{\rho}_{\tc{ab},0}$ in the basis $\{\ket{+_\tc{a}+_\tc{b}},\ket{+_\tc{a}-_\tc{b}}$ $,\ket{-_\tc{a}+_\tc{b}},\ket{-_\tc{a}-_\tc{b}}\}$, the final state of the qubits can then be written as
\begin{equation}\label{eq:rhoABgaplessQC}
{\footnotesize
    \hat{\rho}_\tc{c} = \begin{pmatrix}
        \rho_{11} & e^{ - \ii\Delta_{\tc{ab}}}\rho_{12} & e^{-\ii\Delta_{\tc{ab}}}\rho_{13} & \rho_{14} \\
        e^{\ii\Delta_{\tc{ab}}}\rho_{21} & \rho_{22} & \rho_{23} & e^{\ii\Delta_{\tc{ab}}}\rho_{24} \\
        e^{\ii\Delta_{\tc{ab}}}\rho_{31} & \rho_{32} & \rho_{33} & e^{\ii\Delta_{\tc{ab}}}\rho_{34} \\
        \rho_{41} & e^{-\ii \Delta_{\tc{ab}}}\rho_{42} & e^{- \ii\Delta_{\tc{ab}}}\rho_{43} & \rho_{44} 
    \end{pmatrix}.}
\end{equation}

We chose to write the density matrix explicitly to allow for a straightforward comparison with Eq.~\eqref{eq:rhoABgapless}. Notice that by setting $H$ and $E$ to 0 in Eq.~\eqref{eq:rhoABgaplessQC}, one obtains the quantum-controlled result from its quantum field theory counterpart. While it is clear that $H$ is related to the state dependent terms in the theory, the causal propagator is fundamental in Eq.~\eqref{eq:rhoABgapless} to ensure causality in the interaction. We can see this explicitly by looking at the reduced density operators of A and B. Using that $\Delta +E = 2 G_R$ and $\Delta - E = 2 G_A$, we find 
\begin{align}
    \r_\tc{a} = \begin{pmatrix}
        \rho_{11}+\rho_{22} & e^{-\ii \lambda^2G_R(\Lambda_\tc{a},\Lambda_\tc{b})}\rho_{13} +e^{\ii \lambda^2G_R(\Lambda_\tc{a},\Lambda_\tc{b})}\rho_{24}\\
        e^{\ii \lambda^2G_R(\Lambda_\tc{a},\Lambda_\tc{b})}\rho_{31} +e^{-\ii \lambda^2G_R(\Lambda_\tc{a},\Lambda_\tc{b})}\rho_{42} & \rho_{33}+\rho_{44}
    \end{pmatrix},\\
    \r_\tc{b} = \begin{pmatrix}
        \rho_{11}+\rho_{33} & e^{-\ii \lambda^2G_R(\Lambda_\tc{b},\Lambda_\tc{a})}\rho_{12} +e^{\ii \lambda^2G_R(\Lambda_\tc{b},\Lambda_\tc{a})}\rho_{34}\\
        e^{\ii \lambda^2G_R(\Lambda_\tc{b},\Lambda_\tc{a})}\rho_{21} +e^{-\ii \lambda^2G_R(\Lambda_\tc{b},\Lambda_\tc{a})}\rho_{43} & \rho_{22}+\rho_{44}
    \end{pmatrix},
\end{align}
where the sum and difference of $\Delta_\tc{ab}$ and $E_\tc{ab}$ make sure that $\hat{\rho}_\tc{a}$ only depends on retarded propagation from B and $\r_\tc{b}$ only depends on retarded propagation from A. This also implies that whenever we have $E_\tc{ab} \neq 0$, the qc-model will predict some level of causality violation. This does not imply that the model cannot be applied; it simply means that one has to make sure that the causality violations are below a certain observable threshold. For a detailed discussion of causality in the qc-model, see~\cite{eirini}.

\begin{figure}[h!]
    \centering
    \includegraphics[width=10cm]{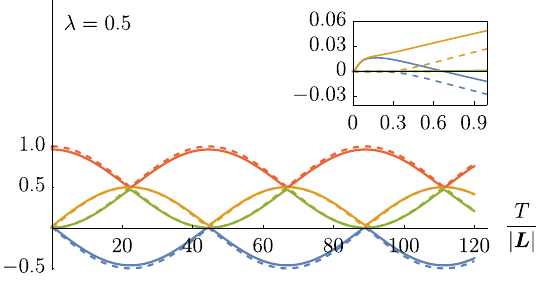}
    \includegraphics[width=10cm]{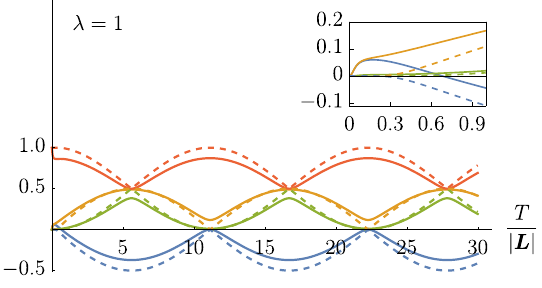}
    \includegraphics[width=10cm]{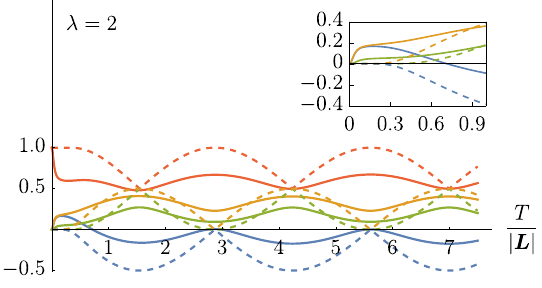}
    \caption{The solid lines correspond to the eigenvalues of the partial transpose of the detectors final state $\hat{\rho}_\tc{ab}^{t_\tc{b}}$, when the detectors start both in their ground state, as a function of the interaction time $T$, scaled by the detectors' separation $|\bm L|$. The dashed lines are the eigenvalues of $\hat{\rho}_\tc{c}^{t_\tc{b}}$, obtained using the qc-model. We picked  $\sigma = 0.05|\bm L|$ for these plots.}
    \label{fig:t}
\end{figure}

\subsubsection*{Comparison Between the Gapless Models}

Let us now consider an explicit example, where $\hat{\mu}_\tc{a} = \hat{\sigma}_\tc{a}^+ + \hat{\sigma}_\tc{a}^-$, $\hat{\mu}_\tc{b} = \hat{\sigma}_\tc{b}^+ + \hat{\sigma}_\tc{b}^-$, the detectors undergo inertial trajectories in Minkowski spacetime, and interact with the vacuum of a massless scalar field in Gaussian spacetime regions. For convenience, we use the same spacetime smearing functions as in~\eqref{eq:prototypeAB} with $t_0 = 0$. We will assume both detectors to start in their ground states, with \mbox{$\hat{\rho}_{\tc{ab},0} = \ket{g_\tc{a}}\!\!\bra{g_\tc{a}}\otimes \ket{g_\tc{b}}\!\!\bra{g_\tc{b}}$}. Notice that in this setup, the spacetime smearing functions differ only by a shift in space, making $E_{\tc{ab}} = 0$, so that the qc-model does not imply any causality violations.

As we did in the example of entanglement harvesting, we will be interested in checking the conditions so that the detectors can end up in an entangled state. To check this, we plot the eigenvalues of the partial transpose of $\hat{\rho}_{\tc{ab}}$ in Fig.~\ref{fig:t} as a function of the effective interaction time $T$ for three different values of the coupling constant (solid lines) as well as the eigenvalues of $\hat{\rho}_{\tc{c}}^{t_\tc{b}}$ (dashed lines).

Keep in mind that the detectors are entangled if and only if the partial transpose of their density operator has a negative eigenvalue. Notice that in this example, for small values of the coupling constant $\lambda$, the behaviour of the state evolved through the qc-model is very similar to the interaction with the quantum field. This is because, for small coupling constants, the detectors are not subject to too much noise, and they are able to communicate through the field without getting too entangled with the field itself.

Also notice that for small values of $T$, the detectors cannot become entangled in either model. This can be seen in the extended plots on the top right of Fig~\ref{fig:t}, which display the region $\frac{T}{|\bm L|} < 1$. The fact that the detectors interacting with a quantum field cannot become entangled when $T\lesssim |\bm L|$ is in agreement with the results of the no-go theorems proven in~\cite{HarvestingQueNemLouko,nogo}, where it was proven that gapless detectors cannot harvest if their interaction regions are spacelike separated. The fact that the detectors start becoming entangled for $T  <|\bm L|$ is merely an artifact of the non-compact support of the Gaussian switching functions considered, in which case there is still enough causal contact for the detectors to become entangled. Importantly, regardless of the coupling constant, we find that the detectors are only able to become entangled for $T \gtrsim 0.7 |\bm L|$, showing that the detectors' inability to communicate is independent of how strongly they couple to the field.

\begin{figure}[h!]
    \centering
    \includegraphics[width=12cm]{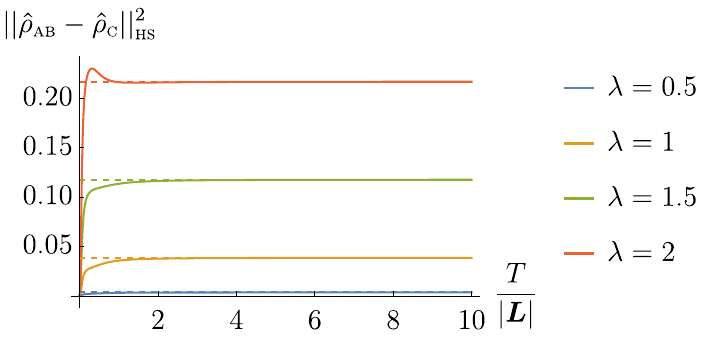}
    \caption{The squared Hilbert-Schmidt norm of the difference between the density operators $\hat{\rho}_\tc{ab}$ and $\hat{\rho}_\tc{c}$ for $\sigma = 0.05 |\bm L|$ considering detectors that start in their ground state. The dashed lines correspond to their asymptotic limit as $T\to\infty$.}
    \label{fig:pass}
\end{figure}

We continue our comparison between the models in this specific example by computing the explicit difference between the final density operator $\hat{\rho}_\tc{ab}$ and the operator $\hat{\rho}_\tc{c}$. In Fig.~\ref{fig:pass}, we plot the squared Hilbert-Schmidt norm\footnote{The Hilbert-Schmidt norm is defined as $||\hat{A}||_\tc{hs} = \sqrt{\Tr(A^\dagger A)}$.} of the difference $\hat{\rho}_\tc{ab} - \hat{\rho}_\tc{c}$ for different values of $\lambda$, using the same setup as we considered before. In the figure, we see that the norm of the difference quickly goes to zero as $\lambda$ decreases. We can also notice that in the limit of $T\to\infty$, the difference between the two evolutions becomes a constant.

\begin{figure}[h!]
    \centering
    \includegraphics[width=10cm]{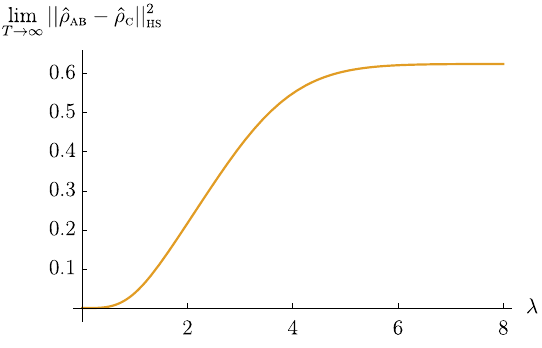}
    \caption{The limit of the asymptotic behaviour of the squared Hilbert-Schmidt norm of the operator $\hat{\rho}_{\tc{ab}} - \hat{\rho}_\tc{c}$. This limit only depends on $\lambda$.}
    \label{fig:passLimit}
\end{figure}

Given that we have access to analytical expressions for both $\hat{\rho}_\tc{ab}$ and $\hat{\rho}_{\tc{c}}$, we can also compute the asymptotic behaviour of $||\hat{\rho}_\tc{ab} - \hat{\rho}_\tc{c}||^2_\tc{hs}$ in the limit where $\sigma\ll |\bm L|$ and $T\gg |\bm L|$ (with the assumption of $\sigma T \ll |\bm L|^2$). In this limit, we find that the result is independent of $\sigma$, $|\bm L|$, and, of course, $T$. We find
\begin{align}
    \lim_{T\to\infty}||\hat{\rho}_\tc{ab} & - \hat{\rho}_\tc{c}||^2_\tc{hs} = \frac{1}{8}\left(5 + e^{-\frac{4 \lambda^2}{\pi}}-2e^{-\frac{2 \lambda^2}{\pi}}+4e^{-\frac{\lambda^2}{\pi}} - 8 e^{-\frac{ \lambda^2}{2\pi}}\right).\nonumber
\end{align}
We plot this result in Fig.~\ref{fig:passLimit}. Also notice that the behaviour of the above limit for $\lambda\to 0$ is given by
\begin{align}
   \lim_{T\to\infty}||\hat{\rho}_\tc{ab} & - \hat{\rho}_\tc{c}||^2_\tc{hs} = \frac{5 \lambda^4}{8 \pi^2} + \mathcal{O}(\lambda^6),
\end{align}
suggesting that, indeed, in the regime of small coupling constants, the evolution is well modelled by the unitary $\hat{U}_\tc{c}$.

From these examples, we reinforce the conclusion that, indeed, the qc-model can be a good approximation for interactions mediated by quantum fields in the limit of long times and small coupling constants.

\subsubsection*{The Role of Quantum Degrees of Freedom of Mediators}

Our initial motivation for the definition of qc-models was to simplify the description of the interactions between quantum systems by neglecting the local degrees of freedom of quantum fields. On the other hand, the qc-model gives rise to a criterion that allows one to evaluate whether or not the quantum degrees of freedom of the mediating field actively participate in the interaction between two localized systems. Explicitly, we can conclude that

\begin{center}
    \textit{``the quantum degrees of freedom of mediators do not play an active role \\in setups that are accurately described by a qc-model.''}
\end{center}

Throughout this Section we saw that experiments that involve weakly coupled sources that are causally connected for long times can be accurately described by quantum-controlled interactions. Hence, these are the regimes where the quantum degrees of freedom of the mediators do not play an active role. In particular, this implies that one can only access the quantum degrees of freedom of a field in the regimes of either strong couplings or short interaction times. While it is to be expected that high energy interactions require quantum field theory for an accurate description, the condition of long interaction times seems to imply that the local degrees of freedom of quantum field theories only manifest themselves in relativistic setups, where the precise spacetime positioning of the probes and their causal contact is relevant.

This quantification of the regimes where quantum field theory is necessary points out that high-energy physics and local operations in quantum field theory are exactly the two relevant cases where the fundamental aspects of quantum field theory play a significant role. On the other hand, in some sense, these two fields of study are opposites: high-energy physics typically considers arbitrary long interaction times in the limit of high energies, while studies of local operations in quantum field theory usually consider finite time interactions in the low-energy regime. As such, the treatment of quantum field theory employed in these two cases also explores different properties of quantum fields.

One could then wonder whether there are any experiments that could be performed outside of these regimes that can be used to probe the quantum degrees of freedom of a mediating field. In other words, it might be possible to devise an experiment that does not explicitly rely on the local quantum degrees of freedom of a field but still depends on its quantum features. Indeed, recent proposals of table-top experiments suggest that it is possible to probe quantum features of the gravitational field in regimes of weak coupling and long interaction times. We devote the next Section to the discussion of these experiments and which assumptions are necessary to bypass the constraints that we discussed in this Section.

\section{Applications to Gravity Mediated Entanglement}\label{sec:GME}

In this Section, we will use the fact that experiments that can be described by the qc-model do not explicitly rely on the quantum degrees of freedom of mediators to discuss the recent proposals of measuring gravity mediated entanglement first presented in~\cite{B,MV}. Even though the experimental proposals are within the regimes of long interaction times and small couplings, these experiments have been claimed to witness quantum behaviour of the gravitational field.

\begin{figure}[h!]
    \centering
    \includegraphics[width=10cm]{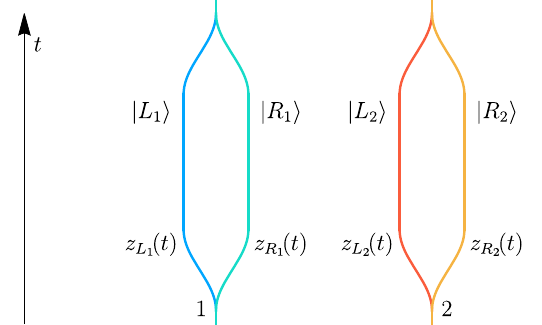}
    \caption{Schematic representation of the gravity mediated entanglement setup, where two particles labelled by $i=1,2$ can undergo a superposition of two trajectories, $\mf z_{L_i}(t)$ and $\mf z_{R_i}(t)$, corresponding to quantum states $\ket{L_i}$ and $\ket{R_i}$.}
    \label{fig:BMV}
\end{figure}

The original proposals of~\cite{B,MV} consider two particles that undergo a superposition of two paths, as shown in Fig.~\ref{fig:BMV} and interact only through the gravitational field. Due to differences in distance between the paths, it is possible that the particles end up in an entangled state after their interaction. For concreteness, one could think of the particles as two neutrons and split their paths by applying a magnetic field with a constant gradient. The neutrons would then interact gravitationally for a finite time and could be recombined through the application of another magnetic pulse. If the two neutrons are sufficiently shielded from other interactions and are measured to be entangled, one could then conclude that the gravitational field was responsible for entangling them. At this stage, the original proposal has been thoroughly studied, and many different setups have been considered~\cite{contBMVOG,contBMVJackson,contBMVBose} in the literature. However, in this Section, we will focus on the original proposal, as it contains the essential features shared by the other proposals.

In a simplified description, one can describe the experiment by considering two pointlike particles labelled by $i\in\{1,2\}$ with masses $m_1$ and $m_2$, whose centres of mass are quantum and can undergo two possible trajectories each, $\mf z_{\tc{R}_i}(t)$ and $\mf z_{\tc{L}_i}(t)$. We associate each possible trajectory to states $\ket{\tc{R}_i}$ and $\ket{\tc{L}_i}$ (see Fig.~\ref{fig:BMV}). During the relevant part of their interaction, the effective non-relativistic Hamiltonian that describes their interaction is simply the gravitational potential between each path:
\begin{equation}
    \hat{H}_I(t) = \sum_{{\substack{p_1\in\{\tc{L}_1,\tc{R}_1\}\\p_2\in\{\tc{L}_2,\tc{R}_2\}}}} -\frac{Gm_1m_2}{|\bm z_{p_1}(t) - \bm z_{p_2}(t)|} \ket{p_1p_2}\!\!\bra{p_1p_2}.
\end{equation}

One then typically considers the following superposed initial state for the two particles:
\begin{equation}\label{eq:initialBMV}
    \ket{\psi_0} = \frac{1}{\sqrt{2}}(\ket{\tc{L}_1} + \ket{\tc{R}_1})\otimes \frac{1}{\sqrt{2}}(\ket{\tc{L}_2} + \ket{\tc{R}_2}) = \frac{1}{2}(\ket{\tc{L}_1\tc{L}_2}+\ket{\tc{L}_1\tc{R}_2}+\ket{\tc{R}_1\tc{L}_2}+\ket{\tc{R}_1\tc{R}_2}).
\end{equation}
Time evolution of this initial state then gives the (in general not separable) final state
\begin{equation}
    \ket{\psi_f} = \frac{1}{2}(e^{-\ii \Phi_{\tc{l}_1\tc{l}_2}}\ket{\tc{L}_1\tc{L}_2}+e^{-\ii \Phi_{\tc{l}_1\tc{r}_2}}\ket{\tc{L}_1\tc{R}_2}+e^{-\ii \Phi_{\tc{r}_1\tc{l}_2}}\ket{\tc{R}_1\tc{L}_2}+e^{-\ii \Phi_{\tc{r}_1\tc{r}_2}}\ket{\tc{R}_1\tc{R}_2}),
\end{equation}
where
\begin{equation}
    \Phi_{p_1p_2} = - \int_0^{T} \dd t \frac{Gm_1m_2}{|\bm z_{p_1}(t) - \bm z_{p_2}(t)|}
\end{equation}
and $T$ denotes the time interval for which the particles remain in a superposition of trajectories. Overall, the final state of the particles can be entangled, with negativity given by
\begin{equation}
    \mathcal{N}(\ket{\psi_f}\!\!\bra{\psi_f}) = \max\big(0,\tfrac{1}{2}\sin(\tfrac{1}{2}\left(\Phi_{\tc{l}_1\tc{l}_2} + \Phi_{\tc{r}_1\tc{r}_2} - \Phi_{\tc{l}_1\tc{r}_2} - \Phi_{\tc{r}_1\tc{l}_2}\right))\big).\label{eq:NegBMVbasic}
\end{equation}
Moreover, in the limit where the paths are approximately inertial while undergoing the superposition of paths and the smallest separations between the paths is between the right branch of particle 1 and the left branch of particle 2 with $|\bm z_{\tc{l}_1} - \bm z_{\tc{l}_2}|, |\bm z_{\tc{r}_1} - \bm z_{\tc{l}_2}|,|\bm z_{\tc{l}_1} - \bm z_{\tc{r}_2}|\gg |\bm z_{\tc{r}_1} - \bm z_{\tc{l}_2}| = r_{12}$ (see Fig.~\eqref{fig:BMV}), we have
\begin{equation}
    \ket{\psi_f} \approx \frac{1}{2}\left(\ket{\tc{L}_1\tc{L}_2}+e^{- \ii \frac{G m_1m_2}{r_{12}}T}\ket{\tc{L}_1\tc{R}_2}+\ket{\tc{R}_1\tc{L}_2}+\ket{\tc{R}_1\tc{R}_2})\right),
\end{equation}
which becomes a maximally entangled state if the interaction time is given by
\begin{equation}
    T =  \frac{\pi r_{12}}{Gm_1m_2}.
\end{equation}

The important remark made in~\cite{B,MV} was that an experiment of this type might be realizable with current technology with reasonable values for the particles' masses, separations and times. The greatest experimental challenge would be to control the decoherence that the particles would experience due to other forces while allowing the masses to interact for the desired time $T$, of the order of seconds. It was then argued that if entanglement was found in the final state of the two particles, this would be evidence of the quantum behaviour of gravity, as

\begin{center}
\textit{``if gravity were a classical field, it could never entangle two particles''.}
\end{center}
\noindent We will discuss the exact assumptions that can lead one to the conclusion that the gravitational field is quantum from the results of the experiments at the end of this Section. In what follows, we will describe the experiment, first with a full quantum field theoretic approach, and then with a quantum-controlled model. From these results, we will discuss the possible implications of the experiment and the role played by the quantum degrees of freedom of the gravitational field in the proposed setup.



\subsubsection*{A Field Theoretic Description of Gravity Mediated Entanglement}

We start by describing a single fermionic particle (such as an electron/neutron/proton) undergoing a superposition of paths within a quantum field theoretic framework. Specifically, we will focus on the specific case where a magnetic field is applied for a finite time $T$, used to split the possible paths of an electron\footnote{We assume that the magnetic field switches on with a positive gradient along the $z$-axis, which splits the paths of spins up and down, then switches to a negative gradient along the $z$-axis, recombining the paths (see Fig.~\ref{fig:BMV}).}. The equation of motion for a Dirac fermion $\psi$ of mass $m$ under the influence of an external electromagnetic potential $A_\mu$ can be written as
\begin{equation}
    (\ii \slashed{\partial} - m + q \slashed{A})\psi = 0.
\end{equation}
We will work under the assumption that the external magnetic field gives rise to localized mode solutions $u_{\bm k,s}(\mf x)$ and $v_{\bm k,s}(\mf x)$, where $s = 1,2$ is the spin polarization and $\bm k$ is a discrete label (due to localized solutions), such that each of the spin polarizations in localized around a different trajectory. The labels $\bm k$ are then associated with the different modes within each path. We can then represent the electron quantum field as
\begin{equation}
    \hat{\psi}(\mf x) = \sum_{s\in\{\tc{l},\tc{r}\}} \sum_{\bm k} u_{\bm k,s}(\mf x) \hat{b}_{\bm k,s} + v_{\bm k,s}(\mf x) \hat{a}_{\bm k,s}^\dagger,
\end{equation}
where $\hat{b}_{\bm k,s}^\dagger$, $\hat{b}_{\bm k,s}$ are the creation and annihilation operators associated with electrons and $\hat{a}_{\bm k,s}^\dagger$, $\hat{a}_{\bm k,s}$ are associated with the positron states. The vacuum of this theory is defined by $\hat{b}_{\bm k,s}\ket{0} =\hat{a}_{\bm k,s}\ket{0} = 0$.

We will then focus on the description of an electron in modes labelled by $\bm k_0$ and $s=\tc{L},\tc{R}$, so that $s=\tc{L}$ corresponds to a trajectory that splits to the left and $s=\tc{R}$ corresponds to a trajectory that splits to the right, and both paths recombine after a timescale $T$. 

The coupling of the fermionic field with gravity is given by the interaction Hamiltonian density
\begin{equation}
    \hat{\mathcal{H}}_I(\mf x) = - \lambda \normord{\hat{T}^{\mu\nu}(\mf x)} \hat{h}_{\mu\nu}(\mf x),
\end{equation}
where $\lambda = \tfrac{1}{2}\sqrt{8 \pi G} = \sqrt{2\pi}\ell_p$ , $\hat{h}_{\mu\nu}(\mf x)$ is the linearized gravitational perturbation, as described in Section~\ref{sec:generalQFT}, and $\normord{\hat{T}_{\mu\nu}}$ is the normal ordered stress-energy momentum tensor of the field $\hat{\psi}(\mf x)$, explicitly given by

\begin{equation}\label{eq:TmunuFermion}
    \normord{\hat{T}^{\mu\nu}(\mf x)} = \frac{1}{2}\left(\normord{\hat{\overline{\psi}}(\mf x) \gamma^{(\mu}i\partial^{\nu)}\hat{\psi}(\mf x)} + \text{H.c.}\right).
\end{equation}
Expanding the fermionic field in terms of the creation and annihilation operators, we find

\begin{align}
    \normord{\hat{T}^{\mu\nu}(\mf x)} \,\,=\!\!\!\!\! \sum_{s,s'\in\{\tc{l},\tc{r}\}}&\sum_{\bm k, \bm k'} \Re(\overline{u}_{\bm k,s}(\mf x)\gamma^{(\mu} i \partial^{\nu)} u_{\bm k',s'}(\mf x)) \hat{b}_{\bm k,s}^\dagger \hat{b}_{\bm k',s'} + \Re(\overline{v}_{\bm k,s}(\mf x)\gamma^{(\mu} i \partial^{\nu)} v_{\bm k',s'}(\mf x)) \hat{a}^\dagger_{\bm k,s} \hat{a}_{\bm k',s'}\nonumber\\
    &+\Re(\overline{u}_{\bm k,s}(\mf x)\gamma^{(\mu} i \partial^{\nu)} v_{\bm k',s'}(\mf x)) \hat{b}_{\bm k,s}^\dagger \hat{a}_{\bm k',s'} + \Re(\overline{v}_{\bm k,s}(\mf x)\gamma^{(\mu} i \partial^{\nu)} u_{\bm k',s'}(\mf x)) \hat{a}^\dagger_{\bm k,s} \hat{b}_{\bm k',s'}
\end{align}
Restricting to the $(\bm k_0,s)$ positive frequency subspace, we obtain the induced observable
\begin{equation}
    \hat{T}_{\bm k_0}^{\mu\nu}(\mf x) \coloneqq \sum_{s,s'\in\{\tc{l},\tc{r}\}}^2 T^{\mu\nu}_{\bm k_0,s,s'}(\mf x)\hat{b}_{\bm k_0,s}^\dagger \hat{b}_{\bm k_0,s'},
\end{equation}
where 
\begin{equation}
    T^{\mu\nu}_{\bm k_0,s,s'}(\mf x) = \Re(\overline{u}_{\bm k_0,s}(\mf x)\gamma^\mu i \partial^\nu u_{\bm k_0,s'}(\mf x)).
\end{equation}
Under the assumption that the two paths determined by the modes $(\bm k_0,1)$ and $(\bm k_0,2)$ are such that the paths are disjoint, we have $T^{\mu\nu}_{\bm k_0,s,s'}(\mf x) = 0$ whenever $s\neq s'$ while the trajectories are split. While the trajectories are the same, we have $T^{\mu\nu}_{\bm k_0,s,s'}(\mf x)$ independent of $s$ or $s'$. During the relevant part of the experiment, we then have

\begin{equation}
     \hat{T}_{\bm k_0}^{\mu\nu}(\mf x) = T^{\mu\nu}_{\bm k_0,\tc{l},\tc{l}}(\mf x)\hat{b}_{\bm k_0,\tc{l}}^\dagger \hat{b}_{\bm k_0,\tc{l}}+T^{\mu\nu}_{\bm k_0,\tc{r},\tc{r}}(\mf x)\hat{b}_{\bm k_0,\tc{r}}^\dagger \hat{b}_{\bm k_0,\tc{r}}.
\end{equation}
For simplicity, we define
\begin{equation}
    \hat{\Pi}_\tc{l} = \hat{b}_{\bm k_0,\tc{l}}^\dagger \hat{b}_{\bm k_0,\tc{l}}, \quad
    \hat{\Pi}_\tc{r} = \hat{b}_{\bm k_0,\tc{r}}^\dagger \hat{b}_{\bm k_0,\tc{r}}, \quad\quad \Lambda_\tc{l}^{\mu\nu}(\mf x) = T^{\mu\nu}_{\bm k_0,\tc{l},\tc{l}}(\mf x), \quad\Lambda_\tc{r}^{\mu\nu}(\mf x) = T^{\mu\nu}_{\bm k_0,\tc{r},\tc{r}}(\mf x), 
\end{equation}
so that the interaction Hamiltonian can be written as
\begin{equation}\label{eq:BMV1part}
    \mathcal{H}_I(\mf x) = - \lambda \Lambda^{\mu\nu}_\tc{l}(\mf x)\hat{\Pi}_\tc{l}\hat{h}_{\mu\nu}(\mf x) - \lambda \Lambda^{\mu\nu}_\tc{r}(\mf x)\hat{\Pi}_\tc{r}\hat{h}_{\mu\nu}(\mf x).
\end{equation}

Not much changes when one considers a fermionic two-particle system, with each particle being able to undergo a superposition of paths in different regions of space, which we conveniently label 1 and 2. In this case the external potential that acts in the fermionic field has two distinct non-interacting parts, such that the mode solutions for the fermionic field of matter can be expanded as
\begin{align}
    \hat{\psi}(\mf x) = &\sum_{s\in\{\tc{l,r}\}} \sum_{\bm k} u^{(1)}_{\bm k,s}(\mf x) \hat{b}^{(1)}_{\bm k,s} + v^{(1)}_{\bm k,s}(\mf x) (\hat{a}^{(1)}_{\bm k,s})^\dagger+ \sum_{s\in\{\tc{l},\tc{r}\}} \sum_{\bm k} u^{(2)}_{\bm k,s}(\mf x) \hat{b}^{(2)}_{\bm k,s} + v^{(2)}_{\bm k,s}(\mf x) (\hat{a}^{2}_{\bm k,s})^\dagger.
\end{align}
The derivation of the interaction Hamiltonian in this case follows a similar approach to that of one single particle. The difference in when two particles are considered is that we will be looking at the two-particle subspace spanned by the states
\begin{align}
    \ket{\tc{L}_1\tc{L}_2} &= (\hat{b}^{(1)}_{\bm k_1,\tc{l}})^\dagger(\hat{b}^{(2)}_{\bm k_2,\tc{l}})^\dagger \ket{0}, &&&
    \ket{\tc{L}_1\tc{R}_2} = (\hat{b}^{(1)}_{\bm k_1,\tc{l}})^\dagger(\hat{b}^{(2)}_{\bm k_2,\tc{r}})^\dagger \ket{0}, \\
    \ket{\tc{R}_1\tc{L}_2} &= (\hat{b}^{(1)}_{\bm k_1,\tc{r}})^\dagger(\hat{b}^{(2)}_{\bm k_2,\tc{l}})^\dagger \ket{0}, &&&
    \ket{\tc{R}_1\tc{R}_2} = (\hat{b}^{(1)}_{\bm k_1,\tc{r}})^\dagger(\hat{b}^{(2)}_{\bm k_2,\tc{r}})^\dagger \ket{0}. 
\end{align}
Here, the labels $1$ and $2$ determine the branch of superposition that the particles undergo, and the labels L and R determine on which side of the setup of Fig.~\ref{fig:BMV} the particle is localized. Also notice that under the assumption that the L and R modes of the field are non-overlapping, we can effectively assign a tensor product structure to the field decomposition in regions L and R\footnote{Formally one cannot do this, but this can be done if one assumes, for instance, that there exists an infinite potential barrier for the field $\hat{\psi}$ between the two regions, which is not an unreasonable assumption in the case of the GME proposals, as one has to shield any other forms of entanglement between the particles.}. This allows us to effectively treat this reduced subspace as a tensor product, with respective vacua $\ket{0} = \ket{0_10_2}$, where the states $\ket{\tc{L}_i} = (\hat{b}^{(i)}_{\bm k_i,\tc{l}})^\dagger \ket{0}$ and $\ket{\tc{R}_i} = (\hat{b}^{(i)}_{\bm k_i,\tc{r}})^\dagger \ket{0}$ span the individual particles' Hilbert spaces.

Expanding the operator $\hat{T}^{\mu\nu}(\mf x)$ given by Eq.~\eqref{eq:TmunuFermion}, and reducing it to the subspace of interest with the same assumptions as in the one-particle case, we obtain an interaction Hamiltonian with the linearized quantum gravitational field that takes the form
\begin{equation}
    \hat{\mathcal{H}}_I(\mf x) = - \lambda \Lambda^{\mu\nu}_{\tc{l}_1}(\mf x)\hat{\Pi}^{(1)}_\tc{l}\hat{h}_{\mu\nu}(\mf x)  - \lambda \Lambda^{\mu\nu}_{\tc{r}_1}(\mf x)\hat{\Pi}^{(1)}_\tc{r}\hat{h}_{\mu\nu}(\mf x)- \lambda \Lambda^{\mu\nu}_{\tc{l}_2}(\mf x)\hat{\Pi}^{(2)}_\tc{l}\hat{h}_{\mu\nu}(\mf x) - \lambda \Lambda^{\mu\nu}_{\tc{r}_2}(\mf x)\hat{\Pi}^{(2)}_\tc{r}\hat{h}_{\mu\nu}(\mf x),
\end{equation}
where the expressions for each one of the terms above is analogous to Eq.~\eqref{eq:BMV1part} with the appropriate indices 1 and 2. In terms of the eigenstates $\ket{\tc{L}_i}$ and $\ket{\tc{R}_i}$, we have $\hat{\Pi}^{(i)}_\tc{l} = \ket{\tc{L}_i}\!\!\bra{\tc{L}_i}$, $\hat{\Pi}^{(i)}_\tc{r} = \ket{\tc{R}_i}\!\!\bra{\tc{R}_i}$.

We can now find the final state of the system of the two particles. Considering the initial state~\eqref{eq:initialBMV}, we can obtain the final state of the system, assuming that the linearized gravitational field starts the interaction in the vacuum state. After the interaction, we trace out the gravitational degrees of freedom to obtain the final state of the two particles after they recombine, $\hat{\rho}_\tc{g}$. This results in a mixed state for the two masses, as the particles become entangled with the gravitational field itself. To simplify the result, we assume that all trajectories are related by rotations and translations in space, so the local vacuum effect in each trajectory is the same. Under this assumption, the 
leading order negativity is computed in Appendix~\ref{app:FinalStates}:
\begin{align}\label{eq:Nq}
    \mathcal{N}(\r_\tc{g}) \!=\! \frac{\lambda^2}{2}\Big(\Big|G_{\tc{l}_1\tc{l}_2}\!\!+\!G_{\tc{r}_1\tc{r}_2}\!\!-\!G_{\tc{l}_1\tc{r}_2}\!\!-\!G_{\tc{r}_1\tc{l}_2}\Big|\!-\!\mathcal{L}\Big) +\mathcal{O}(\lambda^4),
\end{align}
where $\mathcal{L}$ is a local noise term associated with the individual particle's interaction with the gravitational field vacuum---corresponding to gravitational decoherence, given by integrals of the vacuum Wightman function $W_{\mu\nu\alpha'\beta'}(\mf x, \mf x')$ local to each particle---and 
\begin{equation}
    G_{p_1p_2} = G_F(\Lambda_{p_1},\Lambda_{p_j}) = \int \dd V\dd V' \Lambda_{p_1}^{\mu\nu}(\mf x) (G_F)_{\mu\nu\alpha'\beta'}(\mf x,\mf x') \Lambda_{p_2}^{\alpha'\beta'}(\mf x'),
\end{equation}
where $G_F$ is the Feynman propagator for the linearized gravitational field.

\subsubsection*{A Quantum-Controlled model for Gravity Mediated Entanglement}

We will now study the description of the GME experiment in terms of a quantum-controlled model. The associated interaction Hamiltonian density in this case is\footnote{Notice that the qc-interaction Hamiltonian picks up a factor of $2$, as the linearized gravitational field sourced by a stress-energy tensor is given by $h = - \sqrt{8 \pi G} G_RT_{\mu\nu} = - 2 \lambda G_R T_{\mu\nu}$, due to Eq.~\eqref{eq:hmunueom}.}

\begin{equation}\label{eq:Hclass}
    \hat{\mathcal{H}}_{\tc{qc}}(\mf x) = \lambda^2\!\!\!\!\sum_{{\substack{p_1\in\{\tc{L}_1,\tc{R}_1\}\\p_2\in\{\tc{L}_2,\tc{R}_2\}}}} \left(\Lambda^{\mu\nu}_{p_1}(\mf x){\Phi_{\mu\nu}^{p_2}(\mf x)} + \Lambda^{\mu\nu}_{p_2}(\mf x){\Phi_{\mu\nu}^{p_1}(\mf x)}\right)\ket{p_1p_2}\!\!\bra{p_1p_2} ,
\end{equation}
where $\Phi^{p_1}_{\mu\nu}(\mf x)$ denotes the retarded propagation of the stress-energy tensor of each system, i.e.,
\begin{align}
    \Phi^{p_i}_{\mu\nu}(\mf x) = \int \dd V' G_R^{\mu\nu}{}_{\alpha'\beta'}(\mf x,\mf x') \Lambda^{\alpha'\beta'}_{p_i}(\mf x').
\end{align}
This approach can be thought of as the relativistic unapproximated version of the interaction \mbox{$Gm_1m_2/|\hat{\bm x}_1 - \hat{\bm x}_2|$}, which, as we previously mentioned, does not explicitly consider the quantum degrees of freedom of the gravitational field.

Since the Hamiltonian density~\eqref{eq:Hclass} commutes with itself at different times. The time-evolution operator is simply given by
\begin{align}
    \hat{U}_I &= \exp({-\ii \int\dd V\,\hat{\mathcal{H}}_\tc{qc}(\mf x)}) =  \sum_{{\substack{p_1\in\{L_1,R_1\}\\p_2\in\{L_2,R_2\}}}} e^{{\ii\lambda^2} \Delta_{p_1p_2}}\ket{p_1 p_2}\!\!\bra{p_1p_2},
\end{align}
where
\begin{equation}
    \Delta_{p_1p_2} = \Delta(\Lambda_{p_1},\Lambda_{p_2}) = \displaystyle{\int} \dd V \dd V' \Lambda_{p_1}^{\mu\nu}(\mf x)\Delta_{\mu\nu}{}_{\alpha'\beta'}(\mf x,\mf x') \Lambda_{p_2}^{\alpha'\beta'}(\mf x').
\end{equation}

Using the initial state for the particles~\eqref{eq:initialBMV}, we obtain the following final density operator after the interaction
\begin{equation}
    \hat{\rho}_{\tc{c}} = 
    \frac{1}{4}\sum_{{\substack{p_1\in\{\tc{L}_1,\tc{R}_1\}\\p_2\in\{\tc{L}_2,\tc{R}_2\}}}} \!\!\!e^{ {\ii \lambda^2}( \Delta_{p_1p_2}-\Delta_{q_1q_2})}\ket{p_1p_2}\!\!\bra{q_1q_2}.
\end{equation}
The entanglement between the two particles can be evaluated through the negativity of the state $\hat{\rho}_\tc{c}$, which reads
\begin{align}
    \mathcal{N}(\r_\tc{c}) = \frac{1}{2}\sin(\frac{\lambda^2}{2} \Big|{\Delta_{\tc{l}_1\tc{l}_2}\!\!+\!\Delta_{\tc{r}_1\tc{r}_2} \!\!-\! \Delta_{\tc{l}_1\tc{r}_2} \!\!-\! \Delta_{\tc{r}_1\tc{l}_2}}\Big|)\nonumber\\
    = {\frac{\lambda^2}{4}}\Big|{\Delta_{\tc{l}_1\tc{l}_2}\!\!+\!\!\Delta_{\tc{r}_1\tc{r}_2}\!\! -\!\! \Delta_{\tc{l}_1\tc{r}_2}\!\! -\!\! \Delta_{\tc{r}_1\tc{l}_2}}\Big|+\mathcal{O}(\lambda^4).\label{eq:Nc}
\end{align}
With the typical choice of paths for the GME experiment, the above quantity is non-zero. Moreover, using the relationship of the Feynman propagator with the symmetric propagator~\eqref{eq:GFHDelta}, we find that Eq.~\eqref{eq:Nc} matches~\eqref{eq:Nq} when one ignores the noise terms $\mathcal{L}$ and the state dependent part of the Feynman propagator (encoded in the Hadamard distribution). Comparing with the initial description given in~\eqref{eq:NegBMVbasic}, we also see that $\Phi_{p_1p_2} \sim \Delta_{p_1p_2}$.

\subsubsection*{What can Gravity Mediated Entanglement Tell us about Quantum Gravity?}

For the choices of parameters initially proposed in the setup~\cite{B,MV}, the ratio of the time duration of the interaction $T$ and the distance between the systems is at least of the order of $10^{12}$. This is precisely the limit at which the quantum field theory model can be well described by a quantum-controlled model, as we saw in Section~\ref{sec:QFTapproxQC}. One also has $E_{p_1p_2} = 0$ in this case for $p_i \in \{\tc{L}_i,\tc{R}_i\}$, so that the qc-description is safe from causality violations (see~\cite{eirini} for more on this topic). This essentially implies that the results in this experimental setup are independent of the local quantum degrees of freedom of the gravitational field.

However, as pointed out in a series of works~\cite{B,MV,MVwhen,BMVqft}, the GME experiments can be used to probe quantum aspects of the gravitational interaction using an argument based on an additional assumption regarding \textit{locality}. To fully understand the argument, it is important to distinguish between two fundamentally different notions of locality. The first notion comes from the description of spacetime and is deeply linked with causality. It states that operations happen at events in spacetime, and do not affect other events which are causally disconnected from them. We will call this notion \textit{event locality}. The second notion of locality comes from quantum mechanics and states that operations that independently affect two quantum systems must be separable. We call this notion \textit{system locality}~\cite{ThomasFlaminiaAndJohn}. The notion of system locality alone is agnostic about causal structure or any underlying notion of spacetime. Although these notions of locality are different, there is a particular framework that links the two: in quantum field theory, the postulate of microcausality makes it so that two systems can only become entangled through event-local interactions, ensuring that system locality can only be violated when event locality is satisfied. 

Under the assumption that the gravitational interaction also establishes this link between event locality and system locality, one can conclude more about the results of the experiment. In this case, a mediator for the gravitational interaction is required in order to not violate system locality: we need mass A to couple to the field and then the field to carry quantum information to mass B, otherwise we would have action-at-a-distance. In summary, this assumption then becomes equivalent to assuming (the very reasonable idea) that the gravitational field has local degrees of freedom, which rules out a direct interaction between the masses, such as the one prescribed in the qc-model. In other words,

\begin{center}
    \textit{``If gravity has local degrees of freedom that can entangle two masses, \\then these degrees of freedom cannot be classical.''}
\end{center}

\noindent Without the assumption of locality, the experiment does not, however, rule out the possibility that the gravitational interaction is not described by mediators, and instead works very differently from all other known interactions.

Although the assumption gravity has local degrees of freedom might seem reasonable (and it is), it is important to highlight that the only known framework that successfully links the notion of event locality and system locality is quantum field theory. One could then argue that this assumption ends up being equivalent to assuming that the gravitational field is described as a quantum field in the first place. This ends up configuring a circular reasoning, if the intention is to witness quantum degrees of freedom of gravity.

This, however, does not undervalue the GME experiment as a means to understand the relationship between gravity and quantum matter. There is currently no experimental data about how quantum systems source gravity, or how two quantum systems interact gravitationally\footnote{Although the interaction of quantum systems with the gravitational field sourced by classical matter has been known since the COW experiment~\cite{COW}.}. Regardless of any additional assumption, the GME experiments could determine whether gravity can entangle two quantum systems, which would be, by itself, a great achievement in theoretical and experimental physics. For instance, this result would be enough to rule out semiclassical gravity, or the Di\'osi-Pensrose collapse models~\cite{DiosiGravMeas,PenroseCollapse}, as fundamental descriptions for the gravitational field sourced by quantum matter, which have not yet been confirmed experimentally.

\chapter{The Geometry of Spacetime from Quantum Field Theory}\label{chap:geometry}

In this Chapter, we will discuss the relationship between the Hadamard condition and the geometry of spacetime, as well as the possibility that the gravity might be emergent from quantum field theory. In Section~\ref{sec:geometry}, we will show how the Hadamard condition contains all information about the geometry of spacetime, discussing how to physically access the geometry through quantum measurements in Section~\ref{sec:geometryFromMeas}. Section~\ref{sec:resonable?} is devoted to the discussion of whether a theory where spacetime is completely replaced by the correlations of quantum fields could be defined. Section~\ref{sec:speculation} is devoted to a brief discussion about a formulation of spacetime in terms of entanglement in quantum field theory.

\section{The Hadamard Condition and the Geometry of Spacetime}\label{sec:geometry}

The Hadamard condition imposes a universal UV behaviour to the correlations of a quantum field. If the equivalence principle states that measurements at highly localized regions of spacetime behave as they would in flat space, the Hadamard condition imposes the same for quantum field theories. Given this connection between the Hadamard condition and the equivalence principle, it should be no surprise that the Hadamard condition is deeply linked with general relativity and the geometry of spacetime. In this Section, we will see that not only is that the case, but that the Hadamard condition also implies that the correlations of a quantum field contain full information about the geometry of spacetime, and can perhaps even completely replace the role played by the metric.


In 2015, the then PhD students Mehdi Saravani and Siavash Aslanbeigi took the the course AMATH 875, delivered by Prof. Achim Kempf at the Perimeter Institute. During this period, the three made the discovery that one can write the spacetime metric in terms of the Feynman propagator~\cite{achim}. Equation (10) of~\cite{achim} states\footnote{denoting $G(\mf x, \mf x') = G_F(\mf x, \mf x')$ in a spacetime of dimension $D$.}:
\begin{equation}\label{eq:saravani}
g_{ij}(y) = -\frac{1}{2} \left[ \frac{\Gamma\left(\frac{D}{2} - 1\right)}{4\pi^{D/2}} \right]^{\frac{2}{D-2}} 
\lim_{x \to y} \frac{\partial}{\partial x^i} \frac{\partial}{\partial y^j} 
\left( G(x, y)^{\frac{2}{2 - D}} \right).
\end{equation}
This result showed that, at least in principle, one could replace the description of the geometry by the correlations of quantum fields~\cite{achim2}.

It turns out that Eq.~\eqref{eq:saravani} is a consequence of the Hadamard condition, as first pointed out in~\cite{geometry}. As a matter of fact, one can quickly derive it from the expansion that defines the Hadamard condition~\eqref{eq:Hadamard} by noticing that in the limit of $\mf x' \approx \mf x$, the dominant diverging term of the Wightman function is $1/\sigma$,
\begin{equation}
    W(\mf x, \mf x') \approx \frac{1}{8\pi^2 \sigma(\mf x, \mf x')} \quad \Rightarrow \quad \sigma(\mf x, \mf x') \approx \frac{1}{8\pi^2 W(\mf x,\mf x')}.
\end{equation}
One can then rewrite the spacetime metric in terms of the coincidence limit of Synge's world function, $g_{\mu\nu}(\mf x) = - \lim_{\mf x' \to \mf x} \partial_\mu \partial_{\nu'} \sigma(\mf x,\mf x')$, yielding
\begin{equation}\label{mine}
    g_{\mu\nu}(\mf x) =  -\frac{1}{8 \pi^2} \lim_{\mf x' \to \mf x} \left(\pdv{}{x^\mu}\pdv{}{x^{\nu'}}\frac{1}{W(\mf x, \mf x')}\right).
\end{equation}
Given that the Hadamard condition implies that the correlations of a quantum field behave inversely proportional to the geodesic separation between events, it should be no surprise that one can recover the metric from these correlations. Indeed, the essential information encoded by the metric is exactly how distances behave infinitesimally. Equation~\eqref{mine} then tells us how to connect the correlation functions of a quantum field with the background metric. 

In a sense, any physical property that has a scaling behaviour that depends on the distance between two points can be used to recover (at least some) information about the geometry. For instance, one could consider a classical charged particle in spacetime. The charge would then source a Coulomb field that would decay with the inverse of the geodesic separation between a test charge and the source. By measuring the Coulomb field, one would then be able to recover $\sigma(\mf x, \mf x')$ locally around the source, and thus the (spatial) metric. One could then wonder whether there is anything special about Eq.~\eqref{mine}.

It is important to stress that any measurement of distances and time separations gives direct information about the geometry of spacetime, as these measurements provide values of $\sigma(\mf x, \mf x')$ between events. The collection of all possible measurements of space and time will then always allow one to recover spacetime, regardless of how these measurements are performed. However, there are two important features of Eq.~\eqref{mine} that make it significantly more appealing than utilizing other effective ``rulers'' and ``clocks''. The first feature is that the Wightman function can measure both space separations and time intervals: one single object directly gives us both quantities. And the second, and perhaps most important, feature is the fact that whether we choose to measure $W(\mf x, \mf x')$ or not, it is always there. From the point that a quantum field theory is defined, its correlation function is defined everywhere in spacetime, regardless of the (Hadamard) state that the field is at. The fact that the UV behaviour of quantum correlations is universal implies that the distances between all events in spacetime are automatically encoded in the quantum field theory itself.

\subsubsection*{Can One Replace the Geometry by the Correlations of Quantum Fields?}

In 1687 Newton revolutionized physics, starting the tools necessary to begin to study dynamics. Newton showed us that gravity was a force, and we learned that the Moon is falling to the Earth in the same way that an apple falls from a tree. It took three centuries for our understanding of gravity to change. The next revolutionizing discovery came in 1915, when Einstein showed that gravity was not a force, but instead the concept of spacetime itself, and it so happens that spacetime can be curved. Indeed, after General Relativity, the concept of ``measuring gravity'' fundamentally changed. What Newton would have called ``the gravity on Earth's surface'', $9.8\,\text{m/s}^2$, ended up being the acceleration of a static observer that is uniformly accelerated away from the center of the Earth due to \textit{electromagnetic repulsion}. Instead, in the context of General Relativity,

\begin{center}
    ``\textit{gravity is simply where things are, in both space and time,}''
\end{center}

\noindent and gravity measurements are then any measurements of relative distances and time intervals, entirely encoded in $\sigma(\mf x, \mf x')$.


When Einstein first conceived the concept of what we now call Minkowski spacetime, it was in an attempt to put the theory of electromagnetism and Galilean mechanics together. It turned out that the only way of making them compatible was by considering a deep link between space and time, which resulted in Special Relativity. In modern times, we see quantum field theory and general relativity being incompatible, and naturally leading us to the Hadamard condition. One could then wonder whether the Hadamard condition can tell us more about gravity, similar to the way in which Maxwell's equations turned out to imply relativity.

Indeed, as first suggested in~\cite{achim2}, one could potentially phrase spacetime as a differential manifold $\mathcal{M}$ and a correlation function $W(\mf x, \mf x')$, obtaining the corresponding metric from the limit~\eqref{mine}. However, this argument might seem circular: the Wightman function is defined by a formulation of quantum field theory in terms of smeared field operators $\hat{\phi}(f)$, which itself depends on equations of motion in a background spacetime. However, this does not need to be the case if one instead considers some type of background-independent quantum field theory. For instance, one could start with spacetime as a differential manifold $\mathcal{M}$ (without an a-priori metric), and define a quantum field theory as a linear association $f\mapsto \hat{\phi}(f)$ satisfying $\hat{\phi}(f)^\dagger = \hat{\phi}(f)$, and the local algebras $\mathcal{A}(\mathcal{O})$ generated by $\hat{\phi}(f)$ and $\openone$ with $f\in C_0^\infty(\mathcal{O})$, without explicitly imposing the equations of motion or commutation relations. Instead, one would rebuild the causal structure of spacetime from the commutating algebras. In this case, one would still have a Wightman function $W(f,g) = \omega(\hat{\phi}(f)\hat{\phi}(g))$ and its subsequent kernel $W(\mf x, \mf x')$. From $W(\mf x, \mf x')$, one could then define a metric for spacetime, based on~\eqref{mine}, which would be compatible with the causal structure obtained from the commutation of the local algebras. At this stage, it is not clear which restrictions would have to be imposed on the algebras $\mathcal{A}(\mathcal{O})$ in this sort of background independent theory to recover a standard quantum field theory on a Lorentzian manifold. However, it is certainly the case that if $\mathcal{M}$ admits a metric that makes it globally hyperbolic, then there exist associations of local algebras (such as the ones built in Section~\ref{sec:QFT}) that would allow for such construction. 

Following this idea, one could potentially start with a background-independent quantum field theory and a differentiable manifold $\mathcal{M}$ and \textit{define} a Lorentzian metric from Eq.~\eqref{mine}. Valid states in this theory would then be the states that yield the same background metric, implying that they satisfy the Hadamard condition with respect to the corresponding background geometry. In this formulation, gravity would be an effective theory, and its classical degrees of freedom would instead be associated with the ``vacuum'' degrees of freedom of the field $\hat{\phi}(\mf x)$. For instance, a gravitational wave would correspond to differences in the short-scale behaviour of $W(\mf x, \mf x')$. By construction, these short-scale differences would be present in the correlations of all states of the quantum field theory.

The idea of defining the spacetime from correlations of quantum fields is, at the very least, an idea worth pursuing. However, as we will discuss in Section~\ref{sec:resonable?}, it seems that the simple formulation based on Eq.~\eqref{mine} fails to satisfy some desired features for a theory that describes the coupling of quantum fields with gravity. Before addressing the consequences of considering a formulation for gravity emergent from quantum correlations, we will first study an explicit setup where one can recover the geometry of spacetime from measurements of quantum fields.





\section{The spacetime geometry from quantum measurements}\label{sec:geometryFromMeas}

In this Section, we will discuss the results of~\cite{geometry}, which provide a concrete method for determining the geometry of spacetime when one only has access to local measurements of a quantum field. This is an operational approach to the concept of replacing the metric of spacetime with the correlations of quantum fields, showing that one can measure distances and time separations through localized measurements of short-scale quantum correlations.

\subsubsection*{Measuring Two-Point Correlations of a quantum field}

    The first step will be to show how two-level Unruh-DeWitt detectors can be used to recover the correlations of the field. We start by considering a family of functions $\Lambda_{\mf p,\sigma}(\mf x)\in C^\infty(\M)$ that approximates a Dirac delta centered at $\mf p$ in the limit of $\sigma\to 0$, in the sense that
    \begin{equation}
        \lim_{\sigma \to 0}\Lambda_{\mf p, \sigma}(f) = \lim_{\sigma \to 0}\int \dd V \Lambda_{\mf p, \sigma}(\mf x) f(\mf x) = f(\mf p)\quad \forall\,f\in C^\infty(\M).
    \end{equation}
    For instance, in Minkowski spacetime, an example of such a function would be
    \begin{equation}
        \Lambda_{\mf p,\sigma}(\mf x) = \frac{e^{ - \frac{(\mf x - \mf p)^2}{2\sigma^2}}}{(2\pi \sigma^2)^2},
    \end{equation}
    where $(\mf x - \mf p)^2 = (x^0 - p^0)^2 + (\bm x - \bm p)^2$ (notice that this is \textit{not} a Lorentz contraction) and $x^\mu$, $p^\mu$ denote the coordinates of $\mf x$ and $\mf p$ in an inertial coordinate system. In this case, if $\mf p \neq \mf q$ are two events in spacetime that are not in null separation, one can recover the Wightman function $W(\mf p ,\mf q)$ as the limit
    \begin{equation}\label{eq:Wpqlimit}
        W(\mf p, \mf q) = \lim_{\sigma \to 0} W(\Lambda_{\mf p, \sigma},\Lambda_{\mf q, \sigma}),
    \end{equation}
    where we note that $W(\mf p,\mf q)$ is a finite number whenever $\sigma(\mf p,\mf q)\neq 0$ due to the Hadamard condition~\eqref{eq:Hadamard}. 

    The result~\eqref{eq:Wpqlimit} suggests a simple method for approximating the Wightman function between two points using particle detectors. Consider two two-level Unruh-DeWitt detectors that interact with a real scalar field according to the interaction Hamiltonian~\eqref{eq:HItwoUDWs}, with spacetime smearing functions\footnote{Notice that with this choice of spacetime smearing functions, the coupling constant has units of energy.} given by
    \begin{equation}
        \Lambda_\tc{a}(\mf x) = \Lambda_{\mf x_\tc{a},\sigma}(\mf x),\quad\quad 
        \Lambda_\tc{b}(\mf x) = \Lambda_{\mf x_\tc{b},\sigma}(\mf x),
    \end{equation}
    for two \textit{spacelike separated} events $\mf x_\tc{a}$ and $\mf x_\tc{b}$, and sufficiently small $\sigma$, controlling the extension of the interaction both in space and time. If the detectors start their interaction with the field in their ground state and the field is in a quasifree state, the leading order final state of the detectors $\hat{\rho}_\tc{ab}$ will be given by Eq.~\eqref{eq:UDWabFinal}. We can then compute the correlations between the monopole moments $\hat{\mu}_\tc{a}$ and $\hat{\mu}_\tc{b}$ when one performs measurements of detector A at $\tau_\tc{a} = t_\tc{a}$ and of detector B at $\tau_\tc{b} = t_\tc{b}$. The correlation is given by
    \begin{align}\label{eq:corrmuAmuB}
        \langle \hat{\mu}_\tc{a}(t_\tc{a}) \hat{\mu}_\tc{b}(t_\tc{b})\rangle_{\r_{\tc{ab}}} &=  2 \Re(e^{- \ii \Omega_\tc{a} t_\tc{a} + \ii \Omega_\tc{b} t_\tc{a}}\mathcal{L}_\tc{ab}^-) + 2 \Re(e^{+\ii \Omega_\tc{a} t_\tc{a} + \ii \Omega_\tc{b} t_\tc{a}}\mathcal{M}) +\mathcal{O}(\lambda^4)\\
        &= 2\lambda^2 \Re\big(e^{- \ii \Omega_\tc{a} t_\tc{a} + \ii \Omega_\tc{b} t_\tc{a}}W(\Lambda_\tc{a}^-,\Lambda_\tc{b}^+)-e^{\ii \Omega_\tc{a} t_\tc{a} + \ii \Omega_\tc{b} t_\tc{a}}G_F(\Lambda_\tc{a}^+,\Lambda_\tc{b}^+)\big) + \mathcal{O}(\lambda^4),\nonumber
    \end{align}
    where $\Lambda_\tc{a}^\pm(\mf x) = e^{\pm \ii \Omega_\tc{a}\tau_\tc{a}}\Lambda_{\mf x_\tc{a},\sigma}(\mf x)$, $\Lambda_\tc{b}^\pm(\mf x) = e^{\pm\ii \Omega_\tc{b}\tau_\tc{b}}\Lambda_{\mf x_\tc{b},\sigma}(\mf x)$. In the limit $\sigma\to 0$ of the terms in the expressions above we will obtain the Wightman function and Feynman propagator, as well as the exponentials $e^{\pm \ii \Omega_\tc{a}\tau_\tc{a}(\mf x)}$, $e^{\pm \ii \Omega_\tc{b}\tau_\tc{b}(\mf x)}$, evaluated at the center points of the interactions: $\mf x = \mf x_\tc{a}$ and $\mf x = \mf x_\tc{b}$. For convenience, denote $\tau_\tc{a}(\mf x_\tc{a}) = \tau_{\mf x_\tc{a}}$ and $\tau_\tc{b}(\mf x_\tc{b}) = \tau_{\mf x_\tc{b}}$ (not to be confused with the measurement times $t_\tc{a}$ and $t_\tc{b}$). Taking the limit $\sigma\to 0$, we have
    \begin{align}\label{eq:sigmato0WGF}
        \lim_{\sigma\to 0} W(\Lambda_\tc{a}^-,\Lambda_\tc{b}^+) = e^{- \ii \Omega_\tc{a} \tau_{\mf x_\tc{a}} + \ii \Omega_\tc{b} \tau_{\mf x_\tc{b}}}W(\mf x_\tc{a},\mf x_\tc{b}),\quad 
        \lim_{\sigma\to 0} G_F(\Lambda_\tc{a}^-,\Lambda_\tc{b}^+) = e^{\ii \Omega_\tc{a} \tau_{\mf x_\tc{a}} + \ii \Omega_\tc{b} \tau_{\mf x_\tc{b}}}W(\mf x_\tc{a},\mf x_\tc{b}),
    \end{align}
    where we used that both the causal propagator and the symmetric propagator vanish for spacelike separated events, implying $W(\mf x_\tc{a},\mf x_\tc{b}) = G(\mf x_\tc{a},\mf x_\tc{b}) = \tfrac{1}{2}H(\mf x_\tc{a},\mf x_\tc{b})$. Plugging Eq.~\eqref{eq:sigmato0WGF} into Eq.~\eqref{eq:corrmuAmuB} we then find
    \begin{align}
        \langle \hat{\mu}_\tc{a}(t_\tc{a}) \hat{\mu}_\tc{b}(t_\tc{b})\rangle_{\r_{\tc{ab}}} = 4\lambda^2 \sin(\Omega_\tc{a}(\tau_{\mf x_\tc{a}} + t_\tc{a}))\sin(\Omega_\tc{b}(\tau_{\mf x_\tc{b}} +t_\tc{b})) W(\mf x_\tc{a}, \mf x_\tc{b}) + \mathcal{O}(\lambda^4).
    \end{align}
    It is then clear that one can recover $W(\mf x_\tc{a},\mf x_\tc{b})$ from the correlations of two detectors by choosing appropriate measurement times. In particular, by choosing $t_\tc{a} = (2n+\frac{1}{2})\pi - \tau_\tc{a}$, $t_\tc{b} = (2m+\frac{1}{2})\pi - \tau_\tc{b}$, we can simply write
    \begin{equation}\label{eq:Wxaxb}
        W(\mf x_\tc{a}, \mf x_\tc{b}) = \frac{1}{4 \lambda^2} \langle \hat{\mu}_\tc{a}(t_\tc{a}) \hat{\mu}_\tc{b}(t_\tc{b})\rangle_{\r_{\tc{ab}}} + \mathcal{O}(\lambda^2).
    \end{equation}

    It is important to disclaim that, in principle, the Unruh-DeWitt model is not well defined in the limit $\sigma\to 0$, which corresponds to an interaction at a given event in spacetime. Indeed, in this case, each detector's individual excitation probability diverges. However, for sufficiently small $\sigma$, the result of Eq.~\eqref{eq:Wxaxb} approximately holds. In summary, when two Unruh-DeWitt detectors interact with a scalar quantum field in small regions localized around spacelike separated events $\mf x_\tc{a}$ and $\mf x_\tc{b}$, the correlations between their monopole moments become proportional to the field correlations $W(\mf x_\tc{a}, \mf x_\tc{b})$.

    We can also recover the Wightman function between two timelike separated events by considering a single Unruh-DeWitt detector. This result was first shown in~\cite{pipo} and then applied in~\cite{geometry}. To see this explicitly, consider a detector with spacetime smearing function of the form
    \begin{equation}
        \Lambda(\mf x) = \Lambda_{\mf x_1,\sigma}(\mf x)+\Lambda_{\mf x_2,\sigma}(\mf x)
    \end{equation}
    with \textit{timelike} separated $\mf x_1$ and $\mf x_2$, and for convenience, let us assume that $\mf x_1$ is in the causal past of $\mf x_2$. Essentially, this corresponds to a detector that interacts with the field twice, once locally around $\mf x_1$ and the second time around $\mf x_2$. It is then natural to assume that the timelike coordinate that defines the Unruh-DeWitt detector is the Fermi normal coordinate time associated with a trajectory $\mf z(\tau)$ that passes through $\mf x_1$ at $\tau = \tau_1$ and at $\mf x_2$ at $\tau = \tau_2$. 
    
    Let us assume that the detector is initially in its ground state and that the field is in a quasifree state. Then, the detector's leading order excitation probability after interacting with the field at $\mf x_1$ and $\mf x_2$ is
    \begin{equation}
        P_{12} = \lambda^2 W(\Lambda^-,\Lambda^+) = \lambda^2 \big(W(\Lambda_1^-,\Lambda_1^+)+W(\Lambda_1^-,\Lambda_2^+)+W(\Lambda_2^-,\Lambda_1^+)+W(\Lambda_2^-,\Lambda_2^+)\big),
    \end{equation}
    with $\Lambda_i^\pm(\mf x) = e^{\pm \ii \Omega \tau}\Lambda_{\mf x_i,\sigma}(\mf x)$. For convenience, define $P_i = \lambda^2W(\Lambda_{i}^-,\Lambda_{i}^+)$, corresponding to the excitation probabilities if the detector had interacted with the field only in the region around $\mf x_i$. We can then rewrite
    \begin{equation}
        W(\Lambda_1^-,\Lambda_2^+)+W(\Lambda_2^-,\Lambda_1^+) = \frac{1}{\lambda^2}(P_{12} - P_1 - P_2).
    \end{equation}
    Although in the limit $\sigma\to 0$, the probabilities $P_{12}$, $P_1$ and $P_2$ are divergent, the expression on the left-hand side is finite for all $\sigma$, implying that the difference $P_{12} - P_1 - P_2$ is also finite even in the limit of $\sigma\to 0$. In this limit we obtain $W(\Lambda_1^-,\Lambda_2^+)\to e^{-\ii \Omega(\tau_1-\tau_2)}W(\mf x_1, \mf x_2)$ and $W(\Lambda_2^-,\Lambda_1^+)\to e^{\ii \Omega(\tau_1-\tau_2)}W(\mf x_2, \mf x_1)$, so that we find
    \begin{equation}\label{eq:Wx1x2}
        \cos(\Omega \Delta\tau)\Re(W(\mf x_1,\mf x_2)) + \sin(\Omega \Delta\tau)\Im(W(\mf x_1,\mf x_2)) =  \lim_{\sigma \to 0}\frac{1}{2\lambda^2}(P_{12} - P_1 - P_2),
    \end{equation}
    with $\Delta\tau = \tau_2 - \tau_1$. Equation~\eqref{eq:Wx1x2} shows that one can recover both the real and the imaginary parts of the Wightman function between two timelike separated points by comparing $P_{12}$ and $P_1$, $P_2$. Although the limit $\sigma\to 0$ is divergent, Eq.~\eqref{eq:Wx1x2} holds approximately for sufficiently small $\sigma$. Overall, we showed that it is possible to use particle detectors to recover $W(\mf x, \mf x')$ for any two events that are either timelike or spacelike separated.

    \subsubsection*{The setup}\label{sub:general}
    
    The results of the previous Segment show that it is possible to obtain the exact form of the Wightman function in different points of {spacetime} if we have precise enough measurements of the correlations between sufficiently localized Unruh-DeWitt detectors. Combining these results with the fact that the background metric can be recovered from the correlations of quantum fields, we can now show how it is possible to recover the spacetime metric from local measurements of quantum fields.
    
    
    We propose the following setup to recover the spacetime metric by locally measuring a quantum field:
        \begin{enumerate}
            \item Couple local probes to a quantum field.
            \item Measure the correlation between the probes at different {spacetime points}.
            \item From the correlation between the detectors, compute the field two-point function between the corresponding events.
            \item Compute the metric by taking the coincidence limit in Eq. \eqref{mine}.
        \end{enumerate} 
    Although these steps might, in principle, seem simple, one must be careful with their implementation. Indeed, to obtain the spacetime metric with some precision, the limit of step four needs to be taken with enough precision. This relies on coupling probes separated by small enough spacetime intervals, so that the limit can be approximated well enough. In practice, this requires significant control of the probe systems. However, in principle, it is possible to recover the spacetime metric with arbitrary precision using this procedure, provided that the probes are small enough and their coupling with the field can be controlled with enough precision.
    
    For the remainder of this Section, we will consider Unruh-DeWitt detectors in different spacetimes. We will consider detectors that are either pointlike or sufficiently small, that couple fast enough to the quantum field. This allows us to use the approximations \eqref{eq:Wxaxb} and \eqref{eq:Wx1x2}. We will compute the (experimentally accessible) correlation function between detectors placed in a local region of spacetime separated by a coordinate separation $L$ and adapt Eq.~\eqref{mine} to this discrete setup. In essence, given that the interaction happens very approximately pointlikewise in spacetime, we will effectively have access to the Wightman function associated with a discrete lattice of points. We then take the discrete derivative of the Wightman function using the points in the lattice given by the center of the interaction of the detectors\footnote{We will also explicitly analyze the effect of a finite region of interaction in our last example}.

    In our setup, we will assume that the experimentalist can use a coordinate system $x^\mu = (x^0,x^i)$ to label the events in spacetime where the measurements take place, but does not have access to any local notion of space or time separation. In other words, with this information, it is possible to label events, but it is not possible to compute any physical spacetime distance.  Mathematically, this is saying that spacetime can locally be regarded as a $4$-dimensional manifold, but that there is no known spacetime metric. Our goal is to use particle detectors that interact at events which are labelled with values of $x^\mu = (x^0,x^i)$, and from the readouts of those detectors, infer spacetime metric components in the lab coordinates.

    We consider a set of $N^3$ detectors parametrized by $(\text{j}_1,\text{j}_2,\text{j}_3)$ where each $\text{j}_i$ runs from $1$ to $N$. For simplicity, let us work under the assumption that the detectors undergo trajectories associated with the coordinate system $x^\mu$, so that {they} move along the curves \mbox{$x^i = x^i_{\text{j}_i} = \text{const}$}. Then, $x^i_{\text{j}_i}$ are the constants that determine the spatial coordinates of each detector. This simplifying assumption will allow us to easily compute the metric in the $x^\mu$ coordinates. 
        
    \begin{figure}
        \centering
        \includegraphics[scale=0.43]{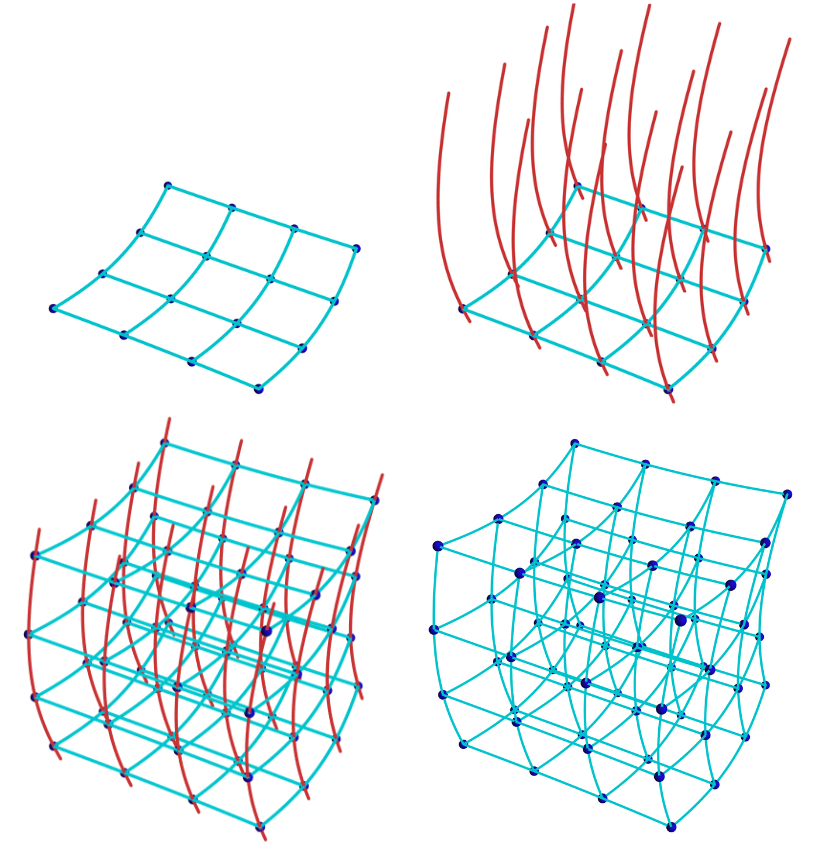}
        \caption{The setup described in Subsection \ref{sub:general}, where a lattice of particle detectors evolves through time and interacts with the field, effectively creating a spacetime lattice from the centres of interaction regions. }
        \label{fig:scheme}
    \end{figure}
    
    We consider that each detector interacts $N_0$ times with the field. The time coordinate of the center points of the interactions will be given by $x^0 = x^0_{\text{j}_0}$, with ${\text{j}_0}$ running from $1$ to $N_0$, corresponding to the values of time where the interactions happen. In this setup, we obtain a $4$-dimensional lattice of points labelled by $({\text{j}_0},\text{j}_1,\text{j}_2,\text{j}_3)$ associated to the events in which the detector interactions take place. A schematic representation of the setup can be found in Fig. \ref{fig:scheme}. We can then measure the detectors simultaneously at spacelike surfaces determined by $x^0 = \text{const}$. This allows one to obtain (from the readouts of the detectors) an approximation to the correlation function of the quantum field $W(\mf x,\mf x')$ when $\mf x$ and $\mf x'$ are events in the lattice.

    As discussed, to recover the spacetime metric, we will employ the discrete derivative of the Wightman function. We can obtain it directly from experimentally measurable detector data from the local measurements centred at the points parametrized by $\mathfrak{j}\coloneqq({\text{j}_0},\text{j}_1,\text{j}_2,\text{j}_3)$. We denote the coordinates of the interaction point $(x^0_{\text{j}_0},x^1_{\text{j}_1},x^2_{\text{j}_2},x^3_{\text{j}_3})$ by $x^\mu_{\mathfrak{j}}$, so that after measuring the quantum field with the particle detectors, we obtain $W(\mf x_{\mathfrak{j}},\mf x_{\mathfrak{l}})$ for all values of the multi-indices $\mathfrak{j}$ and $\mathfrak{l}$.

    To write the discrete derivative in a simple way, we define the object
    \begin{equation}
        \bm 1_\mu = (\underbrace{0,...,0}_{{\mu-1}},1,0,...,0).
    \end{equation}
    With this convention and the labelling $x_{\mathfrak{j}}^\mu$ for the coordinates of the events, given a scalar function $f(\mf x)$, it is possible to write its discrete derivative as
    \begin{equation}\label{eq:wtf}
        \left. \pdv{f}{x^\mu} \right|_{\mf x =\mf x_{\mathfrak{j}}}  \approx \frac{f(\mf x_{\mathfrak{j}+\bm 1_\mu})-f(\mf x_{\mathfrak{j}})}{x^\mu_{\mathfrak{j}+\bm 1_\mu}-x^\mu_{\mathfrak{j}}}.
    \end{equation}
    Intuitively, Eq. \eqref{eq:wtf} compares the value of the function $f(\mf x)$ at nearby points and divides it by the coordinate lattice separation. To recover the spacetime metric, we will compute the derivatives of the function $(W(\mf x,\mf x'))^{-1}$. Its discrete derivative at $(\mf x_{\mathfrak{j}},\mf x_{\mathfrak{l}})$ with respect to its different arguments can be written as

    {\footnotesize
        \begin{equation}\label{eq:Wapprox}
            \left.\pdv{}{x^{\mu'}}\pdv{}{x^{\nu}}W^{-1}(\mf x,\mf x')\right|_{\overset{\displaystyle{\mf x = \mf x_{\mathfrak{j}}}}{\displaystyle{\,\mf x'\!\!=\mf x_{\mathfrak{l}}}}}\approx\frac{ W^{-1}(\mf x_{\mathfrak{j}+\bm 1_\nu},\mf x_{\mathfrak{l}+\bm 1_\mu})-W^{-1}(\mf x_{\mathfrak{j}},\mf x_{\mathfrak{l}+\bm 1_\mu})-W^{-1}(\mf x_{\mathfrak{j}+\bm 1_\nu},\mf x_{\mathfrak{l}})+W^{-1}(\mf x_{\mathfrak{j}},\mf x_{\mathfrak{l}})}{(x^\mu_{\mathfrak{j}+\bm 1_\mu}-x^\mu_{\mathfrak{j}})(x^\nu_{\mathfrak{l}+\bm 1_\nu}-x^\nu_{\mathfrak{l}})}.
        \end{equation}}
    This expression should give an approximate form for Eq. \eqref{mine}, so that we expect to recover the spacetime metric in the case where the detectors are separated by small enough values of {the coordinate separation} $x^\mu_{\mathfrak{j}+\bm 1_\mu}-x^\mu_{\mathfrak{j}}$.
    
    {To simplify the formalism for a proof of principle, we will assume the detectors to be separated by a coordinate distance $L$ in all directions (including the time direction). We can then rewrite Eq. \eqref{eq:Wapprox} using that the coordinates of $\mf x_{\mathfrak{j}+\bm 1_\mu}$ are $x_{\mathfrak{j}}^\mu+L\,\bm 1_\mu$. }
 It is important to remark that in this case, the parameter $L$ does not represent physical spacetime interval separation; it is merely a coordinate parameter. However, continuity ensures that when the coordinate separation between events goes to zero, so does the spacetime interval between them. For this reason, Eq. \eqref{eq:Wapprox} will be used in the examples we study below, so we assume that the detector coordinate separation is $L$ in all directions in the coordinate system that determines their trajectories. It is then expected that if $L$ is small enough, Eq. \eqref{eq:Wapprox} will yield a good approximation for the spacetime metric once the numerical factor from Eq. \eqref{mine} is included. In fact, as we will see in the following examples, for pointlike detectors the metric will be precisely recovered when $L\rightarrow 0$, and very approximately recovered for smeared detectors when the distance between detectors approaches the detectors size.
    
\subsubsection*{Inertial Pointlike Detectors in Minkowski Spacetime}\label{sub:mink}
    
    In this first example, we consider the spacetime (unknown to the experimenter) to Minkowski, with a real massive scalar quantum field in its vacuum state.
        
    \begin{figure}[h!]
        \centering
        \includegraphics[width=10cm]{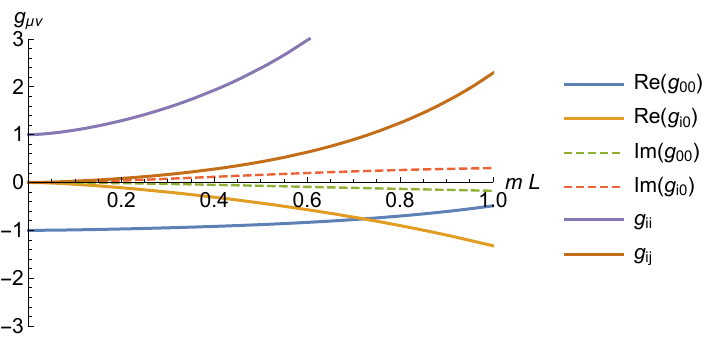}
        \caption{Estimation of the metric coefficients in terms of the coordinate distance between detectors, $L$, for inertial comoving detectors in Minkowski spacetime.}\label{fig:mink}
    \end{figure}
    
    We then consider the inertial coordinate system \mbox{$\mf x = (t,\bm x) {= (t,x,y,z)}$}, and build the lattice of particle detectors in a local region of spacetime according to our setup. For simplicity, in this first example, we consider detectors that interact via delta couplings in space and time. In this case, it is possible to recover the Wightman function of the quantum field exactly. We can then use Eq. \eqref{eq:Wapprox} to approximate the spacetime metric. We obtain the estimates for the metric shown in Figure \ref{fig:mink}. It is possible to see that the readouts of the detector approximate the metric coefficients as the distance between the detectors decreases. Moreover, the imaginary part of the approximate (experimentally obtained) metric goes to zero faster than the real components as $L\rightarrow 0$, so that we are only left with real expressions, which yield the expected value $g_{\mu\nu} = \text{diag}(-1,1,1,1)$.
    
    \subsubsection*{Uniformly Accelerated Pointlike Detectors in Minkowski Spacetime}\label{sub:minkacc}
    
    We now consider uniformly accelerated pointlike detectors probing the Minkowski vacuum of a massless scalar field. The goal of this example is to see whether it is still possible to recover the spacetime metric in different coordinate systems built from particle detectors in different states of motion.  We then consider Rindler coordinates $(T,X,y,z)$ in Minkowski spacetime, associated to the inertial coordinates $(t,\bm x)$ from Subsection \ref{sub:mink} by
    \begin{equation}
        \begin{cases}
            t = X \sinh(a T),\\
            x = X \cosh(a T),
        \end{cases}
    \end{equation}
    with $X>0$ and $T\in\mathbb{R}$. The Minkowski line element in this coordinate system then reads
    \begin{equation}
        \dd s^2 = - a^2 X^2 \dd T^2 + \dd X^2 + \dd y^2 + \dd z^2.
    \end{equation}

    \begin{figure}[h!]
        \centering
        \includegraphics[width=10cm]{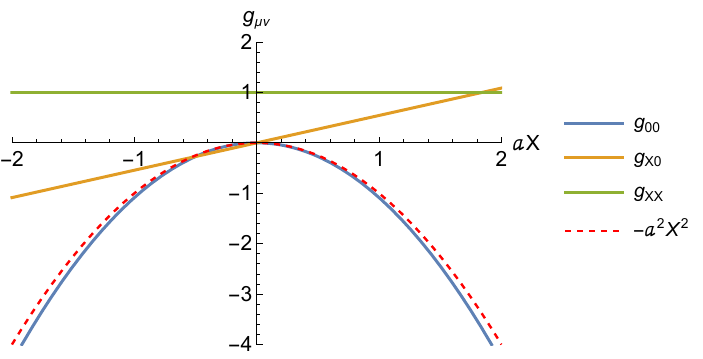}
        \includegraphics[width=10cm]{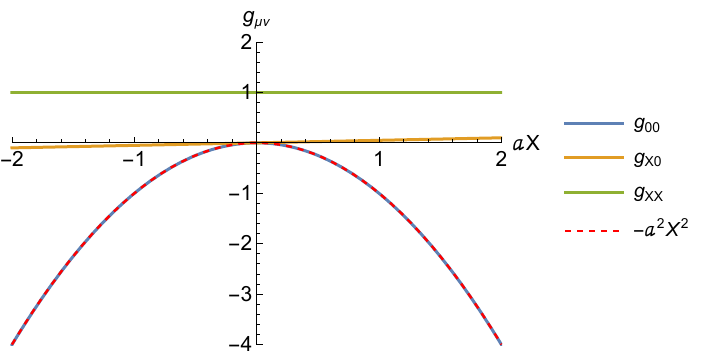}
        \caption{{Metric coefficients obtained from the correlation function of the quantum field in Minkowski spacetime with accelerated detectors. The metric coefficients are plotted as a function of the coordinate $a X$ of the detectors. The detectors were separated by a coordinate distance {$L = a^{-1}$} in the top plot and $L = 0.1a^{-1}$ in the bottom plot.}}\label{fig:MinkAcc}
    \end{figure}

    The lattice of detectors which is associated to this coordinate system is such that each detector follows a trajectory defined by $X = \text{const.}$ with constant values of $y$ and $z$. That is, each detector is uniformly accelerated with different proper accelerations. We consider a massless field, and detectors interacting along different Dirac deltas situated along the corresponding motions of the Rindler flow. Performing the computation in Eq. \eqref{eq:Wapprox}, we find the estimates for the spacetime metric shown in Fig. \ref{fig:MinkAcc} as a function of the coordinate $X$ for detector separations of $L = a^{-1}$ in the top plot and $L = 0.1 a^{-1}$ in the bottom plot. 

    In the limit of $L \to 0$ we recover the metric exactly, as would be expected. Overall, we recover the expected behaviour of the metric components with the coordinate distance $X$ between the detectors. The smaller the value of $L$, the better the fit between the curves. Also notice that for higher values of $aX$, we find more discrepancies between the estimated metric components and the actual Minkowski metric. This is due to the fact that the time separation between the interactions is proportional to $aX$. Overall, we find that it is possible to recover the spacetime metric even when the detectors are in different states of motion, giving rise to different coordinate systems which express the same spacetime metric. This is a general feature of the setup we have considered: it is generally covariant, so that regardless of the relative motion of the detectors, one can recover the metric in the coordinate system associated with their trajectories.

    \subsubsection*{Hyperbolic Static Robertson-Walker Spacetime}
    
    Consider the hyperbolic cosmological spacetime with a constant scale factor $a$. Then the metric in comoving coordinates coordinates can be written as
    \begin{equation}
        \dd s^2 = -\dd t^2 + a^2(\dd \chi^2 + \sinh^2(\chi)(\dd \theta^2 + \sin^2\theta\dd \phi^2)),
    \end{equation}
    We then reparametrize it using the conformal time parameter $\eta = t/a$, so that the coordinates read
    \begin{equation}
        \dd s^2 = a^2(-\dd \eta^2 + \dd \chi^2 + \sinh^2(\chi)(\dd \theta^2 + \sin^2\theta\dd \phi^2)).
    \end{equation}
    Quantizing a conformally coupled real scalar quantum field $\hat{\phi}(\mf x)$ with respect to the conformal time, we can expand it in terms of creation and annihilation operators,
    \begin{align}
        \hat{\phi}(\mf x) = \sum_{k=1}^\infty \sum_{l=0}^{k-1}& \sum_{m=-l}^l \frac{1}{\sqrt{2a^2\omega_{\bm k}}}\left(e^{-\ii\omega_{\bm k}\eta}\Pi_{kl}^{(-)}(\chi)Y_l^m(\theta,\phi)\hat{a}_{\bm k}+\text{H.c.}\right),
    \end{align}
    where $\omega_{\bm k} = k^2 + \mu^2$ and $\mu^2 = a^2(m^2 + (\xi -\frac{1}{6})R)$, where $R$ is the Ricci scalar, which is constant in this spacetime. The explicit expression for $\Pi^{(-)}_{kl}(\chi)$ can be found in e.g. \cite{birrell_davies}. The vacuum Wightman function can then be explicitly computed and reads
    \begin{equation}
        W(\mf x,\mf x') = \frac{\ii\mu(\chi-\chi')H_1^{(2)}(\mu[(\eta-\eta')^2 - (\chi-\chi')^2])}{8\pi a^2\sinh(\chi-\chi')[(\eta-\eta')^2 - (\chi-\chi')^2]},
    \end{equation}
    where $H_1^{(2)}$ is the Hankel function.
        
    \begin{figure}[h!]
        \centering
        \includegraphics[width=10cm]{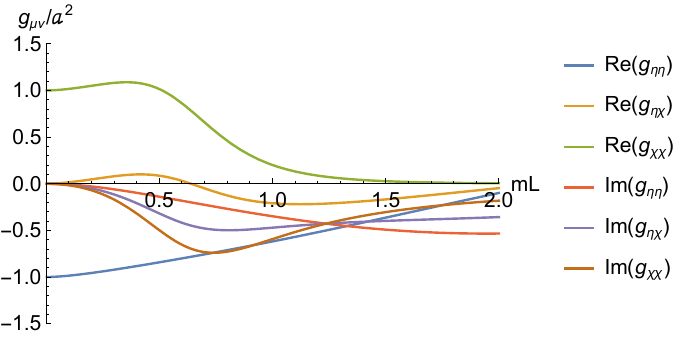}
        \caption{Metric coefficients $g_{\eta\eta}$, $g_{\eta \chi}$, $g_{\chi\eta}$ and $g_{\chi\chi}$ in terms of the coordinate distance between detectors, $L$ in the hyperbolic static Robertson Walker spacetime with choices of mass and conformal parameters such that $\mu = a m$.}\label{fig:cosmo}
    \end{figure}
    
    The results of particle detectors separated by a coordinate distance $L$ in the $(\eta, \chi)$ coordinates coupled to this spacetime can be found in Figure \ref{fig:cosmo}. We then see that in the limit of $L\rightarrow 0$, we recover the exact metric coefficients.

    \subsubsection*{deSitter Spacetime}
    
    In this example, we recover the metric of four-dimensional deSitter spacetime by probing it with particle detectors. deSitter spacetime has a constant curvature with scalar curvature, $R = \text{const.}>0$. It is then possible to write the Riemann curvature tensor as
    \begin{equation}
        R_{\mu \nu \rho \sigma} = \frac{1}{\ell^2}\qty(g_{\mu \rho} g_{\nu \sigma} - g_{\mu \sigma} g_{\nu \rho}),
    \end{equation}
    where $\ell$ is the curvature radius of the spacetime. This will be the first example we investigate where the metric components explicitly depend on the coordinates we use to prescribe the detector's trajectories. 
    \begin{figure}[h]
        \includegraphics[scale=0.65]{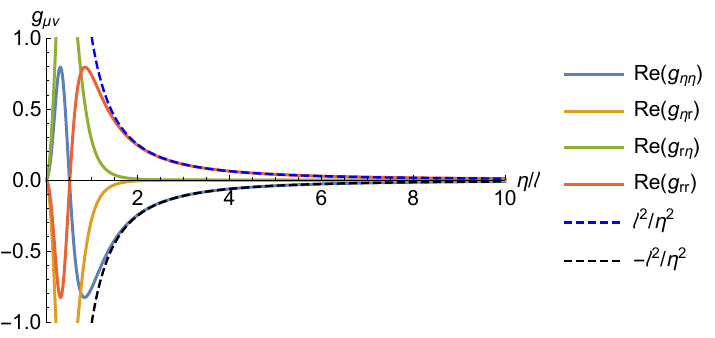}
        \includegraphics[scale=0.65]{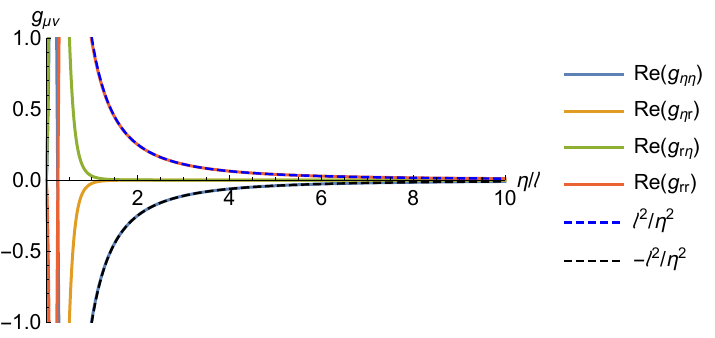}
        \centering
        \caption{Metric coefficients calculated from the correlation function of the quantum field in deSitter spacetime. The metric coefficients are plotted as a function of the coordinate $\eta/\ell$ of the detectors and we choose $\nu = 9/4$. The detectors were separated by a coordinate distance {$L = e^{-7.5}\ell$} in the left plot and $L = e^{-7.5} \ell/2$ in the right plot.}\label{fig:deSitter}
    \end{figure}
    
    We consider conformal coordinates in deSitter spacetime, so that the metric can be written as
    \begin{equation}
        \dd s^2 = \frac{\ell^2}{\eta^2}\left(-\dd \eta^2 + \dd x^2 +\dd y^2 + \dd z^2\right).
    \end{equation}
    The quantization of a real scalar field with respect to the modes adapted to this coordinate system yields the following vacuum Wightman function~\cite{birrell_davies}:
    \begin{align}
        W(\mf x,\mf x') = &\frac{1}{16\pi \ell^2}\left(\frac{1}{4}-\nu^2\right)\sec(\pi \nu){}_2F_1\left(\frac{3}{2}+\nu,\frac{3}{2}-\nu,2,1+\frac{(\Delta \eta)^2-|\Delta \bm x|^2}{4 \eta'\eta}\right),
    \end{align}
    where ${}_2F_1$ is the Hypergeometric function and we write $\mf x = (\eta,\bm x)$ and $\mf x'= (\eta',\bm x')$, defining $\Delta\eta = \eta - \eta'$ and $\Delta \bm x = \bm x - \bm x'$. The parameter $\nu$ contains the information regarding the mass of the field and its coupling to curvature. It is explicitly given by
    \begin{equation}
        \nu^2 = \frac{9}{4}-12\left(\frac{m^2}{R}+\xi \right).
    \end{equation}
    
    To recover the metric in this spacetime, we consider delta-coupled particle detectors that undergo trajectories defined by $\bm x = \text{const}.$, separated by a coordinate distance $L$. We consider these detectors to interact at conformal times that are multiples of $L$. In Fig. \ref{fig:deSitter}, we plot the metric approximation for two values of $L$ as a function of $\eta$. As expected, when $L\rightarrow 0$, we approximate the function $\pm \ell^2/\eta^2$ with high precision. Also notice that the method yields better approximations for larger values of $\eta/\ell$. This is due to the fact that at a given fixed value of conformal time $\eta$, the proper space separation between neighbouring detector trajectories is given by $\frac{\ell}{\eta} L$, which is smaller for larger values of $\eta/\ell$.

    \subsubsection*{The Half Minkowski Space with Dirichlet Boundary Conditions}
    
    In this example we study the effect of boundary conditions in our protocol for recovering the spacetime metric. We analyze a massless Klein-Gordon field in the half Minkowski space $\mf x = (t,x,y,z)$ with $z\geq 0$ and Dirichlet boundary conditions at $z = 0$. This effectively restricts the basis of solutions for the Klein-Gordon equation and changes the field's two-point function. The vacuum state that respects the symmetries of this spacetime then yields the Wightman function~\cite{birrell_davies,alex}
    \begin{equation}
        W(\mf x,\mf x') = \frac{1}{8\pi^2}\frac{1}{\sigma} - \frac{1}{8\pi^2}\frac{1}{\sigma_*},
    \end{equation}
    where $\sigma = \sigma(\mf x,\mf x')$ and $\sigma_* = \sigma_*(\mf x,\mf x')$ are given by
    \begin{align}
        \sigma &= \frac{1}{2}\left(-(t-t')^2 + (x-x')^2 + (y-y')^2 + (z-z')^2\right),\nonumber\\
        \sigma_* &= \frac{1}{2}\left(-(t-t')^2 + (x-x')^2 + (y-y')^2 + (z+z')^2\right).
    \end{align}
    Then, it is possible to verify that whenever $\mf x$ or $\mf x'$ lies at the plane $z = 0$, $W(\mf x,\mf x') = 0$, as expected due to the Dirichlet boundary conditions.
    
    \begin{figure}[h!]
        \includegraphics[width=10cm]{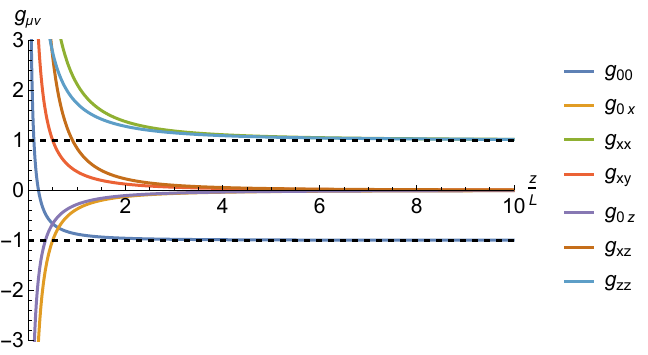}
        \centering
        \caption{{Metric coefficients calculated from the correlation function of a massless quantum field in the half Minkowski space with Dirichlet boundary conditions at $z=0$. The metric coefficients are plotted as a function of the coordinate ratio between their $z$ coordinate and the separation between the detectors, $L$.}}\label{fig:Minkz}
    \end{figure}
    
    We then consider pointlike particle detectors at rest with respect to the coordinate system $\mf x = (t,x,y,z)$ separated by a coordinate distance $L$, which interact with the quantum field at events separated in coordinate time by $L$. Following our general procedure, we estimate the metric coefficients using Eq. \eqref{eq:Wapprox}. In Fig. \ref{fig:Minkz} we plot the obtained metric coefficients as a function of the ratio between the coordinate distance $z$ and the separation between the detectors. As we see, the further away from the boundary, the better the metric estimation is. Moreover, due to the fact that $W(\mf x,\mf x') = 0$ at the boundary, the computation of Eq. \eqref{eq:Wapprox} yields a divergent result at $z = 0$, showing that at the boundary, it is not possible to estimate the metric coefficients. Nevertheless, we highlight that for any $z>0$, the limit $L \to 0$ yields the exact Minkowski metric coefficients. 
    
    Overall, we see that the presence of a boundary disturbs the metric estimation, which fails at the boundary itself. Nevertheless, for any point that is not at the boundary, the correlation function of particle detectors can be used to accurately yield the metric of spacetime, same as in the previous cases.

    \subsubsection*{One-particle Fock states in Minkowski spacetime}
    
    In this example, we consider one-particle Fock wavepackets in Minkowski spacetime to show with an example how the recovery of the spacetime geometry is independent of the field state. We consider the same setup used when probing the Minkowski vacuum, where the detectors undergo inertial motion in a frame $(t,\bm x)$. With respect to these modes, a general normalized one-particle state $\ket{\psi}$ can be written as
    \begin{equation}
        \ket{\psi} = \int \dd^3 \bm k \,f(\bm k) \hat{a}^\dagger_{\bm k} \ket{0},
    \end{equation}
    where $f$ is an $L^2(\mathbb{R}^3)$ normalized function. The two-point function of the field in the state $\ket{\psi}$ will be given by (see, e.g.,~\cite{antiparticles})
    \begin{equation}
        W_\psi(\mf x,\mf x') =  W_0(\mf x,\mf x') + F(\mf x) F^*(\mf x') + F(\mf x') F^*(\mf x),
    \end{equation}
    where $W_0(\mf x,\mf x')$ is the Minkowski vacuum Wightman function and
    \begin{equation}
        F(\mf x) = \frac{1}{(2\pi)^{\frac{3}{2}}}\int \frac{\dd^3 \bm k}{\sqrt{2 \omega_{\bm k}}} \, f(\bm k) e^{\ii \mf k \cdot\mf x}.
    \end{equation}

    \begin{figure}[h]
        \includegraphics[width=10cm]{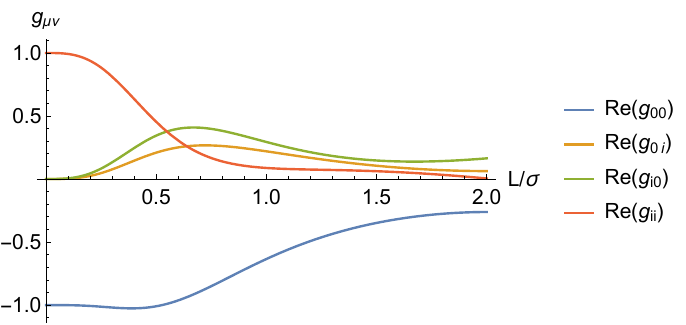}
        \centering
        \caption{{Estimation of metric} coefficients {obtained using} particle detectors in Minkowski spacetime when a massless field is in a Gaussian one-particle state. The metric coefficients are plotted in terms of the coordinate distance between detectors, $L$.}\label{fig:part}
    \end{figure}
    
    For a concrete example, we consider the field to be massless ($m=0$) and prescribe the momentum profile function that defines $\psi$ as a Gaussian centred at $\bm k = \bm 0$ with standard deviation $\sigma$,
    \begin{equation}
        f(\bm k) = \frac{1}{(\pi\sigma)^{3/4}} e^{-\frac{\bm k^2}{2 \sigma^2}}.
    \end{equation}
    We use delta-coupled detectors interacting in events separated by a time/space coordinate separation of $L$. Figure \ref{fig:part} shows the value of the approximated metric coefficients obtained from the detector measurements as a function of the separation between detector interaction events. This allows us to recover the Minkowski metric in the limit where $L \rightarrow 0$. We also find that for larger values of $L$, the results {begin to show} state dependence (compare Figs. \ref{fig:mink} and \ref{fig:part}). This is expected since it is only when the detectors are close to each other that the measurements converge to the metric components independently of the state.

    \subsubsection*{Smeared detectors probing the Minkowski vacuum}\label{sub:smeared}

    As a final example, we consider non-pointlike inertial detectors probing the vacuum of Minkowski spacetime in order to recover the spacetime metric. Unlike the point-like case, it is not possible to recover the Wightman function of the quantum field exactly using smeared particle detectors. However, if the detectors are small, we can resort to the approximation pointed out in Eqs.~\eqref{eq:Wxaxb} and~\eqref{eq:Wx1x2}. Although it is expected that smeared detectors will provide a less accurate measurement of the spacetime metric, these models represent realistic physical systems that are not infinitely localized.
     \begin{figure}[h!]
        \centering
        \includegraphics[width=10cm]{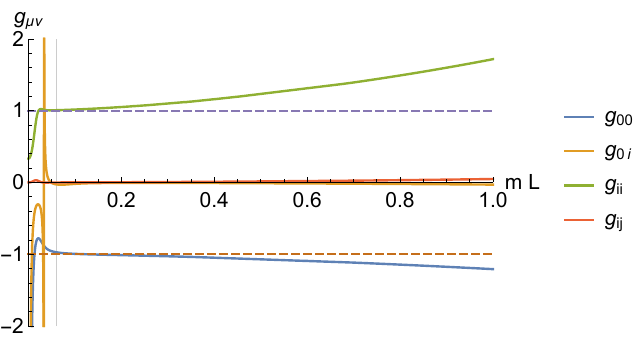}
        \caption{Metric coefficients extracted by Gaussian smeared particle detectors in Minkowski spacetime when the field is in the vacuum state. We have chosen $\Omega = m$, where $m$ is the mass of the field and $\sigma = 10^{-2} \Omega^{-1}$. The metric coefficients are plotted in terms of the proper distance between detectors, $L$. The vertical line on the left indicates $L = 6\sigma$}\label{fig:ohmy}
    \end{figure}
    \begin{figure}[h!]
        \centering
        \includegraphics[width=10cm]{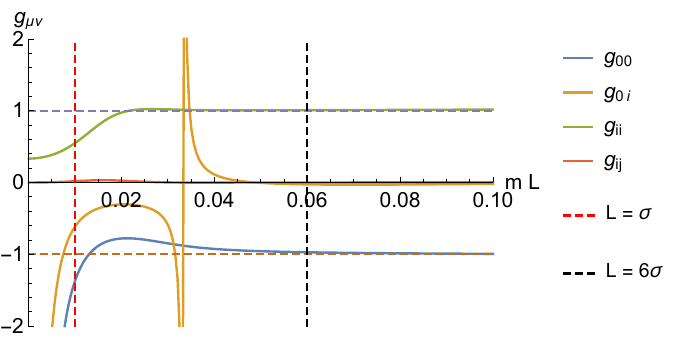}
        \caption{Metric coefficients extracted by Gaussian smeared particle detectors in Minkowski spacetime when the field is in the vacuum state. We have chosen $\Omega = m$, where $m$ is the mass of the field and $\sigma = 10^{-2} \Omega^{-1}$. The metric coefficients are plotted in terms of the proper distance between detectors, $L$.}\label{fig:ohmyzoom}
    \end{figure}
    
    In this example, we consider a lattice of inertial Gaussian-smeared detectors labelled by $\mathfrak{j}$, whose interactions are centred at sites $\mf x_{\mathfrak{j}}$ such that spacetime smearing function can be written as
    \begin{equation}
        \Lambda_{\text{j}}(\mf x) = \sum_{\text{j}_0=1}^{N_0}\frac{e^{-\frac{(\mf x - \mf x_{\mathfrak{j}})^2}{2\sigma^2}}}{(2\pi\sigma^2)^{2}},
    \end{equation}
    where $(\mf x - \mf x_{\mathfrak{j}})^2 = \delta_{\mu\nu} (x^\mu - x_{\mathfrak{j}}^\mu)(x^\nu - x_{\mathfrak{j}}^\nu)$. For each detector trajectory, we sum over the different interaction times $\text{j}_0$. Notice that this corresponds to an interaction that lasts for a time equal to the light crossing time of the detector's spatial profile.
    
    In Fig. \ref{fig:ohmy}, we plot the approximated metric obtained from detector measurements as a function of their coordinate separation $L$, when one considers the approximate Wightman function from Eq.~\eqref{eq:Wxaxb}. For detector separations smaller or comparable to $5\sigma$, the detectors' spacetime smearings have significant overlap, and our method does not apply. In Fig.~\ref{fig:ohmyzoom}, we show a scaled version of the plot, where the approximated metric is shown for smaller values of $L$, and the spurious behaviour for small $L$. Nevertheless, the metric is accurately recovered when $L$ is between $5\sigma$ and $10\sigma$ even when one considers smeared detectors.

\subsubsection*{Conclusions}

The examples discussed above are all able to recover the background metric, regardless of the state of motion of the detectors or the state considered for the field. These results show that one can indeed measure distances and times through localized measurements of the correlations of quantum fields. Regardless of whether it is possible to formulate a consistent theory of gravity from the Hadamard condition, we conclude that quantum fields do locally store complete information about the geometry of spacetime. Moreover, we saw here that this information can, at least in principle, be accessed by physical probes.

\section{Challenges in  Emergent Geometries from Correlations}\label{sec:resonable?}

This section is devoted to discussing the challenges and consequences in formulating spacetime as emerging from the correlations of quantum fields defined through Eq.~\eqref{mine}. We will start discussing how one could potentially obtain dynamics for the metric in this formulation, then discuss whether it is possible to recover the semiclassical Einstein's equations by considering small changes in the state of a quantum field. We will also discuss what would happen if one were to consider more than one scalar field. Overall, this section will show the challenges in obtaining a model for the geometry of spacetime from~\eqref{mine} that could replace general relativity.

\subsubsection*{Trivial Dynamics and Non-Locality}

A physical theory must produce predictions (that hopefully can be tested). The first step towards this goal is to describe the dynamics of a theory. In the context of using the correlations of quantum fields to replace the metric through Eq.~\eqref{mine}, let us assume that we have a theory for a scalar field $\hat{\phi}(\mf x)$ at a Hadamard state $\omega$ with Wightman function $W(\mf x, \mf x')$. Then, the background metric of spacetime can be written as
\begin{equation}
    g_{\mu\nu} = - \frac{1}{8\pi^2} \lim_{\mf x' \to \mf x} \partial_\mu \partial_{\nu'}W^{-1}.
\end{equation}
However, any other state $\tilde{\omega}$ that can be represented in the GNS representation of $\omega$ will result in the same metric $g_{\mu\nu}(\mf x)$. Indeed, any Hadamard state in this spacetime will result in the same background metric. That is to say that the prescription of Eq.~\eqref{mine} does not give any dynamics for the metric: all states produce the same background geometry. This implies that, as is, this framework would not allow one to describe how quantum fields affect spacetime.

One way of resolving this issue (but to introduce many others) would be to \textit{not} take the limit of $\mf x' \to \mf x$ in Eq.~\eqref{mine}, instead defining a bitensor
\begin{equation}\label{eq:qnonlocal}
    q_{\mu\nu'}(\mf x, \mf x') =  - \frac{1}{8\pi^2} \partial_\mu \partial_{\nu'}(W(\mf x, \mf x'))^{-1}.
\end{equation}
This bitensor is certainly not a metric, but it satisfies $\lim_{\mf x' \to \mf x}q_{\mu\nu'} = g_{\mu\nu}$ by construction. Our goal is to allow for a non-local version of the metric through $q_{\mu\nu}$. The idea that quantum field theory, or even quantum gravity, may be non-local close to the Plank scale has been widely debated (see, e.g.~\cite{nonlocal1995,nonlocal2011,nonlocal2024}). We then assume that there is a parameter $\ell$, with dimensions of length and of the order of the Planck length, such that for $|\sigma(\mf x,\mf x')| = \tfrac{\ell^2}{2}$, $q_{\mu\nu'}$ defines a ``non-local spacetime metric''. Intuitively, one can think of $q_{\mu\nu'}(\mf x, \mf x')$ as indicating how the tangent vector of the geodesic that connects $\mf x$ to $\mf x'$ changes as one varies $\mf x'$. This interpretation comes from an analogy with the bitensor $- \sigma_{\mu\nu'}$.

One can obtain a standard tensor $q_{\mu\nu}$ by, for instance, averaging $q_{\mu\nu'}$ over a region $V_{\ell} \subset \{\mf x', \mf x\in \M:|\sigma| = \tfrac{\ell^2}{2}\}$. If $e_{\mu}$ is a basis at $\mf x$, let $e_{\mu'}$ denote its parallel transport to $\mf x'$, and define
\begin{equation}\label{eq:average}
    q_{\mu\nu}(\mf x)e^\mu e^{\nu} := \frac{1}{|V_{\ell}|}\int_{V_\ell} \dd\Sigma'\, q_{\mu\nu'}(\mf x, \mf x')e^\mu e^{\nu'},
\end{equation}
where $|V_\ell|$ denotes the volume of the region, computed with the induced metric\footnote{This volume ca be infinite, but the average of Eq.~\eqref{eq:average} can still often be taken by considering a suitable limit.}. The tensor $q_{\mu\nu}$ would then be symmetric due to the conjugate symmetry of the Wightman function and non-generate if $W(\mf x, \mf x')$. Non-degeneracy is not guaranteed if $W(\mf x, \mf x')$ oscillates with a frequency higher than $1/\ell$, but we will assume that this is not the case and that $q_{\mu\nu}$ indeed defines a metric for the analysis that follows.

The effective metric $q_{\mu\nu}$ would then explicitly depend on the specific field state considered. Indeed, if the state $\tilde{\omega}$ has the Wightman function 
\begin{equation}
    \tilde{W}(\mf x, \mf x') = W(\mf x, \mf x') + w(\mf x, \mf x'),
\end{equation}
then, to leading order in $w$, we have
\begin{equation}
    \tilde{q}_{\mu\nu'} = - \frac{1}{8\pi^2}\partial_\mu \partial_{\nu'} \left(W^{-1} - \frac{w}{8 \pi^2W^2}\right) + \mathcal{O}(w^2)= q_{\mu\nu'} - \partial_\mu \partial_{\nu'} \left(\frac{w}{(8 \pi^2 W)^2}\right)  + \mathcal{O}(w^2),
\end{equation}
which, when averaged according to~\eqref{eq:average} will generally yield a different $\tilde{q}_{\mu\nu}$. On the other hand, the fact that each state generates a different metric effectively implies that one has to consider a different quantum field theory at a different spacetime for each update of $q_{\mu\nu}$, and consider how to relate the different states of the different theories would be an issue all in itself. We will not focus on these issues for now, as this construction will prove ineffective before we can get to this debate.

The Minkowski vacuum of a massless scalar field is, however, stable. Indeed, even without taking the coincidence limit, the Wightman function of the Minkowski vacuum satisfies
\begin{equation}
     - \frac{1}{8\pi^2}\partial_\mu \partial_{\nu'}W_0^{-1} = \eta_{\mu\nu'}, 
\end{equation}
where $\eta_{\mu\nu'}$ is the parallel transport of the Minkowski metric over the second index. In particular, this means that $q_{\mu\nu'} = \eta_{\mu\nu'}$ and $q_{\mu\nu} = \eta_{\mu\nu}$. We conclude that in this formulation, the Minkowski vacuum of a scalar quantum field would not gravitate. Notice, however, that this would not be the case for a massive scalar field or for any other state in the massless theory.

\subsubsection*{Einstein's Equations?}

If the attempt at defining the metric from correlations is to bear any value, the dynamics introduced by changing the state of the field must correspond to the dynamics prescribed by Einstein's equations, at least in some regimes. To test whether this formulation can reproduce Einstein's equations, we consider a real massless scalar field that is at first in its vacuum state, with Wightman function $W_0(\mf x, \mf x')$. We then assume that an operation is applied to the field, updating the state and giving rise to the Wightman function
\begin{equation}
    W(\mf x, \mf x') = W_0(\mf x, \mf x') + w(\mf x, \mf x').
\end{equation}
This modification to the vacuum Wightman function then creates both a modification to the metric $q_{\mu\nu}$, as well as a change in the stress-energy tensor of the field, $\normord{\hat{T}_{\mu\nu}}$, where we use the Minkowski vacuum as a reference state for the normal ordering. For convenience, we employ inertial coordinates $(t,\bm x)$ and define $\sigma_0(\mf x, \mf x') = -\tfrac{1}{2}(t-t')^2 + \frac{1}{2}(\bm x - \bm x')^2$ as Minkowski's Synge's world function, so that $W_0 = 1/8\pi^2\sigma_0$.

The expected value of the stress-energy tensor will be given by
\begin{equation}\label{eq:Tmunuw}
    \langle \normord{\hat{T}_{\mu\nu}}(\mf x)\rangle  = \lim_{\mf x'\to\mf x}\big(\partial_\mu \partial_{\nu'} - \tfrac{1}{2}\eta_{\mu\nu}  \partial_\alpha \partial^{\alpha'}\big) w(\mf x, \mf x').
\end{equation}
The updated non-local metric will be given by Eq.~\eqref{eq:qnonlocal}:
\begin{equation}\label{eq:Gmunutwopts}
    q_{\mu\nu'} = - \frac{1}{8\pi^2}\partial_\mu \partial_{\nu'}(W_0 + w)^{-1} = \eta_{\mu\nu'} - 16\pi^2 \sigma_0 w \,\eta_{\mu\nu'} + \mathcal{O}(|\sigma_0|^{2}),
\end{equation}
where we used that $\sigma_\mu$ is of order $\mathcal{O}(|\sigma|^{1/2})$. Notice that the leading order term is conformal, which matches the fact that a massless real scalar field is a conformal theory. The corrections are also of order $\sigma_0$, which is assumed to be of the order of $\ell_p^2 = G$ in the averaged metric. Our next goal would then be to compute the averaged metric $q_{\mu\nu}$ and check whether it satisfies the linearized Einstein's equations. However, we can see at this stage that if one were to average $q_{\mu\nu}$, we would not obtain terms that involve the derivatives of $w$ with respect to both arguments, so that the associated linearized Einstein tensor would not match $\langle \normord{\hat{T}_{\mu\nu}(\mf x)}\rangle$ in~\eqref{eq:Tmunuw}. This should already be enough evidence that this formalism cannot recover the semiclassical Einstein's equations even in this simple regime.

Our final attempt will be to assume that it is possible to define a bitensor $G_{\mu\nu'}$ whose coincidence limit coincides with $G_{\mu\nu}$. Although there are infinitely many possible bitensors that coincide with $G_{\mu\nu}$ at coincidence, as an example, we pick the linearized $G_{\mu\nu'}$ acting in a bitensor ``metric fluctuation'' $\delta g_{\mu\nu'}$ according to
\begin{equation}
    G_{\mu\nu'} \coloneqq \partial^\alpha \partial_{(\mu} \delta g_{\nu')\alpha} - \tfrac{1}{2} \partial_\mu \partial_{\nu'} \delta g - \tfrac{1}{2} \Box \delta g_{\mu\nu'}  - \tfrac{1}{2}\eta_{\mu\nu'}(\partial^\alpha \partial^{\beta'}\delta g_{\alpha\beta'} - \partial_\alpha\partial^\alpha \delta g), 
\end{equation}
analogous to Eq.~\eqref{eq:Gmunulin}. However, simply applying the definition above to Eq.~\eqref{eq:Gmunutwopts} would get rid of $\sigma_0$ and produce no leading order results for $h_{\mu\nu}$. In a final final attempt, we consider that $\sigma_0$ is kept constant in $q_{\mu\nu'}$. We then find
\begin{align}\label{eq:thecatastrophe}
    G_{\mu\nu'}  &= -8 \pi^2 \sigma_0^2 \big(\partial_{(\mu}\partial_{\nu')} - \eta_{\mu\nu'}\partial_\alpha \partial^{\alpha'}\big) w(\mf x, \mf x'),
\end{align}
where we used that $\partial_\alpha\partial^\alpha w(\mf x,\mf x') = 0$, as the Wightman function is a bi-solution of the equations of motion. Even with all of the (not necessarily reasonable) assumptions made to reach this point, comparing Eq.~\eqref{eq:thecatastrophe} and~\eqref{eq:Tmunuw}, we see that these are \textit{not proportional}. Moreover, no other reasonable choice of $G_{\mu\nu'}$ would yield the desired result. Overall, even attempting to fine-tune the dynamics, we see that we cannot recover the semiclassical Einstein's equations.

\subsubsection*{Generalization to Other Fields}

We end this section by commenting on a final issue regarding the idea of attempting to interpret Eq.~\eqref{mine} as prescribing the metric of spacetime: there are many other fields in the universe, not merely a scalar field. The Hadamard condition takes an alternative form for each type of field, so that fields of all spins have a leading order divergence in their Wightman function that is controlled by Synge's world function. This allows one to recast the metric as a coincidence limit of a power of the Wightman function for each type of field. However, the Wightman function of non-scalar fields are bitensors, which would require one to either contract or trace these. While contractions would unequivocally lead to privileging a frame, the trace operation would involve parallel transport, which could potentially be implemented, allowing one to recast the spacetime metric in terms of a coincidence limit of Wightman functions for fields of more general spin.

However, we would once again face the issue of prescribing the dynamics from limits of the Wightman function. Taking the coincidence limit would yield the same metric for different fields, however, if different fields are present and affect the metric differently, how would combine their effects? At this stage, this answer is not clear. The overall conclusion of this section is that as much as it is interesting that all information about the geometry is encoded in the UV behaviour of quantum fields, it does not look like one can use this fact to determine the influence that quantum fields have in their background spacetime.

\section{Emergent Geometry from Entanglement?}\label{sec:speculation}

The fact that we were unable to formulate a theory for emergent gravity from the correlations of quantum fields using Eq.~\eqref{mine} does not imply that it is not possible to describe gravity as an emergent phenomenon from quantum field theory. Indeed, there has been an ongoing debate about the possibility of the geometry of spacetime emerging from entanglement in quantum field theories~\cite{VanRaamsdonkEmergent2010,ADSCFTemergent,Cao1,Cao2,emergent2021,Gui2023}. In this Section, we will briefly review results that were obtained related to the emergence of spacetime and Einstein's equations and speculate on how our discussions about entanglement and emergence could fit in within this topic.

Perhaps the most clear connection between entanglement and gravity was provided by Ted Jacobson in~\cite{TedOG,TedNew}, where it was shown that it is possible to obtain Einstein's equations by imposing thermodynamic equilibrium involving the entanglement entropy. The derivation in~\cite{TedNew} concerns entropy variations in a causal diamond in Minkowski spacetime (or any maximally symmetric spacetimes), defined as $D(\Sigma_\ell)$, where $\Sigma_\ell$ is a spacelike surface produced by geodesics orthogonal to a timelike vector $n^\mu$ with geodesic length $\ell$. Essentially, the von Neumann entropy of a state in $D(\Sigma_\ell)$ can be split into two contributions,
\begin{equation}
    S = S_{\text{area}} + S_\text{state},
\end{equation}
where $S_\text{area}$ is proportional to the area of the boundary of $\Sigma_\ell$, and is formally divergent, requiring the introduction of a UV cutoff $\epsilon$ (this is the entropy associated with the divergent vacuum entropy), and $S_\text{state}$ is associated with the entropy of the state, neglecting the vacuum contribution. One can show that keeping the volume of $\Sigma_\ell$ constant, while considering small perturbations of the background geometry and of the vacuum,
\begin{align}
    \delta S_\text{area} &= - \frac{\alpha}{4 \epsilon^2}G_{\mu\nu} n^\mu n^\nu, \\\delta S_\text{state} &= 2 \pi \alpha \delta T_{\mu\nu} n^\mu n^\nu,\label{eq:conformal}
\end{align}
where $\alpha$ is a geometrical factor, $\delta T_{\mu\nu}$ is the variation in $\langle \normord{T_{\mu\nu}}\rangle$ due to the perturbations in the quantum state, and $G_{\mu\nu}$ is the Einstein tensor of the perturbed geometry. Eq.~\eqref{eq:conformal} relies on the quantum field theory being conformal, but can be generalized under reasonable assumptions~\cite{TedNew}. Imposing that $\delta S = 0$ for all diamonds $D(\Sigma_\ell)$ of this form then yields the semiclassical Einstein's equations when the cutoff is taken to be the Plank length, $\epsilon = \ell_p = \sqrt{G}$. The connection between entanglement and gravity is then established by the fact that the entropy $S$ corresponds to the von Neumann entropy of the state in $D(\Sigma_\ell)$. 

This fact has led to Cao and Carroll to start a research program where the geometry of spacetime is emergent from the mutual information between subsystems~\cite{Cao1,Cao2}. Specifically, the proposal of~\cite{Cao1,Cao2} is that given a state $\hat{\rho}$ described in a Hilbert space of the form $\mathscr{H} = \otimes_{p=1}^n\mathscr{H}_p$, it is possible to define effective distance $d$ between the subsystems labelled by $p$ and $q$ as a function of their mutual information:
\begin{equation}
    I(p\!:\!q) = S(\hat{\rho}_p)+S(\hat{\rho}_q) - S(\hat{\rho}_{pq}),
\end{equation}
assigning distance $d(p,q) = 0$ if the mutual information is maximal, and $d(p,q) = \infty$ if their mutual information is $0$. In~\cite{Cao2}, it was shown that it is sometimes possible to embed the systems $\mathscr{H}_p$ in a smooth manifold, where the distance defined through the mutual information corresponds to the geodesic distance within this manifold. Specifically, for states that satisfy an analogue of an area law (redundancy constrained states~\cite{AreaLawsCirac2008}), a discrete version of Jacobson's argument applies, allowing for this framework to, in principle, yield both spacetime emergence and dynamics for the geometry, compatible with Einstein's equations. However, this program is still in its infancy and has numerous challenges to overcome, such as the conditions that allow a system to be embedded in a 3-dimensional surface, as well as the Lorentz structure of the associated spacetime.

One particular issue with the program proposed in~\cite{Cao2} is that the mutual information does not quantify the entanglement between two subsystems, unless the combined system at $p$ and $q$ is pure. This would imply that classical statistical mixtures would affect the geometry of spacetime non-trivially. One could potentially fix this issue by instead considering that the distance between quantum systems is a function of the entanglement between them. Indeed, we have seen that there are good arguments for why the entanglement between two regions decays with their separation and that a polynomial decay of entanglement can be achieved by considering measurements in their complement. This seems to imply that the entanglement between two regions could replace the role played by the mutual information in the emergence of spacetime. Indeed, it would not be surprising if one could rewrite an analogue of Eq.~\eqref{mine} in terms of the entanglement between two sufficiently small regions that are separated by distances of the order of the Planck length, allowing one to explicitly recover the background geometry of spacetime. This is, of course, a challenge, given how non-trivial the task of quantifying entanglement in quantum field theory, and even in simpler quantum systems, is.

On the other hand, the idea that the entanglement between two regions can yield their separation in space would also automatically solve the issue that appears when one considers many fields in spacetime: a collection of quantum field theories is still described by a single state, and one could quantify the entanglement of this state between two finite regions of space to define spacetime separations. The method of replacing the role played by the mutual information in~\cite{Cao2} by an entanglement quantifier between the two regions would also be fit for Jacobson's argument, where one could argue that Einstein's equations would follow from a thermodynamical equilibrium argument. It would remain to be seen whether small perturbations on the field's state would be compatible with this formulation, yielding Einstein's equations in a linearized form, as we attempted in Section~\ref{sec:resonable?}.

This approach would also naturally relate to the ER=EPR conjecture, proposed by Maldacena and Susskind in~\cite{MaldacenaSusskind}. Essentially, if the fabric of spacetime indeed emerges from the entanglement structure of quantum fields, in~\cite{MaldacenaSusskind}, it was suggested that one can interpret what are usually seen as nonlocal quantum correlations between two subsystems instead as ``local'' correlations in an effective spacetime that possesses a microscopical Einstein-Rosen bridge connecting them. Quoting~\cite{MaldacenaSusskind}, the ER=EPR conjecture states that ``for an entangled pair of particles, in a quantum theory of gravity there must be a Planckian bridge between them, albeit a very quantum mechanical bridge which probably cannot be described by classical geometry''. If the geometry of spacetime could be described as emergent from entanglement in quantum field theory, and the separations between two regions were defined as inversely proportional to the entanglement within them, a maximally entangled pair between two separated regions would indeed require one to embed these systems in a spacetime that has a non-trivial topology, where the systems seem distant in the external space, but are very close through the Einstein-Rosen bridge.

At this point it should be clear that the first step to tackle the problem of the emergence of spacetime from entanglement present in quantum fields is to find a precise way to determine the entanglement between two regions of a quantum field theory. Thanks to the results of Klco in~\cite{KlcoUVIR,KlcoEntStrQFTI,KlcoEntStrQFTII} (reviewed in Section~\ref{sec:modeEntanglement}), there is a clear path to achieve this goal and to attempt to formulate a theory where gravity is a consequence of the fundamental structure of quantum correlations in quantum field theory. Indeed, most of the discussions held in Section~\ref{sec:geometry} also apply to these theories, where the vacuum correlations of the field close to the Planck scale would give rise to spacetime as a consequence of the Hadamard condition.

%

\chapter{Summary and Conclusions}\label{chap:conclusions}

We discussed four main topics in this thesis: how to probe quantum fields in Chapter~\ref{chap:meas}, how to quantify entanglement in quantum field theory in Chapter~\ref{chap:ent}, how to determine when the degrees of freedom of quantum fields are relevant in Chapter~\ref{chap:qc}, and how quantum fields contain full information about the geometry of spacetime in Chapter~\ref{chap:geometry}. We will now summarize the knowledge acquired about each of these topics and point out directions of future work.

\subsubsection*{Local Probes of Quantum Fields}

It is usual to consider effective non-relativistic systems when implementing operations and measurements in quantum fields. This approach not only simplifies the description of measurements in quantum field theory but also directly connects to physically accessible systems that can be used in realistic setups to measure and operate quantum fields. The price to pay for these simplifications is that the probes will usually be incompatible with relativistic principles. In Section~\ref{sec:LocalizedQuantumFields}, we described precisely how the effective models of the probes arise from an entirely quantum field theoretic description, by explicitly connecting the Fewster-Verch measurement framework to the usual models of particle detectors. We did so by considering a compactly supported quantum field as a probe, and reducing this localized field to modes that can be realistically accessed. 

These effective models that only consider finite modes of the probe naturally led us to the definition of particle detector models, or Unruh-DeWitt detectors, in Section~\ref{sec:UDW}. While studying the dynamics of these effective models, we found that the reason for the incompatibilities with relativity that are present in effective descriptions is that no individual mode of a probe field is local: the modes are smeared along the entire region where the probe is localized, corresponding to one degree of freedom distributed along multiple spacelike separated points. Given that no finite number of field modes satisfy the microcausality condition, this leads to non-covariant dynamics that privilege a time direction with respect to which the effective model is defined.

We then studied how to describe non-relativistic quantum systems undergoing a given trajectory in curved spacetimes in Section~\ref{sec:NRLQS}. There, it became evident that there is one privileged time direction with respect to which one should compute time evolution: the time coordinate that allows the system to be described non-relativistically in the first place. Using this description, we saw how a particle detector model can be defined from a physical system coupled to a relativistic quantum field, connecting this formulation with the typical two-level Unruh-DeWitt detector and with a non-relativistic atom probing the electromagnetic field. Rather than starting from quantum field theory, this formulation allows one to start from a system described by non-relativistic quantum mechanics and embed it into spacetime, allowing one to formulate its interaction with a quantum field within the particle detector model formalism. 

On the other hand, in Section~\ref{sec:MoreRealisticProbes}, we showed that it is not always necessary to employ non-relativistic models to describe physically realistic probes, as we can instead consider more general localized fields corresponding to explicit physical systems. In particular, we discussed fields localized by finite potentials that give rise to non-compactly supported modes and provided an explicit basis-independent definition of localized fields in terms of limits of expectation values of observables at spatial infinity. Along the lines of studying more realistic localized fields, we discussed fields under the influence of not only finite, but bounded potentials, which generally contain a mixture of localized and scattering modes. For a concrete example of a physically realistic localized probe, we discussed a quantum field theoretic description of a hydrogen atom in terms of the electron field bound by a Coulomb potential, showing how it gives rise to an effective probe of an external magnetic field.

Finally, we discussed the implications of general covariance for the models of localized fields used throughout the chapter in Section~\ref{sec:Tmunu}. In essence, we showed that considering a non-dynamical potential for the localization of the probes is incompatible with general covariance: general relativity requires one to describe all matter through dynamical fields. On the other hand, incorporating the dynamics of the external potential also adds their contribution to the stress-energy tensor, which, strictly speaking, rules out any compactly supported fields, as these require an infinite trapping potential. Analogously, any unbounded external potential cannot be consistently described in a background Minkowski spacetime, as these would amount to increasing energies at spatial infinity, which would result in non-trivial effects on the geometry. We then showed an explicit example of a localized quantum field that is trapped by its interaction with a dynamical quantum field and a perfect fluid, giving rise to a finite and bounded stress-energy tensor that satisfies all the energy conditions.

Overall, Chapter~\ref{chap:meas} consisted of a detailed study of the local probes that can be used to measure a quantum field. We connected effective and fundamental models and studied the intricacies of defining localized systems when taking both quantum field theory and general relativity into account. The natural next step regarding the approach taken in this Chapter is to formulate a general covariant model for the hydrogen atom. After all, a non-dynamical pointlike charge and a Coulomb potential both yield infinite contributions to the energy-momentum tensor of the system. If such a model is found, it would also allow one to explicitly describe the stress-energy tensor of a hydrogen atom and, potentially, to study the geometry of spacetime inside the atom. Overall, it would be the first example of a general covariant model that can describe a realistic localized physical system in terms of quantum field theory.

\subsubsection*{Entanglement in Quantum Field Theory}

There are many things about quantum fields that are well understood at this stage. Arguably, entanglement is not one of them. Although specific results are known, such as the fact that the vacuum possesses an arbitrarily large amount of entanglement between a region and its complement, and that this entanglement is directly related to the area between the regions, not many results are available regarding the entanglement between non-complementary regions. In Chapter~\ref{chap:ent}, we discussed why it is non-trivial to quantify entanglement between two finite regions in quantum field theory, and discussed two methods to approach this problem.

The first method was described in Section~\ref{sec:modeEntanglement}, and consisted of analyzing entanglement between finite sets of field modes that are localized in two finite causally disconnected regions. The fact that the vacuum is a Gaussian state allowed us to define independent field modes in each region and to treat these degrees of freedom using standard techniques of Gaussian quantum mechanics. Specifically, we considered the modes defined by canonically commuting pairs of field and momentum operators $(\hat{\Phi}(F_i),\hat{\Pi}(F_i))$ smeared against test functions localized in two non-overlapping subsets of a Cauchy slice. By mapping the vacuum degrees of freedom associated to each mode to the corresponding covariance matrix, we discussed how to compute the entanglement between these degrees of freedom. Using this technique, we considered sets of modes defined by the same smearing functions at different Cauchy slices and, after mapping all modes to a single slice, computed the entanglement between the two regions. We then argued that the entanglement found is a lower bound for the total amount of entanglement between the regions. 

The remainder of Section~\ref{sec:modeEntanglement} was devoted to reviewing the results of~\cite{KlcoUVIR,KlcoEntStrQFTI,KlcoEntStrQFTII,KlcoEntAllDist}, which provided an explicit method to not only compute the entanglement between localized modes, but to explicitly find which modes contribute the most to the negativity between the two regions. The application of these techniques to lattice quantum field theory yielded numerical results in~\cite{KlcoUVIR,KlcoEntStrQFTI,KlcoEntStrQFTII,KlcoEntAllDist} that led to the conclusions that 1) the vacuum contains entanglement between any two finite regions and that this entanglement decays exponentially with their separation, 2) the further apart two regions are, the more energetic the entangled modes are---the so-called UV-IR connection---, and 3) the entanglement in quantum field theories is mostly of GHZ type, allowing one to increase the entanglement between two regions by performing measurements in their complement. These conclusions were later compared with our second approach to quantifying the entanglement in quantum field theory.

Our second approach to studying the entanglement in quantum field theory was discussed in Section~\ref{sec:OperationallyAccessingEnt}, where we considered the entanglement that could be acquired by local probes that couple to a target field, in the protocol of entanglement harvesting. We started our discussion of entanglement harvesting by considering two probes modelled by localized quantum fields and discussed two ways in which the probes can become entangled: either through communication or by extracting entanglement from the field. By imposing that the probes must couple to independent field degrees of freedom we concluded that only when the probes are spacelike separated throughout their interactions can the entanglement in them be traced back to entanglement in the target field. When considering more realistic non-compactly supported probes, we defined a condition based on the symmetric propagator that allows one to conclude that the entanglement acquired by the probes was harvested from the field.

Having the conditions necessary for entanglement harvesting to take place, we studied explicit examples, first describing the probes as localized quantum fields, and then considering two-level Unruh-DeWitt detectors. In the example with particle detectors, we provided closed-form expressions for the leading order final state of the detectors and their negativity. 

Section~\ref{sec:GeneralEnt} was devoted to summarizing general results of entanglement harvesting. We started by reviewing the no-go theorems of~\cite{nogo}, and then proceeded to show that the entanglement extracted by local probes in entanglement harvesting are in agreement with the more general results analyzed in~\cite{KlcoUVIR,KlcoEntStrQFTI,KlcoEntStrQFTII,KlcoEntAllDist}. Specifically, we showed that two detectors can become entangled regardless of their spatial separation and showed an analogue of the UV-IR connection in entanglement harvesting, where the optimal energy gap that allows detectors to harvest entanglement is proportional to their separation.

Although progress has been made in quantifying entanglement in quantum field theory, still very few general results are known. In particular, the question of ``how much entanglement is there between two finite regions'' still only has numerical answers in lattice theories. However, there is a clear path to obtain an answer to this question without the lattice approximation: considering two non-overlapping spheres $B_\tc{a}$ and $B_\tc{b}$, separated by a distance $L$, one can consider bases of real orthonormal functions $F_{n,\tc{a}}\in L^2(B_\tc{a})$ and $F_{n,\tc{b}}\in L^2(B_\tc{b})$, so that the degrees of freedom of the field in the two regions is fully described by the canonically commuting modes $(\hat{\Phi}(F_{n,\tc{a}}),\hat{\Pi}(F_{n,\tc{a}}))$ and $(\hat{\Phi}(F_{n,\tc{b}}),\hat{\Pi}(F_{n,\tc{b}}))$. Considering a large finite number of basis elements, one can then 1) quantify the distillable entanglement between the degrees of freedom in $B_\tc{a}$ and $B_\tc{b}$ to arbitrary precision, and 2) find the spatial and momentum profiles that define the most entangled modes through the methods of~\cite{KlcoUVIR,KlcoEntStrQFTI,KlcoEntStrQFTII}. This is ongoing research which, unfortunately, could not be completed before submission of this thesis. With these results, one could also determine the optimal coupling to be considered in the protocol of entanglement harvesting, and perhaps even allow detectors to harvest significant entanglement from the field, opening the doors for practical applications of the protocol.

\subsubsection*{When is Quantum Field Theory Necessary?}

Although quantum field theory is the benchmark for fundamental descriptions, not all setups require quantum field theory to be accurately described. In Chapter~\ref{chap:qc}, we discussed the quantum-controlled model (qc-model), an effective description of interactions that are mediated by quantum fields that neglects the field's degrees of freedom, while still preserving some relativistic aspects of the interaction. We defined the model in Section~\ref{sec:QCmodels} and discussed its basic properties, such as the fact that it promotes unitary evolution between the systems, and that systems only interact through the qc-model when in causal contact.

In Section~\ref{sec:QFTapproxQC}, we discussed when the quantum-controlled model can approximate interactions mediated by quantum fields. To compare the predictions of both models, we considered two two-level systems and compared the evolution obtained using a qc-model and a quantum field description. The analysis of this setup indicated three conditions that must be fulfilled for a quantum field interaction to be well approximated by a quantum-controlled model: 1) the interaction duration must be much longer than the separation between the systems, 2) the interaction duration must be much longer than the characteristic frequency that determines the systems' internal dynamics, and 3) the coupling between the qubits and the field much be sufficiently weak. We then performed an analysis of the case where the qubits have trivial internal dynamics, where we solved for the final state of the qubits in both cases non-perturbatively and confirmed the conditions previously found. 

Beyond simplifying the description of interactions mediated by quantum fields, the qc-model can be used to indicate the limits where the quantum degrees of freedom of mediating fields play an active role. This allowed us to conclude that it is only in the regime of either high energies or short interactions that the degrees of freedom of quantum fields play an active role. This led us to discuss the gravity mediated entanglement experimental proposals of~\cite{B,MV}, where neither condition is fulfilled, but it is claimed that the experiment can witness quantum degrees of freedom of the gravitational field. We described the experimental setup using both a fully featured quantum field theoretic and a quantum-controlled approach, which led us to conclude that both descriptions are essentially equivalent within the proposed experimental parameters. However, if one assumes that the gravitational field has local degrees of freedom, it is possible to infer quantum properties of gravity from the experimental results, as no classical mediators would be able to entangle two masses. However, we also argued that, in the context of the experiment, the assumption that gravity has local degrees of freedom might, by itself, imply that the gravitational field is described by a quantum field theory, leading to circular reasoning.

Although we defined and studied the quantum-controlled interaction between two systems, nothing prevents one from considering more than two systems coupled through qc-interactions. An important consequence of this fact would be that, in this case, the effective evolution of two subsystems would usually not be unitary, as two systems could also become entangled with the remaining probes. It would be interesting to study how well the qc-model would be able to reproduce quantum field theory when more parties are considered. Moreover, in analogy with the Feynman-Wheeler absorber theory~\cite{Wheeler:1945ps,Wheeler:1949hn}, one could consider effective probes at asymptotic infinity, in which case it might be possible to incorporate many, if not all, effects related to the quantum degrees of freedom of quantum fields. Finding conditions under which generalizations of the qc-model could reproduce the predictions of quantum field theory has the potential to yield different approaches that could illuminate some of its aspects.

\subsubsection*{The Geometry of Spacetime from Quantum Fluctuations}

In Chapter~\ref{chap:geometry}, we explored the deep connection between the Hadamard condition and general relativity, showing that the universal UV divergence of correlation functions of quantum fields contains full information about the geometry of spacetime. In Section~\ref{sec:geometryFromMeas}, we constructed an explicit protocol through which one can recover the correlation function between any two events in a lattice in spacetime by considering sufficiently localized particle detectors. Using the data collected from these detectors we showed that it is indeed possible to approximately recover the background metric of spacetime from these measurements. These results show that particle detectors can also be used to measure space and time separations, showcasing that measurements of gravity can be rephrased in terms of quantum measurements of localized probes.

We also discussed the possibility of using the relationship between the Wightman function and the background metric to formulate a theory where gravity is an emergent phenomenon from correlations of quantum fields. In Section~\ref{sec:resonable?}, we discussed the challenges that such a formulation would face when attempting to recover the metric dynamics prescribed by Einstein's equations. If taken at face value, the relationship between quantum correlations and the metric would be unable to predict dynamics for the geometry of spacetime. This is due to the fact that all states share the same UV divergence, yielding the same background spacetime. We attempted to resolve this issue by considering non-localities at the Planck scale, but even with many additional assumptions, we were unable to recover the dynamics prescribed by Einstein's equations. The implementation of dynamics is also hindered when one attempts to consider multiple fields rather than one. Overall, we were unable to formulate a theory where all aspects of the gravitational field could be captured by the correlations of quantum fields.

Finally, in Section~\ref{sec:speculation}, we briefly discussed alternative programs that aim to describe the geometry of spacetime as an emergent phenomenon. Overall, we argued that it might be possible to formulate a theory where gravity emerges, not from the correlations in quantum field theory, but from entanglement therein.

\subsubsection*{Final Remark}

We looked at two complementary aspects of quantum field theory throughout this thesis. From a fundamental perspective, we studied entanglement in quantum field theory, as well as the regimes where quantum degrees of freedom are necessary, and how quantum correlations contain the full information about the spacetime geometry. From a practical side, we discussed how to use quantum field theory to describe physically realistic low-energy systems, and how it can be approximated to yield simpler and more familiar models and interactions. This all shows how rich the theory is: even after a century of studies in quantum field theory, there sure still seems to be much more to explore.

{\color{magenta}

}

\bibliographystyle{plain}

\cleardoublepage

\phantomsection  
\renewcommand*{\bibname}{References}

\addcontentsline{toc}{chapter}{\textbf{References}}

\bibliography{references.bib}
\nocite{*}


\appendix
\chapter*{APPENDICES}
\addcontentsline{toc}{chapter}{APPENDICES}

\chapter{Evaluation of Smeared Field Bi-Distributions on Gaussian Spacetime Functions}\label{app:analytical}

In this appendix we will compute the bi-distributions $W(f_1,f_2)$, $E(f_1,f_2)$ and $H(f_1,f_2)$ evaluated at functions $f_\tc{i}(\mf x)$ defined by parameters $(T_\tc{i},t_\tc{i},\Omega_\tc{i}, \sigma_\tc{i}, \bm L_\tc{i})$:
\begin{equation}
    f_\tc{i}(\mf x) = \frac{e^{-\frac{|\bm x - \bm L_\tc{i}|^2}{2 \sigma_\tc{i}^2}}}{(2\pi \sigma_\tc{i}^2)^{3/2}}e^{- \frac{(t-t_\tc{i})^2}{2T_\tc{i}^2}}e^{\ii \Omega_\tc{i} t},
\end{equation}
for general parameters  $(T_1,t_1,\Omega_1, \sigma_1, \bm L_1)$ and $(T_2,t_2,\Omega_2, \sigma_2, \bm L_2)$. We will also compute $G_R(f_1,f_2)$, $G_A(f_1,f_2)$, $\Delta(f_1,f_2)$, and $G_F(f_1,f_2)$ in the case $\sigma_1 = \sigma_2 = \sigma$. Finally, we will discuss how to generalize our results for the computation of two-point functions that involve the momentum of a real massless quantum field and more general time derivatives of the field.

We start from Eq.~\eqref{eq:W0Fourier}, and notice that for the functions
\begin{equation}
    F_\tc{i}(\bm x) = \frac{e^{-\frac{|\bm x - \bm L_\tc{i}|^2}{2 \sigma_\tc{i}^2}}}{(2\pi \sigma_\tc{i}^2)^{3/2}}, \quad \quad \chi_\tc{i}(t) = e^{- \frac{(t-t_\tc{i})^2}{2T_\tc{i}^2}}e^{\ii \Omega_\tc{i} t},
\end{equation}
we have the Fourier transforms
\begin{equation}
    \tilde{F}(\bm k) = e^{\ii \bm k \cdot \bm L_\tc{i} - \frac{\sigma_\tc{i}^2 |\bm k|^2}{2}}, \quad \quad \tilde{\chi}(\omega) = \sqrt{2 \pi} T_\tc{i} e^{-\ii (\omega-\Omega_\tc{i}) t_\tc{i} - \frac{(\omega-\Omega_\tc{i})^2 T_\tc{i}^2}{2}}.
\end{equation}
Plugging the results above into Eq.~\eqref{eq:W0Fourier}, we find{\footnotesize
\begin{align}
    W(f_1,f_2) &= \frac{1}{(2\pi)^2} \int \frac{\dd^3\bm k}{2 |\bm k|} e^{\ii \bm k \cdot \bm L_1 - \frac{\sigma_1^2 |\bm k|^2}{2}} e^{-\ii \bm k \cdot \bm L_2 - \frac{\sigma_2^2 |\bm k|^2}{2}} T_2 e^{\ii (|\bm k|+\Omega_2) t_2 - \frac{(|\bm k|+\Omega_2)^2 T_2^2}{2}} T_1 e^{-\ii (|\bm k|-\Omega_1) t_1 - \frac{(|\bm k|-\Omega_1)^2 T_1^2}{2}}\nonumber\\
    &= \frac{T_1T_2e^{\ii (\Omega_1 t_1+\Omega_2 t_2)}}{(2\pi)^2} \int \frac{\dd^3\bm k}{2 |\bm k|} e^{\ii \bm k \cdot (\bm L_1 - \bm L_2)} e^{-\ii |\bm k|( t_1-t_2)}e^{ - \frac{(\sigma_1^2+\sigma_2^2) |\bm k|^2}{2}}   e^{- \frac{(|\bm k|+\Omega_2)^2 T_2^2}{2} - \frac{(|\bm k|-\Omega_1)^2 T_1^2}{2}} \nonumber\\
    &= \frac{T_1T_2e^{\ii (\Omega_1 t_1+\Omega_2 t_2)}}{(2\pi)^2} e^{- \frac{\Omega_2^2 T_2^2}{2} - \frac{\Omega_1^2 T_1^2}{2}}\int \frac{\dd^3\bm k}{2 |\bm k|} e^{\ii \bm k \cdot (\bm L_1 - \bm L_2)} e^{-\ii |\bm k|( t_1-t_2)}e^{ - \frac{(\sigma_1^2+\sigma_2^2+T_1^2 + T_2^2) |\bm k|^2}{2}}    e^{- |\bm k| (\Omega_2 T_2^2-\Omega_1 T_1^2)}\nonumber
    \\
    &= \frac{T_1T_2e^{\ii (\Omega_1 t_1+\Omega_2 t_2)}}{2\pi} e^{- \frac{\Omega_2^2 T_2^2}{2} - \frac{\Omega_1^2 T_1^2}{2}}\int \frac{\dd |\bm k|}{2 |\bm k|} |\bm k|^2 2 \,\text{sinc}(|\bm k|  |\bm L|) e^{-\ii |\bm k|t_0}e^{ - \frac{(\sigma_1^2+\sigma_2^2+T_1^2 + T_2^2) |\bm k|^2}{2}}    e^{- |\bm k| (\Omega_2 T_2^2-\Omega_1 T_1^2)}\nonumber\\
    &= \frac{T_1T_2e^{\ii (\Omega_1 t_1+\Omega_2 t_2)}}{2\pi|\bm L|} e^{- \frac{\Omega_2^2 T_2^2}{2} - \frac{\Omega_1^2 T_1^2}{2}}\int \dd |\bm k|\,\text{sin}(|\bm k|  |\bm L|) e^{-\ii |\bm k|t_0}e^{ - \frac{(\sigma_1^2+\sigma_2^2+T_1^2 + T_2^2) |\bm k|^2}{2}}    e^{- |\bm k| (\Omega_2 T_2^2-\Omega_1 T_1^2)},\label{eq:WfinalInt}
\end{align}}
where $\bm L = \bm L_1 - \bm L_2$ and $t_0 = t_1-t_2$. Using the result
\begin{equation}
    \int_0^\infty \dd r e^{- \ii b r}e^{- \frac{a^2r^2}{2}} = \frac{\sqrt{\pi}}{\sqrt{2}a}e^{- \frac{b^2}{2a^2}}\Big(1 - \ii \text{erfi}\left(\frac{b}{\sqrt{2}a}\right)\Big),
\end{equation}
and writing $\sin(|\bm k||\bm L|)$ as exponentials we find{\footnotesize
\begin{align}
     W(f_1,f_2) &= \frac{T_1T_2e^{\ii (\Omega_1 t_1+\Omega_2 t_2)}}{2\pi|\bm L|}  \frac{1}{2\ii}\frac{\sqrt{\pi}e^{- \frac{\Omega_2^2 T_2^2}{2} - \frac{\Omega_1^2 T_1^2}{2}}}{\sqrt{2} \sqrt{T_1^2 + T_2^2 + \sigma_1^2 + \sigma_2^2}}\Bigg(e^{- \frac{(t_0 - |\bm L|+\ii (\Omega_1 T_1^2- \Omega_2 T_2^2))^2}{2(T_1^2 + T_2^2 + \sigma_1^2 + \sigma_2^2)}}\Bigg(1 - \ii \text{erfi}\left(\tfrac{t_0 - |\bm L|+\ii (\Omega_1 T_1^2- \Omega_2 T_2^2)}{\sqrt{2}\sqrt{T_1^2 + T_2^2 + \sigma_1^2 + \sigma_2^2}}\right)\Bigg)\nonumber\\
     &\:\:\:\:\:\:\:\:\:\:\:\:\:\:\:\:\:\:\:\:\:\:\:\:\:\:\:\:\:\:\:\:\:\:\:\:\:\:\:\:\:\:\:\:\:\:\:\:\:\:\:\:\:\:\:\:\:\:\:\:\:\:\:\:\:\:\:\:\:\:\:\:\:\:\:\:\:\:\:\:\:\:\:\:\:\:\:\:\:\:\:\:\:-e^{- \frac{(t_0 + |\bm L|+\ii (\Omega_1 T_1^2- \Omega_2 T_2^2))^2}{2(T_1^2 + T_2^2 + \sigma_1^2 + \sigma_2^2)}}\Bigg(1 - \ii \text{erfi}\left(\tfrac{t_0 + |\bm L|+\ii (\Omega_1 T_1^2- \Omega_2 T_2^2)}{\sqrt{2}\sqrt{T_1^2 + T_2^2 + \sigma_1^2 + \sigma_2^2}}\right)\Bigg)\nonumber\\
     &= \frac{T_1T_2e^{\ii (\Omega_1 t_1+\Omega_2 t_2)}e^{- \frac{\Omega_2^2 T_2^2}{2} - \frac{\Omega_1^2 T_1^2}{2}}}{4\sqrt{2\pi}|\bm L|\sqrt{T_1^2 + T_2^2 + \sigma_1^2 + \sigma_2^2}} \Bigg(e^{- \frac{(t_0 - |\bm L|+\ii (\Omega_1 T_1^2- \Omega_2 T_2^2))^2}{2(T_1^2 + T_2^2 + \sigma_1^2 + \sigma_2^2)}}\Bigg(-\ii - \text{erfi}\left(\tfrac{t_0 - |\bm L|+\ii (\Omega_1 T_1^2- \Omega_2 T_2^2)}{\sqrt{2}\sqrt{T_1^2 + T_2^2 + \sigma_1^2 + \sigma_2^2}}\right)\Bigg)\nonumber\\
     &\:\:\:\:\:\:\:\:\:\:\:\:\:\:\:\:\:\:\:\:\:\:\:\:\:\:\:\:\:\:\:\:\:\:\:\:\:\:\:\:\:\:\:\:\:\:\:\:\:\:\:\:\:\:\:\:\:\:\:\:\:\:\:\:\:\:\:\:\:\:\:\:\:\:\:\:\:\:\:+e^{- \frac{(t_0 + |\bm L|+\ii (\Omega_1 T_1^2- \Omega_2 T_2^2))^2}{2(T_1^2 + T_2^2 + \sigma_1^2 + \sigma_2^2)}}\Bigg(\ii + \text{erfi}\left(\tfrac{t_0 + |\bm L|+\ii (\Omega_1 T_1^2- \Omega_2 T_2^2)}{\sqrt{2}\sqrt{T_1^2 + T_2^2 + \sigma_1^2 + \sigma_2^2}}\right)\Bigg)\Bigg).
\end{align}}
And, as it turns out, we find that{\footnotesize
\begin{align}
    H(f_1,f_2) &= \frac{T_1T_2e^{\ii (\Omega_1 t_1+\Omega_2 t_2)}e^{- \frac{\Omega_2^2 T_2^2}{2} - \frac{\Omega_1^2 T_1^2}{2}}}{2\sqrt{2\pi}|\bm L|\sqrt{T_1^2 + T_2^2 + \sigma_1^2 + \sigma_2^2}} \Bigg(e^{- \frac{(t_0 - |\bm L|+\ii (\Omega_1 T_1^2- \Omega_2 T_2^2))^2}{2(T_1^2 + T_2^2 + \sigma_1^2 + \sigma_2^2)}}\text{erfi}\left(\frac{|\bm L|-t_0-\ii (\Omega_1 T_1^2- \Omega_2 T_2^2)}{\sqrt{2}\sqrt{T_1^2 + T_2^2 + \sigma_1^2 + \sigma_2^2}}\right)\\
     &\:\:\:\:\:\:\:\:\:\:\:\:\:\:\:\:\:\:\:\:\:\:\:\:\:\:\:\:\:\:\:\:\:\:\:\:\:\:\:\:\:\:\:\:\:\:\:\:\:\:\:\:\:\:\:\:\:\:\:\:\:\:\:\:\:\:\:\:\:\:\:\:\:\:\:\:\:\:\:+e^{- \frac{(t_0 + |\bm L|+\ii (\Omega_1 T_1^2- \Omega_2 T_2^2))^2}{2(T_1^2 + T_2^2 + \sigma_1^2 + \sigma_2^2)}}\text{erfi}\left(\frac{|\bm L|+t_0+\ii (\Omega_1 T_1^2- \Omega_2 T_2^2)}{\sqrt{2}\sqrt{T_1^2 + T_2^2 + \sigma_1^2 + \sigma_2^2}}\right)\Bigg),\label{eq:Hbig}\nonumber
\end{align}}
and
\begin{align}
    E(f_1,f_2) &= \frac{T_1T_2e^{\ii (\Omega_1 t_1+\Omega_2 t_2)}e^{- \frac{\Omega_2^2 T_2^2}{2} - \frac{\Omega_1^2 T_1^2}{2}}}{2\sqrt{2\pi}|\bm L|\sqrt{T_1^2 + T_2^2 + \sigma_1^2 + \sigma_2^2}} \Bigg(e^{- \frac{(t_0 + |\bm L|+\ii (\Omega_1 T_1^2- \Omega_2 T_2^2))^2}{2(T_1^2 + T_2^2 + \sigma_1^2 + \sigma_2^2)}}-e^{- \frac{(t_0 - |\bm L|+\ii (\Omega_1 T_1^2- \Omega_2 T_2^2))^2}{2(T_1^2 + T_2^2 + \sigma_1^2 + \sigma_2^2)}}\Bigg).
\end{align}

We now move on to the Green's functions, which will allow us to compute the Feynman propagator and the symmetric propagator. In order to compute these, we will resort to integration in spacetime of the spacetime smearing functions, using the expressions for the Green's functions of Eqs.~\eqref{eq:GRGAdelta}. Unfortunately, we will have to consider a more restricted parameter space where $\sigma_1 = \sigma_2 = \sigma$ in order to solve these integrals. For the retarded Green's function, we have{\footnotesize
\begin{align}
    G_R(f_1,f_2) &= \int \dd V\dd V' f_1(\mf x) f_2(\mf x') G_R(\mf x,\mf x')\\
    & = \frac{1}{(2\pi)^3 \sigma^6} \int \dd t \dd t' \dd^3\bm x \dd^3 \bm x' e^{- \frac{|\bm x - \bm L_1|^2}{2\sigma^2}}e^{- \frac{|\bm x' - \bm L_2|^2}{2\sigma^2}} e^{\ii \Omega_1 t} e^{- \frac{(t-t_1)^2}{2T_1^2}}e^{\ii \Omega_2 t'} e^{- \frac{(t'-t_2)^2}{2T_2^2}}\left(- \frac{\delta(t' - t + |\bm x - \bm x'|)}{4\pi|\bm x-\bm x'|} \right)\nonumber\\
    & = -\frac{1}{2(2\pi)^4 \sigma^6} \int \dd t  \dd^3\bm x \dd^3 \bm x' e^{- \frac{|\bm x - \bm L_1|^2}{2\sigma^2}}e^{- \frac{|\bm x' - \bm L_2|^2}{2\sigma^2}} e^{\ii \Omega_1 t} e^{- \frac{(t-t_1)^2}{2T_1^2}}e^{\ii \Omega_2 (t - |\bm x - \bm x'|)} e^{- \frac{(t - |\bm x - \bm x'|-t_2)^2}{2T_2^2}}\frac{1}{|\bm x - \bm x'|}.\nonumber
\end{align}}
We now perform the change of variables
\begin{align}
    \bm x = \frac{1}{\sqrt{2}}(\bm u +\bm v) + \bm L_1, \quad\quad
    \bm x' = \frac{1}{\sqrt{2}}(\bm u -\bm v) + \bm L_1,
\end{align}
so that $|\bm x - \bm x'| = \sqrt{2} |\bm v|$. Defining $\bm L = \bm L_1 - \bm L_2$, the integral becomes:{\footnotesize
\begin{align}
    G_R(f_1,f_2) & = -\frac{1}{2(2\pi)^4 \sigma^6} \int \dd^3\bm u \dd^3 \bm v e^{- \frac{|\bm u + \bm v|^2}{4\sigma^2}}e^{- \frac{|\bm u - \bm v + \sqrt{2}\bm L|^2}{4\sigma^2}}\frac{1}{\sqrt{2}|\bm v|} \int \dd t e^{\ii \Omega_1 t} e^{- \frac{(t-t_1)^2}{2T_1^2}}e^{\ii \Omega_2 (t - \sqrt{2}|\bm v|)} e^{- \frac{(t - \sqrt{2}|\bm v|-t_2)^2}{2T_2^2}}\\
    & = -\frac{1}{2(2\pi)^4 \sigma^6} \int   \dd^3\bm u \dd^3 \bm v e^{- \frac{|\bm u|^2 + |\bm v|^2}{4\sigma^2}}e^{- \frac{|\bm u|^2 +|\bm v|^2 + 2|\bm L|^2}{4\sigma^2}}e^{-\frac{\sqrt{2}\bm u\cdot \bm L + \sqrt{2}\bm v \cdot \bm L}{2\sigma^2}} \frac{1}{\sqrt{2}|\bm v|} \nonumber\\
    &\:\:\:\:\:\:\:\:\:\:\:\:\:\:\:\:\:\:\:\:\:\:\:\:\:\:\:\:\:\:\:\:\:\:\:\:\:\:\:\:\:\:\:\:\:\:\:\:\:\:\:\:\:\:\:\:\:\:\:\:\:\:\:\:\:\:\:\:\:\:\:\:\:\:\:\:\:\:\:\:\:\:\:\:\:\:\:\:\:\:\:\:\:\:\:\:\:\:\:\:\:\:\:\:\:\times\int \dd t e^{\ii \Omega_1 t} e^{- \frac{(t-t_1)^2}{2T_1^2}}e^{\ii \Omega_2 (t - \sqrt{2}|\bm v|)} e^{- \frac{(t - \sqrt{2}|\bm v|-t_2)^2}{2T_2^2}}\nonumber\\
    &= -\frac{e^{- \frac{|\bm L|^2}{2\sigma^2}}}{2(2\pi)^4 \sigma^6} \int   \dd^3\bm u \dd^3 \bm v e^{-\frac{|\bm u|^2}{2\sigma^2}}e^{-\frac{|\bm v|^2}{2\sigma^2}} e^{-\frac{\bm u\cdot \bm L}{\sqrt{2}\sigma^2}}e^{\frac{\bm v \cdot \bm L}{\sqrt{2}\sigma^2}}\frac{1}{\sqrt{2}|\bm v|} \int \dd t e^{\ii \Omega_1 t} e^{- \frac{(t-t_1)^2}{2T_1^2}}e^{\ii \Omega_2 (t - \sqrt{2}|\bm v|)} e^{- \frac{(t - \sqrt{2}|\bm v|-t_2)^2}{2T_2^2}}.\nonumber
\end{align}}
We can now proceed to first solve for the angular integrals in $\bm u$ and $\bm v$, and then the radial integral in $|\bm u|$, resulting in{\footnotesize
\begin{align}
    G_R(f_1,f_2) & = -\frac{e^{- \frac{|\bm L|^2}{2\sigma^2}}}{2(2\pi)^2 \sigma^6} \int   \dd |\bm u| \dd |\bm v| \,|\bm u|^2 |\bm v|^2 e^{-\frac{|\bm u|^2}{2\sigma^2}}e^{-\frac{|\bm v|^2}{2\sigma^2}} \frac{2\sqrt{2} \sigma^2\sinh(\frac{|\bm L||\bm u|}{\sqrt{2}\sigma^2})}{|\bm u||\bm L|}\frac{2\sqrt{2} \sigma^2\sinh(\frac{|\bm L||\bm v|}{\sqrt{2}\sigma^2})}{|\bm v||\bm L|}\frac{1}{\sqrt{2}|\bm v|}\nonumber \\
    &\:\:\:\:\:\:\:\:\:\:\:\:\:\:\:\:\:\:\:\:\:\:\:\:\:\:\:\:\:\:\:\:\:\:\:\:\:\:\:\:\:\:\:\:\:\:\:\:\:\:\:\:\:\:\:\:\:\:\:\:\:\:\:\:\:\:\:\:\:\:\:\:\:\:\:\:\:\:\:\:\:\:\:\:\:\:\:\:\:\:\:\:\:\:\:\:\:\:\:\:\:\:\:\:\:\times\int \dd t e^{\ii \Omega_1 t} e^{- \frac{(t-t_1)^2}{2T_1^2}}e^{\ii \Omega_2 (t - \sqrt{2}|\bm v|)} e^{- \frac{(t - \sqrt{2}|\bm v|-t_2)^2}{2T_2^2}}\nonumber\\
    & = -\frac{e^{- \frac{|\bm L|^2}{2\sigma^2}}}{\sqrt{2}\pi^2\sigma^2|\bm L|^2} \int  \dd |\bm u| \dd |\bm v| \,|\bm u| e^{-\frac{|\bm u|^2}{2\sigma^2}} \sinh(\frac{|\bm L||\bm u|}{\sqrt{2}\sigma^2})e^{-\frac{|\bm v|^2}{2\sigma^2}}\sinh(\frac{|\bm L||\bm v|}{\sqrt{2}\sigma^2})\nonumber\\
    &\:\:\:\:\:\:\:\:\:\:\:\:\:\:\:\:\:\:\:\:\:\:\:\:\:\:\:\:\:\:\:\:\:\:\:\:\:\:\:\:\:\:\:\:\:\:\:\:\:\:\:\:\:\:\:\:\:\:\:\:\:\:\:\:\:\:\:\:\:\:\:\:\:\:\:\:\:\:\:\:\:\:\:\:\:\:\:\:\:\:\:\:\:\:\:\:\:\:\:\:\:\:\:\:\:\times \int \dd t e^{\ii \Omega_1 t} e^{- \frac{(t-t_1)^2}{2T_1^2}}e^{\ii \Omega_2 (t - \sqrt{2}|\bm v|)} e^{- \frac{(t - \sqrt{2}|\bm v|-t_2)^2}{2T_2^2}}\nonumber\\
    & = -\frac{e^{- \frac{|\bm L|^2}{2\sigma^2}}}{\sqrt{2}\pi^2\sigma^2|\bm L|^2} \int  \dd |\bm v| \, \frac{\sqrt{\pi}\sigma |\bm L|}{2}e^{\frac{|\bm L|^2}{4 \sigma^2}}e^{-\frac{|\bm v|^2}{2\sigma^2}} \sinh(\frac{|\bm L||\bm v|}{\sqrt{2}\sigma^2}) \int \dd te^{\ii \Omega_1 t} e^{- \frac{(t-t_1)^2}{2T_1^2}}e^{\ii \Omega_2 (t - \sqrt{2}|\bm v|)} e^{- \frac{(t - \sqrt{2}|\bm v|-t_2)^2}{2T_2^2}}\nonumber\\
    & = -\frac{e^{- \frac{|\bm L|^2}{4\sigma^2}}}{2\sqrt{2}\pi^{3/2}\sigma|\bm L|} \int \dd |\bm v| \, e^{-\frac{|\bm v|^2}{2\sigma^2}} \sinh(\frac{|\bm L||\bm v|}{\sqrt{2}\sigma^2}) \int \dd t e^{\ii \Omega_1 t} e^{- \frac{(t-t_1)^2}{2T_1^2}}e^{\ii \Omega_2 (t - \sqrt{2}|\bm v|)} e^{- \frac{(t - \sqrt{2}|\bm v|-t_2)^2}{2T_2^2}}.\label{eq:GRfinalInt}
\end{align}}
Defining $t_0 = t_1 - t_2$, the integral in $t$ yields{\footnotesize
\begin{align}
     \int \dd t e^{\ii \Omega_1 t} e^{- \frac{(t-t_1)^2}{2T_1^2}}e^{\ii \Omega_2 (t - \sqrt{2}|\bm v|)}& e^{- \frac{(t - \sqrt{2}|\bm v|-t_2)^2}{2T_2^2}}\nonumber \textcolor{white}{\text{If you noticed the mistake, congratulations! The final result for GA is correct :)}}\\
     &\!\!\!\! = \frac{\sqrt{2\pi}T_1 T_2}{\sqrt{T_1^2 + T_2^2}}e^{- \frac{|\bm v|^2}{T_1^2 + T_2^2}}e^{\frac{\sqrt{2}|\bm v|(t_0- \ii (\Omega_1 T_1^2 - \Omega_2 T_2^2))}{(T_1^2 + T_2^2)}}e^{-\frac{t_0^2}{2(T_1^2 + T_2^2)}-\frac{T_1^2T_2^2(\Omega_1+\Omega_2)^2}{2(T_1^2 + T_2^2)}+\ii(\Omega_1 + \Omega_2)\frac{(t_1 T_2^2 + t_2 T_1^2)}{T_1^2 + T_2^2}}.
\end{align}}
The final integral in $|\bm v|$ is merely a combination of Gaussian integrals. After simplifications, we finally find{\footnotesize
\begin{align}
    G_R(f_1,f_2) = - \frac{T_1T_2e^{\ii (\Omega_1 t_1+\Omega_2 t_2)}e^{- \frac{\Omega_2^2 T_2^2}{2} - \frac{\Omega_1^2 T_1^2}{2}}}{4\sqrt{2\pi}|\bm L|\sqrt{T_1^2 + T_2^2 + 2\sigma^2}}\Bigg(&e^{- \frac{(t_0 - |\bm L|+\ii (\Omega_1 T_1^2- \Omega_2 T_2^2))^2}{2(T_1^2 + T_2^2 + 2\sigma^2)}}\left(1 + \text{erf}\left(\tfrac{|\bm L|(T_1^2+T_2^2)+2\sigma^2(t_0+\ii (\Omega_1 T_1^2- \Omega_2 T_2^2))}{2\sigma\sqrt{T_1^2 + T_2^2}\sqrt{T_1^2 + T_2^2 + 2\sigma^2}}\right)\right)\nonumber\\
    &\!\!\!\!\!\!\!\!\!\!\!\!\!\!\!\!\!\!+e^{- \frac{(t_0 + |\bm L|+\ii (\Omega_1 T_1^2- \Omega_2 T_2^2))^2}{2(T_1^2 + T_2^2 + 2\sigma^2)}}\left(-1+\text{erf}\left(\tfrac{|\bm L|(T_1^2+T_2^2)-2\sigma^2(t_0+\ii (\Omega_1 T_1^2- \Omega_2 T_2^2))}{2\sigma\sqrt{T_1^2 + T_2^2}\sqrt{T_1^2 + T_2^2 + 2\sigma^2}}\right)\right)\Bigg).\label{eq:GRbig}
\end{align}}
Combining $G_R(f_1,f_2)$ and $E(f_1,f_2)$, we can then find $G_A(f_1,f_2)$,{\footnotesize
\begin{align}
    G_A(f_1,f_2) = - \frac{T_1T_2e^{\ii (\Omega_1 t_1+\Omega_2 t_2)}e^{- \frac{\Omega_2^2 T_2^2}{2} - \frac{\Omega_1^2 T_1^2}{2}}}{4\sqrt{2\pi}|\bm L|\sqrt{T_1^2 + T_2^2 + 2\sigma^2}}\Bigg(&e^{- \frac{(t_0 - |\bm L|+\ii (\Omega_1 T_1^2- \Omega_2 T_2^2))^2}{2(T_1^2 + T_2^2 + 2\sigma^2)}}\left(-1 + \text{erf}\left(\tfrac{|\bm L|(T_1^2+T_2^2)+2\sigma^2(t_0+\ii (\Omega_1 T_1^2- \Omega_2 T_2^2))}{2\sigma\sqrt{T_1^2 + T_2^2}\sqrt{T_1^2 + T_2^2 + 2\sigma^2}}\right)\right)\nonumber\\
    &\!\!\!\!\!\!\!\!\!\!\!\!\!\!\!\!\!\!+e^{- \frac{(t_0 + |\bm L|+\ii (\Omega_1 T_1^2- \Omega_2 T_2^2))^2}{2(T_1^2 + T_2^2 + 2\sigma^2)}}\left(1+\text{erf}\left(\tfrac{|\bm L|(T_1^2+T_2^2)-2\sigma^2(t_0+\ii (\Omega_1 T_1^2- \Omega_2 T_2^2))}{2\sigma\sqrt{T_1^2 + T_2^2}\sqrt{T_1^2 + T_2^2 + 2\sigma^2}}\right)\right)\Bigg).\label{eq:GAbig}
\end{align}}
Finally, we find $\Delta(f_1,f_2)$ by adding Eqs.~\eqref{eq:GRbig} and~\eqref{eq:GAbig}:{\footnotesize
\begin{align}
    \Delta(f_1,f_2) = - \frac{T_1T_2e^{\ii (\Omega_1 t_1+\Omega_2 t_2)}e^{- \frac{\Omega_2^2 T_2^2}{2} - \frac{\Omega_1^2 T_1^2}{2}}}{2\sqrt{2\pi}|\bm L|\sqrt{T_1^2 + T_2^2 + 2\sigma^2}}\Bigg(&e^{- \frac{(t_0 - |\bm L|+\ii (\Omega_1 T_1^2- \Omega_2 T_2^2))^2}{2(T_1^2 + T_2^2 + 2\sigma^2)}} \text{erf}\left(\tfrac{|\bm L|(T_1^2+T_2^2)+2\sigma^2(t_0+\ii (\Omega_1 T_1^2- \Omega_2 T_2^2))}{2\sigma\sqrt{T_1^2 + T_2^2}\sqrt{T_1^2 + T_2^2 + 2\sigma^2}}\right)\nonumber\\
    &+e^{- \frac{(t_0 + |\bm L|+\ii (\Omega_1 T_1^2- \Omega_2 T_2^2))^2}{2(T_1^2 + T_2^2 + 2\sigma^2)}}\text{erf}\left(\tfrac{|\bm L|(T_1^2+T_2^2) - 2\sigma^2(t_0+\ii (\Omega_1 T_1^2- \Omega_2 T_2^2))}{2\sigma\sqrt{T_1^2 + T_2^2}\sqrt{T_1^2 + T_2^2 + 2\sigma^2}}\right)\Bigg).\label{eq:Deltabig}
\end{align}}
Using Eqs.~\eqref{eq:Hbig} and~\eqref{eq:Deltabig}, one can then find $G_F(f_1,f_2) = \frac{1}{2}H(f_1,f_2) + \frac{\ii}{2}\Delta(f_1,f_2)$.

In order to obtain the field bi-distributions evaluated at the field's momentum $\hat{\pi}(\mf x) = \partial_t \hat{\phi}(\mf x)$ rather than at the field's amplitude smeared in Gaussian spacetime regions, one can simply differentiate the results of this section with respect to the parameters $\Omega_1$, $\Omega_2$, using
\begin{equation}
    \omega( \hat{\pi}(f_1)\hat{\phi}(f_2)) = \omega( \hat{\phi}(-\partial_tf_1)\hat{\phi}(f_2)),
\end{equation}
and noticing that for the functions used here
\begin{equation}
    \partial_t f_1(\mf x) = \left(- \frac{t}{T_1^2} + \frac{t_1}{T_1^2} + \ii \Omega_1\right) f_1(\mf x) = \frac{\ii}{T_1^2}\dv{}{\Omega_1}f_1(\mf x) + \left(\frac{t_1}{T_1^2} + \ii \Omega_1\right) f_1(\mf x).
\end{equation}
That is, one finds that
\begin{equation}
    \omega( \hat{\pi}(f_1)\hat{\phi}(f_2)) =  - \frac{\ii}{T_1^2}\dv{}{\Omega_1}W(f_1,f_2) -\left(\frac{t_1}{T_1^2} + \ii \Omega_1\right) W(f_1,f_2).
\end{equation}
Analogous expressions are valid for higher derivatives of the field in both arguments, and for the other bi-distributions $H$, $E$, $G_R$, $G_A$, $G_F$ and $\Delta$. For brevity, we will not write these explicit expressions here, but they can be straightforwardly computed.

\chapter{Estimation of the Fermi bound}\label{app:fermi}

In this appendix we present an estimation for the Fermi bound, that is, we find an approximate bound for the maximum radius that a region contained in the rest spaces $\Sigma_\tau$ can have in order to be contained within the normal neighbourhood of the point $\mf z(\tau)$.

In order to obtain our estimate, we use the expansion of Eq. \eqref{eq:expansionFNC} for the
metric in Fermi normal coordinates. It is important to notice that the approximations of Eq. \eqref{eq:expansionFNC} are not only to second order in the distance of points to the curve, but are also the first order in curvature and acceleration expansions of the metric in Fermi normal coordinates. In the regime where these expansions are valid, the Fermi bound can be estimated as the maximum radius such that the metric does not become degenerate. That is, the largest radius such that the metric of Eq. \eqref{eq:expansionFNC}
is invertible. A good estimate for the $\tau$-Fermi bound can be given by considering the largest value of $|\bm x|$ such that 
\begin{equation}
    -(1+a_i x^i) - R_{0i0j}x^ix^j \neq 0.
\end{equation}
Notice that at $\bm x = 0$ this gives $-1$, so that for each $\tau$ we are looking for a bound on the radius $\ell_r$ such that 
\begin{equation}\label{eq:condition}
    |\bm x| \leq \ell_r \Rightarrow -(1+a_i x^i) - R_{0i0j}x^ix^j < 0. 
\end{equation}

In order to find $\ell_r$, let $a = \sqrt{a_ia^i}$ and define
\begin{equation}
    \lambda_R = \max_{|\bm x| = 1}\left(-R_{0i0j}x^ix^j\right).
\end{equation}
That is, $\lambda_R$ corresponds to the largest negative eigenvalue of $R_{0i0j}$, if there are any, else it is zero. We will now show that
\begin{equation}
    \ell_r = \frac{1}{a+\sqrt{\lambda_R}}
\end{equation}
fulfills the condition of Eq. \eqref{eq:condition}. Let $|\bm x|<\ell_r$, then
\begin{align}
    |\bm x|(a + \sqrt{\lambda_R})<1 \:\:\:\Rightarrow\:\:\: 1 - a|\bm x| >\sqrt{\lambda_R}|\bm x|,
\end{align}
so that using $a_i x^i \geq -a |\bm x|$,
\begin{equation}\label{eq:noname}
    1 + a_i x^i> \sqrt{\lambda_R} |\bm x| \:\:\:\Rightarrow\:\:\: (1 + a_i x^i)^2 > \lambda_R |\bm x| \geq - R_{0i0j}x^i x^j.
\end{equation}
This gives the desired result $0>-(1 + a_i x^i)^2- R_{0i0j}x^i x^j$.

We then obtain the approximate lower bound for the $\tau$-Fermi bound,
\begin{equation}\label{eq:tauFermiApprox}
    \ell_\tau \gtrsim \frac{1}{a+\sqrt{\lambda_R}},
\end{equation}
which is valid for points $\mf x$ such that $|\bm x|$ is sufficiently smaller than the curvature radius of spacetime and $1/a$, which is the regime of validity of the approximation of Eq. \eqref{eq:expansionFNC}. Notice that this bound works exactly in the case where there is no spacetime curvature, giving \mbox{$\ell_r = 1/a$}. The bound for the Fermi bound is then obtained by taking the infimum of Eq. \eqref{eq:tauFermiApprox} with respect to $\tau$.



\chapter{The Redshift Factor and the Metric Determinant}\label{app:det}

The metric in Fermi normal coordinates can be written as
\begin{equation}\label{eq:ADM}
    g = g_{\tau\tau}\dd\tau^2 +2g_{\tau i} \dd \tau \dd x^i+h_{ij}\dd x^i \dd x^j.
\end{equation}
Then, the inverse metric reads
\begin{equation}
    g^{-1} = g^{\tau\tau}\partial_\tau\otimes\partial_\tau + g^{\tau i}\partial_\tau \otimes \partial_i+g^{i\tau}\partial_i \otimes \partial_\tau + g^{ij}\partial_i\otimes\partial_j,
\end{equation}
where
\begin{align}
    g^{\tau\tau} &= \frac{1}{g_{\tau\tau} - g_{\tau i}g_{\tau j} h^{ij}},\\
    g^{\tau i} &= -\frac{h^{ij}g_{\tau j}}{g_{\tau\tau} - g_{\tau k}g_{\tau l} h^{kl}},\\
    g^{ij} &= h^{ij} + \frac{h^{ik}g_{\tau k}h^{jl}g_{\tau l}}{g_{\tau\tau} - g_{\tau k}g_{\tau l} h^{kl}},
\end{align}
and $h^{ij}$ is the inverse of the spatial metric, satisfying $h^{ik}h_{kj} = \delta^i_j$. The metric determinant can be written as
\begin{equation}
    \sqrt{-g} = \sqrt{h^{ij}g_{\tau i} g_{\tau j} -g_{\tau\tau}} \sqrt{g_{\Sigma}},
\end{equation}
where $g_{\Sigma} = \det(h_{ij})$.

The redshift factor associated to the foliation \mbox{$\tau = \text{const.}$} is determined by the norm of the 1-form $\dd \tau$, which is given by 
\begin{equation}
    {|\dd \tau|}^2 = g^{\tau\tau} = \frac{1}{g_{\tau\tau} - g_{\tau i}g_{\tau j} h^{ij}}.
\end{equation}
The redshift factor is then given by
\begin{equation}
    \gamma(\mf x) = \frac{1}{|\dd\tau|} = \abs{g_{\tau\tau} - g_{\tau i}g_{\tau j} h^{ij}}^\frac{1}{2},
\end{equation}
so that the invariant volume element of spacetime can be written as
\begin{equation}
    \dd V = \frac{\dd \tau}{|\dd \tau|}\wedge \dd \Sigma = \dd\tau \wedge(\gamma(\mf x) \dd \Sigma),
\end{equation}
where $\dd \Sigma = \sqrt{g_{\Sigma}}\,\dd^n \bm x$ is the measure in the \mbox{$\tau = \text{const}.$} surfaces.

\definecolor{DeepBlue}{RGB}{30,50,100}

\chapter{The Localization Condition for Quantum Fields}\label{app:locCond}

We can show that  Eq.~\eqref{eq:locCond} holds for a localized field in Minkowski spacetime defined by the Lagrangian~\eqref{eq:Lagphid}. The Riemann normal coordinates in Minkowski spacetime are simply inertial coordinates, and we can write the class of functions ${f}_{\mf x_0}$ simply as $f_{\mf x_0}(t,\bm x) = f(t-t_0,\bm x-\bm x_0)$ in the inertial coordinates $(t,\bm x)$ where the potential $V(\bm x)$ is defined. We will essentially show Eq.~\eqref{eq:locCond} when this restriction is imposed over the shape regions. In this case, the limit~\eqref{eq:locCond} corresponds to the limit $|\bm x_0|\to \infty$. 

Notice that for any compactly smooth real function $h$, we have
\begin{equation}\label{eq:auxlocCond}
    \bra{0_\tc{d}}\hat{\phi}_\tc{d}(f_{\mf x_0})\hat{\phi}_\tc{d}(h)\ket{0_\tc{d}} = \sum_{\bm n} u_{\bm n}({f}_{\mf x_0})u_{\bm n}^*(h),
\end{equation}
where the sum is absolutely convergent, as it corresponds to the inner product between the one-particle states $\ket{f_{\mf x_0}}$ and $\ket{h}$ (see~\eqref{eq:Wfg1part}).  We then have
\begin{equation}
    u_{\bm n}(f_{\mf x_0}) = \int \dd t \,\dd^3 \bm x \, f(t-t_0,\bm x - \bm x_0) \Phi_{\bm n}(\bm x) e^{-\ii \omega_{\bm n} t} = \int \dd t \,\dd^3 \bm x \, f(t,\bm x) \Phi_{\bm n}(\bm x + \bm x_0) e^{-\ii \omega_{\bm n} (t+t_0)}. 
\end{equation}
Due to the fact that the potential $V(\bm x)$ goes to infinity, we also have that the asymptotic behaviour of the functions $\Phi_{\bm n}(\bm x)$ at large $\bm x$ is at most exponential~\cite{expDecay}, so that we can write
 \begin{equation}
     \int \dd^3 \bm x f(t,\bm x) \Phi_{\bm n}(\bm x + \bm x_0) = \chi_{\bm n}(t,\bm x_0) e^{-\gamma_{\bm n}|\bm x_0|}, 
 \end{equation}
 with $\gamma_{\bm n}> \gamma^*>0$ for a given $\gamma^*$ and
 \begin{equation}
     \lim_{|\bm x_0| \to \infty}\chi_{\bm n}(t,\bm x_0) e^{-\gamma^*|\bm x_0|} = 0.
 \end{equation}
 We then find
\begin{equation}
    u_{\bm n}(f_{\mf x_0}) = \int \dd t \chi_{\bm n}(t, \bm x_0)e^{-\ii \omega_{\bm n} (t+t_0)}e^{-\gamma_{\bm n}|\bm x_0|} = \tilde{\chi}_{\bm n}(\omega_{\bm n},\bm x_0)e^{-\ii \omega_{\bm n} t_0}e^{-\gamma_{\bm n}|\bm x_0|},
\end{equation}
where $\tilde{\chi}_{\bm n}$ denotes the Fourier transform. Also notice that the eigenvalues of $L = - \nabla^2 + m_\tc{d}^2 + V(\bm x)$ with positive $V(\bm x)$ constitute an unbounded positive sequence, so that $\omega_{\bm n}\to \infty$ as $\bm n \to \infty$, also implying that $\tilde{\chi}(\omega_{\bm n},\bm x_0)\to 0$ for each $\bm x_0$, using the Riemann-Lebesgue Lemma.

We can now take the limit $|\bm x_0|\to \infty$ in Eq.~\eqref{eq:auxlocCond} using the dominated convergence theorem, as we have a product of the bounded terms $u_{\bm n}^*(h)$, $\tilde{\chi}_{\bm n}(\omega_{\bm n},\bm x_0)e^{-\ii \omega_{\bm n} t_0}e^{-\gamma_{\bm n}|\bm x_0|}\leq \tilde{\chi}_{\bm n}(\omega_{\bm n},\bm x_0)e^{-\ii \omega_{\bm n} t_0}e^{-\gamma^*|\bm x_0|}$ in an absolutely convergent series:
\begin{equation}
    \lim_{|\bm x_0|\to \infty }\bra{0_\tc{d}}\hat{\phi}_\tc{d}(f_{\mf x_0})\hat{\phi}_\tc{d}(h)\ket{0_\tc{d}} = \sum_{\bm n} \lim_{|\bm x_0|\to \infty} u_{\bm n}^*(h)\tilde{\chi}_{\bm n}(\omega_{\bm n},\bm x_0) e^{-\ii \omega_{\bm n} t_0} e^{-\gamma^*|\bm x_0|} = 0.
\end{equation}
The result above also holds when $h = f_{\mf x_0}$ with an analogous proof, so that we have shown that
\begin{equation}
    \lim_{|\bm x_0|\to \infty }\bra{0_\tc{d}}\hat{\phi}_\tc{d}(f_{\mf x_0})^2\ket{0_\tc{d}} =  0.
\end{equation}
Now notice that due to the vacuum $\ket{0_\tc{d}}$ being quasifree, all other terms of the form $\bra{0_\tc{d}}\hat{\phi}_\tc{d}(f_{\mf x_0})^n\ket{0_\tc{d}}$ either vanish if $n$ is odd or are proportional to $\bra{0_\tc{d}}\hat{\phi}_\tc{d}(f_{\mf x_0})^2\ket{0_\tc{d}}$, thus showing that the state $\ket{0_\tc{d}}$ is localized with this restriction of shape functions.

Moreover, given any quasifree state in $\mathcal{F}(\mathscr{H}_\tc{d})$, the expected values of operators of the form $\hat{\phi}_\tc{d}(f_{\mf x_0})^2$ will be given by sums and products of $W(h,{f}_{\mf x_0})$, implying that any quasifree state in the GNS representation of $\ket{0_\tc{d}}$ is also a localized state, from Eq.~\eqref{eq:Wfg1part}. For a Gaussian state $\omega_\tc{g}$ that is not quasifree, we can use the fact that $\omega_\tc{g}(\hat{\phi}_\tc{d}(f_{\mf x_0}))^2 \leq \omega_\tc{g}(\hat{\phi}_\tc{d}(f_{\mf x_0})^2)$, which also implies that the limit of $|\bm x_0|\to \infty$ of $\omega_\tc{g}(\hat{\phi}(f_{\mf x_0})^n)$ vanishes for all $n$.

Finally, notice that although we do not provide an explicit generalization of this result in general globally hyperbolic spacetimes, it is clear that the result still holds, provided that the mode functions $\Phi_{\bm n}(\bm x)$ are exponentially bounded by the proper distance of $\mf x_0$ and any other event $\mf y$ in the same Cauchy surface, which is achieved by a wide class of localizing potentials.

\chapter{Corrections to the Zeeman Interaction to Leading Order in the Fine Structure Constant}\label{app:derivation}

The goal of this appendix is to obtain the result of Eq.~\eqref{hflatzeemanexact}, which shows corrections to the Zeeman effect that depend on the shape of the radial function of the electron field modes corresponding to the electron state. This amounts to computing the integral of the function $\upphi(r)$, defined in Eq.~\eqref{eq:phir}, in space.

In Eq.~\eqref{psi-sol-atom} we showed the form of the $s$ orbital modes of an electron in terms of the radial functions $f(r)$ and $g(r)$. When $j=1/2$ and $p=+1$ these radial functions satisfy the differential equations~\cite{RQMgreiner}
\begin{align}
    \dv{g}{r} - (E-V(r)+m_e)f(r) &= 0,\\
    \dv{f}{r} +\frac{2}{r} f(r) + (E-V(r)-m_e)g(r) &= 0,
\end{align}
where $E$ is the energy level of the given orbital. The differential equations above are valid for any central potential $V(r)$. We can obtain an expression for $\upphi(r)$ by multiplying the first equation by $g(r)$ and the second equation by $f(r)$, yielding
\begin{align}
    \frac{1}{2}\dv{g^2}{r} - (E-V(r)+m_e)f(r)g(r) &= 0,\\
    \frac{1}{2}\dv{f^2}{r} +\frac{2}{r} f(r)g(r) + (E-V(r)-m_e)f(r)g(r) &= 0.
\end{align}
Adding the two equations we find
\begin{align}
    2m_e f(r) g(r) = \frac{1}{2}\dv{}{r}\left(f^2(r) + g^2(r)\right)+ \frac{2}{r} f^2(r).
\end{align}
Integrating the result above from $\infty$ to $r$ we find
\begin{equation}
    \upphi(r) = \int_{\infty}^r \dd r' f(r') g(r') = \frac{1}{4m_e}(f^2(r) + g^2(r)) + \frac{1}{m_e} \int_\infty^r \dd r' \frac{f^2(r')}{r'},
\end{equation}
where we directly integrated the total derivative using that $f(\infty) = g(\infty) = 0$. 

The integral of $\upphi(r)$ in space is then given by
\begin{equation}
    \int \dd^3 \bm x \, \upphi(|\bm x|) = \frac{4\pi}{4m_e} \int_0^\infty \dd r \, r^2 (f^2(r) + g^2(r)) - \frac{4\pi}{m_e}\int_0^\infty \dd r  \int_r^\infty  \dd r'\frac{r^2}{r'} f^2(r'),
\end{equation}
where we picked up factors of $4\pi$ due to integration over the angular coordinates and spherical symmetry of the functions involved. Normalization of the spinor spherical harmonics and of the mode functions $\psi_{\bm N}(\bm x)$ implies that
\begin{equation}\label{eq:normfg}
    \int_0^\infty \dd r \, r^2 (f^2(r) + g^2(r)) = 1.
\end{equation}
To handle the double integral, we can reparametrize the region defined by $0<r<\infty$ and $r<r'<\infty$ as $0<r'<\infty$ and $0<r<r'$. This gives
\begin{equation}\label{eq:changerrp}
    \int_0^\infty \dd r  \int_r^\infty  \dd r'\frac{r^2}{r'} f^2(r') = \int_0^\infty \dd r'  \int_0^{r'}  \dd r\frac{r^2}{r'} f^2(r') = \frac{1}{3}\int_0^\infty \dd r' (r')^2 f^2(r') = \frac{1}{3}\int_0^\infty \dd r\, r^2 f^2(r).
\end{equation}
Combining the results of Eqs.~\eqref{eq:normfg} and~\eqref{eq:changerrp}, we find
\begin{equation}
    \int \dd^3 \bm x \, \upphi(|\bm x|) = \frac{\pi}{m_e} - \frac{4\pi}{3m_e}\int_0^\infty \dd r\, r^2 f^2(r).
\end{equation}
Thus, with a constant magnetic field, we find
\begin{equation}
    \hat{H}_I(t) = - \frac{q}{2\pi} \int d^3\bm x\, \upphi(r)\, \hat{\bm{\sigma}}\cdot\bm{B}(t) = - \frac{q}{2m_e}\left(1 - \frac{4}{3}\int_0^\infty \dd r\, r^2 f^2(r)\right)\hat{\bm{\sigma}}\cdot\bm{B}(t).
\end{equation}

Importantly, this result holds true for modes with $j=1/2$ and $p=+1$ defined by any central potential $V(r)$. The corrections are then entirely determined by the function $f(r)$ and are of the order of the inverse of the product of the mass of the electron and the effective localization of the bound states.

{

\chapter{Two Localized Fields Interacting with a Klein-Gordon Field}\label{app:twoQFTs}

Denote the localized fields by $\hat{\phi}_\tc{a}$ and $\hat{\phi}_\tc{b}$, and the free field that they both interact with by $\hat{\phi}$ in the regions defined by the supports of $\zeta_\tc{a}(\mf x)$ and $\zeta_\tc{b}(\mf x)$, respectively. The interaction Hamiltonian density of the theory will be prescribed as
\begin{equation}
    \hat{h}_I(\mf x) = \lambda  (\zeta_\tc{a}(\mf x)\hat{\phi}(\mf x)\hat{\phi}_\tc{a}(\mf x) + \zeta_\tc{b}(\mf x)\phi(\mf x) \hat{\phi}_\tc{b}(\mf x)).
\end{equation}
We now write the $\phi_{\tc{a},\tc{b}}$ fields with the spacetime smearing functions as
\begin{align}
\zeta_\tc{a}(\mf x)\hat{\phi}_\tc{a}(\mf x) &= \sum_{\bm n} \zeta_\tc{a}(\mf x)\left( u_{\bm n}^{\tc{a}}(\mf x) \hat{a}_{\bm n}^\tc{a} +  u^{\tc{a}*}_{\bm n}(\mf x)\hat{a}^{\tc{a}\dagger}_{\bm n} \right) = \sum_{\bm n} \hat{Q}^{\tc{a}}_{\bm n}(\mf x)\\
\zeta_\tc{b}(\mf x)\hat{\phi}_\tc{b}(\mf x) &= \sum_{\bm n} \zeta_\tc{b}(\mf x)\left( u_{\bm n}^{\tc{b}}(\mf x) \hat{a}_{\bm n}^\tc{b} +  u^{\tc{b}*}_{\bm n}(\mf x) \hat{a}_{\bm n}^{\tc{b}\dagger} \right) = \sum_{\bm n}   \hat{Q}_{\bm n}^\tc{b}(\mf x),
\end{align}
where $u_{\bm n}^{\tc{a}}(\mf x) = e^{-\ii \omega_{\bm n}^{\tc{a}} t}\Phi^\tc{a}_{\bm n}(\bm x)$, $u_{\bm n}^{\tc{b}}(\mf x) = e^{-\ii \omega_{\bm n}^{\tc{b}} t}\Phi_{\bm n}^{\tc{b}}(\bm x)$, and the field expansion will depend on the specific boundary conditions and equations of motion. We are working under the assumption that the field has discrete energy levels, implying that the sums above are discrete. The field expansions automatically define states $\ket{0_\tc{a}}$ and $\ket{0_\tc{b}}$, which are annihilated by all operators $\hat{a}_{\bm n}^\tc{a}$ and $\hat{a}_{\bm n}^\tc{b}$, respectively. We can then write the Hamiltonian interaction density as
\begin{equation}
    \hat{h}_I(\mf x) = \lambda\hat{\phi}(\mf x) \sum_{\bm n} \left(\hat{Q}_{\bm n}^{\tc{a}}(\mf x) +  \hat{Q}_{\bm n}^\tc{b}(\mf x)\right).
\end{equation}
In perturbation theory, we then get
\begin{equation}
    \hat{U} = \mathcal{T}\exp(- \ii \int \dd V \hat{h}_I(\mf x)) = \openone + \hat{U}^{(1)} + \hat{U}^{(2)} + \mathcal{O}(\lambda^3),
\end{equation}
where:
\begin{align}
    \hat{U}^{(1)} = - \ii \int \dd V \hat{h}_I(\mf x) = - \ii \lambda \int \dd V  \hat{\phi}(\mf x)  \sum_{\bm n} \left(\hat{Q}^\tc{a}_{\bm n}(\mf x) +  Q_{\bm n}^\tc{b}(\mf x)\right).
\end{align}{\footnotesize
\begin{align}
    \hat{U}^{(2)}& =  - \int \dd V\dd V' \hat{h}_I(\mf x) \hat{h}_I(\mf x')\theta(t-t')\\&= - \int \dd V \dd V' \hat{\phi}(\mf x)\hat{\phi}(\mf x')\theta(t-t') \sum_{nm} \Big( \hat{Q}_{\bm n}^{\tc{a}}(\mf x)\hat{Q}_{\bm m}^{\tc{a}}(\mf x')+ \hat{Q}^\tc{b}_{\bm n}(\mf x)\hat{Q}_{\bm m}^{\tc{b}}(\mf x')+  \hat{Q}^\tc{a}_{\bm n}(\mf x)\hat{Q}^\tc{b}_{\bm m}(\mf x') +  \hat{Q}^\tc{b}_{\bm n}(\mf x) \hat{Q}^\tc{a}_{\bm m}(\mf x')\Big).
\end{align}}
The final state will be given by
\begin{equation}
    \hat{\rho}_f = \hat{U}\hat{\rho}_0\hat{U}^\dagger = \hat{\rho}_0 + \hat{U}^{(1)}\hat{\rho}_0 + \hat{\rho}_0 \hat{U}^{(1)\dagger} + \hat{U}^{(1)}\hat{\rho}_0\hat{U}^{(1)\dagger}+ \hat{U}^{(2)}\hat{\rho}_0 + \hat{\rho}_0 \hat{U}^{(2)\dagger} + \mathcal{O}(\lambda^3).
\end{equation}
We will assume that $\hat{\rho}_0 = \ket{0_\tc{a}0_\tc{b}}\!\!\bra{0_\tc{a}0_\tc{b}}\otimes \hat{\rho}_\phi = \hat{\rho}_{0,\tc{ab}}\otimes \hat{\rho}_\phi$ and that $\tr_\phi(\hat{U}^{(1)}\hat{\rho}_0) = 0$, so that the $\mathcal{O}(\lambda)$ terms do not contribute to the partial state of the cavities A and B. We then only need to compute \mbox{$\tr_\phi(\hat{U}^{(1)}\hat{\rho}_0\hat{U}^{(1)\dagger}+ \hat{U}^{(2)}\hat{\rho}_0 + \hat{\rho}_0 \hat{U}^{(2)\dagger})$}. We have:
{\footnotesize
\begin{align}
    &\tr_\phi(\hat{U}^{(1)}\hat{\rho}_0\hat{U}^{(1)\dagger}) \nonumber\\&= \lambda^2\int \dd V\dd V' W(\mf x', \mf x)  \sum_{nm} \Big( \hat{Q}^{\tc{a}}_{\bm n}(\mf x)\hat{\rho}_{0,\tc{ab}}\hat{Q}^{\tc{a}}_{\bm m}(\mf x')+ \hat{Q}^\tc{b}_{\bm n}(\mf x)\hat{\rho}_{0,\tc{ab}}\hat{Q}_{\bm m}^{\tc{b}}(\mf x')+ \hat{Q}^\tc{a}_{\bm n}(\mf x)\hat{\rho}_{0,\tc{ab}}\hat{Q}^{\tc{b}}_{\bm m}(\mf x') +  \hat{Q}^\tc{b}_{\bm n}(\mf x) \hat{\rho}_{0,\tc{ab}}\hat{Q}^{\tc{a}}_{\bm m}(\mf x')\Big),
\end{align}}
and{\footnotesize
\begin{align}
    \tr_\phi(\hat{U}^{(2)}\hat{\rho}_0) &= - \lambda^2\int \dd V \dd V' W(\mf x, \mf x')\theta(t-t')  \nonumber\\\times&\sum_{nm} \Big( \hat{Q}^{\tc{a}}_{\bm n}(\mf x)\hat{Q}^{\tc{a}}_{\bm m}(\mf x')\hat{\rho}_{0,\tc{ab}}+ \hat{Q}^\tc{b}_{\bm n}(\mf x)\hat{Q}_{\bm m}^{\tc{b}}(\mf x')\hat{\rho}_{0,\tc{ab}}+  \hat{Q}^\tc{a}_{\bm n}(\mf x)\hat{Q}^\tc{b}_{\bm m}(\mf x')\hat{\rho}_{0,\tc{ab}} +  \hat{Q}^\tc{b}_{\bm n}(\mf x) \hat{Q}^\tc{a}_{\bm m}(\mf x')\hat{\rho}_{0,\tc{ab}}\Big),
\end{align}}
where $W(\mf x, \mf x') = \text{tr}(\hat{\rho}_\phi \hat{\phi}(\mf x) \hat{\phi}(\mf x'))$. Notice that 
\begin{equation}
    \hat{\rho}_{0,\tc{ab}} = \ket{0_\tc{a}0_\tc{b}}\!\!\bra{0_\tc{a}0_\tc{b}} = \bigotimes_{nm} \ket{0_{\bm n}^\tc{a}0_{\bm m}^\tc{b}}\!\!\bra{0_{\bm n}^\tc{a}0_{\bm m}^\tc{b}} = \hat{\rho}_{\tc{d},0}\bigotimes_{n,m>1} \ket{0_{\bm n}^\tc{a}0_{\bm m}^\tc{b}}\!\!\bra{0_{\bm n}^\tc{a}0_{\bm m}^\tc{b}},
\end{equation}
where $\ket{0_{\bm n}^{\tc{a}}}$ and $\ket{0_{\bm n}^{\tc{b}}}$ denotes the ground states of each harmonic of each cavity, and 
\begin{equation}
    \hat{\rho}_{\tc{d},0} = \ket{0_1^\tc{a}0_1^\tc{b}}\!\!\bra{0_1^\tc{a}0_1^\tc{b}}
\end{equation}
is the ground state of the first harmonic in each cavity. We then trace out all cavity modes except for the first harmonic. That is, we will compute the density operator \mbox{$\tr_{\phi,H}(\hat{\rho}_f) = \tr_H(\tr_\phi(\hat{U}^{(1)}\hat{\rho}_0\hat{U}^{(1)\dagger}+ \hat{U}^{(2)}\hat{\rho}_0 + \hat{\rho}_0 \hat{U}^{(2)\dagger}))$}, where $\tr_H$ denotes the trace over all cavity harmonics, except the first, for fields A and B.

Overall, we will need to compute the trace of quantities of the form $\tr_H(\hat{Q}^\tc{a}_{\bm n}(\mf x)\hat{Q}^\tc{a}_{\bm m}(\mf x')\hat{\rho}_{0,\tc{ab}})$. We know that since $\hat{Q}^\tc{i}_{\bm n}(\mf x) = (u_{\bm n}^{\tc{i}}(\mf x)\hat{a}_{\bm n}^\tc{i} + u_{\bm n}^{\tc{i}*}(\mf x)\hat{a}^{\tc{i}\dagger}_{\bm n})$ for $\tc{I} = \tc{A},\tc{B}$. Therefore, the products of the form $\hat{Q}^\tc{i}_{\bm n}(\mf x)\hat{Q}_{\bm m}^\tc{i}(\mf x')$ will only give non-diagonal elements if $n = m$. When $n= m \neq 1$, we have:
\begin{equation}
    \tr_\tc{a}(\hat{Q}^\tc{a}_{\bm n}(\mf x)\hat{Q}^\tc{a}_{\bm n}(\mf x')\ket{0^\tc{a}_{\bm n}}\!\!\bra{0^\tc{a}_{\bm n}}) = \bra{0_{\bm n}^\tc{a}}\hat{Q}^\tc{a}_{\bm n}(\mf x)\hat{Q}^\tc{a}_{\bm n}(\mf x')\ket{0_{\bm n}^\tc{a}} = \zeta_\tc{a}(\mf x)\zeta_\tc{a}(\mf x')u_{\bm n}^\tc{a}(\mf x) u_{\bm n}^{\tc{a}*}(\mf x')
\end{equation}
and
\begin{equation}
    \tr_\tc{b}(\hat{Q}^\tc{b}_{\bm n}(\mf x)\hat{Q}^\tc{b}_{\bm n}(\mf x')\ket{0^\tc{b}_{\bm n}}\!\!\bra{0^\tc{b}_{\bm n}}) = \bra{0_{\bm n}^\tc{b}}\hat{Q}^\tc{b}_{\bm n}(\mf x)\hat{Q}^\tc{b}_{\bm n}(\mf x')\ket{0_{\bm n}^\tc{b}} = \zeta_\tc{b}(\mf x) \zeta_\tc{b}(\mf x')u_{\bm n}^\tc{b}(\mf x) u_{\bm n}^{\tc{b}*}(\mf x').
\end{equation}
We find that for $n$ and $m$ different from $1$,
\begin{equation}
    \tr_H(\hat{Q}_{\bm n}^\tc{i}(\mf x)\hat{Q}_{\bm m}^\tc{j}(\mf x')\hat{\rho}_{0,\tc{ab}}) = \delta_{\bm{nm}} \delta_{\tc{i}\tc{j}} \zeta_\tc{i}(\mf x)\zeta_\tc{j}(\mf x')u_{\bm n}^{\tc{i}}(\mf x)u_{\bm m}^{\tc{j}*}(\mf x')\hat{\rho}_{\tc{d},0},
\end{equation}
where $\tr_H(\hat{Q}_{\bm n}^\tc{a}(\mf x)\hat{Q}_{\bm m}^\tc{b}(\mf x')\hat{\rho}_{0,\tc{ab}}) = 0$ automatically, as it factors into expectation values of creation and annihilation operators in $\tc{A}$ and $\tc{B}$ evaluated at the vacuum. Finally, notice that when $n = m = 1$ we do not need to trace over it, because $H$ encompasses every harmonic except for the first one.

Putting the results above together, we then find that{\footnotesize
\begin{align}
    \tr_H(\tr_\phi(\hat{U}^{(1)} \hat{\rho}_0\hat{U}^{(1)\dagger})) = \!\lambda^2\!\!\!\int &\!\dd V \dd V'  W(\mf x' ,\mf x)\\
    & \times \Big( \hat{Q}^{\tc{a}}_1(\mf x)\hat{\rho}_{\tc{d},0}\hat{Q}_1^{\tc{a}}(\mf x')+ \hat{Q}_1^\tc{b}(\mf x)\hat{\rho}_{\tc{d},0}\hat{Q}_1^{\tc{b}}(\mf x')+ \hat{Q}^\tc{a}_1(\mf x)\hat{\rho}_{\tc{d},0}\hat{Q}^{\tc{b}}_1(\mf x') +  \hat{Q}_1^\tc{b}(\mf x) \hat{\rho}_{\tc{d},0}\hat{Q}^{\tc{a}}_1(\mf x') \nonumber\\&+\sum_{n>1}( \zeta_\tc{a}(\mf x)\zeta_\tc{a}(\mf x')u_{\bm n}^{\tc{a}}(\mf x) u_{\bm n}^{\tc{a}*}(\mf x')+\zeta_\tc{b}(\mf x)\zeta_\tc{b}(\mf x')u_{\bm n}^{\tc{b}}(\mf x) u_{\bm n}^{\tc{b}*}(\mf x'))\hat{\rho}_{\tc{d},0}\Big)\nonumber
\end{align}}
and{\footnotesize
\begin{align}
    \!\tr_H(\tr_\phi(\hat{U}^{(2)}\hat{\rho}_0))\! =\! - \lambda^2\!\!\!\int\! \!\dd V \dd V' &W(\mf x, \mf x')\theta(t-t')\\
    &\times\Big( \hat{Q}_1^{\tc{a}}(\mf x)\hat{Q}_1^{\tc{a}}(\mf x')\hat{\rho}_{\tc{d},0}\!+\! \hat{Q}^\tc{b}_1(\mf x)\hat{Q}_1^{\tc{b}}(\mf x')\hat{\rho}_{\tc{d},0}\!+\! \hat{Q}^\tc{a}_1(\mf x)\hat{Q}^\tc{b}_1(\mf x')\hat{\rho}_{\tc{d},0}\! +\! \hat{Q}^\tc{b}_1(\mf x) \hat{Q}^\tc{a}_1(\mf x')\hat{\rho}_{\tc{d},0}\nonumber\\
   &\:\:\:\:\:\:\:\:\:\:\:\:\:\:\:\:\:\:+\sum_{n>1} (\zeta_\tc{a}(\mf x)\zeta_\tc{a}(\mf x')u_{\bm n}^{\tc{a}}(\mf x)  u_{\bm n}^{\tc{a}*}(\mf x')+ \zeta_\tc{b}(\mf x)\zeta_\tc{b}(\mf x')u_{\bm n}^{\tc{b}}(\mf x) u_{\bm n}^{\tc{b}*}(\mf x'))\hat{\rho}_{\tc{d},0}\Big).\nonumber
   \end{align}}
The last term $\tr_H(\tr_\phi(\hat{\rho}_0 \hat{U}^{(2)\dagger}))$ is simply the conjugate of the term above. Notice that the terms proportional to $\hat{\rho}_{\tc{d},0}$ will cancel when all terms are added together. This can be seen from an explicit calculation using $\theta(t-t') + \theta(t'-t) = 1$, or simply by noticing that each term in the Dyson expansion is traceless due to trace preservation. 

The products of terms $\hat{Q}_1^\tc{i}(\mf x)\hat{Q}_1^\tc{j}(\mf x')$ is given by
{\footnotesize
\begin{align}
    \hat{Q}_1^\tc{i}(\mf x)\hat{Q}_1^\tc{j}(\mf x') &= \zeta_{\tc{i}}(\mf x)\zeta_{\tc{j}}(\mf x') (u_1^\tc{i}(\mf x) \hat{a}_{\bm n}^\tc{i} + u_1^{\tc{i}*}(\mf x) \hat{a}_{\bm n}^{\tc{i}\dagger}) (u_1^\tc{j}(\mf x') \hat{a}_{\bm n}^\tc{j} + u_1^{\tc{j}*}(\mf x') \hat{a}_{\bm n}^{\tc{j}\dagger}) \\
    &= \zeta_{\tc{i}}(\mf x)\zeta_{\tc{j}}(\mf x') (u_1^\tc{i}(\mf x) u_1^\tc{j}(\mf x') \hat{a}_{\bm n}^\tc{i}\hat{a}_{\bm n}^\tc{j} + u_1^{\tc{i}*}(\mf x)u_1^\tc{j}(\mf x') \hat{a}_{\bm n}^{\tc{i}\dagger}\hat{a}_{\bm n}^\tc{j} + u_1^\tc{i}(\mf x)u_1^{\tc{j}*}(\mf x') \hat{a}_{\bm n}^\tc{i} \hat{a}_{\bm n}^{\tc{j}\dagger} + u_1^{\tc{i}*}(\mf x) u_1^{\tc{j}*}(\mf x') \hat{a}_{\bm n}^{\tc{i}\dagger}\hat{a}_{\bm n}^{\tc{j}\dagger}).\nonumber
\end{align}}
We then define the spacetime smearing functions
\begin{align}
    \Lambda_\tc{a}(\mf x) &\coloneqq \zeta_\tc{a}(\mf x) u^{\tc{a}}_1(\mf x),\\
    \Lambda_\tc{b}(\mf x) &\coloneqq \zeta_\tc{b}(\mf x) u^{\tc{b}}_1(\mf x),    
\end{align}
so that the $\hat{Q}_1^{\tc{i}}(\mf x)$ terms read simply as
\begin{align}
    \hat{Q}_1^{\tc{a}}(\mf x) = \Lambda_\tc{a}(\mf x) \hat{a}^{\tc{a}}_1+\Lambda^*_\tc{a}(\mf x) \hat{a}^{\tc{a}\dagger}_1,\\
    \hat{Q}_1^{\tc{b}}(\mf x) = \Lambda_\tc{b}(\mf x) \hat{a}^{\tc{b}}_1+\Lambda^*_\tc{b}(\mf x) \hat{a}^{\tc{b}\dagger}_1.
\end{align}
We then see that the final state of the fields $\tc{A}$ and $\tc{B}$ can be written as
\begin{align}
    \hat\rho_\tc{d} = \tr_{\phi,H}(\hat\rho_f) = \hat{\rho}_{\tc{d},0} + \tr_\phi\left(\hat{U}_I^{(1)} \hat{\rho}_{\tc{d},0}\hat{U}_I^{(1)^\dagger} + \hat{U}_I^{(2)} \hat{\rho}_{\tc{d},0}+ \hat{\rho}_{\tc{d},0}\hat{U}_I^{(2)^\dagger}\right) + \mathcal{O}(\lambda^4),
\end{align}
where
\begin{align}
    \hat{U}_I^{(1)} &= -\ii \int \dd V \, \hat{h}_\text{eff}(\mf x),\\
    \hat{U}_I^{(2)} &= -\int \dd V \dd V' \, \hat{h}_\text{eff}(\mf x)\hat{h}_\text{eff}(\mf x') \theta(t-t'),
\end{align}
with
\begin{equation}
    \hat{h}_\text{eff}(\mf x) = \lambda \hat{Q}_1^{\tc{a}}(\mf x) \hat{\phi}(\mf x) +\lambda \hat{Q}_1^{\tc{b}}(\mf x) \hat{\phi}(\mf x)  = \lambda(\Lambda_\tc{a}(\mf x) \hat{a}^{\tc{a}}_1+\Lambda^*_\tc{a}(\mf x) \hat{a}^{\tc{a}\dagger}_1 + \Lambda_\tc{b}(\mf x) \hat{a}^{\tc{b}}_1+\Lambda^*_\tc{b}(\mf x) \hat{a}^{\tc{b}\dagger}_1)\hat{\phi}(\mf x).
\end{equation}
This is exactly the leading order result when one considers the interaction of two harmonic oscillators interacting with a quantum field $\hat{\phi}(\mf x)$. That is, the final state of the modes can be written as
\begin{equation}
    \hat{\rho}_\tc{d} = \tr_\phi(\hat{U}_I (\hat{\rho}_{\tc{d},0} \otimes \hat{\rho}_\phi )\hat{U}_I^\dagger) + \mathcal{O}(\lambda^4), \quad \hat{U}_I = \mathcal{T}\exp\left(-\ii \int \dd V \, \hat{h}_\text{eff}(\mf x)\right).
\end{equation}
The leading order computations can then be carried on analogously to harmonic oscillator particle detectors (for details on this calculation, see e.g.~\cite{EricksonZero}).

\chapter{The retarded propagator of the gravitational field}\label{app:retGrav}

The retarded propagator $G_R^{\mu\nu}{}_{\alpha'\beta'}(\mf x,\mf x')$ can be written as
\begin{align}
    G_R^{\mu\nu}{}_{\alpha'\beta'}(\mf x,\mf x') &= -\frac{1}{2\pi}\theta(t-t') \delta\left((t-t')^2 - |\bm x - \bm x'|^2\right)\mathcal{P}^{\mu\nu}{}_{\alpha'\beta'} \\
    &= -\frac{1}{4\pi | \bm x - \bm x'|} \delta(t-t' - |\bm x  - \bm x'|)\mathcal{P}^{\mu\nu}{}_{\alpha'\beta'},
\end{align}
where $\mathcal{P}$ is a bitensor. We then have the linearized metric given by $g_{\mu\nu} = \eta_{\mu\nu} + \sqrt{8\pi G}h_{\mu\nu}$, where
\begin{equation}
    h^{\mu\nu}(\mf x) = -\sqrt{8 \pi G}\int \dd V' G_R^{\mu\nu}{}_{\alpha' \beta'}(\mf x,\mf x') T^{\alpha' \beta'}(\mf x'),
\end{equation}
where $T_{\alpha\beta}$ denotes the stress-energy tensor of the source. For the case of a pointlike particle undergoing a trajectory $\mf z_1(t)$  with four-velocity $u_1^\mu(t)$, it reads
\begin{equation}
    T_1^{\mu\nu}(\mf x) = m_1 u_1^\mu(t) u_1^\nu(t) \frac{\delta^{(3)}(\bm x - \bm z_1(t))}{u_1^0(t) \sqrt{-g}},
\end{equation}
so that we obtain{\footnotesize
\begin{align}
    \int \dd V' G_R^{\mu\nu}{}_{\alpha'\beta'}(\mf x,\mf x')T_1^{\alpha' \beta'}(\mf x') & = -\frac{m_1}{4\pi} \int \dd t'  \frac{1}{u_1^0(t') |\bm x - \bm z_1(t')|} \delta(t-t' - |\bm x  - \bm z_1(t')|) \mathcal{P}^{\mu\nu}{}_{\alpha'\beta'}u_1^{\alpha'}(t') u_1^{\beta'}(t').
\end{align}}
Now, let $t_r$ be the retarded time such that $t - t_r - |\bm x - \bm z(t_r)| = 0$, so that
\begin{equation}
    \delta(t-t' - |\bm x  - \bm z_1(t')|) = \frac{\delta(t'-t_r)}{1- \hat{\bm r}_1(t_r)\!\cdot\! \dot{\bm z}_1(t_r)},
\end{equation}
where $\hat{\bm r_1}(t) = (\bm x  - \bm z_1(t))/|\bm x  - \bm z_1(t)|$ and we obtain
\begin{align}
    h_{(1)}^{\mu\nu}(\mf x) =- \sqrt{8 \pi G}\int \dd V' G_R^{\mu\nu}{}_{\alpha'\beta'}(\mf x,\mf x')T_1^{\alpha' \beta'}(\mf x') = \frac{ m_1}{4\pi}  \frac{\sqrt{8 \pi G}\mathcal{P}^{\mu\nu}{}_{\alpha'\beta'}u_1^{\alpha'}(t_r) u_1^{\beta'}(t_r)}{u_1^0(t_r)(1- \hat{\bm r}_1(t_r)\!\cdot\! \dot{\bm z}_1(t_r)) |\bm x - \bm z_1(t_r)|}.
\end{align}
The interaction Hamiltonian of a particle labelled $2$ with the gravitational potential sourced by particle $1$ will then be
\begin{align}
    {H}_{12}(t)  = - \frac{1}{2} \sqrt{8\pi G} \int \dd^3 {\bm x}\, h^{(1)}_{\mu\nu}(\mf x) T_2^{\mu\nu}(\mf x) &=- \frac{G m_1 m_2}{|\bm z_2(t) - \bm z_1(t_r)|} \frac{\mathcal{P}_{\mu\nu}{}_{\alpha'\beta'}u_2^{\mu}(t) u_2^{\nu}(t)u_1^{\alpha'}(t_r) u_1^{\beta'}(t_r)}{u_2^0(t)u_1^0(t_{r_{12}})(1- \hat{\bm r}_{12}\!\cdot\! \dot{\bm z}_1(t_{r_{12}}))},\nonumber\\
    &=- \frac{G m_1 m_2}{|\bm z_2(t) - \bm z_1(t_{r_{12}})|} \frac{2(\eta_{\mu\nu}u_2^{\mu}(t) u_1^{\nu}(t_{r_{12}}))^2-1}{u_2^0(t)u_1^0(t_{r_{12}})(1- \hat{\bm r}_{12}\!\cdot\! \dot{\bm z}_1(t_{r_{12}}))},
\end{align}
where $t_{r_{12}}$ is the solution to $t - t_{r_{12}} = |\bm z_2(t) - \bm z_1(t_{r_{12}})|$ and $\hat{\bm r}_{12} = (\bm z_2(t)  - \bm z_1(t_{r_{12}}))/|\bm z_2(t)  - \bm z_1(t_{r_{12}})|$, and we used $\mathcal{P}_{\mu\nu}{}_{\alpha'\beta'} = \eta_{\mu\alpha'}\eta_{\nu \beta'} + \eta_{\mu \beta'} \eta_{\nu \alpha'} - \eta_{\mu\nu} \eta_{\alpha'\beta'}$. The total interaction between the two particles is then given by
\begin{align}
    H_I(t) = \frac{1}{2}\left(H_{12}(t) + H_{21}(t)\right) =&- \frac{G m_1 m_2}{|\bm z_2(t) - \bm z_1(t_{r_{12}})|}\frac{(\eta_{\mu\nu}u_2^{\mu}(t) u_1^{\nu}(t_{r_{12}}))^2-1/2}{u_2^0(t)u_1^0(t_{r_{12}})(1- \hat{\bm r}_{12}\!\cdot\! \dot{\bm z}_1(t_{r_{12}}))}\\&- \frac{G m_1 m_2}{|\bm z_1(t) - \bm z_2(t_{r_{21}})|} \frac{(\eta_{\mu\nu}u_1^{\mu}(t) u_2^{\nu}(t_{r_{21}}))^2-1/2}{u_1^0(t)u_2^0(t_{r_{21}})(1- \hat{\bm r}_{12}\!\cdot\! \dot{\bm z}_2(t_{r_{21}}))},
\end{align}
$t_{r_{21}}$ is the solution to $t - t_{r_{21}} = |\bm z_2(t) - \bm z_1(t_{r_{21}})|$ and $\hat{\bm r}_{21} = (\bm z_1(t)  - \bm z_2(t_{r_{21}}))/|\bm z_1(t)  - \bm z_2(t_{r_{21}})|$, and we used $\mathcal{P}_{\mu\nu}{}_{\alpha'\beta'} = \eta_{\mu\alpha'}\eta_{\nu \beta'} + \eta_{\mu \beta'} \eta_{\nu \alpha'} - \eta_{\mu\nu} \eta_{\alpha'\beta'}$.
Notice that in the non-relativistic limit, we have $t_{r_{12}}\approx t_{r_{21}} \approx t$, $u_1^0(t) \approx u_2^0(t) \approx 1$ and $\eta_{\mu\nu} u_1^\mu(t)u_2^\nu \approx 1$, allowing one to recover the non-local Newtonian interaction between the particles:
\begin{equation}
    H_I(t)\approx - \frac{G m_1 m_2}{|\bm z_1(t) - \bm z_2(t)|}.
\end{equation}

Also notice that
\begin{align}
    \int \dd t\, H_I(t) &= 2 \pi G\int \dd V \dd V'\left( T^1_{\mu\nu}(\mf x) G_R^{\mu\nu}{}_{\alpha'\beta'}(\mf x, \mf x')T_2^{\alpha' \beta'}(\mf x') +  T^2_{\mu\nu}(\mf x) G_R^{\mu\nu}{}_{\alpha'\beta'}(\mf x, \mf x')T_1^{\alpha' \beta'}(\mf x')\right) \nonumber\\
    &= 2 \pi G\int \dd V \dd V'\left( T^1_{\mu\nu}(\mf x) G_R^{\mu\nu}{}_{\alpha'\beta'}(\mf x, \mf x')T_2^{\alpha' \beta'}(\mf x') +  T^2_{\alpha'\beta'}(\mf x') G_R^{\alpha'\beta'}{}_{\mu\nu}(\mf x', \mf x)T_1^{\mu \nu}(\mf x')\right) \nonumber\\
    &= 2 \pi G\int \dd V \dd V' T^1_{\mu\nu}(\mf x) \Delta^{\mu\nu}{}_{\alpha'\beta'}(\mf x, \mf x')T_2^{\alpha' \beta'}(\mf x'),
\end{align}
where $\Delta^{\mu\nu\alpha'\beta'}(\mf x, \mf x') = G_R^{\mu\nu\alpha'\beta'}(\mf x, \mf x') + G_A^{\mu\nu\alpha'\beta'}(\mf x,\mf x')$ and we used $G_R^{\alpha'\beta'\mu\nu}(\mf x',\mf x) = G_A^{\mu\nu\alpha'\beta'}(\mf x,\mf x')$.

\chapter{Final states in the Quantum Field Approach to Gravity Mediated Entanglement}\label{app:FinalStates}

The initial state of the two particles in matrix form in the basis $\{\ket{L_1L_2},\ket{R_1 L_2},\ket{L_1R_2},\ket{R_1R_2}\}$ reads
\begin{equation}
    \hat{\rho} = \frac{1}{4}
    \begin{pmatrix}
        1 & 1 & 1 & 1\\
        1 & 1 & 1 & 1\\
        1 & 1 & 1 & 1\\
        1 & 1 & 1 & 1        
    \end{pmatrix}.
\end{equation}
To leading order in the gravitational field, the updated state of the particles evolving with respect to the qc-interaction can be written as $\hat{\rho} +  \delta\hat{\rho}_c$, where
\begin{equation}
    \delta\hat{\rho}_c = -\frac{\ii \lambda^2}{4}
    \begin{pmatrix}
        0 & \Delta_{R_1L_2}-\Delta_{L_1L_2} &\Delta_{L_1R_2}-\Delta_{L_1L_2} & \Delta_{R_1R_2}-\Delta_{L_1L_2}\\
        -\Delta_{R_1L_2}+\Delta_{L_1L_2} & 0 & \Delta_{L_1R_2}-\Delta_{R_1L_2} & \Delta_{R_1R_2}-\Delta_{R_1L_2}\\
       -\Delta_{L_1L_2}+\Delta_{L_1R_2} & -\Delta_{L_1R_2}+\Delta_{R_1L_2} & 0 & \Delta_{L_1R_2}-\Delta_{R_1R_2}\\
        -\Delta_{L_1L_2}+\Delta_{R_1R_2} & -\Delta_{R_1L_2}+\Delta_{R_1R_2} & -\Delta_{L_1R_2}+\Delta_{R_1R_2} & 0      
    \end{pmatrix}.
\end{equation}
In the quantum case, the final state of the particles to leading order can be written as $\hat{\rho} + (\delta\hat{\rho}_c+\delta\hat{\rho}_q+\delta \hat{\rho}_l)$, where
\begin{equation}
    \delta\hat{\rho}_l = \frac{\lambda^2}{2} (\mathcal{L}_\textsc{v}-\mathcal{L}_{\textsc{i}})
    \begin{pmatrix}
        0  &  1 & 1 & 2\\
        1  &  0 & 2 & 1\\
        1  &  2 & 0 & 1\\
        2  &  1 & 1 & 0
    \end{pmatrix},
\end{equation}
and
\begin{equation}
    \delta\hat{\rho}_q = \frac{\lambda^2}{4}(H_{L_1R_2} \!\!+ \!\!H_{R_1L_2} \!\!-\!\! H_{L_1L_2}\!\!-\!\!H_{R_1R_2})
    \begin{pmatrix}
        0 & 0 & 0 & 1\\
        0 & 0 & -1 & 0\\
        0 & -1 & 0 & 0\\
        1 & 0 & 0 & 0
    \end{pmatrix},
\end{equation}
where $H_{p_1p_2}$ denotes the integrated Hadamard function along each pair of paths for particles 1 and 2,
\begin{equation}
    H_{p_1p_2} = \int \dd V \dd V' T^{\mu\nu}_{p_1}H_{\mu\nu\alpha'\beta'}(\mf x,\mf x')T_{p_2}^{\alpha'\beta'}(\mf x').
\end{equation}
while $\mathcal{L}_\textsc{v}$ and $\mathcal{L}_{\textsc{i}}$ are noise terms due to the interaction of each particle with the vacuum of the field. These are explicitly given by
\begin{align}
    \mathcal{L}_{\textsc{v}} &= \int \dd V \dd V' T^{\mu\nu}_{L_i}\langle \hat{h}_{\mu\nu}(\mf x) \hat{h}_{\alpha'\beta'}(\mf x')\rangle_0 T_{L_i}^{\alpha'\beta'}(\mf x')= \int \dd V \dd V' T^{\mu\nu}_{R_i}\langle \hat{h}_{\mu\nu}(\mf x) \hat{h}_{\alpha'\beta'}(\mf x')\rangle_0 T_{R_i}^{\alpha'\beta'}(\mf x'),\\
    \mathcal{L}_{\textsc{i}} &= \int \dd V \dd V' T^{\mu\nu}_{L_i}\langle \hat{h}_{\mu\nu}(\mf x) \hat{h}_{\alpha'\beta'}(\mf x')\rangle_0 T_{R_i}^{\alpha'\beta'}(\mf x'),
\end{align}
where, due to the choice of paths, the indices $i$ can be $1$ or $2$, and still yield the same result due to the fact that the paths are related to translations and rotations, which are symmetries of the quantum field theory. Then, the $\mathcal{L}_{\textsc{v}}$ term is a local noise associated to each path, while the $\mathcal{L}_{\textsc{i}}$ term represents an interference term associated with each particle undergoing the superposition of paths. Also notice that due to the fact that the propagator decreases with distance, we have $\mathcal{L}_{\textsc{i}}\leq \mathcal{L}_{\textsc{v}}$, with equality holding only if the paths $1$ and $2$ are identical. Moreover, these vacuum noise terms decay fast with the interaction time, so they are negligible for long interaction times. Then we can interpret $\delta{\hat{\rho}}_l$ as a local vacuum contribution, and $\delta\hat{\rho}_q$ can be seen as the additional correlation contribution due to the quantum nature of the field.


\end{document}